\documentclass[a4paper]{amsart}
\usepackage[utf8]{inputenc}
\usepackage{rotating}
\usepackage{amsfonts,amsmath,amssymb}
\usepackage{mathtools}
\usepackage{color}
\usepackage{tikz}
\usepackage{bibentry}
\usepackage{graphicx}
\usepackage{dsfont}

\usepackage{tabularx,booktabs}
\usepackage{pifont}
\usepackage{enumitem}
\usepackage{faktor}
\usepackage{multirow}
\usepackage[framemethod=tikz]{mdframed}
\usepackage[colorlinks,citecolor=blue]{hyperref}
\usepackage[capitalize]{cleveref}
\usepackage[ruled,vlined,linesnumbered]{algorithm2e}
\DontPrintSemicolon

\SetKwInput{KwParameters}{Parameters}

\usepackage[arrow,matrix,curve,color,pdftex]{xy}
\definecolor{lightgray}{gray}{0.8}
\definecolor{lightergray}{gray}{0.95}

\def\emptyset{\{\}}
\def\units#1{{#1}^*}

\newcommand{\softO}{\widetilde{O}}
\newcommand{\ZZ}{\ensuremath{\mathbb{Z}}}
\newcommand{\FF}{\ensuremath{\mathbb{F}}}
\newcommand{\PP}{\ensuremath{\mathbb{P}}}
\newcommand{\EC}{\ensuremath{\mathcal{E}}}
\newcommand{\AV}{\ensuremath{\mathcal{A}}}
\newcommand{\thegroup}{\ensuremath{\mathcal{G}}}
\newcommand{\subgrp}[1]{\ensuremath{\langle{#1}\rangle}}

\DeclareMathOperator{\Res}{Res}

\theoremstyle{plain}
\newtheorem{theorem}[subsection]{Theorem}
\newtheorem{lemma}[subsection]{Lemma}
\newtheorem{proposition}[subsection]{Proposition}
\theoremstyle{definition}
\newtheorem{definition}[subsection]{Definition}
\newtheorem{example}[subsection]{Example}
\theoremstyle{remark}

\usepackage{todonotes}

\title[Faster computation of isogenies of large prime degree]{Faster computation of isogenies\\ of large prime degree}

\author{Daniel J. Bernstein}
\address{Department of Computer Science\\
University of Illinois at Chicago\\
USA}
\address{Horst Görtz Institute for IT Security\\
Ruhr University Bochum\\
Germany}
\email{djb@cr.yp.to}

\author{Luca De Feo}
\address{IBM Research Zürich\\
Switzerland}
\email{Luca.De.Feo@zurich.ibm.com}

\author{Antonin Leroux}
\address{DGA, Inria and École Polytechnique\\
Institut Polytechnique de Paris\\
Palaiseau\\
France}
\email{antonin.leroux@polytechnique.org}

\author{Benjamin Smith}
\address{Inria and École Polytechnique\\
Institut Polytechnique de Paris\\
Palaiseau\\
France}
\email{smith@lix.polytechnique.fr}

\dedicatory{Dedicated to the memory of Peter Lawrence Montgomery}

\date{2020.03.20}

\thanks{Author list in alphabetical order; see
\url{https://www.ams.org/profession/leaders/culture/CultureStatement04.pdf}.
Part of this work was carried out while the first author
was visiting the Simons Institute for the Theory of Computing.
This work was supported
by the Cisco University Research Program,
by DFG Cluster of Excellence 2092
``CASA: Cyber Security in the Age of Large-Scale Adversaries'',
and
by the U.S. National Science Foundation under grant 1913167.
``Any opinions, findings, and conclusions or recommendations expressed in this material are those
of the author(s) and do not necessarily reflect the views of the National Science Foundation''
(or other funding agencies).
Permanent ID of this document:
{\tt 44d5ade1c1778d86a5b035ad20f880c08031a1dc}.}

\begin{document}

\begin{abstract}
    Let \(\EC/\FF_q\) be an elliptic curve,
    and \(P\) a point in \(\EC(\FF_q)\) of prime order~\(\ell\).
    Vélu's formul\ae{} let us compute a quotient curve
    \(\EC' = \EC/\subgrp{P}\)
    and rational maps defining a quotient isogeny
    \(\phi: \EC \to \EC'\) 
    in \(\softO(\ell)\) \(\FF_q\)-operations,
    where the \(\softO\) is uniform in \(q\).
    This article shows how to compute \(\EC'\),
    and \(\phi(Q)\) for \(Q\) in \(\EC(\FF_q)\),
    using only \(\softO(\sqrt{\ell})\) \(\FF_q\)-operations,
    where the \(\softO\) is again uniform in \(q\).
    As an application,
    this article speeds up some computations
    used in the isogeny-based cryptosystems CSIDH and CSURF.
\end{abstract}

\maketitle

\section{%%%%%%%%%%%%%%%%%%%%%%%%%%%%%%%%%%%%%%%%%%%%%%%%%%%%%%%%%%%%%%%%%%%%%%%
    Introduction
}%%%%%%%%%%%%%%%%%%%%%%%%%%%%%%%%%%%%%%%%%%%%%%%%%%%%%%%%%%%%%%%%%%%%%%%%%%%%%%%
\label{sec:intro}

Let \(\EC\) be an elliptic curve 
over a finite field \(\FF_q\) of odd characteristic,
and let \(P\) be a point in \(\EC(\FF_q)\) 
of order \(n\).
The point \(P\) generates 
a cyclic subgroup \(\thegroup \subseteq \EC(\FF_q)\),
and there exists an elliptic curve \(\EC'\) over \(\FF_q\)
and a separable degree-$n$ quotient isogeny
\[
    \phi: \EC \longrightarrow \EC'
    \qquad
    \text{with}
    \qquad
    \ker\phi = \thegroup = \subgrp{P}
    \,;
\]
the isogeny \(\phi\) is also defined over \(\FF_q\).
We want to compute
\(\phi(Q)\) for a point \(Q\) in \(\EC(\FF_q)\)
as efficiently as possible.

If \(n\) is composite, then we can decompose \(\phi\)
into a series of isogenies of prime degree.
Computationally,
this assumes that we can factor \(n\),
but finding a prime factor $\ell$ of $n$
is not a bottleneck compared to
the computation of an $\ell$-isogeny
by the techniques considered here.
We thus reduce to the case where \(n = \ell\) is prime.

Vélu introduced formul\ae{}
for $\phi$ and \(\EC'\)
(see~\cite{V71} and~\cite[\S2.4]{K96}):
for \(\EC\) defined by \(y^2 = x^3 + a_2x^2 + a_4x + a_6\) 
and \(\ell \ge 3\), we have
\[
    \phi : (X,Y) \longmapsto
    \left(
        \frac{\Phi_\thegroup(X)}{\Psi_\thegroup(X)^2}
        ,
        \frac{Y\Omega_\thegroup(X)}{\Psi_\thegroup(X)^3}
    \right)
\]
where
\begin{align*}
    \Psi_\thegroup(X)
    & = 
    \textstyle\prod_{s = 1}^{(\ell-1)/2}\big(X - x([s]P)\big)
    \,,
    \\
    \Phi_\thegroup(X)
    % b_2 = 4a_2
    % b_4 = 2a_4
    % b_6 = 4a_6
    & = 
    4(X^3 + a_2X^2 + a_4X + a_6)
    (\Psi_\thegroup'(X)^2 - \Psi_\thegroup''(X)\Psi_\thegroup(X))
    \\
    & \qquad 
    {} -
    2(3X^2 + 2a_2X + a_4)\Psi_\thegroup'(X)\Psi(X)
    + 
    %(\ell X - 2s_1)\Psi_\thegroup(X)^2
    (\ell X - \textstyle{\sum_{s=1}^{\ell-1}x([s]P)})\Psi_\thegroup(X)^2
    \,,
    \\
    \Omega_\thegroup(X)
    & = 
    \Phi_\thegroup'(X)\Psi_\thegroup(X)
    -
    2\Phi_\thegroup(X)\Psi_\thegroup'(X)
    \,.
\end{align*}

The obvious way to compute $\phi(Q)$ is
to compute the rational functions shown above,
i.e., to compute the coefficients of the
polynomials $\Psi_G,\Phi_G,\Omega_G$;
and then evaluate those polynomials.
This takes \(\softO(\ell)\) operations.
(If we need the defining equation of \(\EC'\),
then we can obtain it by evaluating \(\phi(Q)\) for a few \(Q\)
outside \(\thegroup\), possibly after extending $\FF_q$,
and then interpolating a curve equation through the resulting points.
Alternatively,
V\'elu gives further formulas for the defining equation.)
We emphasize,
however,
that the goal is not to compute the coefficients
of these functions;
the goal is to evaluate the functions at a specified point.

The core algorithmic problem falls naturally into a more general
framework:
the efficient evaluation of polynomials and rational
functions over \(\FF_{q}\) whose roots are values of 
a function from a cyclic group to $\FF_q$.

Fix a cyclic group \(\thegroup\)
(which we will write additively),
a generator \(P\) of \(\thegroup\),
and a function \(f: \thegroup \to \FF_{q}\).
For each finite subset \(S\) of \(\ZZ\),
we define a polynomial
\[
    h_S(X) = \prod_{s \in S}(X - f([s]P))
    \,,
\]
where \([s]P\) denotes the sum of \(s\) copies of \(P\).
The kernel polynomial \(\Psi_\thegroup(x)\) above
is an example of this,
with \(f = x\) and \(S = \{1,\ldots,(\ell-1)/2\}\).
Another example is
the cyclotomic polynomial \(\Phi_n\),
where \(f\) embeds \(\ZZ/n\ZZ\) in the roots of unity of \(\FF_{q}\),
and \(\Phi_n(X) = h_S(X)\) where \(S = \{i \mid 0 \le i < n, \gcd(i,n) = 1\}\).
More generally,
if \(f\) maps \(i \mapsto \zeta^i\) for some \(\zeta\),
then \(h_S(X)\) is a polynomial whose roots are various powers of \(\zeta\);
similarly,
if \(f\) maps \(i \mapsto i\beta\) for some \(\beta\),
then \(h_S(X)\) is a polynomial whose roots are various integer multiples of \(\beta\).

Given \(f\) and \(S\), then,
we want to compute \(h_S(\alpha)\)
for any \(\alpha\) in \(\FF_{q}\).
We can always do this directly in \(O(\#S)\) \(\FF_{q}\)-operations.
But if \(S\) has enough additive structure,
and if \(f\) is sufficiently compatible with the group structure on \(\thegroup\),
then we can do this in~\(\softO(\sqrt{\#S})\) \(\FF_{q}\)-operations,
as we will see in \S\ref{sec:strassen}, \S\ref{sec:powers},
and~\S\ref{sec:elliptic}.
Our main theoretical result is Theorem~\ref{theorem:x-polynomial},
which shows how to achieve this quasi-square-root complexity
for a large class of \(S\) when \(f\) is the \(x\)-coordinate on an
elliptic curve.
We apply this to the special case of efficient \(\ell\)-isogeny
computation in~\S\ref{sec:isogenies}.
We discuss applications in isogeny-based cryptography
in~\S\ref{sec:protocols}.

Most of this paper
focuses on asymptotic exponents,
in particular improving $\ell$-isogeny evaluation
from cost $\softO(\ell)$ to cost $\softO(\sqrt{\ell})$.
However,
this analysis hides polylogarithmic factors
that can swamp the exponent improvement for small $\ell$.
In Appendix~\ref{sec:concrete}
we instead analyze costs for concrete values of $\ell$,
and ask how large $\ell$ needs to be
for the $\softO(\sqrt{\ell})$ algorithms
to outperform conventional algorithms.

\subsection{Model of computation}
    We state our framework for $\FF_q$ for concreteness.
    All time complexities are in \(\FF_q\)-operations,
    with the \(O\) and \(\softO\) uniform over \(q\).

    The ideas are more general.
    The algorithms here are algebraic algorithms in the sense
    of~\cite{Burgisser--Clausen--Shokrollahi},
    and can further be lifted to algorithms defined over $\ZZ[1/2]$
    and in some cases over $\ZZ$.
    In other words,
    the algorithms are agnostic to the choice of $q$ in $\FF_q$,
    except for sometimes requiring $q$ to be odd;
    and the algorithms can also be applied to more general rings,
    as long as all necessary divisions can be carried out.

    Restricting to algebraic algorithms can damage performance.
    For example,
    for most input sizes,
    the fastest known algorithms to multiply polynomials over $\FF_q$
    are faster than the fastest known algebraic algorithms for the same task.
    This speedup is only polylogarithmic
    and hence is not visible at the level of detail of our analysis
    (before Appendix~\ref{sec:concrete}),
    but implementors should be aware that simply performing a sequence of separate $\FF_q$ operations
    is not always the best approach.

%%%%%%%%%%%%%%%%%%%%%%%%%%%%%%%%%%%%%%%%%%%%%%%%%%%%%%%%%%%%%%%%%%%%%%%
%\input strassen
\section{Strassen's deterministic factorization algorithm}\label{sec:strassen}

As a warmup,
we review a deterministic algorithm that provably factors $n$ into primes
in time $\softO(n^{1/4})$.
There are several such algorithms in the literature
using fast polynomial arithmetic,
including
\cite{1976/strassen},
\cite{2007/bostan},
\cite{2014/costa},
and
\cite{2018/hittmeir};
there is also a separate series of lattice-based algorithms
surveyed in, e.g.,~\cite{2008/bernstein-smallheight}.
Strassen's algorithm from~\cite{1976/strassen}
has the virtue of being particularly simple,
and is essentially the algorithm presented in this section.

The state of the art in integer factorization
has advanced far beyond $\softO(n^{1/4})$.
For example,
ECM~\cite{1987/lenstra},
Lenstra's elliptic-curve method
of factorization,
is plausibly conjectured to take time $n^{o(1)}$.
We present Strassen's algorithm
because Strassen's main subroutine
is the simplest example of
a much broader speedup that we use.

\subsection{Factorization via modular factorials}
Computing $\gcd(n,\ell!\bmod n)$
reveals whether $n$ has a prime factor ${\le}\ell$.
Binary search through all $\ell\le\sqrt{n}$
then finds the smallest prime factor of $n$.
Repeating this process completely factors $n$ into primes.

The rest of this section focuses on the problem of computing $\ell!\bmod n$,
given positive integers $\ell$ and $n$.
The algorithm of~\S\ref{strassen-factorial}
uses $\softO(\sqrt{\ell})$
additions, subtractions, and multiplications in $\ZZ/n\ZZ$,
plus negligible overhead.
For comparison,
a straightforward computation
would use $\ell-1$ multiplications modulo $n$.
The $\softO$ here is uniform over $n$.

\subsection{Modular factorials as an example of the main problem}
Define $\thegroup$ as the additive group $\ZZ$,
define $P=1$,
define $f:\thegroup\to\ZZ/n\ZZ$ as $s\mapsto s$,
and define $h_S(X)=\prod_{s\in S}(X-f([s]P))\in(\ZZ/n\ZZ)[X]$.
Then, in particular, $h_S(X)=(X-1)\cdots(X-\ell)$ for $S=\{1,\ldots,\ell\}$,
and one can compute $\ell!\bmod n$ by computing $h_S(\ell+1)$
or, alternatively, by computing $(-1)^{\ell}h_S(0)$.
This fits the modular-factorials problem,
in the special case that $n$ is a prime number $q$,
into the framework of~\S\ref{sec:intro}.

\subsection{An algorithm for modular factorials}\label{strassen-factorial}
Compute $b=\left\lfloor\smash{\sqrt{\ell}}\right\rfloor$,
and define $I=\{0,1,2,\dots,b-1\}$.
Use a product tree
to compute the polynomial $h_I(X)=X(X-1)(X-2)\cdots(X-(b-1))\in (\ZZ/n\ZZ)[X]$.

Define $J=\{b,2b,3b,\dots,b^2\}$.
Compute $h_J(X)$,
and then compute the resultant of $h_J(X)$ and $h_I(X)$.
This resultant is $h_I(b)h_I(2b)h_I(3b)\cdots h_I(b^2)$,
i.e., $(b^2)!\bmod n$.

One can compute the resultant of two polynomials
via continued fractions;
see, e.g.,~\cite{1983/strassen}.
An alternative here,
since $h_J$ is given as a product of linear polynomials,
is to use a remainder tree to compute $h_I(b),h_I(2b),\dots,h_I(b^2)\in\ZZ/n\ZZ$,
and then multiply.
Either approach uses $\softO(\sqrt{\ell})$ operations.

Finally,
multiply by $(b^2+1)(b^2+2)\cdots\ell$ modulo $n$,
obtaining $\ell!\bmod n$.
%%%%%%%%%%%%%%%%%%%%%%%%%%%%%%%%%%%%%%%%%%%%%%%%%%%%%%%%%%%%%%%%%%%%%%%

\section{%%%%%%%%%%%%%%%%%%%%%%%%%%%%%%%%%%%%%%%%%%%%%%%%%%%%%%%%%%%%%%%%%%%%%%%
    Evaluation of polynomials whose roots are powers
}%%%%%%%%%%%%%%%%%%%%%%%%%%%%%%%%%%%%%%%%%%%%%%%%%%%%%%%%%%%%%%%%%%%%%%%%%%%%%%%
\label{sec:powers}

Pollard~\cite{1974/pollard}
introduced a deterministic algorithm that provably factors $n$
into primes in time $O(n^{1/4+\epsilon})$.
Strassen's algorithm from~\cite{1976/strassen}
was a streamlined version of Pollard's algorithm,
replacing $O(n^{1/4+\epsilon})$ with $\softO(n^{1/4})$.

This section reviews Pollard's main subroutine,
a fast method to evaluate a polynomial
whose roots (with multiplicity) form a geometric progression.
For comparison,
Strassen's main subroutine
is a fast method to evaluate a polynomial
whose roots form an arithmetic progression.
See~\S\ref{strassen-factorial} above.

\subsection{A multiplicative version of modular factorials}
Fix $\zeta\in(\ZZ/n\ZZ)^*$.
Define $\thegroup=\ZZ$, define $P=1$, define $f:\thegroup\to(\ZZ/n\ZZ)^*$ as $s\mapsto \zeta^s$,
and define $h_S(X)=\prod_{s\in S}(X-f([s]P))=\prod_{s\in S}(X-\zeta^s)\in(\ZZ/n\ZZ)[X]$.
(For comparison, in~\S\ref{sec:strassen},
$f$ was $s\mapsto s$,
and $h_S(X)$ was $\prod_{s\in S}(X-s)$.)

In particular, $h_S(X)=\prod_{s=1}^{\ell}(X-\zeta^s)$
for $S=\{1,2,3,\dots,\ell\}$.
Given $\alpha\in\ZZ/n\ZZ$,
one can straightforwardly evaluate $h_S(\alpha)$ for this $S$
using $O(\ell)$ algebraic operations in $\ZZ/n\ZZ$.
The method in~\S\ref{pollard-q-factorial}
accomplishes the same result using only $\softO(\sqrt{\ell})$ operations.
The $O$ and $\softO$ are uniform in $n$,
and all of the algorithms here can take~$\zeta$ as an input rather than fixing it.
There are some divisions by powers of~$\zeta$,
but divisions are included in the definition of algebraic operations.

Pollard uses the special case $h_S(1)=\prod_{s=1}^{\ell}(1-\zeta^s)$.
This is $(1-\zeta)^\ell$ times the quantity
$(1+\zeta)(1+\zeta+\zeta^2)\cdots(1+\zeta+\dots+\zeta^{\ell-1})$.
It would be standard to call the latter quantity a ``$q$-factorial''
if the letter ``$q$'' were used in place of ``$\zeta$'';
beware, however,
that it is not standard to call this quantity a ``$\zeta$-factorial''.
For a vast generalization of Pollard's algorithm to $q$-holonomic
sequences, see~\cite{Bostan2020};
in \S\ref{sec:elliptic}, we will
generalize it in a different direction.

\subsection{An algorithm for the multiplicative version of modular factorials}
\label{pollard-q-factorial}
Compute $b=\left\lfloor\smash{\sqrt{\ell}}\right\rfloor$,
and define $I=\{1,2,3,\dots,b\}$.
Use a product tree to compute the polynomial $h_I(X)=\prod_{i=1}^b (X-\zeta^i)$.

Define $J=\{0,b,2b,\dots,(b-1)b\}$,
and use a remainder tree to compute $h_I(\alpha/\zeta^j)$ for all $j\in J$.
Pollard uses the chirp-$z$ transform~\cite{1969/rabiner}
(Bluestein's trick) instead of a remainder tree,
saving a logarithmic factor in the number of operations,
and it is also easy to save a logarithmic factor in computing $h_I(X)$,
but these speedups are not visible at the level of detail of the analysis in this section.

Multiply $\zeta^{jb}$ by $h_I(\alpha/\zeta^j)$
to obtain $\prod_{i=1}^{b} (\alpha-\zeta^{i+j})$ for each $j$,
and then multiply across $j\in J$
to obtain $\prod_{s=1}^{b^2} (\alpha-\zeta^s)$.
Finally,
multiply by $\prod_{s=b^2+1}^{\ell} (\alpha-\zeta^s)$
to obtain the desired $h_S(\alpha)$.

One can view the product $\prod_{s=1}^{b^2} (\alpha-\zeta^s)$ here,
like the product $(b^2)!$ in~\S\ref{sec:strassen},
as the resultant of two degree-$b$ polynomials.
Specifically,
$\prod_j h_I(\alpha/\zeta^j)$
is the resultant of $\prod_j (X-\alpha/\zeta^j)$ and $h_I$;
and $\prod_j \zeta^{jb} h_I(\alpha/\zeta^j)$
is the resultant of $\prod_j (\zeta^j X-\alpha)$ and $h_I$.
One can, if desired,
use continued-fraction resultant algorithms
rather than multipoint evaluation via a remainder tree.

\ifodd0
The univariate polynomial
$\prod_j \zeta^{jb} h_I(Z/\zeta^j)$
is the $X$-resultant of a bivariate homogenization of $h_J$
and $h_I$,
namely $\prod_j (\zeta^j X-Z)=(-X)^b h_J(Z/X)$.
\fi

\subsection{Structures in $S$ and $f$}
We highlight two structures exploited in the above computation
of $\prod_{s=1}^{\ell} (\alpha-\zeta^s)$.
First,
the set $S=\{1,2,\dots,\ell\}$
has enough additive structure
to allow most of it to be decomposed as $I+J$,
where $I$ and $J$ are much smaller sets.
Second,
the map $s\mapsto\zeta^s$ is a group homomorphism,
allowing each $\zeta^{i+j}$ to be computed as the product of $\zeta^i$ and $\zeta^j$;
we will return to this point in~\S\ref{wantmorphism}.

We now formalize the statement regarding additive structure,
focusing on the $\FF_q$ case that we will need later in the paper.
First, some terminology:
we say that sets of integers \(I\) and \(J\) 
have \emph{no common differences}
if \(i_1 - i_2 \not= j_1 - j_2\) for all \(i_1\not=i_2\) in \(I\)
and all \(j_1\not=j_2\) in \(J\).
If \(I\) and \(J\) have no common differences,
then the map \(I\times J \to I+J\)
sending \((i,j)\) to \(i+j\)
is a bijection.

\begin{lemma}
    \label{lemma:multiplicative-sumset}
    Let $q$ be a prime power.
    Let $\zeta$ be an element of $\FF_q^*$.
    Define $h_S(X)=\prod_{s\in S}(X-\zeta^s)\in\FF_q[X]$
    for each finite subset $S$ of $\ZZ$.
    Let \(I\) and \(J\) be finite subsets of \(\ZZ\)
    with no common differences.
    Then
    \[
        h_{I+J}(X) 
        =
        \Res_Z(h_{I}(Z),H_{J}(X,Z))
    \]
    where
    \(\Res_Z(\cdot,\cdot)\) is the bivariate resultant,
    and
    \[
        H_{J}(X,Z) := \prod_{j \in J}(X -\zeta^j Z)
        .
    \]
\end{lemma}
\begin{proof}
    \(
        \Res_Z(h_I(Z), H_J(X,Z)) 
        = 
        \prod_{i \in I}\prod_{j \in J}(X - \zeta^i\zeta^j) 
        = 
        \prod_{(i,j) \in I\times J}(X - \zeta^{i + j})
        =
        h_{I + J}(X)
    \)
    since the map \(I\times J \to I + J\)
    sending \((i,j)\) to \(i + j\) is bijective.
\end{proof}

Algorithm~\ref{alg:multiplicative}
is an algebraic algorithm that outputs \(h_S(\alpha)\)
given \(\alpha\).
The algorithm is parameterized by $\zeta$ and the set $S$,
and also by finite subsets \(I,J\subset\ZZ\)
with no common differences 
such that \(I + J \subseteq S\).
The algorithm and the proof of
Proposition~\ref{prop:multiplicative}
are stated using generic resultant computation
(via continued fractions),
but, as in~\S\ref{strassen-factorial}
and~\S\ref{pollard-q-factorial},
one can alternatively use multipoint evaluation.

\begin{algorithm}
    \caption{Computing \(h_S(\alpha) = \prod_{s \in S}(\alpha - \zeta^s)\)}
    \label{alg:multiplicative}
    \KwParameters{%
        a prime power $q$;
        $\zeta\in\FF_q^*$;
        finite subsets \(I, J, K \subseteq \ZZ\) 
        such that
        \(I\) and \(J\) have no common differences
        and $(I+J)\cap K=\emptyset$
    }
    \KwIn{%
        \(\alpha\) in \(\FF_q\)
    }
    \KwOut{%
        \(h_S(\alpha)\) 
        where \(h_S(X) = \prod_{s \in S}(X - \zeta^s)\)
        and $S=(I+J)\cup K$
    }
    \(h_I \gets \prod_{i\in I}(Z - \zeta^i) \in \FF_q[Z]\)
    \label{alg:multiplicative:h_I}
    \;
    \(H_J \gets \prod_{j \in J}(\alpha - \zeta^jZ) \in \FF_q[Z]\)
    \label{alg:multiplicative:H_J}
    \;
    \(h_{I+J} \gets \Res_Z(h_I,H_J) \in \FF_q\)
    \label{alg:multiplicative:Res}
    \;
    \(h_K \gets \prod_{k \in K}(\alpha - \zeta^k) \in \FF_q\)
    \label{alg:multiplicative:h_K}
    \;
    \Return{\(h_{I+J}\cdot h_K\)}
\end{algorithm}

\begin{proposition}
\label{prop:multiplicative}
    Let $q$ be a prime power.
    Let $\zeta$ be an element of $\FF_q^*$.
    Let $I,J$ be finite subsets of $\ZZ$ with no common differences.
    Let $K$ be a finite subset of $\ZZ$ disjoint from $I+J$.
    Given $\alpha$ in $\FF_q$,
    Algorithm~\ref{alg:multiplicative}
    outputs $\prod_{s\in S}(\alpha-\zeta^s)$
    using \(\softO(\max(\#I,\#J,\#K))\) \(\FF_{q}\)-operations,
    where $S=(I+J)\cup K$.
\end{proposition}

The $\softO$ is uniform in $q$.
The algorithm can also take $\zeta$ as an input,
at the cost of computing $\zeta^i$ for $i\in I$,
$\zeta^j$ for $j\in J$,
and $\zeta^k$ for $k\in K$.
This preserves the time bound if the elements of $I,J,K$
have $\softO(\max(\#I,\#J,\#K))$ bits.

\begin{proof}
    Since \(S\setminus K = I + J\),
    we have \(h_S(\alpha) = h_{I+J}(\alpha)\cdot h_K(\alpha)\),
    and Lemma~\ref{lemma:multiplicative-sumset}
    shows that \(h_{I+J}(\alpha) = \Res_Z(h_I(Z),H_J(\alpha,Z))\).
    Line~\ref{alg:multiplicative:h_I} computes \(h_I(Z)\)
    in \(\softO(\#I)\) \(\FF_q\)-operations;
    Line~\ref{alg:multiplicative:H_J} computes \(H_J(\alpha,Z)\)
    in \(\softO(\#J)\) \(\FF_q\)-operations;
    Line~\ref{alg:multiplicative:Res} computes \(h_{I+J}(\alpha)\)
    in \(\softO(\max(\#I,\#J))\) \(\FF_q\)-operations;
    and Line~\ref{alg:multiplicative:h_K} computes \(h_K(\alpha)\)
    in \(\softO(\#K)\) \(\FF_q\)-operations.
    The total is \(\softO(\max(\#I,\#J,\#K))\) \(\FF_{q}\)-operations.
\end{proof}

\subsection{Optimization}
The best conceivable case for the time bound in Proposition~\ref{prop:multiplicative},
as a function of $\#S$,
is $\softO(\sqrt{\#S})$.
Indeed, \(\#S = \#I\cdot\#J + \#K\),
so $\max(\#I,\#J,\#K)\ge\sqrt{\#S+1/4}-1/2$.

To reach $\softO(\sqrt{\#S})$
for a given set of exponents \(S\),
we need sets \(I\) and \(J\) with no common differences
such that \(I + J \subseteq S\)
with \(\#I\), \(\#J\), and $\#(S\setminus(I+J))$ in \(\softO(\sqrt{\#S})\).
Such \(I\) and \(J\)
exist for many useful sets \(S\).
Example~\ref{ex:S-ap} shows a simple form for \(I\) and \(J\)
when \(S\) is an arithmetic progression.

\begin{example}
    \label{ex:S-ap}
    Suppose \(S\) is an arithmetic progression of length \(n\):
    that is,
    \[
        S = \{m, m + r, m + 2r, \ldots, m + (n-1)r\}
    \]
    for some \(m\) and some nonzero \(r\).
    Let \(b = \lfloor{\sqrt{n}}\rfloor\),
    and set
    \[
        I := \{ir \mid 0 \le i < b\}
        \qquad
        \text{and}
        \qquad
        J := \{m + jbr \mid 0 \le j < b\}
        \,;
    \]
    then \(I\) and \(J\) have no common differences,
    and
    \( I + J 
        = 
        \{m + kr \mid 0 \le k < b^2\} %, m + r, \ldots, m + (b^2 - 1)r\}
    \),
    so
    \[
        I + J = S \setminus K 
        \quad
        \text{where}
        \quad
        K = \{m + kr \mid b^2 \le k < n \}
        \,.
    \]
    Now \(\#I = \#J = b\),
    and \(\#K = n - b^2 \le 2b\),
    so we can use these sets to compute \(h_S(\alpha)\)
    in \(\softO(b) = \softO(\sqrt{n})\) \(\FF_{q}\)-operations,
    following Proposition~\ref{prop:multiplicative}.
    (In the case \(r = 1\), we recognise the index sets driving Shanks'
    baby-step giant-step algorithm.)
\end{example}

\section{%%%%%%%%%%%%%%%%%%%%%%%%%%%%%%%%%%%%%%%%%%%%%%%%%%%%%%%%%%%%%%%%%%%%%%%
    Elliptic resultants
}%%%%%%%%%%%%%%%%%%%%%%%%%%%%%%%%%%%%%%%%%%%%%%%%%%%%%%%%%%%%%%%%%%%%%%%%%%%%%%%
\label{sec:elliptic}

The technique in~\S\ref{sec:powers}
for evaluating polynomials whose roots are powers is
well known.
Our main theoretical contribution 
is to adapt this
to polynomials whose roots are functions of more interesting groups:
in particular, functions of elliptic-curve torsion points.
The most important such function is the \(x\)-coordinate.
The main complication here is that,
unlike in~\S\ref{sec:powers},
the function \(x\) is not a homomorphism.

\subsection{The elliptic setting}
\label{wantmorphism}
\label{sec:elliptic-setting}

Let \(\EC/\FF_{q}\) be an elliptic curve,
let $P\in\EC(\FF_{q})$,
% XXX ben dones not see need for this: be a point of order \(n\),
% XXX dan does not see need for this: odd
% XXX dan does not see need for this: not divisible by the characteristic of \(\FF_{q}\),
and define $\thegroup=\subgrp{P}$.
% FIXME: do even order
%Define \(f:\thegroup\to\FF_{q}\) as follows: $0\mapsto 0$,
%and $Q\mapsto x(Q)$ for all other $Q$.
Let \(S\) be a finite subset of \(\ZZ\).
We want to evaluate
\[
    h_S(X) = \prod_{s \in S}(X - f([s]P))\,,
    \qquad
    \text{where}
    \qquad
    f:
    Q \longmapsto
    \begin{cases}
        0 & \text{ if } Q = 0
        \,,
        \\
        x(Q) & \text{ if } Q \not= 0
        \,,
    \end{cases}
\]
at some \(\alpha\) in \(\FF_{q}\).
Here \(x: \EC \to \EC/\subgrp{\pm1} \cong \PP^1\)
is the usual map to the \(x\)-line.

Adapting Algorithm~\ref{alg:multiplicative}
to this setting is not a simple matter of 
replacing the multiplicative group with an elliptic curve.
Indeed, Algorithm~\ref{alg:multiplicative}
explicitly uses the homomorphic nature of \(f: s \mapsto \zeta^s\)
to represent the roots \(\zeta^s\) 
as \( \zeta^i\zeta^j\) where \(s = i + j\).
This presents an obstacle
when moving to elliptic curves:
\(x([i+j]P)\) is not a rational function of
\(x([i]P)\) and \(x([j]P)\),
so
we cannot apply the same trick of
decomposing most of \(S\) as \(I + J\) 
before taking a resultant of polynomials encoding \(f(I)\) and \(f(J)\).

This obstacle does not matter
in the factorization context.
For example,
in~\S\ref{sec:powers},
a straightforward resultant $\prod_{i,j}(\alpha/\zeta^j-\zeta^i)$
detects collisions between $\alpha/\zeta^j$ and $\zeta^i$;
our rescaling to $\prod_{i,j}(\alpha-\zeta^{i+j})$ was unnecessary.
Similarly,
Montgomery's FFT extension~\cite{1992/montgomery} to ECM
computes a straightforward resultant $\prod_{i,j}(x([i]P)-x([j]P))$,
detecting any collisions between $x([i]P)$ and $x([j]P)$;
this factorization method does not compute, and does not need to compute,
a product of functions of $x([i+j]P)$.
The isogenies context is different:
we need a product of functions of $x([i+j]P)$.

Fortunately,
even if the \(x\)-map is not homomorphic,
%it respects enough of the group structure on \(\EC\) that 
there is an algebraic relation between 
\(x(P)\), \(x(Q)\), \(x(P+Q)\), and \(x(P-Q)\),
which we will review in~\S\ref{sec:biquadratics}.
The introduction of the difference \(x(P-Q)\) as well as the sum \(x(P+Q)\)
requires us to replace the decomposition of most of \(S\) as \(I+J\) 
with a decomposition involving \(I+J\) and \(I-J\),
which we will formalize in~\S\ref{sec:index-systems}.
We define the resultant required to tie all this together
and compute \(h_{I\pm J}(\alpha)\) 
in~\S\ref{sec:elliptic-resultant}.

\subsection{Biquadratic relations on \(x\)-coordinates}
\label{sec:biquadratics}
Lemma~\ref{lemma:elliptic-biquadratics}
recalls the general relationship between $x(P)$, $x(Q)$, $x(P+Q)$, and $x(P-Q)$.
%Examples~\ref{ex:biquadratics-Weierstrass} and
Example~\ref{ex:biquadratics-Montgomery} 
gives explicit formul\ae{} for the case that is most useful in our applications.

\begin{lemma}
    \label{lemma:elliptic-biquadratics}
    Let $q$ be a prime power.
    Let $\EC/\FF_q$ be an elliptic curve.
    There exist biquadratic polynomials \(F_0\), \(F_1\), and \(F_2\)
    in \(\FF_{q}[X_1,X_2]\) such that 
    \[
        (X - x(P+Q))(X - x(P-Q)) 
        =
        X^2
        +
        \frac{F_1(x(P),x(Q))}{F_0(x(P),x(Q))}X
        + 
        \frac{F_2(x(P),x(Q))}{F_0(x(P),x(Q))}
    \]
    for all \(P\) and \(Q\) in \(\EC\)
    such that \(0 \notin \{P, Q, P+Q, P-Q\}\).
\end{lemma}
\begin{proof}
    The existence of \(F_0\), \(F_1\), and \(F_2\) is classical
    (see e.g.~\cite[p.~132]{Cassels} for the \(F_i\) for Weierstrass
    models);
    indeed, the existence of such biquadratic systems is a general
    phenomenon for theta functions of level \(2\) on abelian varieties
    (see e.g.~\cite[\S3]{Mumford66}).
\end{proof}

%\begin{example}[Biquadratics for short Weierstrass models]
%    \label{ex:biquadratics-Weierstrass}
%    If \(\EC\) is defined by an affine equation
%    \(y^2 = x^3 + ax + b\),
%    then the polynomials of Lemma~\ref{lemma:elliptic-biquadratics}
%    are
%    \begin{align*}
%        F_0(X_1,X_2) & = (X_1 - X_2)^2
%        \,,
%        \\
%        F_1(X_1,X_2) & = -2((X_1X_2 + a)(X_1+X_2) + 2b)
%        \,,
%        \\
%        F_2(X_1,X_2) & = (X_1X_2 - a)^2 - 4b(X_1+X_2) 
%        \,.
%    \end{align*}
%\end{example}

\begin{example}[Biquadratics for Montgomery models]
    \label{ex:biquadratics-Montgomery}
    If \(\EC\) is defined by an affine equation
    \(By^2 = x(x^2 + Ax + 1)\),
    then the polynomials of Lemma~\ref{lemma:elliptic-biquadratics}
    are
    \begin{align*}
        F_0(X_1,X_2) & = (X_1 - X_2)^2
        \,,
        \\
        F_1(X_1,X_2) & = -2((X_1X_2 + 1)(X_1 + X_2) + 2AX_1X_2)
        \,,
        \\
        F_2(X_1,X_2) & = (X_1X_2 - 1)^2
        \,.
    \end{align*}
    The symmetric triquadratic polynomial
    $
    (X_0X_1-1)^2+(X_0X_2-1)^2+(X_1X_2-1)^2-2X_0X_1X_2(X_0+X_1+X_2+2A)-2
    $
    is
    $X_0^2F_0(X_1,X_2)
    +X_0F_1(X_1,X_2)
    +F_2(X_1,X_2)$.
\end{example}

\subsection{Index systems}
\label{sec:index-systems}

In~\S\ref{sec:powers}, 
we represented most of \(S\) as \(I+J\);
requiring \(I\) and~\(J\) to have no common differences
ensured this representation had no redundancy.
Now we will represent most elements of \(S\) as elements
of \((I+J)\cup(I - J)\),
so we need a stronger restriction on \(I\) and \(J\)
to avoid redundancy.

\begin{definition}
    \label{def:index-system}
    Let \(I\) and \(J\) be finite sets of integers.
    \begin{enumerate}
        \item
            We say that \((I,J)\) is an \emph{index system}
            if the maps \(I\times J \to \ZZ\)
            defined by \((i,j) \mapsto i + j\)
            and \((i,j) \mapsto i - j\)
            are both injective and have disjoint images.
        \item
            If \(S\) is a finite subset of \(\ZZ\),
            then we say that an index system \((I,J)\) is an 
            \emph{index system for \(S\)}
            if \(I+J\) and \(I-J\) are both contained in \(S\).
    \end{enumerate}
\end{definition}
If \((I,J)\) is an index system,
then the sets \(I + J\) and \(I - J\) are both in bijection with \(I\times J\).
We write $I\pm J$ for the union of $I+J$ and $I-J$.

\begin{example}
    \label{ex:x-only-index-system}
    Let $m$ be an odd positive integer,
    and consider the set
    \[
        S = \{1, 3, 5, \ldots, m\}
    \]
    in arithmetic progession.
%    If we let
%    \[
%        I := \{ (2i+1)b \mid 0 \le i < b \} 
%        \qquad \text{and} \qquad
%        J := \{ j \mid 0 < j < b \}
%        \,,
%    \]
%    where \(b = \lfloor{\sqrt{(m+1)/2}}\rfloor\),
%    then \((I,J)\) is an example of an index system for \(S\)
%    with \(\#I = b\) and \(\#J = b-1\)
%    both in \(O(\sqrt{m})\).
%    In this case
%    \(
%        K := S \setminus (I \pm J) 
%        = 
%        K_1 \cup K_2
%    \),
%    where \(K_1 = \{ b, 2b, 3b, \ldots, (2b-1)b \}\)
%    and \(K_2 = \{ 2b^2, 2b^2 + 1,\ldots, m \} \).
%    We have \(\#K_1 = 2b-1\)
%    and \(\#K_2 < 4b + 2\),
%    so \(\#K\) is also in \(O(\sqrt{m})\).
%\end{example}
%
%\begin{example}
%    A slightly more efficient way to handle
%    $S=\{1,\dots,m\}$
%    is to take
    Let
    $$
    I:=\{2b(2i+1) \mid 0\le i<b'\}
    \qquad\text{and}\qquad
    J:=\{2j+1\mid 0\le j<b\}
    $$
    where $b=\lfloor{\sqrt{m+1}/2}\rfloor$;
    $b'=\lfloor(m+1)/4b\rfloor$ if $b>0$;
    and $b'=0$ if $b=0$.
    Then \((I,J)\) is an index system for~\(S\),
    and $S\setminus(I\pm J)=K$
    where
    $K=\{4bb'+1,\dots,m-2,m\}$.
    If $b>0$ then $\#I=b'\le b+2$,
    $\#J=b$,
    and $\#K \le 2b-1$.
\end{example}

\subsection{Elliptic resultants}
\label{sec:elliptic-resultant}

We are now ready to adapt the results of~\S\ref{sec:powers}
to the setting of~\S\ref{sec:elliptic-setting}.
Our main tool is Lemma~\ref{lemma:elliptic-resultant},
which expresses \(h_{I\pm J}\)
as a resultant of smaller polynomials.

\begin{lemma}
    \label{lemma:elliptic-resultant}
    Let $q$ be a prime power.
    Let $\EC/\FF_q$ be an elliptic curve.
    Let $P$ be an element of $\EC(\FF_q)$.
    Let $n$ be the order of $P$.
    Let \((I,J)\) be an index system
    such that \(I\), \(J\), \(I+J\), and \(I-J\)
    do not contain any elements of $n\ZZ$.
    Then
    \[
        h_{I \pm J}(X)
        =
        \frac{1}{\Delta_{I,J}}
        \cdot
        \Res_Z\left(h_I(Z),E_J(X,Z)\right)
    \]
    where
    \[
        E_J(X,Z)
        := 
        \prod_{j \in J}\left(
            F_0(Z,x([j]P))X^2 + F_1(Z,x([j]P))X + F_2(Z,x([j]P))
        \right)
    \]
    and \(\Delta_{I,J} := \Res_Z\left(h_I(Z),D_J(Z)\right)\)
    where \(
        D_J(Z)
        := 
        \prod_{j \in J} F_0(Z,x([j]P))
    \).
\end{lemma}
\begin{proof}
    Since \((I,J)\) is an index system,
    \(I+J\) and \(I-J\) are disjoint,
    and therefore
    we have
    \(
        h_{I \pm J}(X)
        = 
        h_{I + J}(X)\cdot h_{I - J}(X)
    \).
    Expanding and regrouping terms, we get
    \begin{align*}
        h_{I \pm J}(X)
        %& = 
        %\prod_{(i,j) \in I\times J}
        %\left(X- x([i+j]P)\right)
        %\prod_{(i,j) \in I\times J}
        %\left(X - x([i-j]P)\right)
        %\\
        & = 
        \prod_{(i,j) \in I\times J}
        \left(X- x([i+j]P)\right)\left(X - x([i-j]P)\right)
        \\
        & = 
        \prod_{i \in I}\prod_{j \in J}
        \left(
            X^2 
            + 
            \frac{F_1(x([i]P),x([j]P))}{F_0(x([i]P),x([j]P))}X
            + 
            \frac{F_2(x([i]P),x([j]P))}{F_0(x([i]P),x([j]P))}
        \right)
    \end{align*}
    by Lemma~\ref{lemma:elliptic-biquadratics}.
    Factoring out the denominator, we find
    \[
        h_{I \pm J}(X)
        = 
        \frac{
            \prod_{i \in I}
            E_J(X,x([i]P))
        }{
            \prod_{i \in I}\prod_{j \in J} F_0(x([i]P),x([j]P))
        }
        = 
        \frac{
            \prod_{i \in I}
            E_J(X,x([i]P))
        }{
            \prod_{i \in I} D_J(x([i]P))
        }
        \,;
    \]
    and finally
    \(
        \prod_{i \in I} E_J(X,x([i]P)) = \Res_Z(h_I(Z), E_J(X,Z)) 
    \)
    and
    \(
        \prod_{i \in I} D_J(x([i]P))
        = 
        \Res_Z(h_I(Z),D_J(Z)) = \Delta_{I,J}
    \),
    which yields the result.
\end{proof}

\subsection{Elliptic polynomial evaluation}

Algorithm~\ref{alg:x-polynomial} is
an algebraic algorithm for computing \(h_S(\alpha)\);
it is the elliptic analogue of Algorithm~\ref{alg:multiplicative}.
Theorem~\ref{theorem:x-polynomial} proves its correctness and runtime.

\begin{algorithm}
    \caption{%
        Computing \(h_S(\alpha)=\prod_{s \in S}\big(\alpha - x([s]P)\big)\)
        for \(P \in \EC(\FF_q)\)
    }
    \label{alg:x-polynomial}
    \KwParameters{%
      a prime power $q$;
      an elliptic curve $\EC/\FF_q$;
      $P\in\EC(\FF_q)$;
        a finite subset \(S\subset\ZZ\);
        an index system \((I,J)\) for \(S\)
        such that \(S\cap n\ZZ = I\cap n\ZZ = J\cap n\ZZ = \emptyset\),
        where \(n\) is the order of \(P\)
    }
    \KwIn{%
        \(\alpha\) in \(\FF_q\)
    }
    \KwOut{%
        \(h_S(\alpha)\)
        where \(h_S(X) = \prod_{s \in S}(X - x([s]P))\)
    }
    \(h_I \gets \prod_{i \in I}(Z - x([i]P)) \in \FF_q[Z]\)
    \label{alg:x-polynomial:h_I}
    \;
    \(D_J \gets \prod_{j \in J}F_0(Z,x([j]P)) \in \FF_q[Z]\)
    \label{alg:x-polynomial:D_J}
    \;
    \(\Delta_{I,J} \gets \Res_Z(h_I,D_J) \in \FF_q\)
    \label{alg:x-polynomial:Delta_IJ}
    \;
    \(
        E_J 
        \gets 
        \prod_{j \in J}\left(
            F_0(Z,x([j]P))\alpha^2 + F_1(Z,x([j]P))\alpha + F_2(Z,x([j]P))
        \right)
        \in \FF_q[Z]
    \)
    \label{alg:x-polynomial:E_J}
    \;
    \(R \gets \Res_Z(h_I,E_J) \in \FF_q\)
    \label{alg:x-polynomial:Res}
    \;
    \(h_K \gets \prod_{k \in S\setminus(I\pm J)}(\alpha - x([k]P)) \in \FF_q\)
    \label{alg:x-polynomial:h_K}
    \;
    \Return{\(h_K\cdot R/\Delta_{I,J}\)}
    \label{alg:x-polynomial:return}
    \;
\end{algorithm}

\begin{theorem}
    \label{theorem:x-polynomial}
    Let $q$ be a prime power.
    Let $\EC/\FF_q$ be an elliptic curve.
    Let $P$ be an element of $\EC(\FF_q)$.
    Let $n$ be the order of $P$.
    Let $(I,J)$ be an index system for a finite set $S\subset\ZZ$.
    Assume that
    \(I\), \(J\), and \(S\)
    contain no elements of \(n\ZZ\).
    Given \(\alpha\) in \(\FF_{q}\),
    Algorithm~\ref{alg:x-polynomial}
    computes 
    \[
        h_S(\alpha) = \prod_{s \in S}\big(\alpha - x([s]P)\big)
    \]
    in \(\softO(\max(\#I,\#J,\#K))\)
    \(\FF_{q}\)-operations,
    where \(K= S\setminus(I\pm J)\).
\end{theorem}

In particular,
if \(\#I\), \(\#J\), and $\#K$ are in
\(\softO(\sqrt{\#S})\),
then Algorithm~\ref{alg:x-polynomial} computes \(h_S(\alpha)\) 
in \(\softO(\sqrt{\#S})\) \(\FF_{q}\)-operations.
The $\softO$ is uniform in $q$.
Algorithm~\ref{alg:x-polynomial} can also take the coordinates of $P$ as input,
at the cost of computing the relevant multiples of $P$.

\begin{proof}
    The proof follows that of Proposition~\ref{prop:multiplicative}.
    Since \(S \setminus K = I \pm J\),
    we have \(h_S(\alpha) = h_{I \pm J}(\alpha)\cdot h_K(\alpha)\).
    Using the notation of Lemma~\ref{lemma:elliptic-resultant}:
    Line~\ref{alg:x-polynomial:h_I}
    computes \(h_I(Z)\) in \(\softO(\#I)\) \(\FF_q\)-operations;
    Line~\ref{alg:x-polynomial:D_J}
    computes \(D_J(Z)\) in \(\softO(\#J)\) \(\FF_q\)-operations;
    Line~\ref{alg:x-polynomial:Delta_IJ}
    computes \(\Delta_{I,J}\) in \(\softO(\max(\#I,\#J))\) \(\FF_q\)-operations;
    Line~\ref{alg:x-polynomial:E_J}
    computes \(E_J(\alpha,Z)\) in \(\softO(\#J)\) \(\FF_q\)-operations;
    Line~\ref{alg:x-polynomial:Res}
    computes \(\Res_Z(h_I(Z),E_J(\alpha,Z))\),
    which is the same as \(\Delta_{I,J}h_{I\pm J}(\alpha)\) by
    Lemma~\ref{lemma:elliptic-resultant},
    in \(\softO(\max(\#I,\#J))\) \(\FF_q\)-operations;
    Line~\ref{alg:x-polynomial:h_K}
    computes \(h_K(\alpha)\)
    in \(\softO(\#K)\) \(\FF_q\)-operations;
    and Line~\ref{alg:x-polynomial:return}
    returns \(h_S(\alpha) = h_K(\alpha)\cdot h_{I \pm J}(\alpha)\).
    The total number of \(\FF_{q}\)-operations
    is in \(\softO(\max(\#I,\#J,\#K))\).
\end{proof}

\begin{example}[Evaluating kernel polynomials]
    \label{ex:kernel-polynomial}
    We now address a problem from the introduction:
    evaluating \(\Psi_\thegroup\),
    the radical of the denominators of the rational functions
    defining the \(\ell\)-isogeny \(\phi: \EC \to \EC'\) 
    with kernel \(\thegroup = \subgrp{P}\), for $\ell$ odd.
    Here
    \[
        \Psi_\thegroup(X)
        =
        h_S(X) 
        = 
        \prod_{s \in S}(X - x([s]P))
        \qquad
        \text{where}
        \qquad
        S = \{1, 3, \ldots, \ell-2\}
    \]
    (the set \(S\) may be replaced by any set of
    representatives of \(((\ZZ/\ell\ZZ)\setminus\{0\})/\subgrp{\pm1}\)).
    Following Example~\ref{ex:x-only-index-system},
    let
    \(I =\{2b(2i+1) \mid 0\le i<b'\}\)
    and \(J = \{1, 3, \ldots, 2b-1\}\)
    with
    \(b = \lfloor\sqrt{\ell-1}/2\rfloor\)
    and (for $b>0$) \(b' = \lfloor(\ell-1)/4b\rfloor\);
    then \((I,J)\) is an index system for \(S\),
    and Algorithm~\ref{alg:x-polynomial}
    computes \(h_S(\alpha) = \Psi_\thegroup(\alpha)\)
    for any \(\alpha\) in \(\FF_{q}\)
    in \(\softO(\sqrt{\ell})\) \(\FF_{q}\)-operations.
\end{example}

\begin{example}[Evaluating derivatives of polynomials]
\label{example:derivatives}
Algorithm~\ref{alg:x-polynomial}
can evaluate $h_S$ at points in any $\FF_q$-algebra,
at the cost of a slowdown that depends on how large the algebra is.
These algebras need not be fields.
For example,
we can evaluate $h_S(\alpha+\epsilon)$
in the algebra $\FF_q[\epsilon]/\epsilon^2$ of 1-jets,
obtaining $h_S(\alpha)+\epsilon h_S'(\alpha)$.
We can thus evaluate derivatives, sums over roots, etc.
The algebra of 1-jets
was used the same way in, e.g.,
\cite{1973/morgenstern,2001/malajovich,2011/bernstein-jetlist};
\cite{2011/oberwolfach} also notes Zagier's suggested terminology ``jet plane''.
\end{example}

\subsection{Irrational generators}
\label{sec:rati-subgr-with}
The point \(P\) in Lemma~\ref{lemma:elliptic-resultant},
Algorithm~\ref{alg:x-polynomial},
and Theorem~\ref{theorem:x-polynomial}
need not be in \(\EC(\FF_q)\): 
everything is defined over \(\FF_q\)
if \(x(P)\) is in \(\FF_q\).
More generally, take \(P\) in \(\EC(\FF_{q^e})\)
with \(x(P)\) in \(\FF_{q^e}\)
for some minimal \(e \ge 1\).  
The \(q\)-power Frobenius \(\pi\) on \(\EC\)
maps \(P\) to \(\pi(P) = [\lambda]P\)
for some eigenvalue \(\lambda\) in \(\ZZ/n\ZZ\)
of order \(e\) in \(\units{(\ZZ/n\ZZ)}\).
Let \(L = \{\lambda^a \mid 0 \le a < e\}\).
For \(h_S(X)\) to be in \(\FF_q[X]\),
we need \(S = LS'\) for some \(S' \subseteq \ZZ\)
(modulo \(n\)):
that is, \(S = \{\lambda^as' \mid 0 \le a < e, s' \in S'\}\).
Then
\[
    h_S(X) 
    = 
    \prod_{s' \in S'}\prod_{a = 0}^{e-1}(X - x([\lambda^as']P))
    = 
    \prod_{s' \in S'}g_{s'}(X)
\]
where the polynomial
\[
    g_{s'}(X) 
    =
    \prod_{a = 0}^{e-1}(X - x([\lambda^as']P))
    = 
    \prod_{a = 0}^{e-1}(X - x(\pi^a([s']P)))
    = 
    \prod_{a = 0}^{e-1}(X - x([s']P)^{q^a})
\]
is in \(\FF_q[X]\),
and can be easily computed from \(x([s]P)\).

To write \(h_I\), \(D_J\), and \(E_J\)
as products of polynomials over \(\FF_q\),
we need the index system \((I,J)\) for \(S\)
to satisfy \((I,J) = (LI',LJ')\)
for some index system \((I',J')\) for~\(S'\).
%then everything is defined over \(\FF_q\)
%and the resultants are computed as usual.
While this does not affect the asymptotic complexity
of the resulting evaluation algorithms
at our level of analysis,
it should be noted that the requirement that \((I,J) = (LI',LJ')\)
is quite strong: typically \(e\) is in \(O(\ell)\),
so \(\#L\) is not in \(\softO(\sqrt{\#S})\),
and a suitable index system \((I,J)\) with \(\#I\) and \(\#J\) in
\(\softO(\sqrt{\#S})\) does not exist.

\subsection{Other functions on \(\EC\)}
We can replace \(x\) with more general functions on~\(\EC\),
though for completely general \(f\)
there may be no useful analogue of
Lemma~\ref{lemma:elliptic-biquadratics},
or at least not one that allows a Lemma~\ref{lemma:elliptic-resultant}
with conveniently small index system.
However, everything above adapts easily to
the case where \(x\) is composed with an automorphism of \(\PP^1\)
(that is, \(f = (ax + b)/(cx + d)\) 
with \(a, b, c, d\) in \(\FF_q\) such that \(ad \not= bc\)).
Less trivially,
we can take \(f = \psi_x\)
for any isogeny \(\psi: \EC \to \EC''\).
In this case, the \(F_0\), \(F_1\), and \(F_2\)
of Lemma~\ref{lemma:elliptic-biquadratics}
are derived from the curve \(\EC''\), not \(\EC\).

\subsection{Abelian varieties}
It is tempting to extend our results to
higher-dimensional principally polarized abelian varieties (PPAVs),
replacing \(\EC\) with a PPAV \(\AV/\FF_q\),
and \(x\) with some coordinate on \(\AV\),
but evaluating the resulting \(h_S\) using our methods is more complicated.
The main issue is the analogue of
Lemma~\ref{lemma:elliptic-biquadratics}.
If we choose any even coordinate \(x\) on \(\AV\),
then the classical theory of theta functions yields quadratic
relations
between \(x(P+Q)\), \(x(P-Q)\),
and the coordinates of \(P\) and \(Q\),
but not \emph{only} \(x(P)\) and \(x(Q)\):
they also require the other even coordinates of \(P\) and \(Q\).
(The simplest example of this is seen in the differential addition
formul\ae{} for Kummer surfaces: see~\cite[\S6]{Chudnovsky86},
\cite[\S3.2]{Gaudry}, and~\cite[\S4.4]{Cassels--Flynn}.)
This means that 
an analogue of Algorithm~\ref{alg:x-polynomial} for
PPAVs would require multivariate polynomials and resultants;
an investigation of this
is well beyond the scope of this article.

\section{%%%%%%%%%%%%%%%%%%%%%%%%%%%%%%%%%%%%%%%%%%%%%%%%%%%%%%%%%%%%%%%%%%%%%%%
    Computing elliptic isogenies
}%%%%%%%%%%%%%%%%%%%%%%%%%%%%%%%%%%%%%%%%%%%%%%%%%%%%%%%%%%%%%%%%%%%%%%%%%%%%%%%
\label{sec:isogenies}

We now apply the techniques of~\S\ref{sec:elliptic}
to the problem of efficient isogeny computation.
The task is divided in two parts: evaluating isogenies on points
(\S\ref{sec:sqrtvelu-evaluation}),
and computing codomain curves (\S\ref{sec:sqrtvelu-codomain}).
Our cryptographic applications use
isogenies between Montgomery models of elliptic curves,
and we concentrate exclusively on this case here;
but our methods adapt easily
to Weierstrass and other models.

\subsection{Evaluating isogenies}
\label{sec:sqrtvelu-evaluation}
Let \(\EC/\FF_q: y^2 = x(x^2 + Ax + 1)\) be an elliptic curve in
Montgomery form,
and let \(P\) be a point of prime order \(\ell \not= 2\) in \(\EC(\FF_q)\).
Costello and Hisil
give explicit formul\ae{}
in~\cite{2017/costello}
for
a quotient isogeny \(\phi: \EC \to \EC'\)
with kernel \(\thegroup = \subgrp{P}\)
such that \(\EC'/\FF_q : y^2 = x(x^2 + A'x + 1)\)
is a Montgomery curve:
\[
    \phi: (X,Y)
    \longmapsto
    \left(
        \phi_x(X)
        ,
        c_0Y\phi_x'(X)
    \right)
\]
where $c_0=\prod_{0<s<\ell/2}x([s]P)$
and
\begin{equation}
    \label{eq:conventional}
    \phi_x(X) 
    = 
    X \prod_{0 < s < \ell} \frac{ x([s]P)X - 1 }{ X - x([s]P) }
    \,.
\end{equation}
See~\cite{R18}
for generalizations and a different proof,
and see the earlier paper~\cite{2016/moody}
for analogous Edwards-coordinate formulas.

Our main goal is to evaluate \(\phi\) on the level of \(x\)-coordinates:
that is, to compute $\phi_x(\alpha)$
given \(\alpha = x(Q)\) for \(Q\) in \(\EC(\FF_q)\).
This is sufficient for our cryptographic applications.
Applications that also need the $y$-coordinate of $\phi(Q)$, namely $c_0 y(Q)\phi_x'(\alpha)$,
can compute $c_0$ as $(-1)^{(\ell-1)/2}h_S(0)$,
and can compute $\phi_x'(\alpha)$ together with $\phi_x(\alpha)$
by the technique of Example~\ref{example:derivatives}.
To compute \(\phi_x(\alpha)\),
we rewrite Eq.~\eqref{eq:conventional}
as
\[
    \phi_x(X)
    =
    \frac{ X^\ell\cdot h_S(1/X)^2 }{ h_S(X)^2 }
    \quad
    \text{where}
    \quad
    S = \{1, 3, \ldots, \ell-2\}
    \,.
\]
Computing \(\phi_x(\alpha)\) thus reduces to two applications of 
Algorithm~\ref{theorem:x-polynomial},
using (for example) the index system \((I,J)\) for \(S\)
in Example~\ref{ex:x-only-index-system}.
The constant \(\Delta_{I,J}\) appears with the same
multiplicity in the numerator and denominator,
so we need not compute it.
All divisions in the computation are by nonzero field elements
except in the following cases,
which can be handled separately:
if $Q=0$
then $\phi(Q)=0$;
if $Q\ne 0$ but $h_S(\alpha)=0$ for $\alpha=x(Q)$
then $\phi(Q)=0$;
if $Q=(0,0)$
then $\phi(Q)=(0,0)$.
        %(One can also homogenize the computation of $h_S(1/\alpha)$ to
        %avoid this division.)
        %(This does not happen if we restrict to \(Q\) of odd order.)

\subsection{Computing codomain curves}
\label{sec:sqrtvelu-codomain}
Our other main task is
to determine the coefficient \(A'\) in the 
defining equation of \(\EC'\).

One approach is as follows.
We can now efficiently compute \(\phi(Q)\) for any \(Q\) in \(\EC(\FF_q)\).
Changing the base ring from $\FF_q$
to $R=\FF_q[\alpha]/(\alpha^2+A\alpha+1)$
(losing a small constant factor in the cost of evaluation)
gives us $\phi(Q)$ for any \(Q\) in \(\EC(R)\).
In particular,
\(Q = (\alpha,0)\) is a point in \(\EC[2](R)\),
and computing
\(\phi(Q) = (\alpha',0)\)
reveals \(A' = -(\alpha' + 1/\alpha')\).
An alternative---%
at the expense of taking a square root,
which is no longer a $q$-independent algebraic computation---%
is to find a point
$(\alpha,0)$ in $\EC(\FF_{q^2})$ with $\alpha\ne 0$.
Sometimes \(\alpha\) is in $\FF_q$,
and then extending to $\FF_{q^2}$ is unnecessary.

Another approach is to use explicit formulas for $A'$.
The formulas from~\cite{2017/costello}
give
\(A' = c_0^2(A - 3\sigma)\)
where
\(c_0^2 = \prod_{0 < s < \ell} x([s]P)\)
and \(\sigma = \sum_{0 < s < \ell}(x([s]P) - 1/x([s]P))\).
As pointed out in~\cite{2018/meyer} in the context of CSIDH,
one can instead transform to twisted Edwards form
and use the formulas from~\cite{2016/moody},
obtaining $A'=2(1+d)/(1-d)$
where
$$d=
  \left(\frac{A-2}{A+2}\right)^\ell
  \biggl(\prod_{s\in S}\frac{x([s]P)-1}{x([s]P)+1}\biggr)^8
  =
  \left(\frac{A-2}{A+2}\right)^\ell
  \biggl(\frac{h_S(1)}{h_S(-1)}\biggr)^8
  \,.
$$
We can thus compute \(A'\) using $\softO(\sqrt{\ell})$
operations:
%as
%$((A-2)/(A+2))^\ell (h_S(1)/h_S(-1))^8$.
%\\
%In the end, 
every task we need can be performed by some evaluations of $h_S$ and some (asymptotically negligible) operations. 

\section{%%%%%%%%%%%%%%%%%%%%%%%%%%%%%%%%%%%%%%%%%%%%%%%%%%%%%%%%%%%%%%%%%%%%%%%
    Applications in isogeny-based cryptography
}%%%%%%%%%%%%%%%%%%%%%%%%%%%%%%%%%%%%%%%%%%%%%%%%%%%%%%%%%%%%%%%%%%%%%%%%%%%%%%%
\label{sec:protocols}

With the notable exception of
SIDH/SIKE~\cite{DFJ11,defeo+jao+plut12,SIKE}, 
most isogeny-based cryptographic protocols
need to evaluate large-degree isogenies.
Specifically,
CRS~\cite{RS06,C96},
CSIDH~\cite{CSIDH2018},
CSURF~\cite{CD19},
etc.~use large-degree isogenies,
since not enough keys are fast compositions
of isogenies of a few small prime degrees.
The largest
isogeny degree,
with standard optimizations,
grows quasi-linearly in the pre-quantum security
level.
For the same post-quantum security level,
known quantum attacks
require an asymptotically larger base field
but do not affect the largest isogeny degree;
see \cite[Remark 11]{CSIDH2018}.

Concretely, targeting 128 bits of pre-quantum security, CSIDH-512 fixes
\[
    p = 4\;\cdot\!\! \underbrace{(3\cdot 5\cdots 373)}_{\text{$73$ first odd
    primes}}\!\! \cdot\; 587 - 1
\]
and uses isogenies of all odd prime degrees \(\ell\mid p+1\). %
Similarly, CSURF-512 fixes
\[
    %p = 8\cdot 9\cdot\!\! \underbrace{(5\cdot 7\cdots 389)}_{\substack{\text{73 consecutive primes,}\\ \text{skipping 347 and 359}}} - 1
    p = 
    8\cdot 9\cdot
    \!\!\!\!\!
    \underbrace{(5\cdot 7\cdots 337)}_{\text{66 consecutive primes}}
    \!\!\!\!\!
    \cdot
    349\cdot353
    \cdot
    \!\!\!\!\!
    \underbrace{(367\cdots389)}_{\text{6 consecutive primes}}
    - 1
\]
and uses isogenies of all prime degrees \(\ell\mid p+1\),
including \(\ell = 2\).

The CSIDH and CSURF algorithms repeatedly sample a random point
of order dividing $p+1$ in $\EC/\FF_p$, multiply it
by an appropriate cofactor to get $P$, and then apply Vélu's
formulas for each of the primes $\ell\mid \mathrm{ord}(P)$ to obtain
$\EC' = \EC/\subgrp{P}$. %
Our algorithm seamlessly replaces Vélu's formulas
in both systems. %
Computing \(\EC'\) is easy in CSURF:
all curves involved have rational $2$-torsion, and can thus be
represented by a root of $\alpha^2+A\alpha-1$ in $\FF_p$. %
For CSIDH, we can apply the techniques of~\S\ref{sec:sqrtvelu-codomain};
alternatively, we can walk to the surface and represent curves
as in CSURF. %

B-SIDH~\cite{Costello2019} is an SIDH variant
using smaller prime fields, at the cost of much larger prime
isogeny degrees. %
One participant uses isogenies of degree \(\ell\mid p+1\),
and the other uses \(\ell\mid p-1\). %
Since primes \(p\) such that \(p-1\) and \(p+1\) \emph{both}
have many small prime factors are rare,
some of the \(\ell\) involved in B-SIDH tend to be even larger
than in CSIDH and CSURF.
The B-SIDH algorithm starts from a single point $P$ and computes 
$\EC/\subgrp{P}$ together with the evaluation of
$\phi:\EC\to\EC/\subgrp{P}$ at three points. %
Unlike CSIDH and CSURF, there is no repeated random sampling of points:
a single \(\ell\)-isogeny evaluation for each prime \(\ell\mid p\pm1\)
is needed. %

Our asymptotic speedup in isogeny evaluation
implies asymptotic speedups for CRS, CSIDH, CSURF, and B-SIDH
as the security level increases.
This does not imply, however,
that there is a speedup for (e.g.)~pre-quantum security $2^{128}$.
Appendix~\ref{sec:concrete} addresses the question
of how large $\ell$ needs to be
before our algorithms become faster than the conventional algorithms.

Cryptographic protocols 
that exploit the KLPT algorithm~\cite{KLPT14} for
isogeny path renormalization, such as the signature
scheme~\cite{GPS17} and the encryption scheme SÉTA~\cite{SETA2019},
%Due to the size of the ideals found by KLPT, 
need to work with irrational torsion points.
They may thus benefit from the
technique of~\S\ref{sec:rati-subgr-with}. %
We did not investigate these protocols further.

%%%%%%%%%%%%%%%%%%%%%%%%%%%%%%%%%%%%%%%%%%%%%%%%%%%%%%%%%%%%%%%%%%%%%%%
\bibliographystyle{plain}
\bibliography{references}
%%%%%%%%%%%%%%%%%%%%%%%%%%%%%%%%%%%%%%%%%%%%%%%%%%%%%%%%%%%%%%%%%%%%%%%

\appendix

\section{%%%%%%%%%%%%%%%%%%%%%%%%%%%%%%%%%%%%%%%%%%%%%%%%%%%%%%%%%%%%%%%%%%%%%%%
    Concrete costs and cross-overs
}%%%%%%%%%%%%%%%%%%%%%%%%%%%%%%%%%%%%%%%%%%%%%%%%%%%%%%%%%%%%%%%%%%%%%%%%%%%%%%%
\label{sec:concrete}
Conventional algorithms
use $\Theta(\ell)$ operations
to evaluate an $\ell$-isogeny.
Any $\softO(\sqrt{\ell})$ algorithm
is better than $\Theta(\ell)$ for all sufficiently large $\ell$,
but this does not answer the question of how large
``sufficiently large'' is.
A more precise asymptotic comparison,
such as $\sqrt{\ell}(\log\ell)^{2+o(1)}$ vs.~$\Theta(\ell)$,
also does not answer the question.

This appendix
looks more closely at performance
and quantifies the cross-over point.
In each of the metrics considered here,
the cross-over point is within the range of primes
used in CSIDH-512.
The new $\ell$-isogeny algorithm
sets new speed records for CSIDH-512 and CSIDH-1024
by small but measurable percentages,
and has more effect on protocols
that use larger $\ell$-isogenies.
Our code is available from
\url{https://velusqrt.isogeny.org}.

\subsection{Choice of function to compute}
\label{choice-of-function}
This appendix highlights the following subroutine:
evaluate an $\ell$-isogeny on a point,
and at the same time
compute the target curve.
The inputs are
an odd prime $\ell$,
a coefficient $A$
for a Montgomery curve $By^2=x(x^2+Ax+1)$,
a coordinate $x(P)$
for a curve point $P$ of order $\ell$,
and a coordinate $x(Q)$
for a curve point $Q$ outside $\langle P\rangle$.
The outputs are $A'$ and $x(\phi(Q))$,
where $\phi$ is a separable $\ell$-isogeny
with kernel $\langle P\rangle$ from the curve
to a Montgomery curve $B'y^2=x(x^2+A'x+1)$.
The inputs and outputs
are represented as fractions.
The base field is a parameter,
and sometimes software
makes a specific choice of this parameter,
such as the CSIDH-512 prime field.

This subroutine
is the sole use of isogenies
in, e.g.,
the CSIDH software from~\cite{CSIDH2018}
and~\cite{2018/meyer}.
This does not mean
that speeding up this subroutine
produces the same speedup
(averaged appropriately over $\ell$)
in this software:
the software also spends
some time on other operations.
Furthermore,
it might be useful to push more points
through the same isogeny,
as in \cite{defeo+jao+plut12},
or fewer;
see generally
\cite[Section 8]{qisog}.

Beyond measuring the costs of $\ell$-isogenies,
we put these costs into context
by measuring various implementations
of the following protocols:
CSIDH-512
and
CSIDH-1024
using the primes from~\cite{CSIDH2018};
CSURF-512
using the prime from~\cite{CD19};
and
B-SIDH using a new prime defined here.

The search for B-SIDH-friendly \(p\) is a hard task; it is not
currently known how large the degrees \(\ell\) could be, though we
expect them to be considerably larger than those used in
CSIDH and CSURF. %
A 256-bit B-SIDH-friendly prime
would have a security level comparable to that of SIKE p434, 
but~\cite{Costello2019} does not give any 256-bit~\(p\).
We use our own 256-bit prime (found by accident),
which is the \(p\) such that
\begin{align*}
    p + 1 
    &= 
    2^{32} \cdot 5^{21} \cdot 7 \cdot 11 \cdot 163 \cdot 1181
  \cdot 2389 \cdot 5233 \cdot 8353 \cdot 10139 \cdot 11939 \\
    & \qquad \qquad \qquad \qquad \qquad \qquad \cdot 22003 \cdot 25391 \cdot 41843 \cdot 3726787 \cdot 6548911
    \,,
    \\
    p - 1 
    &= 
    2 \cdot 3^{56} \cdot 31 \cdot 43 \cdot 59 \cdot 271 \cdot 311
  \cdot 353 \cdot 461 \cdot 593 \cdot 607 \cdot 647 \cdot 691\\
    & \qquad \qquad \qquad \qquad \qquad \qquad \cdot 743 \cdot 769 \cdot 877 \cdot 1549 \cdot 4721 \cdot 12433
  \cdot 26449
    \,.
\end{align*}
For B-SIDH with this \(p\), one participant needs to evaluate
isogenies of degree as large as $\approx 2^{23}$, while the other one
only handles isogenies of degree less than $2^{14}$.

\subsection{Choices of cost metric}
This appendix uses several different metrics:
\begin{itemize}
\item
  Time for software
  in the Magma~\cite{magma} computer-algebra system.
  We use this metric for comparison to~\cite{CD19},
  which provides software in this system for CSIDH and CSURF.
\item
  Time for Julia software,
  using Nemo for the underlying arithmetic.
  Julia~\cite{julia}
  is a just-in-time-compiled programming language
  designed for a
  ``combination of productivity and performance'',
  and Nemo~\cite{nemo}
  is a computer-algebra package for this language.
\item
  Time for C software
  using FLINT~\cite{flint},
  specifically the modules
  \texttt{fmpz\_mod} and \texttt{fmpz\_mod\_poly}
  for arithmetic
  on elements of $\FF_p$ and $\FF_p[X]$
  respectively.
\item
  Time for arbitrary machine-language software.
  We use this metric for comparison to
  the CSIDH-512 implementation from \cite{2018/meyer};
  that implementation, in turn,
  is an improved version
  of the CSIDH-512 implementation from~\cite{CSIDH2018},
  and reuses the assembly-language field operations
  from~\cite{CSIDH2018}.
\item
  Number of multiplications
  in an algebraic algorithm
  using only additions, subtractions, and multiplications.
  There are no divisions here:
  recall that
  inputs and outputs are represented projectively.
\end{itemize}

The last metric
is a traditional object of study
in algebraic complexity theory,
and has the virtue of a simple and clear definition.
However,
this metric is perhaps {\it too\/} simple:
it ignores potentially massive overheads
for additions, subtractions, and non-algebraic overhead,
while
it ignores the possibility of speedups from divisions,
non-algebraic polynomial-multiplication algorithms, etc.

The other metrics
have the virtue of being physical time measurements
that include all overheads,
but these metrics have the complication
of depending upon the choice of CPU.
We used one core of a
3.40GHz Intel Core i7-6700 (Skylake)
with Turbo Boost disabled.

The first three metrics
have the further complications
of depending upon
the details of large software libraries.
We used Magma v2.21-6,
Julia 1.3.1,
Nemo 0.16.2,
the FLINT development branch (commit \texttt{dd1021a}),
and GMP 6.2.0
tuned using the provided \texttt{tune} program.
Improvements in these libraries
would change the metrics,
perhaps increasing or decreasing the $\ell$ cross-over points.
Data points in these metrics
are nevertheless useful examples
of tradeoffs between CPU time and programmer time.

Internally,
Nemo uses FLINT for polynomial arithmetic and field arithmetic;
FLINT uses GMP for integer arithmetic;
and GMP is
``carefully designed to be as fast as possible, both for small operands and for huge operands''.
Unfortunately,
the external and internal library APIs
create various overheads.
For example,
for random integers modulo
the CSIDH-512 prime $p$,
GMP's {\tt mpz\_mul}
takes about 127 Skylake cycles
for squaring
and about 173 cycles
for general multiplication,
and GMP's {\tt mpz\_mod}
takes about 517 cycles
to reduce the product modulo $p$,
so straightforward modular multiplication
takes between 600 and 700 cycles.
GMP's {\tt mpz\_powm}
modular-exponentiation function
is much faster than this,
taking only about 302$n$ cycles
to compute a $2^n$th power for large $n$,
because internally
it uses much faster Montgomery reduction.
The {\tt fp\_mul} modular-multiplication function from~\cite{CSIDH2018}
also uses Montgomery reduction,
avoids the overhead
of allowing variable-size inputs,
and takes only about 200 cycles.

There are other metrics of interest.
Parallel performance metrics
would show a disadvantage
of a continued-fraction computation of resultants,
namely that it has depth $\softO(\sqrt{\ell})$;
but,
as noted in \cite[Section 4.2]{1992/montgomery},
remainder trees avoid this disadvantage
and seem to use somewhat fewer operations in any case.
As another example,
realistic hardware-performance metrics
such as the area-time metric in~\cite{1981/brent-areatime}
would increase our cost from $\softO(\sqrt{\ell})$
to $\softO(\ell^{3/4})$.

\subsection{Results for $\ell$-isogenies}
We wrote four implementations
of our $\ell$-isogeny algorithm
to show upper bounds on costs
in each of the first four metrics above.
Costs here are for our highlighted subroutine,
namely
evaluating an $\ell$-isogeny on one point
and computing the new curve coefficient;
see Appendix~\ref{protocol-timings}
for timings of applications.
The base field here is the CSIDH-512 prime field,
except as noted below.

\begin{figure}[t]
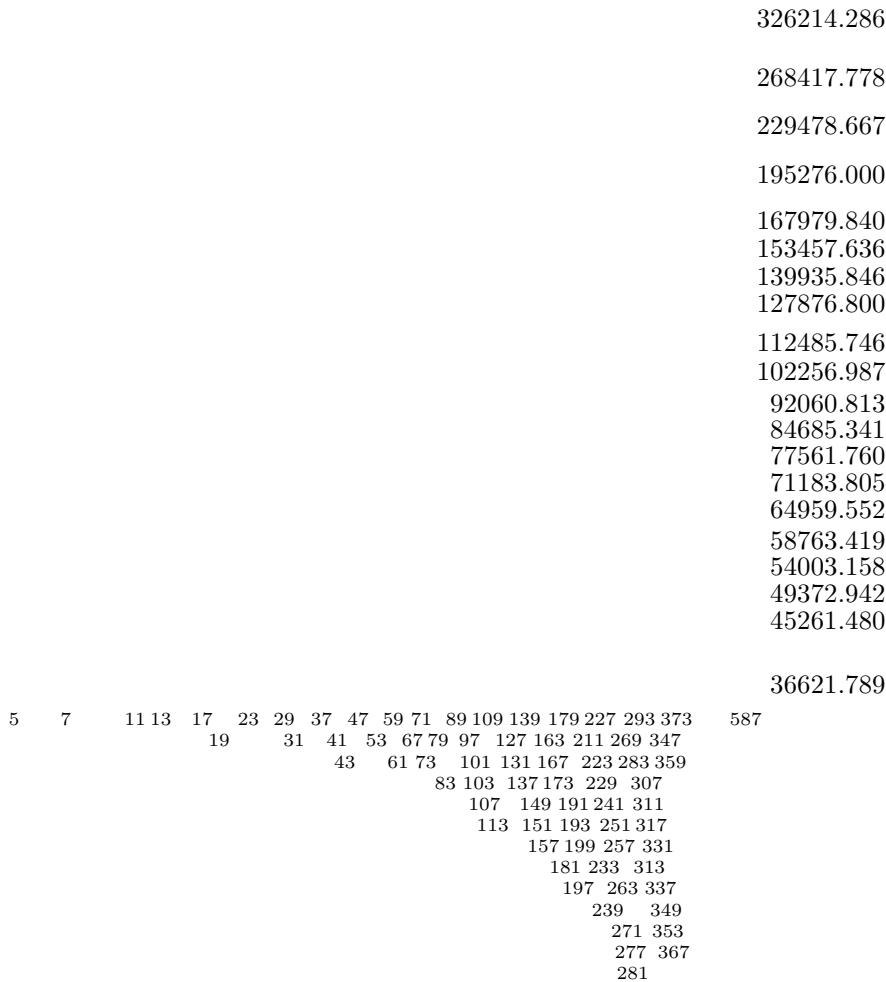

\hbox to\hsize{\hfill
\xy <1.4cm,0cm>:<0cm,2.8cm>::
(2.321928,0.065347); (2.321928,3.220393) **[lightgray]@{-};
(2.321928,-0.100000) *{\scriptstyle 5};
(2.807355,0.065347); (2.807355,3.220393) **[lightgray]@{-};
(2.807355,-0.100000) *{\scriptstyle 7};
(3.459432,0.065347); (3.459432,3.220393) **[lightgray]@{-};
(3.459432,-0.100000) *{\scriptstyle 11};
(3.700440,0.065347); (3.700440,3.220393) **[lightgray]@{-};
(3.700440,-0.100000) *{\scriptstyle 13};
(4.087463,0.065347); (4.087463,3.220393) **[lightgray]@{-};
(4.087463,-0.100000) *{\scriptstyle 17};
(4.247928,0.065347); (4.247928,3.220393) **[lightergray]@{-};
(4.247928,-0.200000) *{\scriptstyle 19};
(4.523562,0.065347); (4.523562,3.220393) **[lightgray]@{-};
(4.523562,-0.100000) *{\scriptstyle 23};
(4.857981,0.065347); (4.857981,3.220393) **[lightgray]@{-};
(4.857981,-0.100000) *{\scriptstyle 29};
(4.954196,0.065347); (4.954196,3.220393) **[lightergray]@{-};
(4.954196,-0.200000) *{\scriptstyle 31};
(5.209453,0.065347); (5.209453,3.220393) **[lightgray]@{-};
(5.209453,-0.100000) *{\scriptstyle 37};
(5.357552,0.065347); (5.357552,3.220393) **[lightergray]@{-};
(5.357552,-0.200000) *{\scriptstyle 41};
(5.426265,0.065347); (5.426265,3.220393) **[lightergray]@{-};
(5.426265,-0.300000) *{\scriptstyle 43};
(5.554589,0.065347); (5.554589,3.220393) **[lightgray]@{-};
(5.554589,-0.100000) *{\scriptstyle 47};
(5.727920,0.065347); (5.727920,3.220393) **[lightergray]@{-};
(5.727920,-0.200000) *{\scriptstyle 53};
(5.882643,0.065347); (5.882643,3.220393) **[lightgray]@{-};
(5.882643,-0.100000) *{\scriptstyle 59};
(5.930737,0.065347); (5.930737,3.220393) **[lightergray]@{-};
(5.930737,-0.300000) *{\scriptstyle 61};
(6.066089,0.065347); (6.066089,3.220393) **[lightergray]@{-};
(6.066089,-0.200000) *{\scriptstyle 67};
(6.149747,0.065347); (6.149747,3.220393) **[lightgray]@{-};
(6.149747,-0.100000) *{\scriptstyle 71};
(6.189825,0.065347); (6.189825,3.220393) **[lightergray]@{-};
(6.189825,-0.300000) *{\scriptstyle 73};
(6.303781,0.065347); (6.303781,3.220393) **[lightergray]@{-};
(6.303781,-0.200000) *{\scriptstyle 79};
(6.375039,0.065347); (6.375039,3.220393) **[lightergray]@{-};
(6.375039,-0.400000) *{\scriptstyle 83};
(6.475733,0.065347); (6.475733,3.220393) **[lightgray]@{-};
(6.475733,-0.100000) *{\scriptstyle 89};
(6.599913,0.065347); (6.599913,3.220393) **[lightergray]@{-};
(6.599913,-0.200000) *{\scriptstyle 97};
(6.658211,0.065347); (6.658211,3.220393) **[lightergray]@{-};
(6.658211,-0.300000) *{\scriptstyle 101};
(6.686501,0.065347); (6.686501,3.220393) **[lightergray]@{-};
(6.686501,-0.400000) *{\scriptstyle 103};
(6.741467,0.065347); (6.741467,3.220393) **[lightergray]@{-};
(6.741467,-0.500000) *{\scriptstyle 107};
(6.768184,0.065347); (6.768184,3.220393) **[lightgray]@{-};
(6.768184,-0.100000) *{\scriptstyle 109};
(6.820179,0.065347); (6.820179,3.220393) **[lightergray]@{-};
(6.820179,-0.600000) *{\scriptstyle 113};
(6.988685,0.065347); (6.988685,3.220393) **[lightergray]@{-};
(6.988685,-0.200000) *{\scriptstyle 127};
(7.033423,0.065347); (7.033423,3.220393) **[lightergray]@{-};
(7.033423,-0.300000) *{\scriptstyle 131};
(7.098032,0.065347); (7.098032,3.220393) **[lightergray]@{-};
(7.098032,-0.400000) *{\scriptstyle 137};
(7.118941,0.065347); (7.118941,3.220393) **[lightgray]@{-};
(7.118941,-0.100000) *{\scriptstyle 139};
(7.219169,0.065347); (7.219169,3.220393) **[lightergray]@{-};
(7.219169,-0.500000) *{\scriptstyle 149};
(7.238405,0.065347); (7.238405,3.220393) **[lightergray]@{-};
(7.238405,-0.600000) *{\scriptstyle 151};
(7.294621,0.065347); (7.294621,3.220393) **[lightergray]@{-};
(7.294621,-0.700000) *{\scriptstyle 157};
(7.348728,0.065347); (7.348728,3.220393) **[lightergray]@{-};
(7.348728,-0.200000) *{\scriptstyle 163};
(7.383704,0.065347); (7.383704,3.220393) **[lightergray]@{-};
(7.383704,-0.300000) *{\scriptstyle 167};
(7.434628,0.065347); (7.434628,3.220393) **[lightergray]@{-};
(7.434628,-0.400000) *{\scriptstyle 173};
(7.483816,0.065347); (7.483816,3.220393) **[lightgray]@{-};
(7.483816,-0.100000) *{\scriptstyle 179};
(7.499846,0.065347); (7.499846,3.220393) **[lightergray]@{-};
(7.499846,-0.800000) *{\scriptstyle 181};
(7.577429,0.065347); (7.577429,3.220393) **[lightergray]@{-};
(7.577429,-0.500000) *{\scriptstyle 191};
(7.592457,0.065347); (7.592457,3.220393) **[lightergray]@{-};
(7.592457,-0.600000) *{\scriptstyle 193};
(7.622052,0.065347); (7.622052,3.220393) **[lightergray]@{-};
(7.622052,-0.900000) *{\scriptstyle 197};
(7.636625,0.065347); (7.636625,3.220393) **[lightergray]@{-};
(7.636625,-0.700000) *{\scriptstyle 199};
(7.721099,0.065347); (7.721099,3.220393) **[lightergray]@{-};
(7.721099,-0.200000) *{\scriptstyle 211};
(7.800900,0.065347); (7.800900,3.220393) **[lightergray]@{-};
(7.800900,-0.300000) *{\scriptstyle 223};
(7.826548,0.065347); (7.826548,3.220393) **[lightgray]@{-};
(7.826548,-0.100000) *{\scriptstyle 227};
(7.839204,0.065347); (7.839204,3.220393) **[lightergray]@{-};
(7.839204,-0.400000) *{\scriptstyle 229};
(7.864186,0.065347); (7.864186,3.220393) **[lightergray]@{-};
(7.864186,-0.800000) *{\scriptstyle 233};
(7.900867,0.065347); (7.900867,3.220393) **[lightergray]@{-};
(7.900867,-1.000000) *{\scriptstyle 239};
(7.912889,0.065347); (7.912889,3.220393) **[lightergray]@{-};
(7.912889,-0.500000) *{\scriptstyle 241};
(7.971544,0.065347); (7.971544,3.220393) **[lightergray]@{-};
(7.971544,-0.600000) *{\scriptstyle 251};
(8.005625,0.065347); (8.005625,3.220393) **[lightergray]@{-};
(8.005625,-0.700000) *{\scriptstyle 257};
(8.038919,0.065347); (8.038919,3.220393) **[lightergray]@{-};
(8.038919,-0.900000) *{\scriptstyle 263};
(8.071462,0.065347); (8.071462,3.220393) **[lightergray]@{-};
(8.071462,-0.200000) *{\scriptstyle 269};
(8.082149,0.065347); (8.082149,3.220393) **[lightergray]@{-};
(8.082149,-1.100000) *{\scriptstyle 271};
(8.113742,0.065347); (8.113742,3.220393) **[lightergray]@{-};
(8.113742,-1.200000) *{\scriptstyle 277};
(8.134426,0.065347); (8.134426,3.220393) **[lightergray]@{-};
(8.134426,-1.300000) *{\scriptstyle 281};
(8.144658,0.065347); (8.144658,3.220393) **[lightergray]@{-};
(8.144658,-0.300000) *{\scriptstyle 283};
(8.194757,0.065347); (8.194757,3.220393) **[lightgray]@{-};
(8.194757,-0.100000) *{\scriptstyle 293};
(8.262095,0.065347); (8.262095,3.220393) **[lightergray]@{-};
(8.262095,-0.400000) *{\scriptstyle 307};
(8.280771,0.065347); (8.280771,3.220393) **[lightergray]@{-};
(8.280771,-0.500000) *{\scriptstyle 311};
(8.290019,0.065347); (8.290019,3.220393) **[lightergray]@{-};
(8.290019,-0.800000) *{\scriptstyle 313};
(8.308339,0.065347); (8.308339,3.220393) **[lightergray]@{-};
(8.308339,-0.600000) *{\scriptstyle 317};
(8.370687,0.065347); (8.370687,3.220393) **[lightergray]@{-};
(8.370687,-0.700000) *{\scriptstyle 331};
(8.396605,0.065347); (8.396605,3.220393) **[lightergray]@{-};
(8.396605,-0.900000) *{\scriptstyle 337};
(8.438792,0.065347); (8.438792,3.220393) **[lightergray]@{-};
(8.438792,-0.200000) *{\scriptstyle 347};
(8.447083,0.065347); (8.447083,3.220393) **[lightergray]@{-};
(8.447083,-1.000000) *{\scriptstyle 349};
(8.463524,0.065347); (8.463524,3.220393) **[lightergray]@{-};
(8.463524,-1.100000) *{\scriptstyle 353};
(8.487840,0.065347); (8.487840,3.220393) **[lightergray]@{-};
(8.487840,-0.300000) *{\scriptstyle 359};
(8.519636,0.065347); (8.519636,3.220393) **[lightergray]@{-};
(8.519636,-1.200000) *{\scriptstyle 367};
(8.543032,0.065347); (8.543032,3.220393) **[lightgray]@{-};
(8.543032,-0.100000) *{\scriptstyle 373};
(9.197217,0.065347); (9.197217,3.220393) **[lightgray]@{-};
(9.197217,-0.100000) *{\scriptstyle 587};
(2.321928,3.220393); (9.197217,3.220393) **@{-};
(10.497217,3.220393) *{\llap{326214.286}};
(2.321928,2.939053); (9.197217,2.939053) **[lightgray]@{-};
(10.497217,2.939053) *{\llap{268417.778}};
(2.321928,2.712933); (9.197217,2.712933) **[lightgray]@{-};
(10.497217,2.712933) *{\llap{229478.667}};
(2.321928,2.480088); (9.197217,2.480088) **[lightgray]@{-};
(10.497217,2.480088) *{\llap{195276.000}};
(2.321928,2.262861); (9.197217,2.262861) **[lightgray]@{-};
(10.497217,2.262861) *{\llap{167979.840}};
(2.321928,2.132414); (9.197217,2.132414) **[lightgray]@{-};
(10.497217,2.132414) *{\llap{153457.636}};
(2.321928,1.999339); (9.197217,1.999339) **[lightgray]@{-};
(10.497217,1.999339) *{\llap{139935.846}};
(2.321928,1.869328); (9.197217,1.869328) **[lightgray]@{-};
(10.497217,1.869328) *{\llap{127876.800}};
(2.321928,1.684315); (9.197217,1.684315) **[lightgray]@{-};
(10.497217,1.684315) *{\llap{112485.746}};
(2.321928,1.546773); (9.197217,1.546773) **[lightgray]@{-};
(10.497217,1.546773) *{\llap{102256.987}};
(2.321928,1.395232); (9.197217,1.395232) **[lightgray]@{-};
(10.497217,1.395232) *{\llap{92060.813}};
(2.321928,1.274757); (9.197217,1.274757) **[lightgray]@{-};
(10.497217,1.274757) *{\llap{84685.341}};
(2.321928,1.147991); (9.197217,1.147991) **[lightgray]@{-};
(10.497217,1.147991) *{\llap{77561.760}};
(2.321928,1.024194); (9.197217,1.024194) **[lightgray]@{-};
(10.497217,1.024194) *{\llap{71183.805}};
(2.321928,0.892187); (9.197217,0.892187) **[lightgray]@{-};
(10.497217,0.892187) *{\llap{64959.552}};
(2.321928,0.747563); (9.197217,0.747563) **[lightgray]@{-};
(10.497217,0.747563) *{\llap{58763.419}};
(2.321928,0.625689); (9.197217,0.625689) **[lightgray]@{-};
(10.497217,0.625689) *{\llap{54003.158}};
(2.321928,0.496366); (9.197217,0.496366) **[lightgray]@{-};
(10.497217,0.496366) *{\llap{49372.942}};
(2.321928,0.370929); (9.197217,0.370929) **[lightgray]@{-};
(10.497217,0.370929) *{\llap{45261.480}};
(2.321928,0.065347); (9.197217,0.065347) **@{-};
(10.497217,0.065347) *{\llap{36621.789}};
(2.241928,0.425788); (2.401928,0.425788) **[blue]@{-};
(2.321928,0.385788); (2.321928,0.465788) **[blue]@{-};
(2.281928,0.415388); (2.361928,0.415388) **[blue]@{-};
(2.281928,0.431310); (2.361928,0.431310) **[blue]@{-};
(2.241928,3.180393); (2.401928,3.260393) **[red]@{-};
(2.401928,3.180393); (2.241928,3.260393) **[red]@{-};
(2.281928,3.192667); (2.361928,3.232667) **[red]@{-};
(2.281928,3.232667); (2.361928,3.192667) **[red]@{-};
(2.281928,3.209820); (2.361928,3.249820) **[red]@{-};
(2.281928,3.249820); (2.361928,3.209820) **[red]@{-};
(2.727355,0.713199); (2.887355,0.713199) **[blue]@{-};
(2.807355,0.673199); (2.807355,0.753199) **[blue]@{-};
(2.767355,0.710268); (2.847355,0.710268) **[blue]@{-};
(2.767355,0.730063); (2.847355,0.730063) **[blue]@{-};
(2.727355,2.899053); (2.887355,2.979053) **[red]@{-};
(2.887355,2.899053); (2.727355,2.979053) **[red]@{-};
(2.767355,2.916919); (2.847355,2.956919) **[red]@{-};
(2.767355,2.956919); (2.847355,2.916919) **[red]@{-};
(2.767355,2.941315); (2.847355,2.981315) **[red]@{-};
(2.767355,2.981315); (2.847355,2.941315) **[red]@{-};
(3.379432,0.945983); (3.539432,0.945983) **[blue]@{-};
(3.459432,0.905983); (3.459432,0.985983) **[blue]@{-};
(3.419432,0.942232); (3.499432,0.942232) **[blue]@{-};
(3.419432,0.957951); (3.499432,0.957951) **[blue]@{-};
(3.379432,2.663195); (3.539432,2.743195) **[red]@{-};
(3.539432,2.663195); (3.379432,2.743195) **[red]@{-};
(3.419432,2.678246); (3.499432,2.718246) **[red]@{-};
(3.419432,2.718246); (3.499432,2.678246) **[red]@{-};
(3.419432,2.686240); (3.499432,2.726240) **[red]@{-};
(3.419432,2.726240); (3.499432,2.686240) **[red]@{-};
(3.620440,1.003604); (3.780440,1.003604) **[blue]@{-};
(3.700440,0.963604); (3.700440,1.043604) **[blue]@{-};
(3.660440,0.999640); (3.740440,0.999640) **[blue]@{-};
(3.660440,1.008486); (3.740440,1.008486) **[blue]@{-};
(3.620440,2.672933); (3.780440,2.752933) **[red]@{-};
(3.780440,2.672933); (3.620440,2.752933) **[red]@{-};
(3.660440,2.683318); (3.740440,2.723318) **[red]@{-};
(3.660440,2.723318); (3.740440,2.683318) **[red]@{-};
(3.660440,2.697337); (3.740440,2.737337) **[red]@{-};
(3.660440,2.737337); (3.740440,2.697337) **[red]@{-};
(4.007463,1.082528); (4.167463,1.082528) **[blue]@{-};
(4.087463,1.042528); (4.087463,1.122528) **[blue]@{-};
(4.047463,1.078735); (4.127463,1.078735) **[blue]@{-};
(4.047463,1.089690); (4.127463,1.089690) **[blue]@{-};
(4.007463,2.440088); (4.167463,2.520088) **[red]@{-};
(4.167463,2.440088); (4.007463,2.520088) **[red]@{-};
(4.047463,2.457381); (4.127463,2.497381) **[red]@{-};
(4.047463,2.497381); (4.127463,2.457381) **[red]@{-};
(4.047463,2.463298); (4.127463,2.503298) **[red]@{-};
(4.047463,2.503298); (4.127463,2.463298) **[red]@{-};
(4.167928,1.108834); (4.327928,1.108834) **[blue]@{-};
(4.247928,1.068834); (4.247928,1.148834) **[blue]@{-};
(4.207928,1.105819); (4.287928,1.105819) **[blue]@{-};
(4.207928,1.113047); (4.287928,1.113047) **[blue]@{-};
(4.167928,2.376788); (4.327928,2.456788) **[red]@{-};
(4.327928,2.376788); (4.167928,2.456788) **[red]@{-};
(4.207928,2.393005); (4.287928,2.433005) **[red]@{-};
(4.207928,2.433005); (4.287928,2.393005) **[red]@{-};
(4.207928,2.407829); (4.287928,2.447829) **[red]@{-};
(4.207928,2.447829); (4.287928,2.407829) **[red]@{-};
(4.443562,1.147991); (4.603562,1.147991) **[blue]@{-};
(4.523562,1.107991); (4.523562,1.187991) **[blue]@{-};
(4.483562,1.142279); (4.563562,1.142279) **[blue]@{-};
(4.483562,1.151592); (4.563562,1.151592) **[blue]@{-};
(4.443562,2.222861); (4.603562,2.302861) **[red]@{-};
(4.603562,2.222861); (4.443562,2.302861) **[red]@{-};
(4.483562,2.237518); (4.563562,2.277518) **[red]@{-};
(4.483562,2.277518); (4.563562,2.237518) **[red]@{-};
(4.483562,2.249236); (4.563562,2.289236) **[red]@{-};
(4.483562,2.289236); (4.563562,2.249236) **[red]@{-};
(4.777981,1.187630); (4.937981,1.187630) **[blue]@{-};
(4.857981,1.147630); (4.857981,1.227630) **[blue]@{-};
(4.817981,1.176656); (4.897981,1.176656) **[blue]@{-};
(4.817981,1.197980); (4.897981,1.197980) **[blue]@{-};
(4.777981,2.132671); (4.937981,2.212671) **[red]@{-};
(4.937981,2.132671); (4.777981,2.212671) **[red]@{-};
(4.817981,2.149449); (4.897981,2.189449) **[red]@{-};
(4.817981,2.189449); (4.897981,2.149449) **[red]@{-};
(4.817981,2.156571); (4.897981,2.196571) **[red]@{-};
(4.817981,2.196571); (4.897981,2.156571) **[red]@{-};
(4.874196,1.189285); (5.034196,1.189285) **[blue]@{-};
(4.954196,1.149285); (4.954196,1.229285) **[blue]@{-};
(4.914196,1.186633); (4.994196,1.186633) **[blue]@{-};
(4.914196,1.201931); (4.994196,1.201931) **[blue]@{-};
(4.874196,2.092414); (5.034196,2.172414) **[red]@{-};
(5.034196,2.092414); (4.874196,2.172414) **[red]@{-};
(4.914196,2.106281); (4.994196,2.146281) **[red]@{-};
(4.914196,2.146281); (4.994196,2.106281) **[red]@{-};
(4.914196,2.117041); (4.994196,2.157041) **[red]@{-};
(4.914196,2.157041); (4.994196,2.117041) **[red]@{-};
(5.129453,1.216070); (5.289453,1.216070) **[blue]@{-};
(5.209453,1.176070); (5.209453,1.256070) **[blue]@{-};
(5.169453,1.207401); (5.249453,1.207401) **[blue]@{-};
(5.169453,1.219588); (5.249453,1.219588) **[blue]@{-};
(5.129453,1.959339); (5.289453,2.039339) **[red]@{-};
(5.289453,1.959339); (5.129453,2.039339) **[red]@{-};
(5.169453,1.975542); (5.249453,2.015542) **[red]@{-};
(5.169453,2.015542); (5.249453,1.975542) **[red]@{-};
(5.169453,1.983480); (5.249453,2.023480) **[red]@{-};
(5.169453,2.023480); (5.249453,1.983480) **[red]@{-};
(5.277552,1.219814); (5.437552,1.219814) **[blue]@{-};
(5.357552,1.179814); (5.357552,1.259814) **[blue]@{-};
(5.317552,1.215905); (5.397552,1.215905) **[blue]@{-};
(5.317552,1.226313); (5.397552,1.226313) **[blue]@{-};
(5.277552,1.974461); (5.437552,2.054461) **[red]@{-};
(5.437552,1.974461); (5.277552,2.054461) **[red]@{-};
(5.317552,1.990198); (5.397552,2.030198) **[red]@{-};
(5.317552,2.030198); (5.397552,1.990198) **[red]@{-};
(5.317552,2.002487); (5.397552,2.042487) **[red]@{-};
(5.317552,2.042487); (5.397552,2.002487) **[red]@{-};
(5.346265,1.236806); (5.506265,1.236806) **[blue]@{-};
(5.426265,1.196806); (5.426265,1.276806) **[blue]@{-};
(5.386265,1.230254); (5.466265,1.230254) **[blue]@{-};
(5.386265,1.239990); (5.466265,1.239990) **[blue]@{-};
(5.346265,1.938174); (5.506265,2.018174) **[red]@{-};
(5.506265,1.938174); (5.346265,2.018174) **[red]@{-};
(5.386265,1.953550); (5.466265,1.993550) **[red]@{-};
(5.386265,1.993550); (5.466265,1.953550) **[red]@{-};
(5.386265,1.962886); (5.466265,2.002886) **[red]@{-};
(5.386265,2.002886); (5.466265,1.962886) **[red]@{-};
(5.474589,1.234941); (5.634589,1.234941) **[blue]@{-};
(5.554589,1.194941); (5.554589,1.274941) **[blue]@{-};
(5.514589,1.227661); (5.594589,1.227661) **[blue]@{-};
(5.514589,1.240889); (5.594589,1.240889) **[blue]@{-};
(5.474589,1.888532); (5.634589,1.968532) **[red]@{-};
(5.634589,1.888532); (5.474589,1.968532) **[red]@{-};
(5.514589,1.901508); (5.594589,1.941508) **[red]@{-};
(5.514589,1.941508); (5.594589,1.901508) **[red]@{-};
(5.514589,1.928584); (5.594589,1.968584) **[red]@{-};
(5.514589,1.968584); (5.594589,1.928584) **[red]@{-};
(5.647920,1.242199); (5.807920,1.242199) **[blue]@{-};
(5.727920,1.202199); (5.727920,1.282199) **[blue]@{-};
(5.687920,1.239473); (5.767920,1.239473) **[blue]@{-};
(5.687920,1.246443); (5.767920,1.246443) **[blue]@{-};
(5.647920,1.829328); (5.807920,1.909328) **[red]@{-};
(5.807920,1.829328); (5.647920,1.909328) **[red]@{-};
(5.687920,1.845979); (5.767920,1.885979) **[red]@{-};
(5.687920,1.885979); (5.767920,1.845979) **[red]@{-};
(5.687920,1.861639); (5.767920,1.901639) **[red]@{-};
(5.687920,1.901639); (5.767920,1.861639) **[red]@{-};
(5.802643,1.252214); (5.962643,1.252214) **[blue]@{-};
(5.882643,1.212214); (5.882643,1.292214) **[blue]@{-};
(5.842643,1.249198); (5.922643,1.249198) **[blue]@{-};
(5.842643,1.255248); (5.922643,1.255248) **[blue]@{-};
(5.802643,1.762720); (5.962643,1.842720) **[red]@{-};
(5.962643,1.762720); (5.802643,1.842720) **[red]@{-};
(5.842643,1.779601); (5.922643,1.819601) **[red]@{-};
(5.842643,1.819601); (5.922643,1.779601) **[red]@{-};
(5.842643,1.788057); (5.922643,1.828057) **[red]@{-};
(5.842643,1.828057); (5.922643,1.788057) **[red]@{-};
(5.850737,1.250030); (6.010737,1.250030) **[blue]@{-};
(5.930737,1.210030); (5.930737,1.290030) **[blue]@{-};
(5.890737,1.248874); (5.970737,1.248874) **[blue]@{-};
(5.890737,1.253942); (5.970737,1.253942) **[blue]@{-};
(5.850737,1.644315); (6.010737,1.724315) **[red]@{-};
(6.010737,1.644315); (5.850737,1.724315) **[red]@{-};
(5.890737,1.659711); (5.970737,1.699711) **[red]@{-};
(5.890737,1.699711); (5.970737,1.659711) **[red]@{-};
(5.890737,1.668422); (5.970737,1.708422) **[red]@{-};
(5.890737,1.708422); (5.970737,1.668422) **[red]@{-};
(5.986089,1.261263); (6.146089,1.261263) **[blue]@{-};
(6.066089,1.221263); (6.066089,1.301263) **[blue]@{-};
(6.026089,1.258437); (6.106089,1.258437) **[blue]@{-};
(6.026089,1.266003); (6.106089,1.266003) **[blue]@{-};
(5.986089,1.618083); (6.146089,1.698083) **[red]@{-};
(6.146089,1.618083); (5.986089,1.698083) **[red]@{-};
(6.026089,1.636020); (6.106089,1.676020) **[red]@{-};
(6.026089,1.676020); (6.106089,1.636020) **[red]@{-};
(6.026089,1.645180); (6.106089,1.685180) **[red]@{-};
(6.026089,1.685180); (6.106089,1.645180) **[red]@{-};
(6.069747,1.261426); (6.229747,1.261426) **[blue]@{-};
(6.149747,1.221426); (6.149747,1.301426) **[blue]@{-};
(6.109747,1.259165); (6.189747,1.259165) **[blue]@{-};
(6.109747,1.265610); (6.189747,1.265610) **[blue]@{-};
(6.069747,1.588347); (6.229747,1.668347) **[red]@{-};
(6.229747,1.588347); (6.069747,1.668347) **[red]@{-};
(6.109747,1.602972); (6.189747,1.642972) **[red]@{-};
(6.109747,1.642972); (6.189747,1.602972) **[red]@{-};
(6.109747,1.614941); (6.189747,1.654941) **[red]@{-};
(6.109747,1.654941); (6.189747,1.614941) **[red]@{-};
(6.109825,1.267158); (6.269825,1.267158) **[blue]@{-};
(6.189825,1.227158); (6.189825,1.307158) **[blue]@{-};
(6.149825,1.263269); (6.229825,1.263269) **[blue]@{-};
(6.149825,1.271714); (6.229825,1.271714) **[blue]@{-};
(6.109825,1.506773); (6.269825,1.586773) **[red]@{-};
(6.269825,1.506773); (6.109825,1.586773) **[red]@{-};
(6.149825,1.524804); (6.229825,1.564804) **[red]@{-};
(6.149825,1.564804); (6.229825,1.524804) **[red]@{-};
(6.149825,1.534327); (6.229825,1.574327) **[red]@{-};
(6.149825,1.574327); (6.229825,1.534327) **[red]@{-};
(6.223781,1.269983); (6.383781,1.269983) **[blue]@{-};
(6.303781,1.229983); (6.303781,1.309983) **[blue]@{-};
(6.263781,1.267665); (6.343781,1.267665) **[blue]@{-};
(6.263781,1.273912); (6.343781,1.273912) **[blue]@{-};
(6.223781,1.481178); (6.383781,1.561178) **[red]@{-};
(6.383781,1.481178); (6.223781,1.561178) **[red]@{-};
(6.263781,1.495184); (6.343781,1.535184) **[red]@{-};
(6.263781,1.535184); (6.343781,1.495184) **[red]@{-};
(6.263781,1.509091); (6.343781,1.549091) **[red]@{-};
(6.263781,1.549091); (6.343781,1.509091) **[red]@{-};
(6.295039,1.271226); (6.455039,1.271226) **[blue]@{-};
(6.375039,1.231226); (6.375039,1.311226) **[blue]@{-};
(6.335039,1.267461); (6.415039,1.267461) **[blue]@{-};
(6.335039,1.283804); (6.415039,1.283804) **[blue]@{-};
(6.295039,1.477559); (6.455039,1.557559) **[red]@{-};
(6.455039,1.477559); (6.295039,1.557559) **[red]@{-};
(6.335039,1.488882); (6.415039,1.528882) **[red]@{-};
(6.335039,1.528882); (6.415039,1.488882) **[red]@{-};
(6.335039,1.504156); (6.415039,1.544156) **[red]@{-};
(6.335039,1.544156); (6.415039,1.504156) **[red]@{-};
(6.395733,1.274757); (6.555733,1.274757) **[blue]@{-};
(6.475733,1.234757); (6.475733,1.314757) **[blue]@{-};
(6.435733,1.273111); (6.515733,1.273111) **[blue]@{-};
(6.435733,1.277897); (6.515733,1.277897) **[blue]@{-};
(6.395733,1.355232); (6.555733,1.435232) **[red]@{-};
(6.555733,1.355232); (6.395733,1.435232) **[red]@{-};
(6.435733,1.369574); (6.515733,1.409574) **[red]@{-};
(6.435733,1.409574); (6.515733,1.369574) **[red]@{-};
(6.435733,1.383110); (6.515733,1.423110) **[red]@{-};
(6.435733,1.423110); (6.515733,1.383110) **[red]@{-};
(6.519913,1.279834); (6.679913,1.279834) **[blue]@{-};
(6.599913,1.239834); (6.599913,1.319834) **[blue]@{-};
(6.559913,1.275885); (6.639913,1.275885) **[blue]@{-};
(6.559913,1.282288); (6.639913,1.282288) **[blue]@{-};
(6.519913,1.339225); (6.679913,1.419225) **[red]@{-};
(6.679913,1.339225); (6.519913,1.419225) **[red]@{-};
(6.559913,1.357691); (6.639913,1.397691) **[red]@{-};
(6.559913,1.397691); (6.639913,1.357691) **[red]@{-};
(6.559913,1.365001); (6.639913,1.405001) **[red]@{-};
(6.559913,1.405001); (6.639913,1.365001) **[red]@{-};
(6.578211,1.276373); (6.738211,1.276373) **[blue]@{-};
(6.658211,1.236373); (6.658211,1.316373) **[blue]@{-};
(6.618211,1.270318); (6.698211,1.270318) **[blue]@{-};
(6.618211,1.280036); (6.698211,1.280036) **[blue]@{-};
(6.578211,1.332117); (6.738211,1.412117) **[red]@{-};
(6.738211,1.332117); (6.578211,1.412117) **[red]@{-};
(6.618211,1.343343); (6.698211,1.383343) **[red]@{-};
(6.618211,1.383343); (6.698211,1.343343) **[red]@{-};
(6.618211,1.359104); (6.698211,1.399104) **[red]@{-};
(6.618211,1.399104); (6.698211,1.359104) **[red]@{-};
(6.606501,1.286822); (6.766501,1.286822) **[blue]@{-};
(6.686501,1.246822); (6.686501,1.326822) **[blue]@{-};
(6.646501,1.280387); (6.726501,1.280387) **[blue]@{-};
(6.646501,1.292753); (6.726501,1.292753) **[blue]@{-};
(6.606501,1.329536); (6.766501,1.409536) **[red]@{-};
(6.766501,1.329536); (6.606501,1.409536) **[red]@{-};
(6.646501,1.342147); (6.726501,1.382147) **[red]@{-};
(6.646501,1.382147); (6.726501,1.342147) **[red]@{-};
(6.646501,1.354111); (6.726501,1.394111) **[red]@{-};
(6.646501,1.394111); (6.726501,1.354111) **[red]@{-};
(6.661467,1.281426); (6.821467,1.281426) **[blue]@{-};
(6.741467,1.241426); (6.741467,1.321426) **[blue]@{-};
(6.701467,1.278806); (6.781467,1.278806) **[blue]@{-};
(6.701467,1.288704); (6.781467,1.288704) **[blue]@{-};
(6.661467,1.313220); (6.821467,1.393220) **[red]@{-};
(6.821467,1.313220); (6.661467,1.393220) **[red]@{-};
(6.701467,1.330674); (6.781467,1.370674) **[red]@{-};
(6.701467,1.370674); (6.781467,1.330674) **[red]@{-};
(6.701467,1.336258); (6.781467,1.376258) **[red]@{-};
(6.701467,1.376258); (6.781467,1.336258) **[red]@{-};
(6.688184,1.282012); (6.848184,1.282012) **[blue]@{-};
(6.768184,1.242012); (6.768184,1.322012) **[blue]@{-};
(6.728184,1.277341); (6.808184,1.277341) **[blue]@{-};
(6.728184,1.284757); (6.808184,1.284757) **[blue]@{-};
(6.688184,1.310409); (6.848184,1.390409) **[red]@{-};
(6.848184,1.310409); (6.688184,1.390409) **[red]@{-};
(6.728184,1.327366); (6.808184,1.367366) **[red]@{-};
(6.728184,1.367366); (6.808184,1.327366) **[red]@{-};
(6.728184,1.335110); (6.808184,1.375110) **[red]@{-};
(6.728184,1.375110); (6.808184,1.335110) **[red]@{-};
(6.740179,1.283977); (6.900179,1.283977) **[blue]@{-};
(6.820179,1.243977); (6.820179,1.323977) **[blue]@{-};
(6.780179,1.277871); (6.860179,1.277871) **[blue]@{-};
(6.780179,1.287896); (6.860179,1.287896) **[blue]@{-};
(6.740179,1.193085); (6.900179,1.273085) **[red]@{-};
(6.900179,1.193085); (6.740179,1.273085) **[red]@{-};
(6.780179,1.209676); (6.860179,1.249676) **[red]@{-};
(6.780179,1.249676); (6.860179,1.209676) **[red]@{-};
(6.780179,1.220934); (6.860179,1.260934) **[red]@{-};
(6.780179,1.260934); (6.860179,1.220934) **[red]@{-};
(6.908685,1.285895); (7.068685,1.285895) **[blue]@{-};
(6.988685,1.245895); (6.988685,1.325895) **[blue]@{-};
(6.948685,1.283915); (7.028685,1.283915) **[blue]@{-};
(6.948685,1.290332); (7.028685,1.290332) **[blue]@{-};
(6.908685,1.200551); (7.068685,1.280551) **[red]@{-};
(7.068685,1.200551); (6.908685,1.280551) **[red]@{-};
(6.948685,1.217322); (7.028685,1.257322) **[red]@{-};
(6.948685,1.257322); (7.028685,1.217322) **[red]@{-};
(6.948685,1.227452); (7.028685,1.267452) **[red]@{-};
(6.948685,1.267452); (7.028685,1.227452) **[red]@{-};
(6.953423,1.288334); (7.113423,1.288334) **[blue]@{-};
(7.033423,1.248334); (7.033423,1.328334) **[blue]@{-};
(6.993423,1.286210); (7.073423,1.286210) **[blue]@{-};
(6.993423,1.289653); (7.073423,1.289653) **[blue]@{-};
(6.953423,0.984194); (7.113423,1.064194) **[red]@{-};
(7.113423,0.984194); (6.953423,1.064194) **[red]@{-};
(6.993423,0.998656); (7.073423,1.038656) **[red]@{-};
(6.993423,1.038656); (7.073423,0.998656) **[red]@{-};
(6.993423,1.010410); (7.073423,1.050410) **[red]@{-};
(6.993423,1.050410); (7.073423,1.010410) **[red]@{-};
(7.018032,1.288688); (7.178032,1.288688) **[blue]@{-};
(7.098032,1.248688); (7.098032,1.328688) **[blue]@{-};
(7.058032,1.282512); (7.138032,1.282512) **[blue]@{-};
(7.058032,1.290987); (7.138032,1.290987) **[blue]@{-};
(7.018032,0.996747); (7.178032,1.076747) **[red]@{-};
(7.178032,0.996747); (7.018032,1.076747) **[red]@{-};
(7.058032,1.010191); (7.138032,1.050191) **[red]@{-};
(7.058032,1.050191); (7.138032,1.010191) **[red]@{-};
(7.058032,1.024563); (7.138032,1.064563) **[red]@{-};
(7.058032,1.064563); (7.138032,1.024563) **[red]@{-};
(7.038941,1.290848); (7.198941,1.290848) **[blue]@{-};
(7.118941,1.250848); (7.118941,1.330848) **[blue]@{-};
(7.078941,1.287903); (7.158941,1.287903) **[blue]@{-};
(7.078941,1.297316); (7.158941,1.297316) **[blue]@{-};
(7.038941,0.994527); (7.198941,1.074527) **[red]@{-};
(7.198941,0.994527); (7.038941,1.074527) **[red]@{-};
(7.078941,1.012832); (7.158941,1.052832) **[red]@{-};
(7.078941,1.052832); (7.158941,1.012832) **[red]@{-};
(7.078941,1.019705); (7.158941,1.059705) **[red]@{-};
(7.078941,1.059705); (7.158941,1.019705) **[red]@{-};
(7.139169,1.292515); (7.299169,1.292515) **[blue]@{-};
(7.219169,1.252515); (7.219169,1.332515) **[blue]@{-};
(7.179169,1.290906); (7.259169,1.290906) **[blue]@{-};
(7.179169,1.294117); (7.259169,1.294117) **[blue]@{-};
(7.139169,0.946994); (7.299169,1.026994) **[red]@{-};
(7.299169,0.946994); (7.139169,1.026994) **[red]@{-};
(7.179169,0.963708); (7.259169,1.003708) **[red]@{-};
(7.179169,1.003708); (7.259169,0.963708) **[red]@{-};
(7.179169,0.974220); (7.259169,1.014220) **[red]@{-};
(7.179169,1.014220); (7.259169,0.974220) **[red]@{-};
(7.158405,1.288731); (7.318405,1.288731) **[blue]@{-};
(7.238405,1.248731); (7.238405,1.328731) **[blue]@{-};
(7.198405,1.284684); (7.278405,1.284684) **[blue]@{-};
(7.198405,1.291667); (7.278405,1.291667) **[blue]@{-};
(7.158405,0.947235); (7.318405,1.027235) **[red]@{-};
(7.318405,0.947235); (7.158405,1.027235) **[red]@{-};
(7.198405,0.963886); (7.278405,1.003886) **[red]@{-};
(7.198405,1.003886); (7.278405,0.963886) **[red]@{-};
(7.198405,0.976316); (7.278405,1.016316) **[red]@{-};
(7.198405,1.016316); (7.278405,0.976316) **[red]@{-};
(7.214621,1.291428); (7.374621,1.291428) **[blue]@{-};
(7.294621,1.251428); (7.294621,1.331428) **[blue]@{-};
(7.254621,1.288180); (7.334621,1.288180) **[blue]@{-};
(7.254621,1.296041); (7.334621,1.296041) **[blue]@{-};
(7.214621,0.963258); (7.374621,1.043258) **[red]@{-};
(7.374621,0.963258); (7.214621,1.043258) **[red]@{-};
(7.254621,0.979923); (7.334621,1.019923) **[red]@{-};
(7.254621,1.019923); (7.334621,0.979923) **[red]@{-};
(7.254621,0.989071); (7.334621,1.029071) **[red]@{-};
(7.254621,1.029071); (7.334621,0.989071) **[red]@{-};
(7.268728,1.289600); (7.428728,1.289600) **[blue]@{-};
(7.348728,1.249600); (7.348728,1.329600) **[blue]@{-};
(7.308728,1.287820); (7.388728,1.287820) **[blue]@{-};
(7.308728,1.291370); (7.388728,1.291370) **[blue]@{-};
(7.268728,0.961426); (7.428728,1.041426) **[red]@{-};
(7.428728,0.961426); (7.268728,1.041426) **[red]@{-};
(7.308728,0.979242); (7.388728,1.019242) **[red]@{-};
(7.308728,1.019242); (7.388728,0.979242) **[red]@{-};
(7.308728,0.987673); (7.388728,1.027673) **[red]@{-};
(7.308728,1.027673); (7.388728,0.987673) **[red]@{-};
(7.303704,1.290319); (7.463704,1.290319) **[blue]@{-};
(7.383704,1.250319); (7.383704,1.330319) **[blue]@{-};
(7.343704,1.288108); (7.423704,1.288108) **[blue]@{-};
(7.343704,1.295006); (7.423704,1.295006) **[blue]@{-};
(7.303704,0.964174); (7.463704,1.044174) **[red]@{-};
(7.463704,0.964174); (7.303704,1.044174) **[red]@{-};
(7.343704,0.979682); (7.423704,1.019682) **[red]@{-};
(7.343704,1.019682); (7.423704,0.979682) **[red]@{-};
(7.343704,0.994650); (7.423704,1.034650) **[red]@{-};
(7.343704,1.034650); (7.423704,0.994650) **[red]@{-};
(7.354628,1.299867); (7.514628,1.299867) **[blue]@{-};
(7.434628,1.259867); (7.434628,1.339867) **[blue]@{-};
(7.394628,1.295332); (7.474628,1.295332) **[blue]@{-};
(7.394628,1.301376); (7.474628,1.301376) **[blue]@{-};
(7.354628,0.988594); (7.514628,1.068594) **[red]@{-};
(7.514628,0.988594); (7.354628,1.068594) **[red]@{-};
(7.394628,1.005301); (7.474628,1.045301) **[red]@{-};
(7.394628,1.045301); (7.474628,1.005301) **[red]@{-};
(7.394628,1.017594); (7.474628,1.057594) **[red]@{-};
(7.394628,1.057594); (7.474628,1.017594) **[red]@{-};
(7.403816,1.292542); (7.563816,1.292542) **[blue]@{-};
(7.483816,1.252542); (7.483816,1.332542) **[blue]@{-};
(7.443816,1.290478); (7.523816,1.290478) **[blue]@{-};
(7.443816,1.296779); (7.523816,1.296779) **[blue]@{-};
(7.403816,0.995784); (7.563816,1.075784) **[red]@{-};
(7.563816,0.995784); (7.403816,1.075784) **[red]@{-};
(7.443816,1.011487); (7.523816,1.051487) **[red]@{-};
(7.443816,1.051487); (7.523816,1.011487) **[red]@{-};
(7.443816,1.020318); (7.523816,1.060318) **[red]@{-};
(7.443816,1.060318); (7.523816,1.020318) **[red]@{-};
(7.419846,1.296830); (7.579846,1.296830) **[blue]@{-};
(7.499846,1.256830); (7.499846,1.336830) **[blue]@{-};
(7.459846,1.291987); (7.539846,1.291987) **[blue]@{-};
(7.459846,1.298968); (7.539846,1.298968) **[blue]@{-};
(7.419846,0.852187); (7.579846,0.932187) **[red]@{-};
(7.579846,0.852187); (7.419846,0.932187) **[red]@{-};
(7.459846,0.869246); (7.539846,0.909246) **[red]@{-};
(7.459846,0.909246); (7.539846,0.869246) **[red]@{-};
(7.459846,0.877673); (7.539846,0.917673) **[red]@{-};
(7.459846,0.917673); (7.539846,0.877673) **[red]@{-};
(7.497429,1.297338); (7.657429,1.297338) **[blue]@{-};
(7.577429,1.257338); (7.577429,1.337338) **[blue]@{-};
(7.537429,1.295048); (7.617429,1.295048) **[blue]@{-};
(7.537429,1.301374); (7.617429,1.301374) **[blue]@{-};
(7.497429,0.883841); (7.657429,0.963841) **[red]@{-};
(7.657429,0.883841); (7.497429,0.963841) **[red]@{-};
(7.537429,0.899053); (7.617429,0.939053) **[red]@{-};
(7.537429,0.939053); (7.617429,0.899053) **[red]@{-};
(7.537429,0.915449); (7.617429,0.955449) **[red]@{-};
(7.537429,0.955449); (7.617429,0.915449) **[red]@{-};
(7.512457,1.296247); (7.672457,1.296247) **[blue]@{-};
(7.592457,1.256247); (7.592457,1.336247) **[blue]@{-};
(7.552457,1.294575); (7.632457,1.294575) **[blue]@{-};
(7.552457,1.302632); (7.632457,1.302632) **[blue]@{-};
(7.512457,0.883686); (7.672457,0.963686) **[red]@{-};
(7.672457,0.883686); (7.512457,0.963686) **[red]@{-};
(7.552457,0.900157); (7.632457,0.940157) **[red]@{-};
(7.552457,0.940157); (7.632457,0.900157) **[red]@{-};
(7.552457,0.913085); (7.632457,0.953085) **[red]@{-};
(7.552457,0.953085); (7.632457,0.913085) **[red]@{-};
(7.542052,1.298424); (7.702052,1.298424) **[blue]@{-};
(7.622052,1.258424); (7.622052,1.338424) **[blue]@{-};
(7.582052,1.295117); (7.662052,1.295117) **[blue]@{-};
(7.582052,1.302195); (7.662052,1.302195) **[blue]@{-};
(7.542052,0.892194); (7.702052,0.972194) **[red]@{-};
(7.702052,0.892194); (7.542052,0.972194) **[red]@{-};
(7.582052,0.907701); (7.662052,0.947701) **[red]@{-};
(7.582052,0.947701); (7.662052,0.907701) **[red]@{-};
(7.582052,0.916985); (7.662052,0.956985) **[red]@{-};
(7.582052,0.956985); (7.662052,0.916985) **[red]@{-};
(7.556625,1.296303); (7.716625,1.296303) **[blue]@{-};
(7.636625,1.256303); (7.636625,1.336303) **[blue]@{-};
(7.596625,1.293933); (7.676625,1.293933) **[blue]@{-};
(7.596625,1.299301); (7.676625,1.299301) **[blue]@{-};
(7.556625,0.890475); (7.716625,0.970475) **[red]@{-};
(7.716625,0.890475); (7.556625,0.970475) **[red]@{-};
(7.596625,0.906541); (7.676625,0.946541) **[red]@{-};
(7.596625,0.946541); (7.676625,0.906541) **[red]@{-};
(7.596625,0.915796); (7.676625,0.955796) **[red]@{-};
(7.596625,0.955796); (7.676625,0.915796) **[red]@{-};
(7.641099,1.297456); (7.801099,1.297456) **[blue]@{-};
(7.721099,1.257456); (7.721099,1.337456) **[blue]@{-};
(7.681099,1.293129); (7.761099,1.293129) **[blue]@{-};
(7.681099,1.300816); (7.761099,1.300816) **[blue]@{-};
(7.641099,0.689957); (7.801099,0.769957) **[red]@{-};
(7.801099,0.689957); (7.641099,0.769957) **[red]@{-};
(7.681099,0.706502); (7.761099,0.746502) **[red]@{-};
(7.681099,0.746502); (7.761099,0.706502) **[red]@{-};
(7.681099,0.716978); (7.761099,0.756978) **[red]@{-};
(7.681099,0.756978); (7.761099,0.716978) **[red]@{-};
(7.720900,1.301533); (7.880900,1.301533) **[blue]@{-};
(7.800900,1.261533); (7.800900,1.341533) **[blue]@{-};
(7.760900,1.298504); (7.840900,1.298504) **[blue]@{-};
(7.760900,1.304840); (7.840900,1.304840) **[blue]@{-};
(7.720900,0.606666); (7.880900,0.686666) **[red]@{-};
(7.880900,0.606666); (7.720900,0.686666) **[red]@{-};
(7.760900,0.621195); (7.840900,0.661195) **[red]@{-};
(7.760900,0.661195); (7.840900,0.621195) **[red]@{-};
(7.760900,0.632564); (7.840900,0.672564) **[red]@{-};
(7.760900,0.672564); (7.840900,0.632564) **[red]@{-};
(7.746548,1.294155); (7.906548,1.294155) **[blue]@{-};
(7.826548,1.254155); (7.826548,1.334155) **[blue]@{-};
(7.786548,1.292605); (7.866548,1.292605) **[blue]@{-};
(7.786548,1.300206); (7.866548,1.300206) **[blue]@{-};
(7.746548,0.619505); (7.906548,0.699505) **[red]@{-};
(7.906548,0.619505); (7.746548,0.699505) **[red]@{-};
(7.786548,0.636973); (7.866548,0.676973) **[red]@{-};
(7.786548,0.676973); (7.866548,0.636973) **[red]@{-};
(7.786548,0.645215); (7.866548,0.685215) **[red]@{-};
(7.786548,0.685215); (7.866548,0.645215) **[red]@{-};
(7.759204,1.295054); (7.919204,1.295054) **[blue]@{-};
(7.839204,1.255054); (7.839204,1.335054) **[blue]@{-};
(7.799204,1.291356); (7.879204,1.291356) **[blue]@{-};
(7.799204,1.299238); (7.879204,1.299238) **[blue]@{-};
(7.759204,0.619627); (7.919204,0.699627) **[red]@{-};
(7.919204,0.619627); (7.759204,0.699627) **[red]@{-};
(7.799204,0.638000); (7.879204,0.678000) **[red]@{-};
(7.799204,0.678000); (7.879204,0.638000) **[red]@{-};
(7.799204,0.647892); (7.879204,0.687892) **[red]@{-};
(7.799204,0.687892); (7.879204,0.647892) **[red]@{-};
(7.784186,1.303287); (7.944186,1.303287) **[blue]@{-};
(7.864186,1.263287); (7.864186,1.343287) **[blue]@{-};
(7.824186,1.298698); (7.904186,1.298698) **[blue]@{-};
(7.824186,1.304445); (7.904186,1.304445) **[blue]@{-};
(7.784186,0.627206); (7.944186,0.707206) **[red]@{-};
(7.944186,0.627206); (7.784186,0.707206) **[red]@{-};
(7.824186,0.645244); (7.904186,0.685244) **[red]@{-};
(7.824186,0.685244); (7.904186,0.645244) **[red]@{-};
(7.824186,0.654450); (7.904186,0.694450) **[red]@{-};
(7.824186,0.694450); (7.904186,0.654450) **[red]@{-};
(7.820867,1.300089); (7.980867,1.300089) **[blue]@{-};
(7.900867,1.260089); (7.900867,1.340089) **[blue]@{-};
(7.860867,1.295594); (7.940867,1.295594) **[blue]@{-};
(7.860867,1.307194); (7.940867,1.307194) **[blue]@{-};
(7.820867,0.644023); (7.980867,0.724023) **[red]@{-};
(7.980867,0.644023); (7.820867,0.724023) **[red]@{-};
(7.860867,0.659801); (7.940867,0.699801) **[red]@{-};
(7.860867,0.699801); (7.940867,0.659801) **[red]@{-};
(7.860867,0.673787); (7.940867,0.713787) **[red]@{-};
(7.860867,0.713787); (7.940867,0.673787) **[red]@{-};
(7.832889,1.300346); (7.992889,1.300346) **[blue]@{-};
(7.912889,1.260346); (7.912889,1.340346) **[blue]@{-};
(7.872889,1.296046); (7.952889,1.296046) **[blue]@{-};
(7.872889,1.302287); (7.952889,1.302287) **[blue]@{-};
(7.832889,0.644430); (7.992889,0.724430) **[red]@{-};
(7.992889,0.644430); (7.832889,0.724430) **[red]@{-};
(7.872889,0.661823); (7.952889,0.701823) **[red]@{-};
(7.872889,0.701823); (7.952889,0.661823) **[red]@{-};
(7.872889,0.667709); (7.952889,0.707709) **[red]@{-};
(7.872889,0.707709); (7.952889,0.667709) **[red]@{-};
(7.891544,1.303837); (8.051544,1.303837) **[blue]@{-};
(7.971544,1.263837); (7.971544,1.343837) **[blue]@{-};
(7.931544,1.300974); (8.011544,1.300974) **[blue]@{-};
(7.931544,1.304831); (8.011544,1.304831) **[blue]@{-};
(7.891544,0.684523); (8.051544,0.764523) **[red]@{-};
(8.051544,0.684523); (7.891544,0.764523) **[red]@{-};
(7.931544,0.703430); (8.011544,0.743430) **[red]@{-};
(7.931544,0.743430); (8.011544,0.703430) **[red]@{-};
(7.931544,0.710408); (8.011544,0.750408) **[red]@{-};
(7.931544,0.750408); (8.011544,0.710408) **[red]@{-};
(7.925625,1.301512); (8.085625,1.301512) **[blue]@{-};
(8.005625,1.261512); (8.005625,1.341512) **[blue]@{-};
(7.965625,1.296726); (8.045625,1.296726) **[blue]@{-};
(7.965625,1.302556); (8.045625,1.302556) **[blue]@{-};
(7.925625,0.688112); (8.085625,0.768112) **[red]@{-};
(8.085625,0.688112); (7.925625,0.768112) **[red]@{-};
(7.965625,0.706454); (8.045625,0.746454) **[red]@{-};
(7.965625,0.746454); (8.045625,0.706454) **[red]@{-};
(7.965625,0.712250); (8.045625,0.752250) **[red]@{-};
(7.965625,0.752250); (8.045625,0.712250) **[red]@{-};
(7.958919,1.301985); (8.118919,1.301985) **[blue]@{-};
(8.038919,1.261985); (8.038919,1.341985) **[blue]@{-};
(7.998919,1.299436); (8.078919,1.299436) **[blue]@{-};
(7.998919,1.303637); (8.078919,1.303637) **[blue]@{-};
(7.958919,0.707563); (8.118919,0.787563) **[red]@{-};
(8.118919,0.707563); (7.958919,0.787563) **[red]@{-};
(7.998919,0.725916); (8.078919,0.765916) **[red]@{-};
(7.998919,0.765916); (8.078919,0.725916) **[red]@{-};
(7.998919,0.730347); (8.078919,0.770347) **[red]@{-};
(7.998919,0.770347); (8.078919,0.730347) **[red]@{-};
(7.991462,1.302223); (8.151462,1.302223) **[blue]@{-};
(8.071462,1.262223); (8.071462,1.342223) **[blue]@{-};
(8.031462,1.299288); (8.111462,1.299288) **[blue]@{-};
(8.031462,1.307448); (8.111462,1.307448) **[blue]@{-};
(7.991462,0.577829); (8.151462,0.657829) **[red]@{-};
(8.151462,0.577829); (7.991462,0.657829) **[red]@{-};
(8.031462,0.596040); (8.111462,0.636040) **[red]@{-};
(8.031462,0.636040); (8.111462,0.596040) **[red]@{-};
(8.031462,0.606229); (8.111462,0.646229) **[red]@{-};
(8.031462,0.646229); (8.111462,0.606229) **[red]@{-};
(8.002149,1.300668); (8.162149,1.300668) **[blue]@{-};
(8.082149,1.260668); (8.082149,1.340668) **[blue]@{-};
(8.042149,1.297929); (8.122149,1.297929) **[blue]@{-};
(8.042149,1.302696); (8.122149,1.302696) **[blue]@{-};
(8.002149,0.585689); (8.162149,0.665689) **[red]@{-};
(8.162149,0.585689); (8.002149,0.665689) **[red]@{-};
(8.042149,0.602353); (8.122149,0.642353) **[red]@{-};
(8.042149,0.642353); (8.122149,0.602353) **[red]@{-};
(8.042149,0.609501); (8.122149,0.649501) **[red]@{-};
(8.042149,0.649501); (8.122149,0.609501) **[red]@{-};
(8.033742,1.304720); (8.193742,1.304720) **[blue]@{-};
(8.113742,1.264720); (8.113742,1.344720) **[blue]@{-};
(8.073742,1.300919); (8.153742,1.300919) **[blue]@{-};
(8.073742,1.307533); (8.153742,1.307533) **[blue]@{-};
(8.033742,0.598774); (8.193742,0.678774) **[red]@{-};
(8.193742,0.598774); (8.033742,0.678774) **[red]@{-};
(8.073742,0.614493); (8.153742,0.654493) **[red]@{-};
(8.073742,0.654493); (8.153742,0.614493) **[red]@{-};
(8.073742,0.622651); (8.153742,0.662651) **[red]@{-};
(8.073742,0.662651); (8.153742,0.622651) **[red]@{-};
(8.054426,1.303803); (8.214426,1.303803) **[blue]@{-};
(8.134426,1.263803); (8.134426,1.343803) **[blue]@{-};
(8.094426,1.298920); (8.174426,1.298920) **[blue]@{-};
(8.094426,1.306231); (8.174426,1.306231) **[blue]@{-};
(8.054426,0.606670); (8.214426,0.686670) **[red]@{-};
(8.214426,0.606670); (8.054426,0.686670) **[red]@{-};
(8.094426,0.623653); (8.174426,0.663653) **[red]@{-};
(8.094426,0.663653); (8.174426,0.623653) **[red]@{-};
(8.094426,0.628730); (8.174426,0.668730) **[red]@{-};
(8.094426,0.668730); (8.174426,0.628730) **[red]@{-};
(8.064658,1.302944); (8.224658,1.302944) **[blue]@{-};
(8.144658,1.262944); (8.144658,1.342944) **[blue]@{-};
(8.104658,1.297886); (8.184658,1.297886) **[blue]@{-};
(8.104658,1.304286); (8.184658,1.304286) **[blue]@{-};
(8.064658,0.618709); (8.224658,0.698709) **[red]@{-};
(8.224658,0.618709); (8.064658,0.698709) **[red]@{-};
(8.104658,0.635470); (8.184658,0.675470) **[red]@{-};
(8.104658,0.675470); (8.184658,0.635470) **[red]@{-};
(8.104658,0.643011); (8.184658,0.683011) **[red]@{-};
(8.104658,0.683011); (8.184658,0.643011) **[red]@{-};
(8.114757,1.301136); (8.274757,1.301136) **[blue]@{-};
(8.194757,1.261136); (8.194757,1.341136) **[blue]@{-};
(8.154757,1.299078); (8.234757,1.299078) **[blue]@{-};
(8.154757,1.303368); (8.234757,1.303368) **[blue]@{-};
(8.114757,0.417506); (8.274757,0.497506) **[red]@{-};
(8.274757,0.417506); (8.114757,0.497506) **[red]@{-};
(8.154757,0.434741); (8.234757,0.474741) **[red]@{-};
(8.154757,0.474741); (8.234757,0.434741) **[red]@{-};
(8.154757,0.442959); (8.234757,0.482959) **[red]@{-};
(8.154757,0.482959); (8.234757,0.442959) **[red]@{-};
(8.182095,1.299559); (8.342095,1.299559) **[blue]@{-};
(8.262095,1.259559); (8.262095,1.339559) **[blue]@{-};
(8.222095,1.296044); (8.302095,1.296044) **[blue]@{-};
(8.222095,1.303780); (8.302095,1.303780) **[blue]@{-};
(8.182095,0.468968); (8.342095,0.548968) **[red]@{-};
(8.342095,0.468968); (8.182095,0.548968) **[red]@{-};
(8.222095,0.486357); (8.302095,0.526357) **[red]@{-};
(8.222095,0.526357); (8.302095,0.486357) **[red]@{-};
(8.222095,0.491380); (8.302095,0.531380) **[red]@{-};
(8.222095,0.531380); (8.302095,0.491380) **[red]@{-};
(8.200771,1.300962); (8.360771,1.300962) **[blue]@{-};
(8.280771,1.260962); (8.280771,1.340962) **[blue]@{-};
(8.240771,1.295853); (8.320771,1.295853) **[blue]@{-};
(8.240771,1.305247); (8.320771,1.305247) **[blue]@{-};
(8.200771,0.476406); (8.360771,0.556406) **[red]@{-};
(8.360771,0.476406); (8.200771,0.556406) **[red]@{-};
(8.240771,0.493863); (8.320771,0.533863) **[red]@{-};
(8.240771,0.533863); (8.320771,0.493863) **[red]@{-};
(8.240771,0.499486); (8.320771,0.539486) **[red]@{-};
(8.240771,0.539486); (8.320771,0.499486) **[red]@{-};
(8.210019,1.302809); (8.370019,1.302809) **[blue]@{-};
(8.290019,1.262809); (8.290019,1.342809) **[blue]@{-};
(8.250019,1.298499); (8.330019,1.298499) **[blue]@{-};
(8.250019,1.304716); (8.330019,1.304716) **[blue]@{-};
(8.210019,0.307805); (8.370019,0.387805) **[red]@{-};
(8.370019,0.307805); (8.210019,0.387805) **[red]@{-};
(8.250019,0.324347); (8.330019,0.364347) **[red]@{-};
(8.250019,0.364347); (8.330019,0.324347) **[red]@{-};
(8.250019,0.334385); (8.330019,0.374385) **[red]@{-};
(8.250019,0.374385); (8.330019,0.334385) **[red]@{-};
(8.228339,1.303058); (8.388339,1.303058) **[blue]@{-};
(8.308339,1.263058); (8.308339,1.343058) **[blue]@{-};
(8.268339,1.298940); (8.348339,1.298940) **[blue]@{-};
(8.268339,1.305663); (8.348339,1.305663) **[blue]@{-};
(8.228339,0.330929); (8.388339,0.410929) **[red]@{-};
(8.388339,0.330929); (8.228339,0.410929) **[red]@{-};
(8.268339,0.346362); (8.348339,0.386362) **[red]@{-};
(8.268339,0.386362); (8.348339,0.346362) **[red]@{-};
(8.268339,0.356500); (8.348339,0.396500) **[red]@{-};
(8.268339,0.396500); (8.348339,0.356500) **[red]@{-};
(8.290687,1.306696); (8.450687,1.306696) **[blue]@{-};
(8.370687,1.266696); (8.370687,1.346696) **[blue]@{-};
(8.330687,1.305125); (8.410687,1.305125) **[blue]@{-};
(8.330687,1.310492); (8.410687,1.310492) **[blue]@{-};
(8.290687,0.366335); (8.450687,0.446335) **[red]@{-};
(8.450687,0.366335); (8.290687,0.446335) **[red]@{-};
(8.330687,0.382844); (8.410687,0.422844) **[red]@{-};
(8.330687,0.422844); (8.410687,0.382844) **[red]@{-};
(8.330687,0.387866); (8.410687,0.427866) **[red]@{-};
(8.330687,0.427866); (8.410687,0.387866) **[red]@{-};
(8.316605,1.304307); (8.476605,1.304307) **[blue]@{-};
(8.396605,1.264307); (8.396605,1.344307) **[blue]@{-};
(8.356605,1.302827); (8.436605,1.302827) **[blue]@{-};
(8.356605,1.307689); (8.436605,1.307689) **[blue]@{-};
(8.316605,0.394702); (8.476605,0.474702) **[red]@{-};
(8.476605,0.394702); (8.316605,0.474702) **[red]@{-};
(8.356605,0.412195); (8.436605,0.452195) **[red]@{-};
(8.356605,0.452195); (8.436605,0.412195) **[red]@{-};
(8.356605,0.416827); (8.436605,0.456827) **[red]@{-};
(8.356605,0.456827); (8.436605,0.416827) **[red]@{-};
(8.358792,1.304994); (8.518792,1.304994) **[blue]@{-};
(8.438792,1.264994); (8.438792,1.344994) **[blue]@{-};
(8.398792,1.300106); (8.478792,1.300106) **[blue]@{-};
(8.398792,1.308013); (8.478792,1.308013) **[blue]@{-};
(8.358792,0.421626); (8.518792,0.501626) **[red]@{-};
(8.518792,0.421626); (8.358792,0.501626) **[red]@{-};
(8.398792,0.439578); (8.478792,0.479578) **[red]@{-};
(8.398792,0.479578); (8.478792,0.439578) **[red]@{-};
(8.398792,0.446119); (8.478792,0.486119) **[red]@{-};
(8.398792,0.486119); (8.478792,0.446119) **[red]@{-};
(8.367083,1.305708); (8.527083,1.305708) **[blue]@{-};
(8.447083,1.265708); (8.447083,1.345708) **[blue]@{-};
(8.407083,1.303349); (8.487083,1.303349) **[blue]@{-};
(8.407083,1.307831); (8.487083,1.307831) **[blue]@{-};
(8.367083,0.427793); (8.527083,0.507793) **[red]@{-};
(8.527083,0.427793); (8.367083,0.507793) **[red]@{-};
(8.407083,0.444963); (8.487083,0.484963) **[red]@{-};
(8.407083,0.484963); (8.487083,0.444963) **[red]@{-};
(8.407083,0.453685); (8.487083,0.493685) **[red]@{-};
(8.407083,0.493685); (8.487083,0.453685) **[red]@{-};
(8.383524,1.303070); (8.543524,1.303070) **[blue]@{-};
(8.463524,1.263070); (8.463524,1.343070) **[blue]@{-};
(8.423524,1.300456); (8.503524,1.300456) **[blue]@{-};
(8.423524,1.305682); (8.503524,1.305682) **[blue]@{-};
(8.383524,0.440580); (8.543524,0.520580) **[red]@{-};
(8.543524,0.440580); (8.383524,0.520580) **[red]@{-};
(8.423524,0.458834); (8.503524,0.498834) **[red]@{-};
(8.423524,0.498834); (8.503524,0.458834) **[red]@{-};
(8.423524,0.462000); (8.503524,0.502000) **[red]@{-};
(8.423524,0.502000); (8.503524,0.462000) **[red]@{-};
(8.407840,1.306927); (8.567840,1.306927) **[blue]@{-};
(8.487840,1.266927); (8.487840,1.346927) **[blue]@{-};
(8.447840,1.303719); (8.527840,1.303719) **[blue]@{-};
(8.447840,1.308692); (8.527840,1.308692) **[blue]@{-};
(8.407840,0.456366); (8.567840,0.536366) **[red]@{-};
(8.567840,0.456366); (8.407840,0.536366) **[red]@{-};
(8.447840,0.472294); (8.527840,0.512294) **[red]@{-};
(8.447840,0.512294); (8.527840,0.472294) **[red]@{-};
(8.447840,0.481044); (8.527840,0.521044) **[red]@{-};
(8.447840,0.521044); (8.527840,0.481044) **[red]@{-};
(8.439636,1.307191); (8.599636,1.307191) **[blue]@{-};
(8.519636,1.267191); (8.519636,1.347191) **[blue]@{-};
(8.479636,1.301063); (8.559636,1.301063) **[blue]@{-};
(8.479636,1.310453); (8.559636,1.310453) **[blue]@{-};
(8.439636,0.329829); (8.599636,0.409829) **[red]@{-};
(8.599636,0.329829); (8.439636,0.409829) **[red]@{-};
(8.479636,0.348658); (8.559636,0.388658) **[red]@{-};
(8.479636,0.388658); (8.559636,0.348658) **[red]@{-};
(8.479636,0.353178); (8.559636,0.393178) **[red]@{-};
(8.479636,0.393178); (8.559636,0.353178) **[red]@{-};
(8.463032,1.299745); (8.623032,1.299745) **[blue]@{-};
(8.543032,1.259745); (8.543032,1.339745) **[blue]@{-};
(8.503032,1.297474); (8.583032,1.297474) **[blue]@{-};
(8.503032,1.303662); (8.583032,1.303662) **[blue]@{-};
(8.463032,0.340931); (8.623032,0.420931) **[red]@{-};
(8.623032,0.340931); (8.463032,0.420931) **[red]@{-};
(8.503032,0.358444); (8.583032,0.398444) **[red]@{-};
(8.503032,0.398444); (8.583032,0.358444) **[red]@{-};
(8.503032,0.362793); (8.583032,0.402793) **[red]@{-};
(8.503032,0.402793); (8.583032,0.362793) **[red]@{-};
(9.117217,1.306383); (9.277217,1.306383) **[blue]@{-};
(9.197217,1.266383); (9.197217,1.346383) **[blue]@{-};
(9.157217,1.303965); (9.237217,1.303965) **[blue]@{-};
(9.157217,1.307434); (9.237217,1.307434) **[blue]@{-};
(9.117217,0.025347); (9.277217,0.105347) **[red]@{-};
(9.277217,0.025347); (9.117217,0.105347) **[red]@{-};
(9.157217,0.039872); (9.237217,0.079872) **[red]@{-};
(9.157217,0.079872); (9.237217,0.039872) **[red]@{-};
(9.157217,0.048286); (9.237217,0.088286) **[red]@{-};
(9.157217,0.088286); (9.237217,0.048286) **[red]@{-};
\endxy
}
\caption{Cost
  to evaluate an $\ell$-isogeny
  on one CSIDH-512 point
  and compute the new curve coefficient.
  Graph shows Skylake cycles
  for {\tt velusqrt-magma}
  divided by $\ell+2$.
}
\label{fig: isogeny-magma}
\end{figure}

\begin{figure}[t]
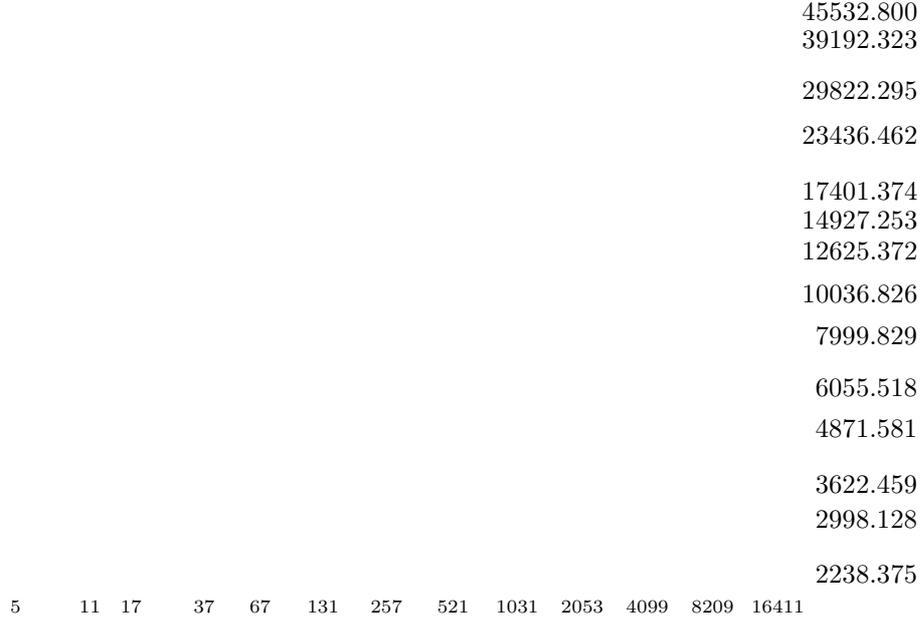

\hbox to\hsize{\hfill
\xy <1.4cm,0cm>:<0cm,2.8cm>::
(1.423413,0.056437); (1.423413,2.720903) **[lightgray]@{-};
(1.423413,-0.100000) *{\scriptstyle 5};
(2.120737,0.056437); (2.120737,2.720903) **[lightgray]@{-};
(2.120737,-0.100000) *{\scriptstyle 11};
(2.505740,0.056437); (2.505740,2.720903) **[lightgray]@{-};
(2.505740,-0.100000) *{\scriptstyle 17};
(3.193554,0.056437); (3.193554,2.720903) **[lightgray]@{-};
(3.193554,-0.100000) *{\scriptstyle 37};
(3.718698,0.056437); (3.718698,2.720903) **[lightgray]@{-};
(3.718698,-0.100000) *{\scriptstyle 67};
(4.311704,0.056437); (4.311704,2.720903) **[lightgray]@{-};
(4.311704,-0.100000) *{\scriptstyle 131};
(4.907693,0.056437); (4.907693,2.720903) **[lightgray]@{-};
(4.907693,-0.100000) *{\scriptstyle 257};
(5.532687,0.056437); (5.532687,2.720903) **[lightgray]@{-};
(5.532687,-0.100000) *{\scriptstyle 521};
(6.136331,0.056437); (6.136331,2.720903) **[lightgray]@{-};
(6.136331,-0.100000) *{\scriptstyle 1031};
(6.745493,0.056437); (6.745493,2.720903) **[lightgray]@{-};
(6.745493,-0.100000) *{\scriptstyle 2053};
(7.357015,0.056437); (7.357015,2.720903) **[lightgray]@{-};
(7.357015,-0.100000) *{\scriptstyle 4099};
(7.971231,0.056437); (7.971231,2.720903) **[lightgray]@{-};
(7.971231,-0.100000) *{\scriptstyle 8209};
(8.583885,0.056437); (8.583885,2.720903) **[lightgray]@{-};
(8.583885,-0.100000) *{\scriptstyle 16411};
(1.423413,2.720903); (8.583885,2.720903) **@{-};
(9.883885,2.720903) *{\llap{45532.800}};
(1.423413,2.588283); (8.583885,2.588283) **[lightgray]@{-};
(9.883885,2.588283) *{\llap{39192.323}};
(1.423413,2.346638); (8.583885,2.346638) **[lightgray]@{-};
(9.883885,2.346638) *{\llap{29822.295}};
(1.423413,2.133526); (8.583885,2.133526) **[lightgray]@{-};
(9.883885,2.133526) *{\llap{23436.462}};
(1.423413,1.870197); (8.583885,1.870197) **[lightgray]@{-};
(9.883885,1.870197) *{\llap{17401.374}};
(1.423413,1.734562); (8.583885,1.734562) **[lightgray]@{-};
(9.883885,1.734562) *{\llap{14927.253}};
(1.423413,1.586440); (8.583885,1.586440) **[lightgray]@{-};
(9.883885,1.586440) *{\llap{12625.372}};
(1.423413,1.383513); (8.583885,1.383513) **[lightgray]@{-};
(9.883885,1.383513) *{\llap{10036.826}};
(1.423413,1.182892); (8.583885,1.182892) **[lightgray]@{-};
(9.883885,1.182892) *{\llap{7999.829}};
(1.423413,0.936626); (8.583885,0.936626) **[lightgray]@{-};
(9.883885,0.936626) *{\llap{6055.518}};
(1.423413,0.744220); (8.583885,0.744220) **[lightgray]@{-};
(9.883885,0.744220) *{\llap{4871.581}};
(1.423413,0.482198); (8.583885,0.482198) **[lightgray]@{-};
(9.883885,0.482198) *{\llap{3622.459}};
(1.423413,0.314897); (8.583885,0.314897) **[lightgray]@{-};
(9.883885,0.314897) *{\llap{2998.128}};
(1.423413,0.056437); (8.583885,0.056437) **@{-};
(9.883885,0.056437) *{\llap{2238.375}};
(1.343413,2.327728); (1.503413,2.327728) **[blue]@{-};
(1.423413,2.287728); (1.423413,2.367728) **[blue]@{-};
(1.383413,2.317486); (1.463413,2.317486) **[blue]@{-};
(1.383413,2.345358); (1.463413,2.345358) **[blue]@{-};
(1.343413,2.680903); (1.503413,2.760903) **[red]@{-};
(1.503413,2.680903); (1.343413,2.760903) **[red]@{-};
(1.383413,2.684544); (1.463413,2.724544) **[red]@{-};
(1.383413,2.724544); (1.463413,2.684544) **[red]@{-};
(1.383413,2.709306); (1.463413,2.749306) **[red]@{-};
(1.383413,2.749306); (1.463413,2.709306) **[red]@{-};
(2.040737,2.133526); (2.200737,2.133526) **[blue]@{-};
(2.120737,2.093526); (2.120737,2.173526) **[blue]@{-};
(2.080737,2.123591); (2.160737,2.123591) **[blue]@{-};
(2.080737,2.140075); (2.160737,2.140075) **[blue]@{-};
(2.040737,2.548283); (2.200737,2.628283) **[red]@{-};
(2.200737,2.548283); (2.040737,2.628283) **[red]@{-};
(2.080737,2.551607); (2.160737,2.591607) **[red]@{-};
(2.080737,2.591607); (2.160737,2.551607) **[red]@{-};
(2.080737,2.574464); (2.160737,2.614464) **[red]@{-};
(2.080737,2.614464); (2.160737,2.574464) **[red]@{-};
(2.425740,2.014924); (2.585740,2.014924) **[blue]@{-};
(2.505740,1.974924); (2.505740,2.054924) **[blue]@{-};
(2.465740,2.006740); (2.545740,2.006740) **[blue]@{-};
(2.465740,2.025028); (2.545740,2.025028) **[blue]@{-};
(2.425740,2.306638); (2.585740,2.386638) **[red]@{-};
(2.585740,2.306638); (2.425740,2.386638) **[red]@{-};
(2.465740,2.315767); (2.545740,2.355767) **[red]@{-};
(2.465740,2.355767); (2.545740,2.315767) **[red]@{-};
(2.465740,2.342115); (2.545740,2.382115) **[red]@{-};
(2.465740,2.382115); (2.545740,2.342115) **[red]@{-};
(3.113554,1.870197); (3.273554,1.870197) **[blue]@{-};
(3.193554,1.830197); (3.193554,1.910197) **[blue]@{-};
(3.153554,1.860771); (3.233554,1.860771) **[blue]@{-};
(3.153554,1.887172); (3.233554,1.887172) **[blue]@{-};
(3.113554,2.017854); (3.273554,2.097854) **[red]@{-};
(3.273554,2.017854); (3.113554,2.097854) **[red]@{-};
(3.153554,2.033644); (3.233554,2.073644) **[red]@{-};
(3.153554,2.073644); (3.233554,2.033644) **[red]@{-};
(3.153554,2.050214); (3.233554,2.090214) **[red]@{-};
(3.153554,2.090214); (3.233554,2.050214) **[red]@{-};
(3.638698,1.821570); (3.798698,1.821570) **[blue]@{-};
(3.718698,1.781570); (3.718698,1.861570) **[blue]@{-};
(3.678698,1.803638); (3.758698,1.803638) **[blue]@{-};
(3.678698,1.826641); (3.758698,1.826641) **[blue]@{-};
(3.638698,1.788362); (3.798698,1.868362) **[red]@{-};
(3.798698,1.788362); (3.638698,1.868362) **[red]@{-};
(3.678698,1.802826); (3.758698,1.842826) **[red]@{-};
(3.678698,1.842826); (3.758698,1.802826) **[red]@{-};
(3.678698,1.817635); (3.758698,1.857635) **[red]@{-};
(3.678698,1.857635); (3.758698,1.817635) **[red]@{-};
(4.231704,1.734562); (4.391704,1.734562) **[blue]@{-};
(4.311704,1.694562); (4.311704,1.774562) **[blue]@{-};
(4.271704,1.730523); (4.351704,1.730523) **[blue]@{-};
(4.271704,1.756815); (4.351704,1.756815) **[blue]@{-};
(4.231704,1.595997); (4.391704,1.675997) **[red]@{-};
(4.391704,1.595997); (4.231704,1.675997) **[red]@{-};
(4.271704,1.601593); (4.351704,1.641593) **[red]@{-};
(4.271704,1.641593); (4.351704,1.601593) **[red]@{-};
(4.271704,1.619982); (4.351704,1.659982) **[red]@{-};
(4.271704,1.659982); (4.351704,1.619982) **[red]@{-};
(4.827693,1.677050); (4.987693,1.677050) **[blue]@{-};
(4.907693,1.637050); (4.907693,1.717050) **[blue]@{-};
(4.867693,1.658894); (4.947693,1.658894) **[blue]@{-};
(4.867693,1.679341); (4.947693,1.679341) **[blue]@{-};
(4.827693,1.343513); (4.987693,1.423513) **[red]@{-};
(4.987693,1.343513); (4.827693,1.423513) **[red]@{-};
(4.867693,1.354454); (4.947693,1.394454) **[red]@{-};
(4.867693,1.394454); (4.947693,1.354454) **[red]@{-};
(4.867693,1.370762); (4.947693,1.410762) **[red]@{-};
(4.867693,1.410762); (4.947693,1.370762) **[red]@{-};
(5.452687,1.625184); (5.612687,1.625184) **[blue]@{-};
(5.532687,1.585184); (5.532687,1.665184) **[blue]@{-};
(5.492687,1.614515); (5.572687,1.614515) **[blue]@{-};
(5.492687,1.628394); (5.572687,1.628394) **[blue]@{-};
(5.452687,1.142892); (5.612687,1.222892) **[red]@{-};
(5.612687,1.142892); (5.452687,1.222892) **[red]@{-};
(5.492687,1.154486); (5.572687,1.194486) **[red]@{-};
(5.492687,1.194486); (5.572687,1.154486) **[red]@{-};
(5.492687,1.169908); (5.572687,1.209908) **[red]@{-};
(5.492687,1.209908); (5.572687,1.169908) **[red]@{-};
(6.056331,1.586440); (6.216331,1.586440) **[blue]@{-};
(6.136331,1.546440); (6.136331,1.626440) **[blue]@{-};
(6.096331,1.577431); (6.176331,1.577431) **[blue]@{-};
(6.096331,1.594839); (6.176331,1.594839) **[blue]@{-};
(6.056331,0.896626); (6.216331,0.976626) **[red]@{-};
(6.216331,0.896626); (6.056331,0.976626) **[red]@{-};
(6.096331,0.903404); (6.176331,0.943404) **[red]@{-};
(6.096331,0.943404); (6.176331,0.903404) **[red]@{-};
(6.096331,0.923048); (6.176331,0.963048) **[red]@{-};
(6.096331,0.963048); (6.176331,0.923048) **[red]@{-};
(6.665493,1.528033); (6.825493,1.528033) **[blue]@{-};
(6.745493,1.488033); (6.745493,1.568033) **[blue]@{-};
(6.705493,1.514877); (6.785493,1.514877) **[blue]@{-};
(6.705493,1.533682); (6.785493,1.533682) **[blue]@{-};
(6.665493,0.704220); (6.825493,0.784220) **[red]@{-};
(6.825493,0.704220); (6.665493,0.784220) **[red]@{-};
(6.705493,0.709346); (6.785493,0.749346) **[red]@{-};
(6.705493,0.749346); (6.785493,0.709346) **[red]@{-};
(6.705493,0.730360); (6.785493,0.770360) **[red]@{-};
(6.705493,0.770360); (6.785493,0.730360) **[red]@{-};
(7.277015,1.503997); (7.437015,1.503997) **[blue]@{-};
(7.357015,1.463997); (7.357015,1.543997) **[blue]@{-};
(7.317015,1.497545); (7.397015,1.497545) **[blue]@{-};
(7.317015,1.510741); (7.397015,1.510741) **[blue]@{-};
(7.277015,0.442198); (7.437015,0.522198) **[red]@{-};
(7.437015,0.442198); (7.277015,0.522198) **[red]@{-};
(7.317015,0.453999); (7.397015,0.493999) **[red]@{-};
(7.317015,0.493999); (7.397015,0.453999) **[red]@{-};
(7.317015,0.465306); (7.397015,0.505306) **[red]@{-};
(7.317015,0.505306); (7.397015,0.465306) **[red]@{-};
(7.891231,1.483990); (8.051231,1.483990) **[blue]@{-};
(7.971231,1.443990); (7.971231,1.523990) **[blue]@{-};
(7.931231,1.483161); (8.011231,1.483161) **[blue]@{-};
(7.931231,1.490626); (8.011231,1.490626) **[blue]@{-};
(7.891231,0.274897); (8.051231,0.354897) **[red]@{-};
(8.051231,0.274897); (7.891231,0.354897) **[red]@{-};
(7.931231,0.281083); (8.011231,0.321083) **[red]@{-};
(7.931231,0.321083); (8.011231,0.281083) **[red]@{-};
(7.931231,0.304517); (8.011231,0.344517) **[red]@{-};
(7.931231,0.344517); (8.011231,0.304517) **[red]@{-};
(8.503885,1.477972); (8.663885,1.477972) **[blue]@{-};
(8.583885,1.437972); (8.583885,1.517972) **[blue]@{-};
(8.543885,1.475357); (8.623885,1.475357) **[blue]@{-};
(8.543885,1.484127); (8.623885,1.484127) **[blue]@{-};
(8.503885,0.016437); (8.663885,0.096437) **[red]@{-};
(8.663885,0.016437); (8.503885,0.096437) **[red]@{-};
(8.543885,0.030650); (8.623885,0.070650) **[red]@{-};
(8.543885,0.070650); (8.623885,0.030650) **[red]@{-};
(8.543885,0.040236); (8.623885,0.080236) **[red]@{-};
(8.543885,0.080236); (8.623885,0.040236) **[red]@{-};
\endxy
}
\caption{Cost
  to evaluate an $\ell$-isogeny
  on one point
  and compute the new curve coefficient.
  Graph shows Skylake cycles
  for {\tt velusqrt-julia}
  divided by $\ell+2$.
}
\label{fig:isogeny-julia}
\end{figure}

\begin{figure}[t]
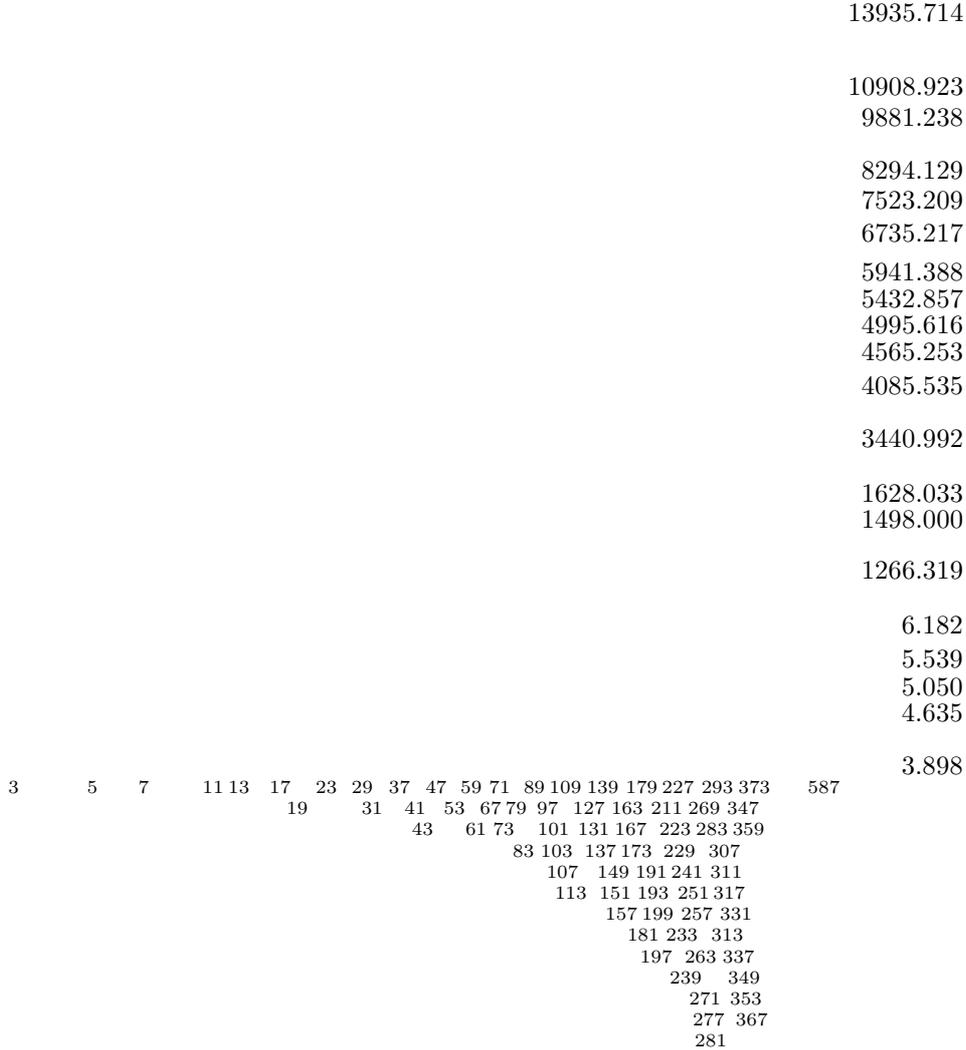

\hbox to\hsize{\hfill
\xy <1.4cm,0cm>:<0cm,2.8cm>::
(1.584963,0.003013); (1.584963,0.668261) **[lightgray]@{-};
(1.584963,0.928374); (1.584963,1.290863) **[lightgray]@{-};
(1.584963,1.553897); (1.584963,3.571788) **[lightgray]@{-};
(1.584963,-0.100000) *{\scriptstyle 3};
(2.321928,0.003013); (2.321928,0.668261) **[lightgray]@{-};
(2.321928,0.928374); (2.321928,1.290863) **[lightgray]@{-};
(2.321928,1.553897); (2.321928,3.571788) **[lightgray]@{-};
(2.321928,-0.100000) *{\scriptstyle 5};
(2.807355,0.003013); (2.807355,0.668261) **[lightgray]@{-};
(2.807355,0.928374); (2.807355,1.290863) **[lightgray]@{-};
(2.807355,1.553897); (2.807355,3.571788) **[lightgray]@{-};
(2.807355,-0.100000) *{\scriptstyle 7};
(3.459432,0.003013); (3.459432,0.668261) **[lightgray]@{-};
(3.459432,0.928374); (3.459432,1.290863) **[lightgray]@{-};
(3.459432,1.553897); (3.459432,3.571788) **[lightgray]@{-};
(3.459432,-0.100000) *{\scriptstyle 11};
(3.700440,0.003013); (3.700440,0.668261) **[lightgray]@{-};
(3.700440,0.928374); (3.700440,1.290863) **[lightgray]@{-};
(3.700440,1.553897); (3.700440,3.571788) **[lightgray]@{-};
(3.700440,-0.100000) *{\scriptstyle 13};
(4.087463,0.003013); (4.087463,0.668261) **[lightgray]@{-};
(4.087463,0.928374); (4.087463,1.290863) **[lightgray]@{-};
(4.087463,1.553897); (4.087463,3.571788) **[lightgray]@{-};
(4.087463,-0.100000) *{\scriptstyle 17};
(4.247928,0.003013); (4.247928,0.668261) **[lightergray]@{-};
(4.247928,0.928374); (4.247928,1.290863) **[lightergray]@{-};
(4.247928,1.553897); (4.247928,3.571788) **[lightergray]@{-};
(4.247928,-0.200000) *{\scriptstyle 19};
(4.523562,0.003013); (4.523562,0.668261) **[lightgray]@{-};
(4.523562,0.928374); (4.523562,1.290863) **[lightgray]@{-};
(4.523562,1.553897); (4.523562,3.571788) **[lightgray]@{-};
(4.523562,-0.100000) *{\scriptstyle 23};
(4.857981,0.003013); (4.857981,0.668261) **[lightgray]@{-};
(4.857981,0.928374); (4.857981,1.290863) **[lightgray]@{-};
(4.857981,1.553897); (4.857981,3.571788) **[lightgray]@{-};
(4.857981,-0.100000) *{\scriptstyle 29};
(4.954196,0.003013); (4.954196,0.668261) **[lightergray]@{-};
(4.954196,0.928374); (4.954196,1.290863) **[lightergray]@{-};
(4.954196,1.553897); (4.954196,3.571788) **[lightergray]@{-};
(4.954196,-0.200000) *{\scriptstyle 31};
(5.209453,0.003013); (5.209453,0.668261) **[lightgray]@{-};
(5.209453,0.928374); (5.209453,1.290863) **[lightgray]@{-};
(5.209453,1.553897); (5.209453,3.571788) **[lightgray]@{-};
(5.209453,-0.100000) *{\scriptstyle 37};
(5.357552,0.003013); (5.357552,0.668261) **[lightergray]@{-};
(5.357552,0.928374); (5.357552,1.290863) **[lightergray]@{-};
(5.357552,1.553897); (5.357552,3.571788) **[lightergray]@{-};
(5.357552,-0.200000) *{\scriptstyle 41};
(5.426265,0.003013); (5.426265,0.668261) **[lightergray]@{-};
(5.426265,0.928374); (5.426265,1.290863) **[lightergray]@{-};
(5.426265,1.553897); (5.426265,3.571788) **[lightergray]@{-};
(5.426265,-0.300000) *{\scriptstyle 43};
(5.554589,0.003013); (5.554589,0.668261) **[lightgray]@{-};
(5.554589,0.928374); (5.554589,1.290863) **[lightgray]@{-};
(5.554589,1.553897); (5.554589,3.571788) **[lightgray]@{-};
(5.554589,-0.100000) *{\scriptstyle 47};
(5.727920,0.003013); (5.727920,0.668261) **[lightergray]@{-};
(5.727920,0.928374); (5.727920,1.290863) **[lightergray]@{-};
(5.727920,1.553897); (5.727920,3.571788) **[lightergray]@{-};
(5.727920,-0.200000) *{\scriptstyle 53};
(5.882643,0.003013); (5.882643,0.668261) **[lightgray]@{-};
(5.882643,0.928374); (5.882643,1.290863) **[lightgray]@{-};
(5.882643,1.553897); (5.882643,3.571788) **[lightgray]@{-};
(5.882643,-0.100000) *{\scriptstyle 59};
(5.930737,0.003013); (5.930737,0.668261) **[lightergray]@{-};
(5.930737,0.928374); (5.930737,1.290863) **[lightergray]@{-};
(5.930737,1.553897); (5.930737,3.571788) **[lightergray]@{-};
(5.930737,-0.300000) *{\scriptstyle 61};
(6.066089,0.003013); (6.066089,0.668261) **[lightergray]@{-};
(6.066089,0.928374); (6.066089,1.290863) **[lightergray]@{-};
(6.066089,1.553897); (6.066089,3.571788) **[lightergray]@{-};
(6.066089,-0.200000) *{\scriptstyle 67};
(6.149747,0.003013); (6.149747,0.668261) **[lightgray]@{-};
(6.149747,0.928374); (6.149747,1.290863) **[lightgray]@{-};
(6.149747,1.553897); (6.149747,3.571788) **[lightgray]@{-};
(6.149747,-0.100000) *{\scriptstyle 71};
(6.189825,0.003013); (6.189825,0.668261) **[lightergray]@{-};
(6.189825,0.928374); (6.189825,1.290863) **[lightergray]@{-};
(6.189825,1.553897); (6.189825,3.571788) **[lightergray]@{-};
(6.189825,-0.300000) *{\scriptstyle 73};
(6.303781,0.003013); (6.303781,0.668261) **[lightergray]@{-};
(6.303781,0.928374); (6.303781,1.290863) **[lightergray]@{-};
(6.303781,1.553897); (6.303781,3.571788) **[lightergray]@{-};
(6.303781,-0.200000) *{\scriptstyle 79};
(6.375039,0.003013); (6.375039,0.668261) **[lightergray]@{-};
(6.375039,0.928374); (6.375039,1.290863) **[lightergray]@{-};
(6.375039,1.553897); (6.375039,3.571788) **[lightergray]@{-};
(6.375039,-0.400000) *{\scriptstyle 83};
(6.475733,0.003013); (6.475733,0.668261) **[lightgray]@{-};
(6.475733,0.928374); (6.475733,1.290863) **[lightgray]@{-};
(6.475733,1.553897); (6.475733,3.571788) **[lightgray]@{-};
(6.475733,-0.100000) *{\scriptstyle 89};
(6.599913,0.003013); (6.599913,0.668261) **[lightergray]@{-};
(6.599913,0.928374); (6.599913,1.290863) **[lightergray]@{-};
(6.599913,1.553897); (6.599913,3.571788) **[lightergray]@{-};
(6.599913,-0.200000) *{\scriptstyle 97};
(6.658211,0.003013); (6.658211,0.668261) **[lightergray]@{-};
(6.658211,0.928374); (6.658211,1.290863) **[lightergray]@{-};
(6.658211,1.553897); (6.658211,3.571788) **[lightergray]@{-};
(6.658211,-0.300000) *{\scriptstyle 101};
(6.686501,0.003013); (6.686501,0.668261) **[lightergray]@{-};
(6.686501,0.928374); (6.686501,1.290863) **[lightergray]@{-};
(6.686501,1.553897); (6.686501,3.571788) **[lightergray]@{-};
(6.686501,-0.400000) *{\scriptstyle 103};
(6.741467,0.003013); (6.741467,0.668261) **[lightergray]@{-};
(6.741467,0.928374); (6.741467,1.290863) **[lightergray]@{-};
(6.741467,1.553897); (6.741467,3.571788) **[lightergray]@{-};
(6.741467,-0.500000) *{\scriptstyle 107};
(6.768184,0.003013); (6.768184,0.668261) **[lightgray]@{-};
(6.768184,0.928374); (6.768184,1.290863) **[lightgray]@{-};
(6.768184,1.553897); (6.768184,3.571788) **[lightgray]@{-};
(6.768184,-0.100000) *{\scriptstyle 109};
(6.820179,0.003013); (6.820179,0.668261) **[lightergray]@{-};
(6.820179,0.928374); (6.820179,1.290863) **[lightergray]@{-};
(6.820179,1.553897); (6.820179,3.571788) **[lightergray]@{-};
(6.820179,-0.600000) *{\scriptstyle 113};
(6.988685,0.003013); (6.988685,0.668261) **[lightergray]@{-};
(6.988685,0.928374); (6.988685,1.290863) **[lightergray]@{-};
(6.988685,1.553897); (6.988685,3.571788) **[lightergray]@{-};
(6.988685,-0.200000) *{\scriptstyle 127};
(7.033423,0.003013); (7.033423,0.668261) **[lightergray]@{-};
(7.033423,0.928374); (7.033423,1.290863) **[lightergray]@{-};
(7.033423,1.553897); (7.033423,3.571788) **[lightergray]@{-};
(7.033423,-0.300000) *{\scriptstyle 131};
(7.098032,0.003013); (7.098032,0.668261) **[lightergray]@{-};
(7.098032,0.928374); (7.098032,1.290863) **[lightergray]@{-};
(7.098032,1.553897); (7.098032,3.571788) **[lightergray]@{-};
(7.098032,-0.400000) *{\scriptstyle 137};
(7.118941,0.003013); (7.118941,0.668261) **[lightgray]@{-};
(7.118941,0.928374); (7.118941,1.290863) **[lightgray]@{-};
(7.118941,1.553897); (7.118941,3.571788) **[lightgray]@{-};
(7.118941,-0.100000) *{\scriptstyle 139};
(7.219169,0.003013); (7.219169,0.668261) **[lightergray]@{-};
(7.219169,0.928374); (7.219169,1.290863) **[lightergray]@{-};
(7.219169,1.553897); (7.219169,3.571788) **[lightergray]@{-};
(7.219169,-0.500000) *{\scriptstyle 149};
(7.238405,0.003013); (7.238405,0.668261) **[lightergray]@{-};
(7.238405,0.928374); (7.238405,1.290863) **[lightergray]@{-};
(7.238405,1.553897); (7.238405,3.571788) **[lightergray]@{-};
(7.238405,-0.600000) *{\scriptstyle 151};
(7.294621,0.003013); (7.294621,0.668261) **[lightergray]@{-};
(7.294621,0.928374); (7.294621,1.290863) **[lightergray]@{-};
(7.294621,1.553897); (7.294621,3.571788) **[lightergray]@{-};
(7.294621,-0.700000) *{\scriptstyle 157};
(7.348728,0.003013); (7.348728,0.668261) **[lightergray]@{-};
(7.348728,0.928374); (7.348728,1.290863) **[lightergray]@{-};
(7.348728,1.553897); (7.348728,3.571788) **[lightergray]@{-};
(7.348728,-0.200000) *{\scriptstyle 163};
(7.383704,0.003013); (7.383704,0.668261) **[lightergray]@{-};
(7.383704,0.928374); (7.383704,1.290863) **[lightergray]@{-};
(7.383704,1.553897); (7.383704,3.571788) **[lightergray]@{-};
(7.383704,-0.300000) *{\scriptstyle 167};
(7.434628,0.003013); (7.434628,0.668261) **[lightergray]@{-};
(7.434628,0.928374); (7.434628,1.290863) **[lightergray]@{-};
(7.434628,1.553897); (7.434628,3.571788) **[lightergray]@{-};
(7.434628,-0.400000) *{\scriptstyle 173};
(7.483816,0.003013); (7.483816,0.668261) **[lightgray]@{-};
(7.483816,0.928374); (7.483816,1.290863) **[lightgray]@{-};
(7.483816,1.553897); (7.483816,3.571788) **[lightgray]@{-};
(7.483816,-0.100000) *{\scriptstyle 179};
(7.499846,0.003013); (7.499846,0.668261) **[lightergray]@{-};
(7.499846,0.928374); (7.499846,1.290863) **[lightergray]@{-};
(7.499846,1.553897); (7.499846,3.571788) **[lightergray]@{-};
(7.499846,-0.800000) *{\scriptstyle 181};
(7.577429,0.003013); (7.577429,0.668261) **[lightergray]@{-};
(7.577429,0.928374); (7.577429,1.290863) **[lightergray]@{-};
(7.577429,1.553897); (7.577429,3.571788) **[lightergray]@{-};
(7.577429,-0.500000) *{\scriptstyle 191};
(7.592457,0.003013); (7.592457,0.668261) **[lightergray]@{-};
(7.592457,0.928374); (7.592457,1.290863) **[lightergray]@{-};
(7.592457,1.553897); (7.592457,3.571788) **[lightergray]@{-};
(7.592457,-0.600000) *{\scriptstyle 193};
(7.622052,0.003013); (7.622052,0.668261) **[lightergray]@{-};
(7.622052,0.928374); (7.622052,1.290863) **[lightergray]@{-};
(7.622052,1.553897); (7.622052,3.571788) **[lightergray]@{-};
(7.622052,-0.900000) *{\scriptstyle 197};
(7.636625,0.003013); (7.636625,0.668261) **[lightergray]@{-};
(7.636625,0.928374); (7.636625,1.290863) **[lightergray]@{-};
(7.636625,1.553897); (7.636625,3.571788) **[lightergray]@{-};
(7.636625,-0.700000) *{\scriptstyle 199};
(7.721099,0.003013); (7.721099,0.668261) **[lightergray]@{-};
(7.721099,0.928374); (7.721099,1.290863) **[lightergray]@{-};
(7.721099,1.553897); (7.721099,3.571788) **[lightergray]@{-};
(7.721099,-0.200000) *{\scriptstyle 211};
(7.800900,0.003013); (7.800900,0.668261) **[lightergray]@{-};
(7.800900,0.928374); (7.800900,1.290863) **[lightergray]@{-};
(7.800900,1.553897); (7.800900,3.571788) **[lightergray]@{-};
(7.800900,-0.300000) *{\scriptstyle 223};
(7.826548,0.003013); (7.826548,0.668261) **[lightgray]@{-};
(7.826548,0.928374); (7.826548,1.290863) **[lightgray]@{-};
(7.826548,1.553897); (7.826548,3.571788) **[lightgray]@{-};
(7.826548,-0.100000) *{\scriptstyle 227};
(7.839204,0.003013); (7.839204,0.668261) **[lightergray]@{-};
(7.839204,0.928374); (7.839204,1.290863) **[lightergray]@{-};
(7.839204,1.553897); (7.839204,3.571788) **[lightergray]@{-};
(7.839204,-0.400000) *{\scriptstyle 229};
(7.864186,0.003013); (7.864186,0.668261) **[lightergray]@{-};
(7.864186,0.928374); (7.864186,1.290863) **[lightergray]@{-};
(7.864186,1.553897); (7.864186,3.571788) **[lightergray]@{-};
(7.864186,-0.800000) *{\scriptstyle 233};
(7.900867,0.003013); (7.900867,0.668261) **[lightergray]@{-};
(7.900867,0.928374); (7.900867,1.290863) **[lightergray]@{-};
(7.900867,1.553897); (7.900867,3.571788) **[lightergray]@{-};
(7.900867,-1.000000) *{\scriptstyle 239};
(7.912889,0.003013); (7.912889,0.668261) **[lightergray]@{-};
(7.912889,0.928374); (7.912889,1.290863) **[lightergray]@{-};
(7.912889,1.553897); (7.912889,3.571788) **[lightergray]@{-};
(7.912889,-0.500000) *{\scriptstyle 241};
(7.971544,0.003013); (7.971544,0.668261) **[lightergray]@{-};
(7.971544,0.928374); (7.971544,1.290863) **[lightergray]@{-};
(7.971544,1.553897); (7.971544,3.571788) **[lightergray]@{-};
(7.971544,-0.600000) *{\scriptstyle 251};
(8.005625,0.003013); (8.005625,0.668261) **[lightergray]@{-};
(8.005625,0.928374); (8.005625,1.290863) **[lightergray]@{-};
(8.005625,1.553897); (8.005625,3.571788) **[lightergray]@{-};
(8.005625,-0.700000) *{\scriptstyle 257};
(8.038919,0.003013); (8.038919,0.668261) **[lightergray]@{-};
(8.038919,0.928374); (8.038919,1.290863) **[lightergray]@{-};
(8.038919,1.553897); (8.038919,3.571788) **[lightergray]@{-};
(8.038919,-0.900000) *{\scriptstyle 263};
(8.071462,0.003013); (8.071462,0.668261) **[lightergray]@{-};
(8.071462,0.928374); (8.071462,1.290863) **[lightergray]@{-};
(8.071462,1.553897); (8.071462,3.571788) **[lightergray]@{-};
(8.071462,-0.200000) *{\scriptstyle 269};
(8.082149,0.003013); (8.082149,0.668261) **[lightergray]@{-};
(8.082149,0.928374); (8.082149,1.290863) **[lightergray]@{-};
(8.082149,1.553897); (8.082149,3.571788) **[lightergray]@{-};
(8.082149,-1.100000) *{\scriptstyle 271};
(8.113742,0.003013); (8.113742,0.668261) **[lightergray]@{-};
(8.113742,0.928374); (8.113742,1.290863) **[lightergray]@{-};
(8.113742,1.553897); (8.113742,3.571788) **[lightergray]@{-};
(8.113742,-1.200000) *{\scriptstyle 277};
(8.134426,0.003013); (8.134426,0.668261) **[lightergray]@{-};
(8.134426,0.928374); (8.134426,1.290863) **[lightergray]@{-};
(8.134426,1.553897); (8.134426,3.571788) **[lightergray]@{-};
(8.134426,-1.300000) *{\scriptstyle 281};
(8.144658,0.003013); (8.144658,0.668261) **[lightergray]@{-};
(8.144658,0.928374); (8.144658,1.290863) **[lightergray]@{-};
(8.144658,1.553897); (8.144658,3.571788) **[lightergray]@{-};
(8.144658,-0.300000) *{\scriptstyle 283};
(8.194757,0.003013); (8.194757,0.668261) **[lightgray]@{-};
(8.194757,0.928374); (8.194757,1.290863) **[lightgray]@{-};
(8.194757,1.553897); (8.194757,3.571788) **[lightgray]@{-};
(8.194757,-0.100000) *{\scriptstyle 293};
(8.262095,0.003013); (8.262095,0.668261) **[lightergray]@{-};
(8.262095,0.928374); (8.262095,1.290863) **[lightergray]@{-};
(8.262095,1.553897); (8.262095,3.571788) **[lightergray]@{-};
(8.262095,-0.400000) *{\scriptstyle 307};
(8.280771,0.003013); (8.280771,0.668261) **[lightergray]@{-};
(8.280771,0.928374); (8.280771,1.290863) **[lightergray]@{-};
(8.280771,1.553897); (8.280771,3.571788) **[lightergray]@{-};
(8.280771,-0.500000) *{\scriptstyle 311};
(8.290019,0.003013); (8.290019,0.668261) **[lightergray]@{-};
(8.290019,0.928374); (8.290019,1.290863) **[lightergray]@{-};
(8.290019,1.553897); (8.290019,3.571788) **[lightergray]@{-};
(8.290019,-0.800000) *{\scriptstyle 313};
(8.308339,0.003013); (8.308339,0.668261) **[lightergray]@{-};
(8.308339,0.928374); (8.308339,1.290863) **[lightergray]@{-};
(8.308339,1.553897); (8.308339,3.571788) **[lightergray]@{-};
(8.308339,-0.600000) *{\scriptstyle 317};
(8.370687,0.003013); (8.370687,0.668261) **[lightergray]@{-};
(8.370687,0.928374); (8.370687,1.290863) **[lightergray]@{-};
(8.370687,1.553897); (8.370687,3.571788) **[lightergray]@{-};
(8.370687,-0.700000) *{\scriptstyle 331};
(8.396605,0.003013); (8.396605,0.668261) **[lightergray]@{-};
(8.396605,0.928374); (8.396605,1.290863) **[lightergray]@{-};
(8.396605,1.553897); (8.396605,3.571788) **[lightergray]@{-};
(8.396605,-0.900000) *{\scriptstyle 337};
(8.438792,0.003013); (8.438792,0.668261) **[lightergray]@{-};
(8.438792,0.928374); (8.438792,1.290863) **[lightergray]@{-};
(8.438792,1.553897); (8.438792,3.571788) **[lightergray]@{-};
(8.438792,-0.200000) *{\scriptstyle 347};
(8.447083,0.003013); (8.447083,0.668261) **[lightergray]@{-};
(8.447083,0.928374); (8.447083,1.290863) **[lightergray]@{-};
(8.447083,1.553897); (8.447083,3.571788) **[lightergray]@{-};
(8.447083,-1.000000) *{\scriptstyle 349};
(8.463524,0.003013); (8.463524,0.668261) **[lightergray]@{-};
(8.463524,0.928374); (8.463524,1.290863) **[lightergray]@{-};
(8.463524,1.553897); (8.463524,3.571788) **[lightergray]@{-};
(8.463524,-1.100000) *{\scriptstyle 353};
(8.487840,0.003013); (8.487840,0.668261) **[lightergray]@{-};
(8.487840,0.928374); (8.487840,1.290863) **[lightergray]@{-};
(8.487840,1.553897); (8.487840,3.571788) **[lightergray]@{-};
(8.487840,-0.300000) *{\scriptstyle 359};
(8.519636,0.003013); (8.519636,0.668261) **[lightergray]@{-};
(8.519636,0.928374); (8.519636,1.290863) **[lightergray]@{-};
(8.519636,1.553897); (8.519636,3.571788) **[lightergray]@{-};
(8.519636,-1.200000) *{\scriptstyle 367};
(8.543032,0.003013); (8.543032,0.668261) **[lightgray]@{-};
(8.543032,0.928374); (8.543032,1.290863) **[lightgray]@{-};
(8.543032,1.553897); (8.543032,3.571788) **[lightgray]@{-};
(8.543032,-0.100000) *{\scriptstyle 373};
(9.197217,0.003013); (9.197217,0.668261) **[lightgray]@{-};
(9.197217,0.928374); (9.197217,1.290863) **[lightgray]@{-};
(9.197217,1.553897); (9.197217,3.571788) **[lightgray]@{-};
(9.197217,-0.100000) *{\scriptstyle 587};
(1.584963,0.668261); (9.197217,0.668261) **@{-};
(10.497217,0.668261) *{\llap{6.182}};
(1.584963,0.509819); (9.197217,0.509819) **[lightgray]@{-};
(10.497217,0.509819) *{\llap{5.539}};
(1.584963,0.376564); (9.197217,0.376564) **[lightgray]@{-};
(10.497217,0.376564) *{\llap{5.050}};
(1.584963,0.252695); (9.197217,0.252695) **[lightgray]@{-};
(10.497217,0.252695) *{\llap{4.635}};
(1.584963,0.003013); (9.197217,0.003013) **@{-};
(10.497217,0.003013) *{\llap{3.898}};
(1.584963,1.290863); (9.197217,1.290863) **@{-};
(10.497217,1.290863) *{\llap{1628.033}};
(1.584963,1.170771); (9.197217,1.170771) **[lightgray]@{-};
(10.497217,1.170771) *{\llap{1498.000}};
(1.584963,0.928374); (9.197217,0.928374) **@{-};
(10.497217,0.928374) *{\llap{1266.319}};
(1.584963,3.571788); (9.197217,3.571788) **@{-};
(10.497217,3.571788) *{\llap{13935.714}};
(1.584963,3.218510); (9.197217,3.218510) **[lightgray]@{-};
(10.497217,3.218510) *{\llap{10908.923}};
(1.584963,3.075765); (9.197217,3.075765) **[lightgray]@{-};
(10.497217,3.075765) *{\llap{9881.238}};
(1.584963,2.823163); (9.197217,2.823163) **[lightgray]@{-};
(10.497217,2.823163) *{\llap{8294.129}};
(1.584963,2.682421); (9.197217,2.682421) **[lightgray]@{-};
(10.497217,2.682421) *{\llap{7523.209}};
(1.584963,2.522797); (9.197217,2.522797) **[lightgray]@{-};
(10.497217,2.522797) *{\llap{6735.217}};
(1.584963,2.341873); (9.197217,2.341873) **[lightgray]@{-};
(10.497217,2.341873) *{\llap{5941.388}};
(1.584963,2.212784); (9.197217,2.212784) **[lightgray]@{-};
(10.497217,2.212784) *{\llap{5432.857}};
(1.584963,2.091736); (9.197217,2.091736) **[lightgray]@{-};
(10.497217,2.091736) *{\llap{4995.616}};
(1.584963,1.961768); (9.197217,1.961768) **[lightgray]@{-};
(10.497217,1.961768) *{\llap{4565.253}};
(1.584963,1.801598); (9.197217,1.801598) **[lightgray]@{-};
(10.497217,1.801598) *{\llap{4085.535}};
(1.584963,1.553897); (9.197217,1.553897) **@{-};
(10.497217,1.553897) *{\llap{3440.992}};
(1.504963,0.625192); (1.664963,0.625192) **[blue]@{-};
(1.584963,0.585192); (1.584963,0.665192) **[blue]@{-};
(1.544963,0.625192); (1.624963,0.625192) **[blue]@{-};
(1.544963,0.625192); (1.624963,0.625192) **[blue]@{-};
(1.504963,0.585192); (1.664963,0.665192) **[red]@{-};
(1.664963,0.585192); (1.504963,0.665192) **[red]@{-};
(1.544963,0.605192); (1.624963,0.645192) **[red]@{-};
(1.544963,0.645192); (1.624963,0.605192) **[red]@{-};
(1.544963,0.605192); (1.624963,0.645192) **[red]@{-};
(1.544963,0.645192); (1.624963,0.605192) **[red]@{-};
(2.241928,0.480802); (2.401928,0.480802) **[blue]@{-};
(2.321928,0.440802); (2.321928,0.520802) **[blue]@{-};
(2.281928,0.480802); (2.361928,0.480802) **[blue]@{-};
(2.281928,0.480802); (2.361928,0.480802) **[blue]@{-};
(2.241928,0.440802); (2.401928,0.520802) **[red]@{-};
(2.401928,0.440802); (2.241928,0.520802) **[red]@{-};
(2.281928,0.460802); (2.361928,0.500802) **[red]@{-};
(2.281928,0.500802); (2.361928,0.460802) **[red]@{-};
(2.281928,0.460802); (2.361928,0.500802) **[red]@{-};
(2.281928,0.500802); (2.361928,0.460802) **[red]@{-};
(2.727355,0.570745); (2.887355,0.570745) **[blue]@{-};
(2.807355,0.530745); (2.807355,0.610745) **[blue]@{-};
(2.767355,0.570745); (2.847355,0.570745) **[blue]@{-};
(2.767355,0.570745); (2.847355,0.570745) **[blue]@{-};
(2.727355,0.530745); (2.887355,0.610745) **[red]@{-};
(2.887355,0.530745); (2.727355,0.610745) **[red]@{-};
(2.767355,0.550745); (2.847355,0.590745) **[red]@{-};
(2.767355,0.590745); (2.847355,0.550745) **[red]@{-};
(2.767355,0.550745); (2.847355,0.590745) **[red]@{-};
(2.767355,0.590745); (2.847355,0.550745) **[red]@{-};
(3.379432,0.625192); (3.539432,0.625192) **[blue]@{-};
(3.459432,0.585192); (3.459432,0.665192) **[blue]@{-};
(3.419432,0.625192); (3.499432,0.625192) **[blue]@{-};
(3.419432,0.625192); (3.499432,0.625192) **[blue]@{-};
(3.379432,0.585192); (3.539432,0.665192) **[red]@{-};
(3.539432,0.585192); (3.379432,0.665192) **[red]@{-};
(3.419432,0.605192); (3.499432,0.645192) **[red]@{-};
(3.419432,0.645192); (3.499432,0.605192) **[red]@{-};
(3.419432,0.605192); (3.499432,0.645192) **[red]@{-};
(3.419432,0.645192); (3.499432,0.605192) **[red]@{-};
(3.620440,0.625192); (3.780440,0.625192) **[blue]@{-};
(3.700440,0.585192); (3.700440,0.665192) **[blue]@{-};
(3.660440,0.625192); (3.740440,0.625192) **[blue]@{-};
(3.660440,0.625192); (3.740440,0.625192) **[blue]@{-};
(3.620440,0.585192); (3.780440,0.665192) **[red]@{-};
(3.780440,0.585192); (3.620440,0.665192) **[red]@{-};
(3.660440,0.605192); (3.740440,0.645192) **[red]@{-};
(3.660440,0.645192); (3.740440,0.605192) **[red]@{-};
(3.660440,0.605192); (3.740440,0.645192) **[red]@{-};
(3.660440,0.645192); (3.740440,0.605192) **[red]@{-};
(4.007463,0.625192); (4.167463,0.625192) **[blue]@{-};
(4.087463,0.585192); (4.087463,0.665192) **[blue]@{-};
(4.047463,0.625192); (4.127463,0.625192) **[blue]@{-};
(4.047463,0.625192); (4.127463,0.625192) **[blue]@{-};
(4.007463,0.585192); (4.167463,0.665192) **[red]@{-};
(4.167463,0.585192); (4.007463,0.665192) **[red]@{-};
(4.047463,0.605192); (4.127463,0.645192) **[red]@{-};
(4.047463,0.645192); (4.127463,0.605192) **[red]@{-};
(4.047463,0.605192); (4.127463,0.645192) **[red]@{-};
(4.047463,0.645192); (4.127463,0.605192) **[red]@{-};
(4.167928,0.647912); (4.327928,0.647912) **[blue]@{-};
(4.247928,0.607912); (4.247928,0.687912) **[blue]@{-};
(4.207928,0.647912); (4.287928,0.647912) **[blue]@{-};
(4.207928,0.647912); (4.287928,0.647912) **[blue]@{-};
(4.167928,0.607912); (4.327928,0.687912) **[red]@{-};
(4.327928,0.607912); (4.167928,0.687912) **[red]@{-};
(4.207928,0.627912); (4.287928,0.667912) **[red]@{-};
(4.207928,0.667912); (4.287928,0.627912) **[red]@{-};
(4.207928,0.627912); (4.287928,0.667912) **[red]@{-};
(4.207928,0.667912); (4.287928,0.627912) **[red]@{-};
(4.443562,0.663160); (4.603562,0.663160) **[blue]@{-};
(4.523562,0.623160); (4.523562,0.703160) **[blue]@{-};
(4.483562,0.663160); (4.563562,0.663160) **[blue]@{-};
(4.483562,0.663160); (4.563562,0.663160) **[blue]@{-};
(4.443562,0.623160); (4.603562,0.703160) **[red]@{-};
(4.603562,0.623160); (4.443562,0.703160) **[red]@{-};
(4.483562,0.643160); (4.563562,0.683160) **[red]@{-};
(4.483562,0.683160); (4.563562,0.643160) **[red]@{-};
(4.483562,0.643160); (4.563562,0.683160) **[red]@{-};
(4.483562,0.683160); (4.563562,0.643160) **[red]@{-};
(4.777981,0.655889); (4.937981,0.655889) **[blue]@{-};
(4.857981,0.615889); (4.857981,0.695889) **[blue]@{-};
(4.817981,0.655889); (4.897981,0.655889) **[blue]@{-};
(4.817981,0.655889); (4.897981,0.655889) **[blue]@{-};
(4.777981,0.615889); (4.937981,0.695889) **[red]@{-};
(4.937981,0.615889); (4.777981,0.695889) **[red]@{-};
(4.817981,0.635889); (4.897981,0.675889) **[red]@{-};
(4.817981,0.675889); (4.897981,0.635889) **[red]@{-};
(4.817981,0.635889); (4.897981,0.675889) **[red]@{-};
(4.817981,0.675889); (4.897981,0.635889) **[red]@{-};
(4.874196,0.668261); (5.034196,0.668261) **[blue]@{-};
(4.954196,0.628261); (4.954196,0.708261) **[blue]@{-};
(4.914196,0.668261); (4.994196,0.668261) **[blue]@{-};
(4.914196,0.668261); (4.994196,0.668261) **[blue]@{-};
(4.874196,0.628261); (5.034196,0.708261) **[red]@{-};
(5.034196,0.628261); (4.874196,0.708261) **[red]@{-};
(4.914196,0.648261); (4.994196,0.688261) **[red]@{-};
(4.914196,0.688261); (4.994196,0.648261) **[red]@{-};
(4.914196,0.648261); (4.994196,0.688261) **[red]@{-};
(4.914196,0.688261); (4.994196,0.648261) **[red]@{-};
(5.129453,0.649645); (5.289453,0.649645) **[blue]@{-};
(5.209453,0.609645); (5.209453,0.689645) **[blue]@{-};
(5.169453,0.649645); (5.249453,0.649645) **[blue]@{-};
(5.169453,0.649645); (5.249453,0.649645) **[blue]@{-};
(5.129453,0.609645); (5.289453,0.689645) **[red]@{-};
(5.289453,0.609645); (5.129453,0.689645) **[red]@{-};
(5.169453,0.629645); (5.249453,0.669645) **[red]@{-};
(5.169453,0.669645); (5.249453,0.629645) **[red]@{-};
(5.169453,0.629645); (5.249453,0.669645) **[red]@{-};
(5.169453,0.669645); (5.249453,0.629645) **[red]@{-};
(5.277552,0.647388); (5.437552,0.647388) **[blue]@{-};
(5.357552,0.607388); (5.357552,0.687388) **[blue]@{-};
(5.317552,0.647388); (5.397552,0.647388) **[blue]@{-};
(5.317552,0.647388); (5.397552,0.647388) **[blue]@{-};
(5.277552,0.607388); (5.437552,0.687388) **[red]@{-};
(5.437552,0.607388); (5.277552,0.687388) **[red]@{-};
(5.317552,0.627388); (5.397552,0.667388) **[red]@{-};
(5.317552,0.667388); (5.397552,0.627388) **[red]@{-};
(5.317552,0.627388); (5.397552,0.667388) **[red]@{-};
(5.317552,0.667388); (5.397552,0.627388) **[red]@{-};
(5.346265,0.656901); (5.506265,0.656901) **[blue]@{-};
(5.426265,0.616901); (5.426265,0.696901) **[blue]@{-};
(5.386265,0.656901); (5.466265,0.656901) **[blue]@{-};
(5.386265,0.656901); (5.466265,0.656901) **[blue]@{-};
(5.346265,0.616901); (5.506265,0.696901) **[red]@{-};
(5.506265,0.616901); (5.346265,0.696901) **[red]@{-};
(5.386265,0.636901); (5.466265,0.676901) **[red]@{-};
(5.386265,0.676901); (5.466265,0.636901) **[red]@{-};
(5.386265,0.636901); (5.466265,0.676901) **[red]@{-};
(5.386265,0.676901); (5.466265,0.636901) **[red]@{-};
(5.474589,0.663925); (5.634589,0.663925) **[blue]@{-};
(5.554589,0.623925); (5.554589,0.703925) **[blue]@{-};
(5.514589,0.663925); (5.594589,0.663925) **[blue]@{-};
(5.514589,0.663925); (5.594589,0.663925) **[blue]@{-};
(5.474589,0.623925); (5.634589,0.703925) **[red]@{-};
(5.634589,0.623925); (5.474589,0.703925) **[red]@{-};
(5.514589,0.643925); (5.594589,0.683925) **[red]@{-};
(5.514589,0.683925); (5.594589,0.643925) **[red]@{-};
(5.514589,0.643925); (5.594589,0.683925) **[red]@{-};
(5.514589,0.683925); (5.594589,0.643925) **[red]@{-};
(5.647920,0.651188); (5.807920,0.651188) **[blue]@{-};
(5.727920,0.611188); (5.727920,0.691188) **[blue]@{-};
(5.687920,0.651188); (5.767920,0.651188) **[blue]@{-};
(5.687920,0.651188); (5.767920,0.651188) **[blue]@{-};
(5.647920,0.611188); (5.807920,0.691188) **[red]@{-};
(5.807920,0.611188); (5.647920,0.691188) **[red]@{-};
(5.687920,0.631188); (5.767920,0.671188) **[red]@{-};
(5.687920,0.671188); (5.767920,0.631188) **[red]@{-};
(5.687920,0.631188); (5.767920,0.671188) **[red]@{-};
(5.687920,0.671188); (5.767920,0.631188) **[red]@{-};
(5.802643,0.656387); (5.962643,0.656387) **[blue]@{-};
(5.882643,0.616387); (5.882643,0.696387) **[blue]@{-};
(5.842643,0.656387); (5.922643,0.656387) **[blue]@{-};
(5.842643,0.656387); (5.922643,0.656387) **[blue]@{-};
(5.802643,0.616387); (5.962643,0.696387) **[red]@{-};
(5.962643,0.616387); (5.802643,0.696387) **[red]@{-};
(5.842643,0.636387); (5.922643,0.676387) **[red]@{-};
(5.842643,0.676387); (5.922643,0.636387) **[red]@{-};
(5.842643,0.636387); (5.922643,0.676387) **[red]@{-};
(5.842643,0.676387); (5.922643,0.636387) **[red]@{-};
(5.850737,0.655407); (6.010737,0.655407) **[blue]@{-};
(5.930737,0.615407); (5.930737,0.695407) **[blue]@{-};
(5.890737,0.655407); (5.970737,0.655407) **[blue]@{-};
(5.890737,0.655407); (5.970737,0.655407) **[blue]@{-};
(5.850737,0.615407); (6.010737,0.695407) **[red]@{-};
(6.010737,0.615407); (5.850737,0.695407) **[red]@{-};
(5.890737,0.635407); (5.970737,0.675407) **[red]@{-};
(5.890737,0.675407); (5.970737,0.635407) **[red]@{-};
(5.890737,0.635407); (5.970737,0.675407) **[red]@{-};
(5.890737,0.675407); (5.970737,0.635407) **[red]@{-};
(5.986089,0.645951); (6.146089,0.645951) **[blue]@{-};
(6.066089,0.605951); (6.066089,0.685951) **[blue]@{-};
(6.026089,0.645951); (6.106089,0.645951) **[blue]@{-};
(6.026089,0.645951); (6.106089,0.645951) **[blue]@{-};
(5.986089,0.605951); (6.146089,0.685951) **[red]@{-};
(6.146089,0.605951); (5.986089,0.685951) **[red]@{-};
(6.026089,0.625951); (6.106089,0.665951) **[red]@{-};
(6.026089,0.665951); (6.106089,0.625951) **[red]@{-};
(6.026089,0.625951); (6.106089,0.665951) **[red]@{-};
(6.026089,0.665951); (6.106089,0.625951) **[red]@{-};
(6.069747,0.651305); (6.229747,0.651305) **[blue]@{-};
(6.149747,0.611305); (6.149747,0.691305) **[blue]@{-};
(6.109747,0.651305); (6.189747,0.651305) **[blue]@{-};
(6.109747,0.651305); (6.189747,0.651305) **[blue]@{-};
(6.069747,0.611305); (6.229747,0.691305) **[red]@{-};
(6.229747,0.611305); (6.069747,0.691305) **[red]@{-};
(6.109747,0.631305); (6.189747,0.671305) **[red]@{-};
(6.109747,0.671305); (6.189747,0.631305) **[red]@{-};
(6.109747,0.631305); (6.189747,0.671305) **[red]@{-};
(6.109747,0.671305); (6.189747,0.631305) **[red]@{-};
(6.109825,0.644301); (6.269825,0.644301) **[blue]@{-};
(6.189825,0.604301); (6.189825,0.684301) **[blue]@{-};
(6.149825,0.644301); (6.229825,0.644301) **[blue]@{-};
(6.149825,0.644301); (6.229825,0.644301) **[blue]@{-};
(6.109825,0.604301); (6.269825,0.684301) **[red]@{-};
(6.269825,0.604301); (6.109825,0.684301) **[red]@{-};
(6.149825,0.624301); (6.229825,0.664301) **[red]@{-};
(6.149825,0.664301); (6.229825,0.624301) **[red]@{-};
(6.149825,0.624301); (6.229825,0.664301) **[red]@{-};
(6.149825,0.664301); (6.229825,0.624301) **[red]@{-};
(6.223781,0.654576); (6.383781,0.654576) **[blue]@{-};
(6.303781,0.614576); (6.303781,0.694576) **[blue]@{-};
(6.263781,0.654576); (6.343781,0.654576) **[blue]@{-};
(6.263781,0.654576); (6.343781,0.654576) **[blue]@{-};
(6.223781,0.614576); (6.383781,0.694576) **[red]@{-};
(6.383781,0.614576); (6.223781,0.694576) **[red]@{-};
(6.263781,0.634576); (6.343781,0.674576) **[red]@{-};
(6.263781,0.674576); (6.343781,0.634576) **[red]@{-};
(6.263781,0.634576); (6.343781,0.674576) **[red]@{-};
(6.263781,0.674576); (6.343781,0.634576) **[red]@{-};
(6.295039,0.647647); (6.455039,0.647647) **[blue]@{-};
(6.375039,0.607647); (6.375039,0.687647) **[blue]@{-};
(6.335039,0.647647); (6.415039,0.647647) **[blue]@{-};
(6.335039,0.647647); (6.415039,0.647647) **[blue]@{-};
(6.295039,0.607647); (6.455039,0.687647) **[red]@{-};
(6.455039,0.607647); (6.295039,0.687647) **[red]@{-};
(6.335039,0.627647); (6.415039,0.667647) **[red]@{-};
(6.335039,0.667647); (6.415039,0.627647) **[red]@{-};
(6.335039,0.627647); (6.415039,0.667647) **[red]@{-};
(6.335039,0.667647); (6.415039,0.627647) **[red]@{-};
(6.395733,0.646177); (6.555733,0.646177) **[blue]@{-};
(6.475733,0.606177); (6.475733,0.686177) **[blue]@{-};
(6.435733,0.646177); (6.515733,0.646177) **[blue]@{-};
(6.435733,0.646177); (6.515733,0.646177) **[blue]@{-};
(6.395733,0.606177); (6.555733,0.686177) **[red]@{-};
(6.555733,0.606177); (6.395733,0.686177) **[red]@{-};
(6.435733,0.626177); (6.515733,0.666177) **[red]@{-};
(6.435733,0.666177); (6.515733,0.626177) **[red]@{-};
(6.435733,0.626177); (6.515733,0.666177) **[red]@{-};
(6.435733,0.666177); (6.515733,0.626177) **[red]@{-};
(6.519913,0.639692); (6.679913,0.639692) **[blue]@{-};
(6.599913,0.599692); (6.599913,0.679692) **[blue]@{-};
(6.559913,0.639692); (6.639913,0.639692) **[blue]@{-};
(6.559913,0.639692); (6.639913,0.639692) **[blue]@{-};
(6.519913,0.599692); (6.679913,0.679692) **[red]@{-};
(6.679913,0.599692); (6.519913,0.679692) **[red]@{-};
(6.559913,0.619692); (6.639913,0.659692) **[red]@{-};
(6.559913,0.659692); (6.639913,0.619692) **[red]@{-};
(6.559913,0.619692); (6.639913,0.659692) **[red]@{-};
(6.559913,0.659692); (6.639913,0.619692) **[red]@{-};
(6.578211,0.643748); (6.738211,0.643748) **[blue]@{-};
(6.658211,0.603748); (6.658211,0.683748) **[blue]@{-};
(6.618211,0.643748); (6.698211,0.643748) **[blue]@{-};
(6.618211,0.643748); (6.698211,0.643748) **[blue]@{-};
(6.578211,0.603748); (6.738211,0.683748) **[red]@{-};
(6.738211,0.603748); (6.578211,0.683748) **[red]@{-};
(6.618211,0.623748); (6.698211,0.663748) **[red]@{-};
(6.618211,0.663748); (6.698211,0.623748) **[red]@{-};
(6.618211,0.623748); (6.698211,0.663748) **[red]@{-};
(6.618211,0.663748); (6.698211,0.623748) **[red]@{-};
(6.606501,0.647912); (6.766501,0.647912) **[blue]@{-};
(6.686501,0.607912); (6.686501,0.687912) **[blue]@{-};
(6.646501,0.647912); (6.726501,0.647912) **[blue]@{-};
(6.646501,0.647912); (6.726501,0.647912) **[blue]@{-};
(6.606501,0.607912); (6.766501,0.687912) **[red]@{-};
(6.766501,0.607912); (6.606501,0.687912) **[red]@{-};
(6.646501,0.627912); (6.726501,0.667912) **[red]@{-};
(6.646501,0.667912); (6.726501,0.627912) **[red]@{-};
(6.646501,0.627912); (6.726501,0.667912) **[red]@{-};
(6.646501,0.667912); (6.726501,0.627912) **[red]@{-};
(6.661467,0.647085); (6.821467,0.647085) **[blue]@{-};
(6.741467,0.607085); (6.741467,0.687085) **[blue]@{-};
(6.701467,0.647085); (6.781467,0.647085) **[blue]@{-};
(6.701467,0.647085); (6.781467,0.647085) **[blue]@{-};
(6.661467,0.607085); (6.821467,0.687085) **[red]@{-};
(6.821467,0.607085); (6.661467,0.687085) **[red]@{-};
(6.701467,0.627085); (6.781467,0.667085) **[red]@{-};
(6.701467,0.667085); (6.781467,0.627085) **[red]@{-};
(6.701467,0.627085); (6.781467,0.667085) **[red]@{-};
(6.701467,0.667085); (6.781467,0.627085) **[red]@{-};
(6.688184,0.646693); (6.848184,0.646693) **[blue]@{-};
(6.768184,0.606693); (6.768184,0.686693) **[blue]@{-};
(6.728184,0.646693); (6.808184,0.646693) **[blue]@{-};
(6.728184,0.646693); (6.808184,0.646693) **[blue]@{-};
(6.688184,0.606693); (6.848184,0.686693) **[red]@{-};
(6.848184,0.606693); (6.688184,0.686693) **[red]@{-};
(6.728184,0.626693); (6.808184,0.666693) **[red]@{-};
(6.728184,0.666693); (6.808184,0.626693) **[red]@{-};
(6.728184,0.626693); (6.808184,0.666693) **[red]@{-};
(6.728184,0.666693); (6.808184,0.626693) **[red]@{-};
(6.740179,0.641823); (6.900179,0.641823) **[blue]@{-};
(6.820179,0.601823); (6.820179,0.681823) **[blue]@{-};
(6.780179,0.641823); (6.860179,0.641823) **[blue]@{-};
(6.780179,0.641823); (6.860179,0.641823) **[blue]@{-};
(6.740179,0.601823); (6.900179,0.681823) **[red]@{-};
(6.900179,0.601823); (6.740179,0.681823) **[red]@{-};
(6.780179,0.621823); (6.860179,0.661823) **[red]@{-};
(6.780179,0.661823); (6.860179,0.621823) **[red]@{-};
(6.780179,0.621823); (6.860179,0.661823) **[red]@{-};
(6.780179,0.661823); (6.860179,0.621823) **[red]@{-};
(6.908685,0.651054); (7.068685,0.651054) **[blue]@{-};
(6.988685,0.611054); (6.988685,0.691054) **[blue]@{-};
(6.948685,0.651054); (7.028685,0.651054) **[blue]@{-};
(6.948685,0.651054); (7.028685,0.651054) **[blue]@{-};
(6.908685,0.600027); (7.068685,0.680027) **[red]@{-};
(7.068685,0.600027); (6.908685,0.680027) **[red]@{-};
(6.948685,0.620027); (7.028685,0.660027) **[red]@{-};
(6.948685,0.660027); (7.028685,0.620027) **[red]@{-};
(6.948685,0.620027); (7.028685,0.660027) **[red]@{-};
(6.948685,0.660027); (7.028685,0.620027) **[red]@{-};
(6.953423,0.639583); (7.113423,0.639583) **[blue]@{-};
(7.033423,0.599583); (7.033423,0.679583) **[blue]@{-};
(6.993423,0.639583); (7.073423,0.639583) **[blue]@{-};
(6.993423,0.639583); (7.073423,0.639583) **[blue]@{-};
(6.953423,0.524263); (7.113423,0.604263) **[red]@{-};
(7.113423,0.524263); (6.953423,0.604263) **[red]@{-};
(6.993423,0.544263); (7.073423,0.584263) **[red]@{-};
(6.993423,0.584263); (7.073423,0.544263) **[red]@{-};
(6.993423,0.544263); (7.073423,0.584263) **[red]@{-};
(6.993423,0.584263); (7.073423,0.544263) **[red]@{-};
(7.018032,0.638965); (7.178032,0.638965) **[blue]@{-};
(7.098032,0.598965); (7.098032,0.678965) **[blue]@{-};
(7.058032,0.638965); (7.138032,0.638965) **[blue]@{-};
(7.058032,0.638965); (7.138032,0.638965) **[blue]@{-};
(7.018032,0.526947); (7.178032,0.606947) **[red]@{-};
(7.178032,0.526947); (7.018032,0.606947) **[red]@{-};
(7.058032,0.546947); (7.138032,0.586947) **[red]@{-};
(7.058032,0.586947); (7.138032,0.546947) **[red]@{-};
(7.058032,0.546947); (7.138032,0.586947) **[red]@{-};
(7.058032,0.586947); (7.138032,0.546947) **[red]@{-};
(7.038941,0.642145); (7.198941,0.642145) **[blue]@{-};
(7.118941,0.602145); (7.118941,0.682145) **[blue]@{-};
(7.078941,0.642145); (7.158941,0.642145) **[blue]@{-};
(7.078941,0.642145); (7.158941,0.642145) **[blue]@{-};
(7.038941,0.531335); (7.198941,0.611335) **[red]@{-};
(7.198941,0.531335); (7.038941,0.611335) **[red]@{-};
(7.078941,0.551335); (7.158941,0.591335) **[red]@{-};
(7.078941,0.591335); (7.158941,0.551335) **[red]@{-};
(7.078941,0.551335); (7.158941,0.591335) **[red]@{-};
(7.078941,0.591335); (7.158941,0.551335) **[red]@{-};
(7.139169,0.641029); (7.299169,0.641029) **[blue]@{-};
(7.219169,0.601029); (7.219169,0.681029) **[blue]@{-};
(7.179169,0.641029); (7.259169,0.641029) **[blue]@{-};
(7.179169,0.641029); (7.259169,0.641029) **[blue]@{-};
(7.139169,0.525037); (7.299169,0.605037) **[red]@{-};
(7.299169,0.525037); (7.139169,0.605037) **[red]@{-};
(7.179169,0.545037); (7.259169,0.585037) **[red]@{-};
(7.179169,0.585037); (7.259169,0.545037) **[red]@{-};
(7.179169,0.545037); (7.259169,0.585037) **[red]@{-};
(7.179169,0.585037); (7.259169,0.545037) **[red]@{-};
(7.158405,0.643929); (7.318405,0.643929) **[blue]@{-};
(7.238405,0.603929); (7.238405,0.683929) **[blue]@{-};
(7.198405,0.643929); (7.278405,0.643929) **[blue]@{-};
(7.198405,0.643929); (7.278405,0.643929) **[blue]@{-};
(7.158405,0.529112); (7.318405,0.609112) **[red]@{-};
(7.318405,0.529112); (7.158405,0.609112) **[red]@{-};
(7.198405,0.549112); (7.278405,0.589112) **[red]@{-};
(7.198405,0.589112); (7.278405,0.549112) **[red]@{-};
(7.198405,0.549112); (7.278405,0.589112) **[red]@{-};
(7.198405,0.589112); (7.278405,0.549112) **[red]@{-};
(7.214621,0.643226); (7.374621,0.643226) **[blue]@{-};
(7.294621,0.603226); (7.294621,0.683226) **[blue]@{-};
(7.254621,0.643226); (7.334621,0.643226) **[blue]@{-};
(7.254621,0.643226); (7.334621,0.643226) **[blue]@{-};
(7.214621,0.531268); (7.374621,0.611268) **[red]@{-};
(7.374621,0.531268); (7.214621,0.611268) **[red]@{-};
(7.254621,0.551268); (7.334621,0.591268) **[red]@{-};
(7.254621,0.591268); (7.334621,0.551268) **[red]@{-};
(7.254621,0.551268); (7.334621,0.591268) **[red]@{-};
(7.254621,0.591268); (7.334621,0.551268) **[red]@{-};
(7.268728,0.639692); (7.428728,0.639692) **[blue]@{-};
(7.348728,0.599692); (7.348728,0.679692) **[blue]@{-};
(7.308728,0.639692); (7.388728,0.639692) **[blue]@{-};
(7.308728,0.639692); (7.388728,0.639692) **[blue]@{-};
(7.268728,0.449291); (7.428728,0.529291) **[red]@{-};
(7.428728,0.449291); (7.268728,0.529291) **[red]@{-};
(7.308728,0.469291); (7.388728,0.509291) **[red]@{-};
(7.308728,0.509291); (7.388728,0.469291) **[red]@{-};
(7.308728,0.469291); (7.388728,0.509291) **[red]@{-};
(7.308728,0.509291); (7.388728,0.469291) **[red]@{-};
(7.303704,0.642165); (7.463704,0.642165) **[blue]@{-};
(7.383704,0.602165); (7.383704,0.682165) **[blue]@{-};
(7.343704,0.642165); (7.423704,0.642165) **[blue]@{-};
(7.343704,0.642165); (7.423704,0.642165) **[blue]@{-};
(7.303704,0.455776); (7.463704,0.535776) **[red]@{-};
(7.463704,0.455776); (7.303704,0.535776) **[red]@{-};
(7.343704,0.475776); (7.423704,0.515776) **[red]@{-};
(7.343704,0.515776); (7.423704,0.475776) **[red]@{-};
(7.343704,0.475776); (7.423704,0.515776) **[red]@{-};
(7.343704,0.515776); (7.423704,0.475776) **[red]@{-};
(7.354628,0.641587); (7.514628,0.641587) **[blue]@{-};
(7.434628,0.601587); (7.434628,0.681587) **[blue]@{-};
(7.394628,0.641587); (7.474628,0.641587) **[blue]@{-};
(7.394628,0.641587); (7.474628,0.641587) **[blue]@{-};
(7.354628,0.460411); (7.514628,0.540411) **[red]@{-};
(7.514628,0.460411); (7.354628,0.540411) **[red]@{-};
(7.394628,0.480411); (7.474628,0.520411) **[red]@{-};
(7.394628,0.520411); (7.474628,0.480411) **[red]@{-};
(7.394628,0.480411); (7.474628,0.520411) **[red]@{-};
(7.394628,0.520411); (7.474628,0.480411) **[red]@{-};
(7.403816,0.641046); (7.563816,0.641046) **[blue]@{-};
(7.483816,0.601046); (7.483816,0.681046) **[blue]@{-};
(7.443816,0.641046); (7.523816,0.641046) **[blue]@{-};
(7.443816,0.641046); (7.523816,0.641046) **[blue]@{-};
(7.403816,0.456034); (7.563816,0.536034) **[red]@{-};
(7.563816,0.456034); (7.403816,0.536034) **[red]@{-};
(7.443816,0.476034); (7.523816,0.516034) **[red]@{-};
(7.443816,0.516034); (7.523816,0.476034) **[red]@{-};
(7.443816,0.476034); (7.523816,0.516034) **[red]@{-};
(7.443816,0.516034); (7.523816,0.476034) **[red]@{-};
(7.419846,0.640874); (7.579846,0.640874) **[blue]@{-};
(7.499846,0.600874); (7.499846,0.680874) **[blue]@{-};
(7.459846,0.640874); (7.539846,0.640874) **[blue]@{-};
(7.459846,0.640874); (7.539846,0.640874) **[blue]@{-};
(7.419846,0.457510); (7.579846,0.537510) **[red]@{-};
(7.579846,0.457510); (7.419846,0.537510) **[red]@{-};
(7.459846,0.477510); (7.539846,0.517510) **[red]@{-};
(7.459846,0.517510); (7.539846,0.477510) **[red]@{-};
(7.459846,0.477510); (7.539846,0.517510) **[red]@{-};
(7.459846,0.517510); (7.539846,0.477510) **[red]@{-};
(7.497429,0.644990); (7.657429,0.644990) **[blue]@{-};
(7.577429,0.604990); (7.577429,0.684990) **[blue]@{-};
(7.537429,0.644990); (7.617429,0.644990) **[blue]@{-};
(7.537429,0.644990); (7.617429,0.644990) **[blue]@{-};
(7.497429,0.469819); (7.657429,0.549819) **[red]@{-};
(7.657429,0.469819); (7.497429,0.549819) **[red]@{-};
(7.537429,0.489819); (7.617429,0.529819) **[red]@{-};
(7.537429,0.529819); (7.617429,0.489819) **[red]@{-};
(7.537429,0.489819); (7.617429,0.529819) **[red]@{-};
(7.537429,0.529819); (7.617429,0.489819) **[red]@{-};
(7.512457,0.635023); (7.672457,0.635023) **[blue]@{-};
(7.592457,0.595023); (7.592457,0.675023) **[blue]@{-};
(7.552457,0.635023); (7.632457,0.635023) **[blue]@{-};
(7.552457,0.635023); (7.632457,0.635023) **[blue]@{-};
(7.512457,0.378741); (7.672457,0.458741) **[red]@{-};
(7.672457,0.378741); (7.512457,0.458741) **[red]@{-};
(7.552457,0.398741); (7.632457,0.438741) **[red]@{-};
(7.552457,0.438741); (7.632457,0.398741) **[red]@{-};
(7.552457,0.398741); (7.632457,0.438741) **[red]@{-};
(7.552457,0.438741); (7.632457,0.398741) **[red]@{-};
(7.542052,0.637225); (7.702052,0.637225) **[blue]@{-};
(7.622052,0.597225); (7.622052,0.677225) **[blue]@{-};
(7.582052,0.637225); (7.662052,0.637225) **[blue]@{-};
(7.582052,0.637225); (7.662052,0.637225) **[blue]@{-};
(7.542052,0.377626); (7.702052,0.457626) **[red]@{-};
(7.702052,0.377626); (7.542052,0.457626) **[red]@{-};
(7.582052,0.397626); (7.662052,0.437626) **[red]@{-};
(7.582052,0.437626); (7.662052,0.397626) **[red]@{-};
(7.582052,0.397626); (7.662052,0.437626) **[red]@{-};
(7.582052,0.437626); (7.662052,0.397626) **[red]@{-};
(7.556625,0.639477); (7.716625,0.639477) **[blue]@{-};
(7.636625,0.599477); (7.636625,0.679477) **[blue]@{-};
(7.596625,0.639477); (7.676625,0.639477) **[blue]@{-};
(7.596625,0.639477); (7.676625,0.639477) **[blue]@{-};
(7.556625,0.382601); (7.716625,0.462601) **[red]@{-};
(7.716625,0.382601); (7.556625,0.462601) **[red]@{-};
(7.596625,0.402601); (7.676625,0.442601) **[red]@{-};
(7.596625,0.442601); (7.676625,0.402601) **[red]@{-};
(7.596625,0.402601); (7.676625,0.442601) **[red]@{-};
(7.596625,0.442601); (7.676625,0.402601) **[red]@{-};
(7.641099,0.638676); (7.801099,0.638676) **[blue]@{-};
(7.721099,0.598676); (7.721099,0.678676) **[blue]@{-};
(7.681099,0.638676); (7.761099,0.638676) **[blue]@{-};
(7.681099,0.638676); (7.761099,0.638676) **[blue]@{-};
(7.641099,0.387054); (7.801099,0.467054) **[red]@{-};
(7.801099,0.387054); (7.641099,0.467054) **[red]@{-};
(7.681099,0.407054); (7.761099,0.447054) **[red]@{-};
(7.681099,0.447054); (7.761099,0.407054) **[red]@{-};
(7.681099,0.407054); (7.761099,0.447054) **[red]@{-};
(7.681099,0.447054); (7.761099,0.407054) **[red]@{-};
(7.720900,0.642190); (7.880900,0.642190) **[blue]@{-};
(7.800900,0.602190); (7.800900,0.682190) **[blue]@{-};
(7.760900,0.642190); (7.840900,0.642190) **[blue]@{-};
(7.760900,0.642190); (7.840900,0.642190) **[blue]@{-};
(7.720900,0.403195); (7.880900,0.483195) **[red]@{-};
(7.880900,0.403195); (7.720900,0.483195) **[red]@{-};
(7.760900,0.423195); (7.840900,0.463195) **[red]@{-};
(7.760900,0.463195); (7.840900,0.423195) **[red]@{-};
(7.760900,0.423195); (7.840900,0.463195) **[red]@{-};
(7.760900,0.463195); (7.840900,0.423195) **[red]@{-};
(7.746548,0.637738); (7.906548,0.637738) **[blue]@{-};
(7.826548,0.597738); (7.826548,0.677738) **[blue]@{-};
(7.786548,0.637738); (7.866548,0.637738) **[blue]@{-};
(7.786548,0.637738); (7.866548,0.637738) **[blue]@{-};
(7.746548,0.393447); (7.906548,0.473447) **[red]@{-};
(7.906548,0.393447); (7.746548,0.473447) **[red]@{-};
(7.786548,0.413447); (7.866548,0.453447) **[red]@{-};
(7.786548,0.453447); (7.866548,0.413447) **[red]@{-};
(7.786548,0.413447); (7.866548,0.453447) **[red]@{-};
(7.786548,0.453447); (7.866548,0.413447) **[red]@{-};
(7.759204,0.637629); (7.919204,0.637629) **[blue]@{-};
(7.839204,0.597629); (7.839204,0.677629) **[blue]@{-};
(7.799204,0.637629); (7.879204,0.637629) **[blue]@{-};
(7.799204,0.637629); (7.879204,0.637629) **[blue]@{-};
(7.759204,0.395221); (7.919204,0.475221) **[red]@{-};
(7.919204,0.395221); (7.759204,0.475221) **[red]@{-};
(7.799204,0.415221); (7.879204,0.455221) **[red]@{-};
(7.799204,0.455221); (7.879204,0.415221) **[red]@{-};
(7.799204,0.415221); (7.879204,0.455221) **[red]@{-};
(7.799204,0.455221); (7.879204,0.415221) **[red]@{-};
(7.784186,0.637419); (7.944186,0.637419) **[blue]@{-};
(7.864186,0.597419); (7.864186,0.677419) **[blue]@{-};
(7.824186,0.637419); (7.904186,0.637419) **[blue]@{-};
(7.824186,0.637419); (7.904186,0.637419) **[blue]@{-};
(7.784186,0.398673); (7.944186,0.478673) **[red]@{-};
(7.944186,0.398673); (7.784186,0.478673) **[red]@{-};
(7.824186,0.418673); (7.904186,0.458673) **[red]@{-};
(7.824186,0.458673); (7.904186,0.418673) **[red]@{-};
(7.824186,0.418673); (7.904186,0.458673) **[red]@{-};
(7.824186,0.458673); (7.904186,0.418673) **[red]@{-};
(7.820867,0.641068); (7.980867,0.641068) **[blue]@{-};
(7.900867,0.601068); (7.900867,0.681068) **[blue]@{-};
(7.860867,0.641068); (7.940867,0.641068) **[blue]@{-};
(7.860867,0.641068); (7.940867,0.641068) **[blue]@{-};
(7.820867,0.408141); (7.980867,0.488141) **[red]@{-};
(7.980867,0.408141); (7.820867,0.488141) **[red]@{-};
(7.860867,0.428141); (7.940867,0.468141) **[red]@{-};
(7.860867,0.468141); (7.940867,0.428141) **[red]@{-};
(7.860867,0.428141); (7.940867,0.468141) **[red]@{-};
(7.860867,0.468141); (7.940867,0.428141) **[red]@{-};
(7.832889,0.637018); (7.992889,0.637018) **[blue]@{-};
(7.912889,0.597018); (7.912889,0.677018) **[blue]@{-};
(7.872889,0.637018); (7.952889,0.637018) **[blue]@{-};
(7.872889,0.637018); (7.952889,0.637018) **[blue]@{-};
(7.832889,0.310235); (7.992889,0.390235) **[red]@{-};
(7.992889,0.310235); (7.832889,0.390235) **[red]@{-};
(7.872889,0.330235); (7.952889,0.370235) **[red]@{-};
(7.872889,0.370235); (7.952889,0.330235) **[red]@{-};
(7.872889,0.330235); (7.952889,0.370235) **[red]@{-};
(7.872889,0.370235); (7.952889,0.330235) **[red]@{-};
(7.891544,0.640319); (8.051544,0.640319) **[blue]@{-};
(7.971544,0.600319); (7.971544,0.680319) **[blue]@{-};
(7.931544,0.640319); (8.011544,0.640319) **[blue]@{-};
(7.931544,0.640319); (8.011544,0.640319) **[blue]@{-};
(7.891544,0.319875); (8.051544,0.399875) **[red]@{-};
(8.051544,0.319875); (7.891544,0.399875) **[red]@{-};
(7.931544,0.339875); (8.011544,0.379875) **[red]@{-};
(7.931544,0.379875); (8.011544,0.339875) **[red]@{-};
(7.931544,0.339875); (8.011544,0.379875) **[red]@{-};
(7.931544,0.379875); (8.011544,0.339875) **[red]@{-};
(7.925625,0.632600); (8.085625,0.632600) **[blue]@{-};
(8.005625,0.592600); (8.005625,0.672600) **[blue]@{-};
(7.965625,0.632600); (8.045625,0.632600) **[blue]@{-};
(7.965625,0.632600); (8.045625,0.632600) **[blue]@{-};
(7.925625,0.317695); (8.085625,0.397695) **[red]@{-};
(8.085625,0.317695); (7.925625,0.397695) **[red]@{-};
(7.965625,0.337695); (8.045625,0.377695) **[red]@{-};
(7.965625,0.377695); (8.045625,0.337695) **[red]@{-};
(7.965625,0.337695); (8.045625,0.377695) **[red]@{-};
(7.965625,0.377695); (8.045625,0.337695) **[red]@{-};
(7.958919,0.636040); (8.118919,0.636040) **[blue]@{-};
(8.038919,0.596040); (8.038919,0.676040) **[blue]@{-};
(7.998919,0.636040); (8.078919,0.636040) **[blue]@{-};
(7.998919,0.636040); (8.078919,0.636040) **[blue]@{-};
(7.958919,0.322158); (8.118919,0.402158) **[red]@{-};
(8.118919,0.322158); (7.958919,0.402158) **[red]@{-};
(7.998919,0.342158); (8.078919,0.382158) **[red]@{-};
(7.998919,0.382158); (8.078919,0.342158) **[red]@{-};
(7.998919,0.342158); (8.078919,0.382158) **[red]@{-};
(7.998919,0.382158); (8.078919,0.342158) **[red]@{-};
(7.991462,0.635800); (8.151462,0.635800) **[blue]@{-};
(8.071462,0.595800); (8.071462,0.675800) **[blue]@{-};
(8.031462,0.635800); (8.111462,0.635800) **[blue]@{-};
(8.031462,0.635800); (8.111462,0.635800) **[blue]@{-};
(7.991462,0.328532); (8.151462,0.408532) **[red]@{-};
(8.151462,0.328532); (7.991462,0.408532) **[red]@{-};
(8.031462,0.348532); (8.111462,0.388532) **[red]@{-};
(8.031462,0.388532); (8.111462,0.348532) **[red]@{-};
(8.031462,0.348532); (8.111462,0.388532) **[red]@{-};
(8.031462,0.388532); (8.111462,0.348532) **[red]@{-};
(8.002149,0.637471); (8.162149,0.637471) **[blue]@{-};
(8.082149,0.597471); (8.082149,0.677471) **[blue]@{-};
(8.042149,0.637471); (8.122149,0.637471) **[blue]@{-};
(8.042149,0.637471); (8.122149,0.637471) **[blue]@{-};
(8.002149,0.332689); (8.162149,0.412689) **[red]@{-};
(8.162149,0.332689); (8.002149,0.412689) **[red]@{-};
(8.042149,0.352689); (8.122149,0.392689) **[red]@{-};
(8.042149,0.392689); (8.122149,0.352689) **[red]@{-};
(8.042149,0.352689); (8.122149,0.392689) **[red]@{-};
(8.042149,0.392689); (8.122149,0.352689) **[red]@{-};
(8.033742,0.635497); (8.193742,0.635497) **[blue]@{-};
(8.113742,0.595497); (8.113742,0.675497) **[blue]@{-};
(8.073742,0.635497); (8.153742,0.635497) **[blue]@{-};
(8.073742,0.635497); (8.153742,0.635497) **[blue]@{-};
(8.033742,0.336564); (8.193742,0.416564) **[red]@{-};
(8.193742,0.336564); (8.033742,0.416564) **[red]@{-};
(8.073742,0.356564); (8.153742,0.396564) **[red]@{-};
(8.073742,0.396564); (8.153742,0.356564) **[red]@{-};
(8.073742,0.356564); (8.153742,0.396564) **[red]@{-};
(8.073742,0.396564); (8.153742,0.356564) **[red]@{-};
(8.054426,0.635352); (8.214426,0.635352) **[blue]@{-};
(8.134426,0.595352); (8.134426,0.675352) **[blue]@{-};
(8.094426,0.635352); (8.174426,0.635352) **[blue]@{-};
(8.094426,0.635352); (8.174426,0.635352) **[blue]@{-};
(8.054426,0.290200); (8.214426,0.370200) **[red]@{-};
(8.214426,0.290200); (8.054426,0.370200) **[red]@{-};
(8.094426,0.310200); (8.174426,0.350200) **[red]@{-};
(8.094426,0.350200); (8.174426,0.310200) **[red]@{-};
(8.094426,0.310200); (8.174426,0.350200) **[red]@{-};
(8.094426,0.350200); (8.174426,0.310200) **[red]@{-};
(8.064658,0.636956); (8.224658,0.636956) **[blue]@{-};
(8.144658,0.596956); (8.144658,0.676956) **[blue]@{-};
(8.104658,0.636956); (8.184658,0.636956) **[blue]@{-};
(8.104658,0.636956); (8.184658,0.636956) **[blue]@{-};
(8.064658,0.288355); (8.224658,0.368355) **[red]@{-};
(8.224658,0.288355); (8.064658,0.368355) **[red]@{-};
(8.104658,0.308355); (8.184658,0.348355) **[red]@{-};
(8.104658,0.348355); (8.184658,0.308355) **[red]@{-};
(8.104658,0.308355); (8.184658,0.348355) **[red]@{-};
(8.104658,0.348355); (8.184658,0.308355) **[red]@{-};
(8.114757,0.634940); (8.274757,0.634940) **[blue]@{-};
(8.194757,0.594940); (8.194757,0.674940) **[blue]@{-};
(8.154757,0.634940); (8.234757,0.634940) **[blue]@{-};
(8.154757,0.634940); (8.234757,0.634940) **[blue]@{-};
(8.114757,0.254065); (8.274757,0.334065) **[red]@{-};
(8.274757,0.254065); (8.114757,0.334065) **[red]@{-};
(8.154757,0.274065); (8.234757,0.314065) **[red]@{-};
(8.154757,0.314065); (8.234757,0.274065) **[red]@{-};
(8.154757,0.274065); (8.234757,0.314065) **[red]@{-};
(8.154757,0.314065); (8.234757,0.274065) **[red]@{-};
(8.182095,0.636046); (8.342095,0.636046) **[blue]@{-};
(8.262095,0.596046); (8.262095,0.676046) **[blue]@{-};
(8.222095,0.636046); (8.302095,0.636046) **[blue]@{-};
(8.222095,0.636046); (8.302095,0.636046) **[blue]@{-};
(8.182095,0.272765); (8.342095,0.352765) **[red]@{-};
(8.342095,0.272765); (8.182095,0.352765) **[red]@{-};
(8.222095,0.292765); (8.302095,0.332765) **[red]@{-};
(8.222095,0.332765); (8.302095,0.292765) **[red]@{-};
(8.222095,0.292765); (8.302095,0.332765) **[red]@{-};
(8.222095,0.332765); (8.302095,0.292765) **[red]@{-};
(8.200771,0.637432); (8.360771,0.637432) **[blue]@{-};
(8.280771,0.597432); (8.280771,0.677432) **[blue]@{-};
(8.240771,0.637432); (8.320771,0.637432) **[blue]@{-};
(8.240771,0.637432); (8.320771,0.637432) **[blue]@{-};
(8.200771,0.279117); (8.360771,0.359117) **[red]@{-};
(8.360771,0.279117); (8.200771,0.359117) **[red]@{-};
(8.240771,0.299117); (8.320771,0.339117) **[red]@{-};
(8.240771,0.339117); (8.320771,0.299117) **[red]@{-};
(8.240771,0.299117); (8.320771,0.339117) **[red]@{-};
(8.240771,0.339117); (8.320771,0.299117) **[red]@{-};
(8.210019,0.635840); (8.370019,0.635840) **[blue]@{-};
(8.290019,0.595840); (8.290019,0.675840) **[blue]@{-};
(8.250019,0.635840); (8.330019,0.635840) **[blue]@{-};
(8.250019,0.635840); (8.330019,0.635840) **[blue]@{-};
(8.210019,0.273722); (8.370019,0.353722) **[red]@{-};
(8.370019,0.273722); (8.210019,0.353722) **[red]@{-};
(8.250019,0.293722); (8.330019,0.333722) **[red]@{-};
(8.250019,0.333722); (8.330019,0.293722) **[red]@{-};
(8.250019,0.293722); (8.330019,0.333722) **[red]@{-};
(8.250019,0.333722); (8.330019,0.293722) **[red]@{-};
(8.228339,0.637202); (8.388339,0.637202) **[blue]@{-};
(8.308339,0.597202); (8.308339,0.677202) **[blue]@{-};
(8.268339,0.637202); (8.348339,0.637202) **[blue]@{-};
(8.268339,0.637202); (8.348339,0.637202) **[blue]@{-};
(8.228339,0.279939); (8.388339,0.359939) **[red]@{-};
(8.388339,0.279939); (8.228339,0.359939) **[red]@{-};
(8.268339,0.299939); (8.348339,0.339939) **[red]@{-};
(8.268339,0.339939); (8.348339,0.299939) **[red]@{-};
(8.268339,0.299939); (8.348339,0.339939) **[red]@{-};
(8.268339,0.339939); (8.348339,0.299939) **[red]@{-};
(8.290687,0.635266); (8.450687,0.635266) **[blue]@{-};
(8.370687,0.595266); (8.370687,0.675266) **[blue]@{-};
(8.330687,0.635266); (8.410687,0.635266) **[blue]@{-};
(8.330687,0.635266); (8.410687,0.635266) **[blue]@{-};
(8.290687,0.239237); (8.450687,0.319237) **[red]@{-};
(8.450687,0.239237); (8.290687,0.319237) **[red]@{-};
(8.330687,0.259237); (8.410687,0.299237) **[red]@{-};
(8.330687,0.299237); (8.410687,0.259237) **[red]@{-};
(8.330687,0.259237); (8.410687,0.299237) **[red]@{-};
(8.330687,0.299237); (8.410687,0.259237) **[red]@{-};
(8.316605,0.633679); (8.476605,0.633679) **[blue]@{-};
(8.396605,0.593679); (8.396605,0.673679) **[blue]@{-};
(8.356605,0.633679); (8.436605,0.633679) **[blue]@{-};
(8.356605,0.633679); (8.436605,0.633679) **[blue]@{-};
(8.316605,0.174404); (8.476605,0.254404) **[red]@{-};
(8.476605,0.174404); (8.316605,0.254404) **[red]@{-};
(8.356605,0.194404); (8.436605,0.234404) **[red]@{-};
(8.356605,0.234404); (8.436605,0.194404) **[red]@{-};
(8.356605,0.194404); (8.436605,0.234404) **[red]@{-};
(8.356605,0.234404); (8.436605,0.194404) **[red]@{-};
(8.358792,0.636174); (8.518792,0.636174) **[blue]@{-};
(8.438792,0.596174); (8.438792,0.676174) **[blue]@{-};
(8.398792,0.636174); (8.478792,0.636174) **[blue]@{-};
(8.398792,0.636174); (8.478792,0.636174) **[blue]@{-};
(8.358792,0.186142); (8.518792,0.266142) **[red]@{-};
(8.518792,0.186142); (8.358792,0.266142) **[red]@{-};
(8.398792,0.206142); (8.478792,0.246142) **[red]@{-};
(8.398792,0.246142); (8.478792,0.206142) **[red]@{-};
(8.398792,0.206142); (8.478792,0.246142) **[red]@{-};
(8.398792,0.246142); (8.478792,0.206142) **[red]@{-};
(8.367083,0.636112); (8.527083,0.636112) **[blue]@{-};
(8.447083,0.596112); (8.447083,0.676112) **[blue]@{-};
(8.407083,0.636112); (8.487083,0.636112) **[blue]@{-};
(8.407083,0.636112); (8.487083,0.636112) **[blue]@{-};
(8.367083,0.188759); (8.527083,0.268759) **[red]@{-};
(8.527083,0.188759); (8.367083,0.268759) **[red]@{-};
(8.407083,0.208759); (8.487083,0.248759) **[red]@{-};
(8.407083,0.248759); (8.487083,0.208759) **[red]@{-};
(8.407083,0.208759); (8.487083,0.248759) **[red]@{-};
(8.407083,0.248759); (8.487083,0.208759) **[red]@{-};
(8.383524,0.633297); (8.543524,0.633297) **[blue]@{-};
(8.463524,0.593297); (8.463524,0.673297) **[blue]@{-};
(8.423524,0.633297); (8.503524,0.633297) **[blue]@{-};
(8.423524,0.633297); (8.503524,0.633297) **[blue]@{-};
(8.383524,0.190333); (8.543524,0.270333) **[red]@{-};
(8.543524,0.190333); (8.383524,0.270333) **[red]@{-};
(8.423524,0.210333); (8.503524,0.250333) **[red]@{-};
(8.423524,0.250333); (8.503524,0.210333) **[red]@{-};
(8.423524,0.210333); (8.503524,0.250333) **[red]@{-};
(8.423524,0.250333); (8.503524,0.210333) **[red]@{-};
(8.407840,0.635810); (8.567840,0.635810) **[blue]@{-};
(8.487840,0.595810); (8.487840,0.675810) **[blue]@{-};
(8.447840,0.635810); (8.527840,0.635810) **[blue]@{-};
(8.447840,0.635810); (8.527840,0.635810) **[blue]@{-};
(8.407840,0.201342); (8.567840,0.281342) **[red]@{-};
(8.567840,0.201342); (8.407840,0.281342) **[red]@{-};
(8.447840,0.221342); (8.527840,0.261342) **[red]@{-};
(8.447840,0.261342); (8.527840,0.221342) **[red]@{-};
(8.447840,0.221342); (8.527840,0.261342) **[red]@{-};
(8.447840,0.261342); (8.527840,0.221342) **[red]@{-};
(8.439636,0.636874); (8.599636,0.636874) **[blue]@{-};
(8.519636,0.596874); (8.519636,0.676874) **[blue]@{-};
(8.479636,0.636874); (8.559636,0.636874) **[blue]@{-};
(8.479636,0.636874); (8.559636,0.636874) **[blue]@{-};
(8.439636,0.207462); (8.599636,0.287462) **[red]@{-};
(8.599636,0.207462); (8.439636,0.287462) **[red]@{-};
(8.479636,0.227462); (8.559636,0.267462) **[red]@{-};
(8.479636,0.267462); (8.559636,0.227462) **[red]@{-};
(8.479636,0.227462); (8.559636,0.267462) **[red]@{-};
(8.479636,0.267462); (8.559636,0.227462) **[red]@{-};
(8.463032,0.635415); (8.623032,0.635415) **[blue]@{-};
(8.543032,0.595415); (8.543032,0.675415) **[blue]@{-};
(8.503032,0.635415); (8.583032,0.635415) **[blue]@{-};
(8.503032,0.635415); (8.583032,0.635415) **[blue]@{-};
(8.463032,0.212695); (8.623032,0.292695) **[red]@{-};
(8.623032,0.212695); (8.463032,0.292695) **[red]@{-};
(8.503032,0.232695); (8.583032,0.272695) **[red]@{-};
(8.503032,0.272695); (8.583032,0.232695) **[red]@{-};
(8.503032,0.232695); (8.583032,0.272695) **[red]@{-};
(8.503032,0.272695); (8.583032,0.232695) **[red]@{-};
(9.117217,0.631709); (9.277217,0.631709) **[blue]@{-};
(9.197217,0.591709); (9.197217,0.671709) **[blue]@{-};
(9.157217,0.631709); (9.237217,0.631709) **[blue]@{-};
(9.157217,0.631709); (9.237217,0.631709) **[blue]@{-};
(9.117217,-0.036987); (9.277217,0.043013) **[red]@{-};
(9.277217,-0.036987); (9.117217,0.043013) **[red]@{-};
(9.157217,-0.016987); (9.237217,0.023013) **[red]@{-};
(9.157217,0.023013); (9.237217,-0.016987) **[red]@{-};
(9.157217,-0.016987); (9.237217,0.023013) **[red]@{-};
(9.157217,0.023013); (9.237217,-0.016987) **[red]@{-};
(1.504963,1.258211); (1.664963,1.258211) **[blue]@{-};
(1.584963,1.218211); (1.584963,1.298211) **[blue]@{-};
(1.544963,1.245465); (1.624963,1.245465) **[blue]@{-};
(1.544963,1.271205); (1.624963,1.271205) **[blue]@{-};
(1.504963,1.225444); (1.664963,1.305444) **[red]@{-};
(1.664963,1.225444); (1.504963,1.305444) **[red]@{-};
(1.544963,1.234581); (1.624963,1.274581) **[red]@{-};
(1.544963,1.274581); (1.624963,1.234581) **[red]@{-};
(1.544963,1.255152); (1.624963,1.295152) **[red]@{-};
(1.544963,1.295152); (1.624963,1.255152) **[red]@{-};
(2.241928,1.093334); (2.401928,1.093334) **[blue]@{-};
(2.321928,1.053334); (2.321928,1.133334) **[blue]@{-};
(2.281928,1.082551); (2.361928,1.082551) **[blue]@{-};
(2.281928,1.097393); (2.361928,1.097393) **[blue]@{-};
(2.241928,1.061152); (2.401928,1.141152) **[red]@{-};
(2.401928,1.061152); (2.241928,1.141152) **[red]@{-};
(2.281928,1.069263); (2.361928,1.109263) **[red]@{-};
(2.281928,1.109263); (2.361928,1.069263) **[red]@{-};
(2.281928,1.092656); (2.361928,1.132656) **[red]@{-};
(2.281928,1.132656); (2.361928,1.092656) **[red]@{-};
(2.727355,1.170771); (2.887355,1.170771) **[blue]@{-};
(2.807355,1.130771); (2.807355,1.210771) **[blue]@{-};
(2.767355,1.160030); (2.847355,1.160030) **[blue]@{-};
(2.767355,1.178668); (2.847355,1.178668) **[blue]@{-};
(2.727355,1.141857); (2.887355,1.221857) **[red]@{-};
(2.887355,1.141857); (2.727355,1.221857) **[red]@{-};
(2.767355,1.154831); (2.847355,1.194831) **[red]@{-};
(2.767355,1.194831); (2.847355,1.154831) **[red]@{-};
(2.767355,1.169060); (2.847355,1.209060) **[red]@{-};
(2.767355,1.209060); (2.847355,1.169060) **[red]@{-};
(3.379432,1.225718); (3.539432,1.225718) **[blue]@{-};
(3.459432,1.185718); (3.459432,1.265718) **[blue]@{-};
(3.419432,1.221003); (3.499432,1.221003) **[blue]@{-};
(3.419432,1.234676); (3.499432,1.234676) **[blue]@{-};
(3.379432,1.197508); (3.539432,1.277508) **[red]@{-};
(3.539432,1.197508); (3.379432,1.277508) **[red]@{-};
(3.419432,1.211412); (3.499432,1.251412) **[red]@{-};
(3.419432,1.251412); (3.499432,1.211412) **[red]@{-};
(3.419432,1.220334); (3.499432,1.260334) **[red]@{-};
(3.419432,1.260334); (3.499432,1.220334) **[red]@{-};
(3.620440,1.228662); (3.780440,1.228662) **[blue]@{-};
(3.700440,1.188662); (3.700440,1.268662) **[blue]@{-};
(3.660440,1.221490); (3.740440,1.221490) **[blue]@{-};
(3.660440,1.233097); (3.740440,1.233097) **[blue]@{-};
(3.620440,1.204123); (3.780440,1.284123) **[red]@{-};
(3.780440,1.204123); (3.620440,1.284123) **[red]@{-};
(3.660440,1.221557); (3.740440,1.261557) **[red]@{-};
(3.660440,1.261557); (3.740440,1.221557) **[red]@{-};
(3.660440,1.227171); (3.740440,1.267171) **[red]@{-};
(3.660440,1.267171); (3.740440,1.227171) **[red]@{-};
(4.007463,1.224110); (4.167463,1.224110) **[blue]@{-};
(4.087463,1.184110); (4.087463,1.264110) **[blue]@{-};
(4.047463,1.219021); (4.127463,1.219021) **[blue]@{-};
(4.047463,1.227526); (4.127463,1.227526) **[blue]@{-};
(4.007463,1.202651); (4.167463,1.282651) **[red]@{-};
(4.167463,1.202651); (4.007463,1.282651) **[red]@{-};
(4.047463,1.219851); (4.127463,1.259851) **[red]@{-};
(4.047463,1.259851); (4.127463,1.219851) **[red]@{-};
(4.047463,1.227658); (4.127463,1.267658) **[red]@{-};
(4.047463,1.267658); (4.127463,1.227658) **[red]@{-};
(4.167928,1.240072); (4.327928,1.240072) **[blue]@{-};
(4.247928,1.200072); (4.247928,1.280072) **[blue]@{-};
(4.207928,1.237972); (4.287928,1.237972) **[blue]@{-};
(4.207928,1.248005); (4.287928,1.248005) **[blue]@{-};
(4.167928,1.217278); (4.327928,1.297278) **[red]@{-};
(4.327928,1.217278); (4.167928,1.297278) **[red]@{-};
(4.207928,1.235204); (4.287928,1.275204) **[red]@{-};
(4.207928,1.275204); (4.287928,1.235204) **[red]@{-};
(4.207928,1.243829); (4.287928,1.283829) **[red]@{-};
(4.207928,1.283829); (4.287928,1.243829) **[red]@{-};
(4.443562,1.257776); (4.603562,1.257776) **[blue]@{-};
(4.523562,1.217776); (4.523562,1.297776) **[blue]@{-};
(4.483562,1.254435); (4.563562,1.254435) **[blue]@{-};
(4.483562,1.262411); (4.563562,1.262411) **[blue]@{-};
(4.443562,1.233144); (4.603562,1.313144) **[red]@{-};
(4.603562,1.233144); (4.443562,1.313144) **[red]@{-};
(4.483562,1.250486); (4.563562,1.290486) **[red]@{-};
(4.483562,1.290486); (4.563562,1.250486) **[red]@{-};
(4.483562,1.257014); (4.563562,1.297014) **[red]@{-};
(4.483562,1.297014); (4.563562,1.257014) **[red]@{-};
(4.777981,1.287170); (4.937981,1.287170) **[blue]@{-};
(4.857981,1.247170); (4.857981,1.327170) **[blue]@{-};
(4.817981,1.284933); (4.897981,1.284933) **[blue]@{-};
(4.817981,1.290491); (4.897981,1.290491) **[blue]@{-};
(4.777981,1.229465); (4.937981,1.309465) **[red]@{-};
(4.937981,1.229465); (4.777981,1.309465) **[red]@{-};
(4.817981,1.247433); (4.897981,1.287433) **[red]@{-};
(4.817981,1.287433); (4.897981,1.247433) **[red]@{-};
(4.817981,1.249987); (4.897981,1.289987) **[red]@{-};
(4.817981,1.289987); (4.897981,1.249987) **[red]@{-};
(4.874196,1.273001); (5.034196,1.273001) **[blue]@{-};
(4.954196,1.233001); (4.954196,1.313001) **[blue]@{-};
(4.914196,1.270497); (4.994196,1.270497) **[blue]@{-};
(4.914196,1.277073); (4.994196,1.277073) **[blue]@{-};
(4.874196,1.235228); (5.034196,1.315228) **[red]@{-};
(5.034196,1.235228); (4.874196,1.315228) **[red]@{-};
(4.914196,1.254848); (4.994196,1.294848) **[red]@{-};
(4.914196,1.294848); (4.994196,1.254848) **[red]@{-};
(4.914196,1.257669); (4.994196,1.297669) **[red]@{-};
(4.914196,1.297669); (4.994196,1.257669) **[red]@{-};
(5.129453,1.244329); (5.289453,1.244329) **[blue]@{-};
(5.209453,1.204329); (5.209453,1.284329) **[blue]@{-};
(5.169453,1.241699); (5.249453,1.241699) **[blue]@{-};
(5.169453,1.245737); (5.249453,1.245737) **[blue]@{-};
(5.129453,1.217783); (5.289453,1.297783) **[red]@{-};
(5.289453,1.217783); (5.129453,1.297783) **[red]@{-};
(5.169453,1.236388); (5.249453,1.276388) **[red]@{-};
(5.169453,1.276388); (5.249453,1.236388) **[red]@{-};
(5.169453,1.241498); (5.249453,1.281498) **[red]@{-};
(5.169453,1.281498); (5.249453,1.241498) **[red]@{-};
(5.277552,1.240548); (5.437552,1.240548) **[blue]@{-};
(5.357552,1.200548); (5.357552,1.280548) **[blue]@{-};
(5.317552,1.238070); (5.397552,1.238070) **[blue]@{-};
(5.317552,1.242680); (5.397552,1.242680) **[blue]@{-};
(5.277552,1.215535); (5.437552,1.295535) **[red]@{-};
(5.437552,1.215535); (5.277552,1.295535) **[red]@{-};
(5.317552,1.231898); (5.397552,1.271898) **[red]@{-};
(5.317552,1.271898); (5.397552,1.231898) **[red]@{-};
(5.317552,1.239922); (5.397552,1.279922) **[red]@{-};
(5.317552,1.279922); (5.397552,1.239922) **[red]@{-};
(5.346265,1.248591); (5.506265,1.248591) **[blue]@{-};
(5.426265,1.208591); (5.426265,1.288591) **[blue]@{-};
(5.386265,1.245749); (5.466265,1.245749) **[blue]@{-};
(5.386265,1.251993); (5.466265,1.251993) **[blue]@{-};
(5.346265,1.224883); (5.506265,1.304883) **[red]@{-};
(5.506265,1.224883); (5.346265,1.304883) **[red]@{-};
(5.386265,1.240988); (5.466265,1.280988) **[red]@{-};
(5.386265,1.280988); (5.466265,1.240988) **[red]@{-};
(5.386265,1.245925); (5.466265,1.285925) **[red]@{-};
(5.386265,1.285925); (5.466265,1.245925) **[red]@{-};
(5.474589,1.254017); (5.634589,1.254017) **[blue]@{-};
(5.554589,1.214017); (5.554589,1.294017) **[blue]@{-};
(5.514589,1.251863); (5.594589,1.251863) **[blue]@{-};
(5.514589,1.255871); (5.594589,1.255871) **[blue]@{-};
(5.474589,1.229003); (5.634589,1.309003) **[red]@{-};
(5.634589,1.229003); (5.474589,1.309003) **[red]@{-};
(5.514589,1.247019); (5.594589,1.287019) **[red]@{-};
(5.514589,1.287019); (5.594589,1.247019) **[red]@{-};
(5.514589,1.250251); (5.594589,1.290251) **[red]@{-};
(5.514589,1.290251); (5.594589,1.250251) **[red]@{-};
(5.647920,1.242669); (5.807920,1.242669) **[blue]@{-};
(5.727920,1.202669); (5.727920,1.282669) **[blue]@{-};
(5.687920,1.241902); (5.767920,1.241902) **[blue]@{-};
(5.687920,1.246661); (5.767920,1.246661) **[blue]@{-};
(5.647920,1.219595); (5.807920,1.299595) **[red]@{-};
(5.807920,1.219595); (5.647920,1.299595) **[red]@{-};
(5.687920,1.237453); (5.767920,1.277453) **[red]@{-};
(5.687920,1.277453); (5.767920,1.237453) **[red]@{-};
(5.687920,1.240911); (5.767920,1.280911) **[red]@{-};
(5.687920,1.280911); (5.767920,1.240911) **[red]@{-};
(5.802643,1.283785); (5.962643,1.283785) **[blue]@{-};
(5.882643,1.243785); (5.882643,1.323785) **[blue]@{-};
(5.842643,1.282120); (5.922643,1.282120) **[blue]@{-};
(5.842643,1.285449); (5.922643,1.285449) **[blue]@{-};
(5.802643,1.250863); (5.962643,1.330863) **[red]@{-};
(5.962643,1.250863); (5.802643,1.330863) **[red]@{-};
(5.842643,1.268682); (5.922643,1.308682) **[red]@{-};
(5.842643,1.308682); (5.922643,1.268682) **[red]@{-};
(5.842643,1.271792); (5.922643,1.311792) **[red]@{-};
(5.842643,1.311792); (5.922643,1.271792) **[red]@{-};
(5.850737,1.282963); (6.010737,1.282963) **[blue]@{-};
(5.930737,1.242963); (5.930737,1.322963) **[blue]@{-};
(5.890737,1.281775); (5.970737,1.281775) **[blue]@{-};
(5.890737,1.286184); (5.970737,1.286184) **[blue]@{-};
(5.850737,1.221706); (6.010737,1.301706) **[red]@{-};
(6.010737,1.221706); (5.850737,1.301706) **[red]@{-};
(5.890737,1.240212); (5.970737,1.280212) **[red]@{-};
(5.890737,1.280212); (5.970737,1.240212) **[red]@{-};
(5.890737,1.242595); (5.970737,1.282595) **[red]@{-};
(5.890737,1.282595); (5.970737,1.242595) **[red]@{-};
(5.986089,1.238737); (6.146089,1.238737) **[blue]@{-};
(6.066089,1.198737); (6.066089,1.278737) **[blue]@{-};
(6.026089,1.236365); (6.106089,1.236365) **[blue]@{-};
(6.026089,1.241451); (6.106089,1.241451) **[blue]@{-};
(5.986089,1.211965); (6.146089,1.291965) **[red]@{-};
(6.146089,1.211965); (5.986089,1.291965) **[red]@{-};
(6.026089,1.230381); (6.106089,1.270381) **[red]@{-};
(6.026089,1.270381); (6.106089,1.230381) **[red]@{-};
(6.026089,1.233126); (6.106089,1.273126) **[red]@{-};
(6.026089,1.273126); (6.106089,1.233126) **[red]@{-};
(6.069747,1.242495); (6.229747,1.242495) **[blue]@{-};
(6.149747,1.202495); (6.149747,1.282495) **[blue]@{-};
(6.109747,1.241465); (6.189747,1.241465) **[blue]@{-};
(6.109747,1.245304); (6.189747,1.245304) **[blue]@{-};
(6.069747,1.218772); (6.229747,1.298772) **[red]@{-};
(6.229747,1.218772); (6.069747,1.298772) **[red]@{-};
(6.109747,1.237158); (6.189747,1.277158) **[red]@{-};
(6.109747,1.277158); (6.189747,1.237158) **[red]@{-};
(6.109747,1.240534); (6.189747,1.280534) **[red]@{-};
(6.109747,1.280534); (6.189747,1.240534) **[red]@{-};
(6.109825,1.236781); (6.269825,1.236781) **[blue]@{-};
(6.189825,1.196781); (6.189825,1.276781) **[blue]@{-};
(6.149825,1.232457); (6.229825,1.232457) **[blue]@{-};
(6.149825,1.239869); (6.229825,1.239869) **[blue]@{-};
(6.109825,1.210407); (6.269825,1.290407) **[red]@{-};
(6.269825,1.210407); (6.109825,1.290407) **[red]@{-};
(6.149825,1.228753); (6.229825,1.268753) **[red]@{-};
(6.149825,1.268753); (6.229825,1.228753) **[red]@{-};
(6.149825,1.231937); (6.229825,1.271937) **[red]@{-};
(6.149825,1.271937); (6.229825,1.231937) **[red]@{-};
(6.223781,1.244566); (6.383781,1.244566) **[blue]@{-};
(6.303781,1.204566); (6.303781,1.284566) **[blue]@{-};
(6.263781,1.243322); (6.343781,1.243322) **[blue]@{-};
(6.263781,1.246620); (6.343781,1.246620) **[blue]@{-};
(6.223781,1.219625); (6.383781,1.299625) **[red]@{-};
(6.383781,1.219625); (6.223781,1.299625) **[red]@{-};
(6.263781,1.238081); (6.343781,1.278081) **[red]@{-};
(6.263781,1.278081); (6.343781,1.238081) **[red]@{-};
(6.263781,1.240675); (6.343781,1.280675) **[red]@{-};
(6.263781,1.280675); (6.343781,1.240675) **[red]@{-};
(6.295039,1.238411); (6.455039,1.238411) **[blue]@{-};
(6.375039,1.198411); (6.375039,1.278411) **[blue]@{-};
(6.335039,1.237892); (6.415039,1.237892) **[blue]@{-};
(6.335039,1.239773); (6.415039,1.239773) **[blue]@{-};
(6.295039,1.218083); (6.455039,1.298083) **[red]@{-};
(6.455039,1.218083); (6.295039,1.298083) **[red]@{-};
(6.335039,1.236674); (6.415039,1.276674) **[red]@{-};
(6.335039,1.276674); (6.415039,1.236674) **[red]@{-};
(6.335039,1.239831); (6.415039,1.279831) **[red]@{-};
(6.335039,1.279831); (6.415039,1.239831) **[red]@{-};
(6.395733,1.237386); (6.555733,1.237386) **[blue]@{-};
(6.475733,1.197386); (6.475733,1.277386) **[blue]@{-};
(6.435733,1.235303); (6.515733,1.235303) **[blue]@{-};
(6.435733,1.239891); (6.515733,1.239891) **[blue]@{-};
(6.395733,1.240239); (6.555733,1.320239) **[red]@{-};
(6.555733,1.240239); (6.395733,1.320239) **[red]@{-};
(6.435733,1.259336); (6.515733,1.299336) **[red]@{-};
(6.435733,1.299336); (6.515733,1.259336) **[red]@{-};
(6.435733,1.261984); (6.515733,1.301984) **[red]@{-};
(6.435733,1.301984); (6.515733,1.261984) **[red]@{-};
(6.519913,1.231836); (6.679913,1.231836) **[blue]@{-};
(6.599913,1.191836); (6.599913,1.271836) **[blue]@{-};
(6.559913,1.230381); (6.639913,1.230381) **[blue]@{-};
(6.559913,1.234184); (6.639913,1.234184) **[blue]@{-};
(6.519913,1.205775); (6.679913,1.285775) **[red]@{-};
(6.679913,1.205775); (6.519913,1.285775) **[red]@{-};
(6.559913,1.223908); (6.639913,1.263908) **[red]@{-};
(6.559913,1.263908); (6.639913,1.223908) **[red]@{-};
(6.559913,1.226809); (6.639913,1.266809) **[red]@{-};
(6.559913,1.266809); (6.639913,1.226809) **[red]@{-};
(6.578211,1.235872); (6.738211,1.235872) **[blue]@{-};
(6.658211,1.195872); (6.658211,1.275872) **[blue]@{-};
(6.618211,1.233224); (6.698211,1.233224) **[blue]@{-};
(6.618211,1.236748); (6.698211,1.236748) **[blue]@{-};
(6.578211,1.208189); (6.738211,1.288189) **[red]@{-};
(6.738211,1.208189); (6.578211,1.288189) **[red]@{-};
(6.618211,1.227640); (6.698211,1.267640) **[red]@{-};
(6.618211,1.267640); (6.698211,1.227640) **[red]@{-};
(6.618211,1.229536); (6.698211,1.269536) **[red]@{-};
(6.618211,1.269536); (6.698211,1.229536) **[red]@{-};
(6.606501,1.239495); (6.766501,1.239495) **[blue]@{-};
(6.686501,1.199495); (6.686501,1.279495) **[blue]@{-};
(6.646501,1.236922); (6.726501,1.236922) **[blue]@{-};
(6.646501,1.241977); (6.726501,1.241977) **[blue]@{-};
(6.606501,1.212606); (6.766501,1.292606) **[red]@{-};
(6.766501,1.212606); (6.606501,1.292606) **[red]@{-};
(6.646501,1.230924); (6.726501,1.270924) **[red]@{-};
(6.646501,1.270924); (6.726501,1.230924) **[red]@{-};
(6.646501,1.234996); (6.726501,1.274996) **[red]@{-};
(6.646501,1.274996); (6.726501,1.234996) **[red]@{-};
(6.661467,1.237941); (6.821467,1.237941) **[blue]@{-};
(6.741467,1.197941); (6.741467,1.277941) **[blue]@{-};
(6.701467,1.237081); (6.781467,1.237081) **[blue]@{-};
(6.701467,1.238565); (6.781467,1.238565) **[blue]@{-};
(6.661467,1.212157); (6.821467,1.292157) **[red]@{-};
(6.821467,1.212157); (6.661467,1.292157) **[red]@{-};
(6.701467,1.229918); (6.781467,1.269918) **[red]@{-};
(6.701467,1.269918); (6.781467,1.229918) **[red]@{-};
(6.701467,1.233443); (6.781467,1.273443) **[red]@{-};
(6.701467,1.273443); (6.781467,1.233443) **[red]@{-};
(6.688184,1.238398); (6.848184,1.238398) **[blue]@{-};
(6.768184,1.198398); (6.768184,1.278398) **[blue]@{-};
(6.728184,1.237007); (6.808184,1.237007) **[blue]@{-};
(6.728184,1.241838); (6.808184,1.241838) **[blue]@{-};
(6.688184,1.209910); (6.848184,1.289910) **[red]@{-};
(6.848184,1.209910); (6.688184,1.289910) **[red]@{-};
(6.728184,1.229023); (6.808184,1.269023) **[red]@{-};
(6.728184,1.269023); (6.808184,1.229023) **[red]@{-};
(6.728184,1.231519); (6.808184,1.271519) **[red]@{-};
(6.728184,1.271519); (6.808184,1.231519) **[red]@{-};
(6.740179,1.233824); (6.900179,1.233824) **[blue]@{-};
(6.820179,1.193824); (6.820179,1.273824) **[blue]@{-};
(6.780179,1.232091); (6.860179,1.232091) **[blue]@{-};
(6.780179,1.235906); (6.860179,1.235906) **[blue]@{-};
(6.740179,1.211305); (6.900179,1.291305) **[red]@{-};
(6.900179,1.211305); (6.740179,1.291305) **[red]@{-};
(6.780179,1.229435); (6.860179,1.269435) **[red]@{-};
(6.780179,1.269435); (6.860179,1.229435) **[red]@{-};
(6.780179,1.232176); (6.860179,1.272176) **[red]@{-};
(6.780179,1.272176); (6.860179,1.232176) **[red]@{-};
(6.908685,1.256464); (7.068685,1.256464) **[blue]@{-};
(6.988685,1.216464); (6.988685,1.296464) **[blue]@{-};
(6.948685,1.254578); (7.028685,1.254578) **[blue]@{-};
(6.948685,1.257913); (7.028685,1.257913) **[blue]@{-};
(6.908685,1.214113); (7.068685,1.294113) **[red]@{-};
(7.068685,1.214113); (6.908685,1.294113) **[red]@{-};
(6.948685,1.232830); (7.028685,1.272830) **[red]@{-};
(6.948685,1.272830); (7.028685,1.232830) **[red]@{-};
(6.948685,1.236858); (7.028685,1.276858) **[red]@{-};
(6.948685,1.276858); (7.028685,1.236858) **[red]@{-};
(6.953423,1.230016); (7.113423,1.230016) **[blue]@{-};
(7.033423,1.190016); (7.033423,1.270016) **[blue]@{-};
(6.993423,1.228681); (7.073423,1.228681) **[blue]@{-};
(6.993423,1.232655); (7.073423,1.232655) **[blue]@{-};
(6.953423,1.205253); (7.113423,1.285253) **[red]@{-};
(7.113423,1.205253); (6.953423,1.285253) **[red]@{-};
(6.993423,1.224881); (7.073423,1.264881) **[red]@{-};
(6.993423,1.264881); (7.073423,1.224881) **[red]@{-};
(6.993423,1.226325); (7.073423,1.266325) **[red]@{-};
(6.993423,1.266325); (7.073423,1.226325) **[red]@{-};
(7.018032,1.230237); (7.178032,1.230237) **[blue]@{-};
(7.098032,1.190237); (7.098032,1.270237) **[blue]@{-};
(7.058032,1.227975); (7.138032,1.227975) **[blue]@{-};
(7.058032,1.232111); (7.138032,1.232111) **[blue]@{-};
(7.018032,1.203237); (7.178032,1.283237) **[red]@{-};
(7.178032,1.203237); (7.018032,1.283237) **[red]@{-};
(7.058032,1.221694); (7.138032,1.261694) **[red]@{-};
(7.058032,1.261694); (7.138032,1.221694) **[red]@{-};
(7.058032,1.226830); (7.138032,1.266830) **[red]@{-};
(7.058032,1.266830); (7.138032,1.226830) **[red]@{-};
(7.038941,1.232961); (7.198941,1.232961) **[blue]@{-};
(7.118941,1.192961); (7.118941,1.272961) **[blue]@{-};
(7.078941,1.230944); (7.158941,1.230944) **[blue]@{-};
(7.078941,1.233928); (7.158941,1.233928) **[blue]@{-};
(7.038941,1.206530); (7.198941,1.286530) **[red]@{-};
(7.198941,1.206530); (7.038941,1.286530) **[red]@{-};
(7.078941,1.225558); (7.158941,1.265558) **[red]@{-};
(7.078941,1.265558); (7.158941,1.225558) **[red]@{-};
(7.078941,1.227852); (7.158941,1.267852) **[red]@{-};
(7.078941,1.267852); (7.158941,1.227852) **[red]@{-};
(7.139169,1.230638); (7.299169,1.230638) **[blue]@{-};
(7.219169,1.190638); (7.219169,1.270638) **[blue]@{-};
(7.179169,1.229781); (7.259169,1.229781) **[blue]@{-};
(7.179169,1.233560); (7.259169,1.233560) **[blue]@{-};
(7.139169,1.223020); (7.299169,1.303020) **[red]@{-};
(7.299169,1.223020); (7.139169,1.303020) **[red]@{-};
(7.179169,1.241176); (7.259169,1.281176) **[red]@{-};
(7.179169,1.281176); (7.259169,1.241176) **[red]@{-};
(7.179169,1.243893); (7.259169,1.283893) **[red]@{-};
(7.179169,1.283893); (7.259169,1.243893) **[red]@{-};
(7.158405,1.237356); (7.318405,1.237356) **[blue]@{-};
(7.238405,1.197356); (7.238405,1.277356) **[blue]@{-};
(7.198405,1.234649); (7.278405,1.234649) **[blue]@{-};
(7.198405,1.239878); (7.278405,1.239878) **[blue]@{-};
(7.158405,1.239016); (7.318405,1.319016) **[red]@{-};
(7.318405,1.239016); (7.158405,1.319016) **[red]@{-};
(7.198405,1.257497); (7.278405,1.297497) **[red]@{-};
(7.198405,1.297497); (7.278405,1.257497) **[red]@{-};
(7.198405,1.260930); (7.278405,1.300930) **[red]@{-};
(7.198405,1.300930); (7.278405,1.260930) **[red]@{-};
(7.214621,1.271273); (7.374621,1.271273) **[blue]@{-};
(7.294621,1.231273); (7.294621,1.311273) **[blue]@{-};
(7.254621,1.270742); (7.334621,1.270742) **[blue]@{-};
(7.254621,1.272346); (7.334621,1.272346) **[blue]@{-};
(7.214621,1.207279); (7.374621,1.287279) **[red]@{-};
(7.374621,1.207279); (7.214621,1.287279) **[red]@{-};
(7.254621,1.226624); (7.334621,1.266624) **[red]@{-};
(7.254621,1.266624); (7.334621,1.226624) **[red]@{-};
(7.254621,1.228232); (7.334621,1.268232) **[red]@{-};
(7.254621,1.268232); (7.334621,1.228232) **[red]@{-};
(7.268728,1.231172); (7.428728,1.231172) **[blue]@{-};
(7.348728,1.191172); (7.348728,1.271172) **[blue]@{-};
(7.308728,1.230254); (7.388728,1.230254) **[blue]@{-};
(7.308728,1.232112); (7.388728,1.232112) **[blue]@{-};
(7.268728,1.203956); (7.428728,1.283956) **[red]@{-};
(7.428728,1.203956); (7.268728,1.283956) **[red]@{-};
(7.308728,1.223279); (7.388728,1.263279) **[red]@{-};
(7.308728,1.263279); (7.388728,1.223279) **[red]@{-};
(7.308728,1.224877); (7.388728,1.264877) **[red]@{-};
(7.308728,1.264877); (7.388728,1.224877) **[red]@{-};
(7.303704,1.233105); (7.463704,1.233105) **[blue]@{-};
(7.383704,1.193105); (7.383704,1.273105) **[blue]@{-};
(7.343704,1.232351); (7.423704,1.232351) **[blue]@{-};
(7.343704,1.236256); (7.423704,1.236256) **[blue]@{-};
(7.303704,1.206793); (7.463704,1.286793) **[red]@{-};
(7.463704,1.206793); (7.303704,1.286793) **[red]@{-};
(7.343704,1.225235); (7.423704,1.265235) **[red]@{-};
(7.343704,1.265235); (7.423704,1.225235) **[red]@{-};
(7.343704,1.227420); (7.423704,1.267420) **[red]@{-};
(7.343704,1.267420); (7.423704,1.227420) **[red]@{-};
(7.354628,1.232046); (7.514628,1.232046) **[blue]@{-};
(7.434628,1.192046); (7.434628,1.272046) **[blue]@{-};
(7.394628,1.231085); (7.474628,1.231085) **[blue]@{-};
(7.394628,1.232657); (7.474628,1.232657) **[blue]@{-};
(7.354628,1.205130); (7.514628,1.285130) **[red]@{-};
(7.514628,1.205130); (7.354628,1.285130) **[red]@{-};
(7.394628,1.223886); (7.474628,1.263886) **[red]@{-};
(7.394628,1.263886); (7.474628,1.223886) **[red]@{-};
(7.394628,1.226478); (7.474628,1.266478) **[red]@{-};
(7.394628,1.266478); (7.474628,1.226478) **[red]@{-};
(7.403816,1.232015); (7.563816,1.232015) **[blue]@{-};
(7.483816,1.192015); (7.483816,1.272015) **[blue]@{-};
(7.443816,1.230893); (7.523816,1.230893) **[blue]@{-};
(7.443816,1.233361); (7.523816,1.233361) **[blue]@{-};
(7.403816,1.203899); (7.563816,1.283899) **[red]@{-};
(7.563816,1.203899); (7.403816,1.283899) **[red]@{-};
(7.443816,1.223272); (7.523816,1.263272) **[red]@{-};
(7.443816,1.263272); (7.523816,1.223272) **[red]@{-};
(7.443816,1.227214); (7.523816,1.267214) **[red]@{-};
(7.443816,1.267214); (7.523816,1.227214) **[red]@{-};
(7.419846,1.232469); (7.579846,1.232469) **[blue]@{-};
(7.499846,1.192469); (7.499846,1.272469) **[blue]@{-};
(7.459846,1.231491); (7.539846,1.231491) **[blue]@{-};
(7.459846,1.233840); (7.539846,1.233840) **[blue]@{-};
(7.419846,1.226751); (7.579846,1.306751) **[red]@{-};
(7.579846,1.226751); (7.419846,1.306751) **[red]@{-};
(7.459846,1.246002); (7.539846,1.286002) **[red]@{-};
(7.459846,1.286002); (7.539846,1.246002) **[red]@{-};
(7.459846,1.248316); (7.539846,1.288316) **[red]@{-};
(7.459846,1.288316); (7.539846,1.248316) **[red]@{-};
(7.497429,1.250645); (7.657429,1.250645) **[blue]@{-};
(7.577429,1.210645); (7.577429,1.290645) **[blue]@{-};
(7.537429,1.250002); (7.617429,1.250002) **[blue]@{-};
(7.537429,1.251919); (7.617429,1.251919) **[blue]@{-};
(7.497429,1.208301); (7.657429,1.288301) **[red]@{-};
(7.657429,1.208301); (7.497429,1.288301) **[red]@{-};
(7.537429,1.227279); (7.617429,1.267279) **[red]@{-};
(7.537429,1.267279); (7.617429,1.227279) **[red]@{-};
(7.537429,1.229228); (7.617429,1.269228) **[red]@{-};
(7.537429,1.269228); (7.617429,1.229228) **[red]@{-};
(7.512457,1.227647); (7.672457,1.227647) **[blue]@{-};
(7.592457,1.187647); (7.592457,1.267647) **[blue]@{-};
(7.552457,1.225718); (7.632457,1.225718) **[blue]@{-};
(7.552457,1.229270); (7.632457,1.229270) **[blue]@{-};
(7.512457,1.199280); (7.672457,1.279280) **[red]@{-};
(7.672457,1.199280); (7.512457,1.279280) **[red]@{-};
(7.552457,1.218771); (7.632457,1.258771) **[red]@{-};
(7.552457,1.258771); (7.632457,1.218771) **[red]@{-};
(7.552457,1.220494); (7.632457,1.260494) **[red]@{-};
(7.552457,1.260494); (7.632457,1.220494) **[red]@{-};
(7.542052,1.229288); (7.702052,1.229288) **[blue]@{-};
(7.622052,1.189288); (7.622052,1.269288) **[blue]@{-};
(7.582052,1.228349); (7.662052,1.228349) **[blue]@{-};
(7.582052,1.229577); (7.662052,1.229577) **[blue]@{-};
(7.542052,1.200731); (7.702052,1.280731) **[red]@{-};
(7.702052,1.200731); (7.542052,1.280731) **[red]@{-};
(7.582052,1.219864); (7.662052,1.259864) **[red]@{-};
(7.582052,1.259864); (7.662052,1.219864) **[red]@{-};
(7.582052,1.222574); (7.662052,1.262574) **[red]@{-};
(7.582052,1.262574); (7.662052,1.222574) **[red]@{-};
(7.556625,1.231587); (7.716625,1.231587) **[blue]@{-};
(7.636625,1.191587); (7.636625,1.271587) **[blue]@{-};
(7.596625,1.230052); (7.676625,1.230052) **[blue]@{-};
(7.596625,1.232404); (7.676625,1.232404) **[blue]@{-};
(7.556625,1.203900); (7.716625,1.283900) **[red]@{-};
(7.716625,1.203900); (7.556625,1.283900) **[red]@{-};
(7.596625,1.222533); (7.676625,1.262533) **[red]@{-};
(7.596625,1.262533); (7.676625,1.222533) **[red]@{-};
(7.596625,1.225111); (7.676625,1.265111) **[red]@{-};
(7.596625,1.265111); (7.676625,1.225111) **[red]@{-};
(7.641099,1.229496); (7.801099,1.229496) **[blue]@{-};
(7.721099,1.189496); (7.721099,1.269496) **[blue]@{-};
(7.681099,1.228949); (7.761099,1.228949) **[blue]@{-};
(7.681099,1.231795); (7.761099,1.231795) **[blue]@{-};
(7.641099,1.224390); (7.801099,1.304390) **[red]@{-};
(7.801099,1.224390); (7.641099,1.304390) **[red]@{-};
(7.681099,1.243602); (7.761099,1.283602) **[red]@{-};
(7.681099,1.283602); (7.761099,1.243602) **[red]@{-};
(7.681099,1.244721); (7.761099,1.284721) **[red]@{-};
(7.681099,1.284721); (7.761099,1.244721) **[red]@{-};
(7.720900,1.248891); (7.880900,1.248891) **[blue]@{-};
(7.800900,1.208891); (7.800900,1.288891) **[blue]@{-};
(7.760900,1.246895); (7.840900,1.246895) **[blue]@{-};
(7.760900,1.249142); (7.840900,1.249142) **[blue]@{-};
(7.720900,1.205895); (7.880900,1.285895) **[red]@{-};
(7.880900,1.205895); (7.720900,1.285895) **[red]@{-};
(7.760900,1.224546); (7.840900,1.264546) **[red]@{-};
(7.760900,1.264546); (7.840900,1.224546) **[red]@{-};
(7.760900,1.227861); (7.840900,1.267861) **[red]@{-};
(7.760900,1.267861); (7.840900,1.227861) **[red]@{-};
(7.746548,1.228601); (7.906548,1.228601) **[blue]@{-};
(7.826548,1.188601); (7.826548,1.268601) **[blue]@{-};
(7.786548,1.228019); (7.866548,1.228019) **[blue]@{-};
(7.786548,1.230958); (7.866548,1.230958) **[blue]@{-};
(7.746548,1.201232); (7.906548,1.281232) **[red]@{-};
(7.906548,1.201232); (7.746548,1.281232) **[red]@{-};
(7.786548,1.220583); (7.866548,1.260583) **[red]@{-};
(7.786548,1.260583); (7.866548,1.220583) **[red]@{-};
(7.786548,1.221912); (7.866548,1.261912) **[red]@{-};
(7.786548,1.261912); (7.866548,1.221912) **[red]@{-};
(7.759204,1.228703); (7.919204,1.228703) **[blue]@{-};
(7.839204,1.188703); (7.839204,1.268703) **[blue]@{-};
(7.799204,1.227877); (7.879204,1.227877) **[blue]@{-};
(7.799204,1.229431); (7.879204,1.229431) **[blue]@{-};
(7.759204,1.202272); (7.919204,1.282272) **[red]@{-};
(7.919204,1.202272); (7.759204,1.282272) **[red]@{-};
(7.799204,1.221661); (7.879204,1.261661) **[red]@{-};
(7.799204,1.261661); (7.879204,1.221661) **[red]@{-};
(7.799204,1.222661); (7.879204,1.262661) **[red]@{-};
(7.799204,1.262661); (7.879204,1.222661) **[red]@{-};
(7.784186,1.228484); (7.944186,1.228484) **[blue]@{-};
(7.864186,1.188484); (7.864186,1.268484) **[blue]@{-};
(7.824186,1.226861); (7.904186,1.226861) **[blue]@{-};
(7.824186,1.228815); (7.904186,1.228815) **[blue]@{-};
(7.784186,1.201716); (7.944186,1.281716) **[red]@{-};
(7.944186,1.201716); (7.784186,1.281716) **[red]@{-};
(7.824186,1.220202); (7.904186,1.260202) **[red]@{-};
(7.824186,1.260202); (7.904186,1.220202) **[red]@{-};
(7.824186,1.223689); (7.904186,1.263689) **[red]@{-};
(7.824186,1.263689); (7.904186,1.223689) **[red]@{-};
(7.820867,1.232292); (7.980867,1.232292) **[blue]@{-};
(7.900867,1.192292); (7.900867,1.272292) **[blue]@{-};
(7.860867,1.231833); (7.940867,1.231833) **[blue]@{-};
(7.860867,1.233318); (7.940867,1.233318) **[blue]@{-};
(7.820867,1.207809); (7.980867,1.287809) **[red]@{-};
(7.980867,1.207809); (7.820867,1.287809) **[red]@{-};
(7.860867,1.227051); (7.940867,1.267051) **[red]@{-};
(7.860867,1.267051); (7.940867,1.227051) **[red]@{-};
(7.860867,1.228748); (7.940867,1.268748) **[red]@{-};
(7.860867,1.268748); (7.940867,1.228748) **[red]@{-};
(7.832889,1.227893); (7.992889,1.227893) **[blue]@{-};
(7.912889,1.187893); (7.912889,1.267893) **[blue]@{-};
(7.872889,1.227009); (7.952889,1.227009) **[blue]@{-};
(7.872889,1.229287); (7.952889,1.229287) **[blue]@{-};
(7.832889,1.184896); (7.992889,1.264896) **[red]@{-};
(7.992889,1.184896); (7.832889,1.264896) **[red]@{-};
(7.872889,1.203796); (7.952889,1.243796) **[red]@{-};
(7.872889,1.243796); (7.952889,1.203796) **[red]@{-};
(7.872889,1.206277); (7.952889,1.246277) **[red]@{-};
(7.872889,1.246277); (7.952889,1.206277) **[red]@{-};
(7.891544,1.268732); (8.051544,1.268732) **[blue]@{-};
(7.971544,1.228732); (7.971544,1.308732) **[blue]@{-};
(7.931544,1.268056); (8.011544,1.268056) **[blue]@{-};
(7.931544,1.269770); (8.011544,1.269770) **[blue]@{-};
(7.891544,1.182897); (8.051544,1.262897) **[red]@{-};
(8.051544,1.182897); (7.891544,1.262897) **[red]@{-};
(7.931544,1.201480); (8.011544,1.241480) **[red]@{-};
(7.931544,1.241480); (8.011544,1.201480) **[red]@{-};
(7.931544,1.203323); (8.011544,1.243323) **[red]@{-};
(7.931544,1.243323); (8.011544,1.203323) **[red]@{-};
(7.925625,1.225332); (8.085625,1.225332) **[blue]@{-};
(8.005625,1.185332); (8.005625,1.265332) **[blue]@{-};
(7.965625,1.223325); (8.045625,1.223325) **[blue]@{-};
(7.965625,1.227564); (8.045625,1.227564) **[blue]@{-};
(7.925625,1.177642); (8.085625,1.257642) **[red]@{-};
(8.085625,1.177642); (7.925625,1.257642) **[red]@{-};
(7.965625,1.196814); (8.045625,1.236814) **[red]@{-};
(7.965625,1.236814); (8.045625,1.196814) **[red]@{-};
(7.965625,1.198736); (8.045625,1.238736) **[red]@{-};
(7.965625,1.238736); (8.045625,1.198736) **[red]@{-};
(7.958919,1.226939); (8.118919,1.226939) **[blue]@{-};
(8.038919,1.186939); (8.038919,1.266939) **[blue]@{-};
(7.998919,1.226233); (8.078919,1.226233) **[blue]@{-};
(7.998919,1.228358); (8.078919,1.228358) **[blue]@{-};
(7.958919,1.200528); (8.118919,1.280528) **[red]@{-};
(8.118919,1.200528); (7.958919,1.280528) **[red]@{-};
(7.998919,1.219503); (8.078919,1.259503) **[red]@{-};
(7.998919,1.259503); (8.078919,1.219503) **[red]@{-};
(7.998919,1.222155); (8.078919,1.262155) **[red]@{-};
(7.998919,1.262155); (8.078919,1.222155) **[red]@{-};
(7.991462,1.227510); (8.151462,1.227510) **[blue]@{-};
(8.071462,1.187510); (8.071462,1.267510) **[blue]@{-};
(8.031462,1.225931); (8.111462,1.225931) **[blue]@{-};
(8.031462,1.227777); (8.111462,1.227777) **[blue]@{-};
(7.991462,1.179740); (8.151462,1.259740) **[red]@{-};
(8.151462,1.179740); (7.991462,1.259740) **[red]@{-};
(8.031462,1.198228); (8.111462,1.238228) **[red]@{-};
(8.031462,1.238228); (8.111462,1.198228) **[red]@{-};
(8.031462,1.200468); (8.111462,1.240468) **[red]@{-};
(8.031462,1.240468); (8.111462,1.200468) **[red]@{-};
(8.002149,1.228737); (8.162149,1.228737) **[blue]@{-};
(8.082149,1.188737); (8.082149,1.268737) **[blue]@{-};
(8.042149,1.227639); (8.122149,1.227639) **[blue]@{-};
(8.042149,1.230830); (8.122149,1.230830) **[blue]@{-};
(8.002149,1.179019); (8.162149,1.259019) **[red]@{-};
(8.162149,1.179019); (8.002149,1.259019) **[red]@{-};
(8.042149,1.198494); (8.122149,1.238494) **[red]@{-};
(8.042149,1.238494); (8.122149,1.198494) **[red]@{-};
(8.042149,1.200035); (8.122149,1.240035) **[red]@{-};
(8.042149,1.240035); (8.122149,1.200035) **[red]@{-};
(8.033742,1.226705); (8.193742,1.226705) **[blue]@{-};
(8.113742,1.186705); (8.113742,1.266705) **[blue]@{-};
(8.073742,1.225077); (8.153742,1.225077) **[blue]@{-};
(8.073742,1.227647); (8.153742,1.227647) **[blue]@{-};
(8.033742,1.176153); (8.193742,1.256153) **[red]@{-};
(8.193742,1.176153); (8.033742,1.256153) **[red]@{-};
(8.073742,1.194701); (8.153742,1.234701) **[red]@{-};
(8.073742,1.234701); (8.153742,1.194701) **[red]@{-};
(8.073742,1.196902); (8.153742,1.236902) **[red]@{-};
(8.073742,1.236902); (8.153742,1.196902) **[red]@{-};
(8.054426,1.227101); (8.214426,1.227101) **[blue]@{-};
(8.134426,1.187101); (8.134426,1.267101) **[blue]@{-};
(8.094426,1.225700); (8.174426,1.225700) **[blue]@{-};
(8.094426,1.228553); (8.174426,1.228553) **[blue]@{-};
(8.054426,1.199387); (8.214426,1.279387) **[red]@{-};
(8.214426,1.199387); (8.054426,1.279387) **[red]@{-};
(8.094426,1.218361); (8.174426,1.258361) **[red]@{-};
(8.094426,1.258361); (8.174426,1.218361) **[red]@{-};
(8.094426,1.221216); (8.174426,1.261216) **[red]@{-};
(8.094426,1.261216); (8.174426,1.221216) **[red]@{-};
(8.064658,1.265944); (8.224658,1.265944) **[blue]@{-};
(8.144658,1.225944); (8.144658,1.305944) **[blue]@{-};
(8.104658,1.264571); (8.184658,1.264571) **[blue]@{-};
(8.104658,1.266419); (8.184658,1.266419) **[blue]@{-};
(8.064658,1.154706); (8.224658,1.234706) **[red]@{-};
(8.224658,1.154706); (8.064658,1.234706) **[red]@{-};
(8.104658,1.172344); (8.184658,1.212344) **[red]@{-};
(8.104658,1.212344); (8.184658,1.172344) **[red]@{-};
(8.104658,1.175782); (8.184658,1.215782) **[red]@{-};
(8.104658,1.215782); (8.184658,1.175782) **[red]@{-};
(8.114757,1.225933); (8.274757,1.225933) **[blue]@{-};
(8.194757,1.185933); (8.194757,1.265933) **[blue]@{-};
(8.154757,1.225682); (8.234757,1.225682) **[blue]@{-};
(8.154757,1.226687); (8.234757,1.226687) **[blue]@{-};
(8.114757,1.135998); (8.274757,1.215998) **[red]@{-};
(8.274757,1.135998); (8.114757,1.215998) **[red]@{-};
(8.154757,1.155165); (8.234757,1.195165) **[red]@{-};
(8.154757,1.195165); (8.234757,1.155165) **[red]@{-};
(8.154757,1.157383); (8.234757,1.197383) **[red]@{-};
(8.154757,1.197383); (8.234757,1.157383) **[red]@{-};
(8.182095,1.226463); (8.342095,1.226463) **[blue]@{-};
(8.262095,1.186463); (8.262095,1.266463) **[blue]@{-};
(8.222095,1.225581); (8.302095,1.225581) **[blue]@{-};
(8.222095,1.227566); (8.302095,1.227566) **[blue]@{-};
(8.182095,1.139775); (8.342095,1.219775) **[red]@{-};
(8.342095,1.139775); (8.182095,1.219775) **[red]@{-};
(8.222095,1.158957); (8.302095,1.198957) **[red]@{-};
(8.222095,1.198957); (8.302095,1.158957) **[red]@{-};
(8.222095,1.160704); (8.302095,1.200704) **[red]@{-};
(8.222095,1.200704); (8.302095,1.160704) **[red]@{-};
(8.200771,1.230443); (8.360771,1.230443) **[blue]@{-};
(8.280771,1.190443); (8.280771,1.270443) **[blue]@{-};
(8.240771,1.228765); (8.320771,1.228765) **[blue]@{-};
(8.240771,1.232697); (8.320771,1.232697) **[blue]@{-};
(8.200771,1.139654); (8.360771,1.219654) **[red]@{-};
(8.360771,1.139654); (8.200771,1.219654) **[red]@{-};
(8.240771,1.158651); (8.320771,1.198651) **[red]@{-};
(8.240771,1.198651); (8.320771,1.158651) **[red]@{-};
(8.240771,1.160412); (8.320771,1.200412) **[red]@{-};
(8.240771,1.200412); (8.320771,1.160412) **[red]@{-};
(8.210019,1.226834); (8.370019,1.226834) **[blue]@{-};
(8.290019,1.186834); (8.290019,1.266834) **[blue]@{-};
(8.250019,1.226376); (8.330019,1.226376) **[blue]@{-};
(8.250019,1.227052); (8.330019,1.227052) **[blue]@{-};
(8.210019,1.135569); (8.370019,1.215569) **[red]@{-};
(8.370019,1.135569); (8.210019,1.215569) **[red]@{-};
(8.250019,1.153892); (8.330019,1.193892) **[red]@{-};
(8.250019,1.193892); (8.330019,1.153892) **[red]@{-};
(8.250019,1.156641); (8.330019,1.196641) **[red]@{-};
(8.250019,1.196641); (8.330019,1.156641) **[red]@{-};
(8.228339,1.265997); (8.388339,1.265997) **[blue]@{-};
(8.308339,1.225997); (8.308339,1.305997) **[blue]@{-};
(8.268339,1.264991); (8.348339,1.264991) **[blue]@{-};
(8.268339,1.266353); (8.348339,1.266353) **[blue]@{-};
(8.228339,1.136496); (8.388339,1.216496) **[red]@{-};
(8.388339,1.136496); (8.228339,1.216496) **[red]@{-};
(8.268339,1.155136); (8.348339,1.195136) **[red]@{-};
(8.268339,1.195136); (8.348339,1.155136) **[red]@{-};
(8.268339,1.157950); (8.348339,1.197950) **[red]@{-};
(8.268339,1.197950); (8.348339,1.157950) **[red]@{-};
(8.290687,1.226978); (8.450687,1.226978) **[blue]@{-};
(8.370687,1.186978); (8.370687,1.266978) **[blue]@{-};
(8.330687,1.225820); (8.410687,1.225820) **[blue]@{-};
(8.330687,1.228724); (8.410687,1.228724) **[blue]@{-};
(8.290687,1.117860); (8.450687,1.197860) **[red]@{-};
(8.450687,1.117860); (8.290687,1.197860) **[red]@{-};
(8.330687,1.137534); (8.410687,1.177534) **[red]@{-};
(8.330687,1.177534); (8.410687,1.137534) **[red]@{-};
(8.330687,1.139517); (8.410687,1.179517) **[red]@{-};
(8.330687,1.179517); (8.410687,1.139517) **[red]@{-};
(8.316605,1.225477); (8.476605,1.225477) **[blue]@{-};
(8.396605,1.185477); (8.396605,1.265477) **[blue]@{-};
(8.356605,1.225099); (8.436605,1.225099) **[blue]@{-};
(8.356605,1.225783); (8.436605,1.225783) **[blue]@{-};
(8.316605,1.091496); (8.476605,1.171496) **[red]@{-};
(8.476605,1.091496); (8.316605,1.171496) **[red]@{-};
(8.356605,1.110836); (8.436605,1.150836) **[red]@{-};
(8.356605,1.150836); (8.436605,1.110836) **[red]@{-};
(8.356605,1.114529); (8.436605,1.154529) **[red]@{-};
(8.356605,1.154529); (8.436605,1.114529) **[red]@{-};
(8.358792,1.265345); (8.518792,1.265345) **[blue]@{-};
(8.438792,1.225345); (8.438792,1.305345) **[blue]@{-};
(8.398792,1.263690); (8.478792,1.263690) **[blue]@{-};
(8.398792,1.266978); (8.478792,1.266978) **[blue]@{-};
(8.358792,1.093254); (8.518792,1.173254) **[red]@{-};
(8.518792,1.093254); (8.358792,1.173254) **[red]@{-};
(8.398792,1.111174); (8.478792,1.151174) **[red]@{-};
(8.398792,1.151174); (8.478792,1.111174) **[red]@{-};
(8.398792,1.114941); (8.478792,1.154941) **[red]@{-};
(8.398792,1.154941); (8.478792,1.114941) **[red]@{-};
(8.367083,1.264880); (8.527083,1.264880) **[blue]@{-};
(8.447083,1.224880); (8.447083,1.304880) **[blue]@{-};
(8.407083,1.263779); (8.487083,1.263779) **[blue]@{-};
(8.407083,1.265399); (8.487083,1.265399) **[blue]@{-};
(8.367083,1.099585); (8.527083,1.179585) **[red]@{-};
(8.527083,1.099585); (8.367083,1.179585) **[red]@{-};
(8.407083,1.117458); (8.487083,1.157458) **[red]@{-};
(8.407083,1.157458); (8.487083,1.117458) **[red]@{-};
(8.407083,1.120442); (8.487083,1.160442) **[red]@{-};
(8.407083,1.160442); (8.487083,1.120442) **[red]@{-};
(8.383524,1.224666); (8.543524,1.224666) **[blue]@{-};
(8.463524,1.184666); (8.463524,1.264666) **[blue]@{-};
(8.423524,1.223594); (8.503524,1.223594) **[blue]@{-};
(8.423524,1.226155); (8.503524,1.226155) **[blue]@{-};
(8.383524,1.092827); (8.543524,1.172827) **[red]@{-};
(8.543524,1.092827); (8.383524,1.172827) **[red]@{-};
(8.423524,1.111267); (8.503524,1.151267) **[red]@{-};
(8.423524,1.151267); (8.503524,1.111267) **[red]@{-};
(8.423524,1.113735); (8.503524,1.153735) **[red]@{-};
(8.423524,1.153735); (8.503524,1.113735) **[red]@{-};
(8.407840,1.227536); (8.567840,1.227536) **[blue]@{-};
(8.487840,1.187536); (8.487840,1.267536) **[blue]@{-};
(8.447840,1.226330); (8.527840,1.226330) **[blue]@{-};
(8.447840,1.228203); (8.527840,1.228203) **[blue]@{-};
(8.407840,1.094900); (8.567840,1.174900) **[red]@{-};
(8.567840,1.094900); (8.407840,1.174900) **[red]@{-};
(8.447840,1.113942); (8.527840,1.153942) **[red]@{-};
(8.447840,1.153942); (8.527840,1.113942) **[red]@{-};
(8.447840,1.115753); (8.527840,1.155753) **[red]@{-};
(8.447840,1.155753); (8.527840,1.115753) **[red]@{-};
(8.439636,1.227458); (8.599636,1.227458) **[blue]@{-};
(8.519636,1.187458); (8.519636,1.267458) **[blue]@{-};
(8.479636,1.226419); (8.559636,1.226419) **[blue]@{-};
(8.479636,1.228763); (8.559636,1.228763) **[blue]@{-};
(8.439636,1.096580); (8.599636,1.176580) **[red]@{-};
(8.599636,1.096580); (8.439636,1.176580) **[red]@{-};
(8.479636,1.114505); (8.559636,1.154505) **[red]@{-};
(8.479636,1.154505); (8.559636,1.114505) **[red]@{-};
(8.479636,1.117093); (8.559636,1.157093) **[red]@{-};
(8.479636,1.157093); (8.559636,1.117093) **[red]@{-};
(8.463032,1.226035); (8.623032,1.226035) **[blue]@{-};
(8.543032,1.186035); (8.543032,1.266035) **[blue]@{-};
(8.503032,1.225654); (8.583032,1.225654) **[blue]@{-};
(8.503032,1.228386); (8.583032,1.228386) **[blue]@{-};
(8.463032,1.100232); (8.623032,1.180232) **[red]@{-};
(8.623032,1.100232); (8.463032,1.180232) **[red]@{-};
(8.503032,1.119361); (8.583032,1.159361) **[red]@{-};
(8.503032,1.159361); (8.583032,1.119361) **[red]@{-};
(8.503032,1.121444); (8.583032,1.161444) **[red]@{-};
(8.503032,1.161444); (8.583032,1.121444) **[red]@{-};
(9.117217,1.225666); (9.277217,1.225666) **[blue]@{-};
(9.197217,1.185666); (9.197217,1.265666) **[blue]@{-};
(9.157217,1.225043); (9.237217,1.225043) **[blue]@{-};
(9.157217,1.226444); (9.237217,1.226444) **[blue]@{-};
(9.117217,0.888374); (9.277217,0.968374) **[red]@{-};
(9.277217,0.888374); (9.117217,0.968374) **[red]@{-};
(9.157217,0.907213); (9.237217,0.947213) **[red]@{-};
(9.157217,0.947213); (9.237217,0.907213) **[red]@{-};
(9.157217,0.910469); (9.237217,0.950469) **[red]@{-};
(9.157217,0.950469); (9.237217,0.910469) **[red]@{-};
(2.241928,2.212784); (2.401928,2.212784) **[blue]@{-};
(2.321928,2.172784); (2.321928,2.252784) **[blue]@{-};
(2.281928,2.128022); (2.361928,2.128022) **[blue]@{-};
(2.281928,2.232152); (2.361928,2.232152) **[blue]@{-};
(2.241928,3.531788); (2.401928,3.611788) **[red]@{-};
(2.401928,3.531788); (2.241928,3.611788) **[red]@{-};
(2.281928,3.450403); (2.361928,3.490403) **[red]@{-};
(2.281928,3.490403); (2.361928,3.450403) **[red]@{-};
(2.281928,3.575724); (2.361928,3.615724) **[red]@{-};
(2.281928,3.615724); (2.361928,3.575724) **[red]@{-};
(2.727355,1.973752); (2.887355,1.973752) **[blue]@{-};
(2.807355,1.933752); (2.807355,2.013752) **[blue]@{-};
(2.767355,1.961302); (2.847355,1.961302) **[blue]@{-};
(2.767355,1.990577); (2.847355,1.990577) **[blue]@{-};
(2.727355,2.945025); (2.887355,3.025025) **[red]@{-};
(2.887355,2.945025); (2.727355,3.025025) **[red]@{-};
(2.767355,2.956396); (2.847355,2.996396) **[red]@{-};
(2.767355,2.996396); (2.847355,2.956396) **[red]@{-};
(2.767355,2.975971); (2.847355,3.015971) **[red]@{-};
(2.767355,3.015971); (2.847355,2.975971) **[red]@{-};
(3.379432,2.035263); (3.539432,2.035263) **[blue]@{-};
(3.459432,1.995263); (3.459432,2.075263) **[blue]@{-};
(3.419432,2.027294); (3.499432,2.027294) **[blue]@{-};
(3.419432,2.041441); (3.499432,2.041441) **[blue]@{-};
(3.379432,3.178510); (3.539432,3.258510) **[red]@{-};
(3.539432,3.178510); (3.379432,3.258510) **[red]@{-};
(3.419432,3.193373); (3.499432,3.233373) **[red]@{-};
(3.419432,3.233373); (3.499432,3.193373) **[red]@{-};
(3.419432,3.253188); (3.499432,3.293188) **[red]@{-};
(3.419432,3.293188); (3.499432,3.253188) **[red]@{-};
(3.620440,2.036390); (3.780440,2.036390) **[blue]@{-};
(3.700440,1.996390); (3.700440,2.076390) **[blue]@{-};
(3.660440,2.032142); (3.740440,2.032142) **[blue]@{-};
(3.660440,2.041582); (3.740440,2.041582) **[blue]@{-};
(3.620440,3.074500); (3.780440,3.154500) **[red]@{-};
(3.780440,3.074500); (3.620440,3.154500) **[red]@{-};
(3.660440,3.084745); (3.740440,3.124745) **[red]@{-};
(3.660440,3.124745); (3.740440,3.084745) **[red]@{-};
(3.660440,3.101155); (3.740440,3.141155) **[red]@{-};
(3.660440,3.141155); (3.740440,3.101155) **[red]@{-};
(4.007463,2.050716); (4.167463,2.050716) **[blue]@{-};
(4.087463,2.010716); (4.087463,2.090716) **[blue]@{-};
(4.047463,2.044102); (4.127463,2.044102) **[blue]@{-};
(4.047463,2.054402); (4.127463,2.054402) **[blue]@{-};
(4.007463,3.104391); (4.167463,3.184391) **[red]@{-};
(4.167463,3.104391); (4.007463,3.184391) **[red]@{-};
(4.047463,3.116707); (4.127463,3.156707) **[red]@{-};
(4.047463,3.156707); (4.127463,3.116707) **[red]@{-};
(4.047463,3.131248); (4.127463,3.171248) **[red]@{-};
(4.047463,3.171248); (4.127463,3.131248) **[red]@{-};
(4.167928,2.070122); (4.327928,2.070122) **[blue]@{-};
(4.247928,2.030122); (4.247928,2.110122) **[blue]@{-};
(4.207928,2.066460); (4.287928,2.066460) **[blue]@{-};
(4.207928,2.076974); (4.287928,2.076974) **[blue]@{-};
(4.167928,3.035765); (4.327928,3.115765) **[red]@{-};
(4.327928,3.035765); (4.167928,3.115765) **[red]@{-};
(4.207928,3.054165); (4.287928,3.094165) **[red]@{-};
(4.207928,3.094165); (4.287928,3.054165) **[red]@{-};
(4.207928,3.057905); (4.287928,3.097905) **[red]@{-};
(4.207928,3.097905); (4.287928,3.057905) **[red]@{-};
(4.443562,2.070318); (4.603562,2.070318) **[blue]@{-};
(4.523562,2.030318); (4.523562,2.110318) **[blue]@{-};
(4.483562,2.067196); (4.563562,2.067196) **[blue]@{-};
(4.483562,2.073831); (4.563562,2.073831) **[blue]@{-};
(4.443562,2.916356); (4.603562,2.996356) **[red]@{-};
(4.603562,2.916356); (4.443562,2.996356) **[red]@{-};
(4.483562,2.932748); (4.563562,2.972748) **[red]@{-};
(4.483562,2.972748); (4.563562,2.932748) **[red]@{-};
(4.483562,2.944163); (4.563562,2.984163) **[red]@{-};
(4.483562,2.984163); (4.563562,2.944163) **[red]@{-};
(4.777981,2.077015); (4.937981,2.077015) **[blue]@{-};
(4.857981,2.037015); (4.857981,2.117015) **[blue]@{-};
(4.817981,2.068350); (4.897981,2.068350) **[blue]@{-};
(4.817981,2.085274); (4.897981,2.085274) **[blue]@{-};
(4.777981,2.783163); (4.937981,2.863163) **[red]@{-};
(4.937981,2.783163); (4.777981,2.863163) **[red]@{-};
(4.817981,2.788070); (4.897981,2.828070) **[red]@{-};
(4.817981,2.828070); (4.897981,2.788070) **[red]@{-};
(4.817981,2.811924); (4.897981,2.851924) **[red]@{-};
(4.817981,2.851924); (4.897981,2.811924) **[red]@{-};
(4.874196,2.085339); (5.034196,2.085339) **[blue]@{-};
(4.954196,2.045339); (4.954196,2.125339) **[blue]@{-};
(4.914196,2.077194); (4.994196,2.077194) **[blue]@{-};
(4.914196,2.096198); (4.994196,2.096198) **[blue]@{-};
(4.874196,2.744530); (5.034196,2.824530) **[red]@{-};
(5.034196,2.744530); (4.874196,2.824530) **[red]@{-};
(4.914196,2.760887); (4.994196,2.800887) **[red]@{-};
(4.914196,2.800887); (4.994196,2.760887) **[red]@{-};
(4.914196,2.768779); (4.994196,2.808779) **[red]@{-};
(4.914196,2.808779); (4.994196,2.768779) **[red]@{-};
(5.129453,2.071430); (5.289453,2.071430) **[blue]@{-};
(5.209453,2.031430); (5.209453,2.111430) **[blue]@{-};
(5.169453,2.069009); (5.249453,2.069009) **[blue]@{-};
(5.169453,2.082904); (5.249453,2.082904) **[blue]@{-};
(5.129453,2.696793); (5.289453,2.776793) **[red]@{-};
(5.289453,2.696793); (5.129453,2.776793) **[red]@{-};
(5.169453,2.711214); (5.249453,2.751214) **[red]@{-};
(5.169453,2.751214); (5.249453,2.711214) **[red]@{-};
(5.169453,2.724246); (5.249453,2.764246) **[red]@{-};
(5.169453,2.764246); (5.249453,2.724246) **[red]@{-};
(5.277552,2.069404); (5.437552,2.069404) **[blue]@{-};
(5.357552,2.029404); (5.357552,2.109404) **[blue]@{-};
(5.317552,2.064375); (5.397552,2.064375) **[blue]@{-};
(5.317552,2.084413); (5.397552,2.084413) **[blue]@{-};
(5.277552,2.642421); (5.437552,2.722421) **[red]@{-};
(5.437552,2.642421); (5.277552,2.722421) **[red]@{-};
(5.317552,2.660824); (5.397552,2.700824) **[red]@{-};
(5.317552,2.700824); (5.397552,2.660824) **[red]@{-};
(5.317552,2.667159); (5.397552,2.707159) **[red]@{-};
(5.317552,2.707159); (5.397552,2.667159) **[red]@{-};
(5.346265,2.077115); (5.506265,2.077115) **[blue]@{-};
(5.426265,2.037115); (5.426265,2.117115) **[blue]@{-};
(5.386265,2.068519); (5.466265,2.068519) **[blue]@{-};
(5.386265,2.082369); (5.466265,2.082369) **[blue]@{-};
(5.346265,2.622181); (5.506265,2.702181) **[red]@{-};
(5.506265,2.622181); (5.346265,2.702181) **[red]@{-};
(5.386265,2.636578); (5.466265,2.676578) **[red]@{-};
(5.386265,2.676578); (5.466265,2.636578) **[red]@{-};
(5.386265,2.651700); (5.466265,2.691700) **[red]@{-};
(5.386265,2.691700); (5.466265,2.651700) **[red]@{-};
(5.474589,2.079584); (5.634589,2.079584) **[blue]@{-};
(5.554589,2.039584); (5.554589,2.119584) **[blue]@{-};
(5.514589,2.075703); (5.594589,2.075703) **[blue]@{-};
(5.514589,2.090207); (5.594589,2.090207) **[blue]@{-};
(5.474589,2.597412); (5.634589,2.677412) **[red]@{-};
(5.634589,2.597412); (5.474589,2.677412) **[red]@{-};
(5.514589,2.604653); (5.594589,2.644653) **[red]@{-};
(5.514589,2.644653); (5.594589,2.604653) **[red]@{-};
(5.514589,2.626853); (5.594589,2.666853) **[red]@{-};
(5.514589,2.666853); (5.594589,2.626853) **[red]@{-};
(5.647920,2.079105); (5.807920,2.079105) **[blue]@{-};
(5.727920,2.039105); (5.727920,2.119105) **[blue]@{-};
(5.687920,2.073310); (5.767920,2.073310) **[blue]@{-};
(5.687920,2.086376); (5.767920,2.086376) **[blue]@{-};
(5.647920,2.580097); (5.807920,2.660097) **[red]@{-};
(5.807920,2.580097); (5.647920,2.660097) **[red]@{-};
(5.687920,2.588701); (5.767920,2.628701) **[red]@{-};
(5.687920,2.628701); (5.767920,2.588701) **[red]@{-};
(5.687920,2.608916); (5.767920,2.648916) **[red]@{-};
(5.687920,2.648916); (5.767920,2.608916) **[red]@{-};
(5.802643,2.080784); (5.962643,2.080784) **[blue]@{-};
(5.882643,2.040784); (5.882643,2.120784) **[blue]@{-};
(5.842643,2.073716); (5.922643,2.073716) **[blue]@{-};
(5.842643,2.087779); (5.922643,2.087779) **[blue]@{-};
(5.802643,2.533548); (5.962643,2.613548) **[red]@{-};
(5.962643,2.533548); (5.802643,2.613548) **[red]@{-};
(5.842643,2.550582); (5.922643,2.590582) **[red]@{-};
(5.842643,2.590582); (5.922643,2.550582) **[red]@{-};
(5.842643,2.560029); (5.922643,2.600029) **[red]@{-};
(5.842643,2.600029); (5.922643,2.560029) **[red]@{-};
(5.850737,2.082927); (6.010737,2.082927) **[blue]@{-};
(5.930737,2.042927); (5.930737,2.122927) **[blue]@{-};
(5.890737,2.077437); (5.970737,2.077437) **[blue]@{-};
(5.890737,2.091846); (5.970737,2.091846) **[blue]@{-};
(5.850737,2.522685); (6.010737,2.602685) **[red]@{-};
(6.010737,2.522685); (5.850737,2.602685) **[red]@{-};
(5.890737,2.541541); (5.970737,2.581541) **[red]@{-};
(5.890737,2.581541); (5.970737,2.541541) **[red]@{-};
(5.890737,2.551075); (5.970737,2.591075) **[red]@{-};
(5.890737,2.591075); (5.970737,2.551075) **[red]@{-};
(5.986089,2.086236); (6.146089,2.086236) **[blue]@{-};
(6.066089,2.046236); (6.066089,2.126236) **[blue]@{-};
(6.026089,2.077809); (6.106089,2.077809) **[blue]@{-};
(6.026089,2.094013); (6.106089,2.094013) **[blue]@{-};
(5.986089,2.482797); (6.146089,2.562797) **[red]@{-};
(6.146089,2.482797); (5.986089,2.562797) **[red]@{-};
(6.026089,2.494828); (6.106089,2.534828) **[red]@{-};
(6.026089,2.534828); (6.106089,2.494828) **[red]@{-};
(6.026089,2.516442); (6.106089,2.556442) **[red]@{-};
(6.026089,2.556442); (6.106089,2.516442) **[red]@{-};
(6.069747,2.084207); (6.229747,2.084207) **[blue]@{-};
(6.149747,2.044207); (6.149747,2.124207) **[blue]@{-};
(6.109747,2.079364); (6.189747,2.079364) **[blue]@{-};
(6.109747,2.090049); (6.189747,2.090049) **[blue]@{-};
(6.069747,2.464041); (6.229747,2.544041) **[red]@{-};
(6.229747,2.464041); (6.069747,2.544041) **[red]@{-};
(6.109747,2.478925); (6.189747,2.518925) **[red]@{-};
(6.109747,2.518925); (6.189747,2.478925) **[red]@{-};
(6.109747,2.494266); (6.189747,2.534266) **[red]@{-};
(6.109747,2.534266); (6.189747,2.494266) **[red]@{-};
(6.109825,2.082002); (6.269825,2.082002) **[blue]@{-};
(6.189825,2.042002); (6.189825,2.122002) **[blue]@{-};
(6.149825,2.074914); (6.229825,2.074914) **[blue]@{-};
(6.149825,2.088555); (6.229825,2.088555) **[blue]@{-};
(6.109825,2.449120); (6.269825,2.529120) **[red]@{-};
(6.269825,2.449120); (6.109825,2.529120) **[red]@{-};
(6.149825,2.466006); (6.229825,2.506006) **[red]@{-};
(6.149825,2.506006); (6.229825,2.466006) **[red]@{-};
(6.149825,2.481346); (6.229825,2.521346) **[red]@{-};
(6.149825,2.521346); (6.229825,2.481346) **[red]@{-};
(6.223781,2.085923); (6.383781,2.085923) **[blue]@{-};
(6.303781,2.045923); (6.303781,2.125923) **[blue]@{-};
(6.263781,2.078386); (6.343781,2.078386) **[blue]@{-};
(6.263781,2.090976); (6.343781,2.090976) **[blue]@{-};
(6.223781,2.428726); (6.383781,2.508726) **[red]@{-};
(6.383781,2.428726); (6.223781,2.508726) **[red]@{-};
(6.263781,2.444013); (6.343781,2.484013) **[red]@{-};
(6.263781,2.484013); (6.343781,2.444013) **[red]@{-};
(6.263781,2.453741); (6.343781,2.493741) **[red]@{-};
(6.263781,2.493741); (6.343781,2.453741) **[red]@{-};
(6.295039,2.085104); (6.455039,2.085104) **[blue]@{-};
(6.375039,2.045104); (6.375039,2.125104) **[blue]@{-};
(6.335039,2.081735); (6.415039,2.081735) **[blue]@{-};
(6.335039,2.090344); (6.415039,2.090344) **[blue]@{-};
(6.295039,2.301873); (6.455039,2.381873) **[red]@{-};
(6.455039,2.301873); (6.295039,2.381873) **[red]@{-};
(6.335039,2.316332); (6.415039,2.356332) **[red]@{-};
(6.335039,2.356332); (6.415039,2.316332) **[red]@{-};
(6.335039,2.328884); (6.415039,2.368884) **[red]@{-};
(6.335039,2.368884); (6.415039,2.328884) **[red]@{-};
(6.395733,2.084134); (6.555733,2.084134) **[blue]@{-};
(6.475733,2.044134); (6.475733,2.124134) **[blue]@{-};
(6.435733,2.075263); (6.515733,2.075263) **[blue]@{-};
(6.435733,2.088771); (6.515733,2.088771) **[blue]@{-};
(6.395733,2.282406); (6.555733,2.362406) **[red]@{-};
(6.555733,2.282406); (6.395733,2.362406) **[red]@{-};
(6.435733,2.297600); (6.515733,2.337600) **[red]@{-};
(6.435733,2.337600); (6.515733,2.297600) **[red]@{-};
(6.435733,2.308775); (6.515733,2.348775) **[red]@{-};
(6.435733,2.348775); (6.515733,2.308775) **[red]@{-};
(6.519913,2.085222); (6.679913,2.085222) **[blue]@{-};
(6.599913,2.045222); (6.599913,2.125222) **[blue]@{-};
(6.559913,2.077023); (6.639913,2.077023) **[blue]@{-};
(6.559913,2.087956); (6.639913,2.087956) **[blue]@{-};
(6.519913,2.264450); (6.679913,2.344450) **[red]@{-};
(6.679913,2.264450); (6.519913,2.344450) **[red]@{-};
(6.559913,2.280578); (6.639913,2.320578) **[red]@{-};
(6.559913,2.320578); (6.639913,2.280578) **[red]@{-};
(6.559913,2.287638); (6.639913,2.327638) **[red]@{-};
(6.559913,2.327638); (6.639913,2.287638) **[red]@{-};
(6.578211,2.081554); (6.738211,2.081554) **[blue]@{-};
(6.658211,2.041554); (6.658211,2.121554) **[blue]@{-};
(6.618211,2.079763); (6.698211,2.079763) **[blue]@{-};
(6.618211,2.090107); (6.698211,2.090107) **[blue]@{-};
(6.578211,2.204610); (6.738211,2.284610) **[red]@{-};
(6.738211,2.204610); (6.578211,2.284610) **[red]@{-};
(6.618211,2.220448); (6.698211,2.260448) **[red]@{-};
(6.618211,2.260448); (6.698211,2.220448) **[red]@{-};
(6.618211,2.230555); (6.698211,2.270555) **[red]@{-};
(6.618211,2.270555); (6.698211,2.230555) **[red]@{-};
(6.606501,2.085684); (6.766501,2.085684) **[blue]@{-};
(6.686501,2.045684); (6.686501,2.125684) **[blue]@{-};
(6.646501,2.075884); (6.726501,2.075884) **[blue]@{-};
(6.646501,2.089920); (6.726501,2.089920) **[blue]@{-};
(6.606501,2.202198); (6.766501,2.282198) **[red]@{-};
(6.766501,2.202198); (6.606501,2.282198) **[red]@{-};
(6.646501,2.219510); (6.726501,2.259510) **[red]@{-};
(6.646501,2.259510); (6.726501,2.219510) **[red]@{-};
(6.646501,2.226341); (6.726501,2.266341) **[red]@{-};
(6.646501,2.266341); (6.726501,2.226341) **[red]@{-};
(6.661467,2.085740); (6.821467,2.085740) **[blue]@{-};
(6.741467,2.045740); (6.741467,2.125740) **[blue]@{-};
(6.701467,2.074914); (6.781467,2.074914) **[blue]@{-};
(6.701467,2.093573); (6.781467,2.093573) **[blue]@{-};
(6.661467,2.193519); (6.821467,2.273519) **[red]@{-};
(6.821467,2.193519); (6.661467,2.273519) **[red]@{-};
(6.701467,2.210822); (6.781467,2.250822) **[red]@{-};
(6.701467,2.250822); (6.781467,2.210822) **[red]@{-};
(6.701467,2.218573); (6.781467,2.258573) **[red]@{-};
(6.701467,2.258573); (6.781467,2.218573) **[red]@{-};
(6.688184,2.083299); (6.848184,2.083299) **[blue]@{-};
(6.768184,2.043299); (6.768184,2.123299) **[blue]@{-};
(6.728184,2.075174); (6.808184,2.075174) **[blue]@{-};
(6.728184,2.096734); (6.808184,2.096734) **[blue]@{-};
(6.688184,2.207069); (6.848184,2.287069) **[red]@{-};
(6.848184,2.207069); (6.688184,2.287069) **[red]@{-};
(6.728184,2.219522); (6.808184,2.259522) **[red]@{-};
(6.728184,2.259522); (6.808184,2.219522) **[red]@{-};
(6.728184,2.228918); (6.808184,2.268918) **[red]@{-};
(6.728184,2.268918); (6.808184,2.228918) **[red]@{-};
(6.740179,2.078648); (6.900179,2.078648) **[blue]@{-};
(6.820179,2.038648); (6.820179,2.118648) **[blue]@{-};
(6.780179,2.071393); (6.860179,2.071393) **[blue]@{-};
(6.780179,2.085752); (6.860179,2.085752) **[blue]@{-};
(6.740179,2.195323); (6.900179,2.275323) **[red]@{-};
(6.900179,2.195323); (6.740179,2.275323) **[red]@{-};
(6.780179,2.213571); (6.860179,2.253571) **[red]@{-};
(6.780179,2.253571); (6.860179,2.213571) **[red]@{-};
(6.780179,2.221896); (6.860179,2.261896) **[red]@{-};
(6.780179,2.261896); (6.860179,2.221896) **[red]@{-};
(6.908685,2.086284); (7.068685,2.086284) **[blue]@{-};
(6.988685,2.046284); (6.988685,2.126284) **[blue]@{-};
(6.948685,2.078637); (7.028685,2.078637) **[blue]@{-};
(6.948685,2.096517); (7.028685,2.096517) **[blue]@{-};
(6.908685,2.148696); (7.068685,2.228696) **[red]@{-};
(7.068685,2.148696); (6.908685,2.228696) **[red]@{-};
(6.948685,2.165599); (7.028685,2.205599) **[red]@{-};
(6.948685,2.205599); (7.028685,2.165599) **[red]@{-};
(6.948685,2.172605); (7.028685,2.212605) **[red]@{-};
(6.948685,2.212605); (7.028685,2.172605) **[red]@{-};
(6.953423,2.081531); (7.113423,2.081531) **[blue]@{-};
(7.033423,2.041531); (7.033423,2.121531) **[blue]@{-};
(6.993423,2.072515); (7.073423,2.072515) **[blue]@{-};
(6.993423,2.087358); (7.073423,2.087358) **[blue]@{-};
(6.953423,2.140524); (7.113423,2.220524) **[red]@{-};
(7.113423,2.140524); (6.953423,2.220524) **[red]@{-};
(6.993423,2.158579); (7.073423,2.198579) **[red]@{-};
(6.993423,2.198579); (7.073423,2.158579) **[red]@{-};
(6.993423,2.167287); (7.073423,2.207287) **[red]@{-};
(6.993423,2.207287); (7.073423,2.167287) **[red]@{-};
(7.018032,2.087223); (7.178032,2.087223) **[blue]@{-};
(7.098032,2.047223); (7.098032,2.127223) **[blue]@{-};
(7.058032,2.076260); (7.138032,2.076260) **[blue]@{-};
(7.058032,2.093412); (7.138032,2.093412) **[blue]@{-};
(7.018032,2.135533); (7.178032,2.215533) **[red]@{-};
(7.178032,2.135533); (7.018032,2.215533) **[red]@{-};
(7.058032,2.151678); (7.138032,2.191678) **[red]@{-};
(7.058032,2.191678); (7.138032,2.151678) **[red]@{-};
(7.058032,2.156995); (7.138032,2.196995) **[red]@{-};
(7.058032,2.196995); (7.138032,2.156995) **[red]@{-};
(7.038941,2.091441); (7.198941,2.091441) **[blue]@{-};
(7.118941,2.051441); (7.118941,2.131441) **[blue]@{-};
(7.078941,2.079650); (7.158941,2.079650) **[blue]@{-};
(7.078941,2.098817); (7.158941,2.098817) **[blue]@{-};
(7.038941,2.139947); (7.198941,2.219947) **[red]@{-};
(7.198941,2.139947); (7.038941,2.219947) **[red]@{-};
(7.078941,2.156930); (7.158941,2.196930) **[red]@{-};
(7.078941,2.196930); (7.158941,2.156930) **[red]@{-};
(7.078941,2.161953); (7.158941,2.201953) **[red]@{-};
(7.078941,2.201953); (7.158941,2.161953) **[red]@{-};
(7.139169,2.085403); (7.299169,2.085403) **[blue]@{-};
(7.219169,2.045403); (7.219169,2.125403) **[blue]@{-};
(7.179169,2.078791); (7.259169,2.078791) **[blue]@{-};
(7.179169,2.090729); (7.259169,2.090729) **[blue]@{-};
(7.139169,2.051736); (7.299169,2.131736) **[red]@{-};
(7.299169,2.051736); (7.139169,2.131736) **[red]@{-};
(7.179169,2.061818); (7.259169,2.101818) **[red]@{-};
(7.179169,2.101818); (7.259169,2.061818) **[red]@{-};
(7.179169,2.084516); (7.259169,2.124516) **[red]@{-};
(7.179169,2.124516); (7.259169,2.084516) **[red]@{-};
(7.158405,2.087719); (7.318405,2.087719) **[blue]@{-};
(7.238405,2.047719); (7.238405,2.127719) **[blue]@{-};
(7.198405,2.082405); (7.278405,2.082405) **[blue]@{-};
(7.198405,2.091960); (7.278405,2.091960) **[blue]@{-};
(7.158405,2.043062); (7.318405,2.123062) **[red]@{-};
(7.318405,2.043062); (7.158405,2.123062) **[red]@{-};
(7.198405,2.059633); (7.278405,2.099633) **[red]@{-};
(7.198405,2.099633); (7.278405,2.059633) **[red]@{-};
(7.198405,2.066802); (7.278405,2.106802) **[red]@{-};
(7.198405,2.106802); (7.278405,2.066802) **[red]@{-};
(7.214621,2.084793); (7.374621,2.084793) **[blue]@{-};
(7.294621,2.044793); (7.294621,2.124793) **[blue]@{-};
(7.254621,2.078718); (7.334621,2.078718) **[blue]@{-};
(7.254621,2.096771); (7.334621,2.096771) **[blue]@{-};
(7.214621,2.043640); (7.374621,2.123640) **[red]@{-};
(7.374621,2.043640); (7.214621,2.123640) **[red]@{-};
(7.254621,2.061095); (7.334621,2.101095) **[red]@{-};
(7.254621,2.101095); (7.334621,2.061095) **[red]@{-};
(7.254621,2.065841); (7.334621,2.105841) **[red]@{-};
(7.254621,2.105841); (7.334621,2.065841) **[red]@{-};
(7.268728,2.088016); (7.428728,2.088016) **[blue]@{-};
(7.348728,2.048016); (7.348728,2.128016) **[blue]@{-};
(7.308728,2.079452); (7.388728,2.079452) **[blue]@{-};
(7.308728,2.095744); (7.388728,2.095744) **[blue]@{-};
(7.268728,2.045072); (7.428728,2.125072) **[red]@{-};
(7.428728,2.045072); (7.268728,2.125072) **[red]@{-};
(7.308728,2.062234); (7.388728,2.102234) **[red]@{-};
(7.308728,2.102234); (7.388728,2.062234) **[red]@{-};
(7.308728,2.070172); (7.388728,2.110172) **[red]@{-};
(7.308728,2.110172); (7.388728,2.070172) **[red]@{-};
(7.303704,2.083202); (7.463704,2.083202) **[blue]@{-};
(7.383704,2.043202); (7.383704,2.123202) **[blue]@{-};
(7.343704,2.074353); (7.423704,2.074353) **[blue]@{-};
(7.343704,2.089107); (7.423704,2.089107) **[blue]@{-};
(7.303704,2.042978); (7.463704,2.122978) **[red]@{-};
(7.463704,2.042978); (7.303704,2.122978) **[red]@{-};
(7.343704,2.061417); (7.423704,2.101417) **[red]@{-};
(7.343704,2.101417); (7.423704,2.061417) **[red]@{-};
(7.343704,2.065401); (7.423704,2.105401) **[red]@{-};
(7.343704,2.105401); (7.423704,2.065401) **[red]@{-};
(7.354628,2.089683); (7.514628,2.089683) **[blue]@{-};
(7.434628,2.049683); (7.434628,2.129683) **[blue]@{-};
(7.394628,2.080509); (7.474628,2.080509) **[blue]@{-};
(7.394628,2.092266); (7.474628,2.092266) **[blue]@{-};
(7.354628,1.970111); (7.514628,2.050111) **[red]@{-};
(7.514628,1.970111); (7.354628,2.050111) **[red]@{-};
(7.394628,1.987038); (7.474628,2.027038) **[red]@{-};
(7.394628,2.027038); (7.474628,1.987038) **[red]@{-};
(7.394628,1.997417); (7.474628,2.037417) **[red]@{-};
(7.394628,2.037417); (7.474628,1.997417) **[red]@{-};
(7.403816,2.082236); (7.563816,2.082236) **[blue]@{-};
(7.483816,2.042236); (7.483816,2.122236) **[blue]@{-};
(7.443816,2.071004); (7.523816,2.071004) **[blue]@{-};
(7.443816,2.089768); (7.523816,2.089768) **[blue]@{-};
(7.403816,1.966979); (7.563816,2.046979) **[red]@{-};
(7.563816,1.966979); (7.403816,2.046979) **[red]@{-};
(7.443816,1.983479); (7.523816,2.023479) **[red]@{-};
(7.443816,2.023479); (7.523816,1.983479) **[red]@{-};
(7.443816,1.988704); (7.523816,2.028704) **[red]@{-};
(7.443816,2.028704); (7.523816,1.988704) **[red]@{-};
(7.419846,2.083506); (7.579846,2.083506) **[blue]@{-};
(7.499846,2.043506); (7.499846,2.123506) **[blue]@{-};
(7.459846,2.079320); (7.539846,2.079320) **[blue]@{-};
(7.459846,2.089263); (7.539846,2.089263) **[blue]@{-};
(7.419846,1.968102); (7.579846,2.048102) **[red]@{-};
(7.579846,1.968102); (7.419846,2.048102) **[red]@{-};
(7.459846,1.985973); (7.539846,2.025973) **[red]@{-};
(7.459846,2.025973); (7.539846,1.985973) **[red]@{-};
(7.459846,1.990808); (7.539846,2.030808) **[red]@{-};
(7.459846,2.030808); (7.539846,1.990808) **[red]@{-};
(7.497429,2.080325); (7.657429,2.080325) **[blue]@{-};
(7.577429,2.040325); (7.577429,2.120325) **[blue]@{-};
(7.537429,2.074476); (7.617429,2.074476) **[blue]@{-};
(7.537429,2.087375); (7.617429,2.087375) **[blue]@{-};
(7.497429,1.970952); (7.657429,2.050952) **[red]@{-};
(7.657429,1.970952); (7.497429,2.050952) **[red]@{-};
(7.537429,1.989353); (7.617429,2.029353) **[red]@{-};
(7.537429,2.029353); (7.617429,1.989353) **[red]@{-};
(7.537429,1.994035); (7.617429,2.034035) **[red]@{-};
(7.537429,2.034035); (7.617429,1.994035) **[red]@{-};
(7.512457,2.084801); (7.672457,2.084801) **[blue]@{-};
(7.592457,2.044801); (7.592457,2.124801) **[blue]@{-};
(7.552457,2.073042); (7.632457,2.073042) **[blue]@{-};
(7.552457,2.092667); (7.632457,2.092667) **[blue]@{-};
(7.512457,1.971498); (7.672457,2.051498) **[red]@{-};
(7.672457,1.971498); (7.512457,2.051498) **[red]@{-};
(7.552457,1.987372); (7.632457,2.027372) **[red]@{-};
(7.552457,2.027372); (7.632457,1.987372) **[red]@{-};
(7.552457,1.997863); (7.632457,2.037863) **[red]@{-};
(7.552457,2.037863); (7.632457,1.997863) **[red]@{-};
(7.542052,2.076180); (7.702052,2.076180) **[blue]@{-};
(7.622052,2.036180); (7.622052,2.116180) **[blue]@{-};
(7.582052,2.070565); (7.662052,2.070565) **[blue]@{-};
(7.582052,2.084042); (7.662052,2.084042) **[blue]@{-};
(7.542052,1.903451); (7.702052,1.983451) **[red]@{-};
(7.702052,1.903451); (7.542052,1.983451) **[red]@{-};
(7.582052,1.919138); (7.662052,1.959138) **[red]@{-};
(7.582052,1.959138); (7.662052,1.919138) **[red]@{-};
(7.582052,1.925017); (7.662052,1.965017) **[red]@{-};
(7.582052,1.965017); (7.662052,1.925017) **[red]@{-};
(7.556625,2.081593); (7.716625,2.081593) **[blue]@{-};
(7.636625,2.041593); (7.636625,2.121593) **[blue]@{-};
(7.596625,2.072358); (7.676625,2.072358) **[blue]@{-};
(7.596625,2.089560); (7.676625,2.089560) **[blue]@{-};
(7.556625,1.906192); (7.716625,1.986192) **[red]@{-};
(7.716625,1.906192); (7.556625,1.986192) **[red]@{-};
(7.596625,1.923596); (7.676625,1.963596) **[red]@{-};
(7.596625,1.963596); (7.676625,1.923596) **[red]@{-};
(7.596625,1.932332); (7.676625,1.972332) **[red]@{-};
(7.596625,1.972332); (7.676625,1.932332) **[red]@{-};
(7.641099,2.081315); (7.801099,2.081315) **[blue]@{-};
(7.721099,2.041315); (7.721099,2.121315) **[blue]@{-};
(7.681099,2.075584); (7.761099,2.075584) **[blue]@{-};
(7.681099,2.089130); (7.761099,2.089130) **[blue]@{-};
(7.641099,1.914425); (7.801099,1.994425) **[red]@{-};
(7.801099,1.914425); (7.641099,1.994425) **[red]@{-};
(7.681099,1.931762); (7.761099,1.971762) **[red]@{-};
(7.681099,1.971762); (7.761099,1.931762) **[red]@{-};
(7.681099,1.937669); (7.761099,1.977669) **[red]@{-};
(7.681099,1.977669); (7.761099,1.937669) **[red]@{-};
(7.720900,2.082313); (7.880900,2.082313) **[blue]@{-};
(7.800900,2.042313); (7.800900,2.122313) **[blue]@{-};
(7.760900,2.077379); (7.840900,2.077379) **[blue]@{-};
(7.760900,2.089223); (7.840900,2.089223) **[blue]@{-};
(7.720900,1.921768); (7.880900,2.001768) **[red]@{-};
(7.880900,1.921768); (7.720900,2.001768) **[red]@{-};
(7.760900,1.938869); (7.840900,1.978869) **[red]@{-};
(7.760900,1.978869); (7.840900,1.938869) **[red]@{-};
(7.760900,1.946482); (7.840900,1.986482) **[red]@{-};
(7.760900,1.986482); (7.840900,1.946482) **[red]@{-};
(7.746548,2.080859); (7.906548,2.080859) **[blue]@{-};
(7.826548,2.040859); (7.826548,2.120859) **[blue]@{-};
(7.786548,2.073041); (7.866548,2.073041) **[blue]@{-};
(7.786548,2.083345); (7.866548,2.083345) **[blue]@{-};
(7.746548,1.886884); (7.906548,1.966884) **[red]@{-};
(7.906548,1.886884); (7.746548,1.966884) **[red]@{-};
(7.786548,1.901601); (7.866548,1.941601) **[red]@{-};
(7.786548,1.941601); (7.866548,1.901601) **[red]@{-};
(7.786548,1.910697); (7.866548,1.950697) **[red]@{-};
(7.786548,1.950697); (7.866548,1.910697) **[red]@{-};
(7.759204,2.084661); (7.919204,2.084661) **[blue]@{-};
(7.839204,2.044661); (7.839204,2.124661) **[blue]@{-};
(7.799204,2.076057); (7.879204,2.076057) **[blue]@{-};
(7.799204,2.089287); (7.879204,2.089287) **[blue]@{-};
(7.759204,1.887901); (7.919204,1.967901) **[red]@{-};
(7.919204,1.887901); (7.759204,1.967901) **[red]@{-};
(7.799204,1.904581); (7.879204,1.944581) **[red]@{-};
(7.799204,1.944581); (7.879204,1.904581) **[red]@{-};
(7.799204,1.909125); (7.879204,1.949125) **[red]@{-};
(7.799204,1.949125); (7.879204,1.909125) **[red]@{-};
(7.784186,2.081128); (7.944186,2.081128) **[blue]@{-};
(7.864186,2.041128); (7.864186,2.121128) **[blue]@{-};
(7.824186,2.072235); (7.904186,2.072235) **[blue]@{-};
(7.824186,2.086469); (7.904186,2.086469) **[blue]@{-};
(7.784186,1.885127); (7.944186,1.965127) **[red]@{-};
(7.944186,1.885127); (7.784186,1.965127) **[red]@{-};
(7.824186,1.902722); (7.904186,1.942722) **[red]@{-};
(7.824186,1.942722); (7.904186,1.902722) **[red]@{-};
(7.824186,1.907726); (7.904186,1.947726) **[red]@{-};
(7.824186,1.947726); (7.904186,1.907726) **[red]@{-};
(7.820867,2.078726); (7.980867,2.078726) **[blue]@{-};
(7.900867,2.038726); (7.900867,2.118726) **[blue]@{-};
(7.860867,2.073758); (7.940867,2.073758) **[blue]@{-};
(7.860867,2.086856); (7.940867,2.086856) **[blue]@{-};
(7.820867,1.891736); (7.980867,1.971736) **[red]@{-};
(7.980867,1.891736); (7.820867,1.971736) **[red]@{-};
(7.860867,1.909753); (7.940867,1.949753) **[red]@{-};
(7.860867,1.949753); (7.940867,1.909753) **[red]@{-};
(7.860867,1.915256); (7.940867,1.955256) **[red]@{-};
(7.860867,1.955256); (7.940867,1.915256) **[red]@{-};
(7.832889,2.079521); (7.992889,2.079521) **[blue]@{-};
(7.912889,2.039521); (7.912889,2.119521) **[blue]@{-};
(7.872889,2.071731); (7.952889,2.071731) **[blue]@{-};
(7.872889,2.086537); (7.952889,2.086537) **[blue]@{-};
(7.832889,1.891868); (7.992889,1.971868) **[red]@{-};
(7.992889,1.891868); (7.832889,1.971868) **[red]@{-};
(7.872889,1.909652); (7.952889,1.949652) **[red]@{-};
(7.872889,1.949652); (7.952889,1.909652) **[red]@{-};
(7.872889,1.916434); (7.952889,1.956434) **[red]@{-};
(7.872889,1.956434); (7.952889,1.916434) **[red]@{-};
(7.891544,2.078154); (8.051544,2.078154) **[blue]@{-};
(7.971544,2.038154); (7.971544,2.118154) **[blue]@{-};
(7.931544,2.072510); (8.011544,2.072510) **[blue]@{-};
(7.931544,2.089569); (8.011544,2.089569) **[blue]@{-};
(7.891544,1.903520); (8.051544,1.983520) **[red]@{-};
(8.051544,1.903520); (7.891544,1.983520) **[red]@{-};
(7.931544,1.919753); (8.011544,1.959753) **[red]@{-};
(7.931544,1.959753); (8.011544,1.919753) **[red]@{-};
(7.931544,1.926719); (8.011544,1.966719) **[red]@{-};
(7.931544,1.966719); (8.011544,1.926719) **[red]@{-};
(7.925625,2.076511); (8.085625,2.076511) **[blue]@{-};
(8.005625,2.036511); (8.005625,2.116511) **[blue]@{-};
(7.965625,2.073454); (8.045625,2.073454) **[blue]@{-};
(7.965625,2.082823); (8.045625,2.082823) **[blue]@{-};
(7.925625,1.882343); (8.085625,1.962343) **[red]@{-};
(8.085625,1.882343); (7.925625,1.962343) **[red]@{-};
(7.965625,1.898523); (8.045625,1.938523) **[red]@{-};
(7.965625,1.938523); (8.045625,1.898523) **[red]@{-};
(7.965625,1.908237); (8.045625,1.948237) **[red]@{-};
(7.965625,1.948237); (8.045625,1.908237) **[red]@{-};
(7.958919,2.080686); (8.118919,2.080686) **[blue]@{-};
(8.038919,2.040686); (8.038919,2.120686) **[blue]@{-};
(7.998919,2.076954); (8.078919,2.076954) **[blue]@{-};
(7.998919,2.089349); (8.078919,2.089349) **[blue]@{-};
(7.958919,1.888939); (8.118919,1.968939) **[red]@{-};
(8.118919,1.888939); (7.958919,1.968939) **[red]@{-};
(7.998919,1.905204); (8.078919,1.945204) **[red]@{-};
(7.998919,1.945204); (8.078919,1.905204) **[red]@{-};
(7.998919,1.910895); (8.078919,1.950895) **[red]@{-};
(7.998919,1.950895); (8.078919,1.910895) **[red]@{-};
(7.991462,2.087875); (8.151462,2.087875) **[blue]@{-};
(8.071462,2.047875); (8.071462,2.127875) **[blue]@{-};
(8.031462,2.079944); (8.111462,2.079944) **[blue]@{-};
(8.031462,2.091237); (8.111462,2.091237) **[blue]@{-};
(7.991462,1.889670); (8.151462,1.969670) **[red]@{-};
(8.151462,1.889670); (7.991462,1.969670) **[red]@{-};
(8.031462,1.905711); (8.111462,1.945711) **[red]@{-};
(8.031462,1.945711); (8.111462,1.905711) **[red]@{-};
(8.031462,1.913785); (8.111462,1.953785) **[red]@{-};
(8.031462,1.953785); (8.111462,1.913785) **[red]@{-};
(8.002149,2.086242); (8.162149,2.086242) **[blue]@{-};
(8.082149,2.046242); (8.082149,2.126242) **[blue]@{-};
(8.042149,2.079025); (8.122149,2.079025) **[blue]@{-};
(8.042149,2.089488); (8.122149,2.089488) **[blue]@{-};
(8.002149,1.890825); (8.162149,1.970825) **[red]@{-};
(8.162149,1.890825); (8.002149,1.970825) **[red]@{-};
(8.042149,1.908005); (8.122149,1.948005) **[red]@{-};
(8.042149,1.948005); (8.122149,1.908005) **[red]@{-};
(8.042149,1.913316); (8.122149,1.953316) **[red]@{-};
(8.042149,1.953316); (8.122149,1.913316) **[red]@{-};
(8.033742,2.081851); (8.193742,2.081851) **[blue]@{-};
(8.113742,2.041851); (8.113742,2.121851) **[blue]@{-};
(8.073742,2.075578); (8.153742,2.075578) **[blue]@{-};
(8.073742,2.091571); (8.153742,2.091571) **[blue]@{-};
(8.033742,1.893147); (8.193742,1.973147) **[red]@{-};
(8.193742,1.893147); (8.033742,1.973147) **[red]@{-};
(8.073742,1.910670); (8.153742,1.950670) **[red]@{-};
(8.073742,1.950670); (8.153742,1.910670) **[red]@{-};
(8.073742,1.917858); (8.153742,1.957858) **[red]@{-};
(8.073742,1.957858); (8.153742,1.917858) **[red]@{-};
(8.054426,2.083642); (8.214426,2.083642) **[blue]@{-};
(8.134426,2.043642); (8.134426,2.123642) **[blue]@{-};
(8.094426,2.077401); (8.174426,2.077401) **[blue]@{-};
(8.094426,2.088770); (8.174426,2.088770) **[blue]@{-};
(8.054426,1.894733); (8.214426,1.974733) **[red]@{-};
(8.214426,1.894733); (8.054426,1.974733) **[red]@{-};
(8.094426,1.912543); (8.174426,1.952543) **[red]@{-};
(8.094426,1.952543); (8.174426,1.912543) **[red]@{-};
(8.094426,1.922225); (8.174426,1.962225) **[red]@{-};
(8.094426,1.962225); (8.174426,1.922225) **[red]@{-};
(8.064658,2.083257); (8.224658,2.083257) **[blue]@{-};
(8.144658,2.043257); (8.144658,2.123257) **[blue]@{-};
(8.104658,2.077632); (8.184658,2.077632) **[blue]@{-};
(8.104658,2.092681); (8.184658,2.092681) **[blue]@{-};
(8.064658,1.895423); (8.224658,1.975423) **[red]@{-};
(8.224658,1.895423); (8.064658,1.975423) **[red]@{-};
(8.104658,1.914649); (8.184658,1.954649) **[red]@{-};
(8.104658,1.954649); (8.184658,1.914649) **[red]@{-};
(8.104658,1.925513); (8.184658,1.965513) **[red]@{-};
(8.104658,1.965513); (8.184658,1.925513) **[red]@{-};
(8.114757,2.090863); (8.274757,2.090863) **[blue]@{-};
(8.194757,2.050863); (8.194757,2.130863) **[blue]@{-};
(8.154757,2.080131); (8.234757,2.080131) **[blue]@{-};
(8.154757,2.100955); (8.234757,2.100955) **[blue]@{-};
(8.114757,1.830566); (8.274757,1.910566) **[red]@{-};
(8.274757,1.830566); (8.114757,1.910566) **[red]@{-};
(8.154757,1.840137); (8.234757,1.880137) **[red]@{-};
(8.154757,1.880137); (8.234757,1.840137) **[red]@{-};
(8.154757,1.881354); (8.234757,1.921354) **[red]@{-};
(8.154757,1.921354); (8.234757,1.881354) **[red]@{-};
(8.182095,2.080302); (8.342095,2.080302) **[blue]@{-};
(8.262095,2.040302); (8.262095,2.120302) **[blue]@{-};
(8.222095,2.072289); (8.302095,2.072289) **[blue]@{-};
(8.222095,2.091551); (8.302095,2.091551) **[blue]@{-};
(8.182095,1.811433); (8.342095,1.891433) **[red]@{-};
(8.342095,1.811433); (8.182095,1.891433) **[red]@{-};
(8.222095,1.829218); (8.302095,1.869218) **[red]@{-};
(8.222095,1.869218); (8.302095,1.829218) **[red]@{-};
(8.222095,1.835408); (8.302095,1.875408) **[red]@{-};
(8.222095,1.875408); (8.302095,1.835408) **[red]@{-};
(8.200771,2.085809); (8.360771,2.085809) **[blue]@{-};
(8.280771,2.045809); (8.280771,2.125809) **[blue]@{-};
(8.240771,2.081556); (8.320771,2.081556) **[blue]@{-};
(8.240771,2.089300); (8.320771,2.089300) **[blue]@{-};
(8.200771,1.830485); (8.360771,1.910485) **[red]@{-};
(8.360771,1.830485); (8.200771,1.910485) **[red]@{-};
(8.240771,1.835933); (8.320771,1.875933) **[red]@{-};
(8.240771,1.875933); (8.320771,1.835933) **[red]@{-};
(8.240771,1.862772); (8.320771,1.902772) **[red]@{-};
(8.240771,1.902772); (8.320771,1.862772) **[red]@{-};
(8.210019,2.082449); (8.370019,2.082449) **[blue]@{-};
(8.290019,2.042449); (8.290019,2.122449) **[blue]@{-};
(8.250019,2.073162); (8.330019,2.073162) **[blue]@{-};
(8.250019,2.092887); (8.330019,2.092887) **[blue]@{-};
(8.210019,1.825082); (8.370019,1.905082) **[red]@{-};
(8.370019,1.825082); (8.210019,1.905082) **[red]@{-};
(8.250019,1.838200); (8.330019,1.878200) **[red]@{-};
(8.250019,1.878200); (8.330019,1.838200) **[red]@{-};
(8.250019,1.851391); (8.330019,1.891391) **[red]@{-};
(8.250019,1.891391); (8.330019,1.851391) **[red]@{-};
(8.228339,2.084526); (8.388339,2.084526) **[blue]@{-};
(8.308339,2.044526); (8.308339,2.124526) **[blue]@{-};
(8.268339,2.080829); (8.348339,2.080829) **[blue]@{-};
(8.268339,2.092160); (8.348339,2.092160) **[blue]@{-};
(8.228339,1.820441); (8.388339,1.900441) **[red]@{-};
(8.388339,1.820441); (8.228339,1.900441) **[red]@{-};
(8.268339,1.837022); (8.348339,1.877022) **[red]@{-};
(8.268339,1.877022); (8.348339,1.837022) **[red]@{-};
(8.268339,1.842258); (8.348339,1.882258) **[red]@{-};
(8.268339,1.882258); (8.348339,1.842258) **[red]@{-};
(8.290687,2.081125); (8.450687,2.081125) **[blue]@{-};
(8.370687,2.041125); (8.370687,2.121125) **[blue]@{-};
(8.330687,2.078148); (8.410687,2.078148) **[blue]@{-};
(8.330687,2.088391); (8.410687,2.088391) **[blue]@{-};
(8.290687,1.734839); (8.450687,1.814839) **[red]@{-};
(8.450687,1.734839); (8.290687,1.814839) **[red]@{-};
(8.330687,1.751175); (8.410687,1.791175) **[red]@{-};
(8.330687,1.791175); (8.410687,1.751175) **[red]@{-};
(8.330687,1.758163); (8.410687,1.798163) **[red]@{-};
(8.330687,1.798163); (8.410687,1.758163) **[red]@{-};
(8.316605,2.086181); (8.476605,2.086181) **[blue]@{-};
(8.396605,2.046181); (8.396605,2.126181) **[blue]@{-};
(8.356605,2.079952); (8.436605,2.079952) **[blue]@{-};
(8.356605,2.090781); (8.436605,2.090781) **[blue]@{-};
(8.316605,1.743250); (8.476605,1.823250) **[red]@{-};
(8.476605,1.743250); (8.316605,1.823250) **[red]@{-};
(8.356605,1.756345); (8.436605,1.796345) **[red]@{-};
(8.356605,1.796345); (8.436605,1.756345) **[red]@{-};
(8.356605,1.766498); (8.436605,1.806498) **[red]@{-};
(8.356605,1.806498); (8.436605,1.766498) **[red]@{-};
(8.358792,2.081511); (8.518792,2.081511) **[blue]@{-};
(8.438792,2.041511); (8.438792,2.121511) **[blue]@{-};
(8.398792,2.074367); (8.478792,2.074367) **[blue]@{-};
(8.398792,2.086965); (8.478792,2.086965) **[blue]@{-};
(8.358792,1.748952); (8.518792,1.828952) **[red]@{-};
(8.518792,1.748952); (8.358792,1.828952) **[red]@{-};
(8.398792,1.765148); (8.478792,1.805148) **[red]@{-};
(8.398792,1.805148); (8.478792,1.765148) **[red]@{-};
(8.398792,1.772543); (8.478792,1.812543) **[red]@{-};
(8.398792,1.812543); (8.478792,1.772543) **[red]@{-};
(8.367083,2.087673); (8.527083,2.087673) **[blue]@{-};
(8.447083,2.047673); (8.447083,2.127673) **[blue]@{-};
(8.407083,2.079527); (8.487083,2.079527) **[blue]@{-};
(8.407083,2.094799); (8.487083,2.094799) **[blue]@{-};
(8.367083,1.751479); (8.527083,1.831479) **[red]@{-};
(8.527083,1.751479); (8.367083,1.831479) **[red]@{-};
(8.407083,1.769102); (8.487083,1.809102) **[red]@{-};
(8.407083,1.809102); (8.487083,1.769102) **[red]@{-};
(8.407083,1.777501); (8.487083,1.817501) **[red]@{-};
(8.407083,1.817501); (8.487083,1.777501) **[red]@{-};
(8.383524,2.087424); (8.543524,2.087424) **[blue]@{-};
(8.463524,2.047424); (8.463524,2.127424) **[blue]@{-};
(8.423524,2.078217); (8.503524,2.078217) **[blue]@{-};
(8.423524,2.091404); (8.503524,2.091404) **[blue]@{-};
(8.383524,1.754989); (8.543524,1.834989) **[red]@{-};
(8.543524,1.754989); (8.383524,1.834989) **[red]@{-};
(8.423524,1.772729); (8.503524,1.812729) **[red]@{-};
(8.423524,1.812729); (8.503524,1.772729) **[red]@{-};
(8.423524,1.781104); (8.503524,1.821104) **[red]@{-};
(8.423524,1.821104); (8.503524,1.781104) **[red]@{-};
(8.407840,2.080369); (8.567840,2.080369) **[blue]@{-};
(8.487840,2.040369); (8.487840,2.120369) **[blue]@{-};
(8.447840,2.073780); (8.527840,2.073780) **[blue]@{-};
(8.447840,2.085496); (8.527840,2.085496) **[blue]@{-};
(8.407840,1.761598); (8.567840,1.841598) **[red]@{-};
(8.567840,1.761598); (8.407840,1.841598) **[red]@{-};
(8.447840,1.779066); (8.527840,1.819066) **[red]@{-};
(8.447840,1.819066); (8.527840,1.779066) **[red]@{-};
(8.447840,1.786223); (8.527840,1.826223) **[red]@{-};
(8.447840,1.826223); (8.527840,1.786223) **[red]@{-};
(8.439636,2.085321); (8.599636,2.085321) **[blue]@{-};
(8.519636,2.045321); (8.519636,2.125321) **[blue]@{-};
(8.479636,2.080538); (8.559636,2.080538) **[blue]@{-};
(8.479636,2.088592); (8.559636,2.088592) **[blue]@{-};
(8.439636,1.732745); (8.599636,1.812745) **[red]@{-};
(8.599636,1.732745); (8.439636,1.812745) **[red]@{-};
(8.479636,1.748075); (8.559636,1.788075) **[red]@{-};
(8.479636,1.788075); (8.559636,1.748075) **[red]@{-};
(8.479636,1.756540); (8.559636,1.796540) **[red]@{-};
(8.479636,1.796540); (8.559636,1.756540) **[red]@{-};
(8.463032,2.084498); (8.623032,2.084498) **[blue]@{-};
(8.543032,2.044498); (8.543032,2.124498) **[blue]@{-};
(8.503032,2.079737); (8.583032,2.079737) **[blue]@{-};
(8.503032,2.093740); (8.583032,2.093740) **[blue]@{-};
(8.463032,1.737338); (8.623032,1.817338) **[red]@{-};
(8.623032,1.737338); (8.463032,1.817338) **[red]@{-};
(8.503032,1.755297); (8.583032,1.795297) **[red]@{-};
(8.503032,1.795297); (8.583032,1.755297) **[red]@{-};
(8.503032,1.762466); (8.583032,1.802466) **[red]@{-};
(8.503032,1.802466); (8.583032,1.762466) **[red]@{-};
(9.117217,2.088518); (9.277217,2.088518) **[blue]@{-};
(9.197217,2.048518); (9.197217,2.128518) **[blue]@{-};
(9.157217,2.079673); (9.237217,2.079673) **[blue]@{-};
(9.157217,2.094927); (9.237217,2.094927) **[blue]@{-};
(9.117217,1.513897); (9.277217,1.593897) **[red]@{-};
(9.277217,1.513897); (9.117217,1.593897) **[red]@{-};
(9.157217,1.531544); (9.237217,1.571544) **[red]@{-};
(9.157217,1.571544); (9.237217,1.531544) **[red]@{-};
(9.157217,1.538938); (9.237217,1.578938) **[red]@{-};
(9.157217,1.578938); (9.237217,1.538938) **[red]@{-};
\endxy
}
\caption{Cost
  to evaluate an $\ell$-isogeny
  on one CSIDH-512 point
  and compute the new curve coefficient.
  Top graph: Skylake cycles
    for {\tt velusqrt-flint}
    divided by $\ell+2$.
  Middle graph: Skylake cycles
    for {\tt velusqrt-asm}
    divided by $\ell+2$.
  Bottom graph: Multiplications
    inside {\tt velusqrt-asm}
    divided by $\ell+2$.
  }
\label{isogeny-cost512}
\end{figure}

We display the results graphically
in~\cref{fig: isogeny-magma},
\cref{fig:isogeny-julia},
and~\cref{isogeny-cost512} as follows:
\begin{itemize}
\item
The horizontal axis is $\ell$.
We limit these graphs
to the range of degrees $\ell$ used in CSIDH-512,
except as noted below.
\item
The vertical axis is cost divided by $\ell+2$.
(The conventional algorithm
has a main loop of length $(\ell-1)/2$,
but also has operations outside the main loop.
Dividing by $\ell+2$
does a better job of reflecting these costs
than dividing by $\ell-1$.)
\item
Blue plus indicates the cost
of the conventional $\ell$-isogeny algorithm,
and red cross indicates the cost
of the new $\ell$-isogeny algorithm.
\item
Large cross and large plus indicate medians
across 15 experiments.
Small plus and small cross indicate quartiles.
(The quartiles are often
so close to medians as to be invisible in the graphs.)
\item
Each axis has a logarithmic scale,
with a factor $2$ horizontally
occupying the same visual distance
as a factor $\sqrt{2}$ vertically.
(This choice means that a
$\widetilde{\Theta}(\sqrt{\ell})$ speedup
is a 45-degree line asymptotically.)
\end{itemize}

The implementations are as follows.
First,
{\tt velusqrt-magma}
implements the new $\ell$-isogeny algorithm in Magma.
\cref{fig: isogeny-magma}
shows the resulting cycle counts.
The graph shows that the new algorithm
is better than the old algorithm
for $\ell\ge 113$.

Second,
{\tt velusqrt-julia}
implements the new $\ell$-isogeny algorithm in Julia,
on top of
the field arithmetic and polynomial arithmetic
provided by Nemo.
\cref{fig:isogeny-julia}
shows the resulting cycle counts.
This graph goes beyond
the range of $\ell$
used in CSIDH-512,
and uses a smaller base field:
each $\ell$ is measured
with prime $p=8\cdot\ell\cdot f-1$,
where $f$ is
the smallest cofactor
producing a $256$-bit prime.

Third,
{\tt velusqrt-flint}
implements $\ell$-isogenies in C
on top of
the field arithmetic and polynomial arithmetic
provided by FLINT\null.
The sets $I$ and $J$ are constructed
as in Example~\ref{ex:kernel-polynomial},
with the parameter $b$ manually tuned.
This software was compiled using
\texttt{gcc-7.5.0 -O3 -march=native -Wall -Wextra -pedantic -std=c99}.
The top graph in
Figure~\ref{isogeny-cost512}
shows the resulting cycle counts:
% XXX: update this if benchmarks are updated
e.g., $2026744\approx 3440.992(\ell+2)$ cycles for $\ell=587$ for the
new algorithm,
about $45\%$ faster than the old one.

Fourth,
{\tt velusqrt-asm}
implements polynomial arithmetic in C
on top of the CSIDH-512
assembly-language field-arithmetic subroutines from~\cite{CSIDH2018},
and implements $\ell$-isogenies
(with automatic tuning of $\#I$ and $\#J$)
on top of that.
We compiled the C portion
of this software using
{\tt clang-6.0 -O3 -Os -march=native -mtune=native -Wall -Wextra -std=gnu99 -pedantic}.
The middle graph in Figure~\ref{isogeny-cost512}
shows the resulting cycle counts:
% XXX: update this if benchmarks are updated
e.g., $745862\approx 1266.319(\ell+2)$ cycles
for $\ell=587$
for the new software,
about 22\% faster
than the software from~\cite{2018/meyer}.
(For small $\ell$,
the automatic tuning chooses
$\#I=0$ and $\#J=0$,
falling back to the conventional algorithm,
so the red curve overlaps the blue curve.)

The {\tt velusqrt-asm} software
includes an internal multiplication counter,
and we used this
to show an upper bound on cost in the fifth metric.
The bottom graph in Figure~\ref{isogeny-cost512}
shows the resulting multiplication counts
(with $\#I$ and $\#J$ automatically
re-tuned for this metric):
e.g.,
$2296\approx 3.898(\ell+2)$ multiplications
for $\ell=587$
for the new algorithm,
about 55\% faster
than the
$3550\approx 6.027(\ell+2)$ multiplications
for the conventional algorithm.

We also implemented the $\ell$-isogeny algorithm
in SageMath~\cite{sagemath},
but this was only to double-check
the correctness of the algorithm output,
not to evaluate performance in any particular metric.

\subsection{Results for protocols}
\label{protocol-timings}
We integrated our $\ell$-isogeny implementations
into the CSIDH-512 and CSURF-512 code
from~\cite{CD19}
(\url{https://github.com/TDecru/CSURF}),
and into the CSIDH-512 code from~\cite{2018/meyer}.
We also
wrote Julia/Nemo code
for CSIDH-512, CSURF-512, and B-SIDH,
and C/FLINT code for CSIDH-512 and CSURF-512.
We also adapted the~\cite{CSIDH2018}
assembly-language field arithmetic
from CSIDH-512 to CSIDH-1024,
saving about a factor 3
compared to the C code
provided in~\cite{CSIDH2018} for CSIDH-1024.
The Magma implementation switches over
to the old $\ell$-isogeny algorithm
for $\ell<113$,
and the FLINT implementation switches over
to the old $\ell$-isogeny algorithm
for $\ell<150$.
These implementations
of CSIDH, CSURF, and B-SIDH
are included in
{\tt velusqrt-magma},
{\tt velusqrt-julia},
{\tt velusqrt-flint},
and
{\tt velusqrt-asm}.

Beware that none of these protocol implementations are constant-time.
Constant-time implementations of CSIDH
\cite{10.1007/978-3-030-25510-7_17,10.1007/978-3-030-26834-3_2,10.1007/978-3-030-30530-7_9,cryptoeprint:2019:1121}
are an active area of research,
and it is too early to guess what the final performance
of constant-time CSIDH, CSURF, and B-SIDH
will be on top of our $\ell$-isogeny algorithm.

We display CSIDH and CSURF performance
in a series of graphs.
Each graph has horizontal lines
showing 25\% quartile, median, and 75\% quartile.
Each graph shows $EK$ cost measurements,
namely $E$ evaluations for each of $K$ keys.
Keys are sorted horizontally
in increasing order of median cost.
Within each key,
evaluations are sorted horizontally
in increasing order of cost.
Blue plus indicates the cost
of an action using
the conventional $\ell$-isogeny algorithm,
while red cross indicates the cost
of an action that switches over to
the new $\ell$-isogeny algorithm
for sufficiently large $\ell$.
Within each graph,
the blue and red curves use the same sample of keys.

\begin{figure}[t]
\centerline{
\xy <1.1cm,0cm>:<0cm,4cm>::
(0,0.803185); (8,0.803185) **[blue]@{-};
(8.1,0.803185) *[blue]{\rlap{6979795712}};
(0,0.886457); (8,0.886457) **[blue]@{-};
(8.1,0.886457) *[blue]{\rlap{7394523616}};
(0,0.974736); (8,0.974736) **[blue]@{-};
(8.1,0.974736) *[blue]{\rlap{7861127478}};
(0,0.596231); (8,0.596231) **[red]@{-};
(-0.1,0.596231) *[red]{\llap{6047049948}};
(0,0.673773); (8,0.673773) **[red]@{-};
(-0.1,0.673773) *[red]{\llap{6380956104}};
(0,0.778384); (8,0.778384) **[red]@{-};
(-0.1,0.778384) *[red]{\llap{6860835654}};
(-0.009091,0.347576); (-0.009091,1.249344) **[lightgray]@{-};
(0.118182,0.347576); (0.118182,1.249344) **[lightgray]@{-};
(0.245455,0.347576); (0.245455,1.249344) **[lightgray]@{-};
(0.372727,0.347576); (0.372727,1.249344) **[lightgray]@{-};
(0.500000,0.347576); (0.500000,1.249344) **[lightgray]@{-};
(0.627273,0.347576); (0.627273,1.249344) **[lightgray]@{-};
(0.754545,0.347576); (0.754545,1.249344) **[lightgray]@{-};
(0.881818,0.347576); (0.881818,1.249344) **[lightgray]@{-};
(1.009091,0.347576); (1.009091,1.249344) **[lightgray]@{-};
(1.136364,0.347576); (1.136364,1.249344) **[lightgray]@{-};
(1.263636,0.347576); (1.263636,1.249344) **[lightgray]@{-};
(1.390909,0.347576); (1.390909,1.249344) **[lightgray]@{-};
(1.518182,0.347576); (1.518182,1.249344) **[lightgray]@{-};
(1.645455,0.347576); (1.645455,1.249344) **[lightgray]@{-};
(1.772727,0.347576); (1.772727,1.249344) **[lightgray]@{-};
(1.900000,0.347576); (1.900000,1.249344) **[lightgray]@{-};
(2.027273,0.347576); (2.027273,1.249344) **[lightgray]@{-};
(2.154545,0.347576); (2.154545,1.249344) **[lightgray]@{-};
(2.281818,0.347576); (2.281818,1.249344) **[lightgray]@{-};
(2.409091,0.347576); (2.409091,1.249344) **[lightgray]@{-};
(2.536364,0.347576); (2.536364,1.249344) **[lightgray]@{-};
(2.663636,0.347576); (2.663636,1.249344) **[lightgray]@{-};
(2.790909,0.347576); (2.790909,1.249344) **[lightgray]@{-};
(2.918182,0.347576); (2.918182,1.249344) **[lightgray]@{-};
(3.045455,0.347576); (3.045455,1.249344) **[lightgray]@{-};
(3.172727,0.347576); (3.172727,1.249344) **[lightgray]@{-};
(3.300000,0.347576); (3.300000,1.249344) **[lightgray]@{-};
(3.427273,0.347576); (3.427273,1.249344) **[lightgray]@{-};
(3.554545,0.347576); (3.554545,1.249344) **[lightgray]@{-};
(3.681818,0.347576); (3.681818,1.249344) **[lightgray]@{-};
(3.809091,0.347576); (3.809091,1.249344) **[lightgray]@{-};
(3.936364,0.347576); (3.936364,1.249344) **[lightgray]@{-};
(4.063636,0.347576); (4.063636,1.249344) **[lightgray]@{-};
(4.190909,0.347576); (4.190909,1.249344) **[lightgray]@{-};
(4.318182,0.347576); (4.318182,1.249344) **[lightgray]@{-};
(4.445455,0.347576); (4.445455,1.249344) **[lightgray]@{-};
(4.572727,0.347576); (4.572727,1.249344) **[lightgray]@{-};
(4.700000,0.347576); (4.700000,1.249344) **[lightgray]@{-};
(4.827273,0.347576); (4.827273,1.249344) **[lightgray]@{-};
(4.954545,0.347576); (4.954545,1.249344) **[lightgray]@{-};
(5.081818,0.347576); (5.081818,1.249344) **[lightgray]@{-};
(5.209091,0.347576); (5.209091,1.249344) **[lightgray]@{-};
(5.336364,0.347576); (5.336364,1.249344) **[lightgray]@{-};
(5.463636,0.347576); (5.463636,1.249344) **[lightgray]@{-};
(5.590909,0.347576); (5.590909,1.249344) **[lightgray]@{-};
(5.718182,0.347576); (5.718182,1.249344) **[lightgray]@{-};
(5.845455,0.347576); (5.845455,1.249344) **[lightgray]@{-};
(5.972727,0.347576); (5.972727,1.249344) **[lightgray]@{-};
(6.100000,0.347576); (6.100000,1.249344) **[lightgray]@{-};
(6.227273,0.347576); (6.227273,1.249344) **[lightgray]@{-};
(6.354545,0.347576); (6.354545,1.249344) **[lightgray]@{-};
(6.481818,0.347576); (6.481818,1.249344) **[lightgray]@{-};
(6.609091,0.347576); (6.609091,1.249344) **[lightgray]@{-};
(6.736364,0.347576); (6.736364,1.249344) **[lightgray]@{-};
(6.863636,0.347576); (6.863636,1.249344) **[lightgray]@{-};
(6.990909,0.347576); (6.990909,1.249344) **[lightgray]@{-};
(7.118182,0.347576); (7.118182,1.249344) **[lightgray]@{-};
(7.245455,0.347576); (7.245455,1.249344) **[lightgray]@{-};
(7.372727,0.347576); (7.372727,1.249344) **[lightgray]@{-};
(7.500000,0.347576); (7.500000,1.249344) **[lightgray]@{-};
(7.627273,0.347576); (7.627273,1.249344) **[lightgray]@{-};
(7.754545,0.347576); (7.754545,1.249344) **[lightgray]@{-};
(7.881818,0.347576); (7.881818,1.249344) **[lightgray]@{-};
(8.009091,0.347576); (8.009091,1.249344) **[lightgray]@{-};
(-0.009091,0.347576); (-0.009091,1.249344) **[lightgray]@{-};
(0.118182,0.347576); (0.118182,1.249344) **[lightgray]@{-};
(0.245455,0.347576); (0.245455,1.249344) **[lightgray]@{-};
(0.372727,0.347576); (0.372727,1.249344) **[lightgray]@{-};
(0.500000,0.347576); (0.500000,1.249344) **[lightgray]@{-};
(0.627273,0.347576); (0.627273,1.249344) **[lightgray]@{-};
(0.754545,0.347576); (0.754545,1.249344) **[lightgray]@{-};
(0.881818,0.347576); (0.881818,1.249344) **[lightgray]@{-};
(1.009091,0.347576); (1.009091,1.249344) **[lightgray]@{-};
(1.136364,0.347576); (1.136364,1.249344) **[lightgray]@{-};
(1.263636,0.347576); (1.263636,1.249344) **[lightgray]@{-};
(1.390909,0.347576); (1.390909,1.249344) **[lightgray]@{-};
(1.518182,0.347576); (1.518182,1.249344) **[lightgray]@{-};
(1.645455,0.347576); (1.645455,1.249344) **[lightgray]@{-};
(1.772727,0.347576); (1.772727,1.249344) **[lightgray]@{-};
(1.900000,0.347576); (1.900000,1.249344) **[lightgray]@{-};
(2.027273,0.347576); (2.027273,1.249344) **[lightgray]@{-};
(2.154545,0.347576); (2.154545,1.249344) **[lightgray]@{-};
(2.281818,0.347576); (2.281818,1.249344) **[lightgray]@{-};
(2.409091,0.347576); (2.409091,1.249344) **[lightgray]@{-};
(2.536364,0.347576); (2.536364,1.249344) **[lightgray]@{-};
(2.663636,0.347576); (2.663636,1.249344) **[lightgray]@{-};
(2.790909,0.347576); (2.790909,1.249344) **[lightgray]@{-};
(2.918182,0.347576); (2.918182,1.249344) **[lightgray]@{-};
(3.045455,0.347576); (3.045455,1.249344) **[lightgray]@{-};
(3.172727,0.347576); (3.172727,1.249344) **[lightgray]@{-};
(3.300000,0.347576); (3.300000,1.249344) **[lightgray]@{-};
(3.427273,0.347576); (3.427273,1.249344) **[lightgray]@{-};
(3.554545,0.347576); (3.554545,1.249344) **[lightgray]@{-};
(3.681818,0.347576); (3.681818,1.249344) **[lightgray]@{-};
(3.809091,0.347576); (3.809091,1.249344) **[lightgray]@{-};
(3.936364,0.347576); (3.936364,1.249344) **[lightgray]@{-};
(4.063636,0.347576); (4.063636,1.249344) **[lightgray]@{-};
(4.190909,0.347576); (4.190909,1.249344) **[lightgray]@{-};
(4.318182,0.347576); (4.318182,1.249344) **[lightgray]@{-};
(4.445455,0.347576); (4.445455,1.249344) **[lightgray]@{-};
(4.572727,0.347576); (4.572727,1.249344) **[lightgray]@{-};
(4.700000,0.347576); (4.700000,1.249344) **[lightgray]@{-};
(4.827273,0.347576); (4.827273,1.249344) **[lightgray]@{-};
(4.954545,0.347576); (4.954545,1.249344) **[lightgray]@{-};
(5.081818,0.347576); (5.081818,1.249344) **[lightgray]@{-};
(5.209091,0.347576); (5.209091,1.249344) **[lightgray]@{-};
(5.336364,0.347576); (5.336364,1.249344) **[lightgray]@{-};
(5.463636,0.347576); (5.463636,1.249344) **[lightgray]@{-};
(5.590909,0.347576); (5.590909,1.249344) **[lightgray]@{-};
(5.718182,0.347576); (5.718182,1.249344) **[lightgray]@{-};
(5.845455,0.347576); (5.845455,1.249344) **[lightgray]@{-};
(5.972727,0.347576); (5.972727,1.249344) **[lightgray]@{-};
(6.100000,0.347576); (6.100000,1.249344) **[lightgray]@{-};
(6.227273,0.347576); (6.227273,1.249344) **[lightgray]@{-};
(6.354545,0.347576); (6.354545,1.249344) **[lightgray]@{-};
(6.481818,0.347576); (6.481818,1.249344) **[lightgray]@{-};
(6.609091,0.347576); (6.609091,1.249344) **[lightgray]@{-};
(6.736364,0.347576); (6.736364,1.249344) **[lightgray]@{-};
(6.863636,0.347576); (6.863636,1.249344) **[lightgray]@{-};
(6.990909,0.347576); (6.990909,1.249344) **[lightgray]@{-};
(7.118182,0.347576); (7.118182,1.249344) **[lightgray]@{-};
(7.245455,0.347576); (7.245455,1.249344) **[lightgray]@{-};
(7.372727,0.347576); (7.372727,1.249344) **[lightgray]@{-};
(7.500000,0.347576); (7.500000,1.249344) **[lightgray]@{-};
(7.627273,0.347576); (7.627273,1.249344) **[lightgray]@{-};
(7.754545,0.347576); (7.754545,1.249344) **[lightgray]@{-};
(7.881818,0.347576); (7.881818,1.249344) **[lightgray]@{-};
(8.009091,0.347576); (8.009091,1.249344) **[lightgray]@{-};
(0.000000,0.454204) *[blue]{\scriptscriptstyle+};
(0.018182,0.672272) *[blue]{\scriptscriptstyle+};
(0.036364,0.705004) *[blue]{\scriptscriptstyle+};
(0.054545,0.723684) *[blue]{\scriptscriptstyle+};
(0.072727,0.778888) *[blue]{\scriptscriptstyle+};
(0.090909,0.831923) *[blue]{\scriptscriptstyle+};
(0.109091,1.040992) *[blue]{\scriptscriptstyle+};
(0.127273,0.583468) *[blue]{\scriptscriptstyle+};
(0.145455,0.675458) *[blue]{\scriptscriptstyle+};
(0.163636,0.737611) *[blue]{\scriptscriptstyle+};
(0.181818,0.827248) *[blue]{\scriptscriptstyle+};
(0.200000,0.838287) *[blue]{\scriptscriptstyle+};
(0.218182,0.924635) *[blue]{\scriptscriptstyle+};
(0.236364,0.928403) *[blue]{\scriptscriptstyle+};
(0.254545,0.657376) *[blue]{\scriptscriptstyle+};
(0.272727,0.702726) *[blue]{\scriptscriptstyle+};
(0.290909,0.749192) *[blue]{\scriptscriptstyle+};
(0.309091,0.822910) *[blue]{\scriptscriptstyle+};
(0.327273,0.837647) *[blue]{\scriptscriptstyle+};
(0.345455,0.884816) *[blue]{\scriptscriptstyle+};
(0.363636,0.898090) *[blue]{\scriptscriptstyle+};
(0.381818,0.579034) *[blue]{\scriptscriptstyle+};
(0.400000,0.682086) *[blue]{\scriptscriptstyle+};
(0.418182,0.719464) *[blue]{\scriptscriptstyle+};
(0.436364,0.854242) *[blue]{\scriptscriptstyle+};
(0.454545,0.858763) *[blue]{\scriptscriptstyle+};
(0.472727,0.957284) *[blue]{\scriptscriptstyle+};
(0.490909,1.160175) *[blue]{\scriptscriptstyle+};
(0.509091,0.603980) *[blue]{\scriptscriptstyle+};
(0.527273,0.658320) *[blue]{\scriptscriptstyle+};
(0.545455,0.682070) *[blue]{\scriptscriptstyle+};
(0.563636,0.850429) *[blue]{\scriptscriptstyle+};
(0.581818,0.918895) *[blue]{\scriptscriptstyle+};
(0.600000,1.052697) *[blue]{\scriptscriptstyle+};
(0.618182,1.081253) *[blue]{\scriptscriptstyle+};
(0.636364,0.633984) *[blue]{\scriptscriptstyle+};
(0.654545,0.678741) *[blue]{\scriptscriptstyle+};
(0.672727,0.711266) *[blue]{\scriptscriptstyle+};
(0.690909,0.721399) *[blue]{\scriptscriptstyle+};
(0.709091,0.959333) *[blue]{\scriptscriptstyle+};
(0.727273,0.969063) *[blue]{\scriptscriptstyle+};
(0.745455,1.056330) *[blue]{\scriptscriptstyle+};
(0.763636,0.759378) *[blue]{\scriptscriptstyle+};
(0.781818,0.774878) *[blue]{\scriptscriptstyle+};
(0.800000,0.785060) *[blue]{\scriptscriptstyle+};
(0.818182,0.789988) *[blue]{\scriptscriptstyle+};
(0.836364,0.867454) *[blue]{\scriptscriptstyle+};
(0.854545,0.867793) *[blue]{\scriptscriptstyle+};
(0.872727,0.894054) *[blue]{\scriptscriptstyle+};
(0.890909,0.689382) *[blue]{\scriptscriptstyle+};
(0.909091,0.757612) *[blue]{\scriptscriptstyle+};
(0.927273,0.764393) *[blue]{\scriptscriptstyle+};
(0.945455,0.851415) *[blue]{\scriptscriptstyle+};
(0.963636,0.886457) *[blue]{\scriptscriptstyle+};
(0.981818,0.917440) *[blue]{\scriptscriptstyle+};
(1.000000,0.946718) *[blue]{\scriptscriptstyle+};
(1.018182,0.707768) *[blue]{\scriptscriptstyle+};
(1.036364,0.717030) *[blue]{\scriptscriptstyle+};
(1.054545,0.734728) *[blue]{\scriptscriptstyle+};
(1.072727,0.919844) *[blue]{\scriptscriptstyle+};
(1.090909,0.939714) *[blue]{\scriptscriptstyle+};
(1.109091,0.941020) *[blue]{\scriptscriptstyle+};
(1.127273,1.051099) *[blue]{\scriptscriptstyle+};
(1.145455,0.606052) *[blue]{\scriptscriptstyle+};
(1.163636,0.773942) *[blue]{\scriptscriptstyle+};
(1.181818,0.780370) *[blue]{\scriptscriptstyle+};
(1.200000,0.788627) *[blue]{\scriptscriptstyle+};
(1.218182,0.815426) *[blue]{\scriptscriptstyle+};
(1.236364,0.940619) *[blue]{\scriptscriptstyle+};
(1.254545,0.970735) *[blue]{\scriptscriptstyle+};
(1.272727,0.602991) *[blue]{\scriptscriptstyle+};
(1.290909,0.710690) *[blue]{\scriptscriptstyle+};
(1.309091,0.890949) *[blue]{\scriptscriptstyle+};
(1.327273,0.911459) *[blue]{\scriptscriptstyle+};
(1.345455,0.929594) *[blue]{\scriptscriptstyle+};
(1.363636,0.934549) *[blue]{\scriptscriptstyle+};
(1.381818,0.945607) *[blue]{\scriptscriptstyle+};
(1.400000,0.608513) *[blue]{\scriptscriptstyle+};
(1.418182,0.669849) *[blue]{\scriptscriptstyle+};
(1.436364,0.871137) *[blue]{\scriptscriptstyle+};
(1.454545,0.886172) *[blue]{\scriptscriptstyle+};
(1.472727,0.905928) *[blue]{\scriptscriptstyle+};
(1.490909,1.021031) *[blue]{\scriptscriptstyle+};
(1.509091,1.041825) *[blue]{\scriptscriptstyle+};
(1.527273,0.614204) *[blue]{\scriptscriptstyle+};
(1.545455,0.710056) *[blue]{\scriptscriptstyle+};
(1.563636,0.747853) *[blue]{\scriptscriptstyle+};
(1.581818,0.754691) *[blue]{\scriptscriptstyle+};
(1.600000,0.993121) *[blue]{\scriptscriptstyle+};
(1.618182,0.996022) *[blue]{\scriptscriptstyle+};
(1.636364,1.077796) *[blue]{\scriptscriptstyle+};
(1.654545,0.691896) *[blue]{\scriptscriptstyle+};
(1.672727,0.786497) *[blue]{\scriptscriptstyle+};
(1.690909,0.806471) *[blue]{\scriptscriptstyle+};
(1.709091,0.834074) *[blue]{\scriptscriptstyle+};
(1.727273,0.838923) *[blue]{\scriptscriptstyle+};
(1.745455,0.843975) *[blue]{\scriptscriptstyle+};
(1.763636,0.907909) *[blue]{\scriptscriptstyle+};
(1.781818,0.667002) *[blue]{\scriptscriptstyle+};
(1.800000,0.772647) *[blue]{\scriptscriptstyle+};
(1.818182,0.773978) *[blue]{\scriptscriptstyle+};
(1.836364,0.803969) *[blue]{\scriptscriptstyle+};
(1.854545,0.916979) *[blue]{\scriptscriptstyle+};
(1.872727,0.962525) *[blue]{\scriptscriptstyle+};
(1.890909,1.033693) *[blue]{\scriptscriptstyle+};
(1.909091,0.723257) *[blue]{\scriptscriptstyle+};
(1.927273,0.735457) *[blue]{\scriptscriptstyle+};
(1.945455,0.872419) *[blue]{\scriptscriptstyle+};
(1.963636,0.891788) *[blue]{\scriptscriptstyle+};
(1.981818,0.915171) *[blue]{\scriptscriptstyle+};
(2.000000,0.976948) *[blue]{\scriptscriptstyle+};
(2.018182,0.978033) *[blue]{\scriptscriptstyle+};
(2.036364,0.647648) *[blue]{\scriptscriptstyle+};
(2.054545,0.791288) *[blue]{\scriptscriptstyle+};
(2.072727,0.794792) *[blue]{\scriptscriptstyle+};
(2.090909,0.825107) *[blue]{\scriptscriptstyle+};
(2.109091,0.855292) *[blue]{\scriptscriptstyle+};
(2.127273,0.936553) *[blue]{\scriptscriptstyle+};
(2.145455,0.972707) *[blue]{\scriptscriptstyle+};
(2.163636,0.754593) *[blue]{\scriptscriptstyle+};
(2.181818,0.805685) *[blue]{\scriptscriptstyle+};
(2.200000,0.858556) *[blue]{\scriptscriptstyle+};
(2.218182,0.860034) *[blue]{\scriptscriptstyle+};
(2.236364,0.882898) *[blue]{\scriptscriptstyle+};
(2.254545,0.899962) *[blue]{\scriptscriptstyle+};
(2.272727,0.908706) *[blue]{\scriptscriptstyle+};
(2.290909,0.581309) *[blue]{\scriptscriptstyle+};
(2.309091,0.744771) *[blue]{\scriptscriptstyle+};
(2.327273,0.755086) *[blue]{\scriptscriptstyle+};
(2.345455,0.880288) *[blue]{\scriptscriptstyle+};
(2.363636,0.945916) *[blue]{\scriptscriptstyle+};
(2.381818,0.969572) *[blue]{\scriptscriptstyle+};
(2.400000,1.063921) *[blue]{\scriptscriptstyle+};
(2.418182,0.744584) *[blue]{\scriptscriptstyle+};
(2.436364,0.766504) *[blue]{\scriptscriptstyle+};
(2.454545,0.785203) *[blue]{\scriptscriptstyle+};
(2.472727,0.794806) *[blue]{\scriptscriptstyle+};
(2.490909,0.834748) *[blue]{\scriptscriptstyle+};
(2.509091,0.867494) *[blue]{\scriptscriptstyle+};
(2.527273,0.974267) *[blue]{\scriptscriptstyle+};
(2.545455,0.617905) *[blue]{\scriptscriptstyle+};
(2.563636,0.728701) *[blue]{\scriptscriptstyle+};
(2.581818,0.765195) *[blue]{\scriptscriptstyle+};
(2.600000,0.878322) *[blue]{\scriptscriptstyle+};
(2.618182,0.917621) *[blue]{\scriptscriptstyle+};
(2.636364,0.974856) *[blue]{\scriptscriptstyle+};
(2.654545,0.974927) *[blue]{\scriptscriptstyle+};
(2.672727,0.666038) *[blue]{\scriptscriptstyle+};
(2.690909,0.764663) *[blue]{\scriptscriptstyle+};
(2.709091,0.767980) *[blue]{\scriptscriptstyle+};
(2.727273,0.920689) *[blue]{\scriptscriptstyle+};
(2.745455,0.940500) *[blue]{\scriptscriptstyle+};
(2.763636,1.029518) *[blue]{\scriptscriptstyle+};
(2.781818,1.107387) *[blue]{\scriptscriptstyle+};
(2.800000,0.657965) *[blue]{\scriptscriptstyle+};
(2.818182,0.794112) *[blue]{\scriptscriptstyle+};
(2.836364,0.830861) *[blue]{\scriptscriptstyle+};
(2.854545,0.880744) *[blue]{\scriptscriptstyle+};
(2.872727,0.913438) *[blue]{\scriptscriptstyle+};
(2.890909,0.918880) *[blue]{\scriptscriptstyle+};
(2.909091,1.054922) *[blue]{\scriptscriptstyle+};
(2.927273,0.674647) *[blue]{\scriptscriptstyle+};
(2.945455,0.883914) *[blue]{\scriptscriptstyle+};
(2.963636,0.891902) *[blue]{\scriptscriptstyle+};
(2.981818,0.917856) *[blue]{\scriptscriptstyle+};
(3.000000,0.934940) *[blue]{\scriptscriptstyle+};
(3.018182,0.964719) *[blue]{\scriptscriptstyle+};
(3.036364,1.054821) *[blue]{\scriptscriptstyle+};
(3.054545,0.653652) *[blue]{\scriptscriptstyle+};
(3.072727,0.735577) *[blue]{\scriptscriptstyle+};
(3.090909,0.799060) *[blue]{\scriptscriptstyle+};
(3.109091,0.872294) *[blue]{\scriptscriptstyle+};
(3.127273,1.021504) *[blue]{\scriptscriptstyle+};
(3.145455,1.032760) *[blue]{\scriptscriptstyle+};
(3.163636,1.091146) *[blue]{\scriptscriptstyle+};
(3.181818,0.796570) *[blue]{\scriptscriptstyle+};
(3.200000,0.798022) *[blue]{\scriptscriptstyle+};
(3.218182,0.863879) *[blue]{\scriptscriptstyle+};
(3.236364,0.871629) *[blue]{\scriptscriptstyle+};
(3.254545,0.909044) *[blue]{\scriptscriptstyle+};
(3.272727,0.918055) *[blue]{\scriptscriptstyle+};
(3.290909,1.010359) *[blue]{\scriptscriptstyle+};
(3.309091,0.672252) *[blue]{\scriptscriptstyle+};
(3.327273,0.762955) *[blue]{\scriptscriptstyle+};
(3.345455,0.862852) *[blue]{\scriptscriptstyle+};
(3.363636,0.886143) *[blue]{\scriptscriptstyle+};
(3.381818,0.926999) *[blue]{\scriptscriptstyle+};
(3.400000,0.948227) *[blue]{\scriptscriptstyle+};
(3.418182,1.094405) *[blue]{\scriptscriptstyle+};
(3.436364,0.782105) *[blue]{\scriptscriptstyle+};
(3.454545,0.815724) *[blue]{\scriptscriptstyle+};
(3.472727,0.820113) *[blue]{\scriptscriptstyle+};
(3.490909,0.876498) *[blue]{\scriptscriptstyle+};
(3.509091,0.886130) *[blue]{\scriptscriptstyle+};
(3.527273,0.893314) *[blue]{\scriptscriptstyle+};
(3.545455,1.076633) *[blue]{\scriptscriptstyle+};
(3.563636,0.670736) *[blue]{\scriptscriptstyle+};
(3.581818,0.775845) *[blue]{\scriptscriptstyle+};
(3.600000,0.806459) *[blue]{\scriptscriptstyle+};
(3.618182,0.846375) *[blue]{\scriptscriptstyle+};
(3.636364,0.980821) *[blue]{\scriptscriptstyle+};
(3.654545,1.013445) *[blue]{\scriptscriptstyle+};
(3.672727,1.018035) *[blue]{\scriptscriptstyle+};
(3.690909,0.803185) *[blue]{\scriptscriptstyle+};
(3.709091,0.838353) *[blue]{\scriptscriptstyle+};
(3.727273,0.867325) *[blue]{\scriptscriptstyle+};
(3.745455,0.920801) *[blue]{\scriptscriptstyle+};
(3.763636,0.954110) *[blue]{\scriptscriptstyle+};
(3.781818,0.963635) *[blue]{\scriptscriptstyle+};
(3.800000,1.034671) *[blue]{\scriptscriptstyle+};
(3.818182,0.729167) *[blue]{\scriptscriptstyle+};
(3.836364,0.838109) *[blue]{\scriptscriptstyle+};
(3.854545,0.909996) *[blue]{\scriptscriptstyle+};
(3.872727,0.910440) *[blue]{\scriptscriptstyle+};
(3.890909,0.930205) *[blue]{\scriptscriptstyle+};
(3.909091,0.997981) *[blue]{\scriptscriptstyle+};
(3.927273,1.026951) *[blue]{\scriptscriptstyle+};
(3.945455,0.713102) *[blue]{\scriptscriptstyle+};
(3.963636,0.809908) *[blue]{\scriptscriptstyle+};
(3.981818,0.821398) *[blue]{\scriptscriptstyle+};
(4.000000,0.835658) *[blue]{\scriptscriptstyle+};
(4.018182,0.899942) *[blue]{\scriptscriptstyle+};
(4.036364,0.978473) *[blue]{\scriptscriptstyle+};
(4.054545,1.007451) *[blue]{\scriptscriptstyle+};
(4.072727,0.668641) *[blue]{\scriptscriptstyle+};
(4.090909,0.817189) *[blue]{\scriptscriptstyle+};
(4.109091,0.836177) *[blue]{\scriptscriptstyle+};
(4.127273,0.920054) *[blue]{\scriptscriptstyle+};
(4.145455,0.938141) *[blue]{\scriptscriptstyle+};
(4.163636,0.993668) *[blue]{\scriptscriptstyle+};
(4.181818,1.130928) *[blue]{\scriptscriptstyle+};
(4.200000,0.642400) *[blue]{\scriptscriptstyle+};
(4.218182,0.775365) *[blue]{\scriptscriptstyle+};
(4.236364,0.881261) *[blue]{\scriptscriptstyle+};
(4.254545,0.911375) *[blue]{\scriptscriptstyle+};
(4.272727,0.934059) *[blue]{\scriptscriptstyle+};
(4.290909,0.944709) *[blue]{\scriptscriptstyle+};
(4.309091,1.006920) *[blue]{\scriptscriptstyle+};
(4.327273,0.778123) *[blue]{\scriptscriptstyle+};
(4.345455,0.811729) *[blue]{\scriptscriptstyle+};
(4.363636,0.815764) *[blue]{\scriptscriptstyle+};
(4.381818,0.855563) *[blue]{\scriptscriptstyle+};
(4.400000,0.920386) *[blue]{\scriptscriptstyle+};
(4.418182,1.031336) *[blue]{\scriptscriptstyle+};
(4.436364,1.063920) *[blue]{\scriptscriptstyle+};
(4.454545,0.761176) *[blue]{\scriptscriptstyle+};
(4.472727,0.934646) *[blue]{\scriptscriptstyle+};
(4.490909,0.939439) *[blue]{\scriptscriptstyle+};
(4.509091,0.946166) *[blue]{\scriptscriptstyle+};
(4.527273,0.949784) *[blue]{\scriptscriptstyle+};
(4.545455,0.972831) *[blue]{\scriptscriptstyle+};
(4.563636,1.003063) *[blue]{\scriptscriptstyle+};
(4.581818,0.727349) *[blue]{\scriptscriptstyle+};
(4.600000,0.801026) *[blue]{\scriptscriptstyle+};
(4.618182,0.804104) *[blue]{\scriptscriptstyle+};
(4.636364,0.825427) *[blue]{\scriptscriptstyle+};
(4.654545,0.850963) *[blue]{\scriptscriptstyle+};
(4.672727,1.050342) *[blue]{\scriptscriptstyle+};
(4.690909,1.139570) *[blue]{\scriptscriptstyle+};
(4.709091,0.741359) *[blue]{\scriptscriptstyle+};
(4.727273,0.813840) *[blue]{\scriptscriptstyle+};
(4.745455,0.814305) *[blue]{\scriptscriptstyle+};
(4.763636,0.852738) *[blue]{\scriptscriptstyle+};
(4.781818,0.963829) *[blue]{\scriptscriptstyle+};
(4.800000,0.995358) *[blue]{\scriptscriptstyle+};
(4.818182,1.029387) *[blue]{\scriptscriptstyle+};
(4.836364,0.753270) *[blue]{\scriptscriptstyle+};
(4.854545,0.871582) *[blue]{\scriptscriptstyle+};
(4.872727,0.874736) *[blue]{\scriptscriptstyle+};
(4.890909,0.889667) *[blue]{\scriptscriptstyle+};
(4.909091,1.017492) *[blue]{\scriptscriptstyle+};
(4.927273,1.040561) *[blue]{\scriptscriptstyle+};
(4.945455,1.086289) *[blue]{\scriptscriptstyle+};
(4.963636,0.769024) *[blue]{\scriptscriptstyle+};
(4.981818,0.835375) *[blue]{\scriptscriptstyle+};
(5.000000,0.925312) *[blue]{\scriptscriptstyle+};
(5.018182,0.948296) *[blue]{\scriptscriptstyle+};
(5.036364,0.992156) *[blue]{\scriptscriptstyle+};
(5.054545,1.006978) *[blue]{\scriptscriptstyle+};
(5.072727,1.023741) *[blue]{\scriptscriptstyle+};
(5.090909,0.631552) *[blue]{\scriptscriptstyle+};
(5.109091,0.838659) *[blue]{\scriptscriptstyle+};
(5.127273,0.844423) *[blue]{\scriptscriptstyle+};
(5.145455,0.880772) *[blue]{\scriptscriptstyle+};
(5.163636,0.893903) *[blue]{\scriptscriptstyle+};
(5.181818,0.996294) *[blue]{\scriptscriptstyle+};
(5.200000,1.019014) *[blue]{\scriptscriptstyle+};
(5.218182,0.716671) *[blue]{\scriptscriptstyle+};
(5.236364,0.756132) *[blue]{\scriptscriptstyle+};
(5.254545,0.843911) *[blue]{\scriptscriptstyle+};
(5.272727,0.897220) *[blue]{\scriptscriptstyle+};
(5.290909,0.984435) *[blue]{\scriptscriptstyle+};
(5.309091,1.023194) *[blue]{\scriptscriptstyle+};
(5.327273,1.071547) *[blue]{\scriptscriptstyle+};
(5.345455,0.804281) *[blue]{\scriptscriptstyle+};
(5.363636,0.808242) *[blue]{\scriptscriptstyle+};
(5.381818,0.835667) *[blue]{\scriptscriptstyle+};
(5.400000,0.875830) *[blue]{\scriptscriptstyle+};
(5.418182,0.887708) *[blue]{\scriptscriptstyle+};
(5.436364,0.983329) *[blue]{\scriptscriptstyle+};
(5.454545,0.998805) *[blue]{\scriptscriptstyle+};
(5.472727,0.804008) *[blue]{\scriptscriptstyle+};
(5.490909,0.809630) *[blue]{\scriptscriptstyle+};
(5.509091,0.814584) *[blue]{\scriptscriptstyle+};
(5.527273,0.938829) *[blue]{\scriptscriptstyle+};
(5.545455,1.000069) *[blue]{\scriptscriptstyle+};
(5.563636,1.021345) *[blue]{\scriptscriptstyle+};
(5.581818,1.054721) *[blue]{\scriptscriptstyle+};
(5.600000,0.761891) *[blue]{\scriptscriptstyle+};
(5.618182,0.810421) *[blue]{\scriptscriptstyle+};
(5.636364,0.844208) *[blue]{\scriptscriptstyle+};
(5.654545,0.865086) *[blue]{\scriptscriptstyle+};
(5.672727,0.880376) *[blue]{\scriptscriptstyle+};
(5.690909,0.974736) *[blue]{\scriptscriptstyle+};
(5.709091,1.082316) *[blue]{\scriptscriptstyle+};
(5.727273,0.757315) *[blue]{\scriptscriptstyle+};
(5.745455,0.771693) *[blue]{\scriptscriptstyle+};
(5.763636,0.908556) *[blue]{\scriptscriptstyle+};
(5.781818,1.006462) *[blue]{\scriptscriptstyle+};
(5.800000,1.013146) *[blue]{\scriptscriptstyle+};
(5.818182,1.019633) *[blue]{\scriptscriptstyle+};
(5.836364,1.088603) *[blue]{\scriptscriptstyle+};
(5.854545,0.830168) *[blue]{\scriptscriptstyle+};
(5.872727,0.860315) *[blue]{\scriptscriptstyle+};
(5.890909,0.871672) *[blue]{\scriptscriptstyle+};
(5.909091,0.886383) *[blue]{\scriptscriptstyle+};
(5.927273,0.913910) *[blue]{\scriptscriptstyle+};
(5.945455,0.989212) *[blue]{\scriptscriptstyle+};
(5.963636,0.992467) *[blue]{\scriptscriptstyle+};
(5.981818,0.717561) *[blue]{\scriptscriptstyle+};
(6.000000,0.828694) *[blue]{\scriptscriptstyle+};
(6.018182,0.853843) *[blue]{\scriptscriptstyle+};
(6.036364,0.894115) *[blue]{\scriptscriptstyle+};
(6.054545,0.968718) *[blue]{\scriptscriptstyle+};
(6.072727,1.045951) *[blue]{\scriptscriptstyle+};
(6.090909,1.177060) *[blue]{\scriptscriptstyle+};
(6.109091,0.789502) *[blue]{\scriptscriptstyle+};
(6.127273,0.809190) *[blue]{\scriptscriptstyle+};
(6.145455,0.838078) *[blue]{\scriptscriptstyle+};
(6.163636,0.872135) *[blue]{\scriptscriptstyle+};
(6.181818,0.881203) *[blue]{\scriptscriptstyle+};
(6.200000,1.093075) *[blue]{\scriptscriptstyle+};
(6.218182,1.116345) *[blue]{\scriptscriptstyle+};
(6.236364,0.666295) *[blue]{\scriptscriptstyle+};
(6.254545,0.851699) *[blue]{\scriptscriptstyle+};
(6.272727,0.943310) *[blue]{\scriptscriptstyle+};
(6.290909,0.960510) *[blue]{\scriptscriptstyle+};
(6.309091,0.961659) *[blue]{\scriptscriptstyle+};
(6.327273,0.973588) *[blue]{\scriptscriptstyle+};
(6.345455,1.070695) *[blue]{\scriptscriptstyle+};
(6.363636,0.823556) *[blue]{\scriptscriptstyle+};
(6.381818,0.830819) *[blue]{\scriptscriptstyle+};
(6.400000,0.831384) *[blue]{\scriptscriptstyle+};
(6.418182,0.855141) *[blue]{\scriptscriptstyle+};
(6.436364,0.881183) *[blue]{\scriptscriptstyle+};
(6.454545,1.007633) *[blue]{\scriptscriptstyle+};
(6.472727,1.119595) *[blue]{\scriptscriptstyle+};
(6.490909,0.821926) *[blue]{\scriptscriptstyle+};
(6.509091,0.835744) *[blue]{\scriptscriptstyle+};
(6.527273,0.857737) *[blue]{\scriptscriptstyle+};
(6.545455,0.894599) *[blue]{\scriptscriptstyle+};
(6.563636,0.896814) *[blue]{\scriptscriptstyle+};
(6.581818,0.956672) *[blue]{\scriptscriptstyle+};
(6.600000,1.007997) *[blue]{\scriptscriptstyle+};
(6.618182,0.735165) *[blue]{\scriptscriptstyle+};
(6.636364,0.828599) *[blue]{\scriptscriptstyle+};
(6.654545,0.863965) *[blue]{\scriptscriptstyle+};
(6.672727,1.011657) *[blue]{\scriptscriptstyle+};
(6.690909,1.045084) *[blue]{\scriptscriptstyle+};
(6.709091,1.092625) *[blue]{\scriptscriptstyle+};
(6.727273,1.133138) *[blue]{\scriptscriptstyle+};
(6.745455,0.789049) *[blue]{\scriptscriptstyle+};
(6.763636,0.826221) *[blue]{\scriptscriptstyle+};
(6.781818,0.900956) *[blue]{\scriptscriptstyle+};
(6.800000,0.965199) *[blue]{\scriptscriptstyle+};
(6.818182,0.986605) *[blue]{\scriptscriptstyle+};
(6.836364,1.043968) *[blue]{\scriptscriptstyle+};
(6.854545,1.172288) *[blue]{\scriptscriptstyle+};
(6.872727,0.707670) *[blue]{\scriptscriptstyle+};
(6.890909,0.881381) *[blue]{\scriptscriptstyle+};
(6.909091,0.899506) *[blue]{\scriptscriptstyle+};
(6.927273,0.905635) *[blue]{\scriptscriptstyle+};
(6.945455,0.920995) *[blue]{\scriptscriptstyle+};
(6.963636,0.933220) *[blue]{\scriptscriptstyle+};
(6.981818,1.126273) *[blue]{\scriptscriptstyle+};
(7.000000,0.834221) *[blue]{\scriptscriptstyle+};
(7.018182,0.888559) *[blue]{\scriptscriptstyle+};
(7.036364,0.908329) *[blue]{\scriptscriptstyle+};
(7.054545,0.952979) *[blue]{\scriptscriptstyle+};
(7.072727,0.971717) *[blue]{\scriptscriptstyle+};
(7.090909,1.027146) *[blue]{\scriptscriptstyle+};
(7.109091,1.159637) *[blue]{\scriptscriptstyle+};
(7.127273,0.832087) *[blue]{\scriptscriptstyle+};
(7.145455,0.894348) *[blue]{\scriptscriptstyle+};
(7.163636,0.899312) *[blue]{\scriptscriptstyle+};
(7.181818,0.904354) *[blue]{\scriptscriptstyle+};
(7.200000,0.939748) *[blue]{\scriptscriptstyle+};
(7.218182,1.022681) *[blue]{\scriptscriptstyle+};
(7.236364,1.123830) *[blue]{\scriptscriptstyle+};
(7.254545,0.707370) *[blue]{\scriptscriptstyle+};
(7.272727,0.746881) *[blue]{\scriptscriptstyle+};
(7.290909,0.781864) *[blue]{\scriptscriptstyle+};
(7.309091,1.028336) *[blue]{\scriptscriptstyle+};
(7.327273,1.074177) *[blue]{\scriptscriptstyle+};
(7.345455,1.093033) *[blue]{\scriptscriptstyle+};
(7.363636,1.249344) *[blue]{\scriptscriptstyle+};
(7.381818,0.830679) *[blue]{\scriptscriptstyle+};
(7.400000,0.859482) *[blue]{\scriptscriptstyle+};
(7.418182,0.902987) *[blue]{\scriptscriptstyle+};
(7.436364,1.019660) *[blue]{\scriptscriptstyle+};
(7.454545,1.021111) *[blue]{\scriptscriptstyle+};
(7.472727,1.069888) *[blue]{\scriptscriptstyle+};
(7.490909,1.198326) *[blue]{\scriptscriptstyle+};
(7.509091,0.861640) *[blue]{\scriptscriptstyle+};
(7.527273,0.901089) *[blue]{\scriptscriptstyle+};
(7.545455,0.923628) *[blue]{\scriptscriptstyle+};
(7.563636,0.942446) *[blue]{\scriptscriptstyle+};
(7.581818,0.946334) *[blue]{\scriptscriptstyle+};
(7.600000,1.033364) *[blue]{\scriptscriptstyle+};
(7.618182,1.082860) *[blue]{\scriptscriptstyle+};
(7.636364,0.718392) *[blue]{\scriptscriptstyle+};
(7.654545,0.861983) *[blue]{\scriptscriptstyle+};
(7.672727,0.922594) *[blue]{\scriptscriptstyle+};
(7.690909,0.947167) *[blue]{\scriptscriptstyle+};
(7.709091,1.068372) *[blue]{\scriptscriptstyle+};
(7.727273,1.081893) *[blue]{\scriptscriptstyle+};
(7.745455,1.154793) *[blue]{\scriptscriptstyle+};
(7.763636,0.714741) *[blue]{\scriptscriptstyle+};
(7.781818,0.907567) *[blue]{\scriptscriptstyle+};
(7.800000,0.914765) *[blue]{\scriptscriptstyle+};
(7.818182,0.956467) *[blue]{\scriptscriptstyle+};
(7.836364,1.031012) *[blue]{\scriptscriptstyle+};
(7.854545,1.092314) *[blue]{\scriptscriptstyle+};
(7.872727,1.114384) *[blue]{\scriptscriptstyle+};
(7.890909,0.941090) *[blue]{\scriptscriptstyle+};
(7.909091,1.006922) *[blue]{\scriptscriptstyle+};
(7.927273,1.014672) *[blue]{\scriptscriptstyle+};
(7.945455,1.014677) *[blue]{\scriptscriptstyle+};
(7.963636,1.017981) *[blue]{\scriptscriptstyle+};
(7.981818,1.018993) *[blue]{\scriptscriptstyle+};
(8.000000,1.043759) *[blue]{\scriptscriptstyle+};
(0.000000,0.347576) *[red]{\scriptscriptstyle\times};
(0.018182,0.534742) *[red]{\scriptscriptstyle\times};
(0.036364,0.557361) *[red]{\scriptscriptstyle\times};
(0.054545,0.577310) *[red]{\scriptscriptstyle\times};
(0.072727,0.593184) *[red]{\scriptscriptstyle\times};
(0.090909,0.642110) *[red]{\scriptscriptstyle\times};
(0.109091,0.808163) *[red]{\scriptscriptstyle\times};
(0.127273,0.453028) *[red]{\scriptscriptstyle\times};
(0.145455,0.492691) *[red]{\scriptscriptstyle\times};
(0.163636,0.534650) *[red]{\scriptscriptstyle\times};
(0.181818,0.605539) *[red]{\scriptscriptstyle\times};
(0.200000,0.621208) *[red]{\scriptscriptstyle\times};
(0.218182,0.646960) *[red]{\scriptscriptstyle\times};
(0.236364,0.704003) *[red]{\scriptscriptstyle\times};
(0.254545,0.455510) *[red]{\scriptscriptstyle\times};
(0.272727,0.476421) *[red]{\scriptscriptstyle\times};
(0.290909,0.561032) *[red]{\scriptscriptstyle\times};
(0.309091,0.623103) *[red]{\scriptscriptstyle\times};
(0.327273,0.643190) *[red]{\scriptscriptstyle\times};
(0.345455,0.684563) *[red]{\scriptscriptstyle\times};
(0.363636,0.781292) *[red]{\scriptscriptstyle\times};
(0.381818,0.412322) *[red]{\scriptscriptstyle\times};
(0.400000,0.426314) *[red]{\scriptscriptstyle\times};
(0.418182,0.500522) *[red]{\scriptscriptstyle\times};
(0.436364,0.620734) *[red]{\scriptscriptstyle\times};
(0.454545,0.666198) *[red]{\scriptscriptstyle\times};
(0.472727,0.775814) *[red]{\scriptscriptstyle\times};
(0.490909,1.009110) *[red]{\scriptscriptstyle\times};
(0.509091,0.392198) *[red]{\scriptscriptstyle\times};
(0.527273,0.423803) *[red]{\scriptscriptstyle\times};
(0.545455,0.553353) *[red]{\scriptscriptstyle\times};
(0.563636,0.576818) *[red]{\scriptscriptstyle\times};
(0.581818,0.737946) *[red]{\scriptscriptstyle\times};
(0.600000,0.784806) *[red]{\scriptscriptstyle\times};
(0.618182,0.788536) *[red]{\scriptscriptstyle\times};
(0.636364,0.399368) *[red]{\scriptscriptstyle\times};
(0.654545,0.467510) *[red]{\scriptscriptstyle\times};
(0.672727,0.502804) *[red]{\scriptscriptstyle\times};
(0.690909,0.665386) *[red]{\scriptscriptstyle\times};
(0.709091,0.739294) *[red]{\scriptscriptstyle\times};
(0.727273,0.830340) *[red]{\scriptscriptstyle\times};
(0.745455,0.832032) *[red]{\scriptscriptstyle\times};
(0.763636,0.556787) *[red]{\scriptscriptstyle\times};
(0.781818,0.576667) *[red]{\scriptscriptstyle\times};
(0.800000,0.611770) *[red]{\scriptscriptstyle\times};
(0.818182,0.641409) *[red]{\scriptscriptstyle\times};
(0.836364,0.652651) *[red]{\scriptscriptstyle\times};
(0.854545,0.673679) *[red]{\scriptscriptstyle\times};
(0.872727,0.676109) *[red]{\scriptscriptstyle\times};
(0.890909,0.430604) *[red]{\scriptscriptstyle\times};
(0.909091,0.557568) *[red]{\scriptscriptstyle\times};
(0.927273,0.574794) *[red]{\scriptscriptstyle\times};
(0.945455,0.583563) *[red]{\scriptscriptstyle\times};
(0.963636,0.667864) *[red]{\scriptscriptstyle\times};
(0.981818,0.670724) *[red]{\scriptscriptstyle\times};
(1.000000,0.782097) *[red]{\scriptscriptstyle\times};
(1.018182,0.410699) *[red]{\scriptscriptstyle\times};
(1.036364,0.492509) *[red]{\scriptscriptstyle\times};
(1.054545,0.495343) *[red]{\scriptscriptstyle\times};
(1.072727,0.645171) *[red]{\scriptscriptstyle\times};
(1.090909,0.709757) *[red]{\scriptscriptstyle\times};
(1.109091,0.806481) *[red]{\scriptscriptstyle\times};
(1.127273,0.877276) *[red]{\scriptscriptstyle\times};
(1.145455,0.461202) *[red]{\scriptscriptstyle\times};
(1.163636,0.533238) *[red]{\scriptscriptstyle\times};
(1.181818,0.544110) *[red]{\scriptscriptstyle\times};
(1.200000,0.564623) *[red]{\scriptscriptstyle\times};
(1.218182,0.597576) *[red]{\scriptscriptstyle\times};
(1.236364,0.717822) *[red]{\scriptscriptstyle\times};
(1.254545,0.734273) *[red]{\scriptscriptstyle\times};
(1.272727,0.525923) *[red]{\scriptscriptstyle\times};
(1.290909,0.539311) *[red]{\scriptscriptstyle\times};
(1.309091,0.561248) *[red]{\scriptscriptstyle\times};
(1.327273,0.663123) *[red]{\scriptscriptstyle\times};
(1.345455,0.718364) *[red]{\scriptscriptstyle\times};
(1.363636,0.757771) *[red]{\scriptscriptstyle\times};
(1.381818,0.844328) *[red]{\scriptscriptstyle\times};
(1.400000,0.409101) *[red]{\scriptscriptstyle\times};
(1.418182,0.467887) *[red]{\scriptscriptstyle\times};
(1.436364,0.515210) *[red]{\scriptscriptstyle\times};
(1.454545,0.609449) *[red]{\scriptscriptstyle\times};
(1.472727,0.709036) *[red]{\scriptscriptstyle\times};
(1.490909,0.768797) *[red]{\scriptscriptstyle\times};
(1.509091,0.815806) *[red]{\scriptscriptstyle\times};
(1.527273,0.445563) *[red]{\scriptscriptstyle\times};
(1.545455,0.471608) *[red]{\scriptscriptstyle\times};
(1.563636,0.516235) *[red]{\scriptscriptstyle\times};
(1.581818,0.603400) *[red]{\scriptscriptstyle\times};
(1.600000,0.733703) *[red]{\scriptscriptstyle\times};
(1.618182,0.795715) *[red]{\scriptscriptstyle\times};
(1.636364,0.901975) *[red]{\scriptscriptstyle\times};
(1.654545,0.538330) *[red]{\scriptscriptstyle\times};
(1.672727,0.608027) *[red]{\scriptscriptstyle\times};
(1.690909,0.639692) *[red]{\scriptscriptstyle\times};
(1.709091,0.646686) *[red]{\scriptscriptstyle\times};
(1.727273,0.671009) *[red]{\scriptscriptstyle\times};
(1.745455,0.695747) *[red]{\scriptscriptstyle\times};
(1.763636,0.784971) *[red]{\scriptscriptstyle\times};
(1.781818,0.530445) *[red]{\scriptscriptstyle\times};
(1.800000,0.543592) *[red]{\scriptscriptstyle\times};
(1.818182,0.552627) *[red]{\scriptscriptstyle\times};
(1.836364,0.634508) *[red]{\scriptscriptstyle\times};
(1.854545,0.698995) *[red]{\scriptscriptstyle\times};
(1.872727,0.713595) *[red]{\scriptscriptstyle\times};
(1.890909,0.813665) *[red]{\scriptscriptstyle\times};
(1.909091,0.544860) *[red]{\scriptscriptstyle\times};
(1.927273,0.609925) *[red]{\scriptscriptstyle\times};
(1.945455,0.657339) *[red]{\scriptscriptstyle\times};
(1.963636,0.711400) *[red]{\scriptscriptstyle\times};
(1.981818,0.732290) *[red]{\scriptscriptstyle\times};
(2.000000,0.751521) *[red]{\scriptscriptstyle\times};
(2.018182,0.791605) *[red]{\scriptscriptstyle\times};
(2.036364,0.496923) *[red]{\scriptscriptstyle\times};
(2.054545,0.562702) *[red]{\scriptscriptstyle\times};
(2.072727,0.584916) *[red]{\scriptscriptstyle\times};
(2.090909,0.625786) *[red]{\scriptscriptstyle\times};
(2.109091,0.628505) *[red]{\scriptscriptstyle\times};
(2.127273,0.696522) *[red]{\scriptscriptstyle\times};
(2.145455,0.818372) *[red]{\scriptscriptstyle\times};
(2.163636,0.597055) *[red]{\scriptscriptstyle\times};
(2.181818,0.597196) *[red]{\scriptscriptstyle\times};
(2.200000,0.619880) *[red]{\scriptscriptstyle\times};
(2.218182,0.621320) *[red]{\scriptscriptstyle\times};
(2.236364,0.655202) *[red]{\scriptscriptstyle\times};
(2.254545,0.689499) *[red]{\scriptscriptstyle\times};
(2.272727,0.743131) *[red]{\scriptscriptstyle\times};
(2.290909,0.421521) *[red]{\scriptscriptstyle\times};
(2.309091,0.519563) *[red]{\scriptscriptstyle\times};
(2.327273,0.565040) *[red]{\scriptscriptstyle\times};
(2.345455,0.708551) *[red]{\scriptscriptstyle\times};
(2.363636,0.720443) *[red]{\scriptscriptstyle\times};
(2.381818,0.761008) *[red]{\scriptscriptstyle\times};
(2.400000,0.815910) *[red]{\scriptscriptstyle\times};
(2.418182,0.468420) *[red]{\scriptscriptstyle\times};
(2.436364,0.525981) *[red]{\scriptscriptstyle\times};
(2.454545,0.585276) *[red]{\scriptscriptstyle\times};
(2.472727,0.613164) *[red]{\scriptscriptstyle\times};
(2.490909,0.647059) *[red]{\scriptscriptstyle\times};
(2.509091,0.685194) *[red]{\scriptscriptstyle\times};
(2.527273,0.831123) *[red]{\scriptscriptstyle\times};
(2.545455,0.439582) *[red]{\scriptscriptstyle\times};
(2.563636,0.440168) *[red]{\scriptscriptstyle\times};
(2.581818,0.571450) *[red]{\scriptscriptstyle\times};
(2.600000,0.670588) *[red]{\scriptscriptstyle\times};
(2.618182,0.758950) *[red]{\scriptscriptstyle\times};
(2.636364,0.759499) *[red]{\scriptscriptstyle\times};
(2.654545,0.782707) *[red]{\scriptscriptstyle\times};
(2.672727,0.407900) *[red]{\scriptscriptstyle\times};
(2.690909,0.531560) *[red]{\scriptscriptstyle\times};
(2.709091,0.534394) *[red]{\scriptscriptstyle\times};
(2.727273,0.660096) *[red]{\scriptscriptstyle\times};
(2.745455,0.662164) *[red]{\scriptscriptstyle\times};
(2.763636,0.869957) *[red]{\scriptscriptstyle\times};
(2.781818,0.905207) *[red]{\scriptscriptstyle\times};
(2.800000,0.472645) *[red]{\scriptscriptstyle\times};
(2.818182,0.617675) *[red]{\scriptscriptstyle\times};
(2.836364,0.696428) *[red]{\scriptscriptstyle\times};
(2.854545,0.697794) *[red]{\scriptscriptstyle\times};
(2.872727,0.744420) *[red]{\scriptscriptstyle\times};
(2.890909,0.756047) *[red]{\scriptscriptstyle\times};
(2.909091,0.893665) *[red]{\scriptscriptstyle\times};
(2.927273,0.491062) *[red]{\scriptscriptstyle\times};
(2.945455,0.662972) *[red]{\scriptscriptstyle\times};
(2.963636,0.708546) *[red]{\scriptscriptstyle\times};
(2.981818,0.728198) *[red]{\scriptscriptstyle\times};
(3.000000,0.752331) *[red]{\scriptscriptstyle\times};
(3.018182,0.756849) *[red]{\scriptscriptstyle\times};
(3.036364,0.801245) *[red]{\scriptscriptstyle\times};
(3.054545,0.475506) *[red]{\scriptscriptstyle\times};
(3.072727,0.488507) *[red]{\scriptscriptstyle\times};
(3.090909,0.610815) *[red]{\scriptscriptstyle\times};
(3.109091,0.651281) *[red]{\scriptscriptstyle\times};
(3.127273,0.760137) *[red]{\scriptscriptstyle\times};
(3.145455,0.860770) *[red]{\scriptscriptstyle\times};
(3.163636,0.893051) *[red]{\scriptscriptstyle\times};
(3.181818,0.589403) *[red]{\scriptscriptstyle\times};
(3.200000,0.605120) *[red]{\scriptscriptstyle\times};
(3.218182,0.646246) *[red]{\scriptscriptstyle\times};
(3.236364,0.687125) *[red]{\scriptscriptstyle\times};
(3.254545,0.703560) *[red]{\scriptscriptstyle\times};
(3.272727,0.706974) *[red]{\scriptscriptstyle\times};
(3.290909,0.765458) *[red]{\scriptscriptstyle\times};
(3.309091,0.502183) *[red]{\scriptscriptstyle\times};
(3.327273,0.521436) *[red]{\scriptscriptstyle\times};
(3.345455,0.655325) *[red]{\scriptscriptstyle\times};
(3.363636,0.684792) *[red]{\scriptscriptstyle\times};
(3.381818,0.768272) *[red]{\scriptscriptstyle\times};
(3.400000,0.796138) *[red]{\scriptscriptstyle\times};
(3.418182,0.869387) *[red]{\scriptscriptstyle\times};
(3.436364,0.535744) *[red]{\scriptscriptstyle\times};
(3.454545,0.588958) *[red]{\scriptscriptstyle\times};
(3.472727,0.628528) *[red]{\scriptscriptstyle\times};
(3.490909,0.693202) *[red]{\scriptscriptstyle\times};
(3.509091,0.693344) *[red]{\scriptscriptstyle\times};
(3.527273,0.730984) *[red]{\scriptscriptstyle\times};
(3.545455,0.785475) *[red]{\scriptscriptstyle\times};
(3.563636,0.530114) *[red]{\scriptscriptstyle\times};
(3.581818,0.602994) *[red]{\scriptscriptstyle\times};
(3.600000,0.610815) *[red]{\scriptscriptstyle\times};
(3.618182,0.640451) *[red]{\scriptscriptstyle\times};
(3.636364,0.655151) *[red]{\scriptscriptstyle\times};
(3.654545,0.801187) *[red]{\scriptscriptstyle\times};
(3.672727,0.835667) *[red]{\scriptscriptstyle\times};
(3.690909,0.566891) *[red]{\scriptscriptstyle\times};
(3.709091,0.630849) *[red]{\scriptscriptstyle\times};
(3.727273,0.634914) *[red]{\scriptscriptstyle\times};
(3.745455,0.657114) *[red]{\scriptscriptstyle\times};
(3.763636,0.671279) *[red]{\scriptscriptstyle\times};
(3.781818,0.763620) *[red]{\scriptscriptstyle\times};
(3.800000,0.775161) *[red]{\scriptscriptstyle\times};
(3.818182,0.536641) *[red]{\scriptscriptstyle\times};
(3.836364,0.657004) *[red]{\scriptscriptstyle\times};
(3.854545,0.671002) *[red]{\scriptscriptstyle\times};
(3.872727,0.692853) *[red]{\scriptscriptstyle\times};
(3.890909,0.706002) *[red]{\scriptscriptstyle\times};
(3.909091,0.740754) *[red]{\scriptscriptstyle\times};
(3.927273,0.875933) *[red]{\scriptscriptstyle\times};
(3.945455,0.420628) *[red]{\scriptscriptstyle\times};
(3.963636,0.596035) *[red]{\scriptscriptstyle\times};
(3.981818,0.646252) *[red]{\scriptscriptstyle\times};
(4.000000,0.650597) *[red]{\scriptscriptstyle\times};
(4.018182,0.742764) *[red]{\scriptscriptstyle\times};
(4.036364,0.771829) *[red]{\scriptscriptstyle\times};
(4.054545,0.834515) *[red]{\scriptscriptstyle\times};
(4.072727,0.507692) *[red]{\scriptscriptstyle\times};
(4.090909,0.631656) *[red]{\scriptscriptstyle\times};
(4.109091,0.651573) *[red]{\scriptscriptstyle\times};
(4.127273,0.671070) *[red]{\scriptscriptstyle\times};
(4.145455,0.675597) *[red]{\scriptscriptstyle\times};
(4.163636,0.764973) *[red]{\scriptscriptstyle\times};
(4.181818,0.920103) *[red]{\scriptscriptstyle\times};
(4.200000,0.358271) *[red]{\scriptscriptstyle\times};
(4.218182,0.582570) *[red]{\scriptscriptstyle\times};
(4.236364,0.668767) *[red]{\scriptscriptstyle\times};
(4.254545,0.701527) *[red]{\scriptscriptstyle\times};
(4.272727,0.763226) *[red]{\scriptscriptstyle\times};
(4.290909,0.811933) *[red]{\scriptscriptstyle\times};
(4.309091,0.856130) *[red]{\scriptscriptstyle\times};
(4.327273,0.594049) *[red]{\scriptscriptstyle\times};
(4.345455,0.599436) *[red]{\scriptscriptstyle\times};
(4.363636,0.643763) *[red]{\scriptscriptstyle\times};
(4.381818,0.656633) *[red]{\scriptscriptstyle\times};
(4.400000,0.720075) *[red]{\scriptscriptstyle\times};
(4.418182,0.744301) *[red]{\scriptscriptstyle\times};
(4.436364,0.908148) *[red]{\scriptscriptstyle\times};
(4.454545,0.598284) *[red]{\scriptscriptstyle\times};
(4.472727,0.731815) *[red]{\scriptscriptstyle\times};
(4.490909,0.740797) *[red]{\scriptscriptstyle\times};
(4.509091,0.766517) *[red]{\scriptscriptstyle\times};
(4.527273,0.780330) *[red]{\scriptscriptstyle\times};
(4.545455,0.782901) *[red]{\scriptscriptstyle\times};
(4.563636,0.808113) *[red]{\scriptscriptstyle\times};
(4.581818,0.504187) *[red]{\scriptscriptstyle\times};
(4.600000,0.588445) *[red]{\scriptscriptstyle\times};
(4.618182,0.606882) *[red]{\scriptscriptstyle\times};
(4.636364,0.617053) *[red]{\scriptscriptstyle\times};
(4.654545,0.781132) *[red]{\scriptscriptstyle\times};
(4.672727,0.791955) *[red]{\scriptscriptstyle\times};
(4.690909,0.962469) *[red]{\scriptscriptstyle\times};
(4.709091,0.505271) *[red]{\scriptscriptstyle\times};
(4.727273,0.559957) *[red]{\scriptscriptstyle\times};
(4.745455,0.596715) *[red]{\scriptscriptstyle\times};
(4.763636,0.682920) *[red]{\scriptscriptstyle\times};
(4.781818,0.690616) *[red]{\scriptscriptstyle\times};
(4.800000,0.790813) *[red]{\scriptscriptstyle\times};
(4.818182,0.832670) *[red]{\scriptscriptstyle\times};
(4.836364,0.552218) *[red]{\scriptscriptstyle\times};
(4.854545,0.633542) *[red]{\scriptscriptstyle\times};
(4.872727,0.667446) *[red]{\scriptscriptstyle\times};
(4.890909,0.703842) *[red]{\scriptscriptstyle\times};
(4.909091,0.760457) *[red]{\scriptscriptstyle\times};
(4.927273,0.795689) *[red]{\scriptscriptstyle\times};
(4.945455,0.810141) *[red]{\scriptscriptstyle\times};
(4.963636,0.525684) *[red]{\scriptscriptstyle\times};
(4.981818,0.664636) *[red]{\scriptscriptstyle\times};
(5.000000,0.672529) *[red]{\scriptscriptstyle\times};
(5.018182,0.707007) *[red]{\scriptscriptstyle\times};
(5.036364,0.740172) *[red]{\scriptscriptstyle\times};
(5.054545,0.774504) *[red]{\scriptscriptstyle\times};
(5.072727,0.832530) *[red]{\scriptscriptstyle\times};
(5.090909,0.377522) *[red]{\scriptscriptstyle\times};
(5.109091,0.666403) *[red]{\scriptscriptstyle\times};
(5.127273,0.669134) *[red]{\scriptscriptstyle\times};
(5.145455,0.679752) *[red]{\scriptscriptstyle\times};
(5.163636,0.711699) *[red]{\scriptscriptstyle\times};
(5.181818,0.769353) *[red]{\scriptscriptstyle\times};
(5.200000,0.850676) *[red]{\scriptscriptstyle\times};
(5.218182,0.510900) *[red]{\scriptscriptstyle\times};
(5.236364,0.550305) *[red]{\scriptscriptstyle\times};
(5.254545,0.610155) *[red]{\scriptscriptstyle\times};
(5.272727,0.660010) *[red]{\scriptscriptstyle\times};
(5.290909,0.766814) *[red]{\scriptscriptstyle\times};
(5.309091,0.871603) *[red]{\scriptscriptstyle\times};
(5.327273,0.934982) *[red]{\scriptscriptstyle\times};
(5.345455,0.595820) *[red]{\scriptscriptstyle\times};
(5.363636,0.603070) *[red]{\scriptscriptstyle\times};
(5.381818,0.637580) *[red]{\scriptscriptstyle\times};
(5.400000,0.652633) *[red]{\scriptscriptstyle\times};
(5.418182,0.779834) *[red]{\scriptscriptstyle\times};
(5.436364,0.788874) *[red]{\scriptscriptstyle\times};
(5.454545,0.851645) *[red]{\scriptscriptstyle\times};
(5.472727,0.559677) *[red]{\scriptscriptstyle\times};
(5.490909,0.664037) *[red]{\scriptscriptstyle\times};
(5.509091,0.668172) *[red]{\scriptscriptstyle\times};
(5.527273,0.766881) *[red]{\scriptscriptstyle\times};
(5.545455,0.782197) *[red]{\scriptscriptstyle\times};
(5.563636,0.795392) *[red]{\scriptscriptstyle\times};
(5.581818,0.862851) *[red]{\scriptscriptstyle\times};
(5.600000,0.552434) *[red]{\scriptscriptstyle\times};
(5.618182,0.596231) *[red]{\scriptscriptstyle\times};
(5.636364,0.620300) *[red]{\scriptscriptstyle\times};
(5.654545,0.622352) *[red]{\scriptscriptstyle\times};
(5.672727,0.651802) *[red]{\scriptscriptstyle\times};
(5.690909,0.804611) *[red]{\scriptscriptstyle\times};
(5.709091,0.920159) *[red]{\scriptscriptstyle\times};
(5.727273,0.497259) *[red]{\scriptscriptstyle\times};
(5.745455,0.622643) *[red]{\scriptscriptstyle\times};
(5.763636,0.718897) *[red]{\scriptscriptstyle\times};
(5.781818,0.778409) *[red]{\scriptscriptstyle\times};
(5.800000,0.795842) *[red]{\scriptscriptstyle\times};
(5.818182,0.828957) *[red]{\scriptscriptstyle\times};
(5.836364,0.874062) *[red]{\scriptscriptstyle\times};
(5.854545,0.603313) *[red]{\scriptscriptstyle\times};
(5.872727,0.629917) *[red]{\scriptscriptstyle\times};
(5.890909,0.643439) *[red]{\scriptscriptstyle\times};
(5.909091,0.677667) *[red]{\scriptscriptstyle\times};
(5.927273,0.713445) *[red]{\scriptscriptstyle\times};
(5.945455,0.749836) *[red]{\scriptscriptstyle\times};
(5.963636,0.806177) *[red]{\scriptscriptstyle\times};
(5.981818,0.538564) *[red]{\scriptscriptstyle\times};
(6.000000,0.625133) *[red]{\scriptscriptstyle\times};
(6.018182,0.695335) *[red]{\scriptscriptstyle\times};
(6.036364,0.705113) *[red]{\scriptscriptstyle\times};
(6.054545,0.746756) *[red]{\scriptscriptstyle\times};
(6.072727,0.814550) *[red]{\scriptscriptstyle\times};
(6.090909,1.016154) *[red]{\scriptscriptstyle\times};
(6.109091,0.562172) *[red]{\scriptscriptstyle\times};
(6.127273,0.588906) *[red]{\scriptscriptstyle\times};
(6.145455,0.649406) *[red]{\scriptscriptstyle\times};
(6.163636,0.668552) *[red]{\scriptscriptstyle\times};
(6.181818,0.785292) *[red]{\scriptscriptstyle\times};
(6.200000,0.910067) *[red]{\scriptscriptstyle\times};
(6.218182,0.927281) *[red]{\scriptscriptstyle\times};
(6.236364,0.581691) *[red]{\scriptscriptstyle\times};
(6.254545,0.640252) *[red]{\scriptscriptstyle\times};
(6.272727,0.664646) *[red]{\scriptscriptstyle\times};
(6.290909,0.716015) *[red]{\scriptscriptstyle\times};
(6.309091,0.717659) *[red]{\scriptscriptstyle\times};
(6.327273,0.810530) *[red]{\scriptscriptstyle\times};
(6.345455,0.843102) *[red]{\scriptscriptstyle\times};
(6.363636,0.607535) *[red]{\scriptscriptstyle\times};
(6.381818,0.611907) *[red]{\scriptscriptstyle\times};
(6.400000,0.663883) *[red]{\scriptscriptstyle\times};
(6.418182,0.742354) *[red]{\scriptscriptstyle\times};
(6.436364,0.745247) *[red]{\scriptscriptstyle\times};
(6.454545,0.815799) *[red]{\scriptscriptstyle\times};
(6.472727,0.957288) *[red]{\scriptscriptstyle\times};
(6.490909,0.588108) *[red]{\scriptscriptstyle\times};
(6.509091,0.601166) *[red]{\scriptscriptstyle\times};
(6.527273,0.667232) *[red]{\scriptscriptstyle\times};
(6.545455,0.720575) *[red]{\scriptscriptstyle\times};
(6.563636,0.756955) *[red]{\scriptscriptstyle\times};
(6.581818,0.760257) *[red]{\scriptscriptstyle\times};
(6.600000,0.853268) *[red]{\scriptscriptstyle\times};
(6.618182,0.488296) *[red]{\scriptscriptstyle\times};
(6.636364,0.632710) *[red]{\scriptscriptstyle\times};
(6.654545,0.636573) *[red]{\scriptscriptstyle\times};
(6.672727,0.750131) *[red]{\scriptscriptstyle\times};
(6.690909,0.818931) *[red]{\scriptscriptstyle\times};
(6.709091,0.829171) *[red]{\scriptscriptstyle\times};
(6.727273,0.888210) *[red]{\scriptscriptstyle\times};
(6.745455,0.559418) *[red]{\scriptscriptstyle\times};
(6.763636,0.593983) *[red]{\scriptscriptstyle\times};
(6.781818,0.624232) *[red]{\scriptscriptstyle\times};
(6.800000,0.775526) *[red]{\scriptscriptstyle\times};
(6.818182,0.805616) *[red]{\scriptscriptstyle\times};
(6.836364,0.831561) *[red]{\scriptscriptstyle\times};
(6.854545,0.932970) *[red]{\scriptscriptstyle\times};
(6.872727,0.479578) *[red]{\scriptscriptstyle\times};
(6.890909,0.645181) *[red]{\scriptscriptstyle\times};
(6.909091,0.673773) *[red]{\scriptscriptstyle\times};
(6.927273,0.727368) *[red]{\scriptscriptstyle\times};
(6.945455,0.730912) *[red]{\scriptscriptstyle\times};
(6.963636,0.778384) *[red]{\scriptscriptstyle\times};
(6.981818,0.920044) *[red]{\scriptscriptstyle\times};
(7.000000,0.625609) *[red]{\scriptscriptstyle\times};
(7.018182,0.661404) *[red]{\scriptscriptstyle\times};
(7.036364,0.746474) *[red]{\scriptscriptstyle\times};
(7.054545,0.762243) *[red]{\scriptscriptstyle\times};
(7.072727,0.805753) *[red]{\scriptscriptstyle\times};
(7.090909,0.828033) *[red]{\scriptscriptstyle\times};
(7.109091,0.881341) *[red]{\scriptscriptstyle\times};
(7.127273,0.682604) *[red]{\scriptscriptstyle\times};
(7.145455,0.685091) *[red]{\scriptscriptstyle\times};
(7.163636,0.703412) *[red]{\scriptscriptstyle\times};
(7.181818,0.711999) *[red]{\scriptscriptstyle\times};
(7.200000,0.736696) *[red]{\scriptscriptstyle\times};
(7.218182,0.833594) *[red]{\scriptscriptstyle\times};
(7.236364,0.943551) *[red]{\scriptscriptstyle\times};
(7.254545,0.501689) *[red]{\scriptscriptstyle\times};
(7.272727,0.524418) *[red]{\scriptscriptstyle\times};
(7.290909,0.590253) *[red]{\scriptscriptstyle\times};
(7.309091,0.857789) *[red]{\scriptscriptstyle\times};
(7.327273,0.884024) *[red]{\scriptscriptstyle\times};
(7.345455,0.920294) *[red]{\scriptscriptstyle\times};
(7.363636,1.018481) *[red]{\scriptscriptstyle\times};
(7.381818,0.620770) *[red]{\scriptscriptstyle\times};
(7.400000,0.654304) *[red]{\scriptscriptstyle\times};
(7.418182,0.687290) *[red]{\scriptscriptstyle\times};
(7.436364,0.815665) *[red]{\scriptscriptstyle\times};
(7.454545,0.833721) *[red]{\scriptscriptstyle\times};
(7.472727,0.926241) *[red]{\scriptscriptstyle\times};
(7.490909,0.970601) *[red]{\scriptscriptstyle\times};
(7.509091,0.567703) *[red]{\scriptscriptstyle\times};
(7.527273,0.688006) *[red]{\scriptscriptstyle\times};
(7.545455,0.694204) *[red]{\scriptscriptstyle\times};
(7.563636,0.700683) *[red]{\scriptscriptstyle\times};
(7.581818,0.782252) *[red]{\scriptscriptstyle\times};
(7.600000,0.874033) *[red]{\scriptscriptstyle\times};
(7.618182,0.933958) *[red]{\scriptscriptstyle\times};
(7.636364,0.493512) *[red]{\scriptscriptstyle\times};
(7.654545,0.560680) *[red]{\scriptscriptstyle\times};
(7.672727,0.729532) *[red]{\scriptscriptstyle\times};
(7.690909,0.729762) *[red]{\scriptscriptstyle\times};
(7.709091,0.800749) *[red]{\scriptscriptstyle\times};
(7.727273,0.917291) *[red]{\scriptscriptstyle\times};
(7.745455,0.959493) *[red]{\scriptscriptstyle\times};
(7.763636,0.627258) *[red]{\scriptscriptstyle\times};
(7.781818,0.637711) *[red]{\scriptscriptstyle\times};
(7.800000,0.752922) *[red]{\scriptscriptstyle\times};
(7.818182,0.824136) *[red]{\scriptscriptstyle\times};
(7.836364,0.866937) *[red]{\scriptscriptstyle\times};
(7.854545,0.873741) *[red]{\scriptscriptstyle\times};
(7.872727,0.938755) *[red]{\scriptscriptstyle\times};
(7.890909,0.744421) *[red]{\scriptscriptstyle\times};
(7.909091,0.776036) *[red]{\scriptscriptstyle\times};
(7.927273,0.796053) *[red]{\scriptscriptstyle\times};
(7.945455,0.804688) *[red]{\scriptscriptstyle\times};
(7.963636,0.806244) *[red]{\scriptscriptstyle\times};
(7.981818,0.826661) *[red]{\scriptscriptstyle\times};
(8.000000,0.867981) *[red]{\scriptscriptstyle\times};
\endxy
}
\caption{Skylake cycles
  for the CSIDH-512 action
  using {\tt velusqrt-magma}.
}
\label{fig: action-magma}
\end{figure}

\begin{figure}[t]
\centerline{
\xy <1.1cm,0cm>:<0cm,4cm>::
(0,0.735008); (8,0.735008) **[blue]@{-};
(8.1,0.735008) *[blue]{\rlap{6657626870}};
(0,0.827607); (8,0.827607) **[blue]@{-};
(8.1,0.827607) *[blue]{\rlap{7098954330}};
(0,0.921592); (8,0.921592) **[blue]@{-};
(8.1,0.921592) *[blue]{\rlap{7576818552}};
(0,0.513701); (8,0.513701) **[red]@{-};
(-0.1,0.513701) *[red]{\llap{5710830916}};
(0,0.597595); (8,0.597595) **[red]@{-};
(-0.1,0.597595) *[red]{\llap{6052768760}};
(0,0.686522); (8,0.686522) **[red]@{-};
(-0.1,0.686522) *[red]{\llap{6437597336}};
(-0.009091,0.262019); (-0.009091,1.171076) **[lightgray]@{-};
(0.118182,0.262019); (0.118182,1.171076) **[lightgray]@{-};
(0.245455,0.262019); (0.245455,1.171076) **[lightgray]@{-};
(0.372727,0.262019); (0.372727,1.171076) **[lightgray]@{-};
(0.500000,0.262019); (0.500000,1.171076) **[lightgray]@{-};
(0.627273,0.262019); (0.627273,1.171076) **[lightgray]@{-};
(0.754545,0.262019); (0.754545,1.171076) **[lightgray]@{-};
(0.881818,0.262019); (0.881818,1.171076) **[lightgray]@{-};
(1.009091,0.262019); (1.009091,1.171076) **[lightgray]@{-};
(1.136364,0.262019); (1.136364,1.171076) **[lightgray]@{-};
(1.263636,0.262019); (1.263636,1.171076) **[lightgray]@{-};
(1.390909,0.262019); (1.390909,1.171076) **[lightgray]@{-};
(1.518182,0.262019); (1.518182,1.171076) **[lightgray]@{-};
(1.645455,0.262019); (1.645455,1.171076) **[lightgray]@{-};
(1.772727,0.262019); (1.772727,1.171076) **[lightgray]@{-};
(1.900000,0.262019); (1.900000,1.171076) **[lightgray]@{-};
(2.027273,0.262019); (2.027273,1.171076) **[lightgray]@{-};
(2.154545,0.262019); (2.154545,1.171076) **[lightgray]@{-};
(2.281818,0.262019); (2.281818,1.171076) **[lightgray]@{-};
(2.409091,0.262019); (2.409091,1.171076) **[lightgray]@{-};
(2.536364,0.262019); (2.536364,1.171076) **[lightgray]@{-};
(2.663636,0.262019); (2.663636,1.171076) **[lightgray]@{-};
(2.790909,0.262019); (2.790909,1.171076) **[lightgray]@{-};
(2.918182,0.262019); (2.918182,1.171076) **[lightgray]@{-};
(3.045455,0.262019); (3.045455,1.171076) **[lightgray]@{-};
(3.172727,0.262019); (3.172727,1.171076) **[lightgray]@{-};
(3.300000,0.262019); (3.300000,1.171076) **[lightgray]@{-};
(3.427273,0.262019); (3.427273,1.171076) **[lightgray]@{-};
(3.554545,0.262019); (3.554545,1.171076) **[lightgray]@{-};
(3.681818,0.262019); (3.681818,1.171076) **[lightgray]@{-};
(3.809091,0.262019); (3.809091,1.171076) **[lightgray]@{-};
(3.936364,0.262019); (3.936364,1.171076) **[lightgray]@{-};
(4.063636,0.262019); (4.063636,1.171076) **[lightgray]@{-};
(4.190909,0.262019); (4.190909,1.171076) **[lightgray]@{-};
(4.318182,0.262019); (4.318182,1.171076) **[lightgray]@{-};
(4.445455,0.262019); (4.445455,1.171076) **[lightgray]@{-};
(4.572727,0.262019); (4.572727,1.171076) **[lightgray]@{-};
(4.700000,0.262019); (4.700000,1.171076) **[lightgray]@{-};
(4.827273,0.262019); (4.827273,1.171076) **[lightgray]@{-};
(4.954545,0.262019); (4.954545,1.171076) **[lightgray]@{-};
(5.081818,0.262019); (5.081818,1.171076) **[lightgray]@{-};
(5.209091,0.262019); (5.209091,1.171076) **[lightgray]@{-};
(5.336364,0.262019); (5.336364,1.171076) **[lightgray]@{-};
(5.463636,0.262019); (5.463636,1.171076) **[lightgray]@{-};
(5.590909,0.262019); (5.590909,1.171076) **[lightgray]@{-};
(5.718182,0.262019); (5.718182,1.171076) **[lightgray]@{-};
(5.845455,0.262019); (5.845455,1.171076) **[lightgray]@{-};
(5.972727,0.262019); (5.972727,1.171076) **[lightgray]@{-};
(6.100000,0.262019); (6.100000,1.171076) **[lightgray]@{-};
(6.227273,0.262019); (6.227273,1.171076) **[lightgray]@{-};
(6.354545,0.262019); (6.354545,1.171076) **[lightgray]@{-};
(6.481818,0.262019); (6.481818,1.171076) **[lightgray]@{-};
(6.609091,0.262019); (6.609091,1.171076) **[lightgray]@{-};
(6.736364,0.262019); (6.736364,1.171076) **[lightgray]@{-};
(6.863636,0.262019); (6.863636,1.171076) **[lightgray]@{-};
(6.990909,0.262019); (6.990909,1.171076) **[lightgray]@{-};
(7.118182,0.262019); (7.118182,1.171076) **[lightgray]@{-};
(7.245455,0.262019); (7.245455,1.171076) **[lightgray]@{-};
(7.372727,0.262019); (7.372727,1.171076) **[lightgray]@{-};
(7.500000,0.262019); (7.500000,1.171076) **[lightgray]@{-};
(7.627273,0.262019); (7.627273,1.171076) **[lightgray]@{-};
(7.754545,0.262019); (7.754545,1.171076) **[lightgray]@{-};
(7.881818,0.262019); (7.881818,1.171076) **[lightgray]@{-};
(8.009091,0.262019); (8.009091,1.171076) **[lightgray]@{-};
(-0.009091,0.262019); (-0.009091,1.171076) **[lightgray]@{-};
(0.118182,0.262019); (0.118182,1.171076) **[lightgray]@{-};
(0.245455,0.262019); (0.245455,1.171076) **[lightgray]@{-};
(0.372727,0.262019); (0.372727,1.171076) **[lightgray]@{-};
(0.500000,0.262019); (0.500000,1.171076) **[lightgray]@{-};
(0.627273,0.262019); (0.627273,1.171076) **[lightgray]@{-};
(0.754545,0.262019); (0.754545,1.171076) **[lightgray]@{-};
(0.881818,0.262019); (0.881818,1.171076) **[lightgray]@{-};
(1.009091,0.262019); (1.009091,1.171076) **[lightgray]@{-};
(1.136364,0.262019); (1.136364,1.171076) **[lightgray]@{-};
(1.263636,0.262019); (1.263636,1.171076) **[lightgray]@{-};
(1.390909,0.262019); (1.390909,1.171076) **[lightgray]@{-};
(1.518182,0.262019); (1.518182,1.171076) **[lightgray]@{-};
(1.645455,0.262019); (1.645455,1.171076) **[lightgray]@{-};
(1.772727,0.262019); (1.772727,1.171076) **[lightgray]@{-};
(1.900000,0.262019); (1.900000,1.171076) **[lightgray]@{-};
(2.027273,0.262019); (2.027273,1.171076) **[lightgray]@{-};
(2.154545,0.262019); (2.154545,1.171076) **[lightgray]@{-};
(2.281818,0.262019); (2.281818,1.171076) **[lightgray]@{-};
(2.409091,0.262019); (2.409091,1.171076) **[lightgray]@{-};
(2.536364,0.262019); (2.536364,1.171076) **[lightgray]@{-};
(2.663636,0.262019); (2.663636,1.171076) **[lightgray]@{-};
(2.790909,0.262019); (2.790909,1.171076) **[lightgray]@{-};
(2.918182,0.262019); (2.918182,1.171076) **[lightgray]@{-};
(3.045455,0.262019); (3.045455,1.171076) **[lightgray]@{-};
(3.172727,0.262019); (3.172727,1.171076) **[lightgray]@{-};
(3.300000,0.262019); (3.300000,1.171076) **[lightgray]@{-};
(3.427273,0.262019); (3.427273,1.171076) **[lightgray]@{-};
(3.554545,0.262019); (3.554545,1.171076) **[lightgray]@{-};
(3.681818,0.262019); (3.681818,1.171076) **[lightgray]@{-};
(3.809091,0.262019); (3.809091,1.171076) **[lightgray]@{-};
(3.936364,0.262019); (3.936364,1.171076) **[lightgray]@{-};
(4.063636,0.262019); (4.063636,1.171076) **[lightgray]@{-};
(4.190909,0.262019); (4.190909,1.171076) **[lightgray]@{-};
(4.318182,0.262019); (4.318182,1.171076) **[lightgray]@{-};
(4.445455,0.262019); (4.445455,1.171076) **[lightgray]@{-};
(4.572727,0.262019); (4.572727,1.171076) **[lightgray]@{-};
(4.700000,0.262019); (4.700000,1.171076) **[lightgray]@{-};
(4.827273,0.262019); (4.827273,1.171076) **[lightgray]@{-};
(4.954545,0.262019); (4.954545,1.171076) **[lightgray]@{-};
(5.081818,0.262019); (5.081818,1.171076) **[lightgray]@{-};
(5.209091,0.262019); (5.209091,1.171076) **[lightgray]@{-};
(5.336364,0.262019); (5.336364,1.171076) **[lightgray]@{-};
(5.463636,0.262019); (5.463636,1.171076) **[lightgray]@{-};
(5.590909,0.262019); (5.590909,1.171076) **[lightgray]@{-};
(5.718182,0.262019); (5.718182,1.171076) **[lightgray]@{-};
(5.845455,0.262019); (5.845455,1.171076) **[lightgray]@{-};
(5.972727,0.262019); (5.972727,1.171076) **[lightgray]@{-};
(6.100000,0.262019); (6.100000,1.171076) **[lightgray]@{-};
(6.227273,0.262019); (6.227273,1.171076) **[lightgray]@{-};
(6.354545,0.262019); (6.354545,1.171076) **[lightgray]@{-};
(6.481818,0.262019); (6.481818,1.171076) **[lightgray]@{-};
(6.609091,0.262019); (6.609091,1.171076) **[lightgray]@{-};
(6.736364,0.262019); (6.736364,1.171076) **[lightgray]@{-};
(6.863636,0.262019); (6.863636,1.171076) **[lightgray]@{-};
(6.990909,0.262019); (6.990909,1.171076) **[lightgray]@{-};
(7.118182,0.262019); (7.118182,1.171076) **[lightgray]@{-};
(7.245455,0.262019); (7.245455,1.171076) **[lightgray]@{-};
(7.372727,0.262019); (7.372727,1.171076) **[lightgray]@{-};
(7.500000,0.262019); (7.500000,1.171076) **[lightgray]@{-};
(7.627273,0.262019); (7.627273,1.171076) **[lightgray]@{-};
(7.754545,0.262019); (7.754545,1.171076) **[lightgray]@{-};
(7.881818,0.262019); (7.881818,1.171076) **[lightgray]@{-};
(8.009091,0.262019); (8.009091,1.171076) **[lightgray]@{-};
(0.000000,0.528526) *[blue]{\scriptscriptstyle+};
(0.018182,0.568912) *[blue]{\scriptscriptstyle+};
(0.036364,0.642558) *[blue]{\scriptscriptstyle+};
(0.054545,0.727094) *[blue]{\scriptscriptstyle+};
(0.072727,0.775285) *[blue]{\scriptscriptstyle+};
(0.090909,0.839906) *[blue]{\scriptscriptstyle+};
(0.109091,0.941663) *[blue]{\scriptscriptstyle+};
(0.127273,0.612397) *[blue]{\scriptscriptstyle+};
(0.145455,0.645130) *[blue]{\scriptscriptstyle+};
(0.163636,0.675034) *[blue]{\scriptscriptstyle+};
(0.181818,0.733991) *[blue]{\scriptscriptstyle+};
(0.200000,0.793495) *[blue]{\scriptscriptstyle+};
(0.218182,0.793751) *[blue]{\scriptscriptstyle+};
(0.236364,0.795261) *[blue]{\scriptscriptstyle+};
(0.254545,0.630597) *[blue]{\scriptscriptstyle+};
(0.272727,0.700435) *[blue]{\scriptscriptstyle+};
(0.290909,0.731415) *[blue]{\scriptscriptstyle+};
(0.309091,0.807472) *[blue]{\scriptscriptstyle+};
(0.327273,0.814501) *[blue]{\scriptscriptstyle+};
(0.345455,0.828064) *[blue]{\scriptscriptstyle+};
(0.363636,0.868955) *[blue]{\scriptscriptstyle+};
(0.381818,0.612542) *[blue]{\scriptscriptstyle+};
(0.400000,0.630630) *[blue]{\scriptscriptstyle+};
(0.418182,0.731816) *[blue]{\scriptscriptstyle+};
(0.436364,0.735790) *[blue]{\scriptscriptstyle+};
(0.454545,0.864744) *[blue]{\scriptscriptstyle+};
(0.472727,0.916267) *[blue]{\scriptscriptstyle+};
(0.490909,0.997520) *[blue]{\scriptscriptstyle+};
(0.509091,0.563671) *[blue]{\scriptscriptstyle+};
(0.527273,0.647033) *[blue]{\scriptscriptstyle+};
(0.545455,0.702673) *[blue]{\scriptscriptstyle+};
(0.563636,0.783530) *[blue]{\scriptscriptstyle+};
(0.581818,0.844885) *[blue]{\scriptscriptstyle+};
(0.600000,0.866562) *[blue]{\scriptscriptstyle+};
(0.618182,0.904223) *[blue]{\scriptscriptstyle+};
(0.636364,0.459686) *[blue]{\scriptscriptstyle+};
(0.654545,0.676101) *[blue]{\scriptscriptstyle+};
(0.672727,0.738038) *[blue]{\scriptscriptstyle+};
(0.690909,0.815474) *[blue]{\scriptscriptstyle+};
(0.709091,0.827607) *[blue]{\scriptscriptstyle+};
(0.727273,0.881766) *[blue]{\scriptscriptstyle+};
(0.745455,0.923150) *[blue]{\scriptscriptstyle+};
(0.763636,0.669389) *[blue]{\scriptscriptstyle+};
(0.781818,0.682299) *[blue]{\scriptscriptstyle+};
(0.800000,0.727929) *[blue]{\scriptscriptstyle+};
(0.818182,0.758861) *[blue]{\scriptscriptstyle+};
(0.836364,0.793763) *[blue]{\scriptscriptstyle+};
(0.854545,0.804133) *[blue]{\scriptscriptstyle+};
(0.872727,0.864365) *[blue]{\scriptscriptstyle+};
(0.890909,0.591971) *[blue]{\scriptscriptstyle+};
(0.909091,0.661022) *[blue]{\scriptscriptstyle+};
(0.927273,0.704912) *[blue]{\scriptscriptstyle+};
(0.945455,0.724862) *[blue]{\scriptscriptstyle+};
(0.963636,0.864689) *[blue]{\scriptscriptstyle+};
(0.981818,0.873739) *[blue]{\scriptscriptstyle+};
(1.000000,0.904343) *[blue]{\scriptscriptstyle+};
(1.018182,0.608526) *[blue]{\scriptscriptstyle+};
(1.036364,0.655626) *[blue]{\scriptscriptstyle+};
(1.054545,0.735335) *[blue]{\scriptscriptstyle+};
(1.072727,0.818108) *[blue]{\scriptscriptstyle+};
(1.090909,0.852966) *[blue]{\scriptscriptstyle+};
(1.109091,0.865550) *[blue]{\scriptscriptstyle+};
(1.127273,0.871407) *[blue]{\scriptscriptstyle+};
(1.145455,0.665898) *[blue]{\scriptscriptstyle+};
(1.163636,0.667680) *[blue]{\scriptscriptstyle+};
(1.181818,0.710225) *[blue]{\scriptscriptstyle+};
(1.200000,0.711812) *[blue]{\scriptscriptstyle+};
(1.218182,0.840604) *[blue]{\scriptscriptstyle+};
(1.236364,0.902172) *[blue]{\scriptscriptstyle+};
(1.254545,0.953447) *[blue]{\scriptscriptstyle+};
(1.272727,0.477446) *[blue]{\scriptscriptstyle+};
(1.290909,0.734434) *[blue]{\scriptscriptstyle+};
(1.309091,0.765274) *[blue]{\scriptscriptstyle+};
(1.327273,0.782574) *[blue]{\scriptscriptstyle+};
(1.345455,0.861980) *[blue]{\scriptscriptstyle+};
(1.363636,0.876602) *[blue]{\scriptscriptstyle+};
(1.381818,1.015550) *[blue]{\scriptscriptstyle+};
(1.400000,0.599183) *[blue]{\scriptscriptstyle+};
(1.418182,0.670211) *[blue]{\scriptscriptstyle+};
(1.436364,0.774127) *[blue]{\scriptscriptstyle+};
(1.454545,0.795686) *[blue]{\scriptscriptstyle+};
(1.472727,0.840498) *[blue]{\scriptscriptstyle+};
(1.490909,0.970872) *[blue]{\scriptscriptstyle+};
(1.509091,1.028081) *[blue]{\scriptscriptstyle+};
(1.527273,0.657857) *[blue]{\scriptscriptstyle+};
(1.545455,0.701000) *[blue]{\scriptscriptstyle+};
(1.563636,0.723209) *[blue]{\scriptscriptstyle+};
(1.581818,0.730497) *[blue]{\scriptscriptstyle+};
(1.600000,0.789010) *[blue]{\scriptscriptstyle+};
(1.618182,0.862268) *[blue]{\scriptscriptstyle+};
(1.636364,1.017355) *[blue]{\scriptscriptstyle+};
(1.654545,0.676776) *[blue]{\scriptscriptstyle+};
(1.672727,0.682762) *[blue]{\scriptscriptstyle+};
(1.690909,0.683170) *[blue]{\scriptscriptstyle+};
(1.709091,0.790449) *[blue]{\scriptscriptstyle+};
(1.727273,0.826309) *[blue]{\scriptscriptstyle+};
(1.745455,0.886254) *[blue]{\scriptscriptstyle+};
(1.763636,0.924735) *[blue]{\scriptscriptstyle+};
(1.781818,0.680900) *[blue]{\scriptscriptstyle+};
(1.800000,0.685393) *[blue]{\scriptscriptstyle+};
(1.818182,0.860980) *[blue]{\scriptscriptstyle+};
(1.836364,0.864295) *[blue]{\scriptscriptstyle+};
(1.854545,0.894489) *[blue]{\scriptscriptstyle+};
(1.872727,0.940726) *[blue]{\scriptscriptstyle+};
(1.890909,0.990541) *[blue]{\scriptscriptstyle+};
(1.909091,0.719250) *[blue]{\scriptscriptstyle+};
(1.927273,0.745501) *[blue]{\scriptscriptstyle+};
(1.945455,0.771736) *[blue]{\scriptscriptstyle+};
(1.963636,0.774704) *[blue]{\scriptscriptstyle+};
(1.981818,0.836865) *[blue]{\scriptscriptstyle+};
(2.000000,0.840788) *[blue]{\scriptscriptstyle+};
(2.018182,0.886102) *[blue]{\scriptscriptstyle+};
(2.036364,0.620264) *[blue]{\scriptscriptstyle+};
(2.054545,0.708846) *[blue]{\scriptscriptstyle+};
(2.072727,0.721506) *[blue]{\scriptscriptstyle+};
(2.090909,0.740384) *[blue]{\scriptscriptstyle+};
(2.109091,0.889079) *[blue]{\scriptscriptstyle+};
(2.127273,0.892987) *[blue]{\scriptscriptstyle+};
(2.145455,0.974697) *[blue]{\scriptscriptstyle+};
(2.163636,0.532438) *[blue]{\scriptscriptstyle+};
(2.181818,0.636366) *[blue]{\scriptscriptstyle+};
(2.200000,0.733705) *[blue]{\scriptscriptstyle+};
(2.218182,0.801134) *[blue]{\scriptscriptstyle+};
(2.236364,0.881171) *[blue]{\scriptscriptstyle+};
(2.254545,0.973859) *[blue]{\scriptscriptstyle+};
(2.272727,1.141263) *[blue]{\scriptscriptstyle+};
(2.290909,0.669549) *[blue]{\scriptscriptstyle+};
(2.309091,0.793338) *[blue]{\scriptscriptstyle+};
(2.327273,0.797516) *[blue]{\scriptscriptstyle+};
(2.345455,0.803062) *[blue]{\scriptscriptstyle+};
(2.363636,0.819175) *[blue]{\scriptscriptstyle+};
(2.381818,0.823258) *[blue]{\scriptscriptstyle+};
(2.400000,0.940829) *[blue]{\scriptscriptstyle+};
(2.418182,0.603691) *[blue]{\scriptscriptstyle+};
(2.436364,0.702022) *[blue]{\scriptscriptstyle+};
(2.454545,0.724813) *[blue]{\scriptscriptstyle+};
(2.472727,0.831514) *[blue]{\scriptscriptstyle+};
(2.490909,0.894909) *[blue]{\scriptscriptstyle+};
(2.509091,0.923653) *[blue]{\scriptscriptstyle+};
(2.527273,0.930258) *[blue]{\scriptscriptstyle+};
(2.545455,0.657651) *[blue]{\scriptscriptstyle+};
(2.563636,0.712578) *[blue]{\scriptscriptstyle+};
(2.581818,0.747749) *[blue]{\scriptscriptstyle+};
(2.600000,0.875425) *[blue]{\scriptscriptstyle+};
(2.618182,0.923920) *[blue]{\scriptscriptstyle+};
(2.636364,0.933397) *[blue]{\scriptscriptstyle+};
(2.654545,1.060895) *[blue]{\scriptscriptstyle+};
(2.672727,0.654399) *[blue]{\scriptscriptstyle+};
(2.690909,0.704308) *[blue]{\scriptscriptstyle+};
(2.709091,0.783475) *[blue]{\scriptscriptstyle+};
(2.727273,0.796484) *[blue]{\scriptscriptstyle+};
(2.745455,0.871546) *[blue]{\scriptscriptstyle+};
(2.763636,0.877671) *[blue]{\scriptscriptstyle+};
(2.781818,0.937102) *[blue]{\scriptscriptstyle+};
(2.800000,0.759228) *[blue]{\scriptscriptstyle+};
(2.818182,0.761256) *[blue]{\scriptscriptstyle+};
(2.836364,0.769879) *[blue]{\scriptscriptstyle+};
(2.854545,0.784486) *[blue]{\scriptscriptstyle+};
(2.872727,0.786736) *[blue]{\scriptscriptstyle+};
(2.890909,0.821213) *[blue]{\scriptscriptstyle+};
(2.909091,0.846872) *[blue]{\scriptscriptstyle+};
(2.927273,0.546775) *[blue]{\scriptscriptstyle+};
(2.945455,0.695535) *[blue]{\scriptscriptstyle+};
(2.963636,0.725683) *[blue]{\scriptscriptstyle+};
(2.981818,0.801476) *[blue]{\scriptscriptstyle+};
(3.000000,0.827243) *[blue]{\scriptscriptstyle+};
(3.018182,0.903775) *[blue]{\scriptscriptstyle+};
(3.036364,0.944963) *[blue]{\scriptscriptstyle+};
(3.054545,0.588605) *[blue]{\scriptscriptstyle+};
(3.072727,0.693191) *[blue]{\scriptscriptstyle+};
(3.090909,0.721163) *[blue]{\scriptscriptstyle+};
(3.109091,0.812917) *[blue]{\scriptscriptstyle+};
(3.127273,0.870709) *[blue]{\scriptscriptstyle+};
(3.145455,0.938709) *[blue]{\scriptscriptstyle+};
(3.163636,1.026583) *[blue]{\scriptscriptstyle+};
(3.181818,0.680951) *[blue]{\scriptscriptstyle+};
(3.200000,0.683926) *[blue]{\scriptscriptstyle+};
(3.218182,0.697885) *[blue]{\scriptscriptstyle+};
(3.236364,0.738649) *[blue]{\scriptscriptstyle+};
(3.254545,0.916334) *[blue]{\scriptscriptstyle+};
(3.272727,0.917944) *[blue]{\scriptscriptstyle+};
(3.290909,0.996080) *[blue]{\scriptscriptstyle+};
(3.309091,0.650946) *[blue]{\scriptscriptstyle+};
(3.327273,0.722050) *[blue]{\scriptscriptstyle+};
(3.345455,0.870427) *[blue]{\scriptscriptstyle+};
(3.363636,0.879226) *[blue]{\scriptscriptstyle+};
(3.381818,0.903201) *[blue]{\scriptscriptstyle+};
(3.400000,0.947811) *[blue]{\scriptscriptstyle+};
(3.418182,1.044453) *[blue]{\scriptscriptstyle+};
(3.436364,0.699312) *[blue]{\scriptscriptstyle+};
(3.454545,0.706846) *[blue]{\scriptscriptstyle+};
(3.472727,0.735008) *[blue]{\scriptscriptstyle+};
(3.490909,0.803480) *[blue]{\scriptscriptstyle+};
(3.509091,0.842445) *[blue]{\scriptscriptstyle+};
(3.527273,0.901251) *[blue]{\scriptscriptstyle+};
(3.545455,0.982080) *[blue]{\scriptscriptstyle+};
(3.563636,0.669385) *[blue]{\scriptscriptstyle+};
(3.581818,0.757347) *[blue]{\scriptscriptstyle+};
(3.600000,0.813116) *[blue]{\scriptscriptstyle+};
(3.618182,0.818618) *[blue]{\scriptscriptstyle+};
(3.636364,0.834057) *[blue]{\scriptscriptstyle+};
(3.654545,0.900455) *[blue]{\scriptscriptstyle+};
(3.672727,0.960851) *[blue]{\scriptscriptstyle+};
(3.690909,0.705763) *[blue]{\scriptscriptstyle+};
(3.709091,0.828324) *[blue]{\scriptscriptstyle+};
(3.727273,0.855574) *[blue]{\scriptscriptstyle+};
(3.745455,0.888264) *[blue]{\scriptscriptstyle+};
(3.763636,0.943673) *[blue]{\scriptscriptstyle+};
(3.781818,0.968398) *[blue]{\scriptscriptstyle+};
(3.800000,1.016985) *[blue]{\scriptscriptstyle+};
(3.818182,0.664785) *[blue]{\scriptscriptstyle+};
(3.836364,0.690275) *[blue]{\scriptscriptstyle+};
(3.854545,0.723721) *[blue]{\scriptscriptstyle+};
(3.872727,0.795842) *[blue]{\scriptscriptstyle+};
(3.890909,0.845475) *[blue]{\scriptscriptstyle+};
(3.909091,0.876757) *[blue]{\scriptscriptstyle+};
(3.927273,0.983539) *[blue]{\scriptscriptstyle+};
(3.945455,0.509265) *[blue]{\scriptscriptstyle+};
(3.963636,0.701774) *[blue]{\scriptscriptstyle+};
(3.981818,0.800974) *[blue]{\scriptscriptstyle+};
(4.000000,0.812852) *[blue]{\scriptscriptstyle+};
(4.018182,0.866956) *[blue]{\scriptscriptstyle+};
(4.036364,0.953210) *[blue]{\scriptscriptstyle+};
(4.054545,0.981635) *[blue]{\scriptscriptstyle+};
(4.072727,0.725504) *[blue]{\scriptscriptstyle+};
(4.090909,0.820786) *[blue]{\scriptscriptstyle+};
(4.109091,0.823341) *[blue]{\scriptscriptstyle+};
(4.127273,0.869362) *[blue]{\scriptscriptstyle+};
(4.145455,0.891352) *[blue]{\scriptscriptstyle+};
(4.163636,0.924770) *[blue]{\scriptscriptstyle+};
(4.181818,0.933321) *[blue]{\scriptscriptstyle+};
(4.200000,0.678224) *[blue]{\scriptscriptstyle+};
(4.218182,0.730683) *[blue]{\scriptscriptstyle+};
(4.236364,0.839403) *[blue]{\scriptscriptstyle+};
(4.254545,0.868714) *[blue]{\scriptscriptstyle+};
(4.272727,0.872498) *[blue]{\scriptscriptstyle+};
(4.290909,0.914496) *[blue]{\scriptscriptstyle+};
(4.309091,1.034399) *[blue]{\scriptscriptstyle+};
(4.327273,0.741368) *[blue]{\scriptscriptstyle+};
(4.345455,0.751226) *[blue]{\scriptscriptstyle+};
(4.363636,0.791296) *[blue]{\scriptscriptstyle+};
(4.381818,0.803576) *[blue]{\scriptscriptstyle+};
(4.400000,0.817969) *[blue]{\scriptscriptstyle+};
(4.418182,0.888638) *[blue]{\scriptscriptstyle+};
(4.436364,0.933369) *[blue]{\scriptscriptstyle+};
(4.454545,0.596402) *[blue]{\scriptscriptstyle+};
(4.472727,0.660036) *[blue]{\scriptscriptstyle+};
(4.490909,0.842833) *[blue]{\scriptscriptstyle+};
(4.509091,0.870349) *[blue]{\scriptscriptstyle+};
(4.527273,0.911586) *[blue]{\scriptscriptstyle+};
(4.545455,0.987544) *[blue]{\scriptscriptstyle+};
(4.563636,1.125683) *[blue]{\scriptscriptstyle+};
(4.581818,0.611425) *[blue]{\scriptscriptstyle+};
(4.600000,0.747026) *[blue]{\scriptscriptstyle+};
(4.618182,0.775743) *[blue]{\scriptscriptstyle+};
(4.636364,0.801996) *[blue]{\scriptscriptstyle+};
(4.654545,0.876108) *[blue]{\scriptscriptstyle+};
(4.672727,0.939116) *[blue]{\scriptscriptstyle+};
(4.690909,0.962958) *[blue]{\scriptscriptstyle+};
(4.709091,0.646630) *[blue]{\scriptscriptstyle+};
(4.727273,0.745607) *[blue]{\scriptscriptstyle+};
(4.745455,0.777965) *[blue]{\scriptscriptstyle+};
(4.763636,0.805607) *[blue]{\scriptscriptstyle+};
(4.781818,0.864187) *[blue]{\scriptscriptstyle+};
(4.800000,0.946522) *[blue]{\scriptscriptstyle+};
(4.818182,0.972975) *[blue]{\scriptscriptstyle+};
(4.836364,0.641885) *[blue]{\scriptscriptstyle+};
(4.854545,0.716262) *[blue]{\scriptscriptstyle+};
(4.872727,0.757139) *[blue]{\scriptscriptstyle+};
(4.890909,0.843018) *[blue]{\scriptscriptstyle+};
(4.909091,0.867866) *[blue]{\scriptscriptstyle+};
(4.927273,1.030523) *[blue]{\scriptscriptstyle+};
(4.945455,1.105595) *[blue]{\scriptscriptstyle+};
(4.963636,0.755806) *[blue]{\scriptscriptstyle+};
(4.981818,0.774778) *[blue]{\scriptscriptstyle+};
(5.000000,0.783751) *[blue]{\scriptscriptstyle+};
(5.018182,0.787372) *[blue]{\scriptscriptstyle+};
(5.036364,0.812480) *[blue]{\scriptscriptstyle+};
(5.054545,0.886934) *[blue]{\scriptscriptstyle+};
(5.072727,0.899424) *[blue]{\scriptscriptstyle+};
(5.090909,0.597152) *[blue]{\scriptscriptstyle+};
(5.109091,0.776908) *[blue]{\scriptscriptstyle+};
(5.127273,0.899142) *[blue]{\scriptscriptstyle+};
(5.145455,0.922714) *[blue]{\scriptscriptstyle+};
(5.163636,0.976314) *[blue]{\scriptscriptstyle+};
(5.181818,0.982914) *[blue]{\scriptscriptstyle+};
(5.200000,0.993247) *[blue]{\scriptscriptstyle+};
(5.218182,0.640111) *[blue]{\scriptscriptstyle+};
(5.236364,0.665327) *[blue]{\scriptscriptstyle+};
(5.254545,0.782762) *[blue]{\scriptscriptstyle+};
(5.272727,0.832832) *[blue]{\scriptscriptstyle+};
(5.290909,0.958683) *[blue]{\scriptscriptstyle+};
(5.309091,0.964561) *[blue]{\scriptscriptstyle+};
(5.327273,1.039986) *[blue]{\scriptscriptstyle+};
(5.345455,0.699918) *[blue]{\scriptscriptstyle+};
(5.363636,0.752753) *[blue]{\scriptscriptstyle+};
(5.381818,0.767666) *[blue]{\scriptscriptstyle+};
(5.400000,0.774854) *[blue]{\scriptscriptstyle+};
(5.418182,0.885799) *[blue]{\scriptscriptstyle+};
(5.436364,0.908097) *[blue]{\scriptscriptstyle+};
(5.454545,1.168883) *[blue]{\scriptscriptstyle+};
(5.472727,0.720001) *[blue]{\scriptscriptstyle+};
(5.490909,0.735070) *[blue]{\scriptscriptstyle+};
(5.509091,0.781494) *[blue]{\scriptscriptstyle+};
(5.527273,0.934977) *[blue]{\scriptscriptstyle+};
(5.545455,0.935307) *[blue]{\scriptscriptstyle+};
(5.563636,0.964820) *[blue]{\scriptscriptstyle+};
(5.581818,0.977842) *[blue]{\scriptscriptstyle+};
(5.600000,0.654579) *[blue]{\scriptscriptstyle+};
(5.618182,0.716940) *[blue]{\scriptscriptstyle+};
(5.636364,0.741157) *[blue]{\scriptscriptstyle+};
(5.654545,0.778594) *[blue]{\scriptscriptstyle+};
(5.672727,0.797985) *[blue]{\scriptscriptstyle+};
(5.690909,1.114294) *[blue]{\scriptscriptstyle+};
(5.709091,1.137622) *[blue]{\scriptscriptstyle+};
(5.727273,0.636951) *[blue]{\scriptscriptstyle+};
(5.745455,0.832487) *[blue]{\scriptscriptstyle+};
(5.763636,0.890195) *[blue]{\scriptscriptstyle+};
(5.781818,0.896106) *[blue]{\scriptscriptstyle+};
(5.800000,0.918842) *[blue]{\scriptscriptstyle+};
(5.818182,0.985341) *[blue]{\scriptscriptstyle+};
(5.836364,0.992638) *[blue]{\scriptscriptstyle+};
(5.854545,0.717674) *[blue]{\scriptscriptstyle+};
(5.872727,0.762470) *[blue]{\scriptscriptstyle+};
(5.890909,0.782829) *[blue]{\scriptscriptstyle+};
(5.909091,0.801982) *[blue]{\scriptscriptstyle+};
(5.927273,0.988923) *[blue]{\scriptscriptstyle+};
(5.945455,0.993961) *[blue]{\scriptscriptstyle+};
(5.963636,1.023796) *[blue]{\scriptscriptstyle+};
(5.981818,0.746909) *[blue]{\scriptscriptstyle+};
(6.000000,0.768831) *[blue]{\scriptscriptstyle+};
(6.018182,0.785365) *[blue]{\scriptscriptstyle+};
(6.036364,0.869904) *[blue]{\scriptscriptstyle+};
(6.054545,0.907845) *[blue]{\scriptscriptstyle+};
(6.072727,0.912407) *[blue]{\scriptscriptstyle+};
(6.090909,1.015229) *[blue]{\scriptscriptstyle+};
(6.109091,0.625654) *[blue]{\scriptscriptstyle+};
(6.127273,0.732276) *[blue]{\scriptscriptstyle+};
(6.145455,0.816888) *[blue]{\scriptscriptstyle+};
(6.163636,0.843987) *[blue]{\scriptscriptstyle+};
(6.181818,0.925488) *[blue]{\scriptscriptstyle+};
(6.200000,0.925900) *[blue]{\scriptscriptstyle+};
(6.218182,1.035655) *[blue]{\scriptscriptstyle+};
(6.236364,0.594905) *[blue]{\scriptscriptstyle+};
(6.254545,0.680686) *[blue]{\scriptscriptstyle+};
(6.272727,0.790289) *[blue]{\scriptscriptstyle+};
(6.290909,0.916372) *[blue]{\scriptscriptstyle+};
(6.309091,0.922774) *[blue]{\scriptscriptstyle+};
(6.327273,0.985370) *[blue]{\scriptscriptstyle+};
(6.345455,1.109230) *[blue]{\scriptscriptstyle+};
(6.363636,0.732596) *[blue]{\scriptscriptstyle+};
(6.381818,0.755140) *[blue]{\scriptscriptstyle+};
(6.400000,0.800321) *[blue]{\scriptscriptstyle+};
(6.418182,0.824876) *[blue]{\scriptscriptstyle+};
(6.436364,0.854509) *[blue]{\scriptscriptstyle+};
(6.454545,0.911275) *[blue]{\scriptscriptstyle+};
(6.472727,1.058862) *[blue]{\scriptscriptstyle+};
(6.490909,0.748564) *[blue]{\scriptscriptstyle+};
(6.509091,0.754284) *[blue]{\scriptscriptstyle+};
(6.527273,0.769076) *[blue]{\scriptscriptstyle+};
(6.545455,0.830407) *[blue]{\scriptscriptstyle+};
(6.563636,0.855373) *[blue]{\scriptscriptstyle+};
(6.581818,0.895556) *[blue]{\scriptscriptstyle+};
(6.600000,1.068629) *[blue]{\scriptscriptstyle+};
(6.618182,0.743299) *[blue]{\scriptscriptstyle+};
(6.636364,0.748109) *[blue]{\scriptscriptstyle+};
(6.654545,0.806125) *[blue]{\scriptscriptstyle+};
(6.672727,0.871878) *[blue]{\scriptscriptstyle+};
(6.690909,0.911185) *[blue]{\scriptscriptstyle+};
(6.709091,0.932934) *[blue]{\scriptscriptstyle+};
(6.727273,1.107308) *[blue]{\scriptscriptstyle+};
(6.745455,0.710625) *[blue]{\scriptscriptstyle+};
(6.763636,0.793589) *[blue]{\scriptscriptstyle+};
(6.781818,0.839460) *[blue]{\scriptscriptstyle+};
(6.800000,0.892396) *[blue]{\scriptscriptstyle+};
(6.818182,0.946767) *[blue]{\scriptscriptstyle+};
(6.836364,0.953027) *[blue]{\scriptscriptstyle+};
(6.854545,0.964044) *[blue]{\scriptscriptstyle+};
(6.872727,0.706461) *[blue]{\scriptscriptstyle+};
(6.890909,0.827598) *[blue]{\scriptscriptstyle+};
(6.909091,0.840714) *[blue]{\scriptscriptstyle+};
(6.927273,0.921592) *[blue]{\scriptscriptstyle+};
(6.945455,1.049865) *[blue]{\scriptscriptstyle+};
(6.963636,1.058341) *[blue]{\scriptscriptstyle+};
(6.981818,1.062953) *[blue]{\scriptscriptstyle+};
(7.000000,0.616739) *[blue]{\scriptscriptstyle+};
(7.018182,0.747174) *[blue]{\scriptscriptstyle+};
(7.036364,0.880235) *[blue]{\scriptscriptstyle+};
(7.054545,0.945520) *[blue]{\scriptscriptstyle+};
(7.072727,0.952528) *[blue]{\scriptscriptstyle+};
(7.090909,0.953900) *[blue]{\scriptscriptstyle+};
(7.109091,0.991155) *[blue]{\scriptscriptstyle+};
(7.127273,0.590917) *[blue]{\scriptscriptstyle+};
(7.145455,0.768364) *[blue]{\scriptscriptstyle+};
(7.163636,0.809445) *[blue]{\scriptscriptstyle+};
(7.181818,0.833454) *[blue]{\scriptscriptstyle+};
(7.200000,0.852242) *[blue]{\scriptscriptstyle+};
(7.218182,0.946128) *[blue]{\scriptscriptstyle+};
(7.236364,1.120758) *[blue]{\scriptscriptstyle+};
(7.254545,0.759592) *[blue]{\scriptscriptstyle+};
(7.272727,0.812664) *[blue]{\scriptscriptstyle+};
(7.290909,0.821430) *[blue]{\scriptscriptstyle+};
(7.309091,0.837135) *[blue]{\scriptscriptstyle+};
(7.327273,0.975521) *[blue]{\scriptscriptstyle+};
(7.345455,0.979778) *[blue]{\scriptscriptstyle+};
(7.363636,1.079353) *[blue]{\scriptscriptstyle+};
(7.381818,0.694808) *[blue]{\scriptscriptstyle+};
(7.400000,0.777842) *[blue]{\scriptscriptstyle+};
(7.418182,0.849653) *[blue]{\scriptscriptstyle+};
(7.436364,0.866227) *[blue]{\scriptscriptstyle+};
(7.454545,0.868854) *[blue]{\scriptscriptstyle+};
(7.472727,0.954791) *[blue]{\scriptscriptstyle+};
(7.490909,1.123510) *[blue]{\scriptscriptstyle+};
(7.509091,0.622496) *[blue]{\scriptscriptstyle+};
(7.527273,0.801039) *[blue]{\scriptscriptstyle+};
(7.545455,0.844021) *[blue]{\scriptscriptstyle+};
(7.563636,0.858526) *[blue]{\scriptscriptstyle+};
(7.581818,0.988876) *[blue]{\scriptscriptstyle+};
(7.600000,1.032778) *[blue]{\scriptscriptstyle+};
(7.618182,1.046517) *[blue]{\scriptscriptstyle+};
(7.636364,0.774050) *[blue]{\scriptscriptstyle+};
(7.654545,0.836423) *[blue]{\scriptscriptstyle+};
(7.672727,0.857785) *[blue]{\scriptscriptstyle+};
(7.690909,0.926321) *[blue]{\scriptscriptstyle+};
(7.709091,0.946251) *[blue]{\scriptscriptstyle+};
(7.727273,1.009039) *[blue]{\scriptscriptstyle+};
(7.745455,1.011007) *[blue]{\scriptscriptstyle+};
(7.763636,0.750246) *[blue]{\scriptscriptstyle+};
(7.781818,0.839264) *[blue]{\scriptscriptstyle+};
(7.800000,0.860113) *[blue]{\scriptscriptstyle+};
(7.818182,0.916689) *[blue]{\scriptscriptstyle+};
(7.836364,1.009864) *[blue]{\scriptscriptstyle+};
(7.854545,1.043441) *[blue]{\scriptscriptstyle+};
(7.872727,1.171076) *[blue]{\scriptscriptstyle+};
(7.890909,0.851063) *[blue]{\scriptscriptstyle+};
(7.909091,0.924249) *[blue]{\scriptscriptstyle+};
(7.927273,0.959173) *[blue]{\scriptscriptstyle+};
(7.945455,0.981139) *[blue]{\scriptscriptstyle+};
(7.963636,1.024707) *[blue]{\scriptscriptstyle+};
(7.981818,1.036392) *[blue]{\scriptscriptstyle+};
(8.000000,1.056462) *[blue]{\scriptscriptstyle+};
(0.000000,0.329874) *[red]{\scriptscriptstyle\times};
(0.018182,0.331646) *[red]{\scriptscriptstyle\times};
(0.036364,0.416159) *[red]{\scriptscriptstyle\times};
(0.054545,0.443899) *[red]{\scriptscriptstyle\times};
(0.072727,0.490507) *[red]{\scriptscriptstyle\times};
(0.090909,0.670640) *[red]{\scriptscriptstyle\times};
(0.109091,0.691180) *[red]{\scriptscriptstyle\times};
(0.127273,0.374686) *[red]{\scriptscriptstyle\times};
(0.145455,0.397059) *[red]{\scriptscriptstyle\times};
(0.163636,0.540818) *[red]{\scriptscriptstyle\times};
(0.181818,0.560383) *[red]{\scriptscriptstyle\times};
(0.200000,0.562062) *[red]{\scriptscriptstyle\times};
(0.218182,0.617063) *[red]{\scriptscriptstyle\times};
(0.236364,0.640799) *[red]{\scriptscriptstyle\times};
(0.254545,0.465234) *[red]{\scriptscriptstyle\times};
(0.272727,0.473152) *[red]{\scriptscriptstyle\times};
(0.290909,0.507510) *[red]{\scriptscriptstyle\times};
(0.309091,0.539406) *[red]{\scriptscriptstyle\times};
(0.327273,0.595060) *[red]{\scriptscriptstyle\times};
(0.345455,0.615414) *[red]{\scriptscriptstyle\times};
(0.363636,0.629117) *[red]{\scriptscriptstyle\times};
(0.381818,0.365409) *[red]{\scriptscriptstyle\times};
(0.400000,0.477505) *[red]{\scriptscriptstyle\times};
(0.418182,0.515695) *[red]{\scriptscriptstyle\times};
(0.436364,0.568626) *[red]{\scriptscriptstyle\times};
(0.454545,0.611503) *[red]{\scriptscriptstyle\times};
(0.472727,0.654534) *[red]{\scriptscriptstyle\times};
(0.490909,0.755742) *[red]{\scriptscriptstyle\times};
(0.509091,0.342196) *[red]{\scriptscriptstyle\times};
(0.527273,0.447596) *[red]{\scriptscriptstyle\times};
(0.545455,0.467504) *[red]{\scriptscriptstyle\times};
(0.563636,0.563590) *[red]{\scriptscriptstyle\times};
(0.581818,0.587439) *[red]{\scriptscriptstyle\times};
(0.600000,0.642300) *[red]{\scriptscriptstyle\times};
(0.618182,0.687576) *[red]{\scriptscriptstyle\times};
(0.636364,0.262019) *[red]{\scriptscriptstyle\times};
(0.654545,0.450283) *[red]{\scriptscriptstyle\times};
(0.672727,0.550787) *[red]{\scriptscriptstyle\times};
(0.690909,0.588289) *[red]{\scriptscriptstyle\times};
(0.709091,0.598029) *[red]{\scriptscriptstyle\times};
(0.727273,0.634956) *[red]{\scriptscriptstyle\times};
(0.745455,0.654952) *[red]{\scriptscriptstyle\times};
(0.763636,0.443857) *[red]{\scriptscriptstyle\times};
(0.781818,0.498763) *[red]{\scriptscriptstyle\times};
(0.800000,0.523214) *[red]{\scriptscriptstyle\times};
(0.818182,0.575943) *[red]{\scriptscriptstyle\times};
(0.836364,0.586783) *[red]{\scriptscriptstyle\times};
(0.854545,0.594548) *[red]{\scriptscriptstyle\times};
(0.872727,0.621928) *[red]{\scriptscriptstyle\times};
(0.890909,0.371974) *[red]{\scriptscriptstyle\times};
(0.909091,0.412026) *[red]{\scriptscriptstyle\times};
(0.927273,0.413725) *[red]{\scriptscriptstyle\times};
(0.945455,0.565882) *[red]{\scriptscriptstyle\times};
(0.963636,0.592250) *[red]{\scriptscriptstyle\times};
(0.981818,0.657603) *[red]{\scriptscriptstyle\times};
(1.000000,0.697373) *[red]{\scriptscriptstyle\times};
(1.018182,0.286037) *[red]{\scriptscriptstyle\times};
(1.036364,0.389732) *[red]{\scriptscriptstyle\times};
(1.054545,0.501847) *[red]{\scriptscriptstyle\times};
(1.072727,0.547551) *[red]{\scriptscriptstyle\times};
(1.090909,0.645682) *[red]{\scriptscriptstyle\times};
(1.109091,0.671093) *[red]{\scriptscriptstyle\times};
(1.127273,0.677641) *[red]{\scriptscriptstyle\times};
(1.145455,0.464606) *[red]{\scriptscriptstyle\times};
(1.163636,0.484152) *[red]{\scriptscriptstyle\times};
(1.181818,0.533277) *[red]{\scriptscriptstyle\times};
(1.200000,0.569939) *[red]{\scriptscriptstyle\times};
(1.218182,0.573595) *[red]{\scriptscriptstyle\times};
(1.236364,0.627521) *[red]{\scriptscriptstyle\times};
(1.254545,0.664261) *[red]{\scriptscriptstyle\times};
(1.272727,0.273472) *[red]{\scriptscriptstyle\times};
(1.290909,0.369891) *[red]{\scriptscriptstyle\times};
(1.309091,0.454455) *[red]{\scriptscriptstyle\times};
(1.327273,0.521054) *[red]{\scriptscriptstyle\times};
(1.345455,0.575413) *[red]{\scriptscriptstyle\times};
(1.363636,0.608557) *[red]{\scriptscriptstyle\times};
(1.381818,0.754168) *[red]{\scriptscriptstyle\times};
(1.400000,0.388910) *[red]{\scriptscriptstyle\times};
(1.418182,0.482605) *[red]{\scriptscriptstyle\times};
(1.436364,0.492555) *[red]{\scriptscriptstyle\times};
(1.454545,0.584051) *[red]{\scriptscriptstyle\times};
(1.472727,0.632208) *[red]{\scriptscriptstyle\times};
(1.490909,0.688783) *[red]{\scriptscriptstyle\times};
(1.509091,0.781263) *[red]{\scriptscriptstyle\times};
(1.527273,0.438103) *[red]{\scriptscriptstyle\times};
(1.545455,0.496575) *[red]{\scriptscriptstyle\times};
(1.563636,0.523368) *[red]{\scriptscriptstyle\times};
(1.581818,0.529769) *[red]{\scriptscriptstyle\times};
(1.600000,0.545010) *[red]{\scriptscriptstyle\times};
(1.618182,0.568564) *[red]{\scriptscriptstyle\times};
(1.636364,0.754616) *[red]{\scriptscriptstyle\times};
(1.654545,0.460277) *[red]{\scriptscriptstyle\times};
(1.672727,0.490836) *[red]{\scriptscriptstyle\times};
(1.690909,0.581486) *[red]{\scriptscriptstyle\times};
(1.709091,0.592635) *[red]{\scriptscriptstyle\times};
(1.727273,0.607797) *[red]{\scriptscriptstyle\times};
(1.745455,0.654499) *[red]{\scriptscriptstyle\times};
(1.763636,0.705096) *[red]{\scriptscriptstyle\times};
(1.781818,0.441278) *[red]{\scriptscriptstyle\times};
(1.800000,0.500906) *[red]{\scriptscriptstyle\times};
(1.818182,0.629924) *[red]{\scriptscriptstyle\times};
(1.836364,0.645929) *[red]{\scriptscriptstyle\times};
(1.854545,0.668726) *[red]{\scriptscriptstyle\times};
(1.872727,0.675761) *[red]{\scriptscriptstyle\times};
(1.890909,0.841942) *[red]{\scriptscriptstyle\times};
(1.909091,0.446915) *[red]{\scriptscriptstyle\times};
(1.927273,0.530594) *[red]{\scriptscriptstyle\times};
(1.945455,0.570510) *[red]{\scriptscriptstyle\times};
(1.963636,0.588162) *[red]{\scriptscriptstyle\times};
(1.981818,0.613643) *[red]{\scriptscriptstyle\times};
(2.000000,0.634760) *[red]{\scriptscriptstyle\times};
(2.018182,0.649738) *[red]{\scriptscriptstyle\times};
(2.036364,0.391942) *[red]{\scriptscriptstyle\times};
(2.054545,0.487113) *[red]{\scriptscriptstyle\times};
(2.072727,0.493016) *[red]{\scriptscriptstyle\times};
(2.090909,0.499365) *[red]{\scriptscriptstyle\times};
(2.109091,0.631236) *[red]{\scriptscriptstyle\times};
(2.127273,0.662392) *[red]{\scriptscriptstyle\times};
(2.145455,0.739067) *[red]{\scriptscriptstyle\times};
(2.163636,0.293965) *[red]{\scriptscriptstyle\times};
(2.181818,0.334258) *[red]{\scriptscriptstyle\times};
(2.200000,0.447239) *[red]{\scriptscriptstyle\times};
(2.218182,0.595456) *[red]{\scriptscriptstyle\times};
(2.236364,0.614791) *[red]{\scriptscriptstyle\times};
(2.254545,0.807213) *[red]{\scriptscriptstyle\times};
(2.272727,0.893617) *[red]{\scriptscriptstyle\times};
(2.290909,0.403348) *[red]{\scriptscriptstyle\times};
(2.309091,0.532116) *[red]{\scriptscriptstyle\times};
(2.327273,0.535533) *[red]{\scriptscriptstyle\times};
(2.345455,0.554367) *[red]{\scriptscriptstyle\times};
(2.363636,0.577141) *[red]{\scriptscriptstyle\times};
(2.381818,0.677106) *[red]{\scriptscriptstyle\times};
(2.400000,0.699773) *[red]{\scriptscriptstyle\times};
(2.418182,0.347816) *[red]{\scriptscriptstyle\times};
(2.436364,0.502679) *[red]{\scriptscriptstyle\times};
(2.454545,0.570429) *[red]{\scriptscriptstyle\times};
(2.472727,0.582043) *[red]{\scriptscriptstyle\times};
(2.490909,0.597595) *[red]{\scriptscriptstyle\times};
(2.509091,0.681999) *[red]{\scriptscriptstyle\times};
(2.527273,0.698201) *[red]{\scriptscriptstyle\times};
(2.545455,0.368712) *[red]{\scriptscriptstyle\times};
(2.563636,0.491941) *[red]{\scriptscriptstyle\times};
(2.581818,0.555627) *[red]{\scriptscriptstyle\times};
(2.600000,0.613569) *[red]{\scriptscriptstyle\times};
(2.618182,0.668324) *[red]{\scriptscriptstyle\times};
(2.636364,0.671928) *[red]{\scriptscriptstyle\times};
(2.654545,0.770578) *[red]{\scriptscriptstyle\times};
(2.672727,0.415626) *[red]{\scriptscriptstyle\times};
(2.690909,0.465212) *[red]{\scriptscriptstyle\times};
(2.709091,0.482524) *[red]{\scriptscriptstyle\times};
(2.727273,0.598072) *[red]{\scriptscriptstyle\times};
(2.745455,0.667385) *[red]{\scriptscriptstyle\times};
(2.763636,0.685008) *[red]{\scriptscriptstyle\times};
(2.781818,0.737271) *[red]{\scriptscriptstyle\times};
(2.800000,0.515092) *[red]{\scriptscriptstyle\times};
(2.818182,0.525739) *[red]{\scriptscriptstyle\times};
(2.836364,0.565641) *[red]{\scriptscriptstyle\times};
(2.854545,0.609548) *[red]{\scriptscriptstyle\times};
(2.872727,0.612944) *[red]{\scriptscriptstyle\times};
(2.890909,0.613462) *[red]{\scriptscriptstyle\times};
(2.909091,0.631838) *[red]{\scriptscriptstyle\times};
(2.927273,0.330746) *[red]{\scriptscriptstyle\times};
(2.945455,0.469102) *[red]{\scriptscriptstyle\times};
(2.963636,0.493796) *[red]{\scriptscriptstyle\times};
(2.981818,0.566611) *[red]{\scriptscriptstyle\times};
(3.000000,0.626184) *[red]{\scriptscriptstyle\times};
(3.018182,0.699520) *[red]{\scriptscriptstyle\times};
(3.036364,0.761162) *[red]{\scriptscriptstyle\times};
(3.054545,0.341330) *[red]{\scriptscriptstyle\times};
(3.072727,0.455542) *[red]{\scriptscriptstyle\times};
(3.090909,0.478468) *[red]{\scriptscriptstyle\times};
(3.109091,0.586025) *[red]{\scriptscriptstyle\times};
(3.127273,0.696202) *[red]{\scriptscriptstyle\times};
(3.145455,0.699554) *[red]{\scriptscriptstyle\times};
(3.163636,0.782514) *[red]{\scriptscriptstyle\times};
(3.181818,0.429363) *[red]{\scriptscriptstyle\times};
(3.200000,0.482992) *[red]{\scriptscriptstyle\times};
(3.218182,0.510600) *[red]{\scriptscriptstyle\times};
(3.236364,0.511100) *[red]{\scriptscriptstyle\times};
(3.254545,0.706060) *[red]{\scriptscriptstyle\times};
(3.272727,0.717044) *[red]{\scriptscriptstyle\times};
(3.290909,0.847080) *[red]{\scriptscriptstyle\times};
(3.309091,0.502971) *[red]{\scriptscriptstyle\times};
(3.327273,0.587080) *[red]{\scriptscriptstyle\times};
(3.345455,0.648921) *[red]{\scriptscriptstyle\times};
(3.363636,0.657796) *[red]{\scriptscriptstyle\times};
(3.381818,0.657848) *[red]{\scriptscriptstyle\times};
(3.400000,0.683592) *[red]{\scriptscriptstyle\times};
(3.418182,0.727657) *[red]{\scriptscriptstyle\times};
(3.436364,0.460995) *[red]{\scriptscriptstyle\times};
(3.454545,0.507506) *[red]{\scriptscriptstyle\times};
(3.472727,0.550207) *[red]{\scriptscriptstyle\times};
(3.490909,0.554541) *[red]{\scriptscriptstyle\times};
(3.509091,0.612280) *[red]{\scriptscriptstyle\times};
(3.527273,0.685324) *[red]{\scriptscriptstyle\times};
(3.545455,0.794740) *[red]{\scriptscriptstyle\times};
(3.563636,0.426810) *[red]{\scriptscriptstyle\times};
(3.581818,0.494733) *[red]{\scriptscriptstyle\times};
(3.600000,0.502331) *[red]{\scriptscriptstyle\times};
(3.618182,0.557326) *[red]{\scriptscriptstyle\times};
(3.636364,0.616086) *[red]{\scriptscriptstyle\times};
(3.654545,0.651054) *[red]{\scriptscriptstyle\times};
(3.672727,0.739161) *[red]{\scriptscriptstyle\times};
(3.690909,0.516533) *[red]{\scriptscriptstyle\times};
(3.709091,0.580468) *[red]{\scriptscriptstyle\times};
(3.727273,0.602374) *[red]{\scriptscriptstyle\times};
(3.745455,0.621521) *[red]{\scriptscriptstyle\times};
(3.763636,0.692163) *[red]{\scriptscriptstyle\times};
(3.781818,0.692208) *[red]{\scriptscriptstyle\times};
(3.800000,0.704665) *[red]{\scriptscriptstyle\times};
(3.818182,0.520540) *[red]{\scriptscriptstyle\times};
(3.836364,0.522481) *[red]{\scriptscriptstyle\times};
(3.854545,0.554272) *[red]{\scriptscriptstyle\times};
(3.872727,0.560711) *[red]{\scriptscriptstyle\times};
(3.890909,0.598018) *[red]{\scriptscriptstyle\times};
(3.909091,0.720413) *[red]{\scriptscriptstyle\times};
(3.927273,0.722084) *[red]{\scriptscriptstyle\times};
(3.945455,0.343966) *[red]{\scriptscriptstyle\times};
(3.963636,0.501977) *[red]{\scriptscriptstyle\times};
(3.981818,0.553118) *[red]{\scriptscriptstyle\times};
(4.000000,0.563271) *[red]{\scriptscriptstyle\times};
(4.018182,0.607559) *[red]{\scriptscriptstyle\times};
(4.036364,0.713209) *[red]{\scriptscriptstyle\times};
(4.054545,0.731257) *[red]{\scriptscriptstyle\times};
(4.072727,0.513701) *[red]{\scriptscriptstyle\times};
(4.090909,0.515480) *[red]{\scriptscriptstyle\times};
(4.109091,0.592431) *[red]{\scriptscriptstyle\times};
(4.127273,0.610028) *[red]{\scriptscriptstyle\times};
(4.145455,0.631237) *[red]{\scriptscriptstyle\times};
(4.163636,0.648289) *[red]{\scriptscriptstyle\times};
(4.181818,0.693460) *[red]{\scriptscriptstyle\times};
(4.200000,0.375511) *[red]{\scriptscriptstyle\times};
(4.218182,0.521707) *[red]{\scriptscriptstyle\times};
(4.236364,0.561905) *[red]{\scriptscriptstyle\times};
(4.254545,0.575611) *[red]{\scriptscriptstyle\times};
(4.272727,0.649362) *[red]{\scriptscriptstyle\times};
(4.290909,0.688947) *[red]{\scriptscriptstyle\times};
(4.309091,0.748852) *[red]{\scriptscriptstyle\times};
(4.327273,0.465245) *[red]{\scriptscriptstyle\times};
(4.345455,0.542476) *[red]{\scriptscriptstyle\times};
(4.363636,0.546206) *[red]{\scriptscriptstyle\times};
(4.381818,0.551345) *[red]{\scriptscriptstyle\times};
(4.400000,0.557931) *[red]{\scriptscriptstyle\times};
(4.418182,0.652425) *[red]{\scriptscriptstyle\times};
(4.436364,0.683647) *[red]{\scriptscriptstyle\times};
(4.454545,0.380767) *[red]{\scriptscriptstyle\times};
(4.472727,0.410252) *[red]{\scriptscriptstyle\times};
(4.490909,0.601266) *[red]{\scriptscriptstyle\times};
(4.509091,0.619420) *[red]{\scriptscriptstyle\times};
(4.527273,0.712748) *[red]{\scriptscriptstyle\times};
(4.545455,0.723071) *[red]{\scriptscriptstyle\times};
(4.563636,0.881801) *[red]{\scriptscriptstyle\times};
(4.581818,0.372968) *[red]{\scriptscriptstyle\times};
(4.600000,0.564422) *[red]{\scriptscriptstyle\times};
(4.618182,0.569785) *[red]{\scriptscriptstyle\times};
(4.636364,0.613528) *[red]{\scriptscriptstyle\times};
(4.654545,0.685860) *[red]{\scriptscriptstyle\times};
(4.672727,0.702921) *[red]{\scriptscriptstyle\times};
(4.690909,0.732830) *[red]{\scriptscriptstyle\times};
(4.709091,0.447747) *[red]{\scriptscriptstyle\times};
(4.727273,0.529311) *[red]{\scriptscriptstyle\times};
(4.745455,0.535756) *[red]{\scriptscriptstyle\times};
(4.763636,0.561323) *[red]{\scriptscriptstyle\times};
(4.781818,0.698757) *[red]{\scriptscriptstyle\times};
(4.800000,0.717407) *[red]{\scriptscriptstyle\times};
(4.818182,0.720438) *[red]{\scriptscriptstyle\times};
(4.836364,0.350192) *[red]{\scriptscriptstyle\times};
(4.854545,0.462475) *[red]{\scriptscriptstyle\times};
(4.872727,0.544344) *[red]{\scriptscriptstyle\times};
(4.890909,0.546129) *[red]{\scriptscriptstyle\times};
(4.909091,0.668517) *[red]{\scriptscriptstyle\times};
(4.927273,0.722024) *[red]{\scriptscriptstyle\times};
(4.945455,0.915922) *[red]{\scriptscriptstyle\times};
(4.963636,0.496339) *[red]{\scriptscriptstyle\times};
(4.981818,0.507006) *[red]{\scriptscriptstyle\times};
(5.000000,0.559872) *[red]{\scriptscriptstyle\times};
(5.018182,0.565400) *[red]{\scriptscriptstyle\times};
(5.036364,0.607392) *[red]{\scriptscriptstyle\times};
(5.054545,0.643024) *[red]{\scriptscriptstyle\times};
(5.072727,0.683894) *[red]{\scriptscriptstyle\times};
(5.090909,0.387333) *[red]{\scriptscriptstyle\times};
(5.109091,0.538855) *[red]{\scriptscriptstyle\times};
(5.127273,0.658486) *[red]{\scriptscriptstyle\times};
(5.145455,0.661707) *[red]{\scriptscriptstyle\times};
(5.163636,0.715217) *[red]{\scriptscriptstyle\times};
(5.181818,0.723538) *[red]{\scriptscriptstyle\times};
(5.200000,0.723770) *[red]{\scriptscriptstyle\times};
(5.218182,0.340993) *[red]{\scriptscriptstyle\times};
(5.236364,0.447551) *[red]{\scriptscriptstyle\times};
(5.254545,0.567161) *[red]{\scriptscriptstyle\times};
(5.272727,0.574762) *[red]{\scriptscriptstyle\times};
(5.290909,0.721434) *[red]{\scriptscriptstyle\times};
(5.309091,0.730637) *[red]{\scriptscriptstyle\times};
(5.327273,0.769801) *[red]{\scriptscriptstyle\times};
(5.345455,0.471092) *[red]{\scriptscriptstyle\times};
(5.363636,0.536420) *[red]{\scriptscriptstyle\times};
(5.381818,0.554637) *[red]{\scriptscriptstyle\times};
(5.400000,0.632026) *[red]{\scriptscriptstyle\times};
(5.418182,0.634754) *[red]{\scriptscriptstyle\times};
(5.436364,0.668184) *[red]{\scriptscriptstyle\times};
(5.454545,0.920751) *[red]{\scriptscriptstyle\times};
(5.472727,0.514823) *[red]{\scriptscriptstyle\times};
(5.490909,0.556238) *[red]{\scriptscriptstyle\times};
(5.509091,0.598957) *[red]{\scriptscriptstyle\times};
(5.527273,0.690550) *[red]{\scriptscriptstyle\times};
(5.545455,0.704297) *[red]{\scriptscriptstyle\times};
(5.563636,0.712925) *[red]{\scriptscriptstyle\times};
(5.581818,0.753651) *[red]{\scriptscriptstyle\times};
(5.600000,0.385665) *[red]{\scriptscriptstyle\times};
(5.618182,0.467072) *[red]{\scriptscriptstyle\times};
(5.636364,0.475059) *[red]{\scriptscriptstyle\times};
(5.654545,0.491778) *[red]{\scriptscriptstyle\times};
(5.672727,0.609706) *[red]{\scriptscriptstyle\times};
(5.690909,0.760136) *[red]{\scriptscriptstyle\times};
(5.709091,0.846467) *[red]{\scriptscriptstyle\times};
(5.727273,0.371999) *[red]{\scriptscriptstyle\times};
(5.745455,0.558616) *[red]{\scriptscriptstyle\times};
(5.763636,0.576688) *[red]{\scriptscriptstyle\times};
(5.781818,0.667184) *[red]{\scriptscriptstyle\times};
(5.800000,0.682921) *[red]{\scriptscriptstyle\times};
(5.818182,0.711679) *[red]{\scriptscriptstyle\times};
(5.836364,0.746353) *[red]{\scriptscriptstyle\times};
(5.854545,0.439992) *[red]{\scriptscriptstyle\times};
(5.872727,0.501598) *[red]{\scriptscriptstyle\times};
(5.890909,0.533579) *[red]{\scriptscriptstyle\times};
(5.909091,0.643088) *[red]{\scriptscriptstyle\times};
(5.927273,0.712743) *[red]{\scriptscriptstyle\times};
(5.945455,0.715256) *[red]{\scriptscriptstyle\times};
(5.963636,0.742797) *[red]{\scriptscriptstyle\times};
(5.981818,0.469019) *[red]{\scriptscriptstyle\times};
(6.000000,0.530176) *[red]{\scriptscriptstyle\times};
(6.018182,0.575609) *[red]{\scriptscriptstyle\times};
(6.036364,0.634818) *[red]{\scriptscriptstyle\times};
(6.054545,0.644329) *[red]{\scriptscriptstyle\times};
(6.072727,0.675251) *[red]{\scriptscriptstyle\times};
(6.090909,0.716548) *[red]{\scriptscriptstyle\times};
(6.109091,0.323315) *[red]{\scriptscriptstyle\times};
(6.127273,0.450848) *[red]{\scriptscriptstyle\times};
(6.145455,0.508575) *[red]{\scriptscriptstyle\times};
(6.163636,0.517591) *[red]{\scriptscriptstyle\times};
(6.181818,0.624651) *[red]{\scriptscriptstyle\times};
(6.200000,0.750516) *[red]{\scriptscriptstyle\times};
(6.218182,0.753696) *[red]{\scriptscriptstyle\times};
(6.236364,0.388351) *[red]{\scriptscriptstyle\times};
(6.254545,0.465375) *[red]{\scriptscriptstyle\times};
(6.272727,0.515771) *[red]{\scriptscriptstyle\times};
(6.290909,0.694133) *[red]{\scriptscriptstyle\times};
(6.309091,0.694785) *[red]{\scriptscriptstyle\times};
(6.327273,0.840205) *[red]{\scriptscriptstyle\times};
(6.345455,0.896114) *[red]{\scriptscriptstyle\times};
(6.363636,0.452150) *[red]{\scriptscriptstyle\times};
(6.381818,0.616715) *[red]{\scriptscriptstyle\times};
(6.400000,0.627370) *[red]{\scriptscriptstyle\times};
(6.418182,0.653628) *[red]{\scriptscriptstyle\times};
(6.436364,0.664145) *[red]{\scriptscriptstyle\times};
(6.454545,0.686522) *[red]{\scriptscriptstyle\times};
(6.472727,0.848740) *[red]{\scriptscriptstyle\times};
(6.490909,0.508982) *[red]{\scriptscriptstyle\times};
(6.509091,0.521576) *[red]{\scriptscriptstyle\times};
(6.527273,0.537718) *[red]{\scriptscriptstyle\times};
(6.545455,0.560009) *[red]{\scriptscriptstyle\times};
(6.563636,0.571247) *[red]{\scriptscriptstyle\times};
(6.581818,0.651326) *[red]{\scriptscriptstyle\times};
(6.600000,0.814314) *[red]{\scriptscriptstyle\times};
(6.618182,0.498299) *[red]{\scriptscriptstyle\times};
(6.636364,0.527494) *[red]{\scriptscriptstyle\times};
(6.654545,0.536362) *[red]{\scriptscriptstyle\times};
(6.672727,0.712785) *[red]{\scriptscriptstyle\times};
(6.690909,0.731868) *[red]{\scriptscriptstyle\times};
(6.709091,0.755400) *[red]{\scriptscriptstyle\times};
(6.727273,0.866279) *[red]{\scriptscriptstyle\times};
(6.745455,0.516219) *[red]{\scriptscriptstyle\times};
(6.763636,0.563287) *[red]{\scriptscriptstyle\times};
(6.781818,0.615498) *[red]{\scriptscriptstyle\times};
(6.800000,0.650111) *[red]{\scriptscriptstyle\times};
(6.818182,0.671628) *[red]{\scriptscriptstyle\times};
(6.836364,0.706229) *[red]{\scriptscriptstyle\times};
(6.854545,0.806868) *[red]{\scriptscriptstyle\times};
(6.872727,0.378881) *[red]{\scriptscriptstyle\times};
(6.890909,0.581760) *[red]{\scriptscriptstyle\times};
(6.909091,0.608942) *[red]{\scriptscriptstyle\times};
(6.927273,0.672064) *[red]{\scriptscriptstyle\times};
(6.945455,0.722055) *[red]{\scriptscriptstyle\times};
(6.963636,0.728762) *[red]{\scriptscriptstyle\times};
(6.981818,0.782319) *[red]{\scriptscriptstyle\times};
(7.000000,0.423012) *[red]{\scriptscriptstyle\times};
(7.018182,0.532460) *[red]{\scriptscriptstyle\times};
(7.036364,0.621670) *[red]{\scriptscriptstyle\times};
(7.054545,0.686134) *[red]{\scriptscriptstyle\times};
(7.072727,0.715959) *[red]{\scriptscriptstyle\times};
(7.090909,0.771115) *[red]{\scriptscriptstyle\times};
(7.109091,0.774495) *[red]{\scriptscriptstyle\times};
(7.127273,0.319111) *[red]{\scriptscriptstyle\times};
(7.145455,0.488573) *[red]{\scriptscriptstyle\times};
(7.163636,0.526757) *[red]{\scriptscriptstyle\times};
(7.181818,0.667487) *[red]{\scriptscriptstyle\times};
(7.200000,0.747229) *[red]{\scriptscriptstyle\times};
(7.218182,0.752353) *[red]{\scriptscriptstyle\times};
(7.236364,0.819637) *[red]{\scriptscriptstyle\times};
(7.254545,0.578495) *[red]{\scriptscriptstyle\times};
(7.272727,0.611596) *[red]{\scriptscriptstyle\times};
(7.290909,0.634613) *[red]{\scriptscriptstyle\times};
(7.309091,0.637173) *[red]{\scriptscriptstyle\times};
(7.327273,0.719235) *[red]{\scriptscriptstyle\times};
(7.345455,0.740702) *[red]{\scriptscriptstyle\times};
(7.363636,0.784889) *[red]{\scriptscriptstyle\times};
(7.381818,0.482939) *[red]{\scriptscriptstyle\times};
(7.400000,0.514972) *[red]{\scriptscriptstyle\times};
(7.418182,0.520789) *[red]{\scriptscriptstyle\times};
(7.436364,0.672328) *[red]{\scriptscriptstyle\times};
(7.454545,0.685933) *[red]{\scriptscriptstyle\times};
(7.472727,0.772715) *[red]{\scriptscriptstyle\times};
(7.490909,0.822500) *[red]{\scriptscriptstyle\times};
(7.509091,0.390653) *[red]{\scriptscriptstyle\times};
(7.527273,0.530078) *[red]{\scriptscriptstyle\times};
(7.545455,0.643673) *[red]{\scriptscriptstyle\times};
(7.563636,0.689002) *[red]{\scriptscriptstyle\times};
(7.581818,0.789634) *[red]{\scriptscriptstyle\times};
(7.600000,0.802307) *[red]{\scriptscriptstyle\times};
(7.618182,0.870809) *[red]{\scriptscriptstyle\times};
(7.636364,0.540475) *[red]{\scriptscriptstyle\times};
(7.654545,0.576958) *[red]{\scriptscriptstyle\times};
(7.672727,0.597517) *[red]{\scriptscriptstyle\times};
(7.690909,0.611551) *[red]{\scriptscriptstyle\times};
(7.709091,0.619653) *[red]{\scriptscriptstyle\times};
(7.727273,0.812100) *[red]{\scriptscriptstyle\times};
(7.745455,0.871447) *[red]{\scriptscriptstyle\times};
(7.763636,0.535167) *[red]{\scriptscriptstyle\times};
(7.781818,0.556748) *[red]{\scriptscriptstyle\times};
(7.800000,0.613781) *[red]{\scriptscriptstyle\times};
(7.818182,0.705496) *[red]{\scriptscriptstyle\times};
(7.836364,0.740602) *[red]{\scriptscriptstyle\times};
(7.854545,0.834985) *[red]{\scriptscriptstyle\times};
(7.872727,0.901187) *[red]{\scriptscriptstyle\times};
(7.890909,0.630383) *[red]{\scriptscriptstyle\times};
(7.909091,0.646332) *[red]{\scriptscriptstyle\times};
(7.927273,0.749645) *[red]{\scriptscriptstyle\times};
(7.945455,0.762580) *[red]{\scriptscriptstyle\times};
(7.963636,0.781234) *[red]{\scriptscriptstyle\times};
(7.981818,0.798229) *[red]{\scriptscriptstyle\times};
(8.000000,0.830980) *[red]{\scriptscriptstyle\times};
\endxy
}
\caption{Skylake cycles
  for the CSURF-512 action
  using {\tt velusqrt-magma}.
}
\label{fig: csurf-magma}
\end{figure}

\begin{figure}[t]
  \centerline{
\xy <1.1cm,0cm>:<0cm,4cm>::
(0,0.324353); (8,0.324353) **[blue]@{-};
(8.1,0.324353) *[blue]{\rlap{313025614}};
(0,0.421341); (8,0.421341) **[blue]@{-};
(8.1,0.421341) *[blue]{\rlap{334792870}};
(0,0.515067); (8,0.515067) **[blue]@{-};
(8.1,0.515067) *[blue]{\rlap{357265234}};
(0,0.247745); (8,0.247745) **[red]@{-};
(-0.1,0.247745) *[red]{\llap{296837404}};
(0,0.348957); (8,0.348957) **[red]@{-};
(-0.1,0.348957) *[red]{\llap{318409802}};
(0,0.448301); (8,0.448301) **[red]@{-};
(-0.1,0.448301) *[red]{\llap{341108192}};
(-0.004107,0.063183); (-0.004107,0.821265) **[lightgray]@{-};
(0.119097,0.063183); (0.119097,0.821265) **[lightgray]@{-};
(0.242300,0.063183); (0.242300,0.821265) **[lightgray]@{-};
(0.365503,0.063183); (0.365503,0.821265) **[lightgray]@{-};
(0.488706,0.063183); (0.488706,0.821265) **[lightgray]@{-};
(0.611910,0.063183); (0.611910,0.821265) **[lightgray]@{-};
(0.735113,0.063183); (0.735113,0.821265) **[lightgray]@{-};
(0.858316,0.063183); (0.858316,0.821265) **[lightgray]@{-};
(0.981520,0.063183); (0.981520,0.821265) **[lightgray]@{-};
(1.104723,0.063183); (1.104723,0.821265) **[lightgray]@{-};
(1.227926,0.063183); (1.227926,0.821265) **[lightgray]@{-};
(1.351129,0.063183); (1.351129,0.821265) **[lightgray]@{-};
(1.474333,0.063183); (1.474333,0.821265) **[lightgray]@{-};
(1.597536,0.063183); (1.597536,0.821265) **[lightgray]@{-};
(1.720739,0.063183); (1.720739,0.821265) **[lightgray]@{-};
(1.843943,0.063183); (1.843943,0.821265) **[lightgray]@{-};
(1.967146,0.063183); (1.967146,0.821265) **[lightgray]@{-};
(2.090349,0.063183); (2.090349,0.821265) **[lightgray]@{-};
(2.213552,0.063183); (2.213552,0.821265) **[lightgray]@{-};
(2.336756,0.063183); (2.336756,0.821265) **[lightgray]@{-};
(2.459959,0.063183); (2.459959,0.821265) **[lightgray]@{-};
(2.583162,0.063183); (2.583162,0.821265) **[lightgray]@{-};
(2.706366,0.063183); (2.706366,0.821265) **[lightgray]@{-};
(2.829569,0.063183); (2.829569,0.821265) **[lightgray]@{-};
(2.952772,0.063183); (2.952772,0.821265) **[lightgray]@{-};
(3.075975,0.063183); (3.075975,0.821265) **[lightgray]@{-};
(3.199179,0.063183); (3.199179,0.821265) **[lightgray]@{-};
(3.322382,0.063183); (3.322382,0.821265) **[lightgray]@{-};
(3.445585,0.063183); (3.445585,0.821265) **[lightgray]@{-};
(3.568789,0.063183); (3.568789,0.821265) **[lightgray]@{-};
(3.691992,0.063183); (3.691992,0.821265) **[lightgray]@{-};
(3.815195,0.063183); (3.815195,0.821265) **[lightgray]@{-};
(3.938398,0.063183); (3.938398,0.821265) **[lightgray]@{-};
(4.061602,0.063183); (4.061602,0.821265) **[lightgray]@{-};
(4.184805,0.063183); (4.184805,0.821265) **[lightgray]@{-};
(4.308008,0.063183); (4.308008,0.821265) **[lightgray]@{-};
(4.431211,0.063183); (4.431211,0.821265) **[lightgray]@{-};
(4.554415,0.063183); (4.554415,0.821265) **[lightgray]@{-};
(4.677618,0.063183); (4.677618,0.821265) **[lightgray]@{-};
(4.800821,0.063183); (4.800821,0.821265) **[lightgray]@{-};
(4.924025,0.063183); (4.924025,0.821265) **[lightgray]@{-};
(5.047228,0.063183); (5.047228,0.821265) **[lightgray]@{-};
(5.170431,0.063183); (5.170431,0.821265) **[lightgray]@{-};
(5.293634,0.063183); (5.293634,0.821265) **[lightgray]@{-};
(5.416838,0.063183); (5.416838,0.821265) **[lightgray]@{-};
(5.540041,0.063183); (5.540041,0.821265) **[lightgray]@{-};
(5.663244,0.063183); (5.663244,0.821265) **[lightgray]@{-};
(5.786448,0.063183); (5.786448,0.821265) **[lightgray]@{-};
(5.909651,0.063183); (5.909651,0.821265) **[lightgray]@{-};
(6.032854,0.063183); (6.032854,0.821265) **[lightgray]@{-};
(6.156057,0.063183); (6.156057,0.821265) **[lightgray]@{-};
(6.279261,0.063183); (6.279261,0.821265) **[lightgray]@{-};
(6.402464,0.063183); (6.402464,0.821265) **[lightgray]@{-};
(6.525667,0.063183); (6.525667,0.821265) **[lightgray]@{-};
(6.648871,0.063183); (6.648871,0.821265) **[lightgray]@{-};
(6.772074,0.063183); (6.772074,0.821265) **[lightgray]@{-};
(6.895277,0.063183); (6.895277,0.821265) **[lightgray]@{-};
(7.018480,0.063183); (7.018480,0.821265) **[lightgray]@{-};
(7.141684,0.063183); (7.141684,0.821265) **[lightgray]@{-};
(7.264887,0.063183); (7.264887,0.821265) **[lightgray]@{-};
(7.388090,0.063183); (7.388090,0.821265) **[lightgray]@{-};
(7.511294,0.063183); (7.511294,0.821265) **[lightgray]@{-};
(7.634497,0.063183); (7.634497,0.821265) **[lightgray]@{-};
(7.757700,0.063183); (7.757700,0.821265) **[lightgray]@{-};
(7.880903,0.063183); (7.880903,0.821265) **[lightgray]@{-};
(8.004107,0.063183); (8.004107,0.821265) **[lightgray]@{-};
(-0.004107,0.063183); (-0.004107,0.821265) **[lightgray]@{-};
(0.119097,0.063183); (0.119097,0.821265) **[lightgray]@{-};
(0.242300,0.063183); (0.242300,0.821265) **[lightgray]@{-};
(0.365503,0.063183); (0.365503,0.821265) **[lightgray]@{-};
(0.488706,0.063183); (0.488706,0.821265) **[lightgray]@{-};
(0.611910,0.063183); (0.611910,0.821265) **[lightgray]@{-};
(0.735113,0.063183); (0.735113,0.821265) **[lightgray]@{-};
(0.858316,0.063183); (0.858316,0.821265) **[lightgray]@{-};
(0.981520,0.063183); (0.981520,0.821265) **[lightgray]@{-};
(1.104723,0.063183); (1.104723,0.821265) **[lightgray]@{-};
(1.227926,0.063183); (1.227926,0.821265) **[lightgray]@{-};
(1.351129,0.063183); (1.351129,0.821265) **[lightgray]@{-};
(1.474333,0.063183); (1.474333,0.821265) **[lightgray]@{-};
(1.597536,0.063183); (1.597536,0.821265) **[lightgray]@{-};
(1.720739,0.063183); (1.720739,0.821265) **[lightgray]@{-};
(1.843943,0.063183); (1.843943,0.821265) **[lightgray]@{-};
(1.967146,0.063183); (1.967146,0.821265) **[lightgray]@{-};
(2.090349,0.063183); (2.090349,0.821265) **[lightgray]@{-};
(2.213552,0.063183); (2.213552,0.821265) **[lightgray]@{-};
(2.336756,0.063183); (2.336756,0.821265) **[lightgray]@{-};
(2.459959,0.063183); (2.459959,0.821265) **[lightgray]@{-};
(2.583162,0.063183); (2.583162,0.821265) **[lightgray]@{-};
(2.706366,0.063183); (2.706366,0.821265) **[lightgray]@{-};
(2.829569,0.063183); (2.829569,0.821265) **[lightgray]@{-};
(2.952772,0.063183); (2.952772,0.821265) **[lightgray]@{-};
(3.075975,0.063183); (3.075975,0.821265) **[lightgray]@{-};
(3.199179,0.063183); (3.199179,0.821265) **[lightgray]@{-};
(3.322382,0.063183); (3.322382,0.821265) **[lightgray]@{-};
(3.445585,0.063183); (3.445585,0.821265) **[lightgray]@{-};
(3.568789,0.063183); (3.568789,0.821265) **[lightgray]@{-};
(3.691992,0.063183); (3.691992,0.821265) **[lightgray]@{-};
(3.815195,0.063183); (3.815195,0.821265) **[lightgray]@{-};
(3.938398,0.063183); (3.938398,0.821265) **[lightgray]@{-};
(4.061602,0.063183); (4.061602,0.821265) **[lightgray]@{-};
(4.184805,0.063183); (4.184805,0.821265) **[lightgray]@{-};
(4.308008,0.063183); (4.308008,0.821265) **[lightgray]@{-};
(4.431211,0.063183); (4.431211,0.821265) **[lightgray]@{-};
(4.554415,0.063183); (4.554415,0.821265) **[lightgray]@{-};
(4.677618,0.063183); (4.677618,0.821265) **[lightgray]@{-};
(4.800821,0.063183); (4.800821,0.821265) **[lightgray]@{-};
(4.924025,0.063183); (4.924025,0.821265) **[lightgray]@{-};
(5.047228,0.063183); (5.047228,0.821265) **[lightgray]@{-};
(5.170431,0.063183); (5.170431,0.821265) **[lightgray]@{-};
(5.293634,0.063183); (5.293634,0.821265) **[lightgray]@{-};
(5.416838,0.063183); (5.416838,0.821265) **[lightgray]@{-};
(5.540041,0.063183); (5.540041,0.821265) **[lightgray]@{-};
(5.663244,0.063183); (5.663244,0.821265) **[lightgray]@{-};
(5.786448,0.063183); (5.786448,0.821265) **[lightgray]@{-};
(5.909651,0.063183); (5.909651,0.821265) **[lightgray]@{-};
(6.032854,0.063183); (6.032854,0.821265) **[lightgray]@{-};
(6.156057,0.063183); (6.156057,0.821265) **[lightgray]@{-};
(6.279261,0.063183); (6.279261,0.821265) **[lightgray]@{-};
(6.402464,0.063183); (6.402464,0.821265) **[lightgray]@{-};
(6.525667,0.063183); (6.525667,0.821265) **[lightgray]@{-};
(6.648871,0.063183); (6.648871,0.821265) **[lightgray]@{-};
(6.772074,0.063183); (6.772074,0.821265) **[lightgray]@{-};
(6.895277,0.063183); (6.895277,0.821265) **[lightgray]@{-};
(7.018480,0.063183); (7.018480,0.821265) **[lightgray]@{-};
(7.141684,0.063183); (7.141684,0.821265) **[lightgray]@{-};
(7.264887,0.063183); (7.264887,0.821265) **[lightgray]@{-};
(7.388090,0.063183); (7.388090,0.821265) **[lightgray]@{-};
(7.511294,0.063183); (7.511294,0.821265) **[lightgray]@{-};
(7.634497,0.063183); (7.634497,0.821265) **[lightgray]@{-};
(7.757700,0.063183); (7.757700,0.821265) **[lightgray]@{-};
(7.880903,0.063183); (7.880903,0.821265) **[lightgray]@{-};
(8.004107,0.063183); (8.004107,0.821265) **[lightgray]@{-};
(0.000000,0.123750) *[blue]{\scriptscriptstyle+};
(0.008214,0.126409) *[blue]{\scriptscriptstyle+};
(0.016427,0.127358) *[blue]{\scriptscriptstyle+};
(0.024641,0.152259) *[blue]{\scriptscriptstyle+};
(0.032854,0.154841) *[blue]{\scriptscriptstyle+};
(0.041068,0.156439) *[blue]{\scriptscriptstyle+};
(0.049281,0.161415) *[blue]{\scriptscriptstyle+};
(0.057495,0.164974) *[blue]{\scriptscriptstyle+};
(0.065708,0.170122) *[blue]{\scriptscriptstyle+};
(0.073922,0.172707) *[blue]{\scriptscriptstyle+};
(0.082136,0.174469) *[blue]{\scriptscriptstyle+};
(0.090349,0.175251) *[blue]{\scriptscriptstyle+};
(0.098563,0.176454) *[blue]{\scriptscriptstyle+};
(0.106776,0.178205) *[blue]{\scriptscriptstyle+};
(0.114990,0.195008) *[blue]{\scriptscriptstyle+};
(0.123203,0.156006) *[blue]{\scriptscriptstyle+};
(0.131417,0.159233) *[blue]{\scriptscriptstyle+};
(0.139630,0.161251) *[blue]{\scriptscriptstyle+};
(0.147844,0.161583) *[blue]{\scriptscriptstyle+};
(0.156057,0.161936) *[blue]{\scriptscriptstyle+};
(0.164271,0.162053) *[blue]{\scriptscriptstyle+};
(0.172485,0.167265) *[blue]{\scriptscriptstyle+};
(0.180698,0.179616) *[blue]{\scriptscriptstyle+};
(0.188912,0.183654) *[blue]{\scriptscriptstyle+};
(0.197125,0.187961) *[blue]{\scriptscriptstyle+};
(0.205339,0.190592) *[blue]{\scriptscriptstyle+};
(0.213552,0.205842) *[blue]{\scriptscriptstyle+};
(0.221766,0.209877) *[blue]{\scriptscriptstyle+};
(0.229979,0.214114) *[blue]{\scriptscriptstyle+};
(0.238193,0.240381) *[blue]{\scriptscriptstyle+};
(0.246407,0.164207) *[blue]{\scriptscriptstyle+};
(0.254620,0.164369) *[blue]{\scriptscriptstyle+};
(0.262834,0.165636) *[blue]{\scriptscriptstyle+};
(0.271047,0.166372) *[blue]{\scriptscriptstyle+};
(0.279261,0.168053) *[blue]{\scriptscriptstyle+};
(0.287474,0.168402) *[blue]{\scriptscriptstyle+};
(0.295688,0.171705) *[blue]{\scriptscriptstyle+};
(0.303901,0.172392) *[blue]{\scriptscriptstyle+};
(0.312115,0.177049) *[blue]{\scriptscriptstyle+};
(0.320329,0.188086) *[blue]{\scriptscriptstyle+};
(0.328542,0.188818) *[blue]{\scriptscriptstyle+};
(0.336756,0.192300) *[blue]{\scriptscriptstyle+};
(0.344969,0.195114) *[blue]{\scriptscriptstyle+};
(0.353183,0.216424) *[blue]{\scriptscriptstyle+};
(0.361396,0.271029) *[blue]{\scriptscriptstyle+};
(0.369610,0.158483) *[blue]{\scriptscriptstyle+};
(0.377823,0.165122) *[blue]{\scriptscriptstyle+};
(0.386037,0.169348) *[blue]{\scriptscriptstyle+};
(0.394251,0.170668) *[blue]{\scriptscriptstyle+};
(0.402464,0.172478) *[blue]{\scriptscriptstyle+};
(0.410678,0.178285) *[blue]{\scriptscriptstyle+};
(0.418891,0.188545) *[blue]{\scriptscriptstyle+};
(0.427105,0.193860) *[blue]{\scriptscriptstyle+};
(0.435318,0.194830) *[blue]{\scriptscriptstyle+};
(0.443532,0.211631) *[blue]{\scriptscriptstyle+};
(0.451745,0.213561) *[blue]{\scriptscriptstyle+};
(0.459959,0.218410) *[blue]{\scriptscriptstyle+};
(0.468172,0.220072) *[blue]{\scriptscriptstyle+};
(0.476386,0.226028) *[blue]{\scriptscriptstyle+};
(0.484600,0.244150) *[blue]{\scriptscriptstyle+};
(0.492813,0.193848) *[blue]{\scriptscriptstyle+};
(0.501027,0.199748) *[blue]{\scriptscriptstyle+};
(0.509240,0.200538) *[blue]{\scriptscriptstyle+};
(0.517454,0.213330) *[blue]{\scriptscriptstyle+};
(0.525667,0.218282) *[blue]{\scriptscriptstyle+};
(0.533881,0.241412) *[blue]{\scriptscriptstyle+};
(0.542094,0.244387) *[blue]{\scriptscriptstyle+};
(0.550308,0.245093) *[blue]{\scriptscriptstyle+};
(0.558522,0.250703) *[blue]{\scriptscriptstyle+};
(0.566735,0.257949) *[blue]{\scriptscriptstyle+};
(0.574949,0.264678) *[blue]{\scriptscriptstyle+};
(0.583162,0.277468) *[blue]{\scriptscriptstyle+};
(0.591376,0.311608) *[blue]{\scriptscriptstyle+};
(0.599589,0.327207) *[blue]{\scriptscriptstyle+};
(0.607803,0.428878) *[blue]{\scriptscriptstyle+};
(0.616016,0.194517) *[blue]{\scriptscriptstyle+};
(0.624230,0.194773) *[blue]{\scriptscriptstyle+};
(0.632444,0.198418) *[blue]{\scriptscriptstyle+};
(0.640657,0.200607) *[blue]{\scriptscriptstyle+};
(0.648871,0.201399) *[blue]{\scriptscriptstyle+};
(0.657084,0.203848) *[blue]{\scriptscriptstyle+};
(0.665298,0.204476) *[blue]{\scriptscriptstyle+};
(0.673511,0.204871) *[blue]{\scriptscriptstyle+};
(0.681725,0.206959) *[blue]{\scriptscriptstyle+};
(0.689938,0.208531) *[blue]{\scriptscriptstyle+};
(0.698152,0.210239) *[blue]{\scriptscriptstyle+};
(0.706366,0.217211) *[blue]{\scriptscriptstyle+};
(0.714579,0.246316) *[blue]{\scriptscriptstyle+};
(0.722793,0.267128) *[blue]{\scriptscriptstyle+};
(0.731006,0.318795) *[blue]{\scriptscriptstyle+};
(0.739220,0.183992) *[blue]{\scriptscriptstyle+};
(0.747433,0.191643) *[blue]{\scriptscriptstyle+};
(0.755647,0.196239) *[blue]{\scriptscriptstyle+};
(0.763860,0.210215) *[blue]{\scriptscriptstyle+};
(0.772074,0.226547) *[blue]{\scriptscriptstyle+};
(0.780287,0.237842) *[blue]{\scriptscriptstyle+};
(0.788501,0.240197) *[blue]{\scriptscriptstyle+};
(0.796715,0.257161) *[blue]{\scriptscriptstyle+};
(0.804928,0.276322) *[blue]{\scriptscriptstyle+};
(0.813142,0.290968) *[blue]{\scriptscriptstyle+};
(0.821355,0.293533) *[blue]{\scriptscriptstyle+};
(0.829569,0.314846) *[blue]{\scriptscriptstyle+};
(0.837782,0.318825) *[blue]{\scriptscriptstyle+};
(0.845996,0.345253) *[blue]{\scriptscriptstyle+};
(0.854209,0.365289) *[blue]{\scriptscriptstyle+};
(0.862423,0.235069) *[blue]{\scriptscriptstyle+};
(0.870637,0.235114) *[blue]{\scriptscriptstyle+};
(0.878850,0.235990) *[blue]{\scriptscriptstyle+};
(0.887064,0.236768) *[blue]{\scriptscriptstyle+};
(0.895277,0.237908) *[blue]{\scriptscriptstyle+};
(0.903491,0.238774) *[blue]{\scriptscriptstyle+};
(0.911704,0.255750) *[blue]{\scriptscriptstyle+};
(0.919918,0.260083) *[blue]{\scriptscriptstyle+};
(0.928131,0.261655) *[blue]{\scriptscriptstyle+};
(0.936345,0.263963) *[blue]{\scriptscriptstyle+};
(0.944559,0.264324) *[blue]{\scriptscriptstyle+};
(0.952772,0.265428) *[blue]{\scriptscriptstyle+};
(0.960986,0.278017) *[blue]{\scriptscriptstyle+};
(0.969199,0.279116) *[blue]{\scriptscriptstyle+};
(0.977413,0.289606) *[blue]{\scriptscriptstyle+};
(0.985626,0.228034) *[blue]{\scriptscriptstyle+};
(0.993840,0.232181) *[blue]{\scriptscriptstyle+};
(1.002053,0.232786) *[blue]{\scriptscriptstyle+};
(1.010267,0.249478) *[blue]{\scriptscriptstyle+};
(1.018480,0.266315) *[blue]{\scriptscriptstyle+};
(1.026694,0.267673) *[blue]{\scriptscriptstyle+};
(1.034908,0.270934) *[blue]{\scriptscriptstyle+};
(1.043121,0.289040) *[blue]{\scriptscriptstyle+};
(1.051335,0.302150) *[blue]{\scriptscriptstyle+};
(1.059548,0.305317) *[blue]{\scriptscriptstyle+};
(1.067762,0.307403) *[blue]{\scriptscriptstyle+};
(1.075975,0.339174) *[blue]{\scriptscriptstyle+};
(1.084189,0.354207) *[blue]{\scriptscriptstyle+};
(1.092402,0.358473) *[blue]{\scriptscriptstyle+};
(1.100616,0.365168) *[blue]{\scriptscriptstyle+};
(1.108830,0.277531) *[blue]{\scriptscriptstyle+};
(1.117043,0.278473) *[blue]{\scriptscriptstyle+};
(1.125257,0.280720) *[blue]{\scriptscriptstyle+};
(1.133470,0.283069) *[blue]{\scriptscriptstyle+};
(1.141684,0.283934) *[blue]{\scriptscriptstyle+};
(1.149897,0.288346) *[blue]{\scriptscriptstyle+};
(1.158111,0.291298) *[blue]{\scriptscriptstyle+};
(1.166324,0.297564) *[blue]{\scriptscriptstyle+};
(1.174538,0.300441) *[blue]{\scriptscriptstyle+};
(1.182752,0.301887) *[blue]{\scriptscriptstyle+};
(1.190965,0.304087) *[blue]{\scriptscriptstyle+};
(1.199179,0.306588) *[blue]{\scriptscriptstyle+};
(1.207392,0.310969) *[blue]{\scriptscriptstyle+};
(1.215606,0.312943) *[blue]{\scriptscriptstyle+};
(1.223819,0.344840) *[blue]{\scriptscriptstyle+};
(1.232033,0.245627) *[blue]{\scriptscriptstyle+};
(1.240246,0.260282) *[blue]{\scriptscriptstyle+};
(1.248460,0.262366) *[blue]{\scriptscriptstyle+};
(1.256674,0.271905) *[blue]{\scriptscriptstyle+};
(1.264887,0.281616) *[blue]{\scriptscriptstyle+};
(1.273101,0.287181) *[blue]{\scriptscriptstyle+};
(1.281314,0.306186) *[blue]{\scriptscriptstyle+};
(1.289528,0.306703) *[blue]{\scriptscriptstyle+};
(1.297741,0.325737) *[blue]{\scriptscriptstyle+};
(1.305955,0.327864) *[blue]{\scriptscriptstyle+};
(1.314168,0.328373) *[blue]{\scriptscriptstyle+};
(1.322382,0.331766) *[blue]{\scriptscriptstyle+};
(1.330595,0.352530) *[blue]{\scriptscriptstyle+};
(1.338809,0.355145) *[blue]{\scriptscriptstyle+};
(1.347023,0.379282) *[blue]{\scriptscriptstyle+};
(1.355236,0.261716) *[blue]{\scriptscriptstyle+};
(1.363450,0.262966) *[blue]{\scriptscriptstyle+};
(1.371663,0.268131) *[blue]{\scriptscriptstyle+};
(1.379877,0.268747) *[blue]{\scriptscriptstyle+};
(1.388090,0.269774) *[blue]{\scriptscriptstyle+};
(1.396304,0.273487) *[blue]{\scriptscriptstyle+};
(1.404517,0.274796) *[blue]{\scriptscriptstyle+};
(1.412731,0.275752) *[blue]{\scriptscriptstyle+};
(1.420945,0.278622) *[blue]{\scriptscriptstyle+};
(1.429158,0.281157) *[blue]{\scriptscriptstyle+};
(1.437372,0.282530) *[blue]{\scriptscriptstyle+};
(1.445585,0.287745) *[blue]{\scriptscriptstyle+};
(1.453799,0.290901) *[blue]{\scriptscriptstyle+};
(1.462012,0.298303) *[blue]{\scriptscriptstyle+};
(1.470226,0.338403) *[blue]{\scriptscriptstyle+};
(1.478439,0.291688) *[blue]{\scriptscriptstyle+};
(1.486653,0.299290) *[blue]{\scriptscriptstyle+};
(1.494867,0.311960) *[blue]{\scriptscriptstyle+};
(1.503080,0.314148) *[blue]{\scriptscriptstyle+};
(1.511294,0.316178) *[blue]{\scriptscriptstyle+};
(1.519507,0.317156) *[blue]{\scriptscriptstyle+};
(1.527721,0.321427) *[blue]{\scriptscriptstyle+};
(1.535934,0.331381) *[blue]{\scriptscriptstyle+};
(1.544148,0.332284) *[blue]{\scriptscriptstyle+};
(1.552361,0.334918) *[blue]{\scriptscriptstyle+};
(1.560575,0.340160) *[blue]{\scriptscriptstyle+};
(1.568789,0.347931) *[blue]{\scriptscriptstyle+};
(1.577002,0.350942) *[blue]{\scriptscriptstyle+};
(1.585216,0.353587) *[blue]{\scriptscriptstyle+};
(1.593429,0.360465) *[blue]{\scriptscriptstyle+};
(1.601643,0.271217) *[blue]{\scriptscriptstyle+};
(1.609856,0.273053) *[blue]{\scriptscriptstyle+};
(1.618070,0.283449) *[blue]{\scriptscriptstyle+};
(1.626283,0.286893) *[blue]{\scriptscriptstyle+};
(1.634497,0.288592) *[blue]{\scriptscriptstyle+};
(1.642710,0.288968) *[blue]{\scriptscriptstyle+};
(1.650924,0.291845) *[blue]{\scriptscriptstyle+};
(1.659138,0.291924) *[blue]{\scriptscriptstyle+};
(1.667351,0.294371) *[blue]{\scriptscriptstyle+};
(1.675565,0.295326) *[blue]{\scriptscriptstyle+};
(1.683778,0.307574) *[blue]{\scriptscriptstyle+};
(1.691992,0.314898) *[blue]{\scriptscriptstyle+};
(1.700205,0.335714) *[blue]{\scriptscriptstyle+};
(1.708419,0.337924) *[blue]{\scriptscriptstyle+};
(1.716632,0.353306) *[blue]{\scriptscriptstyle+};
(1.724846,0.287400) *[blue]{\scriptscriptstyle+};
(1.733060,0.287891) *[blue]{\scriptscriptstyle+};
(1.741273,0.288419) *[blue]{\scriptscriptstyle+};
(1.749487,0.290702) *[blue]{\scriptscriptstyle+};
(1.757700,0.291329) *[blue]{\scriptscriptstyle+};
(1.765914,0.294454) *[blue]{\scriptscriptstyle+};
(1.774127,0.296207) *[blue]{\scriptscriptstyle+};
(1.782341,0.297634) *[blue]{\scriptscriptstyle+};
(1.790554,0.299641) *[blue]{\scriptscriptstyle+};
(1.798768,0.300377) *[blue]{\scriptscriptstyle+};
(1.806982,0.311966) *[blue]{\scriptscriptstyle+};
(1.815195,0.314441) *[blue]{\scriptscriptstyle+};
(1.823409,0.321460) *[blue]{\scriptscriptstyle+};
(1.831622,0.321940) *[blue]{\scriptscriptstyle+};
(1.839836,0.339509) *[blue]{\scriptscriptstyle+};
(1.848049,0.288981) *[blue]{\scriptscriptstyle+};
(1.856263,0.291662) *[blue]{\scriptscriptstyle+};
(1.864476,0.295636) *[blue]{\scriptscriptstyle+};
(1.872690,0.297934) *[blue]{\scriptscriptstyle+};
(1.880903,0.304941) *[blue]{\scriptscriptstyle+};
(1.889117,0.305575) *[blue]{\scriptscriptstyle+};
(1.897331,0.310851) *[blue]{\scriptscriptstyle+};
(1.905544,0.315347) *[blue]{\scriptscriptstyle+};
(1.913758,0.318595) *[blue]{\scriptscriptstyle+};
(1.921971,0.320895) *[blue]{\scriptscriptstyle+};
(1.930185,0.321282) *[blue]{\scriptscriptstyle+};
(1.938398,0.323266) *[blue]{\scriptscriptstyle+};
(1.946612,0.337684) *[blue]{\scriptscriptstyle+};
(1.954825,0.350857) *[blue]{\scriptscriptstyle+};
(1.963039,0.364451) *[blue]{\scriptscriptstyle+};
(1.971253,0.265398) *[blue]{\scriptscriptstyle+};
(1.979466,0.274098) *[blue]{\scriptscriptstyle+};
(1.987680,0.290999) *[blue]{\scriptscriptstyle+};
(1.995893,0.299268) *[blue]{\scriptscriptstyle+};
(2.004107,0.304509) *[blue]{\scriptscriptstyle+};
(2.012320,0.309776) *[blue]{\scriptscriptstyle+};
(2.020534,0.311922) *[blue]{\scriptscriptstyle+};
(2.028747,0.313985) *[blue]{\scriptscriptstyle+};
(2.036961,0.314446) *[blue]{\scriptscriptstyle+};
(2.045175,0.328899) *[blue]{\scriptscriptstyle+};
(2.053388,0.335690) *[blue]{\scriptscriptstyle+};
(2.061602,0.338937) *[blue]{\scriptscriptstyle+};
(2.069815,0.361691) *[blue]{\scriptscriptstyle+};
(2.078029,0.386688) *[blue]{\scriptscriptstyle+};
(2.086242,0.407173) *[blue]{\scriptscriptstyle+};
(2.094456,0.239655) *[blue]{\scriptscriptstyle+};
(2.102669,0.281712) *[blue]{\scriptscriptstyle+};
(2.110883,0.296259) *[blue]{\scriptscriptstyle+};
(2.119097,0.299049) *[blue]{\scriptscriptstyle+};
(2.127310,0.300916) *[blue]{\scriptscriptstyle+};
(2.135524,0.304110) *[blue]{\scriptscriptstyle+};
(2.143737,0.304905) *[blue]{\scriptscriptstyle+};
(2.151951,0.309459) *[blue]{\scriptscriptstyle+};
(2.160164,0.315935) *[blue]{\scriptscriptstyle+};
(2.168378,0.320259) *[blue]{\scriptscriptstyle+};
(2.176591,0.321058) *[blue]{\scriptscriptstyle+};
(2.184805,0.325433) *[blue]{\scriptscriptstyle+};
(2.193018,0.339309) *[blue]{\scriptscriptstyle+};
(2.201232,0.343164) *[blue]{\scriptscriptstyle+};
(2.209446,0.389882) *[blue]{\scriptscriptstyle+};
(2.217659,0.283962) *[blue]{\scriptscriptstyle+};
(2.225873,0.306063) *[blue]{\scriptscriptstyle+};
(2.234086,0.306283) *[blue]{\scriptscriptstyle+};
(2.242300,0.306968) *[blue]{\scriptscriptstyle+};
(2.250513,0.306973) *[blue]{\scriptscriptstyle+};
(2.258727,0.319821) *[blue]{\scriptscriptstyle+};
(2.266940,0.322038) *[blue]{\scriptscriptstyle+};
(2.275154,0.322665) *[blue]{\scriptscriptstyle+};
(2.283368,0.324353) *[blue]{\scriptscriptstyle+};
(2.291581,0.328813) *[blue]{\scriptscriptstyle+};
(2.299795,0.333995) *[blue]{\scriptscriptstyle+};
(2.308008,0.346851) *[blue]{\scriptscriptstyle+};
(2.316222,0.352470) *[blue]{\scriptscriptstyle+};
(2.324435,0.363739) *[blue]{\scriptscriptstyle+};
(2.332649,0.364254) *[blue]{\scriptscriptstyle+};
(2.340862,0.308113) *[blue]{\scriptscriptstyle+};
(2.349076,0.309564) *[blue]{\scriptscriptstyle+};
(2.357290,0.311205) *[blue]{\scriptscriptstyle+};
(2.365503,0.314648) *[blue]{\scriptscriptstyle+};
(2.373717,0.317290) *[blue]{\scriptscriptstyle+};
(2.381930,0.325987) *[blue]{\scriptscriptstyle+};
(2.390144,0.329847) *[blue]{\scriptscriptstyle+};
(2.398357,0.330347) *[blue]{\scriptscriptstyle+};
(2.406571,0.333668) *[blue]{\scriptscriptstyle+};
(2.414784,0.351320) *[blue]{\scriptscriptstyle+};
(2.422998,0.351672) *[blue]{\scriptscriptstyle+};
(2.431211,0.353429) *[blue]{\scriptscriptstyle+};
(2.439425,0.354762) *[blue]{\scriptscriptstyle+};
(2.447639,0.357414) *[blue]{\scriptscriptstyle+};
(2.455852,0.372228) *[blue]{\scriptscriptstyle+};
(2.464066,0.310026) *[blue]{\scriptscriptstyle+};
(2.472279,0.314523) *[blue]{\scriptscriptstyle+};
(2.480493,0.327022) *[blue]{\scriptscriptstyle+};
(2.488706,0.334065) *[blue]{\scriptscriptstyle+};
(2.496920,0.335360) *[blue]{\scriptscriptstyle+};
(2.505133,0.352628) *[blue]{\scriptscriptstyle+};
(2.513347,0.358399) *[blue]{\scriptscriptstyle+};
(2.521561,0.376809) *[blue]{\scriptscriptstyle+};
(2.529774,0.392003) *[blue]{\scriptscriptstyle+};
(2.537988,0.400089) *[blue]{\scriptscriptstyle+};
(2.546201,0.403867) *[blue]{\scriptscriptstyle+};
(2.554415,0.418165) *[blue]{\scriptscriptstyle+};
(2.562628,0.430769) *[blue]{\scriptscriptstyle+};
(2.570842,0.440151) *[blue]{\scriptscriptstyle+};
(2.579055,0.489800) *[blue]{\scriptscriptstyle+};
(2.587269,0.339906) *[blue]{\scriptscriptstyle+};
(2.595483,0.340364) *[blue]{\scriptscriptstyle+};
(2.603696,0.344743) *[blue]{\scriptscriptstyle+};
(2.611910,0.345258) *[blue]{\scriptscriptstyle+};
(2.620123,0.355357) *[blue]{\scriptscriptstyle+};
(2.628337,0.356208) *[blue]{\scriptscriptstyle+};
(2.636550,0.362152) *[blue]{\scriptscriptstyle+};
(2.644764,0.364303) *[blue]{\scriptscriptstyle+};
(2.652977,0.364527) *[blue]{\scriptscriptstyle+};
(2.661191,0.364662) *[blue]{\scriptscriptstyle+};
(2.669405,0.371501) *[blue]{\scriptscriptstyle+};
(2.677618,0.372983) *[blue]{\scriptscriptstyle+};
(2.685832,0.387400) *[blue]{\scriptscriptstyle+};
(2.694045,0.388375) *[blue]{\scriptscriptstyle+};
(2.702259,0.392559) *[blue]{\scriptscriptstyle+};
(2.710472,0.324676) *[blue]{\scriptscriptstyle+};
(2.718686,0.334719) *[blue]{\scriptscriptstyle+};
(2.726899,0.334740) *[blue]{\scriptscriptstyle+};
(2.735113,0.336064) *[blue]{\scriptscriptstyle+};
(2.743326,0.347898) *[blue]{\scriptscriptstyle+};
(2.751540,0.353585) *[blue]{\scriptscriptstyle+};
(2.759754,0.355925) *[blue]{\scriptscriptstyle+};
(2.767967,0.356515) *[blue]{\scriptscriptstyle+};
(2.776181,0.362148) *[blue]{\scriptscriptstyle+};
(2.784394,0.368033) *[blue]{\scriptscriptstyle+};
(2.792608,0.369475) *[blue]{\scriptscriptstyle+};
(2.800821,0.370679) *[blue]{\scriptscriptstyle+};
(2.809035,0.392363) *[blue]{\scriptscriptstyle+};
(2.817248,0.399658) *[blue]{\scriptscriptstyle+};
(2.825462,0.431211) *[blue]{\scriptscriptstyle+};
(2.833676,0.330994) *[blue]{\scriptscriptstyle+};
(2.841889,0.331586) *[blue]{\scriptscriptstyle+};
(2.850103,0.347935) *[blue]{\scriptscriptstyle+};
(2.858316,0.367281) *[blue]{\scriptscriptstyle+};
(2.866530,0.367318) *[blue]{\scriptscriptstyle+};
(2.874743,0.374366) *[blue]{\scriptscriptstyle+};
(2.882957,0.386050) *[blue]{\scriptscriptstyle+};
(2.891170,0.386303) *[blue]{\scriptscriptstyle+};
(2.899384,0.388146) *[blue]{\scriptscriptstyle+};
(2.907598,0.389080) *[blue]{\scriptscriptstyle+};
(2.915811,0.392892) *[blue]{\scriptscriptstyle+};
(2.924025,0.411078) *[blue]{\scriptscriptstyle+};
(2.932238,0.415400) *[blue]{\scriptscriptstyle+};
(2.940452,0.418427) *[blue]{\scriptscriptstyle+};
(2.948665,0.436386) *[blue]{\scriptscriptstyle+};
(2.956879,0.368489) *[blue]{\scriptscriptstyle+};
(2.965092,0.370096) *[blue]{\scriptscriptstyle+};
(2.973306,0.370597) *[blue]{\scriptscriptstyle+};
(2.981520,0.373685) *[blue]{\scriptscriptstyle+};
(2.989733,0.376795) *[blue]{\scriptscriptstyle+};
(2.997947,0.378805) *[blue]{\scriptscriptstyle+};
(3.006160,0.380342) *[blue]{\scriptscriptstyle+};
(3.014374,0.381398) *[blue]{\scriptscriptstyle+};
(3.022587,0.388559) *[blue]{\scriptscriptstyle+};
(3.030801,0.390298) *[blue]{\scriptscriptstyle+};
(3.039014,0.393537) *[blue]{\scriptscriptstyle+};
(3.047228,0.394312) *[blue]{\scriptscriptstyle+};
(3.055441,0.394770) *[blue]{\scriptscriptstyle+};
(3.063655,0.395503) *[blue]{\scriptscriptstyle+};
(3.071869,0.401905) *[blue]{\scriptscriptstyle+};
(3.080082,0.340806) *[blue]{\scriptscriptstyle+};
(3.088296,0.343348) *[blue]{\scriptscriptstyle+};
(3.096509,0.348824) *[blue]{\scriptscriptstyle+};
(3.104723,0.353335) *[blue]{\scriptscriptstyle+};
(3.112936,0.356875) *[blue]{\scriptscriptstyle+};
(3.121150,0.361908) *[blue]{\scriptscriptstyle+};
(3.129363,0.366692) *[blue]{\scriptscriptstyle+};
(3.137577,0.369560) *[blue]{\scriptscriptstyle+};
(3.145791,0.372340) *[blue]{\scriptscriptstyle+};
(3.154004,0.373889) *[blue]{\scriptscriptstyle+};
(3.162218,0.399908) *[blue]{\scriptscriptstyle+};
(3.170431,0.400609) *[blue]{\scriptscriptstyle+};
(3.178645,0.408972) *[blue]{\scriptscriptstyle+};
(3.186858,0.430787) *[blue]{\scriptscriptstyle+};
(3.195072,0.456892) *[blue]{\scriptscriptstyle+};
(3.203285,0.341394) *[blue]{\scriptscriptstyle+};
(3.211499,0.345922) *[blue]{\scriptscriptstyle+};
(3.219713,0.364961) *[blue]{\scriptscriptstyle+};
(3.227926,0.366545) *[blue]{\scriptscriptstyle+};
(3.236140,0.367803) *[blue]{\scriptscriptstyle+};
(3.244353,0.368315) *[blue]{\scriptscriptstyle+};
(3.252567,0.384169) *[blue]{\scriptscriptstyle+};
(3.260780,0.388398) *[blue]{\scriptscriptstyle+};
(3.268994,0.408834) *[blue]{\scriptscriptstyle+};
(3.277207,0.422921) *[blue]{\scriptscriptstyle+};
(3.285421,0.426985) *[blue]{\scriptscriptstyle+};
(3.293634,0.456115) *[blue]{\scriptscriptstyle+};
(3.301848,0.456307) *[blue]{\scriptscriptstyle+};
(3.310062,0.467565) *[blue]{\scriptscriptstyle+};
(3.318275,0.479846) *[blue]{\scriptscriptstyle+};
(3.326489,0.361600) *[blue]{\scriptscriptstyle+};
(3.334702,0.363809) *[blue]{\scriptscriptstyle+};
(3.342916,0.365463) *[blue]{\scriptscriptstyle+};
(3.351129,0.366076) *[blue]{\scriptscriptstyle+};
(3.359343,0.366339) *[blue]{\scriptscriptstyle+};
(3.367556,0.367953) *[blue]{\scriptscriptstyle+};
(3.375770,0.371009) *[blue]{\scriptscriptstyle+};
(3.383984,0.371536) *[blue]{\scriptscriptstyle+};
(3.392197,0.374557) *[blue]{\scriptscriptstyle+};
(3.400411,0.385758) *[blue]{\scriptscriptstyle+};
(3.408624,0.385937) *[blue]{\scriptscriptstyle+};
(3.416838,0.389032) *[blue]{\scriptscriptstyle+};
(3.425051,0.389088) *[blue]{\scriptscriptstyle+};
(3.433265,0.392471) *[blue]{\scriptscriptstyle+};
(3.441478,0.394307) *[blue]{\scriptscriptstyle+};
(3.449692,0.359664) *[blue]{\scriptscriptstyle+};
(3.457906,0.360415) *[blue]{\scriptscriptstyle+};
(3.466119,0.361135) *[blue]{\scriptscriptstyle+};
(3.474333,0.361739) *[blue]{\scriptscriptstyle+};
(3.482546,0.365352) *[blue]{\scriptscriptstyle+};
(3.490760,0.373801) *[blue]{\scriptscriptstyle+};
(3.498973,0.383443) *[blue]{\scriptscriptstyle+};
(3.507187,0.384032) *[blue]{\scriptscriptstyle+};
(3.515400,0.384162) *[blue]{\scriptscriptstyle+};
(3.523614,0.384904) *[blue]{\scriptscriptstyle+};
(3.531828,0.388289) *[blue]{\scriptscriptstyle+};
(3.540041,0.420712) *[blue]{\scriptscriptstyle+};
(3.548255,0.421341) *[blue]{\scriptscriptstyle+};
(3.556468,0.424449) *[blue]{\scriptscriptstyle+};
(3.564682,0.427370) *[blue]{\scriptscriptstyle+};
(3.572895,0.349198) *[blue]{\scriptscriptstyle+};
(3.581109,0.369626) *[blue]{\scriptscriptstyle+};
(3.589322,0.372716) *[blue]{\scriptscriptstyle+};
(3.597536,0.372720) *[blue]{\scriptscriptstyle+};
(3.605749,0.375772) *[blue]{\scriptscriptstyle+};
(3.613963,0.381947) *[blue]{\scriptscriptstyle+};
(3.622177,0.386735) *[blue]{\scriptscriptstyle+};
(3.630390,0.390449) *[blue]{\scriptscriptstyle+};
(3.638604,0.394898) *[blue]{\scriptscriptstyle+};
(3.646817,0.395116) *[blue]{\scriptscriptstyle+};
(3.655031,0.396553) *[blue]{\scriptscriptstyle+};
(3.663244,0.398251) *[blue]{\scriptscriptstyle+};
(3.671458,0.414527) *[blue]{\scriptscriptstyle+};
(3.679671,0.414737) *[blue]{\scriptscriptstyle+};
(3.687885,0.429251) *[blue]{\scriptscriptstyle+};
(3.696099,0.382755) *[blue]{\scriptscriptstyle+};
(3.704312,0.387140) *[blue]{\scriptscriptstyle+};
(3.712526,0.389056) *[blue]{\scriptscriptstyle+};
(3.720739,0.389083) *[blue]{\scriptscriptstyle+};
(3.728953,0.392493) *[blue]{\scriptscriptstyle+};
(3.737166,0.392825) *[blue]{\scriptscriptstyle+};
(3.745380,0.397684) *[blue]{\scriptscriptstyle+};
(3.753593,0.407597) *[blue]{\scriptscriptstyle+};
(3.761807,0.412324) *[blue]{\scriptscriptstyle+};
(3.770021,0.412674) *[blue]{\scriptscriptstyle+};
(3.778234,0.413494) *[blue]{\scriptscriptstyle+};
(3.786448,0.414634) *[blue]{\scriptscriptstyle+};
(3.794661,0.421749) *[blue]{\scriptscriptstyle+};
(3.802875,0.426468) *[blue]{\scriptscriptstyle+};
(3.811088,0.436144) *[blue]{\scriptscriptstyle+};
(3.819302,0.355611) *[blue]{\scriptscriptstyle+};
(3.827515,0.368443) *[blue]{\scriptscriptstyle+};
(3.835729,0.379880) *[blue]{\scriptscriptstyle+};
(3.843943,0.384722) *[blue]{\scriptscriptstyle+};
(3.852156,0.386642) *[blue]{\scriptscriptstyle+};
(3.860370,0.393380) *[blue]{\scriptscriptstyle+};
(3.868583,0.395178) *[blue]{\scriptscriptstyle+};
(3.876797,0.400651) *[blue]{\scriptscriptstyle+};
(3.885010,0.410801) *[blue]{\scriptscriptstyle+};
(3.893224,0.421985) *[blue]{\scriptscriptstyle+};
(3.901437,0.425326) *[blue]{\scriptscriptstyle+};
(3.909651,0.426787) *[blue]{\scriptscriptstyle+};
(3.917864,0.433302) *[blue]{\scriptscriptstyle+};
(3.926078,0.447491) *[blue]{\scriptscriptstyle+};
(3.934292,0.455469) *[blue]{\scriptscriptstyle+};
(3.942505,0.371233) *[blue]{\scriptscriptstyle+};
(3.950719,0.371905) *[blue]{\scriptscriptstyle+};
(3.958932,0.382243) *[blue]{\scriptscriptstyle+};
(3.967146,0.382477) *[blue]{\scriptscriptstyle+};
(3.975359,0.393518) *[blue]{\scriptscriptstyle+};
(3.983573,0.393596) *[blue]{\scriptscriptstyle+};
(3.991786,0.393956) *[blue]{\scriptscriptstyle+};
(4.000000,0.404097) *[blue]{\scriptscriptstyle+};
(4.008214,0.406730) *[blue]{\scriptscriptstyle+};
(4.016427,0.410981) *[blue]{\scriptscriptstyle+};
(4.024641,0.414814) *[blue]{\scriptscriptstyle+};
(4.032854,0.417765) *[blue]{\scriptscriptstyle+};
(4.041068,0.443209) *[blue]{\scriptscriptstyle+};
(4.049281,0.443380) *[blue]{\scriptscriptstyle+};
(4.057495,0.470603) *[blue]{\scriptscriptstyle+};
(4.065708,0.401969) *[blue]{\scriptscriptstyle+};
(4.073922,0.408182) *[blue]{\scriptscriptstyle+};
(4.082136,0.408377) *[blue]{\scriptscriptstyle+};
(4.090349,0.418437) *[blue]{\scriptscriptstyle+};
(4.098563,0.424193) *[blue]{\scriptscriptstyle+};
(4.106776,0.426003) *[blue]{\scriptscriptstyle+};
(4.114990,0.427133) *[blue]{\scriptscriptstyle+};
(4.123203,0.428747) *[blue]{\scriptscriptstyle+};
(4.131417,0.428988) *[blue]{\scriptscriptstyle+};
(4.139630,0.429655) *[blue]{\scriptscriptstyle+};
(4.147844,0.430118) *[blue]{\scriptscriptstyle+};
(4.156057,0.432234) *[blue]{\scriptscriptstyle+};
(4.164271,0.433924) *[blue]{\scriptscriptstyle+};
(4.172485,0.435089) *[blue]{\scriptscriptstyle+};
(4.180698,0.491734) *[blue]{\scriptscriptstyle+};
(4.188912,0.395233) *[blue]{\scriptscriptstyle+};
(4.197125,0.398085) *[blue]{\scriptscriptstyle+};
(4.205339,0.403565) *[blue]{\scriptscriptstyle+};
(4.213552,0.410831) *[blue]{\scriptscriptstyle+};
(4.221766,0.416314) *[blue]{\scriptscriptstyle+};
(4.229979,0.422889) *[blue]{\scriptscriptstyle+};
(4.238193,0.423415) *[blue]{\scriptscriptstyle+};
(4.246407,0.427939) *[blue]{\scriptscriptstyle+};
(4.254620,0.438094) *[blue]{\scriptscriptstyle+};
(4.262834,0.440935) *[blue]{\scriptscriptstyle+};
(4.271047,0.449149) *[blue]{\scriptscriptstyle+};
(4.279261,0.462371) *[blue]{\scriptscriptstyle+};
(4.287474,0.476925) *[blue]{\scriptscriptstyle+};
(4.295688,0.480514) *[blue]{\scriptscriptstyle+};
(4.303901,0.495980) *[blue]{\scriptscriptstyle+};
(4.312115,0.404716) *[blue]{\scriptscriptstyle+};
(4.320329,0.405009) *[blue]{\scriptscriptstyle+};
(4.328542,0.413483) *[blue]{\scriptscriptstyle+};
(4.336756,0.423206) *[blue]{\scriptscriptstyle+};
(4.344969,0.423347) *[blue]{\scriptscriptstyle+};
(4.353183,0.426887) *[blue]{\scriptscriptstyle+};
(4.361396,0.429349) *[blue]{\scriptscriptstyle+};
(4.369610,0.441751) *[blue]{\scriptscriptstyle+};
(4.377823,0.443197) *[blue]{\scriptscriptstyle+};
(4.386037,0.464673) *[blue]{\scriptscriptstyle+};
(4.394251,0.466882) *[blue]{\scriptscriptstyle+};
(4.402464,0.470205) *[blue]{\scriptscriptstyle+};
(4.410678,0.483306) *[blue]{\scriptscriptstyle+};
(4.418891,0.485984) *[blue]{\scriptscriptstyle+};
(4.427105,0.520656) *[blue]{\scriptscriptstyle+};
(4.435318,0.414217) *[blue]{\scriptscriptstyle+};
(4.443532,0.418086) *[blue]{\scriptscriptstyle+};
(4.451745,0.419047) *[blue]{\scriptscriptstyle+};
(4.459959,0.425701) *[blue]{\scriptscriptstyle+};
(4.468172,0.433737) *[blue]{\scriptscriptstyle+};
(4.476386,0.439251) *[blue]{\scriptscriptstyle+};
(4.484600,0.458195) *[blue]{\scriptscriptstyle+};
(4.492813,0.459071) *[blue]{\scriptscriptstyle+};
(4.501027,0.460258) *[blue]{\scriptscriptstyle+};
(4.509240,0.460659) *[blue]{\scriptscriptstyle+};
(4.517454,0.473067) *[blue]{\scriptscriptstyle+};
(4.525667,0.475728) *[blue]{\scriptscriptstyle+};
(4.533881,0.483891) *[blue]{\scriptscriptstyle+};
(4.542094,0.493793) *[blue]{\scriptscriptstyle+};
(4.550308,0.536642) *[blue]{\scriptscriptstyle+};
(4.558522,0.428809) *[blue]{\scriptscriptstyle+};
(4.566735,0.430972) *[blue]{\scriptscriptstyle+};
(4.574949,0.432225) *[blue]{\scriptscriptstyle+};
(4.583162,0.432575) *[blue]{\scriptscriptstyle+};
(4.591376,0.432590) *[blue]{\scriptscriptstyle+};
(4.599589,0.434989) *[blue]{\scriptscriptstyle+};
(4.607803,0.435905) *[blue]{\scriptscriptstyle+};
(4.616016,0.438611) *[blue]{\scriptscriptstyle+};
(4.624230,0.449207) *[blue]{\scriptscriptstyle+};
(4.632444,0.450048) *[blue]{\scriptscriptstyle+};
(4.640657,0.457210) *[blue]{\scriptscriptstyle+};
(4.648871,0.459211) *[blue]{\scriptscriptstyle+};
(4.657084,0.459554) *[blue]{\scriptscriptstyle+};
(4.665298,0.473476) *[blue]{\scriptscriptstyle+};
(4.673511,0.475431) *[blue]{\scriptscriptstyle+};
(4.681725,0.408887) *[blue]{\scriptscriptstyle+};
(4.689938,0.411476) *[blue]{\scriptscriptstyle+};
(4.698152,0.415165) *[blue]{\scriptscriptstyle+};
(4.706366,0.417161) *[blue]{\scriptscriptstyle+};
(4.714579,0.417977) *[blue]{\scriptscriptstyle+};
(4.722793,0.422336) *[blue]{\scriptscriptstyle+};
(4.731006,0.430534) *[blue]{\scriptscriptstyle+};
(4.739220,0.443568) *[blue]{\scriptscriptstyle+};
(4.747433,0.449459) *[blue]{\scriptscriptstyle+};
(4.755647,0.455218) *[blue]{\scriptscriptstyle+};
(4.763860,0.494794) *[blue]{\scriptscriptstyle+};
(4.772074,0.505326) *[blue]{\scriptscriptstyle+};
(4.780287,0.519182) *[blue]{\scriptscriptstyle+};
(4.788501,0.542079) *[blue]{\scriptscriptstyle+};
(4.796715,0.561438) *[blue]{\scriptscriptstyle+};
(4.804928,0.425628) *[blue]{\scriptscriptstyle+};
(4.813142,0.428886) *[blue]{\scriptscriptstyle+};
(4.821355,0.442101) *[blue]{\scriptscriptstyle+};
(4.829569,0.445171) *[blue]{\scriptscriptstyle+};
(4.837782,0.445892) *[blue]{\scriptscriptstyle+};
(4.845996,0.446367) *[blue]{\scriptscriptstyle+};
(4.854209,0.453791) *[blue]{\scriptscriptstyle+};
(4.862423,0.457601) *[blue]{\scriptscriptstyle+};
(4.870637,0.461843) *[blue]{\scriptscriptstyle+};
(4.878850,0.467913) *[blue]{\scriptscriptstyle+};
(4.887064,0.470438) *[blue]{\scriptscriptstyle+};
(4.895277,0.478127) *[blue]{\scriptscriptstyle+};
(4.903491,0.487667) *[blue]{\scriptscriptstyle+};
(4.911704,0.509239) *[blue]{\scriptscriptstyle+};
(4.919918,0.528489) *[blue]{\scriptscriptstyle+};
(4.928131,0.428898) *[blue]{\scriptscriptstyle+};
(4.936345,0.432657) *[blue]{\scriptscriptstyle+};
(4.944559,0.435280) *[blue]{\scriptscriptstyle+};
(4.952772,0.436285) *[blue]{\scriptscriptstyle+};
(4.960986,0.438873) *[blue]{\scriptscriptstyle+};
(4.969199,0.440001) *[blue]{\scriptscriptstyle+};
(4.977413,0.450314) *[blue]{\scriptscriptstyle+};
(4.985626,0.451617) *[blue]{\scriptscriptstyle+};
(4.993840,0.452609) *[blue]{\scriptscriptstyle+};
(5.002053,0.453207) *[blue]{\scriptscriptstyle+};
(5.010267,0.453416) *[blue]{\scriptscriptstyle+};
(5.018480,0.458344) *[blue]{\scriptscriptstyle+};
(5.026694,0.469277) *[blue]{\scriptscriptstyle+};
(5.034908,0.474720) *[blue]{\scriptscriptstyle+};
(5.043121,0.510399) *[blue]{\scriptscriptstyle+};
(5.051335,0.411322) *[blue]{\scriptscriptstyle+};
(5.059548,0.415783) *[blue]{\scriptscriptstyle+};
(5.067762,0.435548) *[blue]{\scriptscriptstyle+};
(5.075975,0.438215) *[blue]{\scriptscriptstyle+};
(5.084189,0.439421) *[blue]{\scriptscriptstyle+};
(5.092402,0.448654) *[blue]{\scriptscriptstyle+};
(5.100616,0.455200) *[blue]{\scriptscriptstyle+};
(5.108830,0.457028) *[blue]{\scriptscriptstyle+};
(5.117043,0.469587) *[blue]{\scriptscriptstyle+};
(5.125257,0.471619) *[blue]{\scriptscriptstyle+};
(5.133470,0.473301) *[blue]{\scriptscriptstyle+};
(5.141684,0.479109) *[blue]{\scriptscriptstyle+};
(5.149897,0.486264) *[blue]{\scriptscriptstyle+};
(5.158111,0.506361) *[blue]{\scriptscriptstyle+};
(5.166324,0.508388) *[blue]{\scriptscriptstyle+};
(5.174538,0.442669) *[blue]{\scriptscriptstyle+};
(5.182752,0.444964) *[blue]{\scriptscriptstyle+};
(5.190965,0.450905) *[blue]{\scriptscriptstyle+};
(5.199179,0.455711) *[blue]{\scriptscriptstyle+};
(5.207392,0.456979) *[blue]{\scriptscriptstyle+};
(5.215606,0.457047) *[blue]{\scriptscriptstyle+};
(5.223819,0.464725) *[blue]{\scriptscriptstyle+};
(5.232033,0.465806) *[blue]{\scriptscriptstyle+};
(5.240246,0.474408) *[blue]{\scriptscriptstyle+};
(5.248460,0.478971) *[blue]{\scriptscriptstyle+};
(5.256674,0.479087) *[blue]{\scriptscriptstyle+};
(5.264887,0.483511) *[blue]{\scriptscriptstyle+};
(5.273101,0.484446) *[blue]{\scriptscriptstyle+};
(5.281314,0.517110) *[blue]{\scriptscriptstyle+};
(5.289528,0.524194) *[blue]{\scriptscriptstyle+};
(5.297741,0.449600) *[blue]{\scriptscriptstyle+};
(5.305955,0.449678) *[blue]{\scriptscriptstyle+};
(5.314168,0.450468) *[blue]{\scriptscriptstyle+};
(5.322382,0.450984) *[blue]{\scriptscriptstyle+};
(5.330595,0.457072) *[blue]{\scriptscriptstyle+};
(5.338809,0.459699) *[blue]{\scriptscriptstyle+};
(5.347023,0.462399) *[blue]{\scriptscriptstyle+};
(5.355236,0.476592) *[blue]{\scriptscriptstyle+};
(5.363450,0.486465) *[blue]{\scriptscriptstyle+};
(5.371663,0.487579) *[blue]{\scriptscriptstyle+};
(5.379877,0.495839) *[blue]{\scriptscriptstyle+};
(5.388090,0.512093) *[blue]{\scriptscriptstyle+};
(5.396304,0.521964) *[blue]{\scriptscriptstyle+};
(5.404517,0.529010) *[blue]{\scriptscriptstyle+};
(5.412731,0.535191) *[blue]{\scriptscriptstyle+};
(5.420945,0.466547) *[blue]{\scriptscriptstyle+};
(5.429158,0.467787) *[blue]{\scriptscriptstyle+};
(5.437372,0.472940) *[blue]{\scriptscriptstyle+};
(5.445585,0.476070) *[blue]{\scriptscriptstyle+};
(5.453799,0.476237) *[blue]{\scriptscriptstyle+};
(5.462012,0.478539) *[blue]{\scriptscriptstyle+};
(5.470226,0.479970) *[blue]{\scriptscriptstyle+};
(5.478439,0.484880) *[blue]{\scriptscriptstyle+};
(5.486653,0.486082) *[blue]{\scriptscriptstyle+};
(5.494867,0.487090) *[blue]{\scriptscriptstyle+};
(5.503080,0.488960) *[blue]{\scriptscriptstyle+};
(5.511294,0.489885) *[blue]{\scriptscriptstyle+};
(5.519507,0.490092) *[blue]{\scriptscriptstyle+};
(5.527721,0.491058) *[blue]{\scriptscriptstyle+};
(5.535934,0.518012) *[blue]{\scriptscriptstyle+};
(5.544148,0.467175) *[blue]{\scriptscriptstyle+};
(5.552361,0.469179) *[blue]{\scriptscriptstyle+};
(5.560575,0.469464) *[blue]{\scriptscriptstyle+};
(5.568789,0.471084) *[blue]{\scriptscriptstyle+};
(5.577002,0.471461) *[blue]{\scriptscriptstyle+};
(5.585216,0.474802) *[blue]{\scriptscriptstyle+};
(5.593429,0.475906) *[blue]{\scriptscriptstyle+};
(5.601643,0.476647) *[blue]{\scriptscriptstyle+};
(5.609856,0.481440) *[blue]{\scriptscriptstyle+};
(5.618070,0.485181) *[blue]{\scriptscriptstyle+};
(5.626283,0.485965) *[blue]{\scriptscriptstyle+};
(5.634497,0.490572) *[blue]{\scriptscriptstyle+};
(5.642710,0.500673) *[blue]{\scriptscriptstyle+};
(5.650924,0.511431) *[blue]{\scriptscriptstyle+};
(5.659138,0.539842) *[blue]{\scriptscriptstyle+};
(5.667351,0.493890) *[blue]{\scriptscriptstyle+};
(5.675565,0.493976) *[blue]{\scriptscriptstyle+};
(5.683778,0.494811) *[blue]{\scriptscriptstyle+};
(5.691992,0.494847) *[blue]{\scriptscriptstyle+};
(5.700205,0.498104) *[blue]{\scriptscriptstyle+};
(5.708419,0.503377) *[blue]{\scriptscriptstyle+};
(5.716632,0.506561) *[blue]{\scriptscriptstyle+};
(5.724846,0.515256) *[blue]{\scriptscriptstyle+};
(5.733060,0.517650) *[blue]{\scriptscriptstyle+};
(5.741273,0.518737) *[blue]{\scriptscriptstyle+};
(5.749487,0.523323) *[blue]{\scriptscriptstyle+};
(5.757700,0.526996) *[blue]{\scriptscriptstyle+};
(5.765914,0.527018) *[blue]{\scriptscriptstyle+};
(5.774127,0.537194) *[blue]{\scriptscriptstyle+};
(5.782341,0.552087) *[blue]{\scriptscriptstyle+};
(5.790554,0.477489) *[blue]{\scriptscriptstyle+};
(5.798768,0.478549) *[blue]{\scriptscriptstyle+};
(5.806982,0.479272) *[blue]{\scriptscriptstyle+};
(5.815195,0.479759) *[blue]{\scriptscriptstyle+};
(5.823409,0.486591) *[blue]{\scriptscriptstyle+};
(5.831622,0.486653) *[blue]{\scriptscriptstyle+};
(5.839836,0.488335) *[blue]{\scriptscriptstyle+};
(5.848049,0.491537) *[blue]{\scriptscriptstyle+};
(5.856263,0.493616) *[blue]{\scriptscriptstyle+};
(5.864476,0.506152) *[blue]{\scriptscriptstyle+};
(5.872690,0.506793) *[blue]{\scriptscriptstyle+};
(5.880903,0.507667) *[blue]{\scriptscriptstyle+};
(5.889117,0.516494) *[blue]{\scriptscriptstyle+};
(5.897331,0.520857) *[blue]{\scriptscriptstyle+};
(5.905544,0.577771) *[blue]{\scriptscriptstyle+};
(5.913758,0.479252) *[blue]{\scriptscriptstyle+};
(5.921971,0.488575) *[blue]{\scriptscriptstyle+};
(5.930185,0.490428) *[blue]{\scriptscriptstyle+};
(5.938398,0.492737) *[blue]{\scriptscriptstyle+};
(5.946612,0.493992) *[blue]{\scriptscriptstyle+};
(5.954825,0.494921) *[blue]{\scriptscriptstyle+};
(5.963039,0.497459) *[blue]{\scriptscriptstyle+};
(5.971253,0.498439) *[blue]{\scriptscriptstyle+};
(5.979466,0.499164) *[blue]{\scriptscriptstyle+};
(5.987680,0.502896) *[blue]{\scriptscriptstyle+};
(5.995893,0.516601) *[blue]{\scriptscriptstyle+};
(6.004107,0.528671) *[blue]{\scriptscriptstyle+};
(6.012320,0.536653) *[blue]{\scriptscriptstyle+};
(6.020534,0.545907) *[blue]{\scriptscriptstyle+};
(6.028747,0.550892) *[blue]{\scriptscriptstyle+};
(6.036961,0.486266) *[blue]{\scriptscriptstyle+};
(6.045175,0.488519) *[blue]{\scriptscriptstyle+};
(6.053388,0.497894) *[blue]{\scriptscriptstyle+};
(6.061602,0.504957) *[blue]{\scriptscriptstyle+};
(6.069815,0.505936) *[blue]{\scriptscriptstyle+};
(6.078029,0.509724) *[blue]{\scriptscriptstyle+};
(6.086242,0.514758) *[blue]{\scriptscriptstyle+};
(6.094456,0.518564) *[blue]{\scriptscriptstyle+};
(6.102669,0.521816) *[blue]{\scriptscriptstyle+};
(6.110883,0.522916) *[blue]{\scriptscriptstyle+};
(6.119097,0.523366) *[blue]{\scriptscriptstyle+};
(6.127310,0.533542) *[blue]{\scriptscriptstyle+};
(6.135524,0.534831) *[blue]{\scriptscriptstyle+};
(6.143737,0.535869) *[blue]{\scriptscriptstyle+};
(6.151951,0.540410) *[blue]{\scriptscriptstyle+};
(6.160164,0.490836) *[blue]{\scriptscriptstyle+};
(6.168378,0.495987) *[blue]{\scriptscriptstyle+};
(6.176591,0.496074) *[blue]{\scriptscriptstyle+};
(6.184805,0.499931) *[blue]{\scriptscriptstyle+};
(6.193018,0.500279) *[blue]{\scriptscriptstyle+};
(6.201232,0.500618) *[blue]{\scriptscriptstyle+};
(6.209446,0.509884) *[blue]{\scriptscriptstyle+};
(6.217659,0.514917) *[blue]{\scriptscriptstyle+};
(6.225873,0.514966) *[blue]{\scriptscriptstyle+};
(6.234086,0.515067) *[blue]{\scriptscriptstyle+};
(6.242300,0.523494) *[blue]{\scriptscriptstyle+};
(6.250513,0.534583) *[blue]{\scriptscriptstyle+};
(6.258727,0.539677) *[blue]{\scriptscriptstyle+};
(6.266940,0.571856) *[blue]{\scriptscriptstyle+};
(6.275154,0.625369) *[blue]{\scriptscriptstyle+};
(6.283368,0.491987) *[blue]{\scriptscriptstyle+};
(6.291581,0.492217) *[blue]{\scriptscriptstyle+};
(6.299795,0.493033) *[blue]{\scriptscriptstyle+};
(6.308008,0.494718) *[blue]{\scriptscriptstyle+};
(6.316222,0.496459) *[blue]{\scriptscriptstyle+};
(6.324435,0.496672) *[blue]{\scriptscriptstyle+};
(6.332649,0.513923) *[blue]{\scriptscriptstyle+};
(6.340862,0.514135) *[blue]{\scriptscriptstyle+};
(6.349076,0.514932) *[blue]{\scriptscriptstyle+};
(6.357290,0.515270) *[blue]{\scriptscriptstyle+};
(6.365503,0.515881) *[blue]{\scriptscriptstyle+};
(6.373717,0.517683) *[blue]{\scriptscriptstyle+};
(6.381930,0.520457) *[blue]{\scriptscriptstyle+};
(6.390144,0.568898) *[blue]{\scriptscriptstyle+};
(6.398357,0.645567) *[blue]{\scriptscriptstyle+};
(6.406571,0.515225) *[blue]{\scriptscriptstyle+};
(6.414784,0.515282) *[blue]{\scriptscriptstyle+};
(6.422998,0.518195) *[blue]{\scriptscriptstyle+};
(6.431211,0.518303) *[blue]{\scriptscriptstyle+};
(6.439425,0.524729) *[blue]{\scriptscriptstyle+};
(6.447639,0.525441) *[blue]{\scriptscriptstyle+};
(6.455852,0.527768) *[blue]{\scriptscriptstyle+};
(6.464066,0.532471) *[blue]{\scriptscriptstyle+};
(6.472279,0.532821) *[blue]{\scriptscriptstyle+};
(6.480493,0.533486) *[blue]{\scriptscriptstyle+};
(6.488706,0.540567) *[blue]{\scriptscriptstyle+};
(6.496920,0.542028) *[blue]{\scriptscriptstyle+};
(6.505133,0.565219) *[blue]{\scriptscriptstyle+};
(6.513347,0.565626) *[blue]{\scriptscriptstyle+};
(6.521561,0.572466) *[blue]{\scriptscriptstyle+};
(6.529774,0.519408) *[blue]{\scriptscriptstyle+};
(6.537988,0.521762) *[blue]{\scriptscriptstyle+};
(6.546201,0.532546) *[blue]{\scriptscriptstyle+};
(6.554415,0.535430) *[blue]{\scriptscriptstyle+};
(6.562628,0.536681) *[blue]{\scriptscriptstyle+};
(6.570842,0.546009) *[blue]{\scriptscriptstyle+};
(6.579055,0.552932) *[blue]{\scriptscriptstyle+};
(6.587269,0.553417) *[blue]{\scriptscriptstyle+};
(6.595483,0.567947) *[blue]{\scriptscriptstyle+};
(6.603696,0.571407) *[blue]{\scriptscriptstyle+};
(6.611910,0.586283) *[blue]{\scriptscriptstyle+};
(6.620123,0.586318) *[blue]{\scriptscriptstyle+};
(6.628337,0.608622) *[blue]{\scriptscriptstyle+};
(6.636550,0.619751) *[blue]{\scriptscriptstyle+};
(6.644764,0.623274) *[blue]{\scriptscriptstyle+};
(6.652977,0.521984) *[blue]{\scriptscriptstyle+};
(6.661191,0.526195) *[blue]{\scriptscriptstyle+};
(6.669405,0.537109) *[blue]{\scriptscriptstyle+};
(6.677618,0.537227) *[blue]{\scriptscriptstyle+};
(6.685832,0.550638) *[blue]{\scriptscriptstyle+};
(6.694045,0.553449) *[blue]{\scriptscriptstyle+};
(6.702259,0.571190) *[blue]{\scriptscriptstyle+};
(6.710472,0.584421) *[blue]{\scriptscriptstyle+};
(6.718686,0.585644) *[blue]{\scriptscriptstyle+};
(6.726899,0.589653) *[blue]{\scriptscriptstyle+};
(6.735113,0.604281) *[blue]{\scriptscriptstyle+};
(6.743326,0.605929) *[blue]{\scriptscriptstyle+};
(6.751540,0.623277) *[blue]{\scriptscriptstyle+};
(6.759754,0.643257) *[blue]{\scriptscriptstyle+};
(6.767967,0.727823) *[blue]{\scriptscriptstyle+};
(6.776181,0.525310) *[blue]{\scriptscriptstyle+};
(6.784394,0.527782) *[blue]{\scriptscriptstyle+};
(6.792608,0.546614) *[blue]{\scriptscriptstyle+};
(6.800821,0.547154) *[blue]{\scriptscriptstyle+};
(6.809035,0.548549) *[blue]{\scriptscriptstyle+};
(6.817248,0.549816) *[blue]{\scriptscriptstyle+};
(6.825462,0.551250) *[blue]{\scriptscriptstyle+};
(6.833676,0.554956) *[blue]{\scriptscriptstyle+};
(6.841889,0.565056) *[blue]{\scriptscriptstyle+};
(6.850103,0.571327) *[blue]{\scriptscriptstyle+};
(6.858316,0.585149) *[blue]{\scriptscriptstyle+};
(6.866530,0.587522) *[blue]{\scriptscriptstyle+};
(6.874743,0.587861) *[blue]{\scriptscriptstyle+};
(6.882957,0.591973) *[blue]{\scriptscriptstyle+};
(6.891170,0.612829) *[blue]{\scriptscriptstyle+};
(6.899384,0.543939) *[blue]{\scriptscriptstyle+};
(6.907598,0.547983) *[blue]{\scriptscriptstyle+};
(6.915811,0.554898) *[blue]{\scriptscriptstyle+};
(6.924025,0.558192) *[blue]{\scriptscriptstyle+};
(6.932238,0.564620) *[blue]{\scriptscriptstyle+};
(6.940452,0.565101) *[blue]{\scriptscriptstyle+};
(6.948665,0.567015) *[blue]{\scriptscriptstyle+};
(6.956879,0.570175) *[blue]{\scriptscriptstyle+};
(6.965092,0.572195) *[blue]{\scriptscriptstyle+};
(6.973306,0.579657) *[blue]{\scriptscriptstyle+};
(6.981520,0.583434) *[blue]{\scriptscriptstyle+};
(6.989733,0.588130) *[blue]{\scriptscriptstyle+};
(6.997947,0.593264) *[blue]{\scriptscriptstyle+};
(7.006160,0.604128) *[blue]{\scriptscriptstyle+};
(7.014374,0.609975) *[blue]{\scriptscriptstyle+};
(7.022587,0.567277) *[blue]{\scriptscriptstyle+};
(7.030801,0.570998) *[blue]{\scriptscriptstyle+};
(7.039014,0.580310) *[blue]{\scriptscriptstyle+};
(7.047228,0.582124) *[blue]{\scriptscriptstyle+};
(7.055441,0.583718) *[blue]{\scriptscriptstyle+};
(7.063655,0.590761) *[blue]{\scriptscriptstyle+};
(7.071869,0.594532) *[blue]{\scriptscriptstyle+};
(7.080082,0.599225) *[blue]{\scriptscriptstyle+};
(7.088296,0.607841) *[blue]{\scriptscriptstyle+};
(7.096509,0.612830) *[blue]{\scriptscriptstyle+};
(7.104723,0.620077) *[blue]{\scriptscriptstyle+};
(7.112936,0.620410) *[blue]{\scriptscriptstyle+};
(7.121150,0.620498) *[blue]{\scriptscriptstyle+};
(7.129363,0.621592) *[blue]{\scriptscriptstyle+};
(7.137577,0.657617) *[blue]{\scriptscriptstyle+};
(7.145791,0.617460) *[blue]{\scriptscriptstyle+};
(7.154004,0.617959) *[blue]{\scriptscriptstyle+};
(7.162218,0.627918) *[blue]{\scriptscriptstyle+};
(7.170431,0.631192) *[blue]{\scriptscriptstyle+};
(7.178645,0.637016) *[blue]{\scriptscriptstyle+};
(7.186858,0.639909) *[blue]{\scriptscriptstyle+};
(7.195072,0.641297) *[blue]{\scriptscriptstyle+};
(7.203285,0.641471) *[blue]{\scriptscriptstyle+};
(7.211499,0.642199) *[blue]{\scriptscriptstyle+};
(7.219713,0.642686) *[blue]{\scriptscriptstyle+};
(7.227926,0.642753) *[blue]{\scriptscriptstyle+};
(7.236140,0.646358) *[blue]{\scriptscriptstyle+};
(7.244353,0.648285) *[blue]{\scriptscriptstyle+};
(7.252567,0.681248) *[blue]{\scriptscriptstyle+};
(7.260780,0.695859) *[blue]{\scriptscriptstyle+};
(7.268994,0.613365) *[blue]{\scriptscriptstyle+};
(7.277207,0.626476) *[blue]{\scriptscriptstyle+};
(7.285421,0.632899) *[blue]{\scriptscriptstyle+};
(7.293634,0.638546) *[blue]{\scriptscriptstyle+};
(7.301848,0.640214) *[blue]{\scriptscriptstyle+};
(7.310062,0.647397) *[blue]{\scriptscriptstyle+};
(7.318275,0.657001) *[blue]{\scriptscriptstyle+};
(7.326489,0.662201) *[blue]{\scriptscriptstyle+};
(7.334702,0.669166) *[blue]{\scriptscriptstyle+};
(7.342916,0.677403) *[blue]{\scriptscriptstyle+};
(7.351129,0.680843) *[blue]{\scriptscriptstyle+};
(7.359343,0.689881) *[blue]{\scriptscriptstyle+};
(7.367556,0.696880) *[blue]{\scriptscriptstyle+};
(7.375770,0.697056) *[blue]{\scriptscriptstyle+};
(7.383984,0.716716) *[blue]{\scriptscriptstyle+};
(7.392197,0.614534) *[blue]{\scriptscriptstyle+};
(7.400411,0.618337) *[blue]{\scriptscriptstyle+};
(7.408624,0.620040) *[blue]{\scriptscriptstyle+};
(7.416838,0.620571) *[blue]{\scriptscriptstyle+};
(7.425051,0.634089) *[blue]{\scriptscriptstyle+};
(7.433265,0.635360) *[blue]{\scriptscriptstyle+};
(7.441478,0.636897) *[blue]{\scriptscriptstyle+};
(7.449692,0.645108) *[blue]{\scriptscriptstyle+};
(7.457906,0.651054) *[blue]{\scriptscriptstyle+};
(7.466119,0.652473) *[blue]{\scriptscriptstyle+};
(7.474333,0.653180) *[blue]{\scriptscriptstyle+};
(7.482546,0.672179) *[blue]{\scriptscriptstyle+};
(7.490760,0.684015) *[blue]{\scriptscriptstyle+};
(7.498973,0.700700) *[blue]{\scriptscriptstyle+};
(7.507187,0.730210) *[blue]{\scriptscriptstyle+};
(7.515400,0.627812) *[blue]{\scriptscriptstyle+};
(7.523614,0.629626) *[blue]{\scriptscriptstyle+};
(7.531828,0.631712) *[blue]{\scriptscriptstyle+};
(7.540041,0.641834) *[blue]{\scriptscriptstyle+};
(7.548255,0.644819) *[blue]{\scriptscriptstyle+};
(7.556468,0.649474) *[blue]{\scriptscriptstyle+};
(7.564682,0.654194) *[blue]{\scriptscriptstyle+};
(7.572895,0.654983) *[blue]{\scriptscriptstyle+};
(7.581109,0.655722) *[blue]{\scriptscriptstyle+};
(7.589322,0.658946) *[blue]{\scriptscriptstyle+};
(7.597536,0.660486) *[blue]{\scriptscriptstyle+};
(7.605749,0.660611) *[blue]{\scriptscriptstyle+};
(7.613963,0.669741) *[blue]{\scriptscriptstyle+};
(7.622177,0.682972) *[blue]{\scriptscriptstyle+};
(7.630390,0.685922) *[blue]{\scriptscriptstyle+};
(7.638604,0.629551) *[blue]{\scriptscriptstyle+};
(7.646817,0.631865) *[blue]{\scriptscriptstyle+};
(7.655031,0.637614) *[blue]{\scriptscriptstyle+};
(7.663244,0.643732) *[blue]{\scriptscriptstyle+};
(7.671458,0.643873) *[blue]{\scriptscriptstyle+};
(7.679671,0.647318) *[blue]{\scriptscriptstyle+};
(7.687885,0.648281) *[blue]{\scriptscriptstyle+};
(7.696099,0.656361) *[blue]{\scriptscriptstyle+};
(7.704312,0.666056) *[blue]{\scriptscriptstyle+};
(7.712526,0.666174) *[blue]{\scriptscriptstyle+};
(7.720739,0.666756) *[blue]{\scriptscriptstyle+};
(7.728953,0.667547) *[blue]{\scriptscriptstyle+};
(7.737166,0.668224) *[blue]{\scriptscriptstyle+};
(7.745380,0.668587) *[blue]{\scriptscriptstyle+};
(7.753593,0.685317) *[blue]{\scriptscriptstyle+};
(7.761807,0.626695) *[blue]{\scriptscriptstyle+};
(7.770021,0.630164) *[blue]{\scriptscriptstyle+};
(7.778234,0.633055) *[blue]{\scriptscriptstyle+};
(7.786448,0.644783) *[blue]{\scriptscriptstyle+};
(7.794661,0.645421) *[blue]{\scriptscriptstyle+};
(7.802875,0.647957) *[blue]{\scriptscriptstyle+};
(7.811088,0.648228) *[blue]{\scriptscriptstyle+};
(7.819302,0.648369) *[blue]{\scriptscriptstyle+};
(7.827515,0.653387) *[blue]{\scriptscriptstyle+};
(7.835729,0.660001) *[blue]{\scriptscriptstyle+};
(7.843943,0.663379) *[blue]{\scriptscriptstyle+};
(7.852156,0.663919) *[blue]{\scriptscriptstyle+};
(7.860370,0.671283) *[blue]{\scriptscriptstyle+};
(7.868583,0.674241) *[blue]{\scriptscriptstyle+};
(7.876797,0.694487) *[blue]{\scriptscriptstyle+};
(7.885010,0.641475) *[blue]{\scriptscriptstyle+};
(7.893224,0.648839) *[blue]{\scriptscriptstyle+};
(7.901437,0.650199) *[blue]{\scriptscriptstyle+};
(7.909651,0.652455) *[blue]{\scriptscriptstyle+};
(7.917864,0.659419) *[blue]{\scriptscriptstyle+};
(7.926078,0.664946) *[blue]{\scriptscriptstyle+};
(7.934292,0.665754) *[blue]{\scriptscriptstyle+};
(7.942505,0.665850) *[blue]{\scriptscriptstyle+};
(7.950719,0.667620) *[blue]{\scriptscriptstyle+};
(7.958932,0.668909) *[blue]{\scriptscriptstyle+};
(7.967146,0.672410) *[blue]{\scriptscriptstyle+};
(7.975359,0.689879) *[blue]{\scriptscriptstyle+};
(7.983573,0.693947) *[blue]{\scriptscriptstyle+};
(7.991786,0.698444) *[blue]{\scriptscriptstyle+};
(8.000000,0.700709) *[blue]{\scriptscriptstyle+};
(0.000000,0.063183) *[red]{\scriptscriptstyle\times};
(0.008214,0.064997) *[red]{\scriptscriptstyle\times};
(0.016427,0.069737) *[red]{\scriptscriptstyle\times};
(0.024641,0.079190) *[red]{\scriptscriptstyle\times};
(0.032854,0.083503) *[red]{\scriptscriptstyle\times};
(0.041068,0.090763) *[red]{\scriptscriptstyle\times};
(0.049281,0.092575) *[red]{\scriptscriptstyle\times};
(0.057495,0.094480) *[red]{\scriptscriptstyle\times};
(0.065708,0.094493) *[red]{\scriptscriptstyle\times};
(0.073922,0.099030) *[red]{\scriptscriptstyle\times};
(0.082136,0.118341) *[red]{\scriptscriptstyle\times};
(0.090349,0.121825) *[red]{\scriptscriptstyle\times};
(0.098563,0.122400) *[red]{\scriptscriptstyle\times};
(0.106776,0.125576) *[red]{\scriptscriptstyle\times};
(0.114990,0.139494) *[red]{\scriptscriptstyle\times};
(0.123203,0.091906) *[red]{\scriptscriptstyle\times};
(0.131417,0.093663) *[red]{\scriptscriptstyle\times};
(0.139630,0.098280) *[red]{\scriptscriptstyle\times};
(0.147844,0.105615) *[red]{\scriptscriptstyle\times};
(0.156057,0.115605) *[red]{\scriptscriptstyle\times};
(0.164271,0.117594) *[red]{\scriptscriptstyle\times};
(0.172485,0.117861) *[red]{\scriptscriptstyle\times};
(0.180698,0.119886) *[red]{\scriptscriptstyle\times};
(0.188912,0.120495) *[red]{\scriptscriptstyle\times};
(0.197125,0.122493) *[red]{\scriptscriptstyle\times};
(0.205339,0.123754) *[red]{\scriptscriptstyle\times};
(0.213552,0.126135) *[red]{\scriptscriptstyle\times};
(0.221766,0.127575) *[red]{\scriptscriptstyle\times};
(0.229979,0.151580) *[red]{\scriptscriptstyle\times};
(0.238193,0.215614) *[red]{\scriptscriptstyle\times};
(0.246407,0.098174) *[red]{\scriptscriptstyle\times};
(0.254620,0.098565) *[red]{\scriptscriptstyle\times};
(0.262834,0.100392) *[red]{\scriptscriptstyle\times};
(0.271047,0.100618) *[red]{\scriptscriptstyle\times};
(0.279261,0.100819) *[red]{\scriptscriptstyle\times};
(0.287474,0.101665) *[red]{\scriptscriptstyle\times};
(0.295688,0.101995) *[red]{\scriptscriptstyle\times};
(0.303901,0.106028) *[red]{\scriptscriptstyle\times};
(0.312115,0.119089) *[red]{\scriptscriptstyle\times};
(0.320329,0.123671) *[red]{\scriptscriptstyle\times};
(0.328542,0.128041) *[red]{\scriptscriptstyle\times};
(0.336756,0.130132) *[red]{\scriptscriptstyle\times};
(0.344969,0.131142) *[red]{\scriptscriptstyle\times};
(0.353183,0.133579) *[red]{\scriptscriptstyle\times};
(0.361396,0.189353) *[red]{\scriptscriptstyle\times};
(0.369610,0.074052) *[red]{\scriptscriptstyle\times};
(0.377823,0.079598) *[red]{\scriptscriptstyle\times};
(0.386037,0.088097) *[red]{\scriptscriptstyle\times};
(0.394251,0.096111) *[red]{\scriptscriptstyle\times};
(0.402464,0.096309) *[red]{\scriptscriptstyle\times};
(0.410678,0.099956) *[red]{\scriptscriptstyle\times};
(0.418891,0.100147) *[red]{\scriptscriptstyle\times};
(0.427105,0.112430) *[red]{\scriptscriptstyle\times};
(0.435318,0.113897) *[red]{\scriptscriptstyle\times};
(0.443532,0.128183) *[red]{\scriptscriptstyle\times};
(0.451745,0.134739) *[red]{\scriptscriptstyle\times};
(0.459959,0.149449) *[red]{\scriptscriptstyle\times};
(0.468172,0.168594) *[red]{\scriptscriptstyle\times};
(0.476386,0.219638) *[red]{\scriptscriptstyle\times};
(0.484600,0.244920) *[red]{\scriptscriptstyle\times};
(0.492813,0.126428) *[red]{\scriptscriptstyle\times};
(0.501027,0.127487) *[red]{\scriptscriptstyle\times};
(0.509240,0.134816) *[red]{\scriptscriptstyle\times};
(0.517454,0.139801) *[red]{\scriptscriptstyle\times};
(0.525667,0.148684) *[red]{\scriptscriptstyle\times};
(0.533881,0.154849) *[red]{\scriptscriptstyle\times};
(0.542094,0.155463) *[red]{\scriptscriptstyle\times};
(0.550308,0.162424) *[red]{\scriptscriptstyle\times};
(0.558522,0.163517) *[red]{\scriptscriptstyle\times};
(0.566735,0.175679) *[red]{\scriptscriptstyle\times};
(0.574949,0.183144) *[red]{\scriptscriptstyle\times};
(0.583162,0.193456) *[red]{\scriptscriptstyle\times};
(0.591376,0.194285) *[red]{\scriptscriptstyle\times};
(0.599589,0.203137) *[red]{\scriptscriptstyle\times};
(0.607803,0.276738) *[red]{\scriptscriptstyle\times};
(0.616016,0.123596) *[red]{\scriptscriptstyle\times};
(0.624230,0.124818) *[red]{\scriptscriptstyle\times};
(0.632444,0.126623) *[red]{\scriptscriptstyle\times};
(0.640657,0.129164) *[red]{\scriptscriptstyle\times};
(0.648871,0.146194) *[red]{\scriptscriptstyle\times};
(0.657084,0.147782) *[red]{\scriptscriptstyle\times};
(0.665298,0.150566) *[red]{\scriptscriptstyle\times};
(0.673511,0.165969) *[red]{\scriptscriptstyle\times};
(0.681725,0.168086) *[red]{\scriptscriptstyle\times};
(0.689938,0.177689) *[red]{\scriptscriptstyle\times};
(0.698152,0.181807) *[red]{\scriptscriptstyle\times};
(0.706366,0.219280) *[red]{\scriptscriptstyle\times};
(0.714579,0.222932) *[red]{\scriptscriptstyle\times};
(0.722793,0.238067) *[red]{\scriptscriptstyle\times};
(0.731006,0.265167) *[red]{\scriptscriptstyle\times};
(0.739220,0.114282) *[red]{\scriptscriptstyle\times};
(0.747433,0.119744) *[red]{\scriptscriptstyle\times};
(0.755647,0.137222) *[red]{\scriptscriptstyle\times};
(0.763860,0.148985) *[red]{\scriptscriptstyle\times};
(0.772074,0.167547) *[red]{\scriptscriptstyle\times};
(0.780287,0.173113) *[red]{\scriptscriptstyle\times};
(0.788501,0.186380) *[red]{\scriptscriptstyle\times};
(0.796715,0.186596) *[red]{\scriptscriptstyle\times};
(0.804928,0.187366) *[red]{\scriptscriptstyle\times};
(0.813142,0.190871) *[red]{\scriptscriptstyle\times};
(0.821355,0.196869) *[red]{\scriptscriptstyle\times};
(0.829569,0.216130) *[red]{\scriptscriptstyle\times};
(0.837782,0.217160) *[red]{\scriptscriptstyle\times};
(0.845996,0.266669) *[red]{\scriptscriptstyle\times};
(0.854209,0.290902) *[red]{\scriptscriptstyle\times};
(0.862423,0.166161) *[red]{\scriptscriptstyle\times};
(0.870637,0.169771) *[red]{\scriptscriptstyle\times};
(0.878850,0.174398) *[red]{\scriptscriptstyle\times};
(0.887064,0.178210) *[red]{\scriptscriptstyle\times};
(0.895277,0.192365) *[red]{\scriptscriptstyle\times};
(0.903491,0.198265) *[red]{\scriptscriptstyle\times};
(0.911704,0.201976) *[red]{\scriptscriptstyle\times};
(0.919918,0.203851) *[red]{\scriptscriptstyle\times};
(0.928131,0.211181) *[red]{\scriptscriptstyle\times};
(0.936345,0.213685) *[red]{\scriptscriptstyle\times};
(0.944559,0.217056) *[red]{\scriptscriptstyle\times};
(0.952772,0.221299) *[red]{\scriptscriptstyle\times};
(0.960986,0.230154) *[red]{\scriptscriptstyle\times};
(0.969199,0.236178) *[red]{\scriptscriptstyle\times};
(0.977413,0.245510) *[red]{\scriptscriptstyle\times};
(0.985626,0.150247) *[red]{\scriptscriptstyle\times};
(0.993840,0.168547) *[red]{\scriptscriptstyle\times};
(1.002053,0.170497) *[red]{\scriptscriptstyle\times};
(1.010267,0.199267) *[red]{\scriptscriptstyle\times};
(1.018480,0.200802) *[red]{\scriptscriptstyle\times};
(1.026694,0.200957) *[red]{\scriptscriptstyle\times};
(1.034908,0.204431) *[red]{\scriptscriptstyle\times};
(1.043121,0.211735) *[red]{\scriptscriptstyle\times};
(1.051335,0.217272) *[red]{\scriptscriptstyle\times};
(1.059548,0.224437) *[red]{\scriptscriptstyle\times};
(1.067762,0.239982) *[red]{\scriptscriptstyle\times};
(1.075975,0.241043) *[red]{\scriptscriptstyle\times};
(1.084189,0.263735) *[red]{\scriptscriptstyle\times};
(1.092402,0.270289) *[red]{\scriptscriptstyle\times};
(1.100616,0.272001) *[red]{\scriptscriptstyle\times};
(1.108830,0.204461) *[red]{\scriptscriptstyle\times};
(1.117043,0.206990) *[red]{\scriptscriptstyle\times};
(1.125257,0.207142) *[red]{\scriptscriptstyle\times};
(1.133470,0.208530) *[red]{\scriptscriptstyle\times};
(1.141684,0.211366) *[red]{\scriptscriptstyle\times};
(1.149897,0.212099) *[red]{\scriptscriptstyle\times};
(1.158111,0.213738) *[red]{\scriptscriptstyle\times};
(1.166324,0.214236) *[red]{\scriptscriptstyle\times};
(1.174538,0.216781) *[red]{\scriptscriptstyle\times};
(1.182752,0.218198) *[red]{\scriptscriptstyle\times};
(1.190965,0.218530) *[red]{\scriptscriptstyle\times};
(1.199179,0.218968) *[red]{\scriptscriptstyle\times};
(1.207392,0.221076) *[red]{\scriptscriptstyle\times};
(1.215606,0.236679) *[red]{\scriptscriptstyle\times};
(1.223819,0.239720) *[red]{\scriptscriptstyle\times};
(1.232033,0.154132) *[red]{\scriptscriptstyle\times};
(1.240246,0.169693) *[red]{\scriptscriptstyle\times};
(1.248460,0.179191) *[red]{\scriptscriptstyle\times};
(1.256674,0.198350) *[red]{\scriptscriptstyle\times};
(1.264887,0.200376) *[red]{\scriptscriptstyle\times};
(1.273101,0.205535) *[red]{\scriptscriptstyle\times};
(1.281314,0.207265) *[red]{\scriptscriptstyle\times};
(1.289528,0.210304) *[red]{\scriptscriptstyle\times};
(1.297741,0.217420) *[red]{\scriptscriptstyle\times};
(1.305955,0.219737) *[red]{\scriptscriptstyle\times};
(1.314168,0.226025) *[red]{\scriptscriptstyle\times};
(1.322382,0.240343) *[red]{\scriptscriptstyle\times};
(1.330595,0.241831) *[red]{\scriptscriptstyle\times};
(1.338809,0.263602) *[red]{\scriptscriptstyle\times};
(1.347023,0.329492) *[red]{\scriptscriptstyle\times};
(1.355236,0.200241) *[red]{\scriptscriptstyle\times};
(1.363450,0.200301) *[red]{\scriptscriptstyle\times};
(1.371663,0.202148) *[red]{\scriptscriptstyle\times};
(1.379877,0.206277) *[red]{\scriptscriptstyle\times};
(1.388090,0.206564) *[red]{\scriptscriptstyle\times};
(1.396304,0.208950) *[red]{\scriptscriptstyle\times};
(1.404517,0.209983) *[red]{\scriptscriptstyle\times};
(1.412731,0.217788) *[red]{\scriptscriptstyle\times};
(1.420945,0.225698) *[red]{\scriptscriptstyle\times};
(1.429158,0.225771) *[red]{\scriptscriptstyle\times};
(1.437372,0.227944) *[red]{\scriptscriptstyle\times};
(1.445585,0.231466) *[red]{\scriptscriptstyle\times};
(1.453799,0.232887) *[red]{\scriptscriptstyle\times};
(1.462012,0.246810) *[red]{\scriptscriptstyle\times};
(1.470226,0.265415) *[red]{\scriptscriptstyle\times};
(1.478439,0.207473) *[red]{\scriptscriptstyle\times};
(1.486653,0.225384) *[red]{\scriptscriptstyle\times};
(1.494867,0.227149) *[red]{\scriptscriptstyle\times};
(1.503080,0.228079) *[red]{\scriptscriptstyle\times};
(1.511294,0.230356) *[red]{\scriptscriptstyle\times};
(1.519507,0.232102) *[red]{\scriptscriptstyle\times};
(1.527721,0.232332) *[red]{\scriptscriptstyle\times};
(1.535934,0.233098) *[red]{\scriptscriptstyle\times};
(1.544148,0.234218) *[red]{\scriptscriptstyle\times};
(1.552361,0.240789) *[red]{\scriptscriptstyle\times};
(1.560575,0.242021) *[red]{\scriptscriptstyle\times};
(1.568789,0.243940) *[red]{\scriptscriptstyle\times};
(1.577002,0.250458) *[red]{\scriptscriptstyle\times};
(1.585216,0.252221) *[red]{\scriptscriptstyle\times};
(1.593429,0.262634) *[red]{\scriptscriptstyle\times};
(1.601643,0.208858) *[red]{\scriptscriptstyle\times};
(1.609856,0.211510) *[red]{\scriptscriptstyle\times};
(1.618070,0.223193) *[red]{\scriptscriptstyle\times};
(1.626283,0.223719) *[red]{\scriptscriptstyle\times};
(1.634497,0.225226) *[red]{\scriptscriptstyle\times};
(1.642710,0.226647) *[red]{\scriptscriptstyle\times};
(1.650924,0.230424) *[red]{\scriptscriptstyle\times};
(1.659138,0.231232) *[red]{\scriptscriptstyle\times};
(1.667351,0.234429) *[red]{\scriptscriptstyle\times};
(1.675565,0.247485) *[red]{\scriptscriptstyle\times};
(1.683778,0.247569) *[red]{\scriptscriptstyle\times};
(1.691992,0.255157) *[red]{\scriptscriptstyle\times};
(1.700205,0.268905) *[red]{\scriptscriptstyle\times};
(1.708419,0.292575) *[red]{\scriptscriptstyle\times};
(1.716632,0.299405) *[red]{\scriptscriptstyle\times};
(1.724846,0.212857) *[red]{\scriptscriptstyle\times};
(1.733060,0.215526) *[red]{\scriptscriptstyle\times};
(1.741273,0.218467) *[red]{\scriptscriptstyle\times};
(1.749487,0.220350) *[red]{\scriptscriptstyle\times};
(1.757700,0.229075) *[red]{\scriptscriptstyle\times};
(1.765914,0.232964) *[red]{\scriptscriptstyle\times};
(1.774127,0.235799) *[red]{\scriptscriptstyle\times};
(1.782341,0.235866) *[red]{\scriptscriptstyle\times};
(1.790554,0.239153) *[red]{\scriptscriptstyle\times};
(1.798768,0.239810) *[red]{\scriptscriptstyle\times};
(1.806982,0.245635) *[red]{\scriptscriptstyle\times};
(1.815195,0.247285) *[red]{\scriptscriptstyle\times};
(1.823409,0.251614) *[red]{\scriptscriptstyle\times};
(1.831622,0.259823) *[red]{\scriptscriptstyle\times};
(1.839836,0.273905) *[red]{\scriptscriptstyle\times};
(1.848049,0.221359) *[red]{\scriptscriptstyle\times};
(1.856263,0.222208) *[red]{\scriptscriptstyle\times};
(1.864476,0.223420) *[red]{\scriptscriptstyle\times};
(1.872690,0.224289) *[red]{\scriptscriptstyle\times};
(1.880903,0.224559) *[red]{\scriptscriptstyle\times};
(1.889117,0.226918) *[red]{\scriptscriptstyle\times};
(1.897331,0.231398) *[red]{\scriptscriptstyle\times};
(1.905544,0.243783) *[red]{\scriptscriptstyle\times};
(1.913758,0.246055) *[red]{\scriptscriptstyle\times};
(1.921971,0.247745) *[red]{\scriptscriptstyle\times};
(1.930185,0.250075) *[red]{\scriptscriptstyle\times};
(1.938398,0.263862) *[red]{\scriptscriptstyle\times};
(1.946612,0.268768) *[red]{\scriptscriptstyle\times};
(1.954825,0.283257) *[red]{\scriptscriptstyle\times};
(1.963039,0.289786) *[red]{\scriptscriptstyle\times};
(1.971253,0.186758) *[red]{\scriptscriptstyle\times};
(1.979466,0.211163) *[red]{\scriptscriptstyle\times};
(1.987680,0.227381) *[red]{\scriptscriptstyle\times};
(1.995893,0.229227) *[red]{\scriptscriptstyle\times};
(2.004107,0.234140) *[red]{\scriptscriptstyle\times};
(2.012320,0.234799) *[red]{\scriptscriptstyle\times};
(2.020534,0.236404) *[red]{\scriptscriptstyle\times};
(2.028747,0.255582) *[red]{\scriptscriptstyle\times};
(2.036961,0.259800) *[red]{\scriptscriptstyle\times};
(2.045175,0.277183) *[red]{\scriptscriptstyle\times};
(2.053388,0.279377) *[red]{\scriptscriptstyle\times};
(2.061602,0.284587) *[red]{\scriptscriptstyle\times};
(2.069815,0.295296) *[red]{\scriptscriptstyle\times};
(2.078029,0.326310) *[red]{\scriptscriptstyle\times};
(2.086242,0.350869) *[red]{\scriptscriptstyle\times};
(2.094456,0.188386) *[red]{\scriptscriptstyle\times};
(2.102669,0.198097) *[red]{\scriptscriptstyle\times};
(2.110883,0.214759) *[red]{\scriptscriptstyle\times};
(2.119097,0.216611) *[red]{\scriptscriptstyle\times};
(2.127310,0.218504) *[red]{\scriptscriptstyle\times};
(2.135524,0.230126) *[red]{\scriptscriptstyle\times};
(2.143737,0.243511) *[red]{\scriptscriptstyle\times};
(2.151951,0.245857) *[red]{\scriptscriptstyle\times};
(2.160164,0.259596) *[red]{\scriptscriptstyle\times};
(2.168378,0.265631) *[red]{\scriptscriptstyle\times};
(2.176591,0.281384) *[red]{\scriptscriptstyle\times};
(2.184805,0.291420) *[red]{\scriptscriptstyle\times};
(2.193018,0.297084) *[red]{\scriptscriptstyle\times};
(2.201232,0.308015) *[red]{\scriptscriptstyle\times};
(2.209446,0.357377) *[red]{\scriptscriptstyle\times};
(2.217659,0.219309) *[red]{\scriptscriptstyle\times};
(2.225873,0.232020) *[red]{\scriptscriptstyle\times};
(2.234086,0.238843) *[red]{\scriptscriptstyle\times};
(2.242300,0.239087) *[red]{\scriptscriptstyle\times};
(2.250513,0.240621) *[red]{\scriptscriptstyle\times};
(2.258727,0.242809) *[red]{\scriptscriptstyle\times};
(2.266940,0.243977) *[red]{\scriptscriptstyle\times};
(2.275154,0.259246) *[red]{\scriptscriptstyle\times};
(2.283368,0.261661) *[red]{\scriptscriptstyle\times};
(2.291581,0.263531) *[red]{\scriptscriptstyle\times};
(2.299795,0.267197) *[red]{\scriptscriptstyle\times};
(2.308008,0.281971) *[red]{\scriptscriptstyle\times};
(2.316222,0.285855) *[red]{\scriptscriptstyle\times};
(2.324435,0.326685) *[red]{\scriptscriptstyle\times};
(2.332649,0.385773) *[red]{\scriptscriptstyle\times};
(2.340862,0.254589) *[red]{\scriptscriptstyle\times};
(2.349076,0.257674) *[red]{\scriptscriptstyle\times};
(2.357290,0.259875) *[red]{\scriptscriptstyle\times};
(2.365503,0.262161) *[red]{\scriptscriptstyle\times};
(2.373717,0.264825) *[red]{\scriptscriptstyle\times};
(2.381930,0.266997) *[red]{\scriptscriptstyle\times};
(2.390144,0.276980) *[red]{\scriptscriptstyle\times};
(2.398357,0.300439) *[red]{\scriptscriptstyle\times};
(2.406571,0.301372) *[red]{\scriptscriptstyle\times};
(2.414784,0.303989) *[red]{\scriptscriptstyle\times};
(2.422998,0.305256) *[red]{\scriptscriptstyle\times};
(2.431211,0.307580) *[red]{\scriptscriptstyle\times};
(2.439425,0.308377) *[red]{\scriptscriptstyle\times};
(2.447639,0.342651) *[red]{\scriptscriptstyle\times};
(2.455852,0.353566) *[red]{\scriptscriptstyle\times};
(2.464066,0.222465) *[red]{\scriptscriptstyle\times};
(2.472279,0.242764) *[red]{\scriptscriptstyle\times};
(2.480493,0.244555) *[red]{\scriptscriptstyle\times};
(2.488706,0.247235) *[red]{\scriptscriptstyle\times};
(2.496920,0.250716) *[red]{\scriptscriptstyle\times};
(2.505133,0.251860) *[red]{\scriptscriptstyle\times};
(2.513347,0.254422) *[red]{\scriptscriptstyle\times};
(2.521561,0.255746) *[red]{\scriptscriptstyle\times};
(2.529774,0.259193) *[red]{\scriptscriptstyle\times};
(2.537988,0.272166) *[red]{\scriptscriptstyle\times};
(2.546201,0.274029) *[red]{\scriptscriptstyle\times};
(2.554415,0.280644) *[red]{\scriptscriptstyle\times};
(2.562628,0.297036) *[red]{\scriptscriptstyle\times};
(2.570842,0.299108) *[red]{\scriptscriptstyle\times};
(2.579055,0.331562) *[red]{\scriptscriptstyle\times};
(2.587269,0.268623) *[red]{\scriptscriptstyle\times};
(2.595483,0.269060) *[red]{\scriptscriptstyle\times};
(2.603696,0.270488) *[red]{\scriptscriptstyle\times};
(2.611910,0.272224) *[red]{\scriptscriptstyle\times};
(2.620123,0.277579) *[red]{\scriptscriptstyle\times};
(2.628337,0.280154) *[red]{\scriptscriptstyle\times};
(2.636550,0.285663) *[red]{\scriptscriptstyle\times};
(2.644764,0.286706) *[red]{\scriptscriptstyle\times};
(2.652977,0.292242) *[red]{\scriptscriptstyle\times};
(2.661191,0.294168) *[red]{\scriptscriptstyle\times};
(2.669405,0.297999) *[red]{\scriptscriptstyle\times};
(2.677618,0.298301) *[red]{\scriptscriptstyle\times};
(2.685832,0.307293) *[red]{\scriptscriptstyle\times};
(2.694045,0.313881) *[red]{\scriptscriptstyle\times};
(2.702259,0.317424) *[red]{\scriptscriptstyle\times};
(2.710472,0.246852) *[red]{\scriptscriptstyle\times};
(2.718686,0.250124) *[red]{\scriptscriptstyle\times};
(2.726899,0.256301) *[red]{\scriptscriptstyle\times};
(2.735113,0.261886) *[red]{\scriptscriptstyle\times};
(2.743326,0.263591) *[red]{\scriptscriptstyle\times};
(2.751540,0.273277) *[red]{\scriptscriptstyle\times};
(2.759754,0.273361) *[red]{\scriptscriptstyle\times};
(2.767967,0.273693) *[red]{\scriptscriptstyle\times};
(2.776181,0.276434) *[red]{\scriptscriptstyle\times};
(2.784394,0.277200) *[red]{\scriptscriptstyle\times};
(2.792608,0.284879) *[red]{\scriptscriptstyle\times};
(2.800821,0.295260) *[red]{\scriptscriptstyle\times};
(2.809035,0.297808) *[red]{\scriptscriptstyle\times};
(2.817248,0.307034) *[red]{\scriptscriptstyle\times};
(2.825462,0.389929) *[red]{\scriptscriptstyle\times};
(2.833676,0.231830) *[red]{\scriptscriptstyle\times};
(2.841889,0.245064) *[red]{\scriptscriptstyle\times};
(2.850103,0.268289) *[red]{\scriptscriptstyle\times};
(2.858316,0.272228) *[red]{\scriptscriptstyle\times};
(2.866530,0.273373) *[red]{\scriptscriptstyle\times};
(2.874743,0.274658) *[red]{\scriptscriptstyle\times};
(2.882957,0.293001) *[red]{\scriptscriptstyle\times};
(2.891170,0.294966) *[red]{\scriptscriptstyle\times};
(2.899384,0.297989) *[red]{\scriptscriptstyle\times};
(2.907598,0.319100) *[red]{\scriptscriptstyle\times};
(2.915811,0.325431) *[red]{\scriptscriptstyle\times};
(2.924025,0.332615) *[red]{\scriptscriptstyle\times};
(2.932238,0.335182) *[red]{\scriptscriptstyle\times};
(2.940452,0.337489) *[red]{\scriptscriptstyle\times};
(2.948665,0.374401) *[red]{\scriptscriptstyle\times};
(2.956879,0.278822) *[red]{\scriptscriptstyle\times};
(2.965092,0.279189) *[red]{\scriptscriptstyle\times};
(2.973306,0.280604) *[red]{\scriptscriptstyle\times};
(2.981520,0.281238) *[red]{\scriptscriptstyle\times};
(2.989733,0.281953) *[red]{\scriptscriptstyle\times};
(2.997947,0.283620) *[red]{\scriptscriptstyle\times};
(3.006160,0.283855) *[red]{\scriptscriptstyle\times};
(3.014374,0.289528) *[red]{\scriptscriptstyle\times};
(3.022587,0.289598) *[red]{\scriptscriptstyle\times};
(3.030801,0.290478) *[red]{\scriptscriptstyle\times};
(3.039014,0.302121) *[red]{\scriptscriptstyle\times};
(3.047228,0.303493) *[red]{\scriptscriptstyle\times};
(3.055441,0.303496) *[red]{\scriptscriptstyle\times};
(3.063655,0.306372) *[red]{\scriptscriptstyle\times};
(3.071869,0.308785) *[red]{\scriptscriptstyle\times};
(3.080082,0.259189) *[red]{\scriptscriptstyle\times};
(3.088296,0.262107) *[red]{\scriptscriptstyle\times};
(3.096509,0.273454) *[red]{\scriptscriptstyle\times};
(3.104723,0.278816) *[red]{\scriptscriptstyle\times};
(3.112936,0.278905) *[red]{\scriptscriptstyle\times};
(3.121150,0.282347) *[red]{\scriptscriptstyle\times};
(3.129363,0.283907) *[red]{\scriptscriptstyle\times};
(3.137577,0.284205) *[red]{\scriptscriptstyle\times};
(3.145791,0.285491) *[red]{\scriptscriptstyle\times};
(3.154004,0.286124) *[red]{\scriptscriptstyle\times};
(3.162218,0.286594) *[red]{\scriptscriptstyle\times};
(3.170431,0.287412) *[red]{\scriptscriptstyle\times};
(3.178645,0.294217) *[red]{\scriptscriptstyle\times};
(3.186858,0.321492) *[red]{\scriptscriptstyle\times};
(3.195072,0.342814) *[red]{\scriptscriptstyle\times};
(3.203285,0.263045) *[red]{\scriptscriptstyle\times};
(3.211499,0.270574) *[red]{\scriptscriptstyle\times};
(3.219713,0.286459) *[red]{\scriptscriptstyle\times};
(3.227926,0.288988) *[red]{\scriptscriptstyle\times};
(3.236140,0.292038) *[red]{\scriptscriptstyle\times};
(3.244353,0.293474) *[red]{\scriptscriptstyle\times};
(3.252567,0.299600) *[red]{\scriptscriptstyle\times};
(3.260780,0.306780) *[red]{\scriptscriptstyle\times};
(3.268994,0.308049) *[red]{\scriptscriptstyle\times};
(3.277207,0.309120) *[red]{\scriptscriptstyle\times};
(3.285421,0.313254) *[red]{\scriptscriptstyle\times};
(3.293634,0.315889) *[red]{\scriptscriptstyle\times};
(3.301848,0.327025) *[red]{\scriptscriptstyle\times};
(3.310062,0.334396) *[red]{\scriptscriptstyle\times};
(3.318275,0.374869) *[red]{\scriptscriptstyle\times};
(3.326489,0.279901) *[red]{\scriptscriptstyle\times};
(3.334702,0.283768) *[red]{\scriptscriptstyle\times};
(3.342916,0.285468) *[red]{\scriptscriptstyle\times};
(3.351129,0.292975) *[red]{\scriptscriptstyle\times};
(3.359343,0.299104) *[red]{\scriptscriptstyle\times};
(3.367556,0.301520) *[red]{\scriptscriptstyle\times};
(3.375770,0.305438) *[red]{\scriptscriptstyle\times};
(3.383984,0.306595) *[red]{\scriptscriptstyle\times};
(3.392197,0.306978) *[red]{\scriptscriptstyle\times};
(3.400411,0.307428) *[red]{\scriptscriptstyle\times};
(3.408624,0.307749) *[red]{\scriptscriptstyle\times};
(3.416838,0.319703) *[red]{\scriptscriptstyle\times};
(3.425051,0.323711) *[red]{\scriptscriptstyle\times};
(3.433265,0.328970) *[red]{\scriptscriptstyle\times};
(3.441478,0.345137) *[red]{\scriptscriptstyle\times};
(3.449692,0.269614) *[red]{\scriptscriptstyle\times};
(3.457906,0.275797) *[red]{\scriptscriptstyle\times};
(3.466119,0.280859) *[red]{\scriptscriptstyle\times};
(3.474333,0.291419) *[red]{\scriptscriptstyle\times};
(3.482546,0.291817) *[red]{\scriptscriptstyle\times};
(3.490760,0.295543) *[red]{\scriptscriptstyle\times};
(3.498973,0.301896) *[red]{\scriptscriptstyle\times};
(3.507187,0.312870) *[red]{\scriptscriptstyle\times};
(3.515400,0.312903) *[red]{\scriptscriptstyle\times};
(3.523614,0.313426) *[red]{\scriptscriptstyle\times};
(3.531828,0.318823) *[red]{\scriptscriptstyle\times};
(3.540041,0.320436) *[red]{\scriptscriptstyle\times};
(3.548255,0.328739) *[red]{\scriptscriptstyle\times};
(3.556468,0.352763) *[red]{\scriptscriptstyle\times};
(3.564682,0.442375) *[red]{\scriptscriptstyle\times};
(3.572895,0.297696) *[red]{\scriptscriptstyle\times};
(3.581109,0.297886) *[red]{\scriptscriptstyle\times};
(3.589322,0.298068) *[red]{\scriptscriptstyle\times};
(3.597536,0.299153) *[red]{\scriptscriptstyle\times};
(3.605749,0.303857) *[red]{\scriptscriptstyle\times};
(3.613963,0.306683) *[red]{\scriptscriptstyle\times};
(3.622177,0.325474) *[red]{\scriptscriptstyle\times};
(3.630390,0.328517) *[red]{\scriptscriptstyle\times};
(3.638604,0.337235) *[red]{\scriptscriptstyle\times};
(3.646817,0.345103) *[red]{\scriptscriptstyle\times};
(3.655031,0.347494) *[red]{\scriptscriptstyle\times};
(3.663244,0.372846) *[red]{\scriptscriptstyle\times};
(3.671458,0.410116) *[red]{\scriptscriptstyle\times};
(3.679671,0.451380) *[red]{\scriptscriptstyle\times};
(3.687885,0.515970) *[red]{\scriptscriptstyle\times};
(3.696099,0.304246) *[red]{\scriptscriptstyle\times};
(3.704312,0.325047) *[red]{\scriptscriptstyle\times};
(3.712526,0.325816) *[red]{\scriptscriptstyle\times};
(3.720739,0.327519) *[red]{\scriptscriptstyle\times};
(3.728953,0.331584) *[red]{\scriptscriptstyle\times};
(3.737166,0.333753) *[red]{\scriptscriptstyle\times};
(3.745380,0.335804) *[red]{\scriptscriptstyle\times};
(3.753593,0.342742) *[red]{\scriptscriptstyle\times};
(3.761807,0.347994) *[red]{\scriptscriptstyle\times};
(3.770021,0.351398) *[red]{\scriptscriptstyle\times};
(3.778234,0.359219) *[red]{\scriptscriptstyle\times};
(3.786448,0.365472) *[red]{\scriptscriptstyle\times};
(3.794661,0.367062) *[red]{\scriptscriptstyle\times};
(3.802875,0.368426) *[red]{\scriptscriptstyle\times};
(3.811088,0.384550) *[red]{\scriptscriptstyle\times};
(3.819302,0.297943) *[red]{\scriptscriptstyle\times};
(3.827515,0.299989) *[red]{\scriptscriptstyle\times};
(3.835729,0.324777) *[red]{\scriptscriptstyle\times};
(3.843943,0.339471) *[red]{\scriptscriptstyle\times};
(3.852156,0.341622) *[red]{\scriptscriptstyle\times};
(3.860370,0.341942) *[red]{\scriptscriptstyle\times};
(3.868583,0.346954) *[red]{\scriptscriptstyle\times};
(3.876797,0.359643) *[red]{\scriptscriptstyle\times};
(3.885010,0.366412) *[red]{\scriptscriptstyle\times};
(3.893224,0.366532) *[red]{\scriptscriptstyle\times};
(3.901437,0.368084) *[red]{\scriptscriptstyle\times};
(3.909651,0.383333) *[red]{\scriptscriptstyle\times};
(3.917864,0.386532) *[red]{\scriptscriptstyle\times};
(3.926078,0.405812) *[red]{\scriptscriptstyle\times};
(3.934292,0.417974) *[red]{\scriptscriptstyle\times};
(3.942505,0.271220) *[red]{\scriptscriptstyle\times};
(3.950719,0.271940) *[red]{\scriptscriptstyle\times};
(3.958932,0.278885) *[red]{\scriptscriptstyle\times};
(3.967146,0.291307) *[red]{\scriptscriptstyle\times};
(3.975359,0.292920) *[red]{\scriptscriptstyle\times};
(3.983573,0.294632) *[red]{\scriptscriptstyle\times};
(3.991786,0.297057) *[red]{\scriptscriptstyle\times};
(4.000000,0.314795) *[red]{\scriptscriptstyle\times};
(4.008214,0.316925) *[red]{\scriptscriptstyle\times};
(4.016427,0.319853) *[red]{\scriptscriptstyle\times};
(4.024641,0.330169) *[red]{\scriptscriptstyle\times};
(4.032854,0.385854) *[red]{\scriptscriptstyle\times};
(4.041068,0.408167) *[red]{\scriptscriptstyle\times};
(4.049281,0.428437) *[red]{\scriptscriptstyle\times};
(4.057495,0.473708) *[red]{\scriptscriptstyle\times};
(4.065708,0.334627) *[red]{\scriptscriptstyle\times};
(4.073922,0.335729) *[red]{\scriptscriptstyle\times};
(4.082136,0.338074) *[red]{\scriptscriptstyle\times};
(4.090349,0.338963) *[red]{\scriptscriptstyle\times};
(4.098563,0.343581) *[red]{\scriptscriptstyle\times};
(4.106776,0.348957) *[red]{\scriptscriptstyle\times};
(4.114990,0.350556) *[red]{\scriptscriptstyle\times};
(4.123203,0.356341) *[red]{\scriptscriptstyle\times};
(4.131417,0.358662) *[red]{\scriptscriptstyle\times};
(4.139630,0.362586) *[red]{\scriptscriptstyle\times};
(4.147844,0.363113) *[red]{\scriptscriptstyle\times};
(4.156057,0.363821) *[red]{\scriptscriptstyle\times};
(4.164271,0.380393) *[red]{\scriptscriptstyle\times};
(4.172485,0.388182) *[red]{\scriptscriptstyle\times};
(4.180698,0.391418) *[red]{\scriptscriptstyle\times};
(4.188912,0.319717) *[red]{\scriptscriptstyle\times};
(4.197125,0.321876) *[red]{\scriptscriptstyle\times};
(4.205339,0.325152) *[red]{\scriptscriptstyle\times};
(4.213552,0.326056) *[red]{\scriptscriptstyle\times};
(4.221766,0.326826) *[red]{\scriptscriptstyle\times};
(4.229979,0.343469) *[red]{\scriptscriptstyle\times};
(4.238193,0.346071) *[red]{\scriptscriptstyle\times};
(4.246407,0.347901) *[red]{\scriptscriptstyle\times};
(4.254620,0.372888) *[red]{\scriptscriptstyle\times};
(4.262834,0.376807) *[red]{\scriptscriptstyle\times};
(4.271047,0.389268) *[red]{\scriptscriptstyle\times};
(4.279261,0.399121) *[red]{\scriptscriptstyle\times};
(4.287474,0.399798) *[red]{\scriptscriptstyle\times};
(4.295688,0.422924) *[red]{\scriptscriptstyle\times};
(4.303901,0.442034) *[red]{\scriptscriptstyle\times};
(4.312115,0.324787) *[red]{\scriptscriptstyle\times};
(4.320329,0.338883) *[red]{\scriptscriptstyle\times};
(4.328542,0.340520) *[red]{\scriptscriptstyle\times};
(4.336756,0.341803) *[red]{\scriptscriptstyle\times};
(4.344969,0.342658) *[red]{\scriptscriptstyle\times};
(4.353183,0.348794) *[red]{\scriptscriptstyle\times};
(4.361396,0.361994) *[red]{\scriptscriptstyle\times};
(4.369610,0.366928) *[red]{\scriptscriptstyle\times};
(4.377823,0.367253) *[red]{\scriptscriptstyle\times};
(4.386037,0.368553) *[red]{\scriptscriptstyle\times};
(4.394251,0.374133) *[red]{\scriptscriptstyle\times};
(4.402464,0.375636) *[red]{\scriptscriptstyle\times};
(4.410678,0.388421) *[red]{\scriptscriptstyle\times};
(4.418891,0.392729) *[red]{\scriptscriptstyle\times};
(4.427105,0.434364) *[red]{\scriptscriptstyle\times};
(4.435318,0.366638) *[red]{\scriptscriptstyle\times};
(4.443532,0.366990) *[red]{\scriptscriptstyle\times};
(4.451745,0.372403) *[red]{\scriptscriptstyle\times};
(4.459959,0.378640) *[red]{\scriptscriptstyle\times};
(4.468172,0.384463) *[red]{\scriptscriptstyle\times};
(4.476386,0.387058) *[red]{\scriptscriptstyle\times};
(4.484600,0.388017) *[red]{\scriptscriptstyle\times};
(4.492813,0.391670) *[red]{\scriptscriptstyle\times};
(4.501027,0.392263) *[red]{\scriptscriptstyle\times};
(4.509240,0.393057) *[red]{\scriptscriptstyle\times};
(4.517454,0.395827) *[red]{\scriptscriptstyle\times};
(4.525667,0.397219) *[red]{\scriptscriptstyle\times};
(4.533881,0.402833) *[red]{\scriptscriptstyle\times};
(4.542094,0.452974) *[red]{\scriptscriptstyle\times};
(4.550308,0.472504) *[red]{\scriptscriptstyle\times};
(4.558522,0.362343) *[red]{\scriptscriptstyle\times};
(4.566735,0.367012) *[red]{\scriptscriptstyle\times};
(4.574949,0.379856) *[red]{\scriptscriptstyle\times};
(4.583162,0.380860) *[red]{\scriptscriptstyle\times};
(4.591376,0.381516) *[red]{\scriptscriptstyle\times};
(4.599589,0.381532) *[red]{\scriptscriptstyle\times};
(4.607803,0.383941) *[red]{\scriptscriptstyle\times};
(4.616016,0.384029) *[red]{\scriptscriptstyle\times};
(4.624230,0.384360) *[red]{\scriptscriptstyle\times};
(4.632444,0.386944) *[red]{\scriptscriptstyle\times};
(4.640657,0.392197) *[red]{\scriptscriptstyle\times};
(4.648871,0.401034) *[red]{\scriptscriptstyle\times};
(4.657084,0.403899) *[red]{\scriptscriptstyle\times};
(4.665298,0.409275) *[red]{\scriptscriptstyle\times};
(4.673511,0.414400) *[red]{\scriptscriptstyle\times};
(4.681725,0.324468) *[red]{\scriptscriptstyle\times};
(4.689938,0.325978) *[red]{\scriptscriptstyle\times};
(4.698152,0.342358) *[red]{\scriptscriptstyle\times};
(4.706366,0.351294) *[red]{\scriptscriptstyle\times};
(4.714579,0.359660) *[red]{\scriptscriptstyle\times};
(4.722793,0.363097) *[red]{\scriptscriptstyle\times};
(4.731006,0.371433) *[red]{\scriptscriptstyle\times};
(4.739220,0.404577) *[red]{\scriptscriptstyle\times};
(4.747433,0.406480) *[red]{\scriptscriptstyle\times};
(4.755647,0.430142) *[red]{\scriptscriptstyle\times};
(4.763860,0.448279) *[red]{\scriptscriptstyle\times};
(4.772074,0.448707) *[red]{\scriptscriptstyle\times};
(4.780287,0.458233) *[red]{\scriptscriptstyle\times};
(4.788501,0.463347) *[red]{\scriptscriptstyle\times};
(4.796715,0.466229) *[red]{\scriptscriptstyle\times};
(4.804928,0.346602) *[red]{\scriptscriptstyle\times};
(4.813142,0.349331) *[red]{\scriptscriptstyle\times};
(4.821355,0.375203) *[red]{\scriptscriptstyle\times};
(4.829569,0.375813) *[red]{\scriptscriptstyle\times};
(4.837782,0.391031) *[red]{\scriptscriptstyle\times};
(4.845996,0.394203) *[red]{\scriptscriptstyle\times};
(4.854209,0.394233) *[red]{\scriptscriptstyle\times};
(4.862423,0.394697) *[red]{\scriptscriptstyle\times};
(4.870637,0.395997) *[red]{\scriptscriptstyle\times};
(4.878850,0.407691) *[red]{\scriptscriptstyle\times};
(4.887064,0.410174) *[red]{\scriptscriptstyle\times};
(4.895277,0.414447) *[red]{\scriptscriptstyle\times};
(4.903491,0.430931) *[red]{\scriptscriptstyle\times};
(4.911704,0.431838) *[red]{\scriptscriptstyle\times};
(4.919918,0.477297) *[red]{\scriptscriptstyle\times};
(4.928131,0.363940) *[red]{\scriptscriptstyle\times};
(4.936345,0.373328) *[red]{\scriptscriptstyle\times};
(4.944559,0.375155) *[red]{\scriptscriptstyle\times};
(4.952772,0.390137) *[red]{\scriptscriptstyle\times};
(4.960986,0.390400) *[red]{\scriptscriptstyle\times};
(4.969199,0.390551) *[red]{\scriptscriptstyle\times};
(4.977413,0.392529) *[red]{\scriptscriptstyle\times};
(4.985626,0.394259) *[red]{\scriptscriptstyle\times};
(4.993840,0.394945) *[red]{\scriptscriptstyle\times};
(5.002053,0.395101) *[red]{\scriptscriptstyle\times};
(5.010267,0.396333) *[red]{\scriptscriptstyle\times};
(5.018480,0.405619) *[red]{\scriptscriptstyle\times};
(5.026694,0.416211) *[red]{\scriptscriptstyle\times};
(5.034908,0.433742) *[red]{\scriptscriptstyle\times};
(5.043121,0.508005) *[red]{\scriptscriptstyle\times};
(5.051335,0.352515) *[red]{\scriptscriptstyle\times};
(5.059548,0.355691) *[red]{\scriptscriptstyle\times};
(5.067762,0.359817) *[red]{\scriptscriptstyle\times};
(5.075975,0.363054) *[red]{\scriptscriptstyle\times};
(5.084189,0.375153) *[red]{\scriptscriptstyle\times};
(5.092402,0.382350) *[red]{\scriptscriptstyle\times};
(5.100616,0.387404) *[red]{\scriptscriptstyle\times};
(5.108830,0.388531) *[red]{\scriptscriptstyle\times};
(5.117043,0.399316) *[red]{\scriptscriptstyle\times};
(5.125257,0.401151) *[red]{\scriptscriptstyle\times};
(5.133470,0.421371) *[red]{\scriptscriptstyle\times};
(5.141684,0.421778) *[red]{\scriptscriptstyle\times};
(5.149897,0.444962) *[red]{\scriptscriptstyle\times};
(5.158111,0.447115) *[red]{\scriptscriptstyle\times};
(5.166324,0.478823) *[red]{\scriptscriptstyle\times};
(5.174538,0.380066) *[red]{\scriptscriptstyle\times};
(5.182752,0.390053) *[red]{\scriptscriptstyle\times};
(5.190965,0.390690) *[red]{\scriptscriptstyle\times};
(5.199179,0.391345) *[red]{\scriptscriptstyle\times};
(5.207392,0.395835) *[red]{\scriptscriptstyle\times};
(5.215606,0.396483) *[red]{\scriptscriptstyle\times};
(5.223819,0.408543) *[red]{\scriptscriptstyle\times};
(5.232033,0.412170) *[red]{\scriptscriptstyle\times};
(5.240246,0.413463) *[red]{\scriptscriptstyle\times};
(5.248460,0.416276) *[red]{\scriptscriptstyle\times};
(5.256674,0.418562) *[red]{\scriptscriptstyle\times};
(5.264887,0.421647) *[red]{\scriptscriptstyle\times};
(5.273101,0.421944) *[red]{\scriptscriptstyle\times};
(5.281314,0.435742) *[red]{\scriptscriptstyle\times};
(5.289528,0.513711) *[red]{\scriptscriptstyle\times};
(5.297741,0.365541) *[red]{\scriptscriptstyle\times};
(5.305955,0.384996) *[red]{\scriptscriptstyle\times};
(5.314168,0.385284) *[red]{\scriptscriptstyle\times};
(5.322382,0.394489) *[red]{\scriptscriptstyle\times};
(5.330595,0.396578) *[red]{\scriptscriptstyle\times};
(5.338809,0.402114) *[red]{\scriptscriptstyle\times};
(5.347023,0.402165) *[red]{\scriptscriptstyle\times};
(5.355236,0.405653) *[red]{\scriptscriptstyle\times};
(5.363450,0.405796) *[red]{\scriptscriptstyle\times};
(5.371663,0.420805) *[red]{\scriptscriptstyle\times};
(5.379877,0.422598) *[red]{\scriptscriptstyle\times};
(5.388090,0.423451) *[red]{\scriptscriptstyle\times};
(5.396304,0.429925) *[red]{\scriptscriptstyle\times};
(5.404517,0.442207) *[red]{\scriptscriptstyle\times};
(5.412731,0.475444) *[red]{\scriptscriptstyle\times};
(5.420945,0.401875) *[red]{\scriptscriptstyle\times};
(5.429158,0.403577) *[red]{\scriptscriptstyle\times};
(5.437372,0.404881) *[red]{\scriptscriptstyle\times};
(5.445585,0.404924) *[red]{\scriptscriptstyle\times};
(5.453799,0.405178) *[red]{\scriptscriptstyle\times};
(5.462012,0.409242) *[red]{\scriptscriptstyle\times};
(5.470226,0.421330) *[red]{\scriptscriptstyle\times};
(5.478439,0.421967) *[red]{\scriptscriptstyle\times};
(5.486653,0.422211) *[red]{\scriptscriptstyle\times};
(5.494867,0.424636) *[red]{\scriptscriptstyle\times};
(5.503080,0.425280) *[red]{\scriptscriptstyle\times};
(5.511294,0.425327) *[red]{\scriptscriptstyle\times};
(5.519507,0.426719) *[red]{\scriptscriptstyle\times};
(5.527721,0.427835) *[red]{\scriptscriptstyle\times};
(5.535934,0.445368) *[red]{\scriptscriptstyle\times};
(5.544148,0.393175) *[red]{\scriptscriptstyle\times};
(5.552361,0.393383) *[red]{\scriptscriptstyle\times};
(5.560575,0.394320) *[red]{\scriptscriptstyle\times};
(5.568789,0.396843) *[red]{\scriptscriptstyle\times};
(5.577002,0.397493) *[red]{\scriptscriptstyle\times};
(5.585216,0.398208) *[red]{\scriptscriptstyle\times};
(5.593429,0.400939) *[red]{\scriptscriptstyle\times};
(5.601643,0.403642) *[red]{\scriptscriptstyle\times};
(5.609856,0.414810) *[red]{\scriptscriptstyle\times};
(5.618070,0.417791) *[red]{\scriptscriptstyle\times};
(5.626283,0.420971) *[red]{\scriptscriptstyle\times};
(5.634497,0.432643) *[red]{\scriptscriptstyle\times};
(5.642710,0.436781) *[red]{\scriptscriptstyle\times};
(5.650924,0.451141) *[red]{\scriptscriptstyle\times};
(5.659138,0.461667) *[red]{\scriptscriptstyle\times};
(5.667351,0.406683) *[red]{\scriptscriptstyle\times};
(5.675565,0.409354) *[red]{\scriptscriptstyle\times};
(5.683778,0.412348) *[red]{\scriptscriptstyle\times};
(5.691992,0.415702) *[red]{\scriptscriptstyle\times};
(5.700205,0.415806) *[red]{\scriptscriptstyle\times};
(5.708419,0.416792) *[red]{\scriptscriptstyle\times};
(5.716632,0.417710) *[red]{\scriptscriptstyle\times};
(5.724846,0.418148) *[red]{\scriptscriptstyle\times};
(5.733060,0.429577) *[red]{\scriptscriptstyle\times};
(5.741273,0.430845) *[red]{\scriptscriptstyle\times};
(5.749487,0.433836) *[red]{\scriptscriptstyle\times};
(5.757700,0.436722) *[red]{\scriptscriptstyle\times};
(5.765914,0.438544) *[red]{\scriptscriptstyle\times};
(5.774127,0.439216) *[red]{\scriptscriptstyle\times};
(5.782341,0.448705) *[red]{\scriptscriptstyle\times};
(5.790554,0.410064) *[red]{\scriptscriptstyle\times};
(5.798768,0.411499) *[red]{\scriptscriptstyle\times};
(5.806982,0.413137) *[red]{\scriptscriptstyle\times};
(5.815195,0.418007) *[red]{\scriptscriptstyle\times};
(5.823409,0.418224) *[red]{\scriptscriptstyle\times};
(5.831622,0.423296) *[red]{\scriptscriptstyle\times};
(5.839836,0.425217) *[red]{\scriptscriptstyle\times};
(5.848049,0.427427) *[red]{\scriptscriptstyle\times};
(5.856263,0.431711) *[red]{\scriptscriptstyle\times};
(5.864476,0.436621) *[red]{\scriptscriptstyle\times};
(5.872690,0.441283) *[red]{\scriptscriptstyle\times};
(5.880903,0.452096) *[red]{\scriptscriptstyle\times};
(5.889117,0.459046) *[red]{\scriptscriptstyle\times};
(5.897331,0.524863) *[red]{\scriptscriptstyle\times};
(5.905544,0.611580) *[red]{\scriptscriptstyle\times};
(5.913758,0.398721) *[red]{\scriptscriptstyle\times};
(5.921971,0.419075) *[red]{\scriptscriptstyle\times};
(5.930185,0.424136) *[red]{\scriptscriptstyle\times};
(5.938398,0.433938) *[red]{\scriptscriptstyle\times};
(5.946612,0.440865) *[red]{\scriptscriptstyle\times};
(5.954825,0.442215) *[red]{\scriptscriptstyle\times};
(5.963039,0.447148) *[red]{\scriptscriptstyle\times};
(5.971253,0.452279) *[red]{\scriptscriptstyle\times};
(5.979466,0.454153) *[red]{\scriptscriptstyle\times};
(5.987680,0.459435) *[red]{\scriptscriptstyle\times};
(5.995893,0.459587) *[red]{\scriptscriptstyle\times};
(6.004107,0.462343) *[red]{\scriptscriptstyle\times};
(6.012320,0.464564) *[red]{\scriptscriptstyle\times};
(6.020534,0.477974) *[red]{\scriptscriptstyle\times};
(6.028747,0.482272) *[red]{\scriptscriptstyle\times};
(6.036961,0.427323) *[red]{\scriptscriptstyle\times};
(6.045175,0.429577) *[red]{\scriptscriptstyle\times};
(6.053388,0.429592) *[red]{\scriptscriptstyle\times};
(6.061602,0.441953) *[red]{\scriptscriptstyle\times};
(6.069815,0.445390) *[red]{\scriptscriptstyle\times};
(6.078029,0.445966) *[red]{\scriptscriptstyle\times};
(6.086242,0.447251) *[red]{\scriptscriptstyle\times};
(6.094456,0.448565) *[red]{\scriptscriptstyle\times};
(6.102669,0.450262) *[red]{\scriptscriptstyle\times};
(6.110883,0.458478) *[red]{\scriptscriptstyle\times};
(6.119097,0.463866) *[red]{\scriptscriptstyle\times};
(6.127310,0.469061) *[red]{\scriptscriptstyle\times};
(6.135524,0.475045) *[red]{\scriptscriptstyle\times};
(6.143737,0.475345) *[red]{\scriptscriptstyle\times};
(6.151951,0.480172) *[red]{\scriptscriptstyle\times};
(6.160164,0.411905) *[red]{\scriptscriptstyle\times};
(6.168378,0.428463) *[red]{\scriptscriptstyle\times};
(6.176591,0.428547) *[red]{\scriptscriptstyle\times};
(6.184805,0.443911) *[red]{\scriptscriptstyle\times};
(6.193018,0.450425) *[red]{\scriptscriptstyle\times};
(6.201232,0.453272) *[red]{\scriptscriptstyle\times};
(6.209446,0.454787) *[red]{\scriptscriptstyle\times};
(6.217659,0.454902) *[red]{\scriptscriptstyle\times};
(6.225873,0.455305) *[red]{\scriptscriptstyle\times};
(6.234086,0.456246) *[red]{\scriptscriptstyle\times};
(6.242300,0.460051) *[red]{\scriptscriptstyle\times};
(6.250513,0.460204) *[red]{\scriptscriptstyle\times};
(6.258727,0.466352) *[red]{\scriptscriptstyle\times};
(6.266940,0.466922) *[red]{\scriptscriptstyle\times};
(6.275154,0.487152) *[red]{\scriptscriptstyle\times};
(6.283368,0.403942) *[red]{\scriptscriptstyle\times};
(6.291581,0.429882) *[red]{\scriptscriptstyle\times};
(6.299795,0.431295) *[red]{\scriptscriptstyle\times};
(6.308008,0.445938) *[red]{\scriptscriptstyle\times};
(6.316222,0.446560) *[red]{\scriptscriptstyle\times};
(6.324435,0.447509) *[red]{\scriptscriptstyle\times};
(6.332649,0.453926) *[red]{\scriptscriptstyle\times};
(6.340862,0.456199) *[red]{\scriptscriptstyle\times};
(6.349076,0.465747) *[red]{\scriptscriptstyle\times};
(6.357290,0.465863) *[red]{\scriptscriptstyle\times};
(6.365503,0.466017) *[red]{\scriptscriptstyle\times};
(6.373717,0.476243) *[red]{\scriptscriptstyle\times};
(6.381930,0.484542) *[red]{\scriptscriptstyle\times};
(6.390144,0.615358) *[red]{\scriptscriptstyle\times};
(6.398357,0.635503) *[red]{\scriptscriptstyle\times};
(6.406571,0.419406) *[red]{\scriptscriptstyle\times};
(6.414784,0.423164) *[red]{\scriptscriptstyle\times};
(6.422998,0.436642) *[red]{\scriptscriptstyle\times};
(6.431211,0.439191) *[red]{\scriptscriptstyle\times};
(6.439425,0.439505) *[red]{\scriptscriptstyle\times};
(6.447639,0.444041) *[red]{\scriptscriptstyle\times};
(6.455852,0.450773) *[red]{\scriptscriptstyle\times};
(6.464066,0.457232) *[red]{\scriptscriptstyle\times};
(6.472279,0.457537) *[red]{\scriptscriptstyle\times};
(6.480493,0.458460) *[red]{\scriptscriptstyle\times};
(6.488706,0.459581) *[red]{\scriptscriptstyle\times};
(6.496920,0.461964) *[red]{\scriptscriptstyle\times};
(6.505133,0.480302) *[red]{\scriptscriptstyle\times};
(6.513347,0.496440) *[red]{\scriptscriptstyle\times};
(6.521561,0.534693) *[red]{\scriptscriptstyle\times};
(6.529774,0.448301) *[red]{\scriptscriptstyle\times};
(6.537988,0.458135) *[red]{\scriptscriptstyle\times};
(6.546201,0.460802) *[red]{\scriptscriptstyle\times};
(6.554415,0.462050) *[red]{\scriptscriptstyle\times};
(6.562628,0.481590) *[red]{\scriptscriptstyle\times};
(6.570842,0.484996) *[red]{\scriptscriptstyle\times};
(6.579055,0.485256) *[red]{\scriptscriptstyle\times};
(6.587269,0.499362) *[red]{\scriptscriptstyle\times};
(6.595483,0.506455) *[red]{\scriptscriptstyle\times};
(6.603696,0.506904) *[red]{\scriptscriptstyle\times};
(6.611910,0.521519) *[red]{\scriptscriptstyle\times};
(6.620123,0.528676) *[red]{\scriptscriptstyle\times};
(6.628337,0.529827) *[red]{\scriptscriptstyle\times};
(6.636550,0.531930) *[red]{\scriptscriptstyle\times};
(6.644764,0.559865) *[red]{\scriptscriptstyle\times};
(6.652977,0.439711) *[red]{\scriptscriptstyle\times};
(6.661191,0.449292) *[red]{\scriptscriptstyle\times};
(6.669405,0.455693) *[red]{\scriptscriptstyle\times};
(6.677618,0.457556) *[red]{\scriptscriptstyle\times};
(6.685832,0.460995) *[red]{\scriptscriptstyle\times};
(6.694045,0.466188) *[red]{\scriptscriptstyle\times};
(6.702259,0.466228) *[red]{\scriptscriptstyle\times};
(6.710472,0.478946) *[red]{\scriptscriptstyle\times};
(6.718686,0.479419) *[red]{\scriptscriptstyle\times};
(6.726899,0.484825) *[red]{\scriptscriptstyle\times};
(6.735113,0.533691) *[red]{\scriptscriptstyle\times};
(6.743326,0.534062) *[red]{\scriptscriptstyle\times};
(6.751540,0.546478) *[red]{\scriptscriptstyle\times};
(6.759754,0.556177) *[red]{\scriptscriptstyle\times};
(6.767967,0.821265) *[red]{\scriptscriptstyle\times};
(6.776181,0.440976) *[red]{\scriptscriptstyle\times};
(6.784394,0.460961) *[red]{\scriptscriptstyle\times};
(6.792608,0.461234) *[red]{\scriptscriptstyle\times};
(6.800821,0.462457) *[red]{\scriptscriptstyle\times};
(6.809035,0.467659) *[red]{\scriptscriptstyle\times};
(6.817248,0.469820) *[red]{\scriptscriptstyle\times};
(6.825462,0.476520) *[red]{\scriptscriptstyle\times};
(6.833676,0.501377) *[red]{\scriptscriptstyle\times};
(6.841889,0.501906) *[red]{\scriptscriptstyle\times};
(6.850103,0.504716) *[red]{\scriptscriptstyle\times};
(6.858316,0.505775) *[red]{\scriptscriptstyle\times};
(6.866530,0.521496) *[red]{\scriptscriptstyle\times};
(6.874743,0.528648) *[red]{\scriptscriptstyle\times};
(6.882957,0.564950) *[red]{\scriptscriptstyle\times};
(6.891170,0.570449) *[red]{\scriptscriptstyle\times};
(6.899384,0.479190) *[red]{\scriptscriptstyle\times};
(6.907598,0.485789) *[red]{\scriptscriptstyle\times};
(6.915811,0.489947) *[red]{\scriptscriptstyle\times};
(6.924025,0.491100) *[red]{\scriptscriptstyle\times};
(6.932238,0.491250) *[red]{\scriptscriptstyle\times};
(6.940452,0.509267) *[red]{\scriptscriptstyle\times};
(6.948665,0.509509) *[red]{\scriptscriptstyle\times};
(6.956879,0.513478) *[red]{\scriptscriptstyle\times};
(6.965092,0.516051) *[red]{\scriptscriptstyle\times};
(6.973306,0.516692) *[red]{\scriptscriptstyle\times};
(6.981520,0.522524) *[red]{\scriptscriptstyle\times};
(6.989733,0.530309) *[red]{\scriptscriptstyle\times};
(6.997947,0.533054) *[red]{\scriptscriptstyle\times};
(7.006160,0.534001) *[red]{\scriptscriptstyle\times};
(7.014374,0.557436) *[red]{\scriptscriptstyle\times};
(7.022587,0.483515) *[red]{\scriptscriptstyle\times};
(7.030801,0.483809) *[red]{\scriptscriptstyle\times};
(7.039014,0.486903) *[red]{\scriptscriptstyle\times};
(7.047228,0.491902) *[red]{\scriptscriptstyle\times};
(7.055441,0.492656) *[red]{\scriptscriptstyle\times};
(7.063655,0.494130) *[red]{\scriptscriptstyle\times};
(7.071869,0.503593) *[red]{\scriptscriptstyle\times};
(7.080082,0.504185) *[red]{\scriptscriptstyle\times};
(7.088296,0.506071) *[red]{\scriptscriptstyle\times};
(7.096509,0.514178) *[red]{\scriptscriptstyle\times};
(7.104723,0.534521) *[red]{\scriptscriptstyle\times};
(7.112936,0.538745) *[red]{\scriptscriptstyle\times};
(7.121150,0.544173) *[red]{\scriptscriptstyle\times};
(7.129363,0.562537) *[red]{\scriptscriptstyle\times};
(7.137577,0.563278) *[red]{\scriptscriptstyle\times};
(7.145791,0.537075) *[red]{\scriptscriptstyle\times};
(7.154004,0.539033) *[red]{\scriptscriptstyle\times};
(7.162218,0.539362) *[red]{\scriptscriptstyle\times};
(7.170431,0.544539) *[red]{\scriptscriptstyle\times};
(7.178645,0.546868) *[red]{\scriptscriptstyle\times};
(7.186858,0.551817) *[red]{\scriptscriptstyle\times};
(7.195072,0.556038) *[red]{\scriptscriptstyle\times};
(7.203285,0.556129) *[red]{\scriptscriptstyle\times};
(7.211499,0.556513) *[red]{\scriptscriptstyle\times};
(7.219713,0.559168) *[red]{\scriptscriptstyle\times};
(7.227926,0.563730) *[red]{\scriptscriptstyle\times};
(7.236140,0.564534) *[red]{\scriptscriptstyle\times};
(7.244353,0.566471) *[red]{\scriptscriptstyle\times};
(7.252567,0.576468) *[red]{\scriptscriptstyle\times};
(7.260780,0.600725) *[red]{\scriptscriptstyle\times};
(7.268994,0.546720) *[red]{\scriptscriptstyle\times};
(7.277207,0.563637) *[red]{\scriptscriptstyle\times};
(7.285421,0.564975) *[red]{\scriptscriptstyle\times};
(7.293634,0.567844) *[red]{\scriptscriptstyle\times};
(7.301848,0.572203) *[red]{\scriptscriptstyle\times};
(7.310062,0.577971) *[red]{\scriptscriptstyle\times};
(7.318275,0.578268) *[red]{\scriptscriptstyle\times};
(7.326489,0.578906) *[red]{\scriptscriptstyle\times};
(7.334702,0.579166) *[red]{\scriptscriptstyle\times};
(7.342916,0.581016) *[red]{\scriptscriptstyle\times};
(7.351129,0.581267) *[red]{\scriptscriptstyle\times};
(7.359343,0.597082) *[red]{\scriptscriptstyle\times};
(7.367556,0.598284) *[red]{\scriptscriptstyle\times};
(7.375770,0.602414) *[red]{\scriptscriptstyle\times};
(7.383984,0.607427) *[red]{\scriptscriptstyle\times};
(7.392197,0.541558) *[red]{\scriptscriptstyle\times};
(7.400411,0.544346) *[red]{\scriptscriptstyle\times};
(7.408624,0.555460) *[red]{\scriptscriptstyle\times};
(7.416838,0.555889) *[red]{\scriptscriptstyle\times};
(7.425051,0.559864) *[red]{\scriptscriptstyle\times};
(7.433265,0.562455) *[red]{\scriptscriptstyle\times};
(7.441478,0.574126) *[red]{\scriptscriptstyle\times};
(7.449692,0.574781) *[red]{\scriptscriptstyle\times};
(7.457906,0.575447) *[red]{\scriptscriptstyle\times};
(7.466119,0.576399) *[red]{\scriptscriptstyle\times};
(7.474333,0.578177) *[red]{\scriptscriptstyle\times};
(7.482546,0.584036) *[red]{\scriptscriptstyle\times};
(7.490760,0.590561) *[red]{\scriptscriptstyle\times};
(7.498973,0.594255) *[red]{\scriptscriptstyle\times};
(7.507187,0.622785) *[red]{\scriptscriptstyle\times};
(7.515400,0.563643) *[red]{\scriptscriptstyle\times};
(7.523614,0.575180) *[red]{\scriptscriptstyle\times};
(7.531828,0.578595) *[red]{\scriptscriptstyle\times};
(7.540041,0.579424) *[red]{\scriptscriptstyle\times};
(7.548255,0.583165) *[red]{\scriptscriptstyle\times};
(7.556468,0.585029) *[red]{\scriptscriptstyle\times};
(7.564682,0.597023) *[red]{\scriptscriptstyle\times};
(7.572895,0.601156) *[red]{\scriptscriptstyle\times};
(7.581109,0.601623) *[red]{\scriptscriptstyle\times};
(7.589322,0.604585) *[red]{\scriptscriptstyle\times};
(7.597536,0.605623) *[red]{\scriptscriptstyle\times};
(7.605749,0.608874) *[red]{\scriptscriptstyle\times};
(7.613963,0.617921) *[red]{\scriptscriptstyle\times};
(7.622177,0.618188) *[red]{\scriptscriptstyle\times};
(7.630390,0.621381) *[red]{\scriptscriptstyle\times};
(7.638604,0.539597) *[red]{\scriptscriptstyle\times};
(7.646817,0.558973) *[red]{\scriptscriptstyle\times};
(7.655031,0.564403) *[red]{\scriptscriptstyle\times};
(7.663244,0.566508) *[red]{\scriptscriptstyle\times};
(7.671458,0.568409) *[red]{\scriptscriptstyle\times};
(7.679671,0.568885) *[red]{\scriptscriptstyle\times};
(7.687885,0.574951) *[red]{\scriptscriptstyle\times};
(7.696099,0.576550) *[red]{\scriptscriptstyle\times};
(7.704312,0.579444) *[red]{\scriptscriptstyle\times};
(7.712526,0.583135) *[red]{\scriptscriptstyle\times};
(7.720739,0.607553) *[red]{\scriptscriptstyle\times};
(7.728953,0.610818) *[red]{\scriptscriptstyle\times};
(7.737166,0.614166) *[red]{\scriptscriptstyle\times};
(7.745380,0.614172) *[red]{\scriptscriptstyle\times};
(7.753593,0.639317) *[red]{\scriptscriptstyle\times};
(7.761807,0.558172) *[red]{\scriptscriptstyle\times};
(7.770021,0.572521) *[red]{\scriptscriptstyle\times};
(7.778234,0.577254) *[red]{\scriptscriptstyle\times};
(7.786448,0.578494) *[red]{\scriptscriptstyle\times};
(7.794661,0.578739) *[red]{\scriptscriptstyle\times};
(7.802875,0.581092) *[red]{\scriptscriptstyle\times};
(7.811088,0.590070) *[red]{\scriptscriptstyle\times};
(7.819302,0.591291) *[red]{\scriptscriptstyle\times};
(7.827515,0.593997) *[red]{\scriptscriptstyle\times};
(7.835729,0.596473) *[red]{\scriptscriptstyle\times};
(7.843943,0.596986) *[red]{\scriptscriptstyle\times};
(7.852156,0.605772) *[red]{\scriptscriptstyle\times};
(7.860370,0.630123) *[red]{\scriptscriptstyle\times};
(7.868583,0.633226) *[red]{\scriptscriptstyle\times};
(7.876797,0.684438) *[red]{\scriptscriptstyle\times};
(7.885010,0.568010) *[red]{\scriptscriptstyle\times};
(7.893224,0.569598) *[red]{\scriptscriptstyle\times};
(7.901437,0.571655) *[red]{\scriptscriptstyle\times};
(7.909651,0.574129) *[red]{\scriptscriptstyle\times};
(7.917864,0.574836) *[red]{\scriptscriptstyle\times};
(7.926078,0.582980) *[red]{\scriptscriptstyle\times};
(7.934292,0.583299) *[red]{\scriptscriptstyle\times};
(7.942505,0.583757) *[red]{\scriptscriptstyle\times};
(7.950719,0.584219) *[red]{\scriptscriptstyle\times};
(7.958932,0.586240) *[red]{\scriptscriptstyle\times};
(7.967146,0.589188) *[red]{\scriptscriptstyle\times};
(7.975359,0.598523) *[red]{\scriptscriptstyle\times};
(7.983573,0.608156) *[red]{\scriptscriptstyle\times};
(7.991786,0.610977) *[red]{\scriptscriptstyle\times};
(8.000000,0.653308) *[red]{\scriptscriptstyle\times};
\endxy
  }
  \caption{Skylake cycles
    for the CSIDH-512 action
    using {\tt velusqrt-flint}.
  }
  \label{fig:csidh-c}
\end{figure}

\begin{figure}[t]
  \centerline{
\xy <1.1cm,0cm>:<0cm,4cm>::
(0,0.291806); (8,0.291806) **[blue]@{-};
(8.1,0.291806) *[blue]{\rlap{293801222}};
(0,0.386163); (8,0.386163) **[blue]@{-};
(8.1,0.386163) *[blue]{\rlap{313659026}};
(0,0.445117); (8,0.445117) **[blue]@{-};
(8.1,0.445117) *[blue]{\rlap{326741956}};
(0,0.252189); (8,0.252189) **[red]@{-};
(-0.1,0.252189) *[red]{\llap{285843060}};
(0,0.347292); (8,0.347292) **[red]@{-};
(-0.1,0.347292) *[red]{\llap{305320816}};
(0,0.403672); (8,0.403672) **[red]@{-};
(-0.1,0.403672) *[red]{\llap{317489062}};
(-0.004107,0.065432); (-0.004107,1.112316) **[lightgray]@{-};
(0.119097,0.065432); (0.119097,1.112316) **[lightgray]@{-};
(0.242300,0.065432); (0.242300,1.112316) **[lightgray]@{-};
(0.365503,0.065432); (0.365503,1.112316) **[lightgray]@{-};
(0.488706,0.065432); (0.488706,1.112316) **[lightgray]@{-};
(0.611910,0.065432); (0.611910,1.112316) **[lightgray]@{-};
(0.735113,0.065432); (0.735113,1.112316) **[lightgray]@{-};
(0.858316,0.065432); (0.858316,1.112316) **[lightgray]@{-};
(0.981520,0.065432); (0.981520,1.112316) **[lightgray]@{-};
(1.104723,0.065432); (1.104723,1.112316) **[lightgray]@{-};
(1.227926,0.065432); (1.227926,1.112316) **[lightgray]@{-};
(1.351129,0.065432); (1.351129,1.112316) **[lightgray]@{-};
(1.474333,0.065432); (1.474333,1.112316) **[lightgray]@{-};
(1.597536,0.065432); (1.597536,1.112316) **[lightgray]@{-};
(1.720739,0.065432); (1.720739,1.112316) **[lightgray]@{-};
(1.843943,0.065432); (1.843943,1.112316) **[lightgray]@{-};
(1.967146,0.065432); (1.967146,1.112316) **[lightgray]@{-};
(2.090349,0.065432); (2.090349,1.112316) **[lightgray]@{-};
(2.213552,0.065432); (2.213552,1.112316) **[lightgray]@{-};
(2.336756,0.065432); (2.336756,1.112316) **[lightgray]@{-};
(2.459959,0.065432); (2.459959,1.112316) **[lightgray]@{-};
(2.583162,0.065432); (2.583162,1.112316) **[lightgray]@{-};
(2.706366,0.065432); (2.706366,1.112316) **[lightgray]@{-};
(2.829569,0.065432); (2.829569,1.112316) **[lightgray]@{-};
(2.952772,0.065432); (2.952772,1.112316) **[lightgray]@{-};
(3.075975,0.065432); (3.075975,1.112316) **[lightgray]@{-};
(3.199179,0.065432); (3.199179,1.112316) **[lightgray]@{-};
(3.322382,0.065432); (3.322382,1.112316) **[lightgray]@{-};
(3.445585,0.065432); (3.445585,1.112316) **[lightgray]@{-};
(3.568789,0.065432); (3.568789,1.112316) **[lightgray]@{-};
(3.691992,0.065432); (3.691992,1.112316) **[lightgray]@{-};
(3.815195,0.065432); (3.815195,1.112316) **[lightgray]@{-};
(3.938398,0.065432); (3.938398,1.112316) **[lightgray]@{-};
(4.061602,0.065432); (4.061602,1.112316) **[lightgray]@{-};
(4.184805,0.065432); (4.184805,1.112316) **[lightgray]@{-};
(4.308008,0.065432); (4.308008,1.112316) **[lightgray]@{-};
(4.431211,0.065432); (4.431211,1.112316) **[lightgray]@{-};
(4.554415,0.065432); (4.554415,1.112316) **[lightgray]@{-};
(4.677618,0.065432); (4.677618,1.112316) **[lightgray]@{-};
(4.800821,0.065432); (4.800821,1.112316) **[lightgray]@{-};
(4.924025,0.065432); (4.924025,1.112316) **[lightgray]@{-};
(5.047228,0.065432); (5.047228,1.112316) **[lightgray]@{-};
(5.170431,0.065432); (5.170431,1.112316) **[lightgray]@{-};
(5.293634,0.065432); (5.293634,1.112316) **[lightgray]@{-};
(5.416838,0.065432); (5.416838,1.112316) **[lightgray]@{-};
(5.540041,0.065432); (5.540041,1.112316) **[lightgray]@{-};
(5.663244,0.065432); (5.663244,1.112316) **[lightgray]@{-};
(5.786448,0.065432); (5.786448,1.112316) **[lightgray]@{-};
(5.909651,0.065432); (5.909651,1.112316) **[lightgray]@{-};
(6.032854,0.065432); (6.032854,1.112316) **[lightgray]@{-};
(6.156057,0.065432); (6.156057,1.112316) **[lightgray]@{-};
(6.279261,0.065432); (6.279261,1.112316) **[lightgray]@{-};
(6.402464,0.065432); (6.402464,1.112316) **[lightgray]@{-};
(6.525667,0.065432); (6.525667,1.112316) **[lightgray]@{-};
(6.648871,0.065432); (6.648871,1.112316) **[lightgray]@{-};
(6.772074,0.065432); (6.772074,1.112316) **[lightgray]@{-};
(6.895277,0.065432); (6.895277,1.112316) **[lightgray]@{-};
(7.018480,0.065432); (7.018480,1.112316) **[lightgray]@{-};
(7.141684,0.065432); (7.141684,1.112316) **[lightgray]@{-};
(7.264887,0.065432); (7.264887,1.112316) **[lightgray]@{-};
(7.388090,0.065432); (7.388090,1.112316) **[lightgray]@{-};
(7.511294,0.065432); (7.511294,1.112316) **[lightgray]@{-};
(7.634497,0.065432); (7.634497,1.112316) **[lightgray]@{-};
(7.757700,0.065432); (7.757700,1.112316) **[lightgray]@{-};
(7.880903,0.065432); (7.880903,1.112316) **[lightgray]@{-};
(8.004107,0.065432); (8.004107,1.112316) **[lightgray]@{-};
(-0.004107,0.065432); (-0.004107,1.112316) **[lightgray]@{-};
(0.119097,0.065432); (0.119097,1.112316) **[lightgray]@{-};
(0.242300,0.065432); (0.242300,1.112316) **[lightgray]@{-};
(0.365503,0.065432); (0.365503,1.112316) **[lightgray]@{-};
(0.488706,0.065432); (0.488706,1.112316) **[lightgray]@{-};
(0.611910,0.065432); (0.611910,1.112316) **[lightgray]@{-};
(0.735113,0.065432); (0.735113,1.112316) **[lightgray]@{-};
(0.858316,0.065432); (0.858316,1.112316) **[lightgray]@{-};
(0.981520,0.065432); (0.981520,1.112316) **[lightgray]@{-};
(1.104723,0.065432); (1.104723,1.112316) **[lightgray]@{-};
(1.227926,0.065432); (1.227926,1.112316) **[lightgray]@{-};
(1.351129,0.065432); (1.351129,1.112316) **[lightgray]@{-};
(1.474333,0.065432); (1.474333,1.112316) **[lightgray]@{-};
(1.597536,0.065432); (1.597536,1.112316) **[lightgray]@{-};
(1.720739,0.065432); (1.720739,1.112316) **[lightgray]@{-};
(1.843943,0.065432); (1.843943,1.112316) **[lightgray]@{-};
(1.967146,0.065432); (1.967146,1.112316) **[lightgray]@{-};
(2.090349,0.065432); (2.090349,1.112316) **[lightgray]@{-};
(2.213552,0.065432); (2.213552,1.112316) **[lightgray]@{-};
(2.336756,0.065432); (2.336756,1.112316) **[lightgray]@{-};
(2.459959,0.065432); (2.459959,1.112316) **[lightgray]@{-};
(2.583162,0.065432); (2.583162,1.112316) **[lightgray]@{-};
(2.706366,0.065432); (2.706366,1.112316) **[lightgray]@{-};
(2.829569,0.065432); (2.829569,1.112316) **[lightgray]@{-};
(2.952772,0.065432); (2.952772,1.112316) **[lightgray]@{-};
(3.075975,0.065432); (3.075975,1.112316) **[lightgray]@{-};
(3.199179,0.065432); (3.199179,1.112316) **[lightgray]@{-};
(3.322382,0.065432); (3.322382,1.112316) **[lightgray]@{-};
(3.445585,0.065432); (3.445585,1.112316) **[lightgray]@{-};
(3.568789,0.065432); (3.568789,1.112316) **[lightgray]@{-};
(3.691992,0.065432); (3.691992,1.112316) **[lightgray]@{-};
(3.815195,0.065432); (3.815195,1.112316) **[lightgray]@{-};
(3.938398,0.065432); (3.938398,1.112316) **[lightgray]@{-};
(4.061602,0.065432); (4.061602,1.112316) **[lightgray]@{-};
(4.184805,0.065432); (4.184805,1.112316) **[lightgray]@{-};
(4.308008,0.065432); (4.308008,1.112316) **[lightgray]@{-};
(4.431211,0.065432); (4.431211,1.112316) **[lightgray]@{-};
(4.554415,0.065432); (4.554415,1.112316) **[lightgray]@{-};
(4.677618,0.065432); (4.677618,1.112316) **[lightgray]@{-};
(4.800821,0.065432); (4.800821,1.112316) **[lightgray]@{-};
(4.924025,0.065432); (4.924025,1.112316) **[lightgray]@{-};
(5.047228,0.065432); (5.047228,1.112316) **[lightgray]@{-};
(5.170431,0.065432); (5.170431,1.112316) **[lightgray]@{-};
(5.293634,0.065432); (5.293634,1.112316) **[lightgray]@{-};
(5.416838,0.065432); (5.416838,1.112316) **[lightgray]@{-};
(5.540041,0.065432); (5.540041,1.112316) **[lightgray]@{-};
(5.663244,0.065432); (5.663244,1.112316) **[lightgray]@{-};
(5.786448,0.065432); (5.786448,1.112316) **[lightgray]@{-};
(5.909651,0.065432); (5.909651,1.112316) **[lightgray]@{-};
(6.032854,0.065432); (6.032854,1.112316) **[lightgray]@{-};
(6.156057,0.065432); (6.156057,1.112316) **[lightgray]@{-};
(6.279261,0.065432); (6.279261,1.112316) **[lightgray]@{-};
(6.402464,0.065432); (6.402464,1.112316) **[lightgray]@{-};
(6.525667,0.065432); (6.525667,1.112316) **[lightgray]@{-};
(6.648871,0.065432); (6.648871,1.112316) **[lightgray]@{-};
(6.772074,0.065432); (6.772074,1.112316) **[lightgray]@{-};
(6.895277,0.065432); (6.895277,1.112316) **[lightgray]@{-};
(7.018480,0.065432); (7.018480,1.112316) **[lightgray]@{-};
(7.141684,0.065432); (7.141684,1.112316) **[lightgray]@{-};
(7.264887,0.065432); (7.264887,1.112316) **[lightgray]@{-};
(7.388090,0.065432); (7.388090,1.112316) **[lightgray]@{-};
(7.511294,0.065432); (7.511294,1.112316) **[lightgray]@{-};
(7.634497,0.065432); (7.634497,1.112316) **[lightgray]@{-};
(7.757700,0.065432); (7.757700,1.112316) **[lightgray]@{-};
(7.880903,0.065432); (7.880903,1.112316) **[lightgray]@{-};
(8.004107,0.065432); (8.004107,1.112316) **[lightgray]@{-};
(0.000000,0.105735) *[blue]{\scriptscriptstyle+};
(0.008214,0.106313) *[blue]{\scriptscriptstyle+};
(0.016427,0.106333) *[blue]{\scriptscriptstyle+};
(0.024641,0.106507) *[blue]{\scriptscriptstyle+};
(0.032854,0.107432) *[blue]{\scriptscriptstyle+};
(0.041068,0.108715) *[blue]{\scriptscriptstyle+};
(0.049281,0.110857) *[blue]{\scriptscriptstyle+};
(0.057495,0.111928) *[blue]{\scriptscriptstyle+};
(0.065708,0.112861) *[blue]{\scriptscriptstyle+};
(0.073922,0.117696) *[blue]{\scriptscriptstyle+};
(0.082136,0.122032) *[blue]{\scriptscriptstyle+};
(0.090349,0.136895) *[blue]{\scriptscriptstyle+};
(0.098563,0.141486) *[blue]{\scriptscriptstyle+};
(0.106776,0.145066) *[blue]{\scriptscriptstyle+};
(0.114990,0.168703) *[blue]{\scriptscriptstyle+};
(0.123203,0.124690) *[blue]{\scriptscriptstyle+};
(0.131417,0.125497) *[blue]{\scriptscriptstyle+};
(0.139630,0.127782) *[blue]{\scriptscriptstyle+};
(0.147844,0.128422) *[blue]{\scriptscriptstyle+};
(0.156057,0.128616) *[blue]{\scriptscriptstyle+};
(0.164271,0.129432) *[blue]{\scriptscriptstyle+};
(0.172485,0.130919) *[blue]{\scriptscriptstyle+};
(0.180698,0.133541) *[blue]{\scriptscriptstyle+};
(0.188912,0.144453) *[blue]{\scriptscriptstyle+};
(0.197125,0.154767) *[blue]{\scriptscriptstyle+};
(0.205339,0.159145) *[blue]{\scriptscriptstyle+};
(0.213552,0.161171) *[blue]{\scriptscriptstyle+};
(0.221766,0.170293) *[blue]{\scriptscriptstyle+};
(0.229979,0.189287) *[blue]{\scriptscriptstyle+};
(0.238193,0.195178) *[blue]{\scriptscriptstyle+};
(0.246407,0.127937) *[blue]{\scriptscriptstyle+};
(0.254620,0.131311) *[blue]{\scriptscriptstyle+};
(0.262834,0.131347) *[blue]{\scriptscriptstyle+};
(0.271047,0.132465) *[blue]{\scriptscriptstyle+};
(0.279261,0.133534) *[blue]{\scriptscriptstyle+};
(0.287474,0.136907) *[blue]{\scriptscriptstyle+};
(0.295688,0.138114) *[blue]{\scriptscriptstyle+};
(0.303901,0.138411) *[blue]{\scriptscriptstyle+};
(0.312115,0.162152) *[blue]{\scriptscriptstyle+};
(0.320329,0.165984) *[blue]{\scriptscriptstyle+};
(0.328542,0.176576) *[blue]{\scriptscriptstyle+};
(0.336756,0.186837) *[blue]{\scriptscriptstyle+};
(0.344969,0.187267) *[blue]{\scriptscriptstyle+};
(0.353183,0.190965) *[blue]{\scriptscriptstyle+};
(0.361396,0.251535) *[blue]{\scriptscriptstyle+};
(0.369610,0.153062) *[blue]{\scriptscriptstyle+};
(0.377823,0.153522) *[blue]{\scriptscriptstyle+};
(0.386037,0.155543) *[blue]{\scriptscriptstyle+};
(0.394251,0.156349) *[blue]{\scriptscriptstyle+};
(0.402464,0.156893) *[blue]{\scriptscriptstyle+};
(0.410678,0.157004) *[blue]{\scriptscriptstyle+};
(0.418891,0.157400) *[blue]{\scriptscriptstyle+};
(0.427105,0.158072) *[blue]{\scriptscriptstyle+};
(0.435318,0.159272) *[blue]{\scriptscriptstyle+};
(0.443532,0.161285) *[blue]{\scriptscriptstyle+};
(0.451745,0.161656) *[blue]{\scriptscriptstyle+};
(0.459959,0.184609) *[blue]{\scriptscriptstyle+};
(0.468172,0.191682) *[blue]{\scriptscriptstyle+};
(0.476386,0.193139) *[blue]{\scriptscriptstyle+};
(0.484600,0.215695) *[blue]{\scriptscriptstyle+};
(0.492813,0.167869) *[blue]{\scriptscriptstyle+};
(0.501027,0.168170) *[blue]{\scriptscriptstyle+};
(0.509240,0.169238) *[blue]{\scriptscriptstyle+};
(0.517454,0.169961) *[blue]{\scriptscriptstyle+};
(0.525667,0.171065) *[blue]{\scriptscriptstyle+};
(0.533881,0.171223) *[blue]{\scriptscriptstyle+};
(0.542094,0.173683) *[blue]{\scriptscriptstyle+};
(0.550308,0.173891) *[blue]{\scriptscriptstyle+};
(0.558522,0.177677) *[blue]{\scriptscriptstyle+};
(0.566735,0.180474) *[blue]{\scriptscriptstyle+};
(0.574949,0.180958) *[blue]{\scriptscriptstyle+};
(0.583162,0.194735) *[blue]{\scriptscriptstyle+};
(0.591376,0.199068) *[blue]{\scriptscriptstyle+};
(0.599589,0.205858) *[blue]{\scriptscriptstyle+};
(0.607803,0.211696) *[blue]{\scriptscriptstyle+};
(0.616016,0.179867) *[blue]{\scriptscriptstyle+};
(0.624230,0.179953) *[blue]{\scriptscriptstyle+};
(0.632444,0.181335) *[blue]{\scriptscriptstyle+};
(0.640657,0.181855) *[blue]{\scriptscriptstyle+};
(0.648871,0.182061) *[blue]{\scriptscriptstyle+};
(0.657084,0.185468) *[blue]{\scriptscriptstyle+};
(0.665298,0.185837) *[blue]{\scriptscriptstyle+};
(0.673511,0.191750) *[blue]{\scriptscriptstyle+};
(0.681725,0.193589) *[blue]{\scriptscriptstyle+};
(0.689938,0.212711) *[blue]{\scriptscriptstyle+};
(0.698152,0.212899) *[blue]{\scriptscriptstyle+};
(0.706366,0.214855) *[blue]{\scriptscriptstyle+};
(0.714579,0.219558) *[blue]{\scriptscriptstyle+};
(0.722793,0.256376) *[blue]{\scriptscriptstyle+};
(0.731006,0.262435) *[blue]{\scriptscriptstyle+};
(0.739220,0.179462) *[blue]{\scriptscriptstyle+};
(0.747433,0.182253) *[blue]{\scriptscriptstyle+};
(0.755647,0.182344) *[blue]{\scriptscriptstyle+};
(0.763860,0.184100) *[blue]{\scriptscriptstyle+};
(0.772074,0.184787) *[blue]{\scriptscriptstyle+};
(0.780287,0.186600) *[blue]{\scriptscriptstyle+};
(0.788501,0.186694) *[blue]{\scriptscriptstyle+};
(0.796715,0.192432) *[blue]{\scriptscriptstyle+};
(0.804928,0.213864) *[blue]{\scriptscriptstyle+};
(0.813142,0.214893) *[blue]{\scriptscriptstyle+};
(0.821355,0.218140) *[blue]{\scriptscriptstyle+};
(0.829569,0.219045) *[blue]{\scriptscriptstyle+};
(0.837782,0.220178) *[blue]{\scriptscriptstyle+};
(0.845996,0.220556) *[blue]{\scriptscriptstyle+};
(0.854209,0.239289) *[blue]{\scriptscriptstyle+};
(0.862423,0.160886) *[blue]{\scriptscriptstyle+};
(0.870637,0.161152) *[blue]{\scriptscriptstyle+};
(0.878850,0.169769) *[blue]{\scriptscriptstyle+};
(0.887064,0.186501) *[blue]{\scriptscriptstyle+};
(0.895277,0.187511) *[blue]{\scriptscriptstyle+};
(0.903491,0.188283) *[blue]{\scriptscriptstyle+};
(0.911704,0.189479) *[blue]{\scriptscriptstyle+};
(0.919918,0.190237) *[blue]{\scriptscriptstyle+};
(0.928131,0.196141) *[blue]{\scriptscriptstyle+};
(0.936345,0.197689) *[blue]{\scriptscriptstyle+};
(0.944559,0.198333) *[blue]{\scriptscriptstyle+};
(0.952772,0.214237) *[blue]{\scriptscriptstyle+};
(0.960986,0.216282) *[blue]{\scriptscriptstyle+};
(0.969199,0.232281) *[blue]{\scriptscriptstyle+};
(0.977413,0.248713) *[blue]{\scriptscriptstyle+};
(0.985626,0.189666) *[blue]{\scriptscriptstyle+};
(0.993840,0.190658) *[blue]{\scriptscriptstyle+};
(1.002053,0.194369) *[blue]{\scriptscriptstyle+};
(1.010267,0.195637) *[blue]{\scriptscriptstyle+};
(1.018480,0.196824) *[blue]{\scriptscriptstyle+};
(1.026694,0.205031) *[blue]{\scriptscriptstyle+};
(1.034908,0.219917) *[blue]{\scriptscriptstyle+};
(1.043121,0.221814) *[blue]{\scriptscriptstyle+};
(1.051335,0.222098) *[blue]{\scriptscriptstyle+};
(1.059548,0.222314) *[blue]{\scriptscriptstyle+};
(1.067762,0.226663) *[blue]{\scriptscriptstyle+};
(1.075975,0.228147) *[blue]{\scriptscriptstyle+};
(1.084189,0.228813) *[blue]{\scriptscriptstyle+};
(1.092402,0.229623) *[blue]{\scriptscriptstyle+};
(1.100616,0.251643) *[blue]{\scriptscriptstyle+};
(1.108830,0.195274) *[blue]{\scriptscriptstyle+};
(1.117043,0.195593) *[blue]{\scriptscriptstyle+};
(1.125257,0.197042) *[blue]{\scriptscriptstyle+};
(1.133470,0.197255) *[blue]{\scriptscriptstyle+};
(1.141684,0.197521) *[blue]{\scriptscriptstyle+};
(1.149897,0.202006) *[blue]{\scriptscriptstyle+};
(1.158111,0.203540) *[blue]{\scriptscriptstyle+};
(1.166324,0.223227) *[blue]{\scriptscriptstyle+};
(1.174538,0.223497) *[blue]{\scriptscriptstyle+};
(1.182752,0.223870) *[blue]{\scriptscriptstyle+};
(1.190965,0.226248) *[blue]{\scriptscriptstyle+};
(1.199179,0.227829) *[blue]{\scriptscriptstyle+};
(1.207392,0.228170) *[blue]{\scriptscriptstyle+};
(1.215606,0.229387) *[blue]{\scriptscriptstyle+};
(1.223819,0.288772) *[blue]{\scriptscriptstyle+};
(1.232033,0.213502) *[blue]{\scriptscriptstyle+};
(1.240246,0.213667) *[blue]{\scriptscriptstyle+};
(1.248460,0.213967) *[blue]{\scriptscriptstyle+};
(1.256674,0.215489) *[blue]{\scriptscriptstyle+};
(1.264887,0.216807) *[blue]{\scriptscriptstyle+};
(1.273101,0.220542) *[blue]{\scriptscriptstyle+};
(1.281314,0.221452) *[blue]{\scriptscriptstyle+};
(1.289528,0.223889) *[blue]{\scriptscriptstyle+};
(1.297741,0.224150) *[blue]{\scriptscriptstyle+};
(1.305955,0.227633) *[blue]{\scriptscriptstyle+};
(1.314168,0.241313) *[blue]{\scriptscriptstyle+};
(1.322382,0.242501) *[blue]{\scriptscriptstyle+};
(1.330595,0.247926) *[blue]{\scriptscriptstyle+};
(1.338809,0.252607) *[blue]{\scriptscriptstyle+};
(1.347023,0.253602) *[blue]{\scriptscriptstyle+};
(1.355236,0.209935) *[blue]{\scriptscriptstyle+};
(1.363450,0.212431) *[blue]{\scriptscriptstyle+};
(1.371663,0.213683) *[blue]{\scriptscriptstyle+};
(1.379877,0.215576) *[blue]{\scriptscriptstyle+};
(1.388090,0.217578) *[blue]{\scriptscriptstyle+};
(1.396304,0.218736) *[blue]{\scriptscriptstyle+};
(1.404517,0.219045) *[blue]{\scriptscriptstyle+};
(1.412731,0.237270) *[blue]{\scriptscriptstyle+};
(1.420945,0.242070) *[blue]{\scriptscriptstyle+};
(1.429158,0.243722) *[blue]{\scriptscriptstyle+};
(1.437372,0.244715) *[blue]{\scriptscriptstyle+};
(1.445585,0.258159) *[blue]{\scriptscriptstyle+};
(1.453799,0.263953) *[blue]{\scriptscriptstyle+};
(1.462012,0.270047) *[blue]{\scriptscriptstyle+};
(1.470226,0.274204) *[blue]{\scriptscriptstyle+};
(1.478439,0.218427) *[blue]{\scriptscriptstyle+};
(1.486653,0.218783) *[blue]{\scriptscriptstyle+};
(1.494867,0.218817) *[blue]{\scriptscriptstyle+};
(1.503080,0.219746) *[blue]{\scriptscriptstyle+};
(1.511294,0.219957) *[blue]{\scriptscriptstyle+};
(1.519507,0.221394) *[blue]{\scriptscriptstyle+};
(1.527721,0.221857) *[blue]{\scriptscriptstyle+};
(1.535934,0.223599) *[blue]{\scriptscriptstyle+};
(1.544148,0.225352) *[blue]{\scriptscriptstyle+};
(1.552361,0.231187) *[blue]{\scriptscriptstyle+};
(1.560575,0.232755) *[blue]{\scriptscriptstyle+};
(1.568789,0.235724) *[blue]{\scriptscriptstyle+};
(1.577002,0.241473) *[blue]{\scriptscriptstyle+};
(1.585216,0.245985) *[blue]{\scriptscriptstyle+};
(1.593429,0.260084) *[blue]{\scriptscriptstyle+};
(1.601643,0.255301) *[blue]{\scriptscriptstyle+};
(1.609856,0.255689) *[blue]{\scriptscriptstyle+};
(1.618070,0.258505) *[blue]{\scriptscriptstyle+};
(1.626283,0.259863) *[blue]{\scriptscriptstyle+};
(1.634497,0.260268) *[blue]{\scriptscriptstyle+};
(1.642710,0.261283) *[blue]{\scriptscriptstyle+};
(1.650924,0.261545) *[blue]{\scriptscriptstyle+};
(1.659138,0.262294) *[blue]{\scriptscriptstyle+};
(1.667351,0.264826) *[blue]{\scriptscriptstyle+};
(1.675565,0.273979) *[blue]{\scriptscriptstyle+};
(1.683778,0.281809) *[blue]{\scriptscriptstyle+};
(1.691992,0.309672) *[blue]{\scriptscriptstyle+};
(1.700205,0.312777) *[blue]{\scriptscriptstyle+};
(1.708419,0.319641) *[blue]{\scriptscriptstyle+};
(1.716632,0.319793) *[blue]{\scriptscriptstyle+};
(1.724846,0.263824) *[blue]{\scriptscriptstyle+};
(1.733060,0.265010) *[blue]{\scriptscriptstyle+};
(1.741273,0.265695) *[blue]{\scriptscriptstyle+};
(1.749487,0.266636) *[blue]{\scriptscriptstyle+};
(1.757700,0.267090) *[blue]{\scriptscriptstyle+};
(1.765914,0.270960) *[blue]{\scriptscriptstyle+};
(1.774127,0.271318) *[blue]{\scriptscriptstyle+};
(1.782341,0.290491) *[blue]{\scriptscriptstyle+};
(1.790554,0.293696) *[blue]{\scriptscriptstyle+};
(1.798768,0.296368) *[blue]{\scriptscriptstyle+};
(1.806982,0.297861) *[blue]{\scriptscriptstyle+};
(1.815195,0.298201) *[blue]{\scriptscriptstyle+};
(1.823409,0.303262) *[blue]{\scriptscriptstyle+};
(1.831622,0.308670) *[blue]{\scriptscriptstyle+};
(1.839836,0.326849) *[blue]{\scriptscriptstyle+};
(1.848049,0.262426) *[blue]{\scriptscriptstyle+};
(1.856263,0.262891) *[blue]{\scriptscriptstyle+};
(1.864476,0.263060) *[blue]{\scriptscriptstyle+};
(1.872690,0.264978) *[blue]{\scriptscriptstyle+};
(1.880903,0.265769) *[blue]{\scriptscriptstyle+};
(1.889117,0.266226) *[blue]{\scriptscriptstyle+};
(1.897331,0.267691) *[blue]{\scriptscriptstyle+};
(1.905544,0.270749) *[blue]{\scriptscriptstyle+};
(1.913758,0.273682) *[blue]{\scriptscriptstyle+};
(1.921971,0.275825) *[blue]{\scriptscriptstyle+};
(1.930185,0.290842) *[blue]{\scriptscriptstyle+};
(1.938398,0.293773) *[blue]{\scriptscriptstyle+};
(1.946612,0.296327) *[blue]{\scriptscriptstyle+};
(1.954825,0.300514) *[blue]{\scriptscriptstyle+};
(1.963039,0.302066) *[blue]{\scriptscriptstyle+};
(1.971253,0.243034) *[blue]{\scriptscriptstyle+};
(1.979466,0.253903) *[blue]{\scriptscriptstyle+};
(1.987680,0.256203) *[blue]{\scriptscriptstyle+};
(1.995893,0.271895) *[blue]{\scriptscriptstyle+};
(2.004107,0.273351) *[blue]{\scriptscriptstyle+};
(2.012320,0.274668) *[blue]{\scriptscriptstyle+};
(2.020534,0.275448) *[blue]{\scriptscriptstyle+};
(2.028747,0.275580) *[blue]{\scriptscriptstyle+};
(2.036961,0.275630) *[blue]{\scriptscriptstyle+};
(2.045175,0.281073) *[blue]{\scriptscriptstyle+};
(2.053388,0.286558) *[blue]{\scriptscriptstyle+};
(2.061602,0.289047) *[blue]{\scriptscriptstyle+};
(2.069815,0.291553) *[blue]{\scriptscriptstyle+};
(2.078029,0.297619) *[blue]{\scriptscriptstyle+};
(2.086242,0.300116) *[blue]{\scriptscriptstyle+};
(2.094456,0.287020) *[blue]{\scriptscriptstyle+};
(2.102669,0.287138) *[blue]{\scriptscriptstyle+};
(2.110883,0.287929) *[blue]{\scriptscriptstyle+};
(2.119097,0.291699) *[blue]{\scriptscriptstyle+};
(2.127310,0.291806) *[blue]{\scriptscriptstyle+};
(2.135524,0.293920) *[blue]{\scriptscriptstyle+};
(2.143737,0.294107) *[blue]{\scriptscriptstyle+};
(2.151951,0.295356) *[blue]{\scriptscriptstyle+};
(2.160164,0.295926) *[blue]{\scriptscriptstyle+};
(2.168378,0.302536) *[blue]{\scriptscriptstyle+};
(2.176591,0.316361) *[blue]{\scriptscriptstyle+};
(2.184805,0.319430) *[blue]{\scriptscriptstyle+};
(2.193018,0.321512) *[blue]{\scriptscriptstyle+};
(2.201232,0.322424) *[blue]{\scriptscriptstyle+};
(2.209446,0.351076) *[blue]{\scriptscriptstyle+};
(2.217659,0.287573) *[blue]{\scriptscriptstyle+};
(2.225873,0.295615) *[blue]{\scriptscriptstyle+};
(2.234086,0.295622) *[blue]{\scriptscriptstyle+};
(2.242300,0.296342) *[blue]{\scriptscriptstyle+};
(2.250513,0.298768) *[blue]{\scriptscriptstyle+};
(2.258727,0.313823) *[blue]{\scriptscriptstyle+};
(2.266940,0.314054) *[blue]{\scriptscriptstyle+};
(2.275154,0.315296) *[blue]{\scriptscriptstyle+};
(2.283368,0.315846) *[blue]{\scriptscriptstyle+};
(2.291581,0.319066) *[blue]{\scriptscriptstyle+};
(2.299795,0.326349) *[blue]{\scriptscriptstyle+};
(2.308008,0.345017) *[blue]{\scriptscriptstyle+};
(2.316222,0.346387) *[blue]{\scriptscriptstyle+};
(2.324435,0.372104) *[blue]{\scriptscriptstyle+};
(2.332649,0.381616) *[blue]{\scriptscriptstyle+};
(2.340862,0.310433) *[blue]{\scriptscriptstyle+};
(2.349076,0.311726) *[blue]{\scriptscriptstyle+};
(2.357290,0.312060) *[blue]{\scriptscriptstyle+};
(2.365503,0.312995) *[blue]{\scriptscriptstyle+};
(2.373717,0.316685) *[blue]{\scriptscriptstyle+};
(2.381930,0.317047) *[blue]{\scriptscriptstyle+};
(2.390144,0.318102) *[blue]{\scriptscriptstyle+};
(2.398357,0.318191) *[blue]{\scriptscriptstyle+};
(2.406571,0.319365) *[blue]{\scriptscriptstyle+};
(2.414784,0.335122) *[blue]{\scriptscriptstyle+};
(2.422998,0.337264) *[blue]{\scriptscriptstyle+};
(2.431211,0.339325) *[blue]{\scriptscriptstyle+};
(2.439425,0.339747) *[blue]{\scriptscriptstyle+};
(2.447639,0.341579) *[blue]{\scriptscriptstyle+};
(2.455852,0.344467) *[blue]{\scriptscriptstyle+};
(2.464066,0.318496) *[blue]{\scriptscriptstyle+};
(2.472279,0.321220) *[blue]{\scriptscriptstyle+};
(2.480493,0.321582) *[blue]{\scriptscriptstyle+};
(2.488706,0.322117) *[blue]{\scriptscriptstyle+};
(2.496920,0.322825) *[blue]{\scriptscriptstyle+};
(2.505133,0.322883) *[blue]{\scriptscriptstyle+};
(2.513347,0.323413) *[blue]{\scriptscriptstyle+};
(2.521561,0.323435) *[blue]{\scriptscriptstyle+};
(2.529774,0.325092) *[blue]{\scriptscriptstyle+};
(2.537988,0.326035) *[blue]{\scriptscriptstyle+};
(2.546201,0.327812) *[blue]{\scriptscriptstyle+};
(2.554415,0.332888) *[blue]{\scriptscriptstyle+};
(2.562628,0.346888) *[blue]{\scriptscriptstyle+};
(2.570842,0.358680) *[blue]{\scriptscriptstyle+};
(2.579055,0.375160) *[blue]{\scriptscriptstyle+};
(2.587269,0.318237) *[blue]{\scriptscriptstyle+};
(2.595483,0.318661) *[blue]{\scriptscriptstyle+};
(2.603696,0.319063) *[blue]{\scriptscriptstyle+};
(2.611910,0.319618) *[blue]{\scriptscriptstyle+};
(2.620123,0.320780) *[blue]{\scriptscriptstyle+};
(2.628337,0.321822) *[blue]{\scriptscriptstyle+};
(2.636550,0.321863) *[blue]{\scriptscriptstyle+};
(2.644764,0.325162) *[blue]{\scriptscriptstyle+};
(2.652977,0.332602) *[blue]{\scriptscriptstyle+};
(2.661191,0.348615) *[blue]{\scriptscriptstyle+};
(2.669405,0.351328) *[blue]{\scriptscriptstyle+};
(2.677618,0.354653) *[blue]{\scriptscriptstyle+};
(2.685832,0.367178) *[blue]{\scriptscriptstyle+};
(2.694045,0.377572) *[blue]{\scriptscriptstyle+};
(2.702259,0.383090) *[blue]{\scriptscriptstyle+};
(2.710472,0.306518) *[blue]{\scriptscriptstyle+};
(2.718686,0.306878) *[blue]{\scriptscriptstyle+};
(2.726899,0.310318) *[blue]{\scriptscriptstyle+};
(2.735113,0.317081) *[blue]{\scriptscriptstyle+};
(2.743326,0.317417) *[blue]{\scriptscriptstyle+};
(2.751540,0.320759) *[blue]{\scriptscriptstyle+};
(2.759754,0.330033) *[blue]{\scriptscriptstyle+};
(2.767967,0.330836) *[blue]{\scriptscriptstyle+};
(2.776181,0.331191) *[blue]{\scriptscriptstyle+};
(2.784394,0.333138) *[blue]{\scriptscriptstyle+};
(2.792608,0.334360) *[blue]{\scriptscriptstyle+};
(2.800821,0.334395) *[blue]{\scriptscriptstyle+};
(2.809035,0.337495) *[blue]{\scriptscriptstyle+};
(2.817248,0.340428) *[blue]{\scriptscriptstyle+};
(2.825462,0.361956) *[blue]{\scriptscriptstyle+};
(2.833676,0.306070) *[blue]{\scriptscriptstyle+};
(2.841889,0.312115) *[blue]{\scriptscriptstyle+};
(2.850103,0.326875) *[blue]{\scriptscriptstyle+};
(2.858316,0.330829) *[blue]{\scriptscriptstyle+};
(2.866530,0.331507) *[blue]{\scriptscriptstyle+};
(2.874743,0.336815) *[blue]{\scriptscriptstyle+};
(2.882957,0.341902) *[blue]{\scriptscriptstyle+};
(2.891170,0.343197) *[blue]{\scriptscriptstyle+};
(2.899384,0.348872) *[blue]{\scriptscriptstyle+};
(2.907598,0.357712) *[blue]{\scriptscriptstyle+};
(2.915811,0.361790) *[blue]{\scriptscriptstyle+};
(2.924025,0.363801) *[blue]{\scriptscriptstyle+};
(2.932238,0.365989) *[blue]{\scriptscriptstyle+};
(2.940452,0.382481) *[blue]{\scriptscriptstyle+};
(2.948665,0.385429) *[blue]{\scriptscriptstyle+};
(2.956879,0.330110) *[blue]{\scriptscriptstyle+};
(2.965092,0.330863) *[blue]{\scriptscriptstyle+};
(2.973306,0.331083) *[blue]{\scriptscriptstyle+};
(2.981520,0.332062) *[blue]{\scriptscriptstyle+};
(2.989733,0.332161) *[blue]{\scriptscriptstyle+};
(2.997947,0.334472) *[blue]{\scriptscriptstyle+};
(3.006160,0.334852) *[blue]{\scriptscriptstyle+};
(3.014374,0.334859) *[blue]{\scriptscriptstyle+};
(3.022587,0.345977) *[blue]{\scriptscriptstyle+};
(3.030801,0.358691) *[blue]{\scriptscriptstyle+};
(3.039014,0.360117) *[blue]{\scriptscriptstyle+};
(3.047228,0.370873) *[blue]{\scriptscriptstyle+};
(3.055441,0.384184) *[blue]{\scriptscriptstyle+};
(3.063655,0.384278) *[blue]{\scriptscriptstyle+};
(3.071869,0.398122) *[blue]{\scriptscriptstyle+};
(3.080082,0.322281) *[blue]{\scriptscriptstyle+};
(3.088296,0.324875) *[blue]{\scriptscriptstyle+};
(3.096509,0.326247) *[blue]{\scriptscriptstyle+};
(3.104723,0.326395) *[blue]{\scriptscriptstyle+};
(3.112936,0.331141) *[blue]{\scriptscriptstyle+};
(3.121150,0.331803) *[blue]{\scriptscriptstyle+};
(3.129363,0.349046) *[blue]{\scriptscriptstyle+};
(3.137577,0.351059) *[blue]{\scriptscriptstyle+};
(3.145791,0.352502) *[blue]{\scriptscriptstyle+};
(3.154004,0.354284) *[blue]{\scriptscriptstyle+};
(3.162218,0.359620) *[blue]{\scriptscriptstyle+};
(3.170431,0.369599) *[blue]{\scriptscriptstyle+};
(3.178645,0.373659) *[blue]{\scriptscriptstyle+};
(3.186858,0.376241) *[blue]{\scriptscriptstyle+};
(3.195072,0.413981) *[blue]{\scriptscriptstyle+};
(3.203285,0.346173) *[blue]{\scriptscriptstyle+};
(3.211499,0.346700) *[blue]{\scriptscriptstyle+};
(3.219713,0.349354) *[blue]{\scriptscriptstyle+};
(3.227926,0.349892) *[blue]{\scriptscriptstyle+};
(3.236140,0.351074) *[blue]{\scriptscriptstyle+};
(3.244353,0.353298) *[blue]{\scriptscriptstyle+};
(3.252567,0.354610) *[blue]{\scriptscriptstyle+};
(3.260780,0.358149) *[blue]{\scriptscriptstyle+};
(3.268994,0.365938) *[blue]{\scriptscriptstyle+};
(3.277207,0.368631) *[blue]{\scriptscriptstyle+};
(3.285421,0.375359) *[blue]{\scriptscriptstyle+};
(3.293634,0.375651) *[blue]{\scriptscriptstyle+};
(3.301848,0.379440) *[blue]{\scriptscriptstyle+};
(3.310062,0.381528) *[blue]{\scriptscriptstyle+};
(3.318275,0.383686) *[blue]{\scriptscriptstyle+};
(3.326489,0.340489) *[blue]{\scriptscriptstyle+};
(3.334702,0.341393) *[blue]{\scriptscriptstyle+};
(3.342916,0.341949) *[blue]{\scriptscriptstyle+};
(3.351129,0.342374) *[blue]{\scriptscriptstyle+};
(3.359343,0.344230) *[blue]{\scriptscriptstyle+};
(3.367556,0.344911) *[blue]{\scriptscriptstyle+};
(3.375770,0.345017) *[blue]{\scriptscriptstyle+};
(3.383984,0.346498) *[blue]{\scriptscriptstyle+};
(3.392197,0.346907) *[blue]{\scriptscriptstyle+};
(3.400411,0.349903) *[blue]{\scriptscriptstyle+};
(3.408624,0.351695) *[blue]{\scriptscriptstyle+};
(3.416838,0.351725) *[blue]{\scriptscriptstyle+};
(3.425051,0.366858) *[blue]{\scriptscriptstyle+};
(3.433265,0.367413) *[blue]{\scriptscriptstyle+};
(3.441478,0.369176) *[blue]{\scriptscriptstyle+};
(3.449692,0.339463) *[blue]{\scriptscriptstyle+};
(3.457906,0.339813) *[blue]{\scriptscriptstyle+};
(3.466119,0.344380) *[blue]{\scriptscriptstyle+};
(3.474333,0.344488) *[blue]{\scriptscriptstyle+};
(3.482546,0.345571) *[blue]{\scriptscriptstyle+};
(3.490760,0.366795) *[blue]{\scriptscriptstyle+};
(3.498973,0.368306) *[blue]{\scriptscriptstyle+};
(3.507187,0.369279) *[blue]{\scriptscriptstyle+};
(3.515400,0.369345) *[blue]{\scriptscriptstyle+};
(3.523614,0.372402) *[blue]{\scriptscriptstyle+};
(3.531828,0.376437) *[blue]{\scriptscriptstyle+};
(3.540041,0.391856) *[blue]{\scriptscriptstyle+};
(3.548255,0.399780) *[blue]{\scriptscriptstyle+};
(3.556468,0.411080) *[blue]{\scriptscriptstyle+};
(3.564682,0.414731) *[blue]{\scriptscriptstyle+};
(3.572895,0.339651) *[blue]{\scriptscriptstyle+};
(3.581109,0.340687) *[blue]{\scriptscriptstyle+};
(3.589322,0.342319) *[blue]{\scriptscriptstyle+};
(3.597536,0.351678) *[blue]{\scriptscriptstyle+};
(3.605749,0.362617) *[blue]{\scriptscriptstyle+};
(3.613963,0.363222) *[blue]{\scriptscriptstyle+};
(3.622177,0.363475) *[blue]{\scriptscriptstyle+};
(3.630390,0.365722) *[blue]{\scriptscriptstyle+};
(3.638604,0.367050) *[blue]{\scriptscriptstyle+};
(3.646817,0.368574) *[blue]{\scriptscriptstyle+};
(3.655031,0.370193) *[blue]{\scriptscriptstyle+};
(3.663244,0.370202) *[blue]{\scriptscriptstyle+};
(3.671458,0.375079) *[blue]{\scriptscriptstyle+};
(3.679671,0.393934) *[blue]{\scriptscriptstyle+};
(3.687885,0.404826) *[blue]{\scriptscriptstyle+};
(3.696099,0.349018) *[blue]{\scriptscriptstyle+};
(3.704312,0.350485) *[blue]{\scriptscriptstyle+};
(3.712526,0.350838) *[blue]{\scriptscriptstyle+};
(3.720739,0.352163) *[blue]{\scriptscriptstyle+};
(3.728953,0.353662) *[blue]{\scriptscriptstyle+};
(3.737166,0.367410) *[blue]{\scriptscriptstyle+};
(3.745380,0.369001) *[blue]{\scriptscriptstyle+};
(3.753593,0.369735) *[blue]{\scriptscriptstyle+};
(3.761807,0.371459) *[blue]{\scriptscriptstyle+};
(3.770021,0.373495) *[blue]{\scriptscriptstyle+};
(3.778234,0.373685) *[blue]{\scriptscriptstyle+};
(3.786448,0.379510) *[blue]{\scriptscriptstyle+};
(3.794661,0.396722) *[blue]{\scriptscriptstyle+};
(3.802875,0.419648) *[blue]{\scriptscriptstyle+};
(3.811088,0.425926) *[blue]{\scriptscriptstyle+};
(3.819302,0.361836) *[blue]{\scriptscriptstyle+};
(3.827515,0.362521) *[blue]{\scriptscriptstyle+};
(3.835729,0.365367) *[blue]{\scriptscriptstyle+};
(3.843943,0.370333) *[blue]{\scriptscriptstyle+};
(3.852156,0.384246) *[blue]{\scriptscriptstyle+};
(3.860370,0.389305) *[blue]{\scriptscriptstyle+};
(3.868583,0.389727) *[blue]{\scriptscriptstyle+};
(3.876797,0.389836) *[blue]{\scriptscriptstyle+};
(3.885010,0.392148) *[blue]{\scriptscriptstyle+};
(3.893224,0.413030) *[blue]{\scriptscriptstyle+};
(3.901437,0.413507) *[blue]{\scriptscriptstyle+};
(3.909651,0.414023) *[blue]{\scriptscriptstyle+};
(3.917864,0.414296) *[blue]{\scriptscriptstyle+};
(3.926078,0.419678) *[blue]{\scriptscriptstyle+};
(3.934292,0.442117) *[blue]{\scriptscriptstyle+};
(3.942505,0.362020) *[blue]{\scriptscriptstyle+};
(3.950719,0.368473) *[blue]{\scriptscriptstyle+};
(3.958932,0.373221) *[blue]{\scriptscriptstyle+};
(3.967146,0.387859) *[blue]{\scriptscriptstyle+};
(3.975359,0.390971) *[blue]{\scriptscriptstyle+};
(3.983573,0.391242) *[blue]{\scriptscriptstyle+};
(3.991786,0.396174) *[blue]{\scriptscriptstyle+};
(4.000000,0.401138) *[blue]{\scriptscriptstyle+};
(4.008214,0.407173) *[blue]{\scriptscriptstyle+};
(4.016427,0.410504) *[blue]{\scriptscriptstyle+};
(4.024641,0.419912) *[blue]{\scriptscriptstyle+};
(4.032854,0.422434) *[blue]{\scriptscriptstyle+};
(4.041068,0.448101) *[blue]{\scriptscriptstyle+};
(4.049281,0.451149) *[blue]{\scriptscriptstyle+};
(4.057495,0.677544) *[blue]{\scriptscriptstyle+};
(4.065708,0.373943) *[blue]{\scriptscriptstyle+};
(4.073922,0.374256) *[blue]{\scriptscriptstyle+};
(4.082136,0.376019) *[blue]{\scriptscriptstyle+};
(4.090349,0.376400) *[blue]{\scriptscriptstyle+};
(4.098563,0.376752) *[blue]{\scriptscriptstyle+};
(4.106776,0.378890) *[blue]{\scriptscriptstyle+};
(4.114990,0.379039) *[blue]{\scriptscriptstyle+};
(4.123203,0.379958) *[blue]{\scriptscriptstyle+};
(4.131417,0.381478) *[blue]{\scriptscriptstyle+};
(4.139630,0.382785) *[blue]{\scriptscriptstyle+};
(4.147844,0.385837) *[blue]{\scriptscriptstyle+};
(4.156057,0.398578) *[blue]{\scriptscriptstyle+};
(4.164271,0.403203) *[blue]{\scriptscriptstyle+};
(4.172485,0.403342) *[blue]{\scriptscriptstyle+};
(4.180698,0.432913) *[blue]{\scriptscriptstyle+};
(4.188912,0.377073) *[blue]{\scriptscriptstyle+};
(4.197125,0.381403) *[blue]{\scriptscriptstyle+};
(4.205339,0.382676) *[blue]{\scriptscriptstyle+};
(4.213552,0.384062) *[blue]{\scriptscriptstyle+};
(4.221766,0.384251) *[blue]{\scriptscriptstyle+};
(4.229979,0.386126) *[blue]{\scriptscriptstyle+};
(4.238193,0.386817) *[blue]{\scriptscriptstyle+};
(4.246407,0.389753) *[blue]{\scriptscriptstyle+};
(4.254620,0.401705) *[blue]{\scriptscriptstyle+};
(4.262834,0.402855) *[blue]{\scriptscriptstyle+};
(4.271047,0.406403) *[blue]{\scriptscriptstyle+};
(4.279261,0.409389) *[blue]{\scriptscriptstyle+};
(4.287474,0.422614) *[blue]{\scriptscriptstyle+};
(4.295688,0.427948) *[blue]{\scriptscriptstyle+};
(4.303901,0.459875) *[blue]{\scriptscriptstyle+};
(4.312115,0.383419) *[blue]{\scriptscriptstyle+};
(4.320329,0.383920) *[blue]{\scriptscriptstyle+};
(4.328542,0.384962) *[blue]{\scriptscriptstyle+};
(4.336756,0.385648) *[blue]{\scriptscriptstyle+};
(4.344969,0.386665) *[blue]{\scriptscriptstyle+};
(4.353183,0.387091) *[blue]{\scriptscriptstyle+};
(4.361396,0.389096) *[blue]{\scriptscriptstyle+};
(4.369610,0.390158) *[blue]{\scriptscriptstyle+};
(4.377823,0.403510) *[blue]{\scriptscriptstyle+};
(4.386037,0.410580) *[blue]{\scriptscriptstyle+};
(4.394251,0.413750) *[blue]{\scriptscriptstyle+};
(4.402464,0.449799) *[blue]{\scriptscriptstyle+};
(4.410678,0.463272) *[blue]{\scriptscriptstyle+};
(4.418891,0.576849) *[blue]{\scriptscriptstyle+};
(4.427105,0.777589) *[blue]{\scriptscriptstyle+};
(4.435318,0.386163) *[blue]{\scriptscriptstyle+};
(4.443532,0.386224) *[blue]{\scriptscriptstyle+};
(4.451745,0.386576) *[blue]{\scriptscriptstyle+};
(4.459959,0.387086) *[blue]{\scriptscriptstyle+};
(4.468172,0.387217) *[blue]{\scriptscriptstyle+};
(4.476386,0.388854) *[blue]{\scriptscriptstyle+};
(4.484600,0.392813) *[blue]{\scriptscriptstyle+};
(4.492813,0.393704) *[blue]{\scriptscriptstyle+};
(4.501027,0.395683) *[blue]{\scriptscriptstyle+};
(4.509240,0.407899) *[blue]{\scriptscriptstyle+};
(4.517454,0.414893) *[blue]{\scriptscriptstyle+};
(4.525667,0.421671) *[blue]{\scriptscriptstyle+};
(4.533881,0.428462) *[blue]{\scriptscriptstyle+};
(4.542094,0.436412) *[blue]{\scriptscriptstyle+};
(4.550308,0.448876) *[blue]{\scriptscriptstyle+};
(4.558522,0.382850) *[blue]{\scriptscriptstyle+};
(4.566735,0.384175) *[blue]{\scriptscriptstyle+};
(4.574949,0.385067) *[blue]{\scriptscriptstyle+};
(4.583162,0.386765) *[blue]{\scriptscriptstyle+};
(4.591376,0.388889) *[blue]{\scriptscriptstyle+};
(4.599589,0.391230) *[blue]{\scriptscriptstyle+};
(4.607803,0.394956) *[blue]{\scriptscriptstyle+};
(4.616016,0.407119) *[blue]{\scriptscriptstyle+};
(4.624230,0.408606) *[blue]{\scriptscriptstyle+};
(4.632444,0.409629) *[blue]{\scriptscriptstyle+};
(4.640657,0.412077) *[blue]{\scriptscriptstyle+};
(4.648871,0.414367) *[blue]{\scriptscriptstyle+};
(4.657084,0.430477) *[blue]{\scriptscriptstyle+};
(4.665298,0.431664) *[blue]{\scriptscriptstyle+};
(4.673511,0.515327) *[blue]{\scriptscriptstyle+};
(4.681725,0.387474) *[blue]{\scriptscriptstyle+};
(4.689938,0.390388) *[blue]{\scriptscriptstyle+};
(4.698152,0.391522) *[blue]{\scriptscriptstyle+};
(4.706366,0.394224) *[blue]{\scriptscriptstyle+};
(4.714579,0.394841) *[blue]{\scriptscriptstyle+};
(4.722793,0.396008) *[blue]{\scriptscriptstyle+};
(4.731006,0.397968) *[blue]{\scriptscriptstyle+};
(4.739220,0.414601) *[blue]{\scriptscriptstyle+};
(4.747433,0.419441) *[blue]{\scriptscriptstyle+};
(4.755647,0.419512) *[blue]{\scriptscriptstyle+};
(4.763860,0.421668) *[blue]{\scriptscriptstyle+};
(4.772074,0.422355) *[blue]{\scriptscriptstyle+};
(4.780287,0.424894) *[blue]{\scriptscriptstyle+};
(4.788501,0.426082) *[blue]{\scriptscriptstyle+};
(4.796715,0.442591) *[blue]{\scriptscriptstyle+};
(4.804928,0.382049) *[blue]{\scriptscriptstyle+};
(4.813142,0.388012) *[blue]{\scriptscriptstyle+};
(4.821355,0.388088) *[blue]{\scriptscriptstyle+};
(4.829569,0.388835) *[blue]{\scriptscriptstyle+};
(4.837782,0.392485) *[blue]{\scriptscriptstyle+};
(4.845996,0.404779) *[blue]{\scriptscriptstyle+};
(4.854209,0.407825) *[blue]{\scriptscriptstyle+};
(4.862423,0.408092) *[blue]{\scriptscriptstyle+};
(4.870637,0.408296) *[blue]{\scriptscriptstyle+};
(4.878850,0.409505) *[blue]{\scriptscriptstyle+};
(4.887064,0.410631) *[blue]{\scriptscriptstyle+};
(4.895277,0.411875) *[blue]{\scriptscriptstyle+};
(4.903491,0.418366) *[blue]{\scriptscriptstyle+};
(4.911704,0.431801) *[blue]{\scriptscriptstyle+};
(4.919918,0.469728) *[blue]{\scriptscriptstyle+};
(4.928131,0.392300) *[blue]{\scriptscriptstyle+};
(4.936345,0.394240) *[blue]{\scriptscriptstyle+};
(4.944559,0.394678) *[blue]{\scriptscriptstyle+};
(4.952772,0.394769) *[blue]{\scriptscriptstyle+};
(4.960986,0.395423) *[blue]{\scriptscriptstyle+};
(4.969199,0.396103) *[blue]{\scriptscriptstyle+};
(4.977413,0.396237) *[blue]{\scriptscriptstyle+};
(4.985626,0.396926) *[blue]{\scriptscriptstyle+};
(4.993840,0.397336) *[blue]{\scriptscriptstyle+};
(5.002053,0.399841) *[blue]{\scriptscriptstyle+};
(5.010267,0.404159) *[blue]{\scriptscriptstyle+};
(5.018480,0.412317) *[blue]{\scriptscriptstyle+};
(5.026694,0.417794) *[blue]{\scriptscriptstyle+};
(5.034908,0.423470) *[blue]{\scriptscriptstyle+};
(5.043121,0.450584) *[blue]{\scriptscriptstyle+};
(5.051335,0.388685) *[blue]{\scriptscriptstyle+};
(5.059548,0.390607) *[blue]{\scriptscriptstyle+};
(5.067762,0.395051) *[blue]{\scriptscriptstyle+};
(5.075975,0.397103) *[blue]{\scriptscriptstyle+};
(5.084189,0.411015) *[blue]{\scriptscriptstyle+};
(5.092402,0.412393) *[blue]{\scriptscriptstyle+};
(5.100616,0.412458) *[blue]{\scriptscriptstyle+};
(5.108830,0.414673) *[blue]{\scriptscriptstyle+};
(5.117043,0.415083) *[blue]{\scriptscriptstyle+};
(5.125257,0.417568) *[blue]{\scriptscriptstyle+};
(5.133470,0.435453) *[blue]{\scriptscriptstyle+};
(5.141684,0.439486) *[blue]{\scriptscriptstyle+};
(5.149897,0.441464) *[blue]{\scriptscriptstyle+};
(5.158111,0.441593) *[blue]{\scriptscriptstyle+};
(5.166324,0.443555) *[blue]{\scriptscriptstyle+};
(5.174538,0.405967) *[blue]{\scriptscriptstyle+};
(5.182752,0.407168) *[blue]{\scriptscriptstyle+};
(5.190965,0.408494) *[blue]{\scriptscriptstyle+};
(5.199179,0.411477) *[blue]{\scriptscriptstyle+};
(5.207392,0.415634) *[blue]{\scriptscriptstyle+};
(5.215606,0.431729) *[blue]{\scriptscriptstyle+};
(5.223819,0.431822) *[blue]{\scriptscriptstyle+};
(5.232033,0.433750) *[blue]{\scriptscriptstyle+};
(5.240246,0.435583) *[blue]{\scriptscriptstyle+};
(5.248460,0.436671) *[blue]{\scriptscriptstyle+};
(5.256674,0.439042) *[blue]{\scriptscriptstyle+};
(5.264887,0.440906) *[blue]{\scriptscriptstyle+};
(5.273101,0.441872) *[blue]{\scriptscriptstyle+};
(5.281314,0.454962) *[blue]{\scriptscriptstyle+};
(5.289528,0.464022) *[blue]{\scriptscriptstyle+};
(5.297741,0.403768) *[blue]{\scriptscriptstyle+};
(5.305955,0.406427) *[blue]{\scriptscriptstyle+};
(5.314168,0.407319) *[blue]{\scriptscriptstyle+};
(5.322382,0.407551) *[blue]{\scriptscriptstyle+};
(5.330595,0.410132) *[blue]{\scriptscriptstyle+};
(5.338809,0.410547) *[blue]{\scriptscriptstyle+};
(5.347023,0.410663) *[blue]{\scriptscriptstyle+};
(5.355236,0.411001) *[blue]{\scriptscriptstyle+};
(5.363450,0.414398) *[blue]{\scriptscriptstyle+};
(5.371663,0.416431) *[blue]{\scriptscriptstyle+};
(5.379877,0.419699) *[blue]{\scriptscriptstyle+};
(5.388090,0.432367) *[blue]{\scriptscriptstyle+};
(5.396304,0.434663) *[blue]{\scriptscriptstyle+};
(5.404517,0.440071) *[blue]{\scriptscriptstyle+};
(5.412731,0.459977) *[blue]{\scriptscriptstyle+};
(5.420945,0.410532) *[blue]{\scriptscriptstyle+};
(5.429158,0.412874) *[blue]{\scriptscriptstyle+};
(5.437372,0.414089) *[blue]{\scriptscriptstyle+};
(5.445585,0.417495) *[blue]{\scriptscriptstyle+};
(5.453799,0.418116) *[blue]{\scriptscriptstyle+};
(5.462012,0.419100) *[blue]{\scriptscriptstyle+};
(5.470226,0.432059) *[blue]{\scriptscriptstyle+};
(5.478439,0.436422) *[blue]{\scriptscriptstyle+};
(5.486653,0.439598) *[blue]{\scriptscriptstyle+};
(5.494867,0.441720) *[blue]{\scriptscriptstyle+};
(5.503080,0.442396) *[blue]{\scriptscriptstyle+};
(5.511294,0.442719) *[blue]{\scriptscriptstyle+};
(5.519507,0.445117) *[blue]{\scriptscriptstyle+};
(5.527721,0.454194) *[blue]{\scriptscriptstyle+};
(5.535934,0.474912) *[blue]{\scriptscriptstyle+};
(5.544148,0.412468) *[blue]{\scriptscriptstyle+};
(5.552361,0.413604) *[blue]{\scriptscriptstyle+};
(5.560575,0.415096) *[blue]{\scriptscriptstyle+};
(5.568789,0.416863) *[blue]{\scriptscriptstyle+};
(5.577002,0.418464) *[blue]{\scriptscriptstyle+};
(5.585216,0.420548) *[blue]{\scriptscriptstyle+};
(5.593429,0.420828) *[blue]{\scriptscriptstyle+};
(5.601643,0.434587) *[blue]{\scriptscriptstyle+};
(5.609856,0.435081) *[blue]{\scriptscriptstyle+};
(5.618070,0.436415) *[blue]{\scriptscriptstyle+};
(5.626283,0.439931) *[blue]{\scriptscriptstyle+};
(5.634497,0.441227) *[blue]{\scriptscriptstyle+};
(5.642710,0.442000) *[blue]{\scriptscriptstyle+};
(5.650924,0.450544) *[blue]{\scriptscriptstyle+};
(5.659138,0.456182) *[blue]{\scriptscriptstyle+};
(5.667351,0.417712) *[blue]{\scriptscriptstyle+};
(5.675565,0.418070) *[blue]{\scriptscriptstyle+};
(5.683778,0.418883) *[blue]{\scriptscriptstyle+};
(5.691992,0.419467) *[blue]{\scriptscriptstyle+};
(5.700205,0.421479) *[blue]{\scriptscriptstyle+};
(5.708419,0.422902) *[blue]{\scriptscriptstyle+};
(5.716632,0.423357) *[blue]{\scriptscriptstyle+};
(5.724846,0.424655) *[blue]{\scriptscriptstyle+};
(5.733060,0.428091) *[blue]{\scriptscriptstyle+};
(5.741273,0.437756) *[blue]{\scriptscriptstyle+};
(5.749487,0.440493) *[blue]{\scriptscriptstyle+};
(5.757700,0.445971) *[blue]{\scriptscriptstyle+};
(5.765914,0.447058) *[blue]{\scriptscriptstyle+};
(5.774127,0.453706) *[blue]{\scriptscriptstyle+};
(5.782341,0.462405) *[blue]{\scriptscriptstyle+};
(5.790554,0.412491) *[blue]{\scriptscriptstyle+};
(5.798768,0.412853) *[blue]{\scriptscriptstyle+};
(5.806982,0.420394) *[blue]{\scriptscriptstyle+};
(5.815195,0.421217) *[blue]{\scriptscriptstyle+};
(5.823409,0.424752) *[blue]{\scriptscriptstyle+};
(5.831622,0.434872) *[blue]{\scriptscriptstyle+};
(5.839836,0.437439) *[blue]{\scriptscriptstyle+};
(5.848049,0.442754) *[blue]{\scriptscriptstyle+};
(5.856263,0.445215) *[blue]{\scriptscriptstyle+};
(5.864476,0.445581) *[blue]{\scriptscriptstyle+};
(5.872690,0.465463) *[blue]{\scriptscriptstyle+};
(5.880903,0.480746) *[blue]{\scriptscriptstyle+};
(5.889117,0.482532) *[blue]{\scriptscriptstyle+};
(5.897331,0.514436) *[blue]{\scriptscriptstyle+};
(5.905544,0.671738) *[blue]{\scriptscriptstyle+};
(5.913758,0.428849) *[blue]{\scriptscriptstyle+};
(5.921971,0.430122) *[blue]{\scriptscriptstyle+};
(5.930185,0.435085) *[blue]{\scriptscriptstyle+};
(5.938398,0.442313) *[blue]{\scriptscriptstyle+};
(5.946612,0.454195) *[blue]{\scriptscriptstyle+};
(5.954825,0.455064) *[blue]{\scriptscriptstyle+};
(5.963039,0.458086) *[blue]{\scriptscriptstyle+};
(5.971253,0.458790) *[blue]{\scriptscriptstyle+};
(5.979466,0.462225) *[blue]{\scriptscriptstyle+};
(5.987680,0.468921) *[blue]{\scriptscriptstyle+};
(5.995893,0.481074) *[blue]{\scriptscriptstyle+};
(6.004107,0.481464) *[blue]{\scriptscriptstyle+};
(6.012320,0.483795) *[blue]{\scriptscriptstyle+};
(6.020534,0.489889) *[blue]{\scriptscriptstyle+};
(6.028747,0.502451) *[blue]{\scriptscriptstyle+};
(6.036961,0.429002) *[blue]{\scriptscriptstyle+};
(6.045175,0.429486) *[blue]{\scriptscriptstyle+};
(6.053388,0.429787) *[blue]{\scriptscriptstyle+};
(6.061602,0.431977) *[blue]{\scriptscriptstyle+};
(6.069815,0.433250) *[blue]{\scriptscriptstyle+};
(6.078029,0.434262) *[blue]{\scriptscriptstyle+};
(6.086242,0.434883) *[blue]{\scriptscriptstyle+};
(6.094456,0.436361) *[blue]{\scriptscriptstyle+};
(6.102669,0.437152) *[blue]{\scriptscriptstyle+};
(6.110883,0.456693) *[blue]{\scriptscriptstyle+};
(6.119097,0.457415) *[blue]{\scriptscriptstyle+};
(6.127310,0.458115) *[blue]{\scriptscriptstyle+};
(6.135524,0.460664) *[blue]{\scriptscriptstyle+};
(6.143737,0.465695) *[blue]{\scriptscriptstyle+};
(6.151951,0.482846) *[blue]{\scriptscriptstyle+};
(6.160164,0.420316) *[blue]{\scriptscriptstyle+};
(6.168378,0.421110) *[blue]{\scriptscriptstyle+};
(6.176591,0.426372) *[blue]{\scriptscriptstyle+};
(6.184805,0.426539) *[blue]{\scriptscriptstyle+};
(6.193018,0.428125) *[blue]{\scriptscriptstyle+};
(6.201232,0.444041) *[blue]{\scriptscriptstyle+};
(6.209446,0.445640) *[blue]{\scriptscriptstyle+};
(6.217659,0.447515) *[blue]{\scriptscriptstyle+};
(6.225873,0.448253) *[blue]{\scriptscriptstyle+};
(6.234086,0.449404) *[blue]{\scriptscriptstyle+};
(6.242300,0.454811) *[blue]{\scriptscriptstyle+};
(6.250513,0.473097) *[blue]{\scriptscriptstyle+};
(6.258727,0.477382) *[blue]{\scriptscriptstyle+};
(6.266940,0.478350) *[blue]{\scriptscriptstyle+};
(6.275154,0.497816) *[blue]{\scriptscriptstyle+};
(6.283368,0.431305) *[blue]{\scriptscriptstyle+};
(6.291581,0.435991) *[blue]{\scriptscriptstyle+};
(6.299795,0.439673) *[blue]{\scriptscriptstyle+};
(6.308008,0.440779) *[blue]{\scriptscriptstyle+};
(6.316222,0.456312) *[blue]{\scriptscriptstyle+};
(6.324435,0.458677) *[blue]{\scriptscriptstyle+};
(6.332649,0.459057) *[blue]{\scriptscriptstyle+};
(6.340862,0.460834) *[blue]{\scriptscriptstyle+};
(6.349076,0.461979) *[blue]{\scriptscriptstyle+};
(6.357290,0.467796) *[blue]{\scriptscriptstyle+};
(6.365503,0.473184) *[blue]{\scriptscriptstyle+};
(6.373717,0.480826) *[blue]{\scriptscriptstyle+};
(6.381930,0.484332) *[blue]{\scriptscriptstyle+};
(6.390144,0.486703) *[blue]{\scriptscriptstyle+};
(6.398357,0.487201) *[blue]{\scriptscriptstyle+};
(6.406571,0.432547) *[blue]{\scriptscriptstyle+};
(6.414784,0.432917) *[blue]{\scriptscriptstyle+};
(6.422998,0.436940) *[blue]{\scriptscriptstyle+};
(6.431211,0.439231) *[blue]{\scriptscriptstyle+};
(6.439425,0.440578) *[blue]{\scriptscriptstyle+};
(6.447639,0.455254) *[blue]{\scriptscriptstyle+};
(6.455852,0.455715) *[blue]{\scriptscriptstyle+};
(6.464066,0.458670) *[blue]{\scriptscriptstyle+};
(6.472279,0.459117) *[blue]{\scriptscriptstyle+};
(6.480493,0.459492) *[blue]{\scriptscriptstyle+};
(6.488706,0.460624) *[blue]{\scriptscriptstyle+};
(6.496920,0.474793) *[blue]{\scriptscriptstyle+};
(6.505133,0.480664) *[blue]{\scriptscriptstyle+};
(6.513347,0.483744) *[blue]{\scriptscriptstyle+};
(6.521561,0.485951) *[blue]{\scriptscriptstyle+};
(6.529774,0.439980) *[blue]{\scriptscriptstyle+};
(6.537988,0.442711) *[blue]{\scriptscriptstyle+};
(6.546201,0.442847) *[blue]{\scriptscriptstyle+};
(6.554415,0.447708) *[blue]{\scriptscriptstyle+};
(6.562628,0.462667) *[blue]{\scriptscriptstyle+};
(6.570842,0.465186) *[blue]{\scriptscriptstyle+};
(6.579055,0.466024) *[blue]{\scriptscriptstyle+};
(6.587269,0.466794) *[blue]{\scriptscriptstyle+};
(6.595483,0.467412) *[blue]{\scriptscriptstyle+};
(6.603696,0.467825) *[blue]{\scriptscriptstyle+};
(6.611910,0.469648) *[blue]{\scriptscriptstyle+};
(6.620123,0.469778) *[blue]{\scriptscriptstyle+};
(6.628337,0.470252) *[blue]{\scriptscriptstyle+};
(6.636550,0.489411) *[blue]{\scriptscriptstyle+};
(6.644764,0.513067) *[blue]{\scriptscriptstyle+};
(6.652977,0.438619) *[blue]{\scriptscriptstyle+};
(6.661191,0.442652) *[blue]{\scriptscriptstyle+};
(6.669405,0.443642) *[blue]{\scriptscriptstyle+};
(6.677618,0.444545) *[blue]{\scriptscriptstyle+};
(6.685832,0.445518) *[blue]{\scriptscriptstyle+};
(6.694045,0.451129) *[blue]{\scriptscriptstyle+};
(6.702259,0.466664) *[blue]{\scriptscriptstyle+};
(6.710472,0.466833) *[blue]{\scriptscriptstyle+};
(6.718686,0.468034) *[blue]{\scriptscriptstyle+};
(6.726899,0.470709) *[blue]{\scriptscriptstyle+};
(6.735113,0.488522) *[blue]{\scriptscriptstyle+};
(6.743326,0.492452) *[blue]{\scriptscriptstyle+};
(6.751540,0.496689) *[blue]{\scriptscriptstyle+};
(6.759754,0.498907) *[blue]{\scriptscriptstyle+};
(6.767967,0.514974) *[blue]{\scriptscriptstyle+};
(6.776181,0.426485) *[blue]{\scriptscriptstyle+};
(6.784394,0.428404) *[blue]{\scriptscriptstyle+};
(6.792608,0.447940) *[blue]{\scriptscriptstyle+};
(6.800821,0.448548) *[blue]{\scriptscriptstyle+};
(6.809035,0.454633) *[blue]{\scriptscriptstyle+};
(6.817248,0.454771) *[blue]{\scriptscriptstyle+};
(6.825462,0.455117) *[blue]{\scriptscriptstyle+};
(6.833676,0.457768) *[blue]{\scriptscriptstyle+};
(6.841889,0.472479) *[blue]{\scriptscriptstyle+};
(6.850103,0.476831) *[blue]{\scriptscriptstyle+};
(6.858316,0.477639) *[blue]{\scriptscriptstyle+};
(6.866530,0.477821) *[blue]{\scriptscriptstyle+};
(6.874743,0.478198) *[blue]{\scriptscriptstyle+};
(6.882957,0.504483) *[blue]{\scriptscriptstyle+};
(6.891170,0.564664) *[blue]{\scriptscriptstyle+};
(6.899384,0.429626) *[blue]{\scriptscriptstyle+};
(6.907598,0.431570) *[blue]{\scriptscriptstyle+};
(6.915811,0.435401) *[blue]{\scriptscriptstyle+};
(6.924025,0.443907) *[blue]{\scriptscriptstyle+};
(6.932238,0.448422) *[blue]{\scriptscriptstyle+};
(6.940452,0.453118) *[blue]{\scriptscriptstyle+};
(6.948665,0.453416) *[blue]{\scriptscriptstyle+};
(6.956879,0.454486) *[blue]{\scriptscriptstyle+};
(6.965092,0.462485) *[blue]{\scriptscriptstyle+};
(6.973306,0.464099) *[blue]{\scriptscriptstyle+};
(6.981520,0.476981) *[blue]{\scriptscriptstyle+};
(6.989733,0.479709) *[blue]{\scriptscriptstyle+};
(6.997947,0.480271) *[blue]{\scriptscriptstyle+};
(7.006160,0.484944) *[blue]{\scriptscriptstyle+};
(7.014374,0.496271) *[blue]{\scriptscriptstyle+};
(7.022587,0.459926) *[blue]{\scriptscriptstyle+};
(7.030801,0.460062) *[blue]{\scriptscriptstyle+};
(7.039014,0.461506) *[blue]{\scriptscriptstyle+};
(7.047228,0.461866) *[blue]{\scriptscriptstyle+};
(7.055441,0.463109) *[blue]{\scriptscriptstyle+};
(7.063655,0.467335) *[blue]{\scriptscriptstyle+};
(7.071869,0.468849) *[blue]{\scriptscriptstyle+};
(7.080082,0.469982) *[blue]{\scriptscriptstyle+};
(7.088296,0.472588) *[blue]{\scriptscriptstyle+};
(7.096509,0.473133) *[blue]{\scriptscriptstyle+};
(7.104723,0.486948) *[blue]{\scriptscriptstyle+};
(7.112936,0.488058) *[blue]{\scriptscriptstyle+};
(7.121150,0.507286) *[blue]{\scriptscriptstyle+};
(7.129363,0.516719) *[blue]{\scriptscriptstyle+};
(7.137577,0.526004) *[blue]{\scriptscriptstyle+};
(7.145791,0.447104) *[blue]{\scriptscriptstyle+};
(7.154004,0.450045) *[blue]{\scriptscriptstyle+};
(7.162218,0.452100) *[blue]{\scriptscriptstyle+};
(7.170431,0.453319) *[blue]{\scriptscriptstyle+};
(7.178645,0.454007) *[blue]{\scriptscriptstyle+};
(7.186858,0.455068) *[blue]{\scriptscriptstyle+};
(7.195072,0.470592) *[blue]{\scriptscriptstyle+};
(7.203285,0.473243) *[blue]{\scriptscriptstyle+};
(7.211499,0.473537) *[blue]{\scriptscriptstyle+};
(7.219713,0.473790) *[blue]{\scriptscriptstyle+};
(7.227926,0.474293) *[blue]{\scriptscriptstyle+};
(7.236140,0.487812) *[blue]{\scriptscriptstyle+};
(7.244353,0.498214) *[blue]{\scriptscriptstyle+};
(7.252567,0.508445) *[blue]{\scriptscriptstyle+};
(7.260780,0.516993) *[blue]{\scriptscriptstyle+};
(7.268994,0.424954) *[blue]{\scriptscriptstyle+};
(7.277207,0.450583) *[blue]{\scriptscriptstyle+};
(7.285421,0.452068) *[blue]{\scriptscriptstyle+};
(7.293634,0.455591) *[blue]{\scriptscriptstyle+};
(7.301848,0.456642) *[blue]{\scriptscriptstyle+};
(7.310062,0.475078) *[blue]{\scriptscriptstyle+};
(7.318275,0.476499) *[blue]{\scriptscriptstyle+};
(7.326489,0.478024) *[blue]{\scriptscriptstyle+};
(7.334702,0.478780) *[blue]{\scriptscriptstyle+};
(7.342916,0.491828) *[blue]{\scriptscriptstyle+};
(7.351129,0.495102) *[blue]{\scriptscriptstyle+};
(7.359343,0.495293) *[blue]{\scriptscriptstyle+};
(7.367556,0.498620) *[blue]{\scriptscriptstyle+};
(7.375770,0.538677) *[blue]{\scriptscriptstyle+};
(7.383984,1.112316) *[blue]{\scriptscriptstyle+};
(7.392197,0.467369) *[blue]{\scriptscriptstyle+};
(7.400411,0.467749) *[blue]{\scriptscriptstyle+};
(7.408624,0.471280) *[blue]{\scriptscriptstyle+};
(7.416838,0.473841) *[blue]{\scriptscriptstyle+};
(7.425051,0.488077) *[blue]{\scriptscriptstyle+};
(7.433265,0.489258) *[blue]{\scriptscriptstyle+};
(7.441478,0.495337) *[blue]{\scriptscriptstyle+};
(7.449692,0.495479) *[blue]{\scriptscriptstyle+};
(7.457906,0.499155) *[blue]{\scriptscriptstyle+};
(7.466119,0.499569) *[blue]{\scriptscriptstyle+};
(7.474333,0.503675) *[blue]{\scriptscriptstyle+};
(7.482546,0.515430) *[blue]{\scriptscriptstyle+};
(7.490760,0.523999) *[blue]{\scriptscriptstyle+};
(7.498973,0.526942) *[blue]{\scriptscriptstyle+};
(7.507187,0.550138) *[blue]{\scriptscriptstyle+};
(7.515400,0.461437) *[blue]{\scriptscriptstyle+};
(7.523614,0.461700) *[blue]{\scriptscriptstyle+};
(7.531828,0.472465) *[blue]{\scriptscriptstyle+};
(7.540041,0.475767) *[blue]{\scriptscriptstyle+};
(7.548255,0.484576) *[blue]{\scriptscriptstyle+};
(7.556468,0.485782) *[blue]{\scriptscriptstyle+};
(7.564682,0.485823) *[blue]{\scriptscriptstyle+};
(7.572895,0.486581) *[blue]{\scriptscriptstyle+};
(7.581109,0.486779) *[blue]{\scriptscriptstyle+};
(7.589322,0.495551) *[blue]{\scriptscriptstyle+};
(7.597536,0.495740) *[blue]{\scriptscriptstyle+};
(7.605749,0.498956) *[blue]{\scriptscriptstyle+};
(7.613963,0.509528) *[blue]{\scriptscriptstyle+};
(7.622177,0.596877) *[blue]{\scriptscriptstyle+};
(7.630390,0.822331) *[blue]{\scriptscriptstyle+};
(7.638604,0.507252) *[blue]{\scriptscriptstyle+};
(7.646817,0.512075) *[blue]{\scriptscriptstyle+};
(7.655031,0.512426) *[blue]{\scriptscriptstyle+};
(7.663244,0.513906) *[blue]{\scriptscriptstyle+};
(7.671458,0.516199) *[blue]{\scriptscriptstyle+};
(7.679671,0.517797) *[blue]{\scriptscriptstyle+};
(7.687885,0.519123) *[blue]{\scriptscriptstyle+};
(7.696099,0.529292) *[blue]{\scriptscriptstyle+};
(7.704312,0.531947) *[blue]{\scriptscriptstyle+};
(7.712526,0.533008) *[blue]{\scriptscriptstyle+};
(7.720739,0.534731) *[blue]{\scriptscriptstyle+};
(7.728953,0.536602) *[blue]{\scriptscriptstyle+};
(7.737166,0.546419) *[blue]{\scriptscriptstyle+};
(7.745380,0.552035) *[blue]{\scriptscriptstyle+};
(7.753593,0.555225) *[blue]{\scriptscriptstyle+};
(7.761807,0.510969) *[blue]{\scriptscriptstyle+};
(7.770021,0.514665) *[blue]{\scriptscriptstyle+};
(7.778234,0.519434) *[blue]{\scriptscriptstyle+};
(7.786448,0.524148) *[blue]{\scriptscriptstyle+};
(7.794661,0.530681) *[blue]{\scriptscriptstyle+};
(7.802875,0.533248) *[blue]{\scriptscriptstyle+};
(7.811088,0.534151) *[blue]{\scriptscriptstyle+};
(7.819302,0.535725) *[blue]{\scriptscriptstyle+};
(7.827515,0.537032) *[blue]{\scriptscriptstyle+};
(7.835729,0.537143) *[blue]{\scriptscriptstyle+};
(7.843943,0.546964) *[blue]{\scriptscriptstyle+};
(7.852156,0.552966) *[blue]{\scriptscriptstyle+};
(7.860370,0.559340) *[blue]{\scriptscriptstyle+};
(7.868583,0.563461) *[blue]{\scriptscriptstyle+};
(7.876797,0.581987) *[blue]{\scriptscriptstyle+};
(7.885010,0.575021) *[blue]{\scriptscriptstyle+};
(7.893224,0.576391) *[blue]{\scriptscriptstyle+};
(7.901437,0.578613) *[blue]{\scriptscriptstyle+};
(7.909651,0.579065) *[blue]{\scriptscriptstyle+};
(7.917864,0.579495) *[blue]{\scriptscriptstyle+};
(7.926078,0.580544) *[blue]{\scriptscriptstyle+};
(7.934292,0.582102) *[blue]{\scriptscriptstyle+};
(7.942505,0.584728) *[blue]{\scriptscriptstyle+};
(7.950719,0.590093) *[blue]{\scriptscriptstyle+};
(7.958932,0.595121) *[blue]{\scriptscriptstyle+};
(7.967146,0.595659) *[blue]{\scriptscriptstyle+};
(7.975359,0.596870) *[blue]{\scriptscriptstyle+};
(7.983573,0.601408) *[blue]{\scriptscriptstyle+};
(7.991786,0.606511) *[blue]{\scriptscriptstyle+};
(8.000000,0.632775) *[blue]{\scriptscriptstyle+};
(0.000000,0.065432) *[red]{\scriptscriptstyle\times};
(0.008214,0.069297) *[red]{\scriptscriptstyle\times};
(0.016427,0.069603) *[red]{\scriptscriptstyle\times};
(0.024641,0.072276) *[red]{\scriptscriptstyle\times};
(0.032854,0.073045) *[red]{\scriptscriptstyle\times};
(0.041068,0.074040) *[red]{\scriptscriptstyle\times};
(0.049281,0.077933) *[red]{\scriptscriptstyle\times};
(0.057495,0.088325) *[red]{\scriptscriptstyle\times};
(0.065708,0.098276) *[red]{\scriptscriptstyle\times};
(0.073922,0.099455) *[red]{\scriptscriptstyle\times};
(0.082136,0.105572) *[red]{\scriptscriptstyle\times};
(0.090349,0.108677) *[red]{\scriptscriptstyle\times};
(0.098563,0.114181) *[red]{\scriptscriptstyle\times};
(0.106776,0.123684) *[red]{\scriptscriptstyle\times};
(0.114990,0.130538) *[red]{\scriptscriptstyle\times};
(0.123203,0.081481) *[red]{\scriptscriptstyle\times};
(0.131417,0.082397) *[red]{\scriptscriptstyle\times};
(0.139630,0.085739) *[red]{\scriptscriptstyle\times};
(0.147844,0.086960) *[red]{\scriptscriptstyle\times};
(0.156057,0.088838) *[red]{\scriptscriptstyle\times};
(0.164271,0.090973) *[red]{\scriptscriptstyle\times};
(0.172485,0.091539) *[red]{\scriptscriptstyle\times};
(0.180698,0.101114) *[red]{\scriptscriptstyle\times};
(0.188912,0.114433) *[red]{\scriptscriptstyle\times};
(0.197125,0.115364) *[red]{\scriptscriptstyle\times};
(0.205339,0.115576) *[red]{\scriptscriptstyle\times};
(0.213552,0.116586) *[red]{\scriptscriptstyle\times};
(0.221766,0.118722) *[red]{\scriptscriptstyle\times};
(0.229979,0.121207) *[red]{\scriptscriptstyle\times};
(0.238193,0.143558) *[red]{\scriptscriptstyle\times};
(0.246407,0.095312) *[red]{\scriptscriptstyle\times};
(0.254620,0.097761) *[red]{\scriptscriptstyle\times};
(0.262834,0.100134) *[red]{\scriptscriptstyle\times};
(0.271047,0.102872) *[red]{\scriptscriptstyle\times};
(0.279261,0.103231) *[red]{\scriptscriptstyle\times};
(0.287474,0.125446) *[red]{\scriptscriptstyle\times};
(0.295688,0.125974) *[red]{\scriptscriptstyle\times};
(0.303901,0.126903) *[red]{\scriptscriptstyle\times};
(0.312115,0.127775) *[red]{\scriptscriptstyle\times};
(0.320329,0.130281) *[red]{\scriptscriptstyle\times};
(0.328542,0.156245) *[red]{\scriptscriptstyle\times};
(0.336756,0.156826) *[red]{\scriptscriptstyle\times};
(0.344969,0.157383) *[red]{\scriptscriptstyle\times};
(0.353183,0.160544) *[red]{\scriptscriptstyle\times};
(0.361396,0.185446) *[red]{\scriptscriptstyle\times};
(0.369610,0.112280) *[red]{\scriptscriptstyle\times};
(0.377823,0.113183) *[red]{\scriptscriptstyle\times};
(0.386037,0.113361) *[red]{\scriptscriptstyle\times};
(0.394251,0.113677) *[red]{\scriptscriptstyle\times};
(0.402464,0.115086) *[red]{\scriptscriptstyle\times};
(0.410678,0.116905) *[red]{\scriptscriptstyle\times};
(0.418891,0.118486) *[red]{\scriptscriptstyle\times};
(0.427105,0.118729) *[red]{\scriptscriptstyle\times};
(0.435318,0.121784) *[red]{\scriptscriptstyle\times};
(0.443532,0.121953) *[red]{\scriptscriptstyle\times};
(0.451745,0.124616) *[red]{\scriptscriptstyle\times};
(0.459959,0.142230) *[red]{\scriptscriptstyle\times};
(0.468172,0.145911) *[red]{\scriptscriptstyle\times};
(0.476386,0.146780) *[red]{\scriptscriptstyle\times};
(0.484600,0.148274) *[red]{\scriptscriptstyle\times};
(0.492813,0.125068) *[red]{\scriptscriptstyle\times};
(0.501027,0.125778) *[red]{\scriptscriptstyle\times};
(0.509240,0.128319) *[red]{\scriptscriptstyle\times};
(0.517454,0.128573) *[red]{\scriptscriptstyle\times};
(0.525667,0.132476) *[red]{\scriptscriptstyle\times};
(0.533881,0.134821) *[red]{\scriptscriptstyle\times};
(0.542094,0.135597) *[red]{\scriptscriptstyle\times};
(0.550308,0.136779) *[red]{\scriptscriptstyle\times};
(0.558522,0.137841) *[red]{\scriptscriptstyle\times};
(0.566735,0.144857) *[red]{\scriptscriptstyle\times};
(0.574949,0.162269) *[red]{\scriptscriptstyle\times};
(0.583162,0.162830) *[red]{\scriptscriptstyle\times};
(0.591376,0.170092) *[red]{\scriptscriptstyle\times};
(0.599589,0.172165) *[red]{\scriptscriptstyle\times};
(0.607803,0.188949) *[red]{\scriptscriptstyle\times};
(0.616016,0.146255) *[red]{\scriptscriptstyle\times};
(0.624230,0.146945) *[red]{\scriptscriptstyle\times};
(0.632444,0.148163) *[red]{\scriptscriptstyle\times};
(0.640657,0.154211) *[red]{\scriptscriptstyle\times};
(0.648871,0.155739) *[red]{\scriptscriptstyle\times};
(0.657084,0.156950) *[red]{\scriptscriptstyle\times};
(0.665298,0.157485) *[red]{\scriptscriptstyle\times};
(0.673511,0.157729) *[red]{\scriptscriptstyle\times};
(0.681725,0.158302) *[red]{\scriptscriptstyle\times};
(0.689938,0.159606) *[red]{\scriptscriptstyle\times};
(0.698152,0.159858) *[red]{\scriptscriptstyle\times};
(0.706366,0.179015) *[red]{\scriptscriptstyle\times};
(0.714579,0.180059) *[red]{\scriptscriptstyle\times};
(0.722793,0.185417) *[red]{\scriptscriptstyle\times};
(0.731006,0.204412) *[red]{\scriptscriptstyle\times};
(0.739220,0.132817) *[red]{\scriptscriptstyle\times};
(0.747433,0.132919) *[red]{\scriptscriptstyle\times};
(0.755647,0.133224) *[red]{\scriptscriptstyle\times};
(0.763860,0.134133) *[red]{\scriptscriptstyle\times};
(0.772074,0.135258) *[red]{\scriptscriptstyle\times};
(0.780287,0.137786) *[red]{\scriptscriptstyle\times};
(0.788501,0.138444) *[red]{\scriptscriptstyle\times};
(0.796715,0.147038) *[red]{\scriptscriptstyle\times};
(0.804928,0.165076) *[red]{\scriptscriptstyle\times};
(0.813142,0.166731) *[red]{\scriptscriptstyle\times};
(0.821355,0.166911) *[red]{\scriptscriptstyle\times};
(0.829569,0.169958) *[red]{\scriptscriptstyle\times};
(0.837782,0.170372) *[red]{\scriptscriptstyle\times};
(0.845996,0.173840) *[red]{\scriptscriptstyle\times};
(0.854209,0.190127) *[red]{\scriptscriptstyle\times};
(0.862423,0.119191) *[red]{\scriptscriptstyle\times};
(0.870637,0.152996) *[red]{\scriptscriptstyle\times};
(0.878850,0.156113) *[red]{\scriptscriptstyle\times};
(0.887064,0.158468) *[red]{\scriptscriptstyle\times};
(0.895277,0.160732) *[red]{\scriptscriptstyle\times};
(0.903491,0.161163) *[red]{\scriptscriptstyle\times};
(0.911704,0.164546) *[red]{\scriptscriptstyle\times};
(0.919918,0.165658) *[red]{\scriptscriptstyle\times};
(0.928131,0.168256) *[red]{\scriptscriptstyle\times};
(0.936345,0.187044) *[red]{\scriptscriptstyle\times};
(0.944559,0.187499) *[red]{\scriptscriptstyle\times};
(0.952772,0.188414) *[red]{\scriptscriptstyle\times};
(0.960986,0.188579) *[red]{\scriptscriptstyle\times};
(0.969199,0.192406) *[red]{\scriptscriptstyle\times};
(0.977413,0.239628) *[red]{\scriptscriptstyle\times};
(0.985626,0.152181) *[red]{\scriptscriptstyle\times};
(0.993840,0.153684) *[red]{\scriptscriptstyle\times};
(1.002053,0.154249) *[red]{\scriptscriptstyle\times};
(1.010267,0.161589) *[red]{\scriptscriptstyle\times};
(1.018480,0.167025) *[red]{\scriptscriptstyle\times};
(1.026694,0.180984) *[red]{\scriptscriptstyle\times};
(1.034908,0.182328) *[red]{\scriptscriptstyle\times};
(1.043121,0.183937) *[red]{\scriptscriptstyle\times};
(1.051335,0.184378) *[red]{\scriptscriptstyle\times};
(1.059548,0.184411) *[red]{\scriptscriptstyle\times};
(1.067762,0.187666) *[red]{\scriptscriptstyle\times};
(1.075975,0.188054) *[red]{\scriptscriptstyle\times};
(1.084189,0.190107) *[red]{\scriptscriptstyle\times};
(1.092402,0.210931) *[red]{\scriptscriptstyle\times};
(1.100616,0.236675) *[red]{\scriptscriptstyle\times};
(1.108830,0.164287) *[red]{\scriptscriptstyle\times};
(1.117043,0.164319) *[red]{\scriptscriptstyle\times};
(1.125257,0.166367) *[red]{\scriptscriptstyle\times};
(1.133470,0.169841) *[red]{\scriptscriptstyle\times};
(1.141684,0.170140) *[red]{\scriptscriptstyle\times};
(1.149897,0.174098) *[red]{\scriptscriptstyle\times};
(1.158111,0.174485) *[red]{\scriptscriptstyle\times};
(1.166324,0.180853) *[red]{\scriptscriptstyle\times};
(1.174538,0.202710) *[red]{\scriptscriptstyle\times};
(1.182752,0.203513) *[red]{\scriptscriptstyle\times};
(1.190965,0.205893) *[red]{\scriptscriptstyle\times};
(1.199179,0.214745) *[red]{\scriptscriptstyle\times};
(1.207392,0.216506) *[red]{\scriptscriptstyle\times};
(1.215606,0.229179) *[red]{\scriptscriptstyle\times};
(1.223819,0.232728) *[red]{\scriptscriptstyle\times};
(1.232033,0.169581) *[red]{\scriptscriptstyle\times};
(1.240246,0.169990) *[red]{\scriptscriptstyle\times};
(1.248460,0.170203) *[red]{\scriptscriptstyle\times};
(1.256674,0.170222) *[red]{\scriptscriptstyle\times};
(1.264887,0.171890) *[red]{\scriptscriptstyle\times};
(1.273101,0.173065) *[red]{\scriptscriptstyle\times};
(1.281314,0.173539) *[red]{\scriptscriptstyle\times};
(1.289528,0.174626) *[red]{\scriptscriptstyle\times};
(1.297741,0.176517) *[red]{\scriptscriptstyle\times};
(1.305955,0.177730) *[red]{\scriptscriptstyle\times};
(1.314168,0.184886) *[red]{\scriptscriptstyle\times};
(1.322382,0.185416) *[red]{\scriptscriptstyle\times};
(1.330595,0.197520) *[red]{\scriptscriptstyle\times};
(1.338809,0.197565) *[red]{\scriptscriptstyle\times};
(1.347023,0.201359) *[red]{\scriptscriptstyle\times};
(1.355236,0.157822) *[red]{\scriptscriptstyle\times};
(1.363450,0.157850) *[red]{\scriptscriptstyle\times};
(1.371663,0.157907) *[red]{\scriptscriptstyle\times};
(1.379877,0.157981) *[red]{\scriptscriptstyle\times};
(1.388090,0.161380) *[red]{\scriptscriptstyle\times};
(1.396304,0.163357) *[red]{\scriptscriptstyle\times};
(1.404517,0.164241) *[red]{\scriptscriptstyle\times};
(1.412731,0.185337) *[red]{\scriptscriptstyle\times};
(1.420945,0.186131) *[red]{\scriptscriptstyle\times};
(1.429158,0.187855) *[red]{\scriptscriptstyle\times};
(1.437372,0.189566) *[red]{\scriptscriptstyle\times};
(1.445585,0.193931) *[red]{\scriptscriptstyle\times};
(1.453799,0.212947) *[red]{\scriptscriptstyle\times};
(1.462012,0.220487) *[red]{\scriptscriptstyle\times};
(1.470226,0.221222) *[red]{\scriptscriptstyle\times};
(1.478439,0.180656) *[red]{\scriptscriptstyle\times};
(1.486653,0.181209) *[red]{\scriptscriptstyle\times};
(1.494867,0.181445) *[red]{\scriptscriptstyle\times};
(1.503080,0.184370) *[red]{\scriptscriptstyle\times};
(1.511294,0.184612) *[red]{\scriptscriptstyle\times};
(1.519507,0.185033) *[red]{\scriptscriptstyle\times};
(1.527721,0.186231) *[red]{\scriptscriptstyle\times};
(1.535934,0.187940) *[red]{\scriptscriptstyle\times};
(1.544148,0.188763) *[red]{\scriptscriptstyle\times};
(1.552361,0.190151) *[red]{\scriptscriptstyle\times};
(1.560575,0.192479) *[red]{\scriptscriptstyle\times};
(1.568789,0.194058) *[red]{\scriptscriptstyle\times};
(1.577002,0.210033) *[red]{\scriptscriptstyle\times};
(1.585216,0.220137) *[red]{\scriptscriptstyle\times};
(1.593429,0.227936) *[red]{\scriptscriptstyle\times};
(1.601643,0.206021) *[red]{\scriptscriptstyle\times};
(1.609856,0.206238) *[red]{\scriptscriptstyle\times};
(1.618070,0.208449) *[red]{\scriptscriptstyle\times};
(1.626283,0.210296) *[red]{\scriptscriptstyle\times};
(1.634497,0.210501) *[red]{\scriptscriptstyle\times};
(1.642710,0.213644) *[red]{\scriptscriptstyle\times};
(1.650924,0.215871) *[red]{\scriptscriptstyle\times};
(1.659138,0.216565) *[red]{\scriptscriptstyle\times};
(1.667351,0.238090) *[red]{\scriptscriptstyle\times};
(1.675565,0.239029) *[red]{\scriptscriptstyle\times};
(1.683778,0.239631) *[red]{\scriptscriptstyle\times};
(1.691992,0.239869) *[red]{\scriptscriptstyle\times};
(1.700205,0.243265) *[red]{\scriptscriptstyle\times};
(1.708419,0.243423) *[red]{\scriptscriptstyle\times};
(1.716632,0.246168) *[red]{\scriptscriptstyle\times};
(1.724846,0.211305) *[red]{\scriptscriptstyle\times};
(1.733060,0.214474) *[red]{\scriptscriptstyle\times};
(1.741273,0.220008) *[red]{\scriptscriptstyle\times};
(1.749487,0.220499) *[red]{\scriptscriptstyle\times};
(1.757700,0.229889) *[red]{\scriptscriptstyle\times};
(1.765914,0.243433) *[red]{\scriptscriptstyle\times};
(1.774127,0.245038) *[red]{\scriptscriptstyle\times};
(1.782341,0.247570) *[red]{\scriptscriptstyle\times};
(1.790554,0.248193) *[red]{\scriptscriptstyle\times};
(1.798768,0.248486) *[red]{\scriptscriptstyle\times};
(1.806982,0.249201) *[red]{\scriptscriptstyle\times};
(1.815195,0.249515) *[red]{\scriptscriptstyle\times};
(1.823409,0.251676) *[red]{\scriptscriptstyle\times};
(1.831622,0.267883) *[red]{\scriptscriptstyle\times};
(1.839836,0.298817) *[red]{\scriptscriptstyle\times};
(1.848049,0.221583) *[red]{\scriptscriptstyle\times};
(1.856263,0.226413) *[red]{\scriptscriptstyle\times};
(1.864476,0.228287) *[red]{\scriptscriptstyle\times};
(1.872690,0.228779) *[red]{\scriptscriptstyle\times};
(1.880903,0.230855) *[red]{\scriptscriptstyle\times};
(1.889117,0.255926) *[red]{\scriptscriptstyle\times};
(1.897331,0.257809) *[red]{\scriptscriptstyle\times};
(1.905544,0.259629) *[red]{\scriptscriptstyle\times};
(1.913758,0.259730) *[red]{\scriptscriptstyle\times};
(1.921971,0.262503) *[red]{\scriptscriptstyle\times};
(1.930185,0.280235) *[red]{\scriptscriptstyle\times};
(1.938398,0.281448) *[red]{\scriptscriptstyle\times};
(1.946612,0.286402) *[red]{\scriptscriptstyle\times};
(1.954825,0.288411) *[red]{\scriptscriptstyle\times};
(1.963039,0.288417) *[red]{\scriptscriptstyle\times};
(1.971253,0.211118) *[red]{\scriptscriptstyle\times};
(1.979466,0.214650) *[red]{\scriptscriptstyle\times};
(1.987680,0.215075) *[red]{\scriptscriptstyle\times};
(1.995893,0.218920) *[red]{\scriptscriptstyle\times};
(2.004107,0.226556) *[red]{\scriptscriptstyle\times};
(2.012320,0.227130) *[red]{\scriptscriptstyle\times};
(2.020534,0.247364) *[red]{\scriptscriptstyle\times};
(2.028747,0.255650) *[red]{\scriptscriptstyle\times};
(2.036961,0.257297) *[red]{\scriptscriptstyle\times};
(2.045175,0.258304) *[red]{\scriptscriptstyle\times};
(2.053388,0.266033) *[red]{\scriptscriptstyle\times};
(2.061602,0.269483) *[red]{\scriptscriptstyle\times};
(2.069815,0.277585) *[red]{\scriptscriptstyle\times};
(2.078029,0.282041) *[red]{\scriptscriptstyle\times};
(2.086242,0.289124) *[red]{\scriptscriptstyle\times};
(2.094456,0.241260) *[red]{\scriptscriptstyle\times};
(2.102669,0.243779) *[red]{\scriptscriptstyle\times};
(2.110883,0.244181) *[red]{\scriptscriptstyle\times};
(2.119097,0.250193) *[red]{\scriptscriptstyle\times};
(2.127310,0.251965) *[red]{\scriptscriptstyle\times};
(2.135524,0.252189) *[red]{\scriptscriptstyle\times};
(2.143737,0.254928) *[red]{\scriptscriptstyle\times};
(2.151951,0.259433) *[red]{\scriptscriptstyle\times};
(2.160164,0.269941) *[red]{\scriptscriptstyle\times};
(2.168378,0.270960) *[red]{\scriptscriptstyle\times};
(2.176591,0.277481) *[red]{\scriptscriptstyle\times};
(2.184805,0.279365) *[red]{\scriptscriptstyle\times};
(2.193018,0.281406) *[red]{\scriptscriptstyle\times};
(2.201232,0.281904) *[red]{\scriptscriptstyle\times};
(2.209446,0.298978) *[red]{\scriptscriptstyle\times};
(2.217659,0.243439) *[red]{\scriptscriptstyle\times};
(2.225873,0.247732) *[red]{\scriptscriptstyle\times};
(2.234086,0.248530) *[red]{\scriptscriptstyle\times};
(2.242300,0.253448) *[red]{\scriptscriptstyle\times};
(2.250513,0.254363) *[red]{\scriptscriptstyle\times};
(2.258727,0.260074) *[red]{\scriptscriptstyle\times};
(2.266940,0.274322) *[red]{\scriptscriptstyle\times};
(2.275154,0.274853) *[red]{\scriptscriptstyle\times};
(2.283368,0.278232) *[red]{\scriptscriptstyle\times};
(2.291581,0.279927) *[red]{\scriptscriptstyle\times};
(2.299795,0.281341) *[red]{\scriptscriptstyle\times};
(2.308008,0.283542) *[red]{\scriptscriptstyle\times};
(2.316222,0.284531) *[red]{\scriptscriptstyle\times};
(2.324435,0.300370) *[red]{\scriptscriptstyle\times};
(2.332649,0.306104) *[red]{\scriptscriptstyle\times};
(2.340862,0.275787) *[red]{\scriptscriptstyle\times};
(2.349076,0.277908) *[red]{\scriptscriptstyle\times};
(2.357290,0.280265) *[red]{\scriptscriptstyle\times};
(2.365503,0.280388) *[red]{\scriptscriptstyle\times};
(2.373717,0.283283) *[red]{\scriptscriptstyle\times};
(2.381930,0.285599) *[red]{\scriptscriptstyle\times};
(2.390144,0.288939) *[red]{\scriptscriptstyle\times};
(2.398357,0.291724) *[red]{\scriptscriptstyle\times};
(2.406571,0.292709) *[red]{\scriptscriptstyle\times};
(2.414784,0.300552) *[red]{\scriptscriptstyle\times};
(2.422998,0.301760) *[red]{\scriptscriptstyle\times};
(2.431211,0.305196) *[red]{\scriptscriptstyle\times};
(2.439425,0.306294) *[red]{\scriptscriptstyle\times};
(2.447639,0.308337) *[red]{\scriptscriptstyle\times};
(2.455852,0.327701) *[red]{\scriptscriptstyle\times};
(2.464066,0.278904) *[red]{\scriptscriptstyle\times};
(2.472279,0.280411) *[red]{\scriptscriptstyle\times};
(2.480493,0.281974) *[red]{\scriptscriptstyle\times};
(2.488706,0.282359) *[red]{\scriptscriptstyle\times};
(2.496920,0.282910) *[red]{\scriptscriptstyle\times};
(2.505133,0.283572) *[red]{\scriptscriptstyle\times};
(2.513347,0.284141) *[red]{\scriptscriptstyle\times};
(2.521561,0.284669) *[red]{\scriptscriptstyle\times};
(2.529774,0.287212) *[red]{\scriptscriptstyle\times};
(2.537988,0.289372) *[red]{\scriptscriptstyle\times};
(2.546201,0.290648) *[red]{\scriptscriptstyle\times};
(2.554415,0.291915) *[red]{\scriptscriptstyle\times};
(2.562628,0.303545) *[red]{\scriptscriptstyle\times};
(2.570842,0.304958) *[red]{\scriptscriptstyle\times};
(2.579055,0.310385) *[red]{\scriptscriptstyle\times};
(2.587269,0.281505) *[red]{\scriptscriptstyle\times};
(2.595483,0.281888) *[red]{\scriptscriptstyle\times};
(2.603696,0.284944) *[red]{\scriptscriptstyle\times};
(2.611910,0.285525) *[red]{\scriptscriptstyle\times};
(2.620123,0.286820) *[red]{\scriptscriptstyle\times};
(2.628337,0.286955) *[red]{\scriptscriptstyle\times};
(2.636550,0.287047) *[red]{\scriptscriptstyle\times};
(2.644764,0.289035) *[red]{\scriptscriptstyle\times};
(2.652977,0.291068) *[red]{\scriptscriptstyle\times};
(2.661191,0.309082) *[red]{\scriptscriptstyle\times};
(2.669405,0.312049) *[red]{\scriptscriptstyle\times};
(2.677618,0.314005) *[red]{\scriptscriptstyle\times};
(2.685832,0.315133) *[red]{\scriptscriptstyle\times};
(2.694045,0.323521) *[red]{\scriptscriptstyle\times};
(2.702259,0.326099) *[red]{\scriptscriptstyle\times};
(2.710472,0.271802) *[red]{\scriptscriptstyle\times};
(2.718686,0.275519) *[red]{\scriptscriptstyle\times};
(2.726899,0.275747) *[red]{\scriptscriptstyle\times};
(2.735113,0.277306) *[red]{\scriptscriptstyle\times};
(2.743326,0.286649) *[red]{\scriptscriptstyle\times};
(2.751540,0.303484) *[red]{\scriptscriptstyle\times};
(2.759754,0.305479) *[red]{\scriptscriptstyle\times};
(2.767967,0.310235) *[red]{\scriptscriptstyle\times};
(2.776181,0.313811) *[red]{\scriptscriptstyle\times};
(2.784394,0.323740) *[red]{\scriptscriptstyle\times};
(2.792608,0.326663) *[red]{\scriptscriptstyle\times};
(2.800821,0.332354) *[red]{\scriptscriptstyle\times};
(2.809035,0.333298) *[red]{\scriptscriptstyle\times};
(2.817248,0.339676) *[red]{\scriptscriptstyle\times};
(2.825462,0.345064) *[red]{\scriptscriptstyle\times};
(2.833676,0.260565) *[red]{\scriptscriptstyle\times};
(2.841889,0.267395) *[red]{\scriptscriptstyle\times};
(2.850103,0.269756) *[red]{\scriptscriptstyle\times};
(2.858316,0.274547) *[red]{\scriptscriptstyle\times};
(2.866530,0.286914) *[red]{\scriptscriptstyle\times};
(2.874743,0.287240) *[red]{\scriptscriptstyle\times};
(2.882957,0.288971) *[red]{\scriptscriptstyle\times};
(2.891170,0.289358) *[red]{\scriptscriptstyle\times};
(2.899384,0.291588) *[red]{\scriptscriptstyle\times};
(2.907598,0.293254) *[red]{\scriptscriptstyle\times};
(2.915811,0.297661) *[red]{\scriptscriptstyle\times};
(2.924025,0.321835) *[red]{\scriptscriptstyle\times};
(2.932238,0.322962) *[red]{\scriptscriptstyle\times};
(2.940452,0.328068) *[red]{\scriptscriptstyle\times};
(2.948665,0.353716) *[red]{\scriptscriptstyle\times};
(2.956879,0.284276) *[red]{\scriptscriptstyle\times};
(2.965092,0.284759) *[red]{\scriptscriptstyle\times};
(2.973306,0.286271) *[red]{\scriptscriptstyle\times};
(2.981520,0.287781) *[red]{\scriptscriptstyle\times};
(2.989733,0.289943) *[red]{\scriptscriptstyle\times};
(2.997947,0.290727) *[red]{\scriptscriptstyle\times};
(3.006160,0.292528) *[red]{\scriptscriptstyle\times};
(3.014374,0.294476) *[red]{\scriptscriptstyle\times};
(3.022587,0.299001) *[red]{\scriptscriptstyle\times};
(3.030801,0.304350) *[red]{\scriptscriptstyle\times};
(3.039014,0.304963) *[red]{\scriptscriptstyle\times};
(3.047228,0.309711) *[red]{\scriptscriptstyle\times};
(3.055441,0.317781) *[red]{\scriptscriptstyle\times};
(3.063655,0.324776) *[red]{\scriptscriptstyle\times};
(3.071869,0.340681) *[red]{\scriptscriptstyle\times};
(3.080082,0.281073) *[red]{\scriptscriptstyle\times};
(3.088296,0.286476) *[red]{\scriptscriptstyle\times};
(3.096509,0.288888) *[red]{\scriptscriptstyle\times};
(3.104723,0.290628) *[red]{\scriptscriptstyle\times};
(3.112936,0.290744) *[red]{\scriptscriptstyle\times};
(3.121150,0.291591) *[red]{\scriptscriptstyle\times};
(3.129363,0.310952) *[red]{\scriptscriptstyle\times};
(3.137577,0.314966) *[red]{\scriptscriptstyle\times};
(3.145791,0.315491) *[red]{\scriptscriptstyle\times};
(3.154004,0.319800) *[red]{\scriptscriptstyle\times};
(3.162218,0.334512) *[red]{\scriptscriptstyle\times};
(3.170431,0.343651) *[red]{\scriptscriptstyle\times};
(3.178645,0.345967) *[red]{\scriptscriptstyle\times};
(3.186858,0.347285) *[red]{\scriptscriptstyle\times};
(3.195072,0.361882) *[red]{\scriptscriptstyle\times};
(3.203285,0.302143) *[red]{\scriptscriptstyle\times};
(3.211499,0.302384) *[red]{\scriptscriptstyle\times};
(3.219713,0.303183) *[red]{\scriptscriptstyle\times};
(3.227926,0.303583) *[red]{\scriptscriptstyle\times};
(3.236140,0.305715) *[red]{\scriptscriptstyle\times};
(3.244353,0.306006) *[red]{\scriptscriptstyle\times};
(3.252567,0.306401) *[red]{\scriptscriptstyle\times};
(3.260780,0.306728) *[red]{\scriptscriptstyle\times};
(3.268994,0.307360) *[red]{\scriptscriptstyle\times};
(3.277207,0.310965) *[red]{\scriptscriptstyle\times};
(3.285421,0.311118) *[red]{\scriptscriptstyle\times};
(3.293634,0.311921) *[red]{\scriptscriptstyle\times};
(3.301848,0.324797) *[red]{\scriptscriptstyle\times};
(3.310062,0.327123) *[red]{\scriptscriptstyle\times};
(3.318275,0.332190) *[red]{\scriptscriptstyle\times};
(3.326489,0.285956) *[red]{\scriptscriptstyle\times};
(3.334702,0.286911) *[red]{\scriptscriptstyle\times};
(3.342916,0.287228) *[red]{\scriptscriptstyle\times};
(3.351129,0.289683) *[red]{\scriptscriptstyle\times};
(3.359343,0.291989) *[red]{\scriptscriptstyle\times};
(3.367556,0.296061) *[red]{\scriptscriptstyle\times};
(3.375770,0.297134) *[red]{\scriptscriptstyle\times};
(3.383984,0.304266) *[red]{\scriptscriptstyle\times};
(3.392197,0.304549) *[red]{\scriptscriptstyle\times};
(3.400411,0.311956) *[red]{\scriptscriptstyle\times};
(3.408624,0.314920) *[red]{\scriptscriptstyle\times};
(3.416838,0.324109) *[red]{\scriptscriptstyle\times};
(3.425051,0.330231) *[red]{\scriptscriptstyle\times};
(3.433265,0.331380) *[red]{\scriptscriptstyle\times};
(3.441478,0.350576) *[red]{\scriptscriptstyle\times};
(3.449692,0.296845) *[red]{\scriptscriptstyle\times};
(3.457906,0.297830) *[red]{\scriptscriptstyle\times};
(3.466119,0.300441) *[red]{\scriptscriptstyle\times};
(3.474333,0.304296) *[red]{\scriptscriptstyle\times};
(3.482546,0.304711) *[red]{\scriptscriptstyle\times};
(3.490760,0.323782) *[red]{\scriptscriptstyle\times};
(3.498973,0.327413) *[red]{\scriptscriptstyle\times};
(3.507187,0.329547) *[red]{\scriptscriptstyle\times};
(3.515400,0.330777) *[red]{\scriptscriptstyle\times};
(3.523614,0.350626) *[red]{\scriptscriptstyle\times};
(3.531828,0.354368) *[red]{\scriptscriptstyle\times};
(3.540041,0.355244) *[red]{\scriptscriptstyle\times};
(3.548255,0.357918) *[red]{\scriptscriptstyle\times};
(3.556468,0.359564) *[red]{\scriptscriptstyle\times};
(3.564682,0.389339) *[red]{\scriptscriptstyle\times};
(3.572895,0.300748) *[red]{\scriptscriptstyle\times};
(3.581109,0.304567) *[red]{\scriptscriptstyle\times};
(3.589322,0.304946) *[red]{\scriptscriptstyle\times};
(3.597536,0.304977) *[red]{\scriptscriptstyle\times};
(3.605749,0.327952) *[red]{\scriptscriptstyle\times};
(3.613963,0.329005) *[red]{\scriptscriptstyle\times};
(3.622177,0.331871) *[red]{\scriptscriptstyle\times};
(3.630390,0.334692) *[red]{\scriptscriptstyle\times};
(3.638604,0.336114) *[red]{\scriptscriptstyle\times};
(3.646817,0.337559) *[red]{\scriptscriptstyle\times};
(3.655031,0.350717) *[red]{\scriptscriptstyle\times};
(3.663244,0.354878) *[red]{\scriptscriptstyle\times};
(3.671458,0.383874) *[red]{\scriptscriptstyle\times};
(3.679671,0.383970) *[red]{\scriptscriptstyle\times};
(3.687885,0.386741) *[red]{\scriptscriptstyle\times};
(3.696099,0.308672) *[red]{\scriptscriptstyle\times};
(3.704312,0.310643) *[red]{\scriptscriptstyle\times};
(3.712526,0.310869) *[red]{\scriptscriptstyle\times};
(3.720739,0.311120) *[red]{\scriptscriptstyle\times};
(3.728953,0.318517) *[red]{\scriptscriptstyle\times};
(3.737166,0.324122) *[red]{\scriptscriptstyle\times};
(3.745380,0.334936) *[red]{\scriptscriptstyle\times};
(3.753593,0.335267) *[red]{\scriptscriptstyle\times};
(3.761807,0.336550) *[red]{\scriptscriptstyle\times};
(3.770021,0.353330) *[red]{\scriptscriptstyle\times};
(3.778234,0.357879) *[red]{\scriptscriptstyle\times};
(3.786448,0.368181) *[red]{\scriptscriptstyle\times};
(3.794661,0.383493) *[red]{\scriptscriptstyle\times};
(3.802875,0.387817) *[red]{\scriptscriptstyle\times};
(3.811088,0.390066) *[red]{\scriptscriptstyle\times};
(3.819302,0.309610) *[red]{\scriptscriptstyle\times};
(3.827515,0.313775) *[red]{\scriptscriptstyle\times};
(3.835729,0.314293) *[red]{\scriptscriptstyle\times};
(3.843943,0.319133) *[red]{\scriptscriptstyle\times};
(3.852156,0.321592) *[red]{\scriptscriptstyle\times};
(3.860370,0.337363) *[red]{\scriptscriptstyle\times};
(3.868583,0.338839) *[red]{\scriptscriptstyle\times};
(3.876797,0.342331) *[red]{\scriptscriptstyle\times};
(3.885010,0.343405) *[red]{\scriptscriptstyle\times};
(3.893224,0.345177) *[red]{\scriptscriptstyle\times};
(3.901437,0.347543) *[red]{\scriptscriptstyle\times};
(3.909651,0.354076) *[red]{\scriptscriptstyle\times};
(3.917864,0.363872) *[red]{\scriptscriptstyle\times};
(3.926078,0.367648) *[red]{\scriptscriptstyle\times};
(3.934292,0.385631) *[red]{\scriptscriptstyle\times};
(3.942505,0.314349) *[red]{\scriptscriptstyle\times};
(3.950719,0.317114) *[red]{\scriptscriptstyle\times};
(3.958932,0.318782) *[red]{\scriptscriptstyle\times};
(3.967146,0.333987) *[red]{\scriptscriptstyle\times};
(3.975359,0.334402) *[red]{\scriptscriptstyle\times};
(3.983573,0.334670) *[red]{\scriptscriptstyle\times};
(3.991786,0.341289) *[red]{\scriptscriptstyle\times};
(4.000000,0.345696) *[red]{\scriptscriptstyle\times};
(4.008214,0.346548) *[red]{\scriptscriptstyle\times};
(4.016427,0.350378) *[red]{\scriptscriptstyle\times};
(4.024641,0.351288) *[red]{\scriptscriptstyle\times};
(4.032854,0.352366) *[red]{\scriptscriptstyle\times};
(4.041068,0.356310) *[red]{\scriptscriptstyle\times};
(4.049281,0.377775) *[red]{\scriptscriptstyle\times};
(4.057495,0.443462) *[red]{\scriptscriptstyle\times};
(4.065708,0.340003) *[red]{\scriptscriptstyle\times};
(4.073922,0.340602) *[red]{\scriptscriptstyle\times};
(4.082136,0.343349) *[red]{\scriptscriptstyle\times};
(4.090349,0.344054) *[red]{\scriptscriptstyle\times};
(4.098563,0.351698) *[red]{\scriptscriptstyle\times};
(4.106776,0.359170) *[red]{\scriptscriptstyle\times};
(4.114990,0.363522) *[red]{\scriptscriptstyle\times};
(4.123203,0.366307) *[red]{\scriptscriptstyle\times};
(4.131417,0.367609) *[red]{\scriptscriptstyle\times};
(4.139630,0.369257) *[red]{\scriptscriptstyle\times};
(4.147844,0.374441) *[red]{\scriptscriptstyle\times};
(4.156057,0.376928) *[red]{\scriptscriptstyle\times};
(4.164271,0.387007) *[red]{\scriptscriptstyle\times};
(4.172485,0.387018) *[red]{\scriptscriptstyle\times};
(4.180698,0.390042) *[red]{\scriptscriptstyle\times};
(4.188912,0.345780) *[red]{\scriptscriptstyle\times};
(4.197125,0.348276) *[red]{\scriptscriptstyle\times};
(4.205339,0.350533) *[red]{\scriptscriptstyle\times};
(4.213552,0.354534) *[red]{\scriptscriptstyle\times};
(4.221766,0.355885) *[red]{\scriptscriptstyle\times};
(4.229979,0.365898) *[red]{\scriptscriptstyle\times};
(4.238193,0.366306) *[red]{\scriptscriptstyle\times};
(4.246407,0.367165) *[red]{\scriptscriptstyle\times};
(4.254620,0.368926) *[red]{\scriptscriptstyle\times};
(4.262834,0.370395) *[red]{\scriptscriptstyle\times};
(4.271047,0.372659) *[red]{\scriptscriptstyle\times};
(4.279261,0.374269) *[red]{\scriptscriptstyle\times};
(4.287474,0.374939) *[red]{\scriptscriptstyle\times};
(4.295688,0.396745) *[red]{\scriptscriptstyle\times};
(4.303901,0.400139) *[red]{\scriptscriptstyle\times};
(4.312115,0.343398) *[red]{\scriptscriptstyle\times};
(4.320329,0.343577) *[red]{\scriptscriptstyle\times};
(4.328542,0.343844) *[red]{\scriptscriptstyle\times};
(4.336756,0.347292) *[red]{\scriptscriptstyle\times};
(4.344969,0.350323) *[red]{\scriptscriptstyle\times};
(4.353183,0.350858) *[red]{\scriptscriptstyle\times};
(4.361396,0.351481) *[red]{\scriptscriptstyle\times};
(4.369610,0.352312) *[red]{\scriptscriptstyle\times};
(4.377823,0.355086) *[red]{\scriptscriptstyle\times};
(4.386037,0.356846) *[red]{\scriptscriptstyle\times};
(4.394251,0.360209) *[red]{\scriptscriptstyle\times};
(4.402464,0.361987) *[red]{\scriptscriptstyle\times};
(4.410678,0.372682) *[red]{\scriptscriptstyle\times};
(4.418891,0.375399) *[red]{\scriptscriptstyle\times};
(4.427105,0.721698) *[red]{\scriptscriptstyle\times};
(4.435318,0.354614) *[red]{\scriptscriptstyle\times};
(4.443532,0.358251) *[red]{\scriptscriptstyle\times};
(4.451745,0.359249) *[red]{\scriptscriptstyle\times};
(4.459959,0.359945) *[red]{\scriptscriptstyle\times};
(4.468172,0.363538) *[red]{\scriptscriptstyle\times};
(4.476386,0.377023) *[red]{\scriptscriptstyle\times};
(4.484600,0.381263) *[red]{\scriptscriptstyle\times};
(4.492813,0.382815) *[red]{\scriptscriptstyle\times};
(4.501027,0.384334) *[red]{\scriptscriptstyle\times};
(4.509240,0.384352) *[red]{\scriptscriptstyle\times};
(4.517454,0.384570) *[red]{\scriptscriptstyle\times};
(4.525667,0.385435) *[red]{\scriptscriptstyle\times};
(4.533881,0.386342) *[red]{\scriptscriptstyle\times};
(4.542094,0.393204) *[red]{\scriptscriptstyle\times};
(4.550308,0.396804) *[red]{\scriptscriptstyle\times};
(4.558522,0.343489) *[red]{\scriptscriptstyle\times};
(4.566735,0.344233) *[red]{\scriptscriptstyle\times};
(4.574949,0.344597) *[red]{\scriptscriptstyle\times};
(4.583162,0.345540) *[red]{\scriptscriptstyle\times};
(4.591376,0.347395) *[red]{\scriptscriptstyle\times};
(4.599589,0.367469) *[red]{\scriptscriptstyle\times};
(4.607803,0.369082) *[red]{\scriptscriptstyle\times};
(4.616016,0.371449) *[red]{\scriptscriptstyle\times};
(4.624230,0.371467) *[red]{\scriptscriptstyle\times};
(4.632444,0.376665) *[red]{\scriptscriptstyle\times};
(4.640657,0.388149) *[red]{\scriptscriptstyle\times};
(4.648871,0.390737) *[red]{\scriptscriptstyle\times};
(4.657084,0.390945) *[red]{\scriptscriptstyle\times};
(4.665298,0.392349) *[red]{\scriptscriptstyle\times};
(4.673511,0.401525) *[red]{\scriptscriptstyle\times};
(4.681725,0.343543) *[red]{\scriptscriptstyle\times};
(4.689938,0.344251) *[red]{\scriptscriptstyle\times};
(4.698152,0.346195) *[red]{\scriptscriptstyle\times};
(4.706366,0.346689) *[red]{\scriptscriptstyle\times};
(4.714579,0.347833) *[red]{\scriptscriptstyle\times};
(4.722793,0.348299) *[red]{\scriptscriptstyle\times};
(4.731006,0.348659) *[red]{\scriptscriptstyle\times};
(4.739220,0.349900) *[red]{\scriptscriptstyle\times};
(4.747433,0.353642) *[red]{\scriptscriptstyle\times};
(4.755647,0.357819) *[red]{\scriptscriptstyle\times};
(4.763860,0.372465) *[red]{\scriptscriptstyle\times};
(4.772074,0.373416) *[red]{\scriptscriptstyle\times};
(4.780287,0.373874) *[red]{\scriptscriptstyle\times};
(4.788501,0.395713) *[red]{\scriptscriptstyle\times};
(4.796715,0.423589) *[red]{\scriptscriptstyle\times};
(4.804928,0.345037) *[red]{\scriptscriptstyle\times};
(4.813142,0.346049) *[red]{\scriptscriptstyle\times};
(4.821355,0.351518) *[red]{\scriptscriptstyle\times};
(4.829569,0.358192) *[red]{\scriptscriptstyle\times};
(4.837782,0.359132) *[red]{\scriptscriptstyle\times};
(4.845996,0.364184) *[red]{\scriptscriptstyle\times};
(4.854209,0.372113) *[red]{\scriptscriptstyle\times};
(4.862423,0.374537) *[red]{\scriptscriptstyle\times};
(4.870637,0.376240) *[red]{\scriptscriptstyle\times};
(4.878850,0.381273) *[red]{\scriptscriptstyle\times};
(4.887064,0.391737) *[red]{\scriptscriptstyle\times};
(4.895277,0.397211) *[red]{\scriptscriptstyle\times};
(4.903491,0.403199) *[red]{\scriptscriptstyle\times};
(4.911704,0.407407) *[red]{\scriptscriptstyle\times};
(4.919918,0.411412) *[red]{\scriptscriptstyle\times};
(4.928131,0.351190) *[red]{\scriptscriptstyle\times};
(4.936345,0.351962) *[red]{\scriptscriptstyle\times};
(4.944559,0.352062) *[red]{\scriptscriptstyle\times};
(4.952772,0.353074) *[red]{\scriptscriptstyle\times};
(4.960986,0.355538) *[red]{\scriptscriptstyle\times};
(4.969199,0.360930) *[red]{\scriptscriptstyle\times};
(4.977413,0.362696) *[red]{\scriptscriptstyle\times};
(4.985626,0.371943) *[red]{\scriptscriptstyle\times};
(4.993840,0.377866) *[red]{\scriptscriptstyle\times};
(5.002053,0.379822) *[red]{\scriptscriptstyle\times};
(5.010267,0.380484) *[red]{\scriptscriptstyle\times};
(5.018480,0.381032) *[red]{\scriptscriptstyle\times};
(5.026694,0.381688) *[red]{\scriptscriptstyle\times};
(5.034908,0.386806) *[red]{\scriptscriptstyle\times};
(5.043121,0.413719) *[red]{\scriptscriptstyle\times};
(5.051335,0.339775) *[red]{\scriptscriptstyle\times};
(5.059548,0.343781) *[red]{\scriptscriptstyle\times};
(5.067762,0.343988) *[red]{\scriptscriptstyle\times};
(5.075975,0.344602) *[red]{\scriptscriptstyle\times};
(5.084189,0.345495) *[red]{\scriptscriptstyle\times};
(5.092402,0.363710) *[red]{\scriptscriptstyle\times};
(5.100616,0.365935) *[red]{\scriptscriptstyle\times};
(5.108830,0.371104) *[red]{\scriptscriptstyle\times};
(5.117043,0.371206) *[red]{\scriptscriptstyle\times};
(5.125257,0.371863) *[red]{\scriptscriptstyle\times};
(5.133470,0.372195) *[red]{\scriptscriptstyle\times};
(5.141684,0.386272) *[red]{\scriptscriptstyle\times};
(5.149897,0.398796) *[red]{\scriptscriptstyle\times};
(5.158111,0.400493) *[red]{\scriptscriptstyle\times};
(5.166324,0.408857) *[red]{\scriptscriptstyle\times};
(5.174538,0.370278) *[red]{\scriptscriptstyle\times};
(5.182752,0.371349) *[red]{\scriptscriptstyle\times};
(5.190965,0.376825) *[red]{\scriptscriptstyle\times};
(5.199179,0.382296) *[red]{\scriptscriptstyle\times};
(5.207392,0.388684) *[red]{\scriptscriptstyle\times};
(5.215606,0.389902) *[red]{\scriptscriptstyle\times};
(5.223819,0.390992) *[red]{\scriptscriptstyle\times};
(5.232033,0.392030) *[red]{\scriptscriptstyle\times};
(5.240246,0.392432) *[red]{\scriptscriptstyle\times};
(5.248460,0.392810) *[red]{\scriptscriptstyle\times};
(5.256674,0.394539) *[red]{\scriptscriptstyle\times};
(5.264887,0.395194) *[red]{\scriptscriptstyle\times};
(5.273101,0.395921) *[red]{\scriptscriptstyle\times};
(5.281314,0.396873) *[red]{\scriptscriptstyle\times};
(5.289528,0.410182) *[red]{\scriptscriptstyle\times};
(5.297741,0.367954) *[red]{\scriptscriptstyle\times};
(5.305955,0.369518) *[red]{\scriptscriptstyle\times};
(5.314168,0.370685) *[red]{\scriptscriptstyle\times};
(5.322382,0.371201) *[red]{\scriptscriptstyle\times};
(5.330595,0.379469) *[red]{\scriptscriptstyle\times};
(5.338809,0.381368) *[red]{\scriptscriptstyle\times};
(5.347023,0.393905) *[red]{\scriptscriptstyle\times};
(5.355236,0.397113) *[red]{\scriptscriptstyle\times};
(5.363450,0.397493) *[red]{\scriptscriptstyle\times};
(5.371663,0.399458) *[red]{\scriptscriptstyle\times};
(5.379877,0.399732) *[red]{\scriptscriptstyle\times};
(5.388090,0.401308) *[red]{\scriptscriptstyle\times};
(5.396304,0.413555) *[red]{\scriptscriptstyle\times};
(5.404517,0.415830) *[red]{\scriptscriptstyle\times};
(5.412731,0.418938) *[red]{\scriptscriptstyle\times};
(5.420945,0.370484) *[red]{\scriptscriptstyle\times};
(5.429158,0.370640) *[red]{\scriptscriptstyle\times};
(5.437372,0.371139) *[red]{\scriptscriptstyle\times};
(5.445585,0.376485) *[red]{\scriptscriptstyle\times};
(5.453799,0.376856) *[red]{\scriptscriptstyle\times};
(5.462012,0.378354) *[red]{\scriptscriptstyle\times};
(5.470226,0.380006) *[red]{\scriptscriptstyle\times};
(5.478439,0.380477) *[red]{\scriptscriptstyle\times};
(5.486653,0.380907) *[red]{\scriptscriptstyle\times};
(5.494867,0.396039) *[red]{\scriptscriptstyle\times};
(5.503080,0.399488) *[red]{\scriptscriptstyle\times};
(5.511294,0.400071) *[red]{\scriptscriptstyle\times};
(5.519507,0.401488) *[red]{\scriptscriptstyle\times};
(5.527721,0.410455) *[red]{\scriptscriptstyle\times};
(5.535934,0.417054) *[red]{\scriptscriptstyle\times};
(5.544148,0.369036) *[red]{\scriptscriptstyle\times};
(5.552361,0.370486) *[red]{\scriptscriptstyle\times};
(5.560575,0.370800) *[red]{\scriptscriptstyle\times};
(5.568789,0.371771) *[red]{\scriptscriptstyle\times};
(5.577002,0.372529) *[red]{\scriptscriptstyle\times};
(5.585216,0.372850) *[red]{\scriptscriptstyle\times};
(5.593429,0.376370) *[red]{\scriptscriptstyle\times};
(5.601643,0.378398) *[red]{\scriptscriptstyle\times};
(5.609856,0.391513) *[red]{\scriptscriptstyle\times};
(5.618070,0.392411) *[red]{\scriptscriptstyle\times};
(5.626283,0.393968) *[red]{\scriptscriptstyle\times};
(5.634497,0.395438) *[red]{\scriptscriptstyle\times};
(5.642710,0.396392) *[red]{\scriptscriptstyle\times};
(5.650924,0.420558) *[red]{\scriptscriptstyle\times};
(5.659138,0.444120) *[red]{\scriptscriptstyle\times};
(5.667351,0.371535) *[red]{\scriptscriptstyle\times};
(5.675565,0.375731) *[red]{\scriptscriptstyle\times};
(5.683778,0.375981) *[red]{\scriptscriptstyle\times};
(5.691992,0.376289) *[red]{\scriptscriptstyle\times};
(5.700205,0.377707) *[red]{\scriptscriptstyle\times};
(5.708419,0.378863) *[red]{\scriptscriptstyle\times};
(5.716632,0.379127) *[red]{\scriptscriptstyle\times};
(5.724846,0.379715) *[red]{\scriptscriptstyle\times};
(5.733060,0.380844) *[red]{\scriptscriptstyle\times};
(5.741273,0.399769) *[red]{\scriptscriptstyle\times};
(5.749487,0.401828) *[red]{\scriptscriptstyle\times};
(5.757700,0.405970) *[red]{\scriptscriptstyle\times};
(5.765914,0.406744) *[red]{\scriptscriptstyle\times};
(5.774127,0.432149) *[red]{\scriptscriptstyle\times};
(5.782341,0.452864) *[red]{\scriptscriptstyle\times};
(5.790554,0.372034) *[red]{\scriptscriptstyle\times};
(5.798768,0.372090) *[red]{\scriptscriptstyle\times};
(5.806982,0.374398) *[red]{\scriptscriptstyle\times};
(5.815195,0.375995) *[red]{\scriptscriptstyle\times};
(5.823409,0.378592) *[red]{\scriptscriptstyle\times};
(5.831622,0.382101) *[red]{\scriptscriptstyle\times};
(5.839836,0.390545) *[red]{\scriptscriptstyle\times};
(5.848049,0.395835) *[red]{\scriptscriptstyle\times};
(5.856263,0.404636) *[red]{\scriptscriptstyle\times};
(5.864476,0.407152) *[red]{\scriptscriptstyle\times};
(5.872690,0.418600) *[red]{\scriptscriptstyle\times};
(5.880903,0.423267) *[red]{\scriptscriptstyle\times};
(5.889117,0.433332) *[red]{\scriptscriptstyle\times};
(5.897331,0.468324) *[red]{\scriptscriptstyle\times};
(5.905544,0.711929) *[red]{\scriptscriptstyle\times};
(5.913758,0.387616) *[red]{\scriptscriptstyle\times};
(5.921971,0.390031) *[red]{\scriptscriptstyle\times};
(5.930185,0.393774) *[red]{\scriptscriptstyle\times};
(5.938398,0.410113) *[red]{\scriptscriptstyle\times};
(5.946612,0.413859) *[red]{\scriptscriptstyle\times};
(5.954825,0.414674) *[red]{\scriptscriptstyle\times};
(5.963039,0.415232) *[red]{\scriptscriptstyle\times};
(5.971253,0.415700) *[red]{\scriptscriptstyle\times};
(5.979466,0.418230) *[red]{\scriptscriptstyle\times};
(5.987680,0.419821) *[red]{\scriptscriptstyle\times};
(5.995893,0.421295) *[red]{\scriptscriptstyle\times};
(6.004107,0.423776) *[red]{\scriptscriptstyle\times};
(6.012320,0.424424) *[red]{\scriptscriptstyle\times};
(6.020534,0.437125) *[red]{\scriptscriptstyle\times};
(6.028747,0.448790) *[red]{\scriptscriptstyle\times};
(6.036961,0.382660) *[red]{\scriptscriptstyle\times};
(6.045175,0.391359) *[red]{\scriptscriptstyle\times};
(6.053388,0.391672) *[red]{\scriptscriptstyle\times};
(6.061602,0.392159) *[red]{\scriptscriptstyle\times};
(6.069815,0.406143) *[red]{\scriptscriptstyle\times};
(6.078029,0.407796) *[red]{\scriptscriptstyle\times};
(6.086242,0.409187) *[red]{\scriptscriptstyle\times};
(6.094456,0.413397) *[red]{\scriptscriptstyle\times};
(6.102669,0.423366) *[red]{\scriptscriptstyle\times};
(6.110883,0.423770) *[red]{\scriptscriptstyle\times};
(6.119097,0.430618) *[red]{\scriptscriptstyle\times};
(6.127310,0.435145) *[red]{\scriptscriptstyle\times};
(6.135524,0.438419) *[red]{\scriptscriptstyle\times};
(6.143737,0.439806) *[red]{\scriptscriptstyle\times};
(6.151951,0.443754) *[red]{\scriptscriptstyle\times};
(6.160164,0.386163) *[red]{\scriptscriptstyle\times};
(6.168378,0.387051) *[red]{\scriptscriptstyle\times};
(6.176591,0.388498) *[red]{\scriptscriptstyle\times};
(6.184805,0.396796) *[red]{\scriptscriptstyle\times};
(6.193018,0.398316) *[red]{\scriptscriptstyle\times};
(6.201232,0.405809) *[red]{\scriptscriptstyle\times};
(6.209446,0.413816) *[red]{\scriptscriptstyle\times};
(6.217659,0.414220) *[red]{\scriptscriptstyle\times};
(6.225873,0.424086) *[red]{\scriptscriptstyle\times};
(6.234086,0.427294) *[red]{\scriptscriptstyle\times};
(6.242300,0.437141) *[red]{\scriptscriptstyle\times};
(6.250513,0.463594) *[red]{\scriptscriptstyle\times};
(6.258727,0.465280) *[red]{\scriptscriptstyle\times};
(6.266940,0.465709) *[red]{\scriptscriptstyle\times};
(6.275154,0.466134) *[red]{\scriptscriptstyle\times};
(6.283368,0.388795) *[red]{\scriptscriptstyle\times};
(6.291581,0.389953) *[red]{\scriptscriptstyle\times};
(6.299795,0.394164) *[red]{\scriptscriptstyle\times};
(6.308008,0.394391) *[red]{\scriptscriptstyle\times};
(6.316222,0.394394) *[red]{\scriptscriptstyle\times};
(6.324435,0.396031) *[red]{\scriptscriptstyle\times};
(6.332649,0.396434) *[red]{\scriptscriptstyle\times};
(6.340862,0.398236) *[red]{\scriptscriptstyle\times};
(6.349076,0.402302) *[red]{\scriptscriptstyle\times};
(6.357290,0.414648) *[red]{\scriptscriptstyle\times};
(6.365503,0.418574) *[red]{\scriptscriptstyle\times};
(6.373717,0.419054) *[red]{\scriptscriptstyle\times};
(6.381930,0.420588) *[red]{\scriptscriptstyle\times};
(6.390144,0.437072) *[red]{\scriptscriptstyle\times};
(6.398357,0.466869) *[red]{\scriptscriptstyle\times};
(6.406571,0.391656) *[red]{\scriptscriptstyle\times};
(6.414784,0.399280) *[red]{\scriptscriptstyle\times};
(6.422998,0.404489) *[red]{\scriptscriptstyle\times};
(6.431211,0.413790) *[red]{\scriptscriptstyle\times};
(6.439425,0.414512) *[red]{\scriptscriptstyle\times};
(6.447639,0.415899) *[red]{\scriptscriptstyle\times};
(6.455852,0.416888) *[red]{\scriptscriptstyle\times};
(6.464066,0.418708) *[red]{\scriptscriptstyle\times};
(6.472279,0.421336) *[red]{\scriptscriptstyle\times};
(6.480493,0.428977) *[red]{\scriptscriptstyle\times};
(6.488706,0.429548) *[red]{\scriptscriptstyle\times};
(6.496920,0.430734) *[red]{\scriptscriptstyle\times};
(6.505133,0.440572) *[red]{\scriptscriptstyle\times};
(6.513347,0.442169) *[red]{\scriptscriptstyle\times};
(6.521561,0.443725) *[red]{\scriptscriptstyle\times};
(6.529774,0.398445) *[red]{\scriptscriptstyle\times};
(6.537988,0.401953) *[red]{\scriptscriptstyle\times};
(6.546201,0.405597) *[red]{\scriptscriptstyle\times};
(6.554415,0.411871) *[red]{\scriptscriptstyle\times};
(6.562628,0.412864) *[red]{\scriptscriptstyle\times};
(6.570842,0.422340) *[red]{\scriptscriptstyle\times};
(6.579055,0.423697) *[red]{\scriptscriptstyle\times};
(6.587269,0.426410) *[red]{\scriptscriptstyle\times};
(6.595483,0.426743) *[red]{\scriptscriptstyle\times};
(6.603696,0.427437) *[red]{\scriptscriptstyle\times};
(6.611910,0.429678) *[red]{\scriptscriptstyle\times};
(6.620123,0.429936) *[red]{\scriptscriptstyle\times};
(6.628337,0.436620) *[red]{\scriptscriptstyle\times};
(6.636550,0.437758) *[red]{\scriptscriptstyle\times};
(6.644764,0.444520) *[red]{\scriptscriptstyle\times};
(6.652977,0.389320) *[red]{\scriptscriptstyle\times};
(6.661191,0.394537) *[red]{\scriptscriptstyle\times};
(6.669405,0.396815) *[red]{\scriptscriptstyle\times};
(6.677618,0.399747) *[red]{\scriptscriptstyle\times};
(6.685832,0.412418) *[red]{\scriptscriptstyle\times};
(6.694045,0.412625) *[red]{\scriptscriptstyle\times};
(6.702259,0.414969) *[red]{\scriptscriptstyle\times};
(6.710472,0.417816) *[red]{\scriptscriptstyle\times};
(6.718686,0.418910) *[red]{\scriptscriptstyle\times};
(6.726899,0.432905) *[red]{\scriptscriptstyle\times};
(6.735113,0.437385) *[red]{\scriptscriptstyle\times};
(6.743326,0.444587) *[red]{\scriptscriptstyle\times};
(6.751540,0.449290) *[red]{\scriptscriptstyle\times};
(6.759754,0.450092) *[red]{\scriptscriptstyle\times};
(6.767967,0.453777) *[red]{\scriptscriptstyle\times};
(6.776181,0.407013) *[red]{\scriptscriptstyle\times};
(6.784394,0.408911) *[red]{\scriptscriptstyle\times};
(6.792608,0.414367) *[red]{\scriptscriptstyle\times};
(6.800821,0.417090) *[red]{\scriptscriptstyle\times};
(6.809035,0.418103) *[red]{\scriptscriptstyle\times};
(6.817248,0.420323) *[red]{\scriptscriptstyle\times};
(6.825462,0.431952) *[red]{\scriptscriptstyle\times};
(6.833676,0.435529) *[red]{\scriptscriptstyle\times};
(6.841889,0.436658) *[red]{\scriptscriptstyle\times};
(6.850103,0.436812) *[red]{\scriptscriptstyle\times};
(6.858316,0.437596) *[red]{\scriptscriptstyle\times};
(6.866530,0.439698) *[red]{\scriptscriptstyle\times};
(6.874743,0.442727) *[red]{\scriptscriptstyle\times};
(6.882957,0.458907) *[red]{\scriptscriptstyle\times};
(6.891170,0.473233) *[red]{\scriptscriptstyle\times};
(6.899384,0.400668) *[red]{\scriptscriptstyle\times};
(6.907598,0.402527) *[red]{\scriptscriptstyle\times};
(6.915811,0.419626) *[red]{\scriptscriptstyle\times};
(6.924025,0.419798) *[red]{\scriptscriptstyle\times};
(6.932238,0.421670) *[red]{\scriptscriptstyle\times};
(6.940452,0.424833) *[red]{\scriptscriptstyle\times};
(6.948665,0.425385) *[red]{\scriptscriptstyle\times};
(6.956879,0.427024) *[red]{\scriptscriptstyle\times};
(6.965092,0.427756) *[red]{\scriptscriptstyle\times};
(6.973306,0.428772) *[red]{\scriptscriptstyle\times};
(6.981520,0.445235) *[red]{\scriptscriptstyle\times};
(6.989733,0.445805) *[red]{\scriptscriptstyle\times};
(6.997947,0.449314) *[red]{\scriptscriptstyle\times};
(7.006160,0.452530) *[red]{\scriptscriptstyle\times};
(7.014374,0.460361) *[red]{\scriptscriptstyle\times};
(7.022587,0.413457) *[red]{\scriptscriptstyle\times};
(7.030801,0.415394) *[red]{\scriptscriptstyle\times};
(7.039014,0.416597) *[red]{\scriptscriptstyle\times};
(7.047228,0.420646) *[red]{\scriptscriptstyle\times};
(7.055441,0.421664) *[red]{\scriptscriptstyle\times};
(7.063655,0.422467) *[red]{\scriptscriptstyle\times};
(7.071869,0.423119) *[red]{\scriptscriptstyle\times};
(7.080082,0.425967) *[red]{\scriptscriptstyle\times};
(7.088296,0.427125) *[red]{\scriptscriptstyle\times};
(7.096509,0.429985) *[red]{\scriptscriptstyle\times};
(7.104723,0.433716) *[red]{\scriptscriptstyle\times};
(7.112936,0.435697) *[red]{\scriptscriptstyle\times};
(7.121150,0.438749) *[red]{\scriptscriptstyle\times};
(7.129363,0.439862) *[red]{\scriptscriptstyle\times};
(7.137577,0.440172) *[red]{\scriptscriptstyle\times};
(7.145791,0.392677) *[red]{\scriptscriptstyle\times};
(7.154004,0.393486) *[red]{\scriptscriptstyle\times};
(7.162218,0.395180) *[red]{\scriptscriptstyle\times};
(7.170431,0.398212) *[red]{\scriptscriptstyle\times};
(7.178645,0.398758) *[red]{\scriptscriptstyle\times};
(7.186858,0.405539) *[red]{\scriptscriptstyle\times};
(7.195072,0.405958) *[red]{\scriptscriptstyle\times};
(7.203285,0.412285) *[red]{\scriptscriptstyle\times};
(7.211499,0.418116) *[red]{\scriptscriptstyle\times};
(7.219713,0.422182) *[red]{\scriptscriptstyle\times};
(7.227926,0.425543) *[red]{\scriptscriptstyle\times};
(7.236140,0.438237) *[red]{\scriptscriptstyle\times};
(7.244353,0.451984) *[red]{\scriptscriptstyle\times};
(7.252567,0.457672) *[red]{\scriptscriptstyle\times};
(7.260780,0.468001) *[red]{\scriptscriptstyle\times};
(7.268994,0.383757) *[red]{\scriptscriptstyle\times};
(7.277207,0.403672) *[red]{\scriptscriptstyle\times};
(7.285421,0.412797) *[red]{\scriptscriptstyle\times};
(7.293634,0.415275) *[red]{\scriptscriptstyle\times};
(7.301848,0.419008) *[red]{\scriptscriptstyle\times};
(7.310062,0.425553) *[red]{\scriptscriptstyle\times};
(7.318275,0.426794) *[red]{\scriptscriptstyle\times};
(7.326489,0.431280) *[red]{\scriptscriptstyle\times};
(7.334702,0.444389) *[red]{\scriptscriptstyle\times};
(7.342916,0.455712) *[red]{\scriptscriptstyle\times};
(7.351129,0.458772) *[red]{\scriptscriptstyle\times};
(7.359343,0.458837) *[red]{\scriptscriptstyle\times};
(7.367556,0.483862) *[red]{\scriptscriptstyle\times};
(7.375770,0.505260) *[red]{\scriptscriptstyle\times};
(7.383984,0.677831) *[red]{\scriptscriptstyle\times};
(7.392197,0.418313) *[red]{\scriptscriptstyle\times};
(7.400411,0.425907) *[red]{\scriptscriptstyle\times};
(7.408624,0.426773) *[red]{\scriptscriptstyle\times};
(7.416838,0.428501) *[red]{\scriptscriptstyle\times};
(7.425051,0.432090) *[red]{\scriptscriptstyle\times};
(7.433265,0.441002) *[red]{\scriptscriptstyle\times};
(7.441478,0.445026) *[red]{\scriptscriptstyle\times};
(7.449692,0.445457) *[red]{\scriptscriptstyle\times};
(7.457906,0.445508) *[red]{\scriptscriptstyle\times};
(7.466119,0.446795) *[red]{\scriptscriptstyle\times};
(7.474333,0.452449) *[red]{\scriptscriptstyle\times};
(7.482546,0.452819) *[red]{\scriptscriptstyle\times};
(7.490760,0.468301) *[red]{\scriptscriptstyle\times};
(7.498973,0.468784) *[red]{\scriptscriptstyle\times};
(7.507187,0.488800) *[red]{\scriptscriptstyle\times};
(7.515400,0.412956) *[red]{\scriptscriptstyle\times};
(7.523614,0.417833) *[red]{\scriptscriptstyle\times};
(7.531828,0.421494) *[red]{\scriptscriptstyle\times};
(7.540041,0.427812) *[red]{\scriptscriptstyle\times};
(7.548255,0.439581) *[red]{\scriptscriptstyle\times};
(7.556468,0.443045) *[red]{\scriptscriptstyle\times};
(7.564682,0.444955) *[red]{\scriptscriptstyle\times};
(7.572895,0.451088) *[red]{\scriptscriptstyle\times};
(7.581109,0.452444) *[red]{\scriptscriptstyle\times};
(7.589322,0.459097) *[red]{\scriptscriptstyle\times};
(7.597536,0.460160) *[red]{\scriptscriptstyle\times};
(7.605749,0.463915) *[red]{\scriptscriptstyle\times};
(7.613963,0.491258) *[red]{\scriptscriptstyle\times};
(7.622177,0.495830) *[red]{\scriptscriptstyle\times};
(7.630390,0.805591) *[red]{\scriptscriptstyle\times};
(7.638604,0.468025) *[red]{\scriptscriptstyle\times};
(7.646817,0.468106) *[red]{\scriptscriptstyle\times};
(7.655031,0.470689) *[red]{\scriptscriptstyle\times};
(7.663244,0.471841) *[red]{\scriptscriptstyle\times};
(7.671458,0.472446) *[red]{\scriptscriptstyle\times};
(7.679671,0.474971) *[red]{\scriptscriptstyle\times};
(7.687885,0.476687) *[red]{\scriptscriptstyle\times};
(7.696099,0.476729) *[red]{\scriptscriptstyle\times};
(7.704312,0.476777) *[red]{\scriptscriptstyle\times};
(7.712526,0.478885) *[red]{\scriptscriptstyle\times};
(7.720739,0.481318) *[red]{\scriptscriptstyle\times};
(7.728953,0.481871) *[red]{\scriptscriptstyle\times};
(7.737166,0.492228) *[red]{\scriptscriptstyle\times};
(7.745380,0.493448) *[red]{\scriptscriptstyle\times};
(7.753593,0.497134) *[red]{\scriptscriptstyle\times};
(7.761807,0.464132) *[red]{\scriptscriptstyle\times};
(7.770021,0.466044) *[red]{\scriptscriptstyle\times};
(7.778234,0.467046) *[red]{\scriptscriptstyle\times};
(7.786448,0.468685) *[red]{\scriptscriptstyle\times};
(7.794661,0.471160) *[red]{\scriptscriptstyle\times};
(7.802875,0.476026) *[red]{\scriptscriptstyle\times};
(7.811088,0.477780) *[red]{\scriptscriptstyle\times};
(7.819302,0.486682) *[red]{\scriptscriptstyle\times};
(7.827515,0.489440) *[red]{\scriptscriptstyle\times};
(7.835729,0.492795) *[red]{\scriptscriptstyle\times};
(7.843943,0.495979) *[red]{\scriptscriptstyle\times};
(7.852156,0.496107) *[red]{\scriptscriptstyle\times};
(7.860370,0.504301) *[red]{\scriptscriptstyle\times};
(7.868583,0.508934) *[red]{\scriptscriptstyle\times};
(7.876797,0.518246) *[red]{\scriptscriptstyle\times};
(7.885010,0.543584) *[red]{\scriptscriptstyle\times};
(7.893224,0.545427) *[red]{\scriptscriptstyle\times};
(7.901437,0.551245) *[red]{\scriptscriptstyle\times};
(7.909651,0.556996) *[red]{\scriptscriptstyle\times};
(7.917864,0.557681) *[red]{\scriptscriptstyle\times};
(7.926078,0.559563) *[red]{\scriptscriptstyle\times};
(7.934292,0.561378) *[red]{\scriptscriptstyle\times};
(7.942505,0.562968) *[red]{\scriptscriptstyle\times};
(7.950719,0.569079) *[red]{\scriptscriptstyle\times};
(7.958932,0.569610) *[red]{\scriptscriptstyle\times};
(7.967146,0.570452) *[red]{\scriptscriptstyle\times};
(7.975359,0.576109) *[red]{\scriptscriptstyle\times};
(7.983573,0.578686) *[red]{\scriptscriptstyle\times};
(7.991786,0.584026) *[red]{\scriptscriptstyle\times};
(8.000000,0.592403) *[red]{\scriptscriptstyle\times};
\endxy
  }
  \caption{Skylake cycles
    for the CSURF-512 action
    using {\tt velusqrt-flint}.
  }
  \label{fig:csurf-c}
\end{figure}

\begin{figure}[t]
\centerline{
\xy <1.1cm,0cm>:<0cm,4cm>::
(0,1.205563); (8,1.205563) **[blue]@{-};
(8.1,1.205563) *[blue]{\rlap{115313652}};
(0,1.279904); (8,1.279904) **[blue]@{-};
(8.1,1.279904) *[blue]{\rlap{121411384}};
(0,1.361763); (8,1.361763) **[blue]@{-};
(8.1,1.361763) *[blue]{\rlap{128499534}};
(0,1.184947); (8,1.184947) **[red]@{-};
(-0.1,1.184947) *[red]{\llap{113677484}};
(0,1.262986); (8,1.262986) **[red]@{-};
(-0.1,1.262986) *[red]{\llap{119995936}};
(0,1.343815); (8,1.343815) **[red]@{-};
(-0.1,1.343815) *[red]{\llap{126910804}};
(-0.004107,0.978379); (-0.004107,1.676527) **[lightgray]@{-};
(0.119097,0.978379); (0.119097,1.676527) **[lightgray]@{-};
(0.242300,0.978379); (0.242300,1.676527) **[lightgray]@{-};
(0.365503,0.978379); (0.365503,1.676527) **[lightgray]@{-};
(0.488706,0.978379); (0.488706,1.676527) **[lightgray]@{-};
(0.611910,0.978379); (0.611910,1.676527) **[lightgray]@{-};
(0.735113,0.978379); (0.735113,1.676527) **[lightgray]@{-};
(0.858316,0.978379); (0.858316,1.676527) **[lightgray]@{-};
(0.981520,0.978379); (0.981520,1.676527) **[lightgray]@{-};
(1.104723,0.978379); (1.104723,1.676527) **[lightgray]@{-};
(1.227926,0.978379); (1.227926,1.676527) **[lightgray]@{-};
(1.351129,0.978379); (1.351129,1.676527) **[lightgray]@{-};
(1.474333,0.978379); (1.474333,1.676527) **[lightgray]@{-};
(1.597536,0.978379); (1.597536,1.676527) **[lightgray]@{-};
(1.720739,0.978379); (1.720739,1.676527) **[lightgray]@{-};
(1.843943,0.978379); (1.843943,1.676527) **[lightgray]@{-};
(1.967146,0.978379); (1.967146,1.676527) **[lightgray]@{-};
(2.090349,0.978379); (2.090349,1.676527) **[lightgray]@{-};
(2.213552,0.978379); (2.213552,1.676527) **[lightgray]@{-};
(2.336756,0.978379); (2.336756,1.676527) **[lightgray]@{-};
(2.459959,0.978379); (2.459959,1.676527) **[lightgray]@{-};
(2.583162,0.978379); (2.583162,1.676527) **[lightgray]@{-};
(2.706366,0.978379); (2.706366,1.676527) **[lightgray]@{-};
(2.829569,0.978379); (2.829569,1.676527) **[lightgray]@{-};
(2.952772,0.978379); (2.952772,1.676527) **[lightgray]@{-};
(3.075975,0.978379); (3.075975,1.676527) **[lightgray]@{-};
(3.199179,0.978379); (3.199179,1.676527) **[lightgray]@{-};
(3.322382,0.978379); (3.322382,1.676527) **[lightgray]@{-};
(3.445585,0.978379); (3.445585,1.676527) **[lightgray]@{-};
(3.568789,0.978379); (3.568789,1.676527) **[lightgray]@{-};
(3.691992,0.978379); (3.691992,1.676527) **[lightgray]@{-};
(3.815195,0.978379); (3.815195,1.676527) **[lightgray]@{-};
(3.938398,0.978379); (3.938398,1.676527) **[lightgray]@{-};
(4.061602,0.978379); (4.061602,1.676527) **[lightgray]@{-};
(4.184805,0.978379); (4.184805,1.676527) **[lightgray]@{-};
(4.308008,0.978379); (4.308008,1.676527) **[lightgray]@{-};
(4.431211,0.978379); (4.431211,1.676527) **[lightgray]@{-};
(4.554415,0.978379); (4.554415,1.676527) **[lightgray]@{-};
(4.677618,0.978379); (4.677618,1.676527) **[lightgray]@{-};
(4.800821,0.978379); (4.800821,1.676527) **[lightgray]@{-};
(4.924025,0.978379); (4.924025,1.676527) **[lightgray]@{-};
(5.047228,0.978379); (5.047228,1.676527) **[lightgray]@{-};
(5.170431,0.978379); (5.170431,1.676527) **[lightgray]@{-};
(5.293634,0.978379); (5.293634,1.676527) **[lightgray]@{-};
(5.416838,0.978379); (5.416838,1.676527) **[lightgray]@{-};
(5.540041,0.978379); (5.540041,1.676527) **[lightgray]@{-};
(5.663244,0.978379); (5.663244,1.676527) **[lightgray]@{-};
(5.786448,0.978379); (5.786448,1.676527) **[lightgray]@{-};
(5.909651,0.978379); (5.909651,1.676527) **[lightgray]@{-};
(6.032854,0.978379); (6.032854,1.676527) **[lightgray]@{-};
(6.156057,0.978379); (6.156057,1.676527) **[lightgray]@{-};
(6.279261,0.978379); (6.279261,1.676527) **[lightgray]@{-};
(6.402464,0.978379); (6.402464,1.676527) **[lightgray]@{-};
(6.525667,0.978379); (6.525667,1.676527) **[lightgray]@{-};
(6.648871,0.978379); (6.648871,1.676527) **[lightgray]@{-};
(6.772074,0.978379); (6.772074,1.676527) **[lightgray]@{-};
(6.895277,0.978379); (6.895277,1.676527) **[lightgray]@{-};
(7.018480,0.978379); (7.018480,1.676527) **[lightgray]@{-};
(7.141684,0.978379); (7.141684,1.676527) **[lightgray]@{-};
(7.264887,0.978379); (7.264887,1.676527) **[lightgray]@{-};
(7.388090,0.978379); (7.388090,1.676527) **[lightgray]@{-};
(7.511294,0.978379); (7.511294,1.676527) **[lightgray]@{-};
(7.634497,0.978379); (7.634497,1.676527) **[lightgray]@{-};
(7.757700,0.978379); (7.757700,1.676527) **[lightgray]@{-};
(7.880903,0.978379); (7.880903,1.676527) **[lightgray]@{-};
(8.004107,0.978379); (8.004107,1.676527) **[lightgray]@{-};
(-0.004107,0.978379); (-0.004107,1.676527) **[lightgray]@{-};
(0.119097,0.978379); (0.119097,1.676527) **[lightgray]@{-};
(0.242300,0.978379); (0.242300,1.676527) **[lightgray]@{-};
(0.365503,0.978379); (0.365503,1.676527) **[lightgray]@{-};
(0.488706,0.978379); (0.488706,1.676527) **[lightgray]@{-};
(0.611910,0.978379); (0.611910,1.676527) **[lightgray]@{-};
(0.735113,0.978379); (0.735113,1.676527) **[lightgray]@{-};
(0.858316,0.978379); (0.858316,1.676527) **[lightgray]@{-};
(0.981520,0.978379); (0.981520,1.676527) **[lightgray]@{-};
(1.104723,0.978379); (1.104723,1.676527) **[lightgray]@{-};
(1.227926,0.978379); (1.227926,1.676527) **[lightgray]@{-};
(1.351129,0.978379); (1.351129,1.676527) **[lightgray]@{-};
(1.474333,0.978379); (1.474333,1.676527) **[lightgray]@{-};
(1.597536,0.978379); (1.597536,1.676527) **[lightgray]@{-};
(1.720739,0.978379); (1.720739,1.676527) **[lightgray]@{-};
(1.843943,0.978379); (1.843943,1.676527) **[lightgray]@{-};
(1.967146,0.978379); (1.967146,1.676527) **[lightgray]@{-};
(2.090349,0.978379); (2.090349,1.676527) **[lightgray]@{-};
(2.213552,0.978379); (2.213552,1.676527) **[lightgray]@{-};
(2.336756,0.978379); (2.336756,1.676527) **[lightgray]@{-};
(2.459959,0.978379); (2.459959,1.676527) **[lightgray]@{-};
(2.583162,0.978379); (2.583162,1.676527) **[lightgray]@{-};
(2.706366,0.978379); (2.706366,1.676527) **[lightgray]@{-};
(2.829569,0.978379); (2.829569,1.676527) **[lightgray]@{-};
(2.952772,0.978379); (2.952772,1.676527) **[lightgray]@{-};
(3.075975,0.978379); (3.075975,1.676527) **[lightgray]@{-};
(3.199179,0.978379); (3.199179,1.676527) **[lightgray]@{-};
(3.322382,0.978379); (3.322382,1.676527) **[lightgray]@{-};
(3.445585,0.978379); (3.445585,1.676527) **[lightgray]@{-};
(3.568789,0.978379); (3.568789,1.676527) **[lightgray]@{-};
(3.691992,0.978379); (3.691992,1.676527) **[lightgray]@{-};
(3.815195,0.978379); (3.815195,1.676527) **[lightgray]@{-};
(3.938398,0.978379); (3.938398,1.676527) **[lightgray]@{-};
(4.061602,0.978379); (4.061602,1.676527) **[lightgray]@{-};
(4.184805,0.978379); (4.184805,1.676527) **[lightgray]@{-};
(4.308008,0.978379); (4.308008,1.676527) **[lightgray]@{-};
(4.431211,0.978379); (4.431211,1.676527) **[lightgray]@{-};
(4.554415,0.978379); (4.554415,1.676527) **[lightgray]@{-};
(4.677618,0.978379); (4.677618,1.676527) **[lightgray]@{-};
(4.800821,0.978379); (4.800821,1.676527) **[lightgray]@{-};
(4.924025,0.978379); (4.924025,1.676527) **[lightgray]@{-};
(5.047228,0.978379); (5.047228,1.676527) **[lightgray]@{-};
(5.170431,0.978379); (5.170431,1.676527) **[lightgray]@{-};
(5.293634,0.978379); (5.293634,1.676527) **[lightgray]@{-};
(5.416838,0.978379); (5.416838,1.676527) **[lightgray]@{-};
(5.540041,0.978379); (5.540041,1.676527) **[lightgray]@{-};
(5.663244,0.978379); (5.663244,1.676527) **[lightgray]@{-};
(5.786448,0.978379); (5.786448,1.676527) **[lightgray]@{-};
(5.909651,0.978379); (5.909651,1.676527) **[lightgray]@{-};
(6.032854,0.978379); (6.032854,1.676527) **[lightgray]@{-};
(6.156057,0.978379); (6.156057,1.676527) **[lightgray]@{-};
(6.279261,0.978379); (6.279261,1.676527) **[lightgray]@{-};
(6.402464,0.978379); (6.402464,1.676527) **[lightgray]@{-};
(6.525667,0.978379); (6.525667,1.676527) **[lightgray]@{-};
(6.648871,0.978379); (6.648871,1.676527) **[lightgray]@{-};
(6.772074,0.978379); (6.772074,1.676527) **[lightgray]@{-};
(6.895277,0.978379); (6.895277,1.676527) **[lightgray]@{-};
(7.018480,0.978379); (7.018480,1.676527) **[lightgray]@{-};
(7.141684,0.978379); (7.141684,1.676527) **[lightgray]@{-};
(7.264887,0.978379); (7.264887,1.676527) **[lightgray]@{-};
(7.388090,0.978379); (7.388090,1.676527) **[lightgray]@{-};
(7.511294,0.978379); (7.511294,1.676527) **[lightgray]@{-};
(7.634497,0.978379); (7.634497,1.676527) **[lightgray]@{-};
(7.757700,0.978379); (7.757700,1.676527) **[lightgray]@{-};
(7.880903,0.978379); (7.880903,1.676527) **[lightgray]@{-};
(8.004107,0.978379); (8.004107,1.676527) **[lightgray]@{-};
(0.000000,1.016639) *[blue]{\scriptscriptstyle+};
(0.008214,1.020422) *[blue]{\scriptscriptstyle+};
(0.016427,1.021154) *[blue]{\scriptscriptstyle+};
(0.024641,1.023992) *[blue]{\scriptscriptstyle+};
(0.032854,1.032530) *[blue]{\scriptscriptstyle+};
(0.041068,1.035310) *[blue]{\scriptscriptstyle+};
(0.049281,1.046346) *[blue]{\scriptscriptstyle+};
(0.057495,1.048383) *[blue]{\scriptscriptstyle+};
(0.065708,1.051206) *[blue]{\scriptscriptstyle+};
(0.073922,1.060484) *[blue]{\scriptscriptstyle+};
(0.082136,1.063728) *[blue]{\scriptscriptstyle+};
(0.090349,1.071510) *[blue]{\scriptscriptstyle+};
(0.098563,1.084226) *[blue]{\scriptscriptstyle+};
(0.106776,1.085705) *[blue]{\scriptscriptstyle+};
(0.114990,1.087449) *[blue]{\scriptscriptstyle+};
(0.123203,1.024299) *[blue]{\scriptscriptstyle+};
(0.131417,1.024832) *[blue]{\scriptscriptstyle+};
(0.139630,1.025210) *[blue]{\scriptscriptstyle+};
(0.147844,1.026007) *[blue]{\scriptscriptstyle+};
(0.156057,1.029313) *[blue]{\scriptscriptstyle+};
(0.164271,1.029782) *[blue]{\scriptscriptstyle+};
(0.172485,1.032225) *[blue]{\scriptscriptstyle+};
(0.180698,1.055975) *[blue]{\scriptscriptstyle+};
(0.188912,1.056677) *[blue]{\scriptscriptstyle+};
(0.197125,1.060691) *[blue]{\scriptscriptstyle+};
(0.205339,1.061909) *[blue]{\scriptscriptstyle+};
(0.213552,1.062376) *[blue]{\scriptscriptstyle+};
(0.221766,1.074535) *[blue]{\scriptscriptstyle+};
(0.229979,1.076124) *[blue]{\scriptscriptstyle+};
(0.238193,1.086662) *[blue]{\scriptscriptstyle+};
(0.246407,1.059697) *[blue]{\scriptscriptstyle+};
(0.254620,1.064794) *[blue]{\scriptscriptstyle+};
(0.262834,1.076528) *[blue]{\scriptscriptstyle+};
(0.271047,1.085973) *[blue]{\scriptscriptstyle+};
(0.279261,1.086638) *[blue]{\scriptscriptstyle+};
(0.287474,1.087101) *[blue]{\scriptscriptstyle+};
(0.295688,1.087277) *[blue]{\scriptscriptstyle+};
(0.303901,1.089439) *[blue]{\scriptscriptstyle+};
(0.312115,1.090420) *[blue]{\scriptscriptstyle+};
(0.320329,1.090503) *[blue]{\scriptscriptstyle+};
(0.328542,1.091289) *[blue]{\scriptscriptstyle+};
(0.336756,1.097834) *[blue]{\scriptscriptstyle+};
(0.344969,1.108229) *[blue]{\scriptscriptstyle+};
(0.353183,1.119883) *[blue]{\scriptscriptstyle+};
(0.361396,1.135146) *[blue]{\scriptscriptstyle+};
(0.369610,1.066932) *[blue]{\scriptscriptstyle+};
(0.377823,1.067488) *[blue]{\scriptscriptstyle+};
(0.386037,1.072249) *[blue]{\scriptscriptstyle+};
(0.394251,1.092332) *[blue]{\scriptscriptstyle+};
(0.402464,1.097012) *[blue]{\scriptscriptstyle+};
(0.410678,1.097871) *[blue]{\scriptscriptstyle+};
(0.418891,1.104846) *[blue]{\scriptscriptstyle+};
(0.427105,1.106237) *[blue]{\scriptscriptstyle+};
(0.435318,1.107627) *[blue]{\scriptscriptstyle+};
(0.443532,1.112515) *[blue]{\scriptscriptstyle+};
(0.451745,1.115236) *[blue]{\scriptscriptstyle+};
(0.459959,1.125597) *[blue]{\scriptscriptstyle+};
(0.468172,1.129387) *[blue]{\scriptscriptstyle+};
(0.476386,1.143840) *[blue]{\scriptscriptstyle+};
(0.484600,1.164646) *[blue]{\scriptscriptstyle+};
(0.492813,1.043397) *[blue]{\scriptscriptstyle+};
(0.501027,1.049245) *[blue]{\scriptscriptstyle+};
(0.509240,1.072193) *[blue]{\scriptscriptstyle+};
(0.517454,1.073257) *[blue]{\scriptscriptstyle+};
(0.525667,1.078625) *[blue]{\scriptscriptstyle+};
(0.533881,1.097238) *[blue]{\scriptscriptstyle+};
(0.542094,1.109361) *[blue]{\scriptscriptstyle+};
(0.550308,1.123262) *[blue]{\scriptscriptstyle+};
(0.558522,1.124475) *[blue]{\scriptscriptstyle+};
(0.566735,1.128030) *[blue]{\scriptscriptstyle+};
(0.574949,1.134886) *[blue]{\scriptscriptstyle+};
(0.583162,1.137001) *[blue]{\scriptscriptstyle+};
(0.591376,1.182086) *[blue]{\scriptscriptstyle+};
(0.599589,1.233635) *[blue]{\scriptscriptstyle+};
(0.607803,1.251231) *[blue]{\scriptscriptstyle+};
(0.616016,1.063392) *[blue]{\scriptscriptstyle+};
(0.624230,1.083074) *[blue]{\scriptscriptstyle+};
(0.632444,1.085576) *[blue]{\scriptscriptstyle+};
(0.640657,1.085926) *[blue]{\scriptscriptstyle+};
(0.648871,1.087906) *[blue]{\scriptscriptstyle+};
(0.657084,1.103574) *[blue]{\scriptscriptstyle+};
(0.665298,1.106159) *[blue]{\scriptscriptstyle+};
(0.673511,1.110480) *[blue]{\scriptscriptstyle+};
(0.681725,1.111315) *[blue]{\scriptscriptstyle+};
(0.689938,1.120249) *[blue]{\scriptscriptstyle+};
(0.698152,1.122716) *[blue]{\scriptscriptstyle+};
(0.706366,1.122846) *[blue]{\scriptscriptstyle+};
(0.714579,1.126624) *[blue]{\scriptscriptstyle+};
(0.722793,1.173277) *[blue]{\scriptscriptstyle+};
(0.731006,1.175924) *[blue]{\scriptscriptstyle+};
(0.739220,1.092753) *[blue]{\scriptscriptstyle+};
(0.747433,1.093539) *[blue]{\scriptscriptstyle+};
(0.755647,1.097669) *[blue]{\scriptscriptstyle+};
(0.763860,1.098078) *[blue]{\scriptscriptstyle+};
(0.772074,1.100738) *[blue]{\scriptscriptstyle+};
(0.780287,1.101129) *[blue]{\scriptscriptstyle+};
(0.788501,1.104083) *[blue]{\scriptscriptstyle+};
(0.796715,1.107925) *[blue]{\scriptscriptstyle+};
(0.804928,1.110985) *[blue]{\scriptscriptstyle+};
(0.813142,1.125756) *[blue]{\scriptscriptstyle+};
(0.821355,1.129003) *[blue]{\scriptscriptstyle+};
(0.829569,1.130679) *[blue]{\scriptscriptstyle+};
(0.837782,1.134867) *[blue]{\scriptscriptstyle+};
(0.845996,1.137529) *[blue]{\scriptscriptstyle+};
(0.854209,1.151371) *[blue]{\scriptscriptstyle+};
(0.862423,1.113684) *[blue]{\scriptscriptstyle+};
(0.870637,1.118107) *[blue]{\scriptscriptstyle+};
(0.878850,1.118329) *[blue]{\scriptscriptstyle+};
(0.887064,1.122377) *[blue]{\scriptscriptstyle+};
(0.895277,1.125283) *[blue]{\scriptscriptstyle+};
(0.903491,1.135337) *[blue]{\scriptscriptstyle+};
(0.911704,1.136918) *[blue]{\scriptscriptstyle+};
(0.919918,1.141025) *[blue]{\scriptscriptstyle+};
(0.928131,1.141031) *[blue]{\scriptscriptstyle+};
(0.936345,1.143553) *[blue]{\scriptscriptstyle+};
(0.944559,1.144161) *[blue]{\scriptscriptstyle+};
(0.952772,1.145389) *[blue]{\scriptscriptstyle+};
(0.960986,1.147251) *[blue]{\scriptscriptstyle+};
(0.969199,1.153898) *[blue]{\scriptscriptstyle+};
(0.977413,1.167716) *[blue]{\scriptscriptstyle+};
(0.985626,1.119066) *[blue]{\scriptscriptstyle+};
(0.993840,1.124409) *[blue]{\scriptscriptstyle+};
(1.002053,1.125333) *[blue]{\scriptscriptstyle+};
(1.010267,1.129511) *[blue]{\scriptscriptstyle+};
(1.018480,1.134857) *[blue]{\scriptscriptstyle+};
(1.026694,1.136289) *[blue]{\scriptscriptstyle+};
(1.034908,1.138755) *[blue]{\scriptscriptstyle+};
(1.043121,1.140737) *[blue]{\scriptscriptstyle+};
(1.051335,1.150823) *[blue]{\scriptscriptstyle+};
(1.059548,1.151527) *[blue]{\scriptscriptstyle+};
(1.067762,1.152305) *[blue]{\scriptscriptstyle+};
(1.075975,1.157188) *[blue]{\scriptscriptstyle+};
(1.084189,1.178698) *[blue]{\scriptscriptstyle+};
(1.092402,1.180261) *[blue]{\scriptscriptstyle+};
(1.100616,1.199884) *[blue]{\scriptscriptstyle+};
(1.108830,1.129189) *[blue]{\scriptscriptstyle+};
(1.117043,1.130878) *[blue]{\scriptscriptstyle+};
(1.125257,1.134148) *[blue]{\scriptscriptstyle+};
(1.133470,1.135512) *[blue]{\scriptscriptstyle+};
(1.141684,1.139027) *[blue]{\scriptscriptstyle+};
(1.149897,1.139348) *[blue]{\scriptscriptstyle+};
(1.158111,1.152857) *[blue]{\scriptscriptstyle+};
(1.166324,1.156843) *[blue]{\scriptscriptstyle+};
(1.174538,1.157755) *[blue]{\scriptscriptstyle+};
(1.182752,1.161091) *[blue]{\scriptscriptstyle+};
(1.190965,1.163942) *[blue]{\scriptscriptstyle+};
(1.199179,1.169327) *[blue]{\scriptscriptstyle+};
(1.207392,1.185349) *[blue]{\scriptscriptstyle+};
(1.215606,1.185565) *[blue]{\scriptscriptstyle+};
(1.223819,1.248170) *[blue]{\scriptscriptstyle+};
(1.232033,1.140242) *[blue]{\scriptscriptstyle+};
(1.240246,1.141082) *[blue]{\scriptscriptstyle+};
(1.248460,1.143830) *[blue]{\scriptscriptstyle+};
(1.256674,1.146256) *[blue]{\scriptscriptstyle+};
(1.264887,1.146943) *[blue]{\scriptscriptstyle+};
(1.273101,1.149706) *[blue]{\scriptscriptstyle+};
(1.281314,1.151843) *[blue]{\scriptscriptstyle+};
(1.289528,1.153117) *[blue]{\scriptscriptstyle+};
(1.297741,1.156742) *[blue]{\scriptscriptstyle+};
(1.305955,1.176736) *[blue]{\scriptscriptstyle+};
(1.314168,1.177447) *[blue]{\scriptscriptstyle+};
(1.322382,1.183954) *[blue]{\scriptscriptstyle+};
(1.330595,1.205269) *[blue]{\scriptscriptstyle+};
(1.338809,1.214024) *[blue]{\scriptscriptstyle+};
(1.347023,1.223494) *[blue]{\scriptscriptstyle+};
(1.355236,1.113465) *[blue]{\scriptscriptstyle+};
(1.363450,1.136074) *[blue]{\scriptscriptstyle+};
(1.371663,1.139444) *[blue]{\scriptscriptstyle+};
(1.379877,1.142873) *[blue]{\scriptscriptstyle+};
(1.388090,1.143816) *[blue]{\scriptscriptstyle+};
(1.396304,1.146630) *[blue]{\scriptscriptstyle+};
(1.404517,1.147419) *[blue]{\scriptscriptstyle+};
(1.412731,1.150232) *[blue]{\scriptscriptstyle+};
(1.420945,1.155004) *[blue]{\scriptscriptstyle+};
(1.429158,1.163709) *[blue]{\scriptscriptstyle+};
(1.437372,1.167933) *[blue]{\scriptscriptstyle+};
(1.445585,1.197104) *[blue]{\scriptscriptstyle+};
(1.453799,1.203785) *[blue]{\scriptscriptstyle+};
(1.462012,1.210695) *[blue]{\scriptscriptstyle+};
(1.470226,1.298060) *[blue]{\scriptscriptstyle+};
(1.478439,1.146355) *[blue]{\scriptscriptstyle+};
(1.486653,1.147161) *[blue]{\scriptscriptstyle+};
(1.494867,1.147684) *[blue]{\scriptscriptstyle+};
(1.503080,1.150725) *[blue]{\scriptscriptstyle+};
(1.511294,1.156373) *[blue]{\scriptscriptstyle+};
(1.519507,1.160590) *[blue]{\scriptscriptstyle+};
(1.527721,1.164376) *[blue]{\scriptscriptstyle+};
(1.535934,1.174986) *[blue]{\scriptscriptstyle+};
(1.544148,1.177835) *[blue]{\scriptscriptstyle+};
(1.552361,1.177939) *[blue]{\scriptscriptstyle+};
(1.560575,1.180303) *[blue]{\scriptscriptstyle+};
(1.568789,1.181454) *[blue]{\scriptscriptstyle+};
(1.577002,1.182262) *[blue]{\scriptscriptstyle+};
(1.585216,1.182314) *[blue]{\scriptscriptstyle+};
(1.593429,1.182793) *[blue]{\scriptscriptstyle+};
(1.601643,1.147997) *[blue]{\scriptscriptstyle+};
(1.609856,1.149420) *[blue]{\scriptscriptstyle+};
(1.618070,1.150300) *[blue]{\scriptscriptstyle+};
(1.626283,1.153454) *[blue]{\scriptscriptstyle+};
(1.634497,1.153564) *[blue]{\scriptscriptstyle+};
(1.642710,1.159357) *[blue]{\scriptscriptstyle+};
(1.650924,1.175496) *[blue]{\scriptscriptstyle+};
(1.659138,1.187338) *[blue]{\scriptscriptstyle+};
(1.667351,1.203485) *[blue]{\scriptscriptstyle+};
(1.675565,1.203769) *[blue]{\scriptscriptstyle+};
(1.683778,1.209658) *[blue]{\scriptscriptstyle+};
(1.691992,1.214620) *[blue]{\scriptscriptstyle+};
(1.700205,1.215053) *[blue]{\scriptscriptstyle+};
(1.708419,1.227192) *[blue]{\scriptscriptstyle+};
(1.716632,1.273583) *[blue]{\scriptscriptstyle+};
(1.724846,1.175019) *[blue]{\scriptscriptstyle+};
(1.733060,1.175599) *[blue]{\scriptscriptstyle+};
(1.741273,1.178377) *[blue]{\scriptscriptstyle+};
(1.749487,1.182920) *[blue]{\scriptscriptstyle+};
(1.757700,1.187703) *[blue]{\scriptscriptstyle+};
(1.765914,1.194868) *[blue]{\scriptscriptstyle+};
(1.774127,1.197742) *[blue]{\scriptscriptstyle+};
(1.782341,1.201617) *[blue]{\scriptscriptstyle+};
(1.790554,1.203031) *[blue]{\scriptscriptstyle+};
(1.798768,1.206339) *[blue]{\scriptscriptstyle+};
(1.806982,1.206618) *[blue]{\scriptscriptstyle+};
(1.815195,1.207734) *[blue]{\scriptscriptstyle+};
(1.823409,1.213307) *[blue]{\scriptscriptstyle+};
(1.831622,1.222050) *[blue]{\scriptscriptstyle+};
(1.839836,1.226581) *[blue]{\scriptscriptstyle+};
(1.848049,1.139772) *[blue]{\scriptscriptstyle+};
(1.856263,1.142305) *[blue]{\scriptscriptstyle+};
(1.864476,1.170149) *[blue]{\scriptscriptstyle+};
(1.872690,1.171106) *[blue]{\scriptscriptstyle+};
(1.880903,1.171878) *[blue]{\scriptscriptstyle+};
(1.889117,1.172407) *[blue]{\scriptscriptstyle+};
(1.897331,1.173268) *[blue]{\scriptscriptstyle+};
(1.905544,1.196195) *[blue]{\scriptscriptstyle+};
(1.913758,1.200170) *[blue]{\scriptscriptstyle+};
(1.921971,1.206376) *[blue]{\scriptscriptstyle+};
(1.930185,1.210264) *[blue]{\scriptscriptstyle+};
(1.938398,1.211766) *[blue]{\scriptscriptstyle+};
(1.946612,1.219137) *[blue]{\scriptscriptstyle+};
(1.954825,1.228631) *[blue]{\scriptscriptstyle+};
(1.963039,1.241784) *[blue]{\scriptscriptstyle+};
(1.971253,1.189009) *[blue]{\scriptscriptstyle+};
(1.979466,1.190204) *[blue]{\scriptscriptstyle+};
(1.987680,1.191583) *[blue]{\scriptscriptstyle+};
(1.995893,1.192812) *[blue]{\scriptscriptstyle+};
(2.004107,1.193215) *[blue]{\scriptscriptstyle+};
(2.012320,1.212214) *[blue]{\scriptscriptstyle+};
(2.020534,1.213305) *[blue]{\scriptscriptstyle+};
(2.028747,1.215039) *[blue]{\scriptscriptstyle+};
(2.036961,1.215612) *[blue]{\scriptscriptstyle+};
(2.045175,1.236558) *[blue]{\scriptscriptstyle+};
(2.053388,1.240808) *[blue]{\scriptscriptstyle+};
(2.061602,1.243574) *[blue]{\scriptscriptstyle+};
(2.069815,1.248202) *[blue]{\scriptscriptstyle+};
(2.078029,1.249351) *[blue]{\scriptscriptstyle+};
(2.086242,1.255309) *[blue]{\scriptscriptstyle+};
(2.094456,1.194558) *[blue]{\scriptscriptstyle+};
(2.102669,1.197555) *[blue]{\scriptscriptstyle+};
(2.110883,1.199844) *[blue]{\scriptscriptstyle+};
(2.119097,1.201074) *[blue]{\scriptscriptstyle+};
(2.127310,1.205563) *[blue]{\scriptscriptstyle+};
(2.135524,1.205756) *[blue]{\scriptscriptstyle+};
(2.143737,1.221071) *[blue]{\scriptscriptstyle+};
(2.151951,1.224409) *[blue]{\scriptscriptstyle+};
(2.160164,1.225755) *[blue]{\scriptscriptstyle+};
(2.168378,1.230054) *[blue]{\scriptscriptstyle+};
(2.176591,1.230615) *[blue]{\scriptscriptstyle+};
(2.184805,1.233379) *[blue]{\scriptscriptstyle+};
(2.193018,1.235441) *[blue]{\scriptscriptstyle+};
(2.201232,1.251559) *[blue]{\scriptscriptstyle+};
(2.209446,1.256741) *[blue]{\scriptscriptstyle+};
(2.217659,1.133368) *[blue]{\scriptscriptstyle+};
(2.225873,1.159064) *[blue]{\scriptscriptstyle+};
(2.234086,1.166185) *[blue]{\scriptscriptstyle+};
(2.242300,1.177842) *[blue]{\scriptscriptstyle+};
(2.250513,1.206388) *[blue]{\scriptscriptstyle+};
(2.258727,1.209193) *[blue]{\scriptscriptstyle+};
(2.266940,1.211089) *[blue]{\scriptscriptstyle+};
(2.275154,1.211674) *[blue]{\scriptscriptstyle+};
(2.283368,1.212142) *[blue]{\scriptscriptstyle+};
(2.291581,1.212246) *[blue]{\scriptscriptstyle+};
(2.299795,1.215418) *[blue]{\scriptscriptstyle+};
(2.308008,1.219232) *[blue]{\scriptscriptstyle+};
(2.316222,1.256916) *[blue]{\scriptscriptstyle+};
(2.324435,1.268381) *[blue]{\scriptscriptstyle+};
(2.332649,1.275175) *[blue]{\scriptscriptstyle+};
(2.340862,1.176709) *[blue]{\scriptscriptstyle+};
(2.349076,1.188152) *[blue]{\scriptscriptstyle+};
(2.357290,1.189981) *[blue]{\scriptscriptstyle+};
(2.365503,1.202232) *[blue]{\scriptscriptstyle+};
(2.373717,1.202247) *[blue]{\scriptscriptstyle+};
(2.381930,1.202647) *[blue]{\scriptscriptstyle+};
(2.390144,1.203904) *[blue]{\scriptscriptstyle+};
(2.398357,1.208876) *[blue]{\scriptscriptstyle+};
(2.406571,1.215862) *[blue]{\scriptscriptstyle+};
(2.414784,1.220516) *[blue]{\scriptscriptstyle+};
(2.422998,1.226653) *[blue]{\scriptscriptstyle+};
(2.431211,1.229548) *[blue]{\scriptscriptstyle+};
(2.439425,1.232593) *[blue]{\scriptscriptstyle+};
(2.447639,1.241701) *[blue]{\scriptscriptstyle+};
(2.455852,1.259016) *[blue]{\scriptscriptstyle+};
(2.464066,1.205893) *[blue]{\scriptscriptstyle+};
(2.472279,1.212357) *[blue]{\scriptscriptstyle+};
(2.480493,1.213780) *[blue]{\scriptscriptstyle+};
(2.488706,1.223199) *[blue]{\scriptscriptstyle+};
(2.496920,1.226495) *[blue]{\scriptscriptstyle+};
(2.505133,1.227298) *[blue]{\scriptscriptstyle+};
(2.513347,1.235919) *[blue]{\scriptscriptstyle+};
(2.521561,1.236513) *[blue]{\scriptscriptstyle+};
(2.529774,1.248026) *[blue]{\scriptscriptstyle+};
(2.537988,1.256259) *[blue]{\scriptscriptstyle+};
(2.546201,1.268072) *[blue]{\scriptscriptstyle+};
(2.554415,1.271420) *[blue]{\scriptscriptstyle+};
(2.562628,1.275024) *[blue]{\scriptscriptstyle+};
(2.570842,1.306899) *[blue]{\scriptscriptstyle+};
(2.579055,1.312549) *[blue]{\scriptscriptstyle+};
(2.587269,1.183772) *[blue]{\scriptscriptstyle+};
(2.595483,1.209166) *[blue]{\scriptscriptstyle+};
(2.603696,1.215360) *[blue]{\scriptscriptstyle+};
(2.611910,1.217546) *[blue]{\scriptscriptstyle+};
(2.620123,1.219235) *[blue]{\scriptscriptstyle+};
(2.628337,1.221829) *[blue]{\scriptscriptstyle+};
(2.636550,1.230224) *[blue]{\scriptscriptstyle+};
(2.644764,1.237053) *[blue]{\scriptscriptstyle+};
(2.652977,1.240458) *[blue]{\scriptscriptstyle+};
(2.661191,1.241204) *[blue]{\scriptscriptstyle+};
(2.669405,1.248301) *[blue]{\scriptscriptstyle+};
(2.677618,1.256886) *[blue]{\scriptscriptstyle+};
(2.685832,1.287055) *[blue]{\scriptscriptstyle+};
(2.694045,1.359950) *[blue]{\scriptscriptstyle+};
(2.702259,1.363792) *[blue]{\scriptscriptstyle+};
(2.710472,1.182589) *[blue]{\scriptscriptstyle+};
(2.718686,1.184561) *[blue]{\scriptscriptstyle+};
(2.726899,1.188041) *[blue]{\scriptscriptstyle+};
(2.735113,1.190572) *[blue]{\scriptscriptstyle+};
(2.743326,1.201799) *[blue]{\scriptscriptstyle+};
(2.751540,1.211623) *[blue]{\scriptscriptstyle+};
(2.759754,1.213967) *[blue]{\scriptscriptstyle+};
(2.767967,1.234417) *[blue]{\scriptscriptstyle+};
(2.776181,1.240286) *[blue]{\scriptscriptstyle+};
(2.784394,1.243868) *[blue]{\scriptscriptstyle+};
(2.792608,1.246802) *[blue]{\scriptscriptstyle+};
(2.800821,1.253866) *[blue]{\scriptscriptstyle+};
(2.809035,1.259030) *[blue]{\scriptscriptstyle+};
(2.817248,1.260982) *[blue]{\scriptscriptstyle+};
(2.825462,1.285722) *[blue]{\scriptscriptstyle+};
(2.833676,1.198445) *[blue]{\scriptscriptstyle+};
(2.841889,1.215389) *[blue]{\scriptscriptstyle+};
(2.850103,1.219820) *[blue]{\scriptscriptstyle+};
(2.858316,1.221795) *[blue]{\scriptscriptstyle+};
(2.866530,1.234858) *[blue]{\scriptscriptstyle+};
(2.874743,1.235671) *[blue]{\scriptscriptstyle+};
(2.882957,1.242583) *[blue]{\scriptscriptstyle+};
(2.891170,1.246090) *[blue]{\scriptscriptstyle+};
(2.899384,1.248426) *[blue]{\scriptscriptstyle+};
(2.907598,1.253551) *[blue]{\scriptscriptstyle+};
(2.915811,1.256841) *[blue]{\scriptscriptstyle+};
(2.924025,1.259635) *[blue]{\scriptscriptstyle+};
(2.932238,1.268492) *[blue]{\scriptscriptstyle+};
(2.940452,1.274276) *[blue]{\scriptscriptstyle+};
(2.948665,1.283458) *[blue]{\scriptscriptstyle+};
(2.956879,1.229680) *[blue]{\scriptscriptstyle+};
(2.965092,1.230278) *[blue]{\scriptscriptstyle+};
(2.973306,1.235193) *[blue]{\scriptscriptstyle+};
(2.981520,1.237885) *[blue]{\scriptscriptstyle+};
(2.989733,1.241894) *[blue]{\scriptscriptstyle+};
(2.997947,1.242691) *[blue]{\scriptscriptstyle+};
(3.006160,1.250653) *[blue]{\scriptscriptstyle+};
(3.014374,1.257084) *[blue]{\scriptscriptstyle+};
(3.022587,1.259652) *[blue]{\scriptscriptstyle+};
(3.030801,1.267960) *[blue]{\scriptscriptstyle+};
(3.039014,1.271059) *[blue]{\scriptscriptstyle+};
(3.047228,1.298611) *[blue]{\scriptscriptstyle+};
(3.055441,1.349124) *[blue]{\scriptscriptstyle+};
(3.063655,1.350352) *[blue]{\scriptscriptstyle+};
(3.071869,1.350813) *[blue]{\scriptscriptstyle+};
(3.080082,1.214722) *[blue]{\scriptscriptstyle+};
(3.088296,1.216530) *[blue]{\scriptscriptstyle+};
(3.096509,1.218831) *[blue]{\scriptscriptstyle+};
(3.104723,1.219115) *[blue]{\scriptscriptstyle+};
(3.112936,1.235634) *[blue]{\scriptscriptstyle+};
(3.121150,1.238232) *[blue]{\scriptscriptstyle+};
(3.129363,1.244831) *[blue]{\scriptscriptstyle+};
(3.137577,1.245353) *[blue]{\scriptscriptstyle+};
(3.145791,1.247445) *[blue]{\scriptscriptstyle+};
(3.154004,1.248424) *[blue]{\scriptscriptstyle+};
(3.162218,1.256597) *[blue]{\scriptscriptstyle+};
(3.170431,1.267782) *[blue]{\scriptscriptstyle+};
(3.178645,1.271136) *[blue]{\scriptscriptstyle+};
(3.186858,1.273104) *[blue]{\scriptscriptstyle+};
(3.195072,1.274555) *[blue]{\scriptscriptstyle+};
(3.203285,1.244841) *[blue]{\scriptscriptstyle+};
(3.211499,1.245765) *[blue]{\scriptscriptstyle+};
(3.219713,1.247979) *[blue]{\scriptscriptstyle+};
(3.227926,1.252059) *[blue]{\scriptscriptstyle+};
(3.236140,1.253557) *[blue]{\scriptscriptstyle+};
(3.244353,1.265971) *[blue]{\scriptscriptstyle+};
(3.252567,1.266302) *[blue]{\scriptscriptstyle+};
(3.260780,1.267176) *[blue]{\scriptscriptstyle+};
(3.268994,1.269338) *[blue]{\scriptscriptstyle+};
(3.277207,1.274868) *[blue]{\scriptscriptstyle+};
(3.285421,1.279882) *[blue]{\scriptscriptstyle+};
(3.293634,1.292073) *[blue]{\scriptscriptstyle+};
(3.301848,1.295792) *[blue]{\scriptscriptstyle+};
(3.310062,1.301809) *[blue]{\scriptscriptstyle+};
(3.318275,1.313111) *[blue]{\scriptscriptstyle+};
(3.326489,1.223278) *[blue]{\scriptscriptstyle+};
(3.334702,1.230647) *[blue]{\scriptscriptstyle+};
(3.342916,1.241161) *[blue]{\scriptscriptstyle+};
(3.351129,1.244440) *[blue]{\scriptscriptstyle+};
(3.359343,1.246395) *[blue]{\scriptscriptstyle+};
(3.367556,1.250436) *[blue]{\scriptscriptstyle+};
(3.375770,1.250880) *[blue]{\scriptscriptstyle+};
(3.383984,1.258792) *[blue]{\scriptscriptstyle+};
(3.392197,1.260750) *[blue]{\scriptscriptstyle+};
(3.400411,1.262888) *[blue]{\scriptscriptstyle+};
(3.408624,1.263797) *[blue]{\scriptscriptstyle+};
(3.416838,1.265180) *[blue]{\scriptscriptstyle+};
(3.425051,1.281019) *[blue]{\scriptscriptstyle+};
(3.433265,1.291805) *[blue]{\scriptscriptstyle+};
(3.441478,1.296230) *[blue]{\scriptscriptstyle+};
(3.449692,1.228090) *[blue]{\scriptscriptstyle+};
(3.457906,1.229627) *[blue]{\scriptscriptstyle+};
(3.466119,1.231186) *[blue]{\scriptscriptstyle+};
(3.474333,1.237027) *[blue]{\scriptscriptstyle+};
(3.482546,1.239180) *[blue]{\scriptscriptstyle+};
(3.490760,1.239228) *[blue]{\scriptscriptstyle+};
(3.498973,1.252866) *[blue]{\scriptscriptstyle+};
(3.507187,1.257978) *[blue]{\scriptscriptstyle+};
(3.515400,1.259501) *[blue]{\scriptscriptstyle+};
(3.523614,1.259801) *[blue]{\scriptscriptstyle+};
(3.531828,1.261051) *[blue]{\scriptscriptstyle+};
(3.540041,1.261599) *[blue]{\scriptscriptstyle+};
(3.548255,1.264103) *[blue]{\scriptscriptstyle+};
(3.556468,1.272738) *[blue]{\scriptscriptstyle+};
(3.564682,1.298884) *[blue]{\scriptscriptstyle+};
(3.572895,1.244058) *[blue]{\scriptscriptstyle+};
(3.581109,1.248490) *[blue]{\scriptscriptstyle+};
(3.589322,1.248577) *[blue]{\scriptscriptstyle+};
(3.597536,1.252961) *[blue]{\scriptscriptstyle+};
(3.605749,1.253922) *[blue]{\scriptscriptstyle+};
(3.613963,1.255082) *[blue]{\scriptscriptstyle+};
(3.622177,1.264820) *[blue]{\scriptscriptstyle+};
(3.630390,1.270241) *[blue]{\scriptscriptstyle+};
(3.638604,1.270339) *[blue]{\scriptscriptstyle+};
(3.646817,1.271285) *[blue]{\scriptscriptstyle+};
(3.655031,1.273889) *[blue]{\scriptscriptstyle+};
(3.663244,1.274277) *[blue]{\scriptscriptstyle+};
(3.671458,1.289603) *[blue]{\scriptscriptstyle+};
(3.679671,1.295219) *[blue]{\scriptscriptstyle+};
(3.687885,1.302137) *[blue]{\scriptscriptstyle+};
(3.696099,1.244979) *[blue]{\scriptscriptstyle+};
(3.704312,1.251043) *[blue]{\scriptscriptstyle+};
(3.712526,1.255686) *[blue]{\scriptscriptstyle+};
(3.720739,1.260562) *[blue]{\scriptscriptstyle+};
(3.728953,1.260647) *[blue]{\scriptscriptstyle+};
(3.737166,1.263761) *[blue]{\scriptscriptstyle+};
(3.745380,1.265678) *[blue]{\scriptscriptstyle+};
(3.753593,1.273656) *[blue]{\scriptscriptstyle+};
(3.761807,1.276069) *[blue]{\scriptscriptstyle+};
(3.770021,1.280473) *[blue]{\scriptscriptstyle+};
(3.778234,1.292081) *[blue]{\scriptscriptstyle+};
(3.786448,1.296818) *[blue]{\scriptscriptstyle+};
(3.794661,1.317580) *[blue]{\scriptscriptstyle+};
(3.802875,1.318101) *[blue]{\scriptscriptstyle+};
(3.811088,1.330146) *[blue]{\scriptscriptstyle+};
(3.819302,1.260831) *[blue]{\scriptscriptstyle+};
(3.827515,1.260963) *[blue]{\scriptscriptstyle+};
(3.835729,1.263177) *[blue]{\scriptscriptstyle+};
(3.843943,1.263251) *[blue]{\scriptscriptstyle+};
(3.852156,1.264265) *[blue]{\scriptscriptstyle+};
(3.860370,1.282964) *[blue]{\scriptscriptstyle+};
(3.868583,1.282966) *[blue]{\scriptscriptstyle+};
(3.876797,1.306425) *[blue]{\scriptscriptstyle+};
(3.885010,1.307657) *[blue]{\scriptscriptstyle+};
(3.893224,1.324068) *[blue]{\scriptscriptstyle+};
(3.901437,1.329654) *[blue]{\scriptscriptstyle+};
(3.909651,1.330464) *[blue]{\scriptscriptstyle+};
(3.917864,1.356277) *[blue]{\scriptscriptstyle+};
(3.926078,1.357956) *[blue]{\scriptscriptstyle+};
(3.934292,1.450837) *[blue]{\scriptscriptstyle+};
(3.942505,1.254096) *[blue]{\scriptscriptstyle+};
(3.950719,1.254608) *[blue]{\scriptscriptstyle+};
(3.958932,1.257705) *[blue]{\scriptscriptstyle+};
(3.967146,1.257902) *[blue]{\scriptscriptstyle+};
(3.975359,1.266227) *[blue]{\scriptscriptstyle+};
(3.983573,1.272883) *[blue]{\scriptscriptstyle+};
(3.991786,1.277285) *[blue]{\scriptscriptstyle+};
(4.000000,1.279773) *[blue]{\scriptscriptstyle+};
(4.008214,1.284149) *[blue]{\scriptscriptstyle+};
(4.016427,1.284644) *[blue]{\scriptscriptstyle+};
(4.024641,1.297525) *[blue]{\scriptscriptstyle+};
(4.032854,1.307682) *[blue]{\scriptscriptstyle+};
(4.041068,1.325282) *[blue]{\scriptscriptstyle+};
(4.049281,1.343817) *[blue]{\scriptscriptstyle+};
(4.057495,1.404097) *[blue]{\scriptscriptstyle+};
(4.065708,1.252672) *[blue]{\scriptscriptstyle+};
(4.073922,1.256654) *[blue]{\scriptscriptstyle+};
(4.082136,1.257791) *[blue]{\scriptscriptstyle+};
(4.090349,1.258172) *[blue]{\scriptscriptstyle+};
(4.098563,1.263254) *[blue]{\scriptscriptstyle+};
(4.106776,1.279904) *[blue]{\scriptscriptstyle+};
(4.114990,1.280814) *[blue]{\scriptscriptstyle+};
(4.123203,1.282567) *[blue]{\scriptscriptstyle+};
(4.131417,1.282577) *[blue]{\scriptscriptstyle+};
(4.139630,1.288249) *[blue]{\scriptscriptstyle+};
(4.147844,1.293551) *[blue]{\scriptscriptstyle+};
(4.156057,1.308567) *[blue]{\scriptscriptstyle+};
(4.164271,1.319656) *[blue]{\scriptscriptstyle+};
(4.172485,1.320155) *[blue]{\scriptscriptstyle+};
(4.180698,1.327551) *[blue]{\scriptscriptstyle+};
(4.188912,1.246988) *[blue]{\scriptscriptstyle+};
(4.197125,1.259865) *[blue]{\scriptscriptstyle+};
(4.205339,1.268906) *[blue]{\scriptscriptstyle+};
(4.213552,1.290136) *[blue]{\scriptscriptstyle+};
(4.221766,1.291620) *[blue]{\scriptscriptstyle+};
(4.229979,1.297134) *[blue]{\scriptscriptstyle+};
(4.238193,1.299627) *[blue]{\scriptscriptstyle+};
(4.246407,1.302355) *[blue]{\scriptscriptstyle+};
(4.254620,1.308756) *[blue]{\scriptscriptstyle+};
(4.262834,1.318968) *[blue]{\scriptscriptstyle+};
(4.271047,1.325170) *[blue]{\scriptscriptstyle+};
(4.279261,1.344307) *[blue]{\scriptscriptstyle+};
(4.287474,1.345837) *[blue]{\scriptscriptstyle+};
(4.295688,1.358333) *[blue]{\scriptscriptstyle+};
(4.303901,1.363785) *[blue]{\scriptscriptstyle+};
(4.312115,1.264879) *[blue]{\scriptscriptstyle+};
(4.320329,1.270237) *[blue]{\scriptscriptstyle+};
(4.328542,1.271450) *[blue]{\scriptscriptstyle+};
(4.336756,1.272410) *[blue]{\scriptscriptstyle+};
(4.344969,1.272590) *[blue]{\scriptscriptstyle+};
(4.353183,1.272784) *[blue]{\scriptscriptstyle+};
(4.361396,1.274127) *[blue]{\scriptscriptstyle+};
(4.369610,1.275643) *[blue]{\scriptscriptstyle+};
(4.377823,1.278067) *[blue]{\scriptscriptstyle+};
(4.386037,1.289552) *[blue]{\scriptscriptstyle+};
(4.394251,1.290962) *[blue]{\scriptscriptstyle+};
(4.402464,1.300059) *[blue]{\scriptscriptstyle+};
(4.410678,1.318802) *[blue]{\scriptscriptstyle+};
(4.418891,1.320277) *[blue]{\scriptscriptstyle+};
(4.427105,1.323159) *[blue]{\scriptscriptstyle+};
(4.435318,1.269165) *[blue]{\scriptscriptstyle+};
(4.443532,1.271296) *[blue]{\scriptscriptstyle+};
(4.451745,1.273502) *[blue]{\scriptscriptstyle+};
(4.459959,1.277362) *[blue]{\scriptscriptstyle+};
(4.468172,1.281374) *[blue]{\scriptscriptstyle+};
(4.476386,1.283702) *[blue]{\scriptscriptstyle+};
(4.484600,1.284042) *[blue]{\scriptscriptstyle+};
(4.492813,1.289750) *[blue]{\scriptscriptstyle+};
(4.501027,1.295633) *[blue]{\scriptscriptstyle+};
(4.509240,1.298773) *[blue]{\scriptscriptstyle+};
(4.517454,1.300729) *[blue]{\scriptscriptstyle+};
(4.525667,1.304408) *[blue]{\scriptscriptstyle+};
(4.533881,1.330607) *[blue]{\scriptscriptstyle+};
(4.542094,1.341932) *[blue]{\scriptscriptstyle+};
(4.550308,1.348138) *[blue]{\scriptscriptstyle+};
(4.558522,1.264317) *[blue]{\scriptscriptstyle+};
(4.566735,1.267581) *[blue]{\scriptscriptstyle+};
(4.574949,1.268172) *[blue]{\scriptscriptstyle+};
(4.583162,1.282946) *[blue]{\scriptscriptstyle+};
(4.591376,1.284850) *[blue]{\scriptscriptstyle+};
(4.599589,1.286228) *[blue]{\scriptscriptstyle+};
(4.607803,1.287789) *[blue]{\scriptscriptstyle+};
(4.616016,1.288869) *[blue]{\scriptscriptstyle+};
(4.624230,1.290581) *[blue]{\scriptscriptstyle+};
(4.632444,1.290662) *[blue]{\scriptscriptstyle+};
(4.640657,1.293072) *[blue]{\scriptscriptstyle+};
(4.648871,1.295033) *[blue]{\scriptscriptstyle+};
(4.657084,1.298443) *[blue]{\scriptscriptstyle+};
(4.665298,1.300137) *[blue]{\scriptscriptstyle+};
(4.673511,1.317334) *[blue]{\scriptscriptstyle+};
(4.681725,1.258955) *[blue]{\scriptscriptstyle+};
(4.689938,1.265373) *[blue]{\scriptscriptstyle+};
(4.698152,1.277110) *[blue]{\scriptscriptstyle+};
(4.706366,1.277359) *[blue]{\scriptscriptstyle+};
(4.714579,1.282040) *[blue]{\scriptscriptstyle+};
(4.722793,1.284559) *[blue]{\scriptscriptstyle+};
(4.731006,1.286533) *[blue]{\scriptscriptstyle+};
(4.739220,1.289604) *[blue]{\scriptscriptstyle+};
(4.747433,1.291009) *[blue]{\scriptscriptstyle+};
(4.755647,1.293404) *[blue]{\scriptscriptstyle+};
(4.763860,1.295026) *[blue]{\scriptscriptstyle+};
(4.772074,1.327709) *[blue]{\scriptscriptstyle+};
(4.780287,1.336170) *[blue]{\scriptscriptstyle+};
(4.788501,1.356923) *[blue]{\scriptscriptstyle+};
(4.796715,1.399407) *[blue]{\scriptscriptstyle+};
(4.804928,1.265586) *[blue]{\scriptscriptstyle+};
(4.813142,1.267242) *[blue]{\scriptscriptstyle+};
(4.821355,1.268971) *[blue]{\scriptscriptstyle+};
(4.829569,1.273205) *[blue]{\scriptscriptstyle+};
(4.837782,1.281121) *[blue]{\scriptscriptstyle+};
(4.845996,1.286693) *[blue]{\scriptscriptstyle+};
(4.854209,1.286778) *[blue]{\scriptscriptstyle+};
(4.862423,1.290573) *[blue]{\scriptscriptstyle+};
(4.870637,1.309779) *[blue]{\scriptscriptstyle+};
(4.878850,1.313731) *[blue]{\scriptscriptstyle+};
(4.887064,1.313745) *[blue]{\scriptscriptstyle+};
(4.895277,1.337500) *[blue]{\scriptscriptstyle+};
(4.903491,1.341593) *[blue]{\scriptscriptstyle+};
(4.911704,1.370576) *[blue]{\scriptscriptstyle+};
(4.919918,1.376791) *[blue]{\scriptscriptstyle+};
(4.928131,1.245259) *[blue]{\scriptscriptstyle+};
(4.936345,1.267663) *[blue]{\scriptscriptstyle+};
(4.944559,1.267759) *[blue]{\scriptscriptstyle+};
(4.952772,1.276692) *[blue]{\scriptscriptstyle+};
(4.960986,1.286933) *[blue]{\scriptscriptstyle+};
(4.969199,1.291814) *[blue]{\scriptscriptstyle+};
(4.977413,1.298064) *[blue]{\scriptscriptstyle+};
(4.985626,1.298416) *[blue]{\scriptscriptstyle+};
(4.993840,1.299423) *[blue]{\scriptscriptstyle+};
(5.002053,1.309510) *[blue]{\scriptscriptstyle+};
(5.010267,1.309576) *[blue]{\scriptscriptstyle+};
(5.018480,1.317987) *[blue]{\scriptscriptstyle+};
(5.026694,1.341722) *[blue]{\scriptscriptstyle+};
(5.034908,1.344615) *[blue]{\scriptscriptstyle+};
(5.043121,1.383933) *[blue]{\scriptscriptstyle+};
(5.051335,1.303022) *[blue]{\scriptscriptstyle+};
(5.059548,1.304635) *[blue]{\scriptscriptstyle+};
(5.067762,1.304908) *[blue]{\scriptscriptstyle+};
(5.075975,1.314312) *[blue]{\scriptscriptstyle+};
(5.084189,1.316039) *[blue]{\scriptscriptstyle+};
(5.092402,1.320885) *[blue]{\scriptscriptstyle+};
(5.100616,1.324707) *[blue]{\scriptscriptstyle+};
(5.108830,1.325709) *[blue]{\scriptscriptstyle+};
(5.117043,1.325947) *[blue]{\scriptscriptstyle+};
(5.125257,1.329071) *[blue]{\scriptscriptstyle+};
(5.133470,1.330615) *[blue]{\scriptscriptstyle+};
(5.141684,1.338188) *[blue]{\scriptscriptstyle+};
(5.149897,1.343093) *[blue]{\scriptscriptstyle+};
(5.158111,1.352337) *[blue]{\scriptscriptstyle+};
(5.166324,1.355946) *[blue]{\scriptscriptstyle+};
(5.174538,1.269841) *[blue]{\scriptscriptstyle+};
(5.182752,1.278515) *[blue]{\scriptscriptstyle+};
(5.190965,1.295517) *[blue]{\scriptscriptstyle+};
(5.199179,1.297579) *[blue]{\scriptscriptstyle+};
(5.207392,1.301416) *[blue]{\scriptscriptstyle+};
(5.215606,1.314887) *[blue]{\scriptscriptstyle+};
(5.223819,1.321005) *[blue]{\scriptscriptstyle+};
(5.232033,1.324166) *[blue]{\scriptscriptstyle+};
(5.240246,1.335864) *[blue]{\scriptscriptstyle+};
(5.248460,1.337522) *[blue]{\scriptscriptstyle+};
(5.256674,1.340386) *[blue]{\scriptscriptstyle+};
(5.264887,1.341625) *[blue]{\scriptscriptstyle+};
(5.273101,1.365670) *[blue]{\scriptscriptstyle+};
(5.281314,1.368801) *[blue]{\scriptscriptstyle+};
(5.289528,1.371443) *[blue]{\scriptscriptstyle+};
(5.297741,1.284991) *[blue]{\scriptscriptstyle+};
(5.305955,1.285415) *[blue]{\scriptscriptstyle+};
(5.314168,1.295785) *[blue]{\scriptscriptstyle+};
(5.322382,1.311606) *[blue]{\scriptscriptstyle+};
(5.330595,1.317121) *[blue]{\scriptscriptstyle+};
(5.338809,1.320682) *[blue]{\scriptscriptstyle+};
(5.347023,1.326333) *[blue]{\scriptscriptstyle+};
(5.355236,1.330143) *[blue]{\scriptscriptstyle+};
(5.363450,1.331339) *[blue]{\scriptscriptstyle+};
(5.371663,1.331458) *[blue]{\scriptscriptstyle+};
(5.379877,1.333623) *[blue]{\scriptscriptstyle+};
(5.388090,1.337277) *[blue]{\scriptscriptstyle+};
(5.396304,1.337350) *[blue]{\scriptscriptstyle+};
(5.404517,1.340629) *[blue]{\scriptscriptstyle+};
(5.412731,1.358757) *[blue]{\scriptscriptstyle+};
(5.420945,1.293960) *[blue]{\scriptscriptstyle+};
(5.429158,1.295331) *[blue]{\scriptscriptstyle+};
(5.437372,1.308267) *[blue]{\scriptscriptstyle+};
(5.445585,1.313271) *[blue]{\scriptscriptstyle+};
(5.453799,1.313802) *[blue]{\scriptscriptstyle+};
(5.462012,1.316583) *[blue]{\scriptscriptstyle+};
(5.470226,1.323985) *[blue]{\scriptscriptstyle+};
(5.478439,1.330398) *[blue]{\scriptscriptstyle+};
(5.486653,1.330428) *[blue]{\scriptscriptstyle+};
(5.494867,1.332139) *[blue]{\scriptscriptstyle+};
(5.503080,1.339137) *[blue]{\scriptscriptstyle+};
(5.511294,1.339196) *[blue]{\scriptscriptstyle+};
(5.519507,1.341296) *[blue]{\scriptscriptstyle+};
(5.527721,1.343440) *[blue]{\scriptscriptstyle+};
(5.535934,1.345812) *[blue]{\scriptscriptstyle+};
(5.544148,1.317548) *[blue]{\scriptscriptstyle+};
(5.552361,1.329275) *[blue]{\scriptscriptstyle+};
(5.560575,1.333968) *[blue]{\scriptscriptstyle+};
(5.568789,1.335429) *[blue]{\scriptscriptstyle+};
(5.577002,1.335483) *[blue]{\scriptscriptstyle+};
(5.585216,1.336424) *[blue]{\scriptscriptstyle+};
(5.593429,1.338492) *[blue]{\scriptscriptstyle+};
(5.601643,1.338507) *[blue]{\scriptscriptstyle+};
(5.609856,1.346451) *[blue]{\scriptscriptstyle+};
(5.618070,1.348290) *[blue]{\scriptscriptstyle+};
(5.626283,1.353960) *[blue]{\scriptscriptstyle+};
(5.634497,1.360140) *[blue]{\scriptscriptstyle+};
(5.642710,1.360266) *[blue]{\scriptscriptstyle+};
(5.650924,1.372865) *[blue]{\scriptscriptstyle+};
(5.659138,1.412367) *[blue]{\scriptscriptstyle+};
(5.667351,1.323810) *[blue]{\scriptscriptstyle+};
(5.675565,1.328999) *[blue]{\scriptscriptstyle+};
(5.683778,1.331549) *[blue]{\scriptscriptstyle+};
(5.691992,1.337794) *[blue]{\scriptscriptstyle+};
(5.700205,1.341332) *[blue]{\scriptscriptstyle+};
(5.708419,1.344849) *[blue]{\scriptscriptstyle+};
(5.716632,1.349874) *[blue]{\scriptscriptstyle+};
(5.724846,1.350497) *[blue]{\scriptscriptstyle+};
(5.733060,1.356074) *[blue]{\scriptscriptstyle+};
(5.741273,1.358788) *[blue]{\scriptscriptstyle+};
(5.749487,1.360989) *[blue]{\scriptscriptstyle+};
(5.757700,1.367264) *[blue]{\scriptscriptstyle+};
(5.765914,1.375868) *[blue]{\scriptscriptstyle+};
(5.774127,1.378695) *[blue]{\scriptscriptstyle+};
(5.782341,1.392488) *[blue]{\scriptscriptstyle+};
(5.790554,1.326355) *[blue]{\scriptscriptstyle+};
(5.798768,1.328142) *[blue]{\scriptscriptstyle+};
(5.806982,1.329974) *[blue]{\scriptscriptstyle+};
(5.815195,1.343019) *[blue]{\scriptscriptstyle+};
(5.823409,1.343186) *[blue]{\scriptscriptstyle+};
(5.831622,1.344118) *[blue]{\scriptscriptstyle+};
(5.839836,1.347913) *[blue]{\scriptscriptstyle+};
(5.848049,1.351770) *[blue]{\scriptscriptstyle+};
(5.856263,1.354261) *[blue]{\scriptscriptstyle+};
(5.864476,1.355412) *[blue]{\scriptscriptstyle+};
(5.872690,1.355810) *[blue]{\scriptscriptstyle+};
(5.880903,1.364672) *[blue]{\scriptscriptstyle+};
(5.889117,1.371210) *[blue]{\scriptscriptstyle+};
(5.897331,1.377952) *[blue]{\scriptscriptstyle+};
(5.905544,1.397692) *[blue]{\scriptscriptstyle+};
(5.913758,1.322739) *[blue]{\scriptscriptstyle+};
(5.921971,1.325835) *[blue]{\scriptscriptstyle+};
(5.930185,1.327606) *[blue]{\scriptscriptstyle+};
(5.938398,1.344555) *[blue]{\scriptscriptstyle+};
(5.946612,1.344721) *[blue]{\scriptscriptstyle+};
(5.954825,1.344849) *[blue]{\scriptscriptstyle+};
(5.963039,1.345964) *[blue]{\scriptscriptstyle+};
(5.971253,1.350908) *[blue]{\scriptscriptstyle+};
(5.979466,1.351082) *[blue]{\scriptscriptstyle+};
(5.987680,1.353845) *[blue]{\scriptscriptstyle+};
(5.995893,1.360068) *[blue]{\scriptscriptstyle+};
(6.004107,1.361763) *[blue]{\scriptscriptstyle+};
(6.012320,1.394623) *[blue]{\scriptscriptstyle+};
(6.020534,1.396992) *[blue]{\scriptscriptstyle+};
(6.028747,1.414808) *[blue]{\scriptscriptstyle+};
(6.036961,1.331277) *[blue]{\scriptscriptstyle+};
(6.045175,1.340571) *[blue]{\scriptscriptstyle+};
(6.053388,1.345000) *[blue]{\scriptscriptstyle+};
(6.061602,1.347615) *[blue]{\scriptscriptstyle+};
(6.069815,1.348814) *[blue]{\scriptscriptstyle+};
(6.078029,1.356522) *[blue]{\scriptscriptstyle+};
(6.086242,1.357057) *[blue]{\scriptscriptstyle+};
(6.094456,1.370042) *[blue]{\scriptscriptstyle+};
(6.102669,1.374820) *[blue]{\scriptscriptstyle+};
(6.110883,1.402911) *[blue]{\scriptscriptstyle+};
(6.119097,1.436727) *[blue]{\scriptscriptstyle+};
(6.127310,1.438010) *[blue]{\scriptscriptstyle+};
(6.135524,1.439372) *[blue]{\scriptscriptstyle+};
(6.143737,1.473629) *[blue]{\scriptscriptstyle+};
(6.151951,1.473984) *[blue]{\scriptscriptstyle+};
(6.160164,1.335050) *[blue]{\scriptscriptstyle+};
(6.168378,1.342088) *[blue]{\scriptscriptstyle+};
(6.176591,1.356178) *[blue]{\scriptscriptstyle+};
(6.184805,1.362742) *[blue]{\scriptscriptstyle+};
(6.193018,1.375379) *[blue]{\scriptscriptstyle+};
(6.201232,1.375993) *[blue]{\scriptscriptstyle+};
(6.209446,1.377298) *[blue]{\scriptscriptstyle+};
(6.217659,1.383156) *[blue]{\scriptscriptstyle+};
(6.225873,1.385685) *[blue]{\scriptscriptstyle+};
(6.234086,1.393787) *[blue]{\scriptscriptstyle+};
(6.242300,1.394795) *[blue]{\scriptscriptstyle+};
(6.250513,1.397440) *[blue]{\scriptscriptstyle+};
(6.258727,1.400292) *[blue]{\scriptscriptstyle+};
(6.266940,1.420583) *[blue]{\scriptscriptstyle+};
(6.275154,1.465105) *[blue]{\scriptscriptstyle+};
(6.283368,1.302843) *[blue]{\scriptscriptstyle+};
(6.291581,1.326449) *[blue]{\scriptscriptstyle+};
(6.299795,1.342806) *[blue]{\scriptscriptstyle+};
(6.308008,1.347705) *[blue]{\scriptscriptstyle+};
(6.316222,1.348671) *[blue]{\scriptscriptstyle+};
(6.324435,1.354928) *[blue]{\scriptscriptstyle+};
(6.332649,1.368559) *[blue]{\scriptscriptstyle+};
(6.340862,1.368962) *[blue]{\scriptscriptstyle+};
(6.349076,1.374785) *[blue]{\scriptscriptstyle+};
(6.357290,1.385035) *[blue]{\scriptscriptstyle+};
(6.365503,1.402741) *[blue]{\scriptscriptstyle+};
(6.373717,1.408958) *[blue]{\scriptscriptstyle+};
(6.381930,1.410759) *[blue]{\scriptscriptstyle+};
(6.390144,1.436710) *[blue]{\scriptscriptstyle+};
(6.398357,1.455513) *[blue]{\scriptscriptstyle+};
(6.406571,1.340583) *[blue]{\scriptscriptstyle+};
(6.414784,1.342808) *[blue]{\scriptscriptstyle+};
(6.422998,1.344255) *[blue]{\scriptscriptstyle+};
(6.431211,1.356328) *[blue]{\scriptscriptstyle+};
(6.439425,1.367275) *[blue]{\scriptscriptstyle+};
(6.447639,1.367474) *[blue]{\scriptscriptstyle+};
(6.455852,1.368188) *[blue]{\scriptscriptstyle+};
(6.464066,1.370213) *[blue]{\scriptscriptstyle+};
(6.472279,1.375851) *[blue]{\scriptscriptstyle+};
(6.480493,1.380654) *[blue]{\scriptscriptstyle+};
(6.488706,1.381910) *[blue]{\scriptscriptstyle+};
(6.496920,1.384829) *[blue]{\scriptscriptstyle+};
(6.505133,1.395607) *[blue]{\scriptscriptstyle+};
(6.513347,1.423414) *[blue]{\scriptscriptstyle+};
(6.521561,1.458964) *[blue]{\scriptscriptstyle+};
(6.529774,1.373231) *[blue]{\scriptscriptstyle+};
(6.537988,1.377549) *[blue]{\scriptscriptstyle+};
(6.546201,1.378450) *[blue]{\scriptscriptstyle+};
(6.554415,1.386102) *[blue]{\scriptscriptstyle+};
(6.562628,1.399007) *[blue]{\scriptscriptstyle+};
(6.570842,1.402052) *[blue]{\scriptscriptstyle+};
(6.579055,1.402371) *[blue]{\scriptscriptstyle+};
(6.587269,1.408191) *[blue]{\scriptscriptstyle+};
(6.595483,1.416464) *[blue]{\scriptscriptstyle+};
(6.603696,1.420942) *[blue]{\scriptscriptstyle+};
(6.611910,1.421115) *[blue]{\scriptscriptstyle+};
(6.620123,1.423974) *[blue]{\scriptscriptstyle+};
(6.628337,1.435238) *[blue]{\scriptscriptstyle+};
(6.636550,1.435260) *[blue]{\scriptscriptstyle+};
(6.644764,1.460648) *[blue]{\scriptscriptstyle+};
(6.652977,1.370458) *[blue]{\scriptscriptstyle+};
(6.661191,1.380148) *[blue]{\scriptscriptstyle+};
(6.669405,1.388129) *[blue]{\scriptscriptstyle+};
(6.677618,1.392502) *[blue]{\scriptscriptstyle+};
(6.685832,1.392810) *[blue]{\scriptscriptstyle+};
(6.694045,1.393281) *[blue]{\scriptscriptstyle+};
(6.702259,1.406510) *[blue]{\scriptscriptstyle+};
(6.710472,1.408158) *[blue]{\scriptscriptstyle+};
(6.718686,1.408801) *[blue]{\scriptscriptstyle+};
(6.726899,1.440031) *[blue]{\scriptscriptstyle+};
(6.735113,1.445705) *[blue]{\scriptscriptstyle+};
(6.743326,1.450610) *[blue]{\scriptscriptstyle+};
(6.751540,1.484846) *[blue]{\scriptscriptstyle+};
(6.759754,1.525821) *[blue]{\scriptscriptstyle+};
(6.767967,1.547960) *[blue]{\scriptscriptstyle+};
(6.776181,1.352286) *[blue]{\scriptscriptstyle+};
(6.784394,1.367241) *[blue]{\scriptscriptstyle+};
(6.792608,1.384048) *[blue]{\scriptscriptstyle+};
(6.800821,1.384598) *[blue]{\scriptscriptstyle+};
(6.809035,1.393346) *[blue]{\scriptscriptstyle+};
(6.817248,1.396808) *[blue]{\scriptscriptstyle+};
(6.825462,1.398857) *[blue]{\scriptscriptstyle+};
(6.833676,1.407440) *[blue]{\scriptscriptstyle+};
(6.841889,1.409880) *[blue]{\scriptscriptstyle+};
(6.850103,1.418828) *[blue]{\scriptscriptstyle+};
(6.858316,1.423293) *[blue]{\scriptscriptstyle+};
(6.866530,1.429622) *[blue]{\scriptscriptstyle+};
(6.874743,1.441209) *[blue]{\scriptscriptstyle+};
(6.882957,1.470587) *[blue]{\scriptscriptstyle+};
(6.891170,1.535133) *[blue]{\scriptscriptstyle+};
(6.899384,1.396968) *[blue]{\scriptscriptstyle+};
(6.907598,1.397987) *[blue]{\scriptscriptstyle+};
(6.915811,1.400149) *[blue]{\scriptscriptstyle+};
(6.924025,1.405121) *[blue]{\scriptscriptstyle+};
(6.932238,1.418972) *[blue]{\scriptscriptstyle+};
(6.940452,1.423870) *[blue]{\scriptscriptstyle+};
(6.948665,1.424659) *[blue]{\scriptscriptstyle+};
(6.956879,1.424880) *[blue]{\scriptscriptstyle+};
(6.965092,1.425051) *[blue]{\scriptscriptstyle+};
(6.973306,1.425371) *[blue]{\scriptscriptstyle+};
(6.981520,1.438874) *[blue]{\scriptscriptstyle+};
(6.989733,1.440301) *[blue]{\scriptscriptstyle+};
(6.997947,1.441962) *[blue]{\scriptscriptstyle+};
(7.006160,1.445911) *[blue]{\scriptscriptstyle+};
(7.014374,1.448012) *[blue]{\scriptscriptstyle+};
(7.022587,1.374877) *[blue]{\scriptscriptstyle+};
(7.030801,1.379375) *[blue]{\scriptscriptstyle+};
(7.039014,1.382483) *[blue]{\scriptscriptstyle+};
(7.047228,1.394130) *[blue]{\scriptscriptstyle+};
(7.055441,1.402008) *[blue]{\scriptscriptstyle+};
(7.063655,1.408584) *[blue]{\scriptscriptstyle+};
(7.071869,1.412469) *[blue]{\scriptscriptstyle+};
(7.080082,1.425017) *[blue]{\scriptscriptstyle+};
(7.088296,1.426175) *[blue]{\scriptscriptstyle+};
(7.096509,1.437328) *[blue]{\scriptscriptstyle+};
(7.104723,1.438512) *[blue]{\scriptscriptstyle+};
(7.112936,1.459726) *[blue]{\scriptscriptstyle+};
(7.121150,1.484979) *[blue]{\scriptscriptstyle+};
(7.129363,1.517245) *[blue]{\scriptscriptstyle+};
(7.137577,1.517463) *[blue]{\scriptscriptstyle+};
(7.145791,1.370485) *[blue]{\scriptscriptstyle+};
(7.154004,1.398165) *[blue]{\scriptscriptstyle+};
(7.162218,1.403084) *[blue]{\scriptscriptstyle+};
(7.170431,1.415708) *[blue]{\scriptscriptstyle+};
(7.178645,1.415787) *[blue]{\scriptscriptstyle+};
(7.186858,1.434328) *[blue]{\scriptscriptstyle+};
(7.195072,1.435329) *[blue]{\scriptscriptstyle+};
(7.203285,1.436141) *[blue]{\scriptscriptstyle+};
(7.211499,1.436769) *[blue]{\scriptscriptstyle+};
(7.219713,1.438739) *[blue]{\scriptscriptstyle+};
(7.227926,1.442393) *[blue]{\scriptscriptstyle+};
(7.236140,1.454593) *[blue]{\scriptscriptstyle+};
(7.244353,1.459239) *[blue]{\scriptscriptstyle+};
(7.252567,1.476191) *[blue]{\scriptscriptstyle+};
(7.260780,1.553110) *[blue]{\scriptscriptstyle+};
(7.268994,1.368797) *[blue]{\scriptscriptstyle+};
(7.277207,1.384142) *[blue]{\scriptscriptstyle+};
(7.285421,1.388763) *[blue]{\scriptscriptstyle+};
(7.293634,1.407000) *[blue]{\scriptscriptstyle+};
(7.301848,1.410986) *[blue]{\scriptscriptstyle+};
(7.310062,1.412845) *[blue]{\scriptscriptstyle+};
(7.318275,1.429137) *[blue]{\scriptscriptstyle+};
(7.326489,1.432704) *[blue]{\scriptscriptstyle+};
(7.334702,1.443814) *[blue]{\scriptscriptstyle+};
(7.342916,1.453340) *[blue]{\scriptscriptstyle+};
(7.351129,1.458658) *[blue]{\scriptscriptstyle+};
(7.359343,1.463077) *[blue]{\scriptscriptstyle+};
(7.367556,1.469057) *[blue]{\scriptscriptstyle+};
(7.375770,1.488247) *[blue]{\scriptscriptstyle+};
(7.383984,1.523229) *[blue]{\scriptscriptstyle+};
(7.392197,1.402361) *[blue]{\scriptscriptstyle+};
(7.400411,1.414214) *[blue]{\scriptscriptstyle+};
(7.408624,1.430426) *[blue]{\scriptscriptstyle+};
(7.416838,1.431823) *[blue]{\scriptscriptstyle+};
(7.425051,1.437175) *[blue]{\scriptscriptstyle+};
(7.433265,1.437594) *[blue]{\scriptscriptstyle+};
(7.441478,1.443815) *[blue]{\scriptscriptstyle+};
(7.449692,1.445909) *[blue]{\scriptscriptstyle+};
(7.457906,1.446863) *[blue]{\scriptscriptstyle+};
(7.466119,1.458833) *[blue]{\scriptscriptstyle+};
(7.474333,1.470849) *[blue]{\scriptscriptstyle+};
(7.482546,1.482029) *[blue]{\scriptscriptstyle+};
(7.490760,1.483608) *[blue]{\scriptscriptstyle+};
(7.498973,1.511009) *[blue]{\scriptscriptstyle+};
(7.507187,1.517061) *[blue]{\scriptscriptstyle+};
(7.515400,1.396077) *[blue]{\scriptscriptstyle+};
(7.523614,1.401671) *[blue]{\scriptscriptstyle+};
(7.531828,1.408182) *[blue]{\scriptscriptstyle+};
(7.540041,1.427976) *[blue]{\scriptscriptstyle+};
(7.548255,1.443837) *[blue]{\scriptscriptstyle+};
(7.556468,1.449209) *[blue]{\scriptscriptstyle+};
(7.564682,1.449737) *[blue]{\scriptscriptstyle+};
(7.572895,1.450044) *[blue]{\scriptscriptstyle+};
(7.581109,1.460923) *[blue]{\scriptscriptstyle+};
(7.589322,1.463279) *[blue]{\scriptscriptstyle+};
(7.597536,1.465197) *[blue]{\scriptscriptstyle+};
(7.605749,1.467490) *[blue]{\scriptscriptstyle+};
(7.613963,1.473628) *[blue]{\scriptscriptstyle+};
(7.622177,1.507126) *[blue]{\scriptscriptstyle+};
(7.630390,1.512164) *[blue]{\scriptscriptstyle+};
(7.638604,1.440006) *[blue]{\scriptscriptstyle+};
(7.646817,1.463012) *[blue]{\scriptscriptstyle+};
(7.655031,1.463572) *[blue]{\scriptscriptstyle+};
(7.663244,1.464237) *[blue]{\scriptscriptstyle+};
(7.671458,1.472679) *[blue]{\scriptscriptstyle+};
(7.679671,1.479381) *[blue]{\scriptscriptstyle+};
(7.687885,1.479898) *[blue]{\scriptscriptstyle+};
(7.696099,1.486556) *[blue]{\scriptscriptstyle+};
(7.704312,1.487674) *[blue]{\scriptscriptstyle+};
(7.712526,1.490880) *[blue]{\scriptscriptstyle+};
(7.720739,1.499133) *[blue]{\scriptscriptstyle+};
(7.728953,1.504049) *[blue]{\scriptscriptstyle+};
(7.737166,1.507862) *[blue]{\scriptscriptstyle+};
(7.745380,1.516257) *[blue]{\scriptscriptstyle+};
(7.753593,1.520336) *[blue]{\scriptscriptstyle+};
(7.761807,1.510578) *[blue]{\scriptscriptstyle+};
(7.770021,1.511211) *[blue]{\scriptscriptstyle+};
(7.778234,1.515985) *[blue]{\scriptscriptstyle+};
(7.786448,1.525317) *[blue]{\scriptscriptstyle+};
(7.794661,1.531787) *[blue]{\scriptscriptstyle+};
(7.802875,1.536226) *[blue]{\scriptscriptstyle+};
(7.811088,1.536489) *[blue]{\scriptscriptstyle+};
(7.819302,1.547389) *[blue]{\scriptscriptstyle+};
(7.827515,1.560682) *[blue]{\scriptscriptstyle+};
(7.835729,1.578675) *[blue]{\scriptscriptstyle+};
(7.843943,1.603007) *[blue]{\scriptscriptstyle+};
(7.852156,1.610471) *[blue]{\scriptscriptstyle+};
(7.860370,1.622630) *[blue]{\scriptscriptstyle+};
(7.868583,1.652975) *[blue]{\scriptscriptstyle+};
(7.876797,1.671759) *[blue]{\scriptscriptstyle+};
(7.885010,1.559491) *[blue]{\scriptscriptstyle+};
(7.893224,1.563327) *[blue]{\scriptscriptstyle+};
(7.901437,1.564212) *[blue]{\scriptscriptstyle+};
(7.909651,1.570302) *[blue]{\scriptscriptstyle+};
(7.917864,1.570934) *[blue]{\scriptscriptstyle+};
(7.926078,1.572874) *[blue]{\scriptscriptstyle+};
(7.934292,1.573877) *[blue]{\scriptscriptstyle+};
(7.942505,1.592784) *[blue]{\scriptscriptstyle+};
(7.950719,1.597229) *[blue]{\scriptscriptstyle+};
(7.958932,1.601774) *[blue]{\scriptscriptstyle+};
(7.967146,1.606399) *[blue]{\scriptscriptstyle+};
(7.975359,1.613835) *[blue]{\scriptscriptstyle+};
(7.983573,1.618685) *[blue]{\scriptscriptstyle+};
(7.991786,1.635682) *[blue]{\scriptscriptstyle+};
(8.000000,1.670206) *[blue]{\scriptscriptstyle+};
(0.000000,0.978379) *[red]{\scriptscriptstyle\times};
(0.008214,0.996207) *[red]{\scriptscriptstyle\times};
(0.016427,1.004992) *[red]{\scriptscriptstyle\times};
(0.024641,1.008742) *[red]{\scriptscriptstyle\times};
(0.032854,1.016631) *[red]{\scriptscriptstyle\times};
(0.041068,1.016883) *[red]{\scriptscriptstyle\times};
(0.049281,1.019549) *[red]{\scriptscriptstyle\times};
(0.057495,1.024479) *[red]{\scriptscriptstyle\times};
(0.065708,1.027285) *[red]{\scriptscriptstyle\times};
(0.073922,1.029698) *[red]{\scriptscriptstyle\times};
(0.082136,1.030891) *[red]{\scriptscriptstyle\times};
(0.090349,1.041967) *[red]{\scriptscriptstyle\times};
(0.098563,1.042710) *[red]{\scriptscriptstyle\times};
(0.106776,1.045932) *[red]{\scriptscriptstyle\times};
(0.114990,1.073388) *[red]{\scriptscriptstyle\times};
(0.123203,1.016636) *[red]{\scriptscriptstyle\times};
(0.131417,1.018568) *[red]{\scriptscriptstyle\times};
(0.139630,1.032277) *[red]{\scriptscriptstyle\times};
(0.147844,1.034154) *[red]{\scriptscriptstyle\times};
(0.156057,1.035084) *[red]{\scriptscriptstyle\times};
(0.164271,1.037860) *[red]{\scriptscriptstyle\times};
(0.172485,1.041193) *[red]{\scriptscriptstyle\times};
(0.180698,1.043342) *[red]{\scriptscriptstyle\times};
(0.188912,1.047355) *[red]{\scriptscriptstyle\times};
(0.197125,1.058494) *[red]{\scriptscriptstyle\times};
(0.205339,1.071180) *[red]{\scriptscriptstyle\times};
(0.213552,1.072602) *[red]{\scriptscriptstyle\times};
(0.221766,1.073228) *[red]{\scriptscriptstyle\times};
(0.229979,1.078610) *[red]{\scriptscriptstyle\times};
(0.238193,1.095241) *[red]{\scriptscriptstyle\times};
(0.246407,1.044281) *[red]{\scriptscriptstyle\times};
(0.254620,1.047691) *[red]{\scriptscriptstyle\times};
(0.262834,1.052818) *[red]{\scriptscriptstyle\times};
(0.271047,1.056721) *[red]{\scriptscriptstyle\times};
(0.279261,1.056912) *[red]{\scriptscriptstyle\times};
(0.287474,1.062837) *[red]{\scriptscriptstyle\times};
(0.295688,1.065033) *[red]{\scriptscriptstyle\times};
(0.303901,1.065781) *[red]{\scriptscriptstyle\times};
(0.312115,1.066953) *[red]{\scriptscriptstyle\times};
(0.320329,1.076504) *[red]{\scriptscriptstyle\times};
(0.328542,1.080076) *[red]{\scriptscriptstyle\times};
(0.336756,1.101173) *[red]{\scriptscriptstyle\times};
(0.344969,1.108167) *[red]{\scriptscriptstyle\times};
(0.353183,1.108530) *[red]{\scriptscriptstyle\times};
(0.361396,1.316956) *[red]{\scriptscriptstyle\times};
(0.369610,1.045255) *[red]{\scriptscriptstyle\times};
(0.377823,1.047940) *[red]{\scriptscriptstyle\times};
(0.386037,1.053397) *[red]{\scriptscriptstyle\times};
(0.394251,1.054680) *[red]{\scriptscriptstyle\times};
(0.402464,1.055037) *[red]{\scriptscriptstyle\times};
(0.410678,1.067300) *[red]{\scriptscriptstyle\times};
(0.418891,1.077937) *[red]{\scriptscriptstyle\times};
(0.427105,1.080206) *[red]{\scriptscriptstyle\times};
(0.435318,1.080620) *[red]{\scriptscriptstyle\times};
(0.443532,1.081232) *[red]{\scriptscriptstyle\times};
(0.451745,1.082687) *[red]{\scriptscriptstyle\times};
(0.459959,1.083614) *[red]{\scriptscriptstyle\times};
(0.468172,1.103073) *[red]{\scriptscriptstyle\times};
(0.476386,1.103314) *[red]{\scriptscriptstyle\times};
(0.484600,1.128827) *[red]{\scriptscriptstyle\times};
(0.492813,1.032296) *[red]{\scriptscriptstyle\times};
(0.501027,1.034722) *[red]{\scriptscriptstyle\times};
(0.509240,1.048889) *[red]{\scriptscriptstyle\times};
(0.517454,1.057291) *[red]{\scriptscriptstyle\times};
(0.525667,1.072473) *[red]{\scriptscriptstyle\times};
(0.533881,1.077026) *[red]{\scriptscriptstyle\times};
(0.542094,1.081816) *[red]{\scriptscriptstyle\times};
(0.550308,1.087045) *[red]{\scriptscriptstyle\times};
(0.558522,1.088002) *[red]{\scriptscriptstyle\times};
(0.566735,1.094057) *[red]{\scriptscriptstyle\times};
(0.574949,1.096758) *[red]{\scriptscriptstyle\times};
(0.583162,1.108402) *[red]{\scriptscriptstyle\times};
(0.591376,1.113073) *[red]{\scriptscriptstyle\times};
(0.599589,1.139101) *[red]{\scriptscriptstyle\times};
(0.607803,1.208427) *[red]{\scriptscriptstyle\times};
(0.616016,1.049599) *[red]{\scriptscriptstyle\times};
(0.624230,1.057661) *[red]{\scriptscriptstyle\times};
(0.632444,1.076100) *[red]{\scriptscriptstyle\times};
(0.640657,1.076134) *[red]{\scriptscriptstyle\times};
(0.648871,1.078065) *[red]{\scriptscriptstyle\times};
(0.657084,1.078280) *[red]{\scriptscriptstyle\times};
(0.665298,1.078488) *[red]{\scriptscriptstyle\times};
(0.673511,1.081962) *[red]{\scriptscriptstyle\times};
(0.681725,1.085427) *[red]{\scriptscriptstyle\times};
(0.689938,1.087713) *[red]{\scriptscriptstyle\times};
(0.698152,1.106191) *[red]{\scriptscriptstyle\times};
(0.706366,1.107880) *[red]{\scriptscriptstyle\times};
(0.714579,1.109755) *[red]{\scriptscriptstyle\times};
(0.722793,1.128208) *[red]{\scriptscriptstyle\times};
(0.731006,1.205689) *[red]{\scriptscriptstyle\times};
(0.739220,1.075888) *[red]{\scriptscriptstyle\times};
(0.747433,1.077513) *[red]{\scriptscriptstyle\times};
(0.755647,1.078512) *[red]{\scriptscriptstyle\times};
(0.763860,1.078981) *[red]{\scriptscriptstyle\times};
(0.772074,1.079526) *[red]{\scriptscriptstyle\times};
(0.780287,1.080690) *[red]{\scriptscriptstyle\times};
(0.788501,1.082828) *[red]{\scriptscriptstyle\times};
(0.796715,1.083127) *[red]{\scriptscriptstyle\times};
(0.804928,1.089721) *[red]{\scriptscriptstyle\times};
(0.813142,1.091593) *[red]{\scriptscriptstyle\times};
(0.821355,1.093870) *[red]{\scriptscriptstyle\times};
(0.829569,1.100042) *[red]{\scriptscriptstyle\times};
(0.837782,1.102414) *[red]{\scriptscriptstyle\times};
(0.845996,1.106957) *[red]{\scriptscriptstyle\times};
(0.854209,1.107964) *[red]{\scriptscriptstyle\times};
(0.862423,1.098731) *[red]{\scriptscriptstyle\times};
(0.870637,1.100477) *[red]{\scriptscriptstyle\times};
(0.878850,1.101161) *[red]{\scriptscriptstyle\times};
(0.887064,1.105675) *[red]{\scriptscriptstyle\times};
(0.895277,1.107827) *[red]{\scriptscriptstyle\times};
(0.903491,1.110918) *[red]{\scriptscriptstyle\times};
(0.911704,1.113059) *[red]{\scriptscriptstyle\times};
(0.919918,1.120010) *[red]{\scriptscriptstyle\times};
(0.928131,1.122801) *[red]{\scriptscriptstyle\times};
(0.936345,1.123678) *[red]{\scriptscriptstyle\times};
(0.944559,1.127525) *[red]{\scriptscriptstyle\times};
(0.952772,1.129310) *[red]{\scriptscriptstyle\times};
(0.960986,1.130569) *[red]{\scriptscriptstyle\times};
(0.969199,1.135167) *[red]{\scriptscriptstyle\times};
(0.977413,1.162882) *[red]{\scriptscriptstyle\times};
(0.985626,1.108932) *[red]{\scriptscriptstyle\times};
(0.993840,1.112428) *[red]{\scriptscriptstyle\times};
(1.002053,1.122857) *[red]{\scriptscriptstyle\times};
(1.010267,1.130926) *[red]{\scriptscriptstyle\times};
(1.018480,1.133794) *[red]{\scriptscriptstyle\times};
(1.026694,1.138152) *[red]{\scriptscriptstyle\times};
(1.034908,1.139473) *[red]{\scriptscriptstyle\times};
(1.043121,1.152070) *[red]{\scriptscriptstyle\times};
(1.051335,1.152261) *[red]{\scriptscriptstyle\times};
(1.059548,1.155166) *[red]{\scriptscriptstyle\times};
(1.067762,1.158694) *[red]{\scriptscriptstyle\times};
(1.075975,1.159112) *[red]{\scriptscriptstyle\times};
(1.084189,1.159408) *[red]{\scriptscriptstyle\times};
(1.092402,1.159579) *[red]{\scriptscriptstyle\times};
(1.100616,1.171688) *[red]{\scriptscriptstyle\times};
(1.108830,1.113757) *[red]{\scriptscriptstyle\times};
(1.117043,1.119605) *[red]{\scriptscriptstyle\times};
(1.125257,1.122195) *[red]{\scriptscriptstyle\times};
(1.133470,1.129683) *[red]{\scriptscriptstyle\times};
(1.141684,1.141780) *[red]{\scriptscriptstyle\times};
(1.149897,1.145030) *[red]{\scriptscriptstyle\times};
(1.158111,1.145157) *[red]{\scriptscriptstyle\times};
(1.166324,1.147275) *[red]{\scriptscriptstyle\times};
(1.174538,1.149487) *[red]{\scriptscriptstyle\times};
(1.182752,1.150828) *[red]{\scriptscriptstyle\times};
(1.190965,1.151297) *[red]{\scriptscriptstyle\times};
(1.199179,1.151367) *[red]{\scriptscriptstyle\times};
(1.207392,1.163690) *[red]{\scriptscriptstyle\times};
(1.215606,1.168837) *[red]{\scriptscriptstyle\times};
(1.223819,1.188851) *[red]{\scriptscriptstyle\times};
(1.232033,1.124915) *[red]{\scriptscriptstyle\times};
(1.240246,1.127900) *[red]{\scriptscriptstyle\times};
(1.248460,1.128724) *[red]{\scriptscriptstyle\times};
(1.256674,1.129413) *[red]{\scriptscriptstyle\times};
(1.264887,1.130722) *[red]{\scriptscriptstyle\times};
(1.273101,1.133785) *[red]{\scriptscriptstyle\times};
(1.281314,1.143469) *[red]{\scriptscriptstyle\times};
(1.289528,1.148461) *[red]{\scriptscriptstyle\times};
(1.297741,1.155035) *[red]{\scriptscriptstyle\times};
(1.305955,1.157086) *[red]{\scriptscriptstyle\times};
(1.314168,1.157121) *[red]{\scriptscriptstyle\times};
(1.322382,1.158294) *[red]{\scriptscriptstyle\times};
(1.330595,1.183286) *[red]{\scriptscriptstyle\times};
(1.338809,1.185040) *[red]{\scriptscriptstyle\times};
(1.347023,1.196638) *[red]{\scriptscriptstyle\times};
(1.355236,1.109203) *[red]{\scriptscriptstyle\times};
(1.363450,1.110407) *[red]{\scriptscriptstyle\times};
(1.371663,1.111524) *[red]{\scriptscriptstyle\times};
(1.379877,1.113770) *[red]{\scriptscriptstyle\times};
(1.388090,1.127781) *[red]{\scriptscriptstyle\times};
(1.396304,1.159911) *[red]{\scriptscriptstyle\times};
(1.404517,1.163418) *[red]{\scriptscriptstyle\times};
(1.412731,1.163854) *[red]{\scriptscriptstyle\times};
(1.420945,1.164906) *[red]{\scriptscriptstyle\times};
(1.429158,1.173653) *[red]{\scriptscriptstyle\times};
(1.437372,1.181520) *[red]{\scriptscriptstyle\times};
(1.445585,1.194659) *[red]{\scriptscriptstyle\times};
(1.453799,1.203990) *[red]{\scriptscriptstyle\times};
(1.462012,1.227866) *[red]{\scriptscriptstyle\times};
(1.470226,1.252678) *[red]{\scriptscriptstyle\times};
(1.478439,1.136352) *[red]{\scriptscriptstyle\times};
(1.486653,1.139306) *[red]{\scriptscriptstyle\times};
(1.494867,1.139469) *[red]{\scriptscriptstyle\times};
(1.503080,1.145670) *[red]{\scriptscriptstyle\times};
(1.511294,1.146508) *[red]{\scriptscriptstyle\times};
(1.519507,1.148408) *[red]{\scriptscriptstyle\times};
(1.527721,1.159403) *[red]{\scriptscriptstyle\times};
(1.535934,1.162959) *[red]{\scriptscriptstyle\times};
(1.544148,1.166894) *[red]{\scriptscriptstyle\times};
(1.552361,1.168526) *[red]{\scriptscriptstyle\times};
(1.560575,1.169519) *[red]{\scriptscriptstyle\times};
(1.568789,1.170786) *[red]{\scriptscriptstyle\times};
(1.577002,1.172491) *[red]{\scriptscriptstyle\times};
(1.585216,1.173012) *[red]{\scriptscriptstyle\times};
(1.593429,1.173309) *[red]{\scriptscriptstyle\times};
(1.601643,1.142723) *[red]{\scriptscriptstyle\times};
(1.609856,1.151572) *[red]{\scriptscriptstyle\times};
(1.618070,1.160482) *[red]{\scriptscriptstyle\times};
(1.626283,1.164047) *[red]{\scriptscriptstyle\times};
(1.634497,1.165773) *[red]{\scriptscriptstyle\times};
(1.642710,1.166095) *[red]{\scriptscriptstyle\times};
(1.650924,1.166800) *[red]{\scriptscriptstyle\times};
(1.659138,1.169354) *[red]{\scriptscriptstyle\times};
(1.667351,1.171429) *[red]{\scriptscriptstyle\times};
(1.675565,1.177748) *[red]{\scriptscriptstyle\times};
(1.683778,1.181585) *[red]{\scriptscriptstyle\times};
(1.691992,1.183368) *[red]{\scriptscriptstyle\times};
(1.700205,1.199354) *[red]{\scriptscriptstyle\times};
(1.708419,1.205832) *[red]{\scriptscriptstyle\times};
(1.716632,1.219008) *[red]{\scriptscriptstyle\times};
(1.724846,1.128970) *[red]{\scriptscriptstyle\times};
(1.733060,1.133800) *[red]{\scriptscriptstyle\times};
(1.741273,1.154965) *[red]{\scriptscriptstyle\times};
(1.749487,1.156944) *[red]{\scriptscriptstyle\times};
(1.757700,1.161822) *[red]{\scriptscriptstyle\times};
(1.765914,1.163615) *[red]{\scriptscriptstyle\times};
(1.774127,1.166435) *[red]{\scriptscriptstyle\times};
(1.782341,1.172267) *[red]{\scriptscriptstyle\times};
(1.790554,1.181243) *[red]{\scriptscriptstyle\times};
(1.798768,1.185303) *[red]{\scriptscriptstyle\times};
(1.806982,1.185952) *[red]{\scriptscriptstyle\times};
(1.815195,1.186086) *[red]{\scriptscriptstyle\times};
(1.823409,1.196267) *[red]{\scriptscriptstyle\times};
(1.831622,1.208276) *[red]{\scriptscriptstyle\times};
(1.839836,1.219263) *[red]{\scriptscriptstyle\times};
(1.848049,1.132178) *[red]{\scriptscriptstyle\times};
(1.856263,1.140508) *[red]{\scriptscriptstyle\times};
(1.864476,1.162896) *[red]{\scriptscriptstyle\times};
(1.872690,1.165430) *[red]{\scriptscriptstyle\times};
(1.880903,1.169429) *[red]{\scriptscriptstyle\times};
(1.889117,1.181445) *[red]{\scriptscriptstyle\times};
(1.897331,1.187062) *[red]{\scriptscriptstyle\times};
(1.905544,1.187964) *[red]{\scriptscriptstyle\times};
(1.913758,1.191921) *[red]{\scriptscriptstyle\times};
(1.921971,1.199073) *[red]{\scriptscriptstyle\times};
(1.930185,1.199497) *[red]{\scriptscriptstyle\times};
(1.938398,1.229114) *[red]{\scriptscriptstyle\times};
(1.946612,1.231649) *[red]{\scriptscriptstyle\times};
(1.954825,1.234484) *[red]{\scriptscriptstyle\times};
(1.963039,1.249722) *[red]{\scriptscriptstyle\times};
(1.971253,1.149500) *[red]{\scriptscriptstyle\times};
(1.979466,1.154379) *[red]{\scriptscriptstyle\times};
(1.987680,1.155667) *[red]{\scriptscriptstyle\times};
(1.995893,1.170591) *[red]{\scriptscriptstyle\times};
(2.004107,1.179193) *[red]{\scriptscriptstyle\times};
(2.012320,1.180237) *[red]{\scriptscriptstyle\times};
(2.020534,1.182645) *[red]{\scriptscriptstyle\times};
(2.028747,1.184947) *[red]{\scriptscriptstyle\times};
(2.036961,1.189244) *[red]{\scriptscriptstyle\times};
(2.045175,1.198935) *[red]{\scriptscriptstyle\times};
(2.053388,1.201782) *[red]{\scriptscriptstyle\times};
(2.061602,1.204615) *[red]{\scriptscriptstyle\times};
(2.069815,1.211175) *[red]{\scriptscriptstyle\times};
(2.078029,1.224782) *[red]{\scriptscriptstyle\times};
(2.086242,1.230077) *[red]{\scriptscriptstyle\times};
(2.094456,1.172070) *[red]{\scriptscriptstyle\times};
(2.102669,1.172165) *[red]{\scriptscriptstyle\times};
(2.110883,1.173927) *[red]{\scriptscriptstyle\times};
(2.119097,1.175521) *[red]{\scriptscriptstyle\times};
(2.127310,1.176855) *[red]{\scriptscriptstyle\times};
(2.135524,1.178153) *[red]{\scriptscriptstyle\times};
(2.143737,1.178560) *[red]{\scriptscriptstyle\times};
(2.151951,1.183926) *[red]{\scriptscriptstyle\times};
(2.160164,1.193719) *[red]{\scriptscriptstyle\times};
(2.168378,1.197211) *[red]{\scriptscriptstyle\times};
(2.176591,1.198395) *[red]{\scriptscriptstyle\times};
(2.184805,1.208141) *[red]{\scriptscriptstyle\times};
(2.193018,1.209181) *[red]{\scriptscriptstyle\times};
(2.201232,1.210422) *[red]{\scriptscriptstyle\times};
(2.209446,1.219489) *[red]{\scriptscriptstyle\times};
(2.217659,1.148824) *[red]{\scriptscriptstyle\times};
(2.225873,1.161424) *[red]{\scriptscriptstyle\times};
(2.234086,1.167946) *[red]{\scriptscriptstyle\times};
(2.242300,1.169227) *[red]{\scriptscriptstyle\times};
(2.250513,1.175037) *[red]{\scriptscriptstyle\times};
(2.258727,1.175101) *[red]{\scriptscriptstyle\times};
(2.266940,1.175834) *[red]{\scriptscriptstyle\times};
(2.275154,1.183157) *[red]{\scriptscriptstyle\times};
(2.283368,1.200729) *[red]{\scriptscriptstyle\times};
(2.291581,1.201005) *[red]{\scriptscriptstyle\times};
(2.299795,1.202313) *[red]{\scriptscriptstyle\times};
(2.308008,1.204394) *[red]{\scriptscriptstyle\times};
(2.316222,1.207326) *[red]{\scriptscriptstyle\times};
(2.324435,1.221391) *[red]{\scriptscriptstyle\times};
(2.332649,1.313732) *[red]{\scriptscriptstyle\times};
(2.340862,1.159750) *[red]{\scriptscriptstyle\times};
(2.349076,1.161178) *[red]{\scriptscriptstyle\times};
(2.357290,1.163041) *[red]{\scriptscriptstyle\times};
(2.365503,1.163427) *[red]{\scriptscriptstyle\times};
(2.373717,1.170258) *[red]{\scriptscriptstyle\times};
(2.381930,1.171749) *[red]{\scriptscriptstyle\times};
(2.390144,1.192230) *[red]{\scriptscriptstyle\times};
(2.398357,1.193574) *[red]{\scriptscriptstyle\times};
(2.406571,1.207934) *[red]{\scriptscriptstyle\times};
(2.414784,1.219317) *[red]{\scriptscriptstyle\times};
(2.422998,1.219470) *[red]{\scriptscriptstyle\times};
(2.431211,1.223381) *[red]{\scriptscriptstyle\times};
(2.439425,1.225231) *[red]{\scriptscriptstyle\times};
(2.447639,1.229701) *[red]{\scriptscriptstyle\times};
(2.455852,1.250576) *[red]{\scriptscriptstyle\times};
(2.464066,1.208924) *[red]{\scriptscriptstyle\times};
(2.472279,1.210628) *[red]{\scriptscriptstyle\times};
(2.480493,1.213450) *[red]{\scriptscriptstyle\times};
(2.488706,1.214447) *[red]{\scriptscriptstyle\times};
(2.496920,1.214569) *[red]{\scriptscriptstyle\times};
(2.505133,1.218164) *[red]{\scriptscriptstyle\times};
(2.513347,1.218404) *[red]{\scriptscriptstyle\times};
(2.521561,1.219063) *[red]{\scriptscriptstyle\times};
(2.529774,1.223395) *[red]{\scriptscriptstyle\times};
(2.537988,1.225795) *[red]{\scriptscriptstyle\times};
(2.546201,1.226742) *[red]{\scriptscriptstyle\times};
(2.554415,1.233224) *[red]{\scriptscriptstyle\times};
(2.562628,1.245563) *[red]{\scriptscriptstyle\times};
(2.570842,1.284693) *[red]{\scriptscriptstyle\times};
(2.579055,1.297924) *[red]{\scriptscriptstyle\times};
(2.587269,1.171516) *[red]{\scriptscriptstyle\times};
(2.595483,1.192356) *[red]{\scriptscriptstyle\times};
(2.603696,1.192922) *[red]{\scriptscriptstyle\times};
(2.611910,1.199343) *[red]{\scriptscriptstyle\times};
(2.620123,1.216988) *[red]{\scriptscriptstyle\times};
(2.628337,1.217155) *[red]{\scriptscriptstyle\times};
(2.636550,1.221067) *[red]{\scriptscriptstyle\times};
(2.644764,1.223006) *[red]{\scriptscriptstyle\times};
(2.652977,1.223064) *[red]{\scriptscriptstyle\times};
(2.661191,1.234604) *[red]{\scriptscriptstyle\times};
(2.669405,1.251333) *[red]{\scriptscriptstyle\times};
(2.677618,1.253194) *[red]{\scriptscriptstyle\times};
(2.685832,1.254248) *[red]{\scriptscriptstyle\times};
(2.694045,1.255820) *[red]{\scriptscriptstyle\times};
(2.702259,1.279222) *[red]{\scriptscriptstyle\times};
(2.710472,1.176915) *[red]{\scriptscriptstyle\times};
(2.718686,1.196406) *[red]{\scriptscriptstyle\times};
(2.726899,1.211100) *[red]{\scriptscriptstyle\times};
(2.735113,1.212810) *[red]{\scriptscriptstyle\times};
(2.743326,1.216318) *[red]{\scriptscriptstyle\times};
(2.751540,1.221439) *[red]{\scriptscriptstyle\times};
(2.759754,1.222970) *[red]{\scriptscriptstyle\times};
(2.767967,1.223661) *[red]{\scriptscriptstyle\times};
(2.776181,1.246709) *[red]{\scriptscriptstyle\times};
(2.784394,1.247834) *[red]{\scriptscriptstyle\times};
(2.792608,1.272934) *[red]{\scriptscriptstyle\times};
(2.800821,1.275730) *[red]{\scriptscriptstyle\times};
(2.809035,1.281567) *[red]{\scriptscriptstyle\times};
(2.817248,1.296128) *[red]{\scriptscriptstyle\times};
(2.825462,1.324679) *[red]{\scriptscriptstyle\times};
(2.833676,1.202724) *[red]{\scriptscriptstyle\times};
(2.841889,1.203872) *[red]{\scriptscriptstyle\times};
(2.850103,1.204820) *[red]{\scriptscriptstyle\times};
(2.858316,1.206634) *[red]{\scriptscriptstyle\times};
(2.866530,1.212662) *[red]{\scriptscriptstyle\times};
(2.874743,1.228437) *[red]{\scriptscriptstyle\times};
(2.882957,1.229188) *[red]{\scriptscriptstyle\times};
(2.891170,1.230098) *[red]{\scriptscriptstyle\times};
(2.899384,1.238277) *[red]{\scriptscriptstyle\times};
(2.907598,1.240540) *[red]{\scriptscriptstyle\times};
(2.915811,1.242819) *[red]{\scriptscriptstyle\times};
(2.924025,1.244026) *[red]{\scriptscriptstyle\times};
(2.932238,1.259218) *[red]{\scriptscriptstyle\times};
(2.940452,1.259744) *[red]{\scriptscriptstyle\times};
(2.948665,1.267325) *[red]{\scriptscriptstyle\times};
(2.956879,1.203219) *[red]{\scriptscriptstyle\times};
(2.965092,1.204616) *[red]{\scriptscriptstyle\times};
(2.973306,1.207352) *[red]{\scriptscriptstyle\times};
(2.981520,1.208913) *[red]{\scriptscriptstyle\times};
(2.989733,1.209357) *[red]{\scriptscriptstyle\times};
(2.997947,1.212945) *[red]{\scriptscriptstyle\times};
(3.006160,1.213307) *[red]{\scriptscriptstyle\times};
(3.014374,1.219575) *[red]{\scriptscriptstyle\times};
(3.022587,1.223021) *[red]{\scriptscriptstyle\times};
(3.030801,1.228971) *[red]{\scriptscriptstyle\times};
(3.039014,1.231961) *[red]{\scriptscriptstyle\times};
(3.047228,1.263879) *[red]{\scriptscriptstyle\times};
(3.055441,1.270987) *[red]{\scriptscriptstyle\times};
(3.063655,1.283699) *[red]{\scriptscriptstyle\times};
(3.071869,1.296125) *[red]{\scriptscriptstyle\times};
(3.080082,1.199904) *[red]{\scriptscriptstyle\times};
(3.088296,1.203265) *[red]{\scriptscriptstyle\times};
(3.096509,1.205192) *[red]{\scriptscriptstyle\times};
(3.104723,1.205525) *[red]{\scriptscriptstyle\times};
(3.112936,1.207852) *[red]{\scriptscriptstyle\times};
(3.121150,1.215603) *[red]{\scriptscriptstyle\times};
(3.129363,1.231625) *[red]{\scriptscriptstyle\times};
(3.137577,1.231633) *[red]{\scriptscriptstyle\times};
(3.145791,1.231960) *[red]{\scriptscriptstyle\times};
(3.154004,1.254839) *[red]{\scriptscriptstyle\times};
(3.162218,1.257196) *[red]{\scriptscriptstyle\times};
(3.170431,1.257871) *[red]{\scriptscriptstyle\times};
(3.178645,1.263465) *[red]{\scriptscriptstyle\times};
(3.186858,1.267540) *[red]{\scriptscriptstyle\times};
(3.195072,1.303294) *[red]{\scriptscriptstyle\times};
(3.203285,1.197327) *[red]{\scriptscriptstyle\times};
(3.211499,1.201129) *[red]{\scriptscriptstyle\times};
(3.219713,1.222942) *[red]{\scriptscriptstyle\times};
(3.227926,1.223962) *[red]{\scriptscriptstyle\times};
(3.236140,1.225987) *[red]{\scriptscriptstyle\times};
(3.244353,1.226562) *[red]{\scriptscriptstyle\times};
(3.252567,1.230040) *[red]{\scriptscriptstyle\times};
(3.260780,1.230427) *[red]{\scriptscriptstyle\times};
(3.268994,1.235048) *[red]{\scriptscriptstyle\times};
(3.277207,1.243482) *[red]{\scriptscriptstyle\times};
(3.285421,1.249028) *[red]{\scriptscriptstyle\times};
(3.293634,1.249553) *[red]{\scriptscriptstyle\times};
(3.301848,1.251070) *[red]{\scriptscriptstyle\times};
(3.310062,1.274177) *[red]{\scriptscriptstyle\times};
(3.318275,1.285318) *[red]{\scriptscriptstyle\times};
(3.326489,1.202397) *[red]{\scriptscriptstyle\times};
(3.334702,1.202436) *[red]{\scriptscriptstyle\times};
(3.342916,1.206006) *[red]{\scriptscriptstyle\times};
(3.351129,1.207356) *[red]{\scriptscriptstyle\times};
(3.359343,1.216420) *[red]{\scriptscriptstyle\times};
(3.367556,1.228647) *[red]{\scriptscriptstyle\times};
(3.375770,1.231607) *[red]{\scriptscriptstyle\times};
(3.383984,1.240003) *[red]{\scriptscriptstyle\times};
(3.392197,1.253046) *[red]{\scriptscriptstyle\times};
(3.400411,1.255598) *[red]{\scriptscriptstyle\times};
(3.408624,1.256590) *[red]{\scriptscriptstyle\times};
(3.416838,1.261223) *[red]{\scriptscriptstyle\times};
(3.425051,1.264278) *[red]{\scriptscriptstyle\times};
(3.433265,1.273770) *[red]{\scriptscriptstyle\times};
(3.441478,1.293463) *[red]{\scriptscriptstyle\times};
(3.449692,1.211824) *[red]{\scriptscriptstyle\times};
(3.457906,1.216493) *[red]{\scriptscriptstyle\times};
(3.466119,1.218559) *[red]{\scriptscriptstyle\times};
(3.474333,1.234950) *[red]{\scriptscriptstyle\times};
(3.482546,1.235782) *[red]{\scriptscriptstyle\times};
(3.490760,1.241379) *[red]{\scriptscriptstyle\times};
(3.498973,1.251437) *[red]{\scriptscriptstyle\times};
(3.507187,1.253319) *[red]{\scriptscriptstyle\times};
(3.515400,1.255028) *[red]{\scriptscriptstyle\times};
(3.523614,1.260748) *[red]{\scriptscriptstyle\times};
(3.531828,1.264399) *[red]{\scriptscriptstyle\times};
(3.540041,1.266327) *[red]{\scriptscriptstyle\times};
(3.548255,1.267520) *[red]{\scriptscriptstyle\times};
(3.556468,1.267985) *[red]{\scriptscriptstyle\times};
(3.564682,1.292572) *[red]{\scriptscriptstyle\times};
(3.572895,1.248855) *[red]{\scriptscriptstyle\times};
(3.581109,1.255396) *[red]{\scriptscriptstyle\times};
(3.589322,1.256659) *[red]{\scriptscriptstyle\times};
(3.597536,1.256921) *[red]{\scriptscriptstyle\times};
(3.605749,1.257352) *[red]{\scriptscriptstyle\times};
(3.613963,1.258880) *[red]{\scriptscriptstyle\times};
(3.622177,1.260326) *[red]{\scriptscriptstyle\times};
(3.630390,1.260407) *[red]{\scriptscriptstyle\times};
(3.638604,1.262348) *[red]{\scriptscriptstyle\times};
(3.646817,1.263975) *[red]{\scriptscriptstyle\times};
(3.655031,1.271590) *[red]{\scriptscriptstyle\times};
(3.663244,1.290837) *[red]{\scriptscriptstyle\times};
(3.671458,1.290868) *[red]{\scriptscriptstyle\times};
(3.679671,1.300091) *[red]{\scriptscriptstyle\times};
(3.687885,1.324819) *[red]{\scriptscriptstyle\times};
(3.696099,1.217546) *[red]{\scriptscriptstyle\times};
(3.704312,1.219954) *[red]{\scriptscriptstyle\times};
(3.712526,1.220043) *[red]{\scriptscriptstyle\times};
(3.720739,1.223220) *[red]{\scriptscriptstyle\times};
(3.728953,1.223514) *[red]{\scriptscriptstyle\times};
(3.737166,1.231300) *[red]{\scriptscriptstyle\times};
(3.745380,1.235549) *[red]{\scriptscriptstyle\times};
(3.753593,1.248110) *[red]{\scriptscriptstyle\times};
(3.761807,1.249744) *[red]{\scriptscriptstyle\times};
(3.770021,1.267906) *[red]{\scriptscriptstyle\times};
(3.778234,1.271099) *[red]{\scriptscriptstyle\times};
(3.786448,1.272465) *[red]{\scriptscriptstyle\times};
(3.794661,1.292702) *[red]{\scriptscriptstyle\times};
(3.802875,1.311363) *[red]{\scriptscriptstyle\times};
(3.811088,1.339973) *[red]{\scriptscriptstyle\times};
(3.819302,1.205561) *[red]{\scriptscriptstyle\times};
(3.827515,1.206027) *[red]{\scriptscriptstyle\times};
(3.835729,1.234004) *[red]{\scriptscriptstyle\times};
(3.843943,1.235294) *[red]{\scriptscriptstyle\times};
(3.852156,1.236926) *[red]{\scriptscriptstyle\times};
(3.860370,1.242002) *[red]{\scriptscriptstyle\times};
(3.868583,1.248999) *[red]{\scriptscriptstyle\times};
(3.876797,1.249719) *[red]{\scriptscriptstyle\times};
(3.885010,1.257386) *[red]{\scriptscriptstyle\times};
(3.893224,1.262825) *[red]{\scriptscriptstyle\times};
(3.901437,1.273160) *[red]{\scriptscriptstyle\times};
(3.909651,1.279580) *[red]{\scriptscriptstyle\times};
(3.917864,1.282047) *[red]{\scriptscriptstyle\times};
(3.926078,1.294288) *[red]{\scriptscriptstyle\times};
(3.934292,1.301449) *[red]{\scriptscriptstyle\times};
(3.942505,1.233734) *[red]{\scriptscriptstyle\times};
(3.950719,1.234138) *[red]{\scriptscriptstyle\times};
(3.958932,1.239460) *[red]{\scriptscriptstyle\times};
(3.967146,1.244720) *[red]{\scriptscriptstyle\times};
(3.975359,1.245687) *[red]{\scriptscriptstyle\times};
(3.983573,1.246911) *[red]{\scriptscriptstyle\times};
(3.991786,1.259190) *[red]{\scriptscriptstyle\times};
(4.000000,1.261431) *[red]{\scriptscriptstyle\times};
(4.008214,1.264311) *[red]{\scriptscriptstyle\times};
(4.016427,1.266399) *[red]{\scriptscriptstyle\times};
(4.024641,1.273134) *[red]{\scriptscriptstyle\times};
(4.032854,1.287439) *[red]{\scriptscriptstyle\times};
(4.041068,1.291670) *[red]{\scriptscriptstyle\times};
(4.049281,1.317316) *[red]{\scriptscriptstyle\times};
(4.057495,1.328695) *[red]{\scriptscriptstyle\times};
(4.065708,1.236050) *[red]{\scriptscriptstyle\times};
(4.073922,1.240766) *[red]{\scriptscriptstyle\times};
(4.082136,1.252064) *[red]{\scriptscriptstyle\times};
(4.090349,1.258150) *[red]{\scriptscriptstyle\times};
(4.098563,1.258596) *[red]{\scriptscriptstyle\times};
(4.106776,1.259355) *[red]{\scriptscriptstyle\times};
(4.114990,1.262986) *[red]{\scriptscriptstyle\times};
(4.123203,1.265429) *[red]{\scriptscriptstyle\times};
(4.131417,1.266652) *[red]{\scriptscriptstyle\times};
(4.139630,1.269951) *[red]{\scriptscriptstyle\times};
(4.147844,1.271092) *[red]{\scriptscriptstyle\times};
(4.156057,1.276098) *[red]{\scriptscriptstyle\times};
(4.164271,1.280726) *[red]{\scriptscriptstyle\times};
(4.172485,1.326365) *[red]{\scriptscriptstyle\times};
(4.180698,1.334106) *[red]{\scriptscriptstyle\times};
(4.188912,1.205559) *[red]{\scriptscriptstyle\times};
(4.197125,1.210923) *[red]{\scriptscriptstyle\times};
(4.205339,1.219224) *[red]{\scriptscriptstyle\times};
(4.213552,1.229795) *[red]{\scriptscriptstyle\times};
(4.221766,1.235452) *[red]{\scriptscriptstyle\times};
(4.229979,1.235780) *[red]{\scriptscriptstyle\times};
(4.238193,1.239453) *[red]{\scriptscriptstyle\times};
(4.246407,1.241347) *[red]{\scriptscriptstyle\times};
(4.254620,1.256546) *[red]{\scriptscriptstyle\times};
(4.262834,1.257398) *[red]{\scriptscriptstyle\times};
(4.271047,1.258820) *[red]{\scriptscriptstyle\times};
(4.279261,1.262448) *[red]{\scriptscriptstyle\times};
(4.287474,1.276309) *[red]{\scriptscriptstyle\times};
(4.295688,1.295283) *[red]{\scriptscriptstyle\times};
(4.303901,1.319261) *[red]{\scriptscriptstyle\times};
(4.312115,1.245987) *[red]{\scriptscriptstyle\times};
(4.320329,1.251587) *[red]{\scriptscriptstyle\times};
(4.328542,1.252737) *[red]{\scriptscriptstyle\times};
(4.336756,1.254774) *[red]{\scriptscriptstyle\times};
(4.344969,1.255404) *[red]{\scriptscriptstyle\times};
(4.353183,1.261394) *[red]{\scriptscriptstyle\times};
(4.361396,1.261944) *[red]{\scriptscriptstyle\times};
(4.369610,1.263042) *[red]{\scriptscriptstyle\times};
(4.377823,1.272169) *[red]{\scriptscriptstyle\times};
(4.386037,1.272207) *[red]{\scriptscriptstyle\times};
(4.394251,1.280616) *[red]{\scriptscriptstyle\times};
(4.402464,1.283129) *[red]{\scriptscriptstyle\times};
(4.410678,1.290949) *[red]{\scriptscriptstyle\times};
(4.418891,1.309719) *[red]{\scriptscriptstyle\times};
(4.427105,1.314212) *[red]{\scriptscriptstyle\times};
(4.435318,1.256335) *[red]{\scriptscriptstyle\times};
(4.443532,1.257933) *[red]{\scriptscriptstyle\times};
(4.451745,1.258307) *[red]{\scriptscriptstyle\times};
(4.459959,1.258966) *[red]{\scriptscriptstyle\times};
(4.468172,1.259675) *[red]{\scriptscriptstyle\times};
(4.476386,1.260174) *[red]{\scriptscriptstyle\times};
(4.484600,1.260428) *[red]{\scriptscriptstyle\times};
(4.492813,1.263684) *[red]{\scriptscriptstyle\times};
(4.501027,1.264948) *[red]{\scriptscriptstyle\times};
(4.509240,1.269219) *[red]{\scriptscriptstyle\times};
(4.517454,1.275191) *[red]{\scriptscriptstyle\times};
(4.525667,1.276698) *[red]{\scriptscriptstyle\times};
(4.533881,1.279427) *[red]{\scriptscriptstyle\times};
(4.542094,1.287746) *[red]{\scriptscriptstyle\times};
(4.550308,1.291260) *[red]{\scriptscriptstyle\times};
(4.558522,1.241599) *[red]{\scriptscriptstyle\times};
(4.566735,1.245709) *[red]{\scriptscriptstyle\times};
(4.574949,1.246724) *[red]{\scriptscriptstyle\times};
(4.583162,1.252230) *[red]{\scriptscriptstyle\times};
(4.591376,1.252766) *[red]{\scriptscriptstyle\times};
(4.599589,1.252897) *[red]{\scriptscriptstyle\times};
(4.607803,1.254595) *[red]{\scriptscriptstyle\times};
(4.616016,1.264199) *[red]{\scriptscriptstyle\times};
(4.624230,1.270779) *[red]{\scriptscriptstyle\times};
(4.632444,1.275748) *[red]{\scriptscriptstyle\times};
(4.640657,1.278456) *[red]{\scriptscriptstyle\times};
(4.648871,1.281072) *[red]{\scriptscriptstyle\times};
(4.657084,1.310099) *[red]{\scriptscriptstyle\times};
(4.665298,1.327711) *[red]{\scriptscriptstyle\times};
(4.673511,1.354029) *[red]{\scriptscriptstyle\times};
(4.681725,1.253833) *[red]{\scriptscriptstyle\times};
(4.689938,1.257174) *[red]{\scriptscriptstyle\times};
(4.698152,1.258003) *[red]{\scriptscriptstyle\times};
(4.706366,1.262818) *[red]{\scriptscriptstyle\times};
(4.714579,1.264147) *[red]{\scriptscriptstyle\times};
(4.722793,1.268436) *[red]{\scriptscriptstyle\times};
(4.731006,1.279510) *[red]{\scriptscriptstyle\times};
(4.739220,1.282409) *[red]{\scriptscriptstyle\times};
(4.747433,1.285384) *[red]{\scriptscriptstyle\times};
(4.755647,1.285403) *[red]{\scriptscriptstyle\times};
(4.763860,1.285500) *[red]{\scriptscriptstyle\times};
(4.772074,1.312200) *[red]{\scriptscriptstyle\times};
(4.780287,1.313189) *[red]{\scriptscriptstyle\times};
(4.788501,1.329786) *[red]{\scriptscriptstyle\times};
(4.796715,1.331152) *[red]{\scriptscriptstyle\times};
(4.804928,1.252893) *[red]{\scriptscriptstyle\times};
(4.813142,1.264904) *[red]{\scriptscriptstyle\times};
(4.821355,1.265405) *[red]{\scriptscriptstyle\times};
(4.829569,1.268307) *[red]{\scriptscriptstyle\times};
(4.837782,1.275556) *[red]{\scriptscriptstyle\times};
(4.845996,1.276522) *[red]{\scriptscriptstyle\times};
(4.854209,1.279515) *[red]{\scriptscriptstyle\times};
(4.862423,1.283613) *[red]{\scriptscriptstyle\times};
(4.870637,1.283682) *[red]{\scriptscriptstyle\times};
(4.878850,1.296370) *[red]{\scriptscriptstyle\times};
(4.887064,1.301944) *[red]{\scriptscriptstyle\times};
(4.895277,1.318061) *[red]{\scriptscriptstyle\times};
(4.903491,1.318529) *[red]{\scriptscriptstyle\times};
(4.911704,1.339019) *[red]{\scriptscriptstyle\times};
(4.919918,1.348495) *[red]{\scriptscriptstyle\times};
(4.928131,1.263066) *[red]{\scriptscriptstyle\times};
(4.936345,1.281840) *[red]{\scriptscriptstyle\times};
(4.944559,1.288911) *[red]{\scriptscriptstyle\times};
(4.952772,1.291118) *[red]{\scriptscriptstyle\times};
(4.960986,1.292419) *[red]{\scriptscriptstyle\times};
(4.969199,1.296577) *[red]{\scriptscriptstyle\times};
(4.977413,1.304793) *[red]{\scriptscriptstyle\times};
(4.985626,1.308424) *[red]{\scriptscriptstyle\times};
(4.993840,1.308527) *[red]{\scriptscriptstyle\times};
(5.002053,1.331930) *[red]{\scriptscriptstyle\times};
(5.010267,1.333832) *[red]{\scriptscriptstyle\times};
(5.018480,1.348704) *[red]{\scriptscriptstyle\times};
(5.026694,1.349380) *[red]{\scriptscriptstyle\times};
(5.034908,1.364503) *[red]{\scriptscriptstyle\times};
(5.043121,1.533074) *[red]{\scriptscriptstyle\times};
(5.051335,1.280368) *[red]{\scriptscriptstyle\times};
(5.059548,1.283182) *[red]{\scriptscriptstyle\times};
(5.067762,1.284140) *[red]{\scriptscriptstyle\times};
(5.075975,1.285152) *[red]{\scriptscriptstyle\times};
(5.084189,1.285488) *[red]{\scriptscriptstyle\times};
(5.092402,1.287873) *[red]{\scriptscriptstyle\times};
(5.100616,1.289400) *[red]{\scriptscriptstyle\times};
(5.108830,1.291210) *[red]{\scriptscriptstyle\times};
(5.117043,1.295410) *[red]{\scriptscriptstyle\times};
(5.125257,1.300902) *[red]{\scriptscriptstyle\times};
(5.133470,1.303427) *[red]{\scriptscriptstyle\times};
(5.141684,1.307481) *[red]{\scriptscriptstyle\times};
(5.149897,1.311122) *[red]{\scriptscriptstyle\times};
(5.158111,1.338737) *[red]{\scriptscriptstyle\times};
(5.166324,1.342102) *[red]{\scriptscriptstyle\times};
(5.174538,1.264318) *[red]{\scriptscriptstyle\times};
(5.182752,1.280442) *[red]{\scriptscriptstyle\times};
(5.190965,1.283249) *[red]{\scriptscriptstyle\times};
(5.199179,1.284643) *[red]{\scriptscriptstyle\times};
(5.207392,1.290103) *[red]{\scriptscriptstyle\times};
(5.215606,1.292687) *[red]{\scriptscriptstyle\times};
(5.223819,1.306869) *[red]{\scriptscriptstyle\times};
(5.232033,1.307302) *[red]{\scriptscriptstyle\times};
(5.240246,1.309214) *[red]{\scriptscriptstyle\times};
(5.248460,1.309216) *[red]{\scriptscriptstyle\times};
(5.256674,1.322372) *[red]{\scriptscriptstyle\times};
(5.264887,1.325673) *[red]{\scriptscriptstyle\times};
(5.273101,1.332694) *[red]{\scriptscriptstyle\times};
(5.281314,1.343060) *[red]{\scriptscriptstyle\times};
(5.289528,1.363184) *[red]{\scriptscriptstyle\times};
(5.297741,1.269553) *[red]{\scriptscriptstyle\times};
(5.305955,1.273118) *[red]{\scriptscriptstyle\times};
(5.314168,1.277649) *[red]{\scriptscriptstyle\times};
(5.322382,1.291453) *[red]{\scriptscriptstyle\times};
(5.330595,1.291709) *[red]{\scriptscriptstyle\times};
(5.338809,1.292086) *[red]{\scriptscriptstyle\times};
(5.347023,1.298221) *[red]{\scriptscriptstyle\times};
(5.355236,1.298629) *[red]{\scriptscriptstyle\times};
(5.363450,1.301277) *[red]{\scriptscriptstyle\times};
(5.371663,1.305204) *[red]{\scriptscriptstyle\times};
(5.379877,1.307551) *[red]{\scriptscriptstyle\times};
(5.388090,1.314407) *[red]{\scriptscriptstyle\times};
(5.396304,1.315737) *[red]{\scriptscriptstyle\times};
(5.404517,1.322349) *[red]{\scriptscriptstyle\times};
(5.412731,1.325137) *[red]{\scriptscriptstyle\times};
(5.420945,1.270172) *[red]{\scriptscriptstyle\times};
(5.429158,1.273421) *[red]{\scriptscriptstyle\times};
(5.437372,1.275694) *[red]{\scriptscriptstyle\times};
(5.445585,1.280340) *[red]{\scriptscriptstyle\times};
(5.453799,1.295142) *[red]{\scriptscriptstyle\times};
(5.462012,1.298912) *[red]{\scriptscriptstyle\times};
(5.470226,1.308041) *[red]{\scriptscriptstyle\times};
(5.478439,1.316112) *[red]{\scriptscriptstyle\times};
(5.486653,1.316416) *[red]{\scriptscriptstyle\times};
(5.494867,1.316743) *[red]{\scriptscriptstyle\times};
(5.503080,1.320511) *[red]{\scriptscriptstyle\times};
(5.511294,1.336638) *[red]{\scriptscriptstyle\times};
(5.519507,1.340668) *[red]{\scriptscriptstyle\times};
(5.527721,1.349891) *[red]{\scriptscriptstyle\times};
(5.535934,1.359949) *[red]{\scriptscriptstyle\times};
(5.544148,1.285557) *[red]{\scriptscriptstyle\times};
(5.552361,1.285770) *[red]{\scriptscriptstyle\times};
(5.560575,1.286736) *[red]{\scriptscriptstyle\times};
(5.568789,1.290351) *[red]{\scriptscriptstyle\times};
(5.577002,1.290869) *[red]{\scriptscriptstyle\times};
(5.585216,1.296875) *[red]{\scriptscriptstyle\times};
(5.593429,1.310289) *[red]{\scriptscriptstyle\times};
(5.601643,1.317456) *[red]{\scriptscriptstyle\times};
(5.609856,1.319535) *[red]{\scriptscriptstyle\times};
(5.618070,1.324508) *[red]{\scriptscriptstyle\times};
(5.626283,1.334927) *[red]{\scriptscriptstyle\times};
(5.634497,1.335705) *[red]{\scriptscriptstyle\times};
(5.642710,1.341130) *[red]{\scriptscriptstyle\times};
(5.650924,1.352751) *[red]{\scriptscriptstyle\times};
(5.659138,1.363536) *[red]{\scriptscriptstyle\times};
(5.667351,1.306563) *[red]{\scriptscriptstyle\times};
(5.675565,1.308705) *[red]{\scriptscriptstyle\times};
(5.683778,1.308917) *[red]{\scriptscriptstyle\times};
(5.691992,1.311147) *[red]{\scriptscriptstyle\times};
(5.700205,1.311795) *[red]{\scriptscriptstyle\times};
(5.708419,1.317153) *[red]{\scriptscriptstyle\times};
(5.716632,1.317477) *[red]{\scriptscriptstyle\times};
(5.724846,1.331721) *[red]{\scriptscriptstyle\times};
(5.733060,1.336804) *[red]{\scriptscriptstyle\times};
(5.741273,1.337307) *[red]{\scriptscriptstyle\times};
(5.749487,1.341362) *[red]{\scriptscriptstyle\times};
(5.757700,1.348423) *[red]{\scriptscriptstyle\times};
(5.765914,1.354502) *[red]{\scriptscriptstyle\times};
(5.774127,1.356698) *[red]{\scriptscriptstyle\times};
(5.782341,1.361167) *[red]{\scriptscriptstyle\times};
(5.790554,1.300536) *[red]{\scriptscriptstyle\times};
(5.798768,1.309107) *[red]{\scriptscriptstyle\times};
(5.806982,1.321915) *[red]{\scriptscriptstyle\times};
(5.815195,1.322986) *[red]{\scriptscriptstyle\times};
(5.823409,1.323770) *[red]{\scriptscriptstyle\times};
(5.831622,1.323992) *[red]{\scriptscriptstyle\times};
(5.839836,1.324612) *[red]{\scriptscriptstyle\times};
(5.848049,1.325448) *[red]{\scriptscriptstyle\times};
(5.856263,1.331106) *[red]{\scriptscriptstyle\times};
(5.864476,1.334765) *[red]{\scriptscriptstyle\times};
(5.872690,1.339304) *[red]{\scriptscriptstyle\times};
(5.880903,1.348011) *[red]{\scriptscriptstyle\times};
(5.889117,1.352335) *[red]{\scriptscriptstyle\times};
(5.897331,1.409137) *[red]{\scriptscriptstyle\times};
(5.905544,1.431873) *[red]{\scriptscriptstyle\times};
(5.913758,1.315793) *[red]{\scriptscriptstyle\times};
(5.921971,1.317486) *[red]{\scriptscriptstyle\times};
(5.930185,1.329393) *[red]{\scriptscriptstyle\times};
(5.938398,1.329486) *[red]{\scriptscriptstyle\times};
(5.946612,1.329752) *[red]{\scriptscriptstyle\times};
(5.954825,1.330776) *[red]{\scriptscriptstyle\times};
(5.963039,1.333498) *[red]{\scriptscriptstyle\times};
(5.971253,1.335368) *[red]{\scriptscriptstyle\times};
(5.979466,1.336375) *[red]{\scriptscriptstyle\times};
(5.987680,1.359201) *[red]{\scriptscriptstyle\times};
(5.995893,1.363810) *[red]{\scriptscriptstyle\times};
(6.004107,1.367731) *[red]{\scriptscriptstyle\times};
(6.012320,1.378824) *[red]{\scriptscriptstyle\times};
(6.020534,1.436941) *[red]{\scriptscriptstyle\times};
(6.028747,1.438308) *[red]{\scriptscriptstyle\times};
(6.036961,1.306537) *[red]{\scriptscriptstyle\times};
(6.045175,1.308230) *[red]{\scriptscriptstyle\times};
(6.053388,1.320258) *[red]{\scriptscriptstyle\times};
(6.061602,1.321719) *[red]{\scriptscriptstyle\times};
(6.069815,1.321790) *[red]{\scriptscriptstyle\times};
(6.078029,1.345621) *[red]{\scriptscriptstyle\times};
(6.086242,1.347736) *[red]{\scriptscriptstyle\times};
(6.094456,1.357479) *[red]{\scriptscriptstyle\times};
(6.102669,1.381927) *[red]{\scriptscriptstyle\times};
(6.110883,1.389648) *[red]{\scriptscriptstyle\times};
(6.119097,1.391437) *[red]{\scriptscriptstyle\times};
(6.127310,1.392456) *[red]{\scriptscriptstyle\times};
(6.135524,1.396653) *[red]{\scriptscriptstyle\times};
(6.143737,1.405711) *[red]{\scriptscriptstyle\times};
(6.151951,1.473055) *[red]{\scriptscriptstyle\times};
(6.160164,1.327702) *[red]{\scriptscriptstyle\times};
(6.168378,1.336309) *[red]{\scriptscriptstyle\times};
(6.176591,1.337188) *[red]{\scriptscriptstyle\times};
(6.184805,1.337404) *[red]{\scriptscriptstyle\times};
(6.193018,1.340377) *[red]{\scriptscriptstyle\times};
(6.201232,1.342446) *[red]{\scriptscriptstyle\times};
(6.209446,1.343491) *[red]{\scriptscriptstyle\times};
(6.217659,1.345772) *[red]{\scriptscriptstyle\times};
(6.225873,1.348625) *[red]{\scriptscriptstyle\times};
(6.234086,1.360966) *[red]{\scriptscriptstyle\times};
(6.242300,1.362465) *[red]{\scriptscriptstyle\times};
(6.250513,1.370160) *[red]{\scriptscriptstyle\times};
(6.258727,1.383383) *[red]{\scriptscriptstyle\times};
(6.266940,1.393927) *[red]{\scriptscriptstyle\times};
(6.275154,1.431052) *[red]{\scriptscriptstyle\times};
(6.283368,1.316340) *[red]{\scriptscriptstyle\times};
(6.291581,1.336334) *[red]{\scriptscriptstyle\times};
(6.299795,1.336805) *[red]{\scriptscriptstyle\times};
(6.308008,1.342626) *[red]{\scriptscriptstyle\times};
(6.316222,1.343815) *[red]{\scriptscriptstyle\times};
(6.324435,1.345267) *[red]{\scriptscriptstyle\times};
(6.332649,1.353670) *[red]{\scriptscriptstyle\times};
(6.340862,1.365459) *[red]{\scriptscriptstyle\times};
(6.349076,1.368097) *[red]{\scriptscriptstyle\times};
(6.357290,1.369058) *[red]{\scriptscriptstyle\times};
(6.365503,1.408674) *[red]{\scriptscriptstyle\times};
(6.373717,1.410665) *[red]{\scriptscriptstyle\times};
(6.381930,1.418563) *[red]{\scriptscriptstyle\times};
(6.390144,1.444858) *[red]{\scriptscriptstyle\times};
(6.398357,1.550779) *[red]{\scriptscriptstyle\times};
(6.406571,1.326102) *[red]{\scriptscriptstyle\times};
(6.414784,1.330769) *[red]{\scriptscriptstyle\times};
(6.422998,1.334518) *[red]{\scriptscriptstyle\times};
(6.431211,1.336510) *[red]{\scriptscriptstyle\times};
(6.439425,1.344524) *[red]{\scriptscriptstyle\times};
(6.447639,1.351340) *[red]{\scriptscriptstyle\times};
(6.455852,1.351919) *[red]{\scriptscriptstyle\times};
(6.464066,1.353025) *[red]{\scriptscriptstyle\times};
(6.472279,1.369189) *[red]{\scriptscriptstyle\times};
(6.480493,1.372554) *[red]{\scriptscriptstyle\times};
(6.488706,1.372875) *[red]{\scriptscriptstyle\times};
(6.496920,1.376405) *[red]{\scriptscriptstyle\times};
(6.505133,1.391009) *[red]{\scriptscriptstyle\times};
(6.513347,1.411741) *[red]{\scriptscriptstyle\times};
(6.521561,1.414535) *[red]{\scriptscriptstyle\times};
(6.529774,1.358622) *[red]{\scriptscriptstyle\times};
(6.537988,1.365463) *[red]{\scriptscriptstyle\times};
(6.546201,1.369582) *[red]{\scriptscriptstyle\times};
(6.554415,1.381209) *[red]{\scriptscriptstyle\times};
(6.562628,1.383648) *[red]{\scriptscriptstyle\times};
(6.570842,1.385385) *[red]{\scriptscriptstyle\times};
(6.579055,1.387377) *[red]{\scriptscriptstyle\times};
(6.587269,1.388092) *[red]{\scriptscriptstyle\times};
(6.595483,1.388771) *[red]{\scriptscriptstyle\times};
(6.603696,1.391796) *[red]{\scriptscriptstyle\times};
(6.611910,1.395415) *[red]{\scriptscriptstyle\times};
(6.620123,1.407197) *[red]{\scriptscriptstyle\times};
(6.628337,1.412627) *[red]{\scriptscriptstyle\times};
(6.636550,1.412998) *[red]{\scriptscriptstyle\times};
(6.644764,1.420014) *[red]{\scriptscriptstyle\times};
(6.652977,1.368828) *[red]{\scriptscriptstyle\times};
(6.661191,1.369817) *[red]{\scriptscriptstyle\times};
(6.669405,1.370419) *[red]{\scriptscriptstyle\times};
(6.677618,1.370557) *[red]{\scriptscriptstyle\times};
(6.685832,1.372800) *[red]{\scriptscriptstyle\times};
(6.694045,1.375004) *[red]{\scriptscriptstyle\times};
(6.702259,1.376557) *[red]{\scriptscriptstyle\times};
(6.710472,1.377115) *[red]{\scriptscriptstyle\times};
(6.718686,1.390542) *[red]{\scriptscriptstyle\times};
(6.726899,1.401267) *[red]{\scriptscriptstyle\times};
(6.735113,1.412721) *[red]{\scriptscriptstyle\times};
(6.743326,1.414689) *[red]{\scriptscriptstyle\times};
(6.751540,1.438438) *[red]{\scriptscriptstyle\times};
(6.759754,1.455872) *[red]{\scriptscriptstyle\times};
(6.767967,1.503100) *[red]{\scriptscriptstyle\times};
(6.776181,1.350705) *[red]{\scriptscriptstyle\times};
(6.784394,1.373440) *[red]{\scriptscriptstyle\times};
(6.792608,1.374957) *[red]{\scriptscriptstyle\times};
(6.800821,1.383184) *[red]{\scriptscriptstyle\times};
(6.809035,1.394406) *[red]{\scriptscriptstyle\times};
(6.817248,1.398787) *[red]{\scriptscriptstyle\times};
(6.825462,1.403598) *[red]{\scriptscriptstyle\times};
(6.833676,1.405937) *[red]{\scriptscriptstyle\times};
(6.841889,1.411950) *[red]{\scriptscriptstyle\times};
(6.850103,1.422182) *[red]{\scriptscriptstyle\times};
(6.858316,1.423463) *[red]{\scriptscriptstyle\times};
(6.866530,1.436411) *[red]{\scriptscriptstyle\times};
(6.874743,1.443585) *[red]{\scriptscriptstyle\times};
(6.882957,1.444101) *[red]{\scriptscriptstyle\times};
(6.891170,1.458959) *[red]{\scriptscriptstyle\times};
(6.899384,1.376643) *[red]{\scriptscriptstyle\times};
(6.907598,1.378942) *[red]{\scriptscriptstyle\times};
(6.915811,1.379792) *[red]{\scriptscriptstyle\times};
(6.924025,1.389735) *[red]{\scriptscriptstyle\times};
(6.932238,1.390406) *[red]{\scriptscriptstyle\times};
(6.940452,1.391073) *[red]{\scriptscriptstyle\times};
(6.948665,1.391462) *[red]{\scriptscriptstyle\times};
(6.956879,1.394285) *[red]{\scriptscriptstyle\times};
(6.965092,1.404000) *[red]{\scriptscriptstyle\times};
(6.973306,1.404050) *[red]{\scriptscriptstyle\times};
(6.981520,1.407568) *[red]{\scriptscriptstyle\times};
(6.989733,1.407786) *[red]{\scriptscriptstyle\times};
(6.997947,1.412047) *[red]{\scriptscriptstyle\times};
(7.006160,1.416323) *[red]{\scriptscriptstyle\times};
(7.014374,1.433129) *[red]{\scriptscriptstyle\times};
(7.022587,1.338831) *[red]{\scriptscriptstyle\times};
(7.030801,1.349934) *[red]{\scriptscriptstyle\times};
(7.039014,1.360301) *[red]{\scriptscriptstyle\times};
(7.047228,1.365150) *[red]{\scriptscriptstyle\times};
(7.055441,1.381069) *[red]{\scriptscriptstyle\times};
(7.063655,1.384756) *[red]{\scriptscriptstyle\times};
(7.071869,1.400768) *[red]{\scriptscriptstyle\times};
(7.080082,1.406632) *[red]{\scriptscriptstyle\times};
(7.088296,1.408763) *[red]{\scriptscriptstyle\times};
(7.096509,1.411057) *[red]{\scriptscriptstyle\times};
(7.104723,1.415827) *[red]{\scriptscriptstyle\times};
(7.112936,1.424557) *[red]{\scriptscriptstyle\times};
(7.121150,1.448583) *[red]{\scriptscriptstyle\times};
(7.129363,1.453173) *[red]{\scriptscriptstyle\times};
(7.137577,1.485854) *[red]{\scriptscriptstyle\times};
(7.145791,1.362060) *[red]{\scriptscriptstyle\times};
(7.154004,1.381452) *[red]{\scriptscriptstyle\times};
(7.162218,1.382387) *[red]{\scriptscriptstyle\times};
(7.170431,1.383675) *[red]{\scriptscriptstyle\times};
(7.178645,1.398441) *[red]{\scriptscriptstyle\times};
(7.186858,1.407628) *[red]{\scriptscriptstyle\times};
(7.195072,1.408809) *[red]{\scriptscriptstyle\times};
(7.203285,1.409080) *[red]{\scriptscriptstyle\times};
(7.211499,1.412929) *[red]{\scriptscriptstyle\times};
(7.219713,1.420985) *[red]{\scriptscriptstyle\times};
(7.227926,1.423183) *[red]{\scriptscriptstyle\times};
(7.236140,1.447195) *[red]{\scriptscriptstyle\times};
(7.244353,1.459978) *[red]{\scriptscriptstyle\times};
(7.252567,1.461457) *[red]{\scriptscriptstyle\times};
(7.260780,1.465309) *[red]{\scriptscriptstyle\times};
(7.268994,1.348527) *[red]{\scriptscriptstyle\times};
(7.277207,1.355815) *[red]{\scriptscriptstyle\times};
(7.285421,1.372922) *[red]{\scriptscriptstyle\times};
(7.293634,1.372966) *[red]{\scriptscriptstyle\times};
(7.301848,1.393470) *[red]{\scriptscriptstyle\times};
(7.310062,1.399577) *[red]{\scriptscriptstyle\times};
(7.318275,1.410840) *[red]{\scriptscriptstyle\times};
(7.326489,1.416771) *[red]{\scriptscriptstyle\times};
(7.334702,1.438527) *[red]{\scriptscriptstyle\times};
(7.342916,1.440848) *[red]{\scriptscriptstyle\times};
(7.351129,1.442280) *[red]{\scriptscriptstyle\times};
(7.359343,1.450082) *[red]{\scriptscriptstyle\times};
(7.367556,1.453179) *[red]{\scriptscriptstyle\times};
(7.375770,1.453537) *[red]{\scriptscriptstyle\times};
(7.383984,1.458367) *[red]{\scriptscriptstyle\times};
(7.392197,1.398658) *[red]{\scriptscriptstyle\times};
(7.400411,1.401724) *[red]{\scriptscriptstyle\times};
(7.408624,1.402150) *[red]{\scriptscriptstyle\times};
(7.416838,1.416660) *[red]{\scriptscriptstyle\times};
(7.425051,1.419736) *[red]{\scriptscriptstyle\times};
(7.433265,1.421015) *[red]{\scriptscriptstyle\times};
(7.441478,1.423917) *[red]{\scriptscriptstyle\times};
(7.449692,1.448386) *[red]{\scriptscriptstyle\times};
(7.457906,1.455580) *[red]{\scriptscriptstyle\times};
(7.466119,1.460842) *[red]{\scriptscriptstyle\times};
(7.474333,1.493423) *[red]{\scriptscriptstyle\times};
(7.482546,1.503092) *[red]{\scriptscriptstyle\times};
(7.490760,1.511787) *[red]{\scriptscriptstyle\times};
(7.498973,1.560622) *[red]{\scriptscriptstyle\times};
(7.507187,1.580604) *[red]{\scriptscriptstyle\times};
(7.515400,1.402209) *[red]{\scriptscriptstyle\times};
(7.523614,1.402684) *[red]{\scriptscriptstyle\times};
(7.531828,1.406869) *[red]{\scriptscriptstyle\times};
(7.540041,1.423671) *[red]{\scriptscriptstyle\times};
(7.548255,1.425447) *[red]{\scriptscriptstyle\times};
(7.556468,1.431395) *[red]{\scriptscriptstyle\times};
(7.564682,1.433209) *[red]{\scriptscriptstyle\times};
(7.572895,1.442311) *[red]{\scriptscriptstyle\times};
(7.581109,1.444784) *[red]{\scriptscriptstyle\times};
(7.589322,1.458156) *[red]{\scriptscriptstyle\times};
(7.597536,1.464742) *[red]{\scriptscriptstyle\times};
(7.605749,1.476614) *[red]{\scriptscriptstyle\times};
(7.613963,1.477266) *[red]{\scriptscriptstyle\times};
(7.622177,1.505725) *[red]{\scriptscriptstyle\times};
(7.630390,1.521068) *[red]{\scriptscriptstyle\times};
(7.638604,1.428956) *[red]{\scriptscriptstyle\times};
(7.646817,1.442728) *[red]{\scriptscriptstyle\times};
(7.655031,1.446549) *[red]{\scriptscriptstyle\times};
(7.663244,1.446991) *[red]{\scriptscriptstyle\times};
(7.671458,1.451515) *[red]{\scriptscriptstyle\times};
(7.679671,1.451528) *[red]{\scriptscriptstyle\times};
(7.687885,1.468248) *[red]{\scriptscriptstyle\times};
(7.696099,1.468517) *[red]{\scriptscriptstyle\times};
(7.704312,1.468769) *[red]{\scriptscriptstyle\times};
(7.712526,1.471011) *[red]{\scriptscriptstyle\times};
(7.720739,1.499972) *[red]{\scriptscriptstyle\times};
(7.728953,1.503356) *[red]{\scriptscriptstyle\times};
(7.737166,1.504488) *[red]{\scriptscriptstyle\times};
(7.745380,1.508969) *[red]{\scriptscriptstyle\times};
(7.753593,1.569344) *[red]{\scriptscriptstyle\times};
(7.761807,1.497252) *[red]{\scriptscriptstyle\times};
(7.770021,1.506931) *[red]{\scriptscriptstyle\times};
(7.778234,1.511833) *[red]{\scriptscriptstyle\times};
(7.786448,1.514644) *[red]{\scriptscriptstyle\times};
(7.794661,1.531485) *[red]{\scriptscriptstyle\times};
(7.802875,1.538110) *[red]{\scriptscriptstyle\times};
(7.811088,1.543260) *[red]{\scriptscriptstyle\times};
(7.819302,1.550152) *[red]{\scriptscriptstyle\times};
(7.827515,1.553661) *[red]{\scriptscriptstyle\times};
(7.835729,1.557721) *[red]{\scriptscriptstyle\times};
(7.843943,1.566841) *[red]{\scriptscriptstyle\times};
(7.852156,1.568453) *[red]{\scriptscriptstyle\times};
(7.860370,1.569741) *[red]{\scriptscriptstyle\times};
(7.868583,1.571025) *[red]{\scriptscriptstyle\times};
(7.876797,1.599735) *[red]{\scriptscriptstyle\times};
(7.885010,1.543050) *[red]{\scriptscriptstyle\times};
(7.893224,1.546914) *[red]{\scriptscriptstyle\times};
(7.901437,1.550378) *[red]{\scriptscriptstyle\times};
(7.909651,1.552383) *[red]{\scriptscriptstyle\times};
(7.917864,1.560247) *[red]{\scriptscriptstyle\times};
(7.926078,1.560442) *[red]{\scriptscriptstyle\times};
(7.934292,1.567450) *[red]{\scriptscriptstyle\times};
(7.942505,1.572148) *[red]{\scriptscriptstyle\times};
(7.950719,1.583381) *[red]{\scriptscriptstyle\times};
(7.958932,1.583568) *[red]{\scriptscriptstyle\times};
(7.967146,1.589920) *[red]{\scriptscriptstyle\times};
(7.975359,1.624949) *[red]{\scriptscriptstyle\times};
(7.983573,1.635231) *[red]{\scriptscriptstyle\times};
(7.991786,1.644348) *[red]{\scriptscriptstyle\times};
(8.000000,1.676527) *[red]{\scriptscriptstyle\times};
\endxy
}
\caption{Skylake cycles
  for the CSIDH-512 action
  using {\tt velusqrt-asm}.
}
\label{action-cycles512}
\end{figure}

\begin{figure}[t]
\centerline{
\xy <1.1cm,0cm>:<0cm,4cm>::
(0,1.321477); (8,1.321477) **[blue]@{-};
(8.1,1.321477) *[blue]{\rlap{749765588}};
(0,1.394655); (8,1.394655) **[blue]@{-};
(8.1,1.394655) *[blue]{\rlap{788777194}};
(0,1.473598); (8,1.473598) **[blue]@{-};
(8.1,1.473598) *[blue]{\rlap{833140632}};
(0,1.193192); (8,1.193192) **[red]@{-};
(-0.1,1.193192) *[red]{\llap{685974488}};
(0,1.261198); (8,1.261198) **[red]@{-};
(-0.1,1.261198) *[red]{\llap{719084288}};
(0,1.341104); (8,1.341104) **[red]@{-};
(-0.1,1.341104) *[red]{\llap{760035568}};
(-0.004107,0.950999); (-0.004107,1.667783) **[lightgray]@{-};
(0.119097,0.950999); (0.119097,1.667783) **[lightgray]@{-};
(0.242300,0.950999); (0.242300,1.667783) **[lightgray]@{-};
(0.365503,0.950999); (0.365503,1.667783) **[lightgray]@{-};
(0.488706,0.950999); (0.488706,1.667783) **[lightgray]@{-};
(0.611910,0.950999); (0.611910,1.667783) **[lightgray]@{-};
(0.735113,0.950999); (0.735113,1.667783) **[lightgray]@{-};
(0.858316,0.950999); (0.858316,1.667783) **[lightgray]@{-};
(0.981520,0.950999); (0.981520,1.667783) **[lightgray]@{-};
(1.104723,0.950999); (1.104723,1.667783) **[lightgray]@{-};
(1.227926,0.950999); (1.227926,1.667783) **[lightgray]@{-};
(1.351129,0.950999); (1.351129,1.667783) **[lightgray]@{-};
(1.474333,0.950999); (1.474333,1.667783) **[lightgray]@{-};
(1.597536,0.950999); (1.597536,1.667783) **[lightgray]@{-};
(1.720739,0.950999); (1.720739,1.667783) **[lightgray]@{-};
(1.843943,0.950999); (1.843943,1.667783) **[lightgray]@{-};
(1.967146,0.950999); (1.967146,1.667783) **[lightgray]@{-};
(2.090349,0.950999); (2.090349,1.667783) **[lightgray]@{-};
(2.213552,0.950999); (2.213552,1.667783) **[lightgray]@{-};
(2.336756,0.950999); (2.336756,1.667783) **[lightgray]@{-};
(2.459959,0.950999); (2.459959,1.667783) **[lightgray]@{-};
(2.583162,0.950999); (2.583162,1.667783) **[lightgray]@{-};
(2.706366,0.950999); (2.706366,1.667783) **[lightgray]@{-};
(2.829569,0.950999); (2.829569,1.667783) **[lightgray]@{-};
(2.952772,0.950999); (2.952772,1.667783) **[lightgray]@{-};
(3.075975,0.950999); (3.075975,1.667783) **[lightgray]@{-};
(3.199179,0.950999); (3.199179,1.667783) **[lightgray]@{-};
(3.322382,0.950999); (3.322382,1.667783) **[lightgray]@{-};
(3.445585,0.950999); (3.445585,1.667783) **[lightgray]@{-};
(3.568789,0.950999); (3.568789,1.667783) **[lightgray]@{-};
(3.691992,0.950999); (3.691992,1.667783) **[lightgray]@{-};
(3.815195,0.950999); (3.815195,1.667783) **[lightgray]@{-};
(3.938398,0.950999); (3.938398,1.667783) **[lightgray]@{-};
(4.061602,0.950999); (4.061602,1.667783) **[lightgray]@{-};
(4.184805,0.950999); (4.184805,1.667783) **[lightgray]@{-};
(4.308008,0.950999); (4.308008,1.667783) **[lightgray]@{-};
(4.431211,0.950999); (4.431211,1.667783) **[lightgray]@{-};
(4.554415,0.950999); (4.554415,1.667783) **[lightgray]@{-};
(4.677618,0.950999); (4.677618,1.667783) **[lightgray]@{-};
(4.800821,0.950999); (4.800821,1.667783) **[lightgray]@{-};
(4.924025,0.950999); (4.924025,1.667783) **[lightgray]@{-};
(5.047228,0.950999); (5.047228,1.667783) **[lightgray]@{-};
(5.170431,0.950999); (5.170431,1.667783) **[lightgray]@{-};
(5.293634,0.950999); (5.293634,1.667783) **[lightgray]@{-};
(5.416838,0.950999); (5.416838,1.667783) **[lightgray]@{-};
(5.540041,0.950999); (5.540041,1.667783) **[lightgray]@{-};
(5.663244,0.950999); (5.663244,1.667783) **[lightgray]@{-};
(5.786448,0.950999); (5.786448,1.667783) **[lightgray]@{-};
(5.909651,0.950999); (5.909651,1.667783) **[lightgray]@{-};
(6.032854,0.950999); (6.032854,1.667783) **[lightgray]@{-};
(6.156057,0.950999); (6.156057,1.667783) **[lightgray]@{-};
(6.279261,0.950999); (6.279261,1.667783) **[lightgray]@{-};
(6.402464,0.950999); (6.402464,1.667783) **[lightgray]@{-};
(6.525667,0.950999); (6.525667,1.667783) **[lightgray]@{-};
(6.648871,0.950999); (6.648871,1.667783) **[lightgray]@{-};
(6.772074,0.950999); (6.772074,1.667783) **[lightgray]@{-};
(6.895277,0.950999); (6.895277,1.667783) **[lightgray]@{-};
(7.018480,0.950999); (7.018480,1.667783) **[lightgray]@{-};
(7.141684,0.950999); (7.141684,1.667783) **[lightgray]@{-};
(7.264887,0.950999); (7.264887,1.667783) **[lightgray]@{-};
(7.388090,0.950999); (7.388090,1.667783) **[lightgray]@{-};
(7.511294,0.950999); (7.511294,1.667783) **[lightgray]@{-};
(7.634497,0.950999); (7.634497,1.667783) **[lightgray]@{-};
(7.757700,0.950999); (7.757700,1.667783) **[lightgray]@{-};
(7.880903,0.950999); (7.880903,1.667783) **[lightgray]@{-};
(8.004107,0.950999); (8.004107,1.667783) **[lightgray]@{-};
(-0.004107,0.950999); (-0.004107,1.667783) **[lightgray]@{-};
(0.119097,0.950999); (0.119097,1.667783) **[lightgray]@{-};
(0.242300,0.950999); (0.242300,1.667783) **[lightgray]@{-};
(0.365503,0.950999); (0.365503,1.667783) **[lightgray]@{-};
(0.488706,0.950999); (0.488706,1.667783) **[lightgray]@{-};
(0.611910,0.950999); (0.611910,1.667783) **[lightgray]@{-};
(0.735113,0.950999); (0.735113,1.667783) **[lightgray]@{-};
(0.858316,0.950999); (0.858316,1.667783) **[lightgray]@{-};
(0.981520,0.950999); (0.981520,1.667783) **[lightgray]@{-};
(1.104723,0.950999); (1.104723,1.667783) **[lightgray]@{-};
(1.227926,0.950999); (1.227926,1.667783) **[lightgray]@{-};
(1.351129,0.950999); (1.351129,1.667783) **[lightgray]@{-};
(1.474333,0.950999); (1.474333,1.667783) **[lightgray]@{-};
(1.597536,0.950999); (1.597536,1.667783) **[lightgray]@{-};
(1.720739,0.950999); (1.720739,1.667783) **[lightgray]@{-};
(1.843943,0.950999); (1.843943,1.667783) **[lightgray]@{-};
(1.967146,0.950999); (1.967146,1.667783) **[lightgray]@{-};
(2.090349,0.950999); (2.090349,1.667783) **[lightgray]@{-};
(2.213552,0.950999); (2.213552,1.667783) **[lightgray]@{-};
(2.336756,0.950999); (2.336756,1.667783) **[lightgray]@{-};
(2.459959,0.950999); (2.459959,1.667783) **[lightgray]@{-};
(2.583162,0.950999); (2.583162,1.667783) **[lightgray]@{-};
(2.706366,0.950999); (2.706366,1.667783) **[lightgray]@{-};
(2.829569,0.950999); (2.829569,1.667783) **[lightgray]@{-};
(2.952772,0.950999); (2.952772,1.667783) **[lightgray]@{-};
(3.075975,0.950999); (3.075975,1.667783) **[lightgray]@{-};
(3.199179,0.950999); (3.199179,1.667783) **[lightgray]@{-};
(3.322382,0.950999); (3.322382,1.667783) **[lightgray]@{-};
(3.445585,0.950999); (3.445585,1.667783) **[lightgray]@{-};
(3.568789,0.950999); (3.568789,1.667783) **[lightgray]@{-};
(3.691992,0.950999); (3.691992,1.667783) **[lightgray]@{-};
(3.815195,0.950999); (3.815195,1.667783) **[lightgray]@{-};
(3.938398,0.950999); (3.938398,1.667783) **[lightgray]@{-};
(4.061602,0.950999); (4.061602,1.667783) **[lightgray]@{-};
(4.184805,0.950999); (4.184805,1.667783) **[lightgray]@{-};
(4.308008,0.950999); (4.308008,1.667783) **[lightgray]@{-};
(4.431211,0.950999); (4.431211,1.667783) **[lightgray]@{-};
(4.554415,0.950999); (4.554415,1.667783) **[lightgray]@{-};
(4.677618,0.950999); (4.677618,1.667783) **[lightgray]@{-};
(4.800821,0.950999); (4.800821,1.667783) **[lightgray]@{-};
(4.924025,0.950999); (4.924025,1.667783) **[lightgray]@{-};
(5.047228,0.950999); (5.047228,1.667783) **[lightgray]@{-};
(5.170431,0.950999); (5.170431,1.667783) **[lightgray]@{-};
(5.293634,0.950999); (5.293634,1.667783) **[lightgray]@{-};
(5.416838,0.950999); (5.416838,1.667783) **[lightgray]@{-};
(5.540041,0.950999); (5.540041,1.667783) **[lightgray]@{-};
(5.663244,0.950999); (5.663244,1.667783) **[lightgray]@{-};
(5.786448,0.950999); (5.786448,1.667783) **[lightgray]@{-};
(5.909651,0.950999); (5.909651,1.667783) **[lightgray]@{-};
(6.032854,0.950999); (6.032854,1.667783) **[lightgray]@{-};
(6.156057,0.950999); (6.156057,1.667783) **[lightgray]@{-};
(6.279261,0.950999); (6.279261,1.667783) **[lightgray]@{-};
(6.402464,0.950999); (6.402464,1.667783) **[lightgray]@{-};
(6.525667,0.950999); (6.525667,1.667783) **[lightgray]@{-};
(6.648871,0.950999); (6.648871,1.667783) **[lightgray]@{-};
(6.772074,0.950999); (6.772074,1.667783) **[lightgray]@{-};
(6.895277,0.950999); (6.895277,1.667783) **[lightgray]@{-};
(7.018480,0.950999); (7.018480,1.667783) **[lightgray]@{-};
(7.141684,0.950999); (7.141684,1.667783) **[lightgray]@{-};
(7.264887,0.950999); (7.264887,1.667783) **[lightgray]@{-};
(7.388090,0.950999); (7.388090,1.667783) **[lightgray]@{-};
(7.511294,0.950999); (7.511294,1.667783) **[lightgray]@{-};
(7.634497,0.950999); (7.634497,1.667783) **[lightgray]@{-};
(7.757700,0.950999); (7.757700,1.667783) **[lightgray]@{-};
(7.880903,0.950999); (7.880903,1.667783) **[lightgray]@{-};
(8.004107,0.950999); (8.004107,1.667783) **[lightgray]@{-};
(0.000000,1.098652) *[blue]{\scriptscriptstyle+};
(0.008214,1.098787) *[blue]{\scriptscriptstyle+};
(0.016427,1.099903) *[blue]{\scriptscriptstyle+};
(0.024641,1.102415) *[blue]{\scriptscriptstyle+};
(0.032854,1.102465) *[blue]{\scriptscriptstyle+};
(0.041068,1.105362) *[blue]{\scriptscriptstyle+};
(0.049281,1.105988) *[blue]{\scriptscriptstyle+};
(0.057495,1.111750) *[blue]{\scriptscriptstyle+};
(0.065708,1.138878) *[blue]{\scriptscriptstyle+};
(0.073922,1.142534) *[blue]{\scriptscriptstyle+};
(0.082136,1.146755) *[blue]{\scriptscriptstyle+};
(0.090349,1.152319) *[blue]{\scriptscriptstyle+};
(0.098563,1.159612) *[blue]{\scriptscriptstyle+};
(0.106776,1.169235) *[blue]{\scriptscriptstyle+};
(0.114990,1.204200) *[blue]{\scriptscriptstyle+};
(0.123203,1.108786) *[blue]{\scriptscriptstyle+};
(0.131417,1.109043) *[blue]{\scriptscriptstyle+};
(0.139630,1.109059) *[blue]{\scriptscriptstyle+};
(0.147844,1.109200) *[blue]{\scriptscriptstyle+};
(0.156057,1.148393) *[blue]{\scriptscriptstyle+};
(0.164271,1.151216) *[blue]{\scriptscriptstyle+};
(0.172485,1.151600) *[blue]{\scriptscriptstyle+};
(0.180698,1.152318) *[blue]{\scriptscriptstyle+};
(0.188912,1.185505) *[blue]{\scriptscriptstyle+};
(0.197125,1.185908) *[blue]{\scriptscriptstyle+};
(0.205339,1.189218) *[blue]{\scriptscriptstyle+};
(0.213552,1.195033) *[blue]{\scriptscriptstyle+};
(0.221766,1.220507) *[blue]{\scriptscriptstyle+};
(0.229979,1.264193) *[blue]{\scriptscriptstyle+};
(0.238193,1.327227) *[blue]{\scriptscriptstyle+};
(0.246407,1.152671) *[blue]{\scriptscriptstyle+};
(0.254620,1.153858) *[blue]{\scriptscriptstyle+};
(0.262834,1.156961) *[blue]{\scriptscriptstyle+};
(0.271047,1.157461) *[blue]{\scriptscriptstyle+};
(0.279261,1.167172) *[blue]{\scriptscriptstyle+};
(0.287474,1.172184) *[blue]{\scriptscriptstyle+};
(0.295688,1.172715) *[blue]{\scriptscriptstyle+};
(0.303901,1.190218) *[blue]{\scriptscriptstyle+};
(0.312115,1.190237) *[blue]{\scriptscriptstyle+};
(0.320329,1.193251) *[blue]{\scriptscriptstyle+};
(0.328542,1.194004) *[blue]{\scriptscriptstyle+};
(0.336756,1.197151) *[blue]{\scriptscriptstyle+};
(0.344969,1.204358) *[blue]{\scriptscriptstyle+};
(0.353183,1.211093) *[blue]{\scriptscriptstyle+};
(0.361396,1.230355) *[blue]{\scriptscriptstyle+};
(0.369610,1.163885) *[blue]{\scriptscriptstyle+};
(0.377823,1.168078) *[blue]{\scriptscriptstyle+};
(0.386037,1.194303) *[blue]{\scriptscriptstyle+};
(0.394251,1.194358) *[blue]{\scriptscriptstyle+};
(0.402464,1.195708) *[blue]{\scriptscriptstyle+};
(0.410678,1.195846) *[blue]{\scriptscriptstyle+};
(0.418891,1.197915) *[blue]{\scriptscriptstyle+};
(0.427105,1.199436) *[blue]{\scriptscriptstyle+};
(0.435318,1.199727) *[blue]{\scriptscriptstyle+};
(0.443532,1.204518) *[blue]{\scriptscriptstyle+};
(0.451745,1.213782) *[blue]{\scriptscriptstyle+};
(0.459959,1.223699) *[blue]{\scriptscriptstyle+};
(0.468172,1.235304) *[blue]{\scriptscriptstyle+};
(0.476386,1.266513) *[blue]{\scriptscriptstyle+};
(0.484600,1.342262) *[blue]{\scriptscriptstyle+};
(0.492813,1.164187) *[blue]{\scriptscriptstyle+};
(0.501027,1.170432) *[blue]{\scriptscriptstyle+};
(0.509240,1.171491) *[blue]{\scriptscriptstyle+};
(0.517454,1.174021) *[blue]{\scriptscriptstyle+};
(0.525667,1.200517) *[blue]{\scriptscriptstyle+};
(0.533881,1.200618) *[blue]{\scriptscriptstyle+};
(0.542094,1.204087) *[blue]{\scriptscriptstyle+};
(0.550308,1.209851) *[blue]{\scriptscriptstyle+};
(0.558522,1.210851) *[blue]{\scriptscriptstyle+};
(0.566735,1.226126) *[blue]{\scriptscriptstyle+};
(0.574949,1.240541) *[blue]{\scriptscriptstyle+};
(0.583162,1.241355) *[blue]{\scriptscriptstyle+};
(0.591376,1.244752) *[blue]{\scriptscriptstyle+};
(0.599589,1.245940) *[blue]{\scriptscriptstyle+};
(0.607803,1.276012) *[blue]{\scriptscriptstyle+};
(0.616016,1.167882) *[blue]{\scriptscriptstyle+};
(0.624230,1.167938) *[blue]{\scriptscriptstyle+};
(0.632444,1.171510) *[blue]{\scriptscriptstyle+};
(0.640657,1.205142) *[blue]{\scriptscriptstyle+};
(0.648871,1.207319) *[blue]{\scriptscriptstyle+};
(0.657084,1.209578) *[blue]{\scriptscriptstyle+};
(0.665298,1.209641) *[blue]{\scriptscriptstyle+};
(0.673511,1.211092) *[blue]{\scriptscriptstyle+};
(0.681725,1.219276) *[blue]{\scriptscriptstyle+};
(0.689938,1.248586) *[blue]{\scriptscriptstyle+};
(0.698152,1.255339) *[blue]{\scriptscriptstyle+};
(0.706366,1.278609) *[blue]{\scriptscriptstyle+};
(0.714579,1.286050) *[blue]{\scriptscriptstyle+};
(0.722793,1.298781) *[blue]{\scriptscriptstyle+};
(0.731006,1.305754) *[blue]{\scriptscriptstyle+};
(0.739220,1.173391) *[blue]{\scriptscriptstyle+};
(0.747433,1.174335) *[blue]{\scriptscriptstyle+};
(0.755647,1.174795) *[blue]{\scriptscriptstyle+};
(0.763860,1.181549) *[blue]{\scriptscriptstyle+};
(0.772074,1.183896) *[blue]{\scriptscriptstyle+};
(0.780287,1.210560) *[blue]{\scriptscriptstyle+};
(0.788501,1.214885) *[blue]{\scriptscriptstyle+};
(0.796715,1.218439) *[blue]{\scriptscriptstyle+};
(0.804928,1.224469) *[blue]{\scriptscriptstyle+};
(0.813142,1.246856) *[blue]{\scriptscriptstyle+};
(0.821355,1.246879) *[blue]{\scriptscriptstyle+};
(0.829569,1.249258) *[blue]{\scriptscriptstyle+};
(0.837782,1.252848) *[blue]{\scriptscriptstyle+};
(0.845996,1.261964) *[blue]{\scriptscriptstyle+};
(0.854209,1.296510) *[blue]{\scriptscriptstyle+};
(0.862423,1.188905) *[blue]{\scriptscriptstyle+};
(0.870637,1.189401) *[blue]{\scriptscriptstyle+};
(0.878850,1.197250) *[blue]{\scriptscriptstyle+};
(0.887064,1.223497) *[blue]{\scriptscriptstyle+};
(0.895277,1.223860) *[blue]{\scriptscriptstyle+};
(0.903491,1.223894) *[blue]{\scriptscriptstyle+};
(0.911704,1.223902) *[blue]{\scriptscriptstyle+};
(0.919918,1.227296) *[blue]{\scriptscriptstyle+};
(0.928131,1.234088) *[blue]{\scriptscriptstyle+};
(0.936345,1.240970) *[blue]{\scriptscriptstyle+};
(0.944559,1.265731) *[blue]{\scriptscriptstyle+};
(0.952772,1.270280) *[blue]{\scriptscriptstyle+};
(0.960986,1.289731) *[blue]{\scriptscriptstyle+};
(0.969199,1.297985) *[blue]{\scriptscriptstyle+};
(0.977413,1.471234) *[blue]{\scriptscriptstyle+};
(0.985626,1.201161) *[blue]{\scriptscriptstyle+};
(0.993840,1.208950) *[blue]{\scriptscriptstyle+};
(1.002053,1.232566) *[blue]{\scriptscriptstyle+};
(1.010267,1.232839) *[blue]{\scriptscriptstyle+};
(1.018480,1.234148) *[blue]{\scriptscriptstyle+};
(1.026694,1.234957) *[blue]{\scriptscriptstyle+};
(1.034908,1.241148) *[blue]{\scriptscriptstyle+};
(1.043121,1.251322) *[blue]{\scriptscriptstyle+};
(1.051335,1.271780) *[blue]{\scriptscriptstyle+};
(1.059548,1.271975) *[blue]{\scriptscriptstyle+};
(1.067762,1.276500) *[blue]{\scriptscriptstyle+};
(1.075975,1.277707) *[blue]{\scriptscriptstyle+};
(1.084189,1.285325) *[blue]{\scriptscriptstyle+};
(1.092402,1.306400) *[blue]{\scriptscriptstyle+};
(1.100616,1.350298) *[blue]{\scriptscriptstyle+};
(1.108830,1.232950) *[blue]{\scriptscriptstyle+};
(1.117043,1.236472) *[blue]{\scriptscriptstyle+};
(1.125257,1.246776) *[blue]{\scriptscriptstyle+};
(1.133470,1.268912) *[blue]{\scriptscriptstyle+};
(1.141684,1.273058) *[blue]{\scriptscriptstyle+};
(1.149897,1.274670) *[blue]{\scriptscriptstyle+};
(1.158111,1.274675) *[blue]{\scriptscriptstyle+};
(1.166324,1.275336) *[blue]{\scriptscriptstyle+};
(1.174538,1.304221) *[blue]{\scriptscriptstyle+};
(1.182752,1.306651) *[blue]{\scriptscriptstyle+};
(1.190965,1.307238) *[blue]{\scriptscriptstyle+};
(1.199179,1.316497) *[blue]{\scriptscriptstyle+};
(1.207392,1.317334) *[blue]{\scriptscriptstyle+};
(1.215606,1.344715) *[blue]{\scriptscriptstyle+};
(1.223819,1.347914) *[blue]{\scriptscriptstyle+};
(1.232033,1.261242) *[blue]{\scriptscriptstyle+};
(1.240246,1.261311) *[blue]{\scriptscriptstyle+};
(1.248460,1.261397) *[blue]{\scriptscriptstyle+};
(1.256674,1.263795) *[blue]{\scriptscriptstyle+};
(1.264887,1.266164) *[blue]{\scriptscriptstyle+};
(1.273101,1.274067) *[blue]{\scriptscriptstyle+};
(1.281314,1.274475) *[blue]{\scriptscriptstyle+};
(1.289528,1.276379) *[blue]{\scriptscriptstyle+};
(1.297741,1.299052) *[blue]{\scriptscriptstyle+};
(1.305955,1.303645) *[blue]{\scriptscriptstyle+};
(1.314168,1.308548) *[blue]{\scriptscriptstyle+};
(1.322382,1.316845) *[blue]{\scriptscriptstyle+};
(1.330595,1.330374) *[blue]{\scriptscriptstyle+};
(1.338809,1.339553) *[blue]{\scriptscriptstyle+};
(1.347023,1.348738) *[blue]{\scriptscriptstyle+};
(1.355236,1.231471) *[blue]{\scriptscriptstyle+};
(1.363450,1.236962) *[blue]{\scriptscriptstyle+};
(1.371663,1.240657) *[blue]{\scriptscriptstyle+};
(1.379877,1.250535) *[blue]{\scriptscriptstyle+};
(1.388090,1.267342) *[blue]{\scriptscriptstyle+};
(1.396304,1.267422) *[blue]{\scriptscriptstyle+};
(1.404517,1.271771) *[blue]{\scriptscriptstyle+};
(1.412731,1.277680) *[blue]{\scriptscriptstyle+};
(1.420945,1.302063) *[blue]{\scriptscriptstyle+};
(1.429158,1.302208) *[blue]{\scriptscriptstyle+};
(1.437372,1.325862) *[blue]{\scriptscriptstyle+};
(1.445585,1.328684) *[blue]{\scriptscriptstyle+};
(1.453799,1.348963) *[blue]{\scriptscriptstyle+};
(1.462012,1.375757) *[blue]{\scriptscriptstyle+};
(1.470226,1.428045) *[blue]{\scriptscriptstyle+};
(1.478439,1.259394) *[blue]{\scriptscriptstyle+};
(1.486653,1.263133) *[blue]{\scriptscriptstyle+};
(1.494867,1.263902) *[blue]{\scriptscriptstyle+};
(1.503080,1.267323) *[blue]{\scriptscriptstyle+};
(1.511294,1.295690) *[blue]{\scriptscriptstyle+};
(1.519507,1.297738) *[blue]{\scriptscriptstyle+};
(1.527721,1.298783) *[blue]{\scriptscriptstyle+};
(1.535934,1.301000) *[blue]{\scriptscriptstyle+};
(1.544148,1.301818) *[blue]{\scriptscriptstyle+};
(1.552361,1.304112) *[blue]{\scriptscriptstyle+};
(1.560575,1.307419) *[blue]{\scriptscriptstyle+};
(1.568789,1.311959) *[blue]{\scriptscriptstyle+};
(1.577002,1.331848) *[blue]{\scriptscriptstyle+};
(1.585216,1.342440) *[blue]{\scriptscriptstyle+};
(1.593429,1.375387) *[blue]{\scriptscriptstyle+};
(1.601643,1.291288) *[blue]{\scriptscriptstyle+};
(1.609856,1.297041) *[blue]{\scriptscriptstyle+};
(1.618070,1.298145) *[blue]{\scriptscriptstyle+};
(1.626283,1.301053) *[blue]{\scriptscriptstyle+};
(1.634497,1.304528) *[blue]{\scriptscriptstyle+};
(1.642710,1.304575) *[blue]{\scriptscriptstyle+};
(1.650924,1.311140) *[blue]{\scriptscriptstyle+};
(1.659138,1.325677) *[blue]{\scriptscriptstyle+};
(1.667351,1.328914) *[blue]{\scriptscriptstyle+};
(1.675565,1.330568) *[blue]{\scriptscriptstyle+};
(1.683778,1.331233) *[blue]{\scriptscriptstyle+};
(1.691992,1.347497) *[blue]{\scriptscriptstyle+};
(1.700205,1.359394) *[blue]{\scriptscriptstyle+};
(1.708419,1.364245) *[blue]{\scriptscriptstyle+};
(1.716632,1.368580) *[blue]{\scriptscriptstyle+};
(1.724846,1.263963) *[blue]{\scriptscriptstyle+};
(1.733060,1.268491) *[blue]{\scriptscriptstyle+};
(1.741273,1.270315) *[blue]{\scriptscriptstyle+};
(1.749487,1.277321) *[blue]{\scriptscriptstyle+};
(1.757700,1.302171) *[blue]{\scriptscriptstyle+};
(1.765914,1.302176) *[blue]{\scriptscriptstyle+};
(1.774127,1.305350) *[blue]{\scriptscriptstyle+};
(1.782341,1.309084) *[blue]{\scriptscriptstyle+};
(1.790554,1.311926) *[blue]{\scriptscriptstyle+};
(1.798768,1.314545) *[blue]{\scriptscriptstyle+};
(1.806982,1.332959) *[blue]{\scriptscriptstyle+};
(1.815195,1.342323) *[blue]{\scriptscriptstyle+};
(1.823409,1.368385) *[blue]{\scriptscriptstyle+};
(1.831622,1.381926) *[blue]{\scriptscriptstyle+};
(1.839836,1.414074) *[blue]{\scriptscriptstyle+};
(1.848049,1.273858) *[blue]{\scriptscriptstyle+};
(1.856263,1.274830) *[blue]{\scriptscriptstyle+};
(1.864476,1.274950) *[blue]{\scriptscriptstyle+};
(1.872690,1.277828) *[blue]{\scriptscriptstyle+};
(1.880903,1.278099) *[blue]{\scriptscriptstyle+};
(1.889117,1.278407) *[blue]{\scriptscriptstyle+};
(1.897331,1.298118) *[blue]{\scriptscriptstyle+};
(1.905544,1.302890) *[blue]{\scriptscriptstyle+};
(1.913758,1.302903) *[blue]{\scriptscriptstyle+};
(1.921971,1.302928) *[blue]{\scriptscriptstyle+};
(1.930185,1.309367) *[blue]{\scriptscriptstyle+};
(1.938398,1.318789) *[blue]{\scriptscriptstyle+};
(1.946612,1.328851) *[blue]{\scriptscriptstyle+};
(1.954825,1.346576) *[blue]{\scriptscriptstyle+};
(1.963039,1.373366) *[blue]{\scriptscriptstyle+};
(1.971253,1.287714) *[blue]{\scriptscriptstyle+};
(1.979466,1.287722) *[blue]{\scriptscriptstyle+};
(1.987680,1.287792) *[blue]{\scriptscriptstyle+};
(1.995893,1.289780) *[blue]{\scriptscriptstyle+};
(2.004107,1.321950) *[blue]{\scriptscriptstyle+};
(2.012320,1.322146) *[blue]{\scriptscriptstyle+};
(2.020534,1.328523) *[blue]{\scriptscriptstyle+};
(2.028747,1.331802) *[blue]{\scriptscriptstyle+};
(2.036961,1.334579) *[blue]{\scriptscriptstyle+};
(2.045175,1.358865) *[blue]{\scriptscriptstyle+};
(2.053388,1.359373) *[blue]{\scriptscriptstyle+};
(2.061602,1.359892) *[blue]{\scriptscriptstyle+};
(2.069815,1.390433) *[blue]{\scriptscriptstyle+};
(2.078029,1.394953) *[blue]{\scriptscriptstyle+};
(2.086242,1.435507) *[blue]{\scriptscriptstyle+};
(2.094456,1.283417) *[blue]{\scriptscriptstyle+};
(2.102669,1.284538) *[blue]{\scriptscriptstyle+};
(2.110883,1.286495) *[blue]{\scriptscriptstyle+};
(2.119097,1.321235) *[blue]{\scriptscriptstyle+};
(2.127310,1.322450) *[blue]{\scriptscriptstyle+};
(2.135524,1.329031) *[blue]{\scriptscriptstyle+};
(2.143737,1.351983) *[blue]{\scriptscriptstyle+};
(2.151951,1.354559) *[blue]{\scriptscriptstyle+};
(2.160164,1.357400) *[blue]{\scriptscriptstyle+};
(2.168378,1.357833) *[blue]{\scriptscriptstyle+};
(2.176591,1.390929) *[blue]{\scriptscriptstyle+};
(2.184805,1.396022) *[blue]{\scriptscriptstyle+};
(2.193018,1.412096) *[blue]{\scriptscriptstyle+};
(2.201232,1.420361) *[blue]{\scriptscriptstyle+};
(2.209446,1.484867) *[blue]{\scriptscriptstyle+};
(2.217659,1.313252) *[blue]{\scriptscriptstyle+};
(2.225873,1.313286) *[blue]{\scriptscriptstyle+};
(2.234086,1.313308) *[blue]{\scriptscriptstyle+};
(2.242300,1.319966) *[blue]{\scriptscriptstyle+};
(2.250513,1.321477) *[blue]{\scriptscriptstyle+};
(2.258727,1.323083) *[blue]{\scriptscriptstyle+};
(2.266940,1.326229) *[blue]{\scriptscriptstyle+};
(2.275154,1.347106) *[blue]{\scriptscriptstyle+};
(2.283368,1.350211) *[blue]{\scriptscriptstyle+};
(2.291581,1.350245) *[blue]{\scriptscriptstyle+};
(2.299795,1.365925) *[blue]{\scriptscriptstyle+};
(2.308008,1.386158) *[blue]{\scriptscriptstyle+};
(2.316222,1.392244) *[blue]{\scriptscriptstyle+};
(2.324435,1.412393) *[blue]{\scriptscriptstyle+};
(2.332649,1.414082) *[blue]{\scriptscriptstyle+};
(2.340862,1.288149) *[blue]{\scriptscriptstyle+};
(2.349076,1.292382) *[blue]{\scriptscriptstyle+};
(2.357290,1.298427) *[blue]{\scriptscriptstyle+};
(2.365503,1.322365) *[blue]{\scriptscriptstyle+};
(2.373717,1.323570) *[blue]{\scriptscriptstyle+};
(2.381930,1.325710) *[blue]{\scriptscriptstyle+};
(2.390144,1.325752) *[blue]{\scriptscriptstyle+};
(2.398357,1.328765) *[blue]{\scriptscriptstyle+};
(2.406571,1.328878) *[blue]{\scriptscriptstyle+};
(2.414784,1.332140) *[blue]{\scriptscriptstyle+};
(2.422998,1.333025) *[blue]{\scriptscriptstyle+};
(2.431211,1.337846) *[blue]{\scriptscriptstyle+};
(2.439425,1.356096) *[blue]{\scriptscriptstyle+};
(2.447639,1.391974) *[blue]{\scriptscriptstyle+};
(2.455852,1.476543) *[blue]{\scriptscriptstyle+};
(2.464066,1.298814) *[blue]{\scriptscriptstyle+};
(2.472279,1.299627) *[blue]{\scriptscriptstyle+};
(2.480493,1.326522) *[blue]{\scriptscriptstyle+};
(2.488706,1.326597) *[blue]{\scriptscriptstyle+};
(2.496920,1.327892) *[blue]{\scriptscriptstyle+};
(2.505133,1.329538) *[blue]{\scriptscriptstyle+};
(2.513347,1.330523) *[blue]{\scriptscriptstyle+};
(2.521561,1.333981) *[blue]{\scriptscriptstyle+};
(2.529774,1.358242) *[blue]{\scriptscriptstyle+};
(2.537988,1.360787) *[blue]{\scriptscriptstyle+};
(2.546201,1.360867) *[blue]{\scriptscriptstyle+};
(2.554415,1.366172) *[blue]{\scriptscriptstyle+};
(2.562628,1.366925) *[blue]{\scriptscriptstyle+};
(2.570842,1.400197) *[blue]{\scriptscriptstyle+};
(2.579055,1.406190) *[blue]{\scriptscriptstyle+};
(2.587269,1.302015) *[blue]{\scriptscriptstyle+};
(2.595483,1.302016) *[blue]{\scriptscriptstyle+};
(2.603696,1.302026) *[blue]{\scriptscriptstyle+};
(2.611910,1.305363) *[blue]{\scriptscriptstyle+};
(2.620123,1.306269) *[blue]{\scriptscriptstyle+};
(2.628337,1.312978) *[blue]{\scriptscriptstyle+};
(2.636550,1.321881) *[blue]{\scriptscriptstyle+};
(2.644764,1.337262) *[blue]{\scriptscriptstyle+};
(2.652977,1.338842) *[blue]{\scriptscriptstyle+};
(2.661191,1.347872) *[blue]{\scriptscriptstyle+};
(2.669405,1.348797) *[blue]{\scriptscriptstyle+};
(2.677618,1.373587) *[blue]{\scriptscriptstyle+};
(2.685832,1.373788) *[blue]{\scriptscriptstyle+};
(2.694045,1.379666) *[blue]{\scriptscriptstyle+};
(2.702259,1.381859) *[blue]{\scriptscriptstyle+};
(2.710472,1.289794) *[blue]{\scriptscriptstyle+};
(2.718686,1.299422) *[blue]{\scriptscriptstyle+};
(2.726899,1.299672) *[blue]{\scriptscriptstyle+};
(2.735113,1.299711) *[blue]{\scriptscriptstyle+};
(2.743326,1.325422) *[blue]{\scriptscriptstyle+};
(2.751540,1.327402) *[blue]{\scriptscriptstyle+};
(2.759754,1.346576) *[blue]{\scriptscriptstyle+};
(2.767967,1.357469) *[blue]{\scriptscriptstyle+};
(2.776181,1.360857) *[blue]{\scriptscriptstyle+};
(2.784394,1.365611) *[blue]{\scriptscriptstyle+};
(2.792608,1.368196) *[blue]{\scriptscriptstyle+};
(2.800821,1.395927) *[blue]{\scriptscriptstyle+};
(2.809035,1.402868) *[blue]{\scriptscriptstyle+};
(2.817248,1.426716) *[blue]{\scriptscriptstyle+};
(2.825462,1.431739) *[blue]{\scriptscriptstyle+};
(2.833676,1.302491) *[blue]{\scriptscriptstyle+};
(2.841889,1.304536) *[blue]{\scriptscriptstyle+};
(2.850103,1.305957) *[blue]{\scriptscriptstyle+};
(2.858316,1.335448) *[blue]{\scriptscriptstyle+};
(2.866530,1.339748) *[blue]{\scriptscriptstyle+};
(2.874743,1.339758) *[blue]{\scriptscriptstyle+};
(2.882957,1.346153) *[blue]{\scriptscriptstyle+};
(2.891170,1.357459) *[blue]{\scriptscriptstyle+};
(2.899384,1.372872) *[blue]{\scriptscriptstyle+};
(2.907598,1.373569) *[blue]{\scriptscriptstyle+};
(2.915811,1.373608) *[blue]{\scriptscriptstyle+};
(2.924025,1.375872) *[blue]{\scriptscriptstyle+};
(2.932238,1.382369) *[blue]{\scriptscriptstyle+};
(2.940452,1.398316) *[blue]{\scriptscriptstyle+};
(2.948665,1.406910) *[blue]{\scriptscriptstyle+};
(2.956879,1.310234) *[blue]{\scriptscriptstyle+};
(2.965092,1.313541) *[blue]{\scriptscriptstyle+};
(2.973306,1.315171) *[blue]{\scriptscriptstyle+};
(2.981520,1.318095) *[blue]{\scriptscriptstyle+};
(2.989733,1.344065) *[blue]{\scriptscriptstyle+};
(2.997947,1.348012) *[blue]{\scriptscriptstyle+};
(3.006160,1.351677) *[blue]{\scriptscriptstyle+};
(3.014374,1.353688) *[blue]{\scriptscriptstyle+};
(3.022587,1.356494) *[blue]{\scriptscriptstyle+};
(3.030801,1.357951) *[blue]{\scriptscriptstyle+};
(3.039014,1.358466) *[blue]{\scriptscriptstyle+};
(3.047228,1.361152) *[blue]{\scriptscriptstyle+};
(3.055441,1.378293) *[blue]{\scriptscriptstyle+};
(3.063655,1.389966) *[blue]{\scriptscriptstyle+};
(3.071869,1.390462) *[blue]{\scriptscriptstyle+};
(3.080082,1.333086) *[blue]{\scriptscriptstyle+};
(3.088296,1.333114) *[blue]{\scriptscriptstyle+};
(3.096509,1.335495) *[blue]{\scriptscriptstyle+};
(3.104723,1.339076) *[blue]{\scriptscriptstyle+};
(3.112936,1.342745) *[blue]{\scriptscriptstyle+};
(3.121150,1.346485) *[blue]{\scriptscriptstyle+};
(3.129363,1.367900) *[blue]{\scriptscriptstyle+};
(3.137577,1.370796) *[blue]{\scriptscriptstyle+};
(3.145791,1.372647) *[blue]{\scriptscriptstyle+};
(3.154004,1.374350) *[blue]{\scriptscriptstyle+};
(3.162218,1.391272) *[blue]{\scriptscriptstyle+};
(3.170431,1.398854) *[blue]{\scriptscriptstyle+};
(3.178645,1.433758) *[blue]{\scriptscriptstyle+};
(3.186858,1.460635) *[blue]{\scriptscriptstyle+};
(3.195072,1.559225) *[blue]{\scriptscriptstyle+};
(3.203285,1.317759) *[blue]{\scriptscriptstyle+};
(3.211499,1.345265) *[blue]{\scriptscriptstyle+};
(3.219713,1.345269) *[blue]{\scriptscriptstyle+};
(3.227926,1.348257) *[blue]{\scriptscriptstyle+};
(3.236140,1.351289) *[blue]{\scriptscriptstyle+};
(3.244353,1.351490) *[blue]{\scriptscriptstyle+};
(3.252567,1.351507) *[blue]{\scriptscriptstyle+};
(3.260780,1.354588) *[blue]{\scriptscriptstyle+};
(3.268994,1.356616) *[blue]{\scriptscriptstyle+};
(3.277207,1.377862) *[blue]{\scriptscriptstyle+};
(3.285421,1.381040) *[blue]{\scriptscriptstyle+};
(3.293634,1.381513) *[blue]{\scriptscriptstyle+};
(3.301848,1.387326) *[blue]{\scriptscriptstyle+};
(3.310062,1.395386) *[blue]{\scriptscriptstyle+};
(3.318275,1.426025) *[blue]{\scriptscriptstyle+};
(3.326489,1.318536) *[blue]{\scriptscriptstyle+};
(3.334702,1.343945) *[blue]{\scriptscriptstyle+};
(3.342916,1.344086) *[blue]{\scriptscriptstyle+};
(3.351129,1.345955) *[blue]{\scriptscriptstyle+};
(3.359343,1.348911) *[blue]{\scriptscriptstyle+};
(3.367556,1.380277) *[blue]{\scriptscriptstyle+};
(3.375770,1.380397) *[blue]{\scriptscriptstyle+};
(3.383984,1.383122) *[blue]{\scriptscriptstyle+};
(3.392197,1.383487) *[blue]{\scriptscriptstyle+};
(3.400411,1.399066) *[blue]{\scriptscriptstyle+};
(3.408624,1.405414) *[blue]{\scriptscriptstyle+};
(3.416838,1.410918) *[blue]{\scriptscriptstyle+};
(3.425051,1.410984) *[blue]{\scriptscriptstyle+};
(3.433265,1.412824) *[blue]{\scriptscriptstyle+};
(3.441478,1.436598) *[blue]{\scriptscriptstyle+};
(3.449692,1.321836) *[blue]{\scriptscriptstyle+};
(3.457906,1.322467) *[blue]{\scriptscriptstyle+};
(3.466119,1.335319) *[blue]{\scriptscriptstyle+};
(3.474333,1.352656) *[blue]{\scriptscriptstyle+};
(3.482546,1.352897) *[blue]{\scriptscriptstyle+};
(3.490760,1.354881) *[blue]{\scriptscriptstyle+};
(3.498973,1.363555) *[blue]{\scriptscriptstyle+};
(3.507187,1.368947) *[blue]{\scriptscriptstyle+};
(3.515400,1.385824) *[blue]{\scriptscriptstyle+};
(3.523614,1.392007) *[blue]{\scriptscriptstyle+};
(3.531828,1.399896) *[blue]{\scriptscriptstyle+};
(3.540041,1.402281) *[blue]{\scriptscriptstyle+};
(3.548255,1.429515) *[blue]{\scriptscriptstyle+};
(3.556468,1.429916) *[blue]{\scriptscriptstyle+};
(3.564682,1.461167) *[blue]{\scriptscriptstyle+};
(3.572895,1.340216) *[blue]{\scriptscriptstyle+};
(3.581109,1.343348) *[blue]{\scriptscriptstyle+};
(3.589322,1.346574) *[blue]{\scriptscriptstyle+};
(3.597536,1.370189) *[blue]{\scriptscriptstyle+};
(3.605749,1.370382) *[blue]{\scriptscriptstyle+};
(3.613963,1.380887) *[blue]{\scriptscriptstyle+};
(3.622177,1.402877) *[blue]{\scriptscriptstyle+};
(3.630390,1.405633) *[blue]{\scriptscriptstyle+};
(3.638604,1.408804) *[blue]{\scriptscriptstyle+};
(3.646817,1.411860) *[blue]{\scriptscriptstyle+};
(3.655031,1.414883) *[blue]{\scriptscriptstyle+};
(3.663244,1.417925) *[blue]{\scriptscriptstyle+};
(3.671458,1.434477) *[blue]{\scriptscriptstyle+};
(3.679671,1.471610) *[blue]{\scriptscriptstyle+};
(3.687885,1.514176) *[blue]{\scriptscriptstyle+};
(3.696099,1.333194) *[blue]{\scriptscriptstyle+};
(3.704312,1.333883) *[blue]{\scriptscriptstyle+};
(3.712526,1.335182) *[blue]{\scriptscriptstyle+};
(3.720739,1.336178) *[blue]{\scriptscriptstyle+};
(3.728953,1.341343) *[blue]{\scriptscriptstyle+};
(3.737166,1.367094) *[blue]{\scriptscriptstyle+};
(3.745380,1.377855) *[blue]{\scriptscriptstyle+};
(3.753593,1.402794) *[blue]{\scriptscriptstyle+};
(3.761807,1.404107) *[blue]{\scriptscriptstyle+};
(3.770021,1.405188) *[blue]{\scriptscriptstyle+};
(3.778234,1.408940) *[blue]{\scriptscriptstyle+};
(3.786448,1.426990) *[blue]{\scriptscriptstyle+};
(3.794661,1.428745) *[blue]{\scriptscriptstyle+};
(3.802875,1.431771) *[blue]{\scriptscriptstyle+};
(3.811088,1.465735) *[blue]{\scriptscriptstyle+};
(3.819302,1.333682) *[blue]{\scriptscriptstyle+};
(3.827515,1.343320) *[blue]{\scriptscriptstyle+};
(3.835729,1.365089) *[blue]{\scriptscriptstyle+};
(3.843943,1.366994) *[blue]{\scriptscriptstyle+};
(3.852156,1.366998) *[blue]{\scriptscriptstyle+};
(3.860370,1.370313) *[blue]{\scriptscriptstyle+};
(3.868583,1.371497) *[blue]{\scriptscriptstyle+};
(3.876797,1.376339) *[blue]{\scriptscriptstyle+};
(3.885010,1.393403) *[blue]{\scriptscriptstyle+};
(3.893224,1.396356) *[blue]{\scriptscriptstyle+};
(3.901437,1.397528) *[blue]{\scriptscriptstyle+};
(3.909651,1.397984) *[blue]{\scriptscriptstyle+};
(3.917864,1.398247) *[blue]{\scriptscriptstyle+};
(3.926078,1.414130) *[blue]{\scriptscriptstyle+};
(3.934292,1.437362) *[blue]{\scriptscriptstyle+};
(3.942505,1.345373) *[blue]{\scriptscriptstyle+};
(3.950719,1.347506) *[blue]{\scriptscriptstyle+};
(3.958932,1.367793) *[blue]{\scriptscriptstyle+};
(3.967146,1.367859) *[blue]{\scriptscriptstyle+};
(3.975359,1.367863) *[blue]{\scriptscriptstyle+};
(3.983573,1.367965) *[blue]{\scriptscriptstyle+};
(3.991786,1.371605) *[blue]{\scriptscriptstyle+};
(4.000000,1.372972) *[blue]{\scriptscriptstyle+};
(4.008214,1.377426) *[blue]{\scriptscriptstyle+};
(4.016427,1.379071) *[blue]{\scriptscriptstyle+};
(4.024641,1.401343) *[blue]{\scriptscriptstyle+};
(4.032854,1.406691) *[blue]{\scriptscriptstyle+};
(4.041068,1.410609) *[blue]{\scriptscriptstyle+};
(4.049281,1.439415) *[blue]{\scriptscriptstyle+};
(4.057495,1.469414) *[blue]{\scriptscriptstyle+};
(4.065708,1.349597) *[blue]{\scriptscriptstyle+};
(4.073922,1.349703) *[blue]{\scriptscriptstyle+};
(4.082136,1.352912) *[blue]{\scriptscriptstyle+};
(4.090349,1.354612) *[blue]{\scriptscriptstyle+};
(4.098563,1.358652) *[blue]{\scriptscriptstyle+};
(4.106776,1.372373) *[blue]{\scriptscriptstyle+};
(4.114990,1.377952) *[blue]{\scriptscriptstyle+};
(4.123203,1.384533) *[blue]{\scriptscriptstyle+};
(4.131417,1.385291) *[blue]{\scriptscriptstyle+};
(4.139630,1.388992) *[blue]{\scriptscriptstyle+};
(4.147844,1.393470) *[blue]{\scriptscriptstyle+};
(4.156057,1.393888) *[blue]{\scriptscriptstyle+};
(4.164271,1.414899) *[blue]{\scriptscriptstyle+};
(4.172485,1.420202) *[blue]{\scriptscriptstyle+};
(4.180698,1.422803) *[blue]{\scriptscriptstyle+};
(4.188912,1.349208) *[blue]{\scriptscriptstyle+};
(4.197125,1.352214) *[blue]{\scriptscriptstyle+};
(4.205339,1.379937) *[blue]{\scriptscriptstyle+};
(4.213552,1.383060) *[blue]{\scriptscriptstyle+};
(4.221766,1.384547) *[blue]{\scriptscriptstyle+};
(4.229979,1.386087) *[blue]{\scriptscriptstyle+};
(4.238193,1.392198) *[blue]{\scriptscriptstyle+};
(4.246407,1.393347) *[blue]{\scriptscriptstyle+};
(4.254620,1.396138) *[blue]{\scriptscriptstyle+};
(4.262834,1.398100) *[blue]{\scriptscriptstyle+};
(4.271047,1.416093) *[blue]{\scriptscriptstyle+};
(4.279261,1.418935) *[blue]{\scriptscriptstyle+};
(4.287474,1.426961) *[blue]{\scriptscriptstyle+};
(4.295688,1.442364) *[blue]{\scriptscriptstyle+};
(4.303901,1.486304) *[blue]{\scriptscriptstyle+};
(4.312115,1.376920) *[blue]{\scriptscriptstyle+};
(4.320329,1.379749) *[blue]{\scriptscriptstyle+};
(4.328542,1.388737) *[blue]{\scriptscriptstyle+};
(4.336756,1.409411) *[blue]{\scriptscriptstyle+};
(4.344969,1.410740) *[blue]{\scriptscriptstyle+};
(4.353183,1.410786) *[blue]{\scriptscriptstyle+};
(4.361396,1.413829) *[blue]{\scriptscriptstyle+};
(4.369610,1.416292) *[blue]{\scriptscriptstyle+};
(4.377823,1.416872) *[blue]{\scriptscriptstyle+};
(4.386037,1.424013) *[blue]{\scriptscriptstyle+};
(4.394251,1.425848) *[blue]{\scriptscriptstyle+};
(4.402464,1.439370) *[blue]{\scriptscriptstyle+};
(4.410678,1.443734) *[blue]{\scriptscriptstyle+};
(4.418891,1.449427) *[blue]{\scriptscriptstyle+};
(4.427105,1.464935) *[blue]{\scriptscriptstyle+};
(4.435318,1.359992) *[blue]{\scriptscriptstyle+};
(4.443532,1.360125) *[blue]{\scriptscriptstyle+};
(4.451745,1.360200) *[blue]{\scriptscriptstyle+};
(4.459959,1.392607) *[blue]{\scriptscriptstyle+};
(4.468172,1.392838) *[blue]{\scriptscriptstyle+};
(4.476386,1.392884) *[blue]{\scriptscriptstyle+};
(4.484600,1.400620) *[blue]{\scriptscriptstyle+};
(4.492813,1.405162) *[blue]{\scriptscriptstyle+};
(4.501027,1.415780) *[blue]{\scriptscriptstyle+};
(4.509240,1.424728) *[blue]{\scriptscriptstyle+};
(4.517454,1.424854) *[blue]{\scriptscriptstyle+};
(4.525667,1.427878) *[blue]{\scriptscriptstyle+};
(4.533881,1.438308) *[blue]{\scriptscriptstyle+};
(4.542094,1.439505) *[blue]{\scriptscriptstyle+};
(4.550308,1.504214) *[blue]{\scriptscriptstyle+};
(4.558522,1.387664) *[blue]{\scriptscriptstyle+};
(4.566735,1.397920) *[blue]{\scriptscriptstyle+};
(4.574949,1.410471) *[blue]{\scriptscriptstyle+};
(4.583162,1.410782) *[blue]{\scriptscriptstyle+};
(4.591376,1.410861) *[blue]{\scriptscriptstyle+};
(4.599589,1.411466) *[blue]{\scriptscriptstyle+};
(4.607803,1.411469) *[blue]{\scriptscriptstyle+};
(4.616016,1.411642) *[blue]{\scriptscriptstyle+};
(4.624230,1.423657) *[blue]{\scriptscriptstyle+};
(4.632444,1.431080) *[blue]{\scriptscriptstyle+};
(4.640657,1.442121) *[blue]{\scriptscriptstyle+};
(4.648871,1.448800) *[blue]{\scriptscriptstyle+};
(4.657084,1.457312) *[blue]{\scriptscriptstyle+};
(4.665298,1.459935) *[blue]{\scriptscriptstyle+};
(4.673511,1.497826) *[blue]{\scriptscriptstyle+};
(4.681725,1.348423) *[blue]{\scriptscriptstyle+};
(4.689938,1.354869) *[blue]{\scriptscriptstyle+};
(4.698152,1.358536) *[blue]{\scriptscriptstyle+};
(4.706366,1.409909) *[blue]{\scriptscriptstyle+};
(4.714579,1.421458) *[blue]{\scriptscriptstyle+};
(4.722793,1.423838) *[blue]{\scriptscriptstyle+};
(4.731006,1.425876) *[blue]{\scriptscriptstyle+};
(4.739220,1.429115) *[blue]{\scriptscriptstyle+};
(4.747433,1.457529) *[blue]{\scriptscriptstyle+};
(4.755647,1.466155) *[blue]{\scriptscriptstyle+};
(4.763860,1.493595) *[blue]{\scriptscriptstyle+};
(4.772074,1.536736) *[blue]{\scriptscriptstyle+};
(4.780287,1.559845) *[blue]{\scriptscriptstyle+};
(4.788501,1.626911) *[blue]{\scriptscriptstyle+};
(4.796715,1.627469) *[blue]{\scriptscriptstyle+};
(4.804928,1.360488) *[blue]{\scriptscriptstyle+};
(4.813142,1.360706) *[blue]{\scriptscriptstyle+};
(4.821355,1.366277) *[blue]{\scriptscriptstyle+};
(4.829569,1.367313) *[blue]{\scriptscriptstyle+};
(4.837782,1.373108) *[blue]{\scriptscriptstyle+};
(4.845996,1.379384) *[blue]{\scriptscriptstyle+};
(4.854209,1.393124) *[blue]{\scriptscriptstyle+};
(4.862423,1.399416) *[blue]{\scriptscriptstyle+};
(4.870637,1.427835) *[blue]{\scriptscriptstyle+};
(4.878850,1.428156) *[blue]{\scriptscriptstyle+};
(4.887064,1.428368) *[blue]{\scriptscriptstyle+};
(4.895277,1.431147) *[blue]{\scriptscriptstyle+};
(4.903491,1.431248) *[blue]{\scriptscriptstyle+};
(4.911704,1.437151) *[blue]{\scriptscriptstyle+};
(4.919918,1.440891) *[blue]{\scriptscriptstyle+};
(4.928131,1.374778) *[blue]{\scriptscriptstyle+};
(4.936345,1.377767) *[blue]{\scriptscriptstyle+};
(4.944559,1.379674) *[blue]{\scriptscriptstyle+};
(4.952772,1.380864) *[blue]{\scriptscriptstyle+};
(4.960986,1.382059) *[blue]{\scriptscriptstyle+};
(4.969199,1.406910) *[blue]{\scriptscriptstyle+};
(4.977413,1.408225) *[blue]{\scriptscriptstyle+};
(4.985626,1.410065) *[blue]{\scriptscriptstyle+};
(4.993840,1.413315) *[blue]{\scriptscriptstyle+};
(5.002053,1.434304) *[blue]{\scriptscriptstyle+};
(5.010267,1.444695) *[blue]{\scriptscriptstyle+};
(5.018480,1.446428) *[blue]{\scriptscriptstyle+};
(5.026694,1.447649) *[blue]{\scriptscriptstyle+};
(5.034908,1.449993) *[blue]{\scriptscriptstyle+};
(5.043121,1.482580) *[blue]{\scriptscriptstyle+};
(5.051335,1.403009) *[blue]{\scriptscriptstyle+};
(5.059548,1.404341) *[blue]{\scriptscriptstyle+};
(5.067762,1.407557) *[blue]{\scriptscriptstyle+};
(5.075975,1.407718) *[blue]{\scriptscriptstyle+};
(5.084189,1.431059) *[blue]{\scriptscriptstyle+};
(5.092402,1.431650) *[blue]{\scriptscriptstyle+};
(5.100616,1.431816) *[blue]{\scriptscriptstyle+};
(5.108830,1.432694) *[blue]{\scriptscriptstyle+};
(5.117043,1.438861) *[blue]{\scriptscriptstyle+};
(5.125257,1.442254) *[blue]{\scriptscriptstyle+};
(5.133470,1.443005) *[blue]{\scriptscriptstyle+};
(5.141684,1.468799) *[blue]{\scriptscriptstyle+};
(5.149897,1.476153) *[blue]{\scriptscriptstyle+};
(5.158111,1.494747) *[blue]{\scriptscriptstyle+};
(5.166324,1.496841) *[blue]{\scriptscriptstyle+};
(5.174538,1.365911) *[blue]{\scriptscriptstyle+};
(5.182752,1.371140) *[blue]{\scriptscriptstyle+};
(5.190965,1.371962) *[blue]{\scriptscriptstyle+};
(5.199179,1.376497) *[blue]{\scriptscriptstyle+};
(5.207392,1.402785) *[blue]{\scriptscriptstyle+};
(5.215606,1.404397) *[blue]{\scriptscriptstyle+};
(5.223819,1.407681) *[blue]{\scriptscriptstyle+};
(5.232033,1.409259) *[blue]{\scriptscriptstyle+};
(5.240246,1.430381) *[blue]{\scriptscriptstyle+};
(5.248460,1.433226) *[blue]{\scriptscriptstyle+};
(5.256674,1.433303) *[blue]{\scriptscriptstyle+};
(5.264887,1.440861) *[blue]{\scriptscriptstyle+};
(5.273101,1.444481) *[blue]{\scriptscriptstyle+};
(5.281314,1.465660) *[blue]{\scriptscriptstyle+};
(5.289528,1.473598) *[blue]{\scriptscriptstyle+};
(5.297741,1.384619) *[blue]{\scriptscriptstyle+};
(5.305955,1.386939) *[blue]{\scriptscriptstyle+};
(5.314168,1.389106) *[blue]{\scriptscriptstyle+};
(5.322382,1.390937) *[blue]{\scriptscriptstyle+};
(5.330595,1.391916) *[blue]{\scriptscriptstyle+};
(5.338809,1.393337) *[blue]{\scriptscriptstyle+};
(5.347023,1.416730) *[blue]{\scriptscriptstyle+};
(5.355236,1.420016) *[blue]{\scriptscriptstyle+};
(5.363450,1.420184) *[blue]{\scriptscriptstyle+};
(5.371663,1.421028) *[blue]{\scriptscriptstyle+};
(5.379877,1.422942) *[blue]{\scriptscriptstyle+};
(5.388090,1.431881) *[blue]{\scriptscriptstyle+};
(5.396304,1.452294) *[blue]{\scriptscriptstyle+};
(5.404517,1.457390) *[blue]{\scriptscriptstyle+};
(5.412731,1.466867) *[blue]{\scriptscriptstyle+};
(5.420945,1.392568) *[blue]{\scriptscriptstyle+};
(5.429158,1.394760) *[blue]{\scriptscriptstyle+};
(5.437372,1.395599) *[blue]{\scriptscriptstyle+};
(5.445585,1.395898) *[blue]{\scriptscriptstyle+};
(5.453799,1.404915) *[blue]{\scriptscriptstyle+};
(5.462012,1.406115) *[blue]{\scriptscriptstyle+};
(5.470226,1.427141) *[blue]{\scriptscriptstyle+};
(5.478439,1.428199) *[blue]{\scriptscriptstyle+};
(5.486653,1.428513) *[blue]{\scriptscriptstyle+};
(5.494867,1.431105) *[blue]{\scriptscriptstyle+};
(5.503080,1.432625) *[blue]{\scriptscriptstyle+};
(5.511294,1.437128) *[blue]{\scriptscriptstyle+};
(5.519507,1.453531) *[blue]{\scriptscriptstyle+};
(5.527721,1.460155) *[blue]{\scriptscriptstyle+};
(5.535934,1.528532) *[blue]{\scriptscriptstyle+};
(5.544148,1.408875) *[blue]{\scriptscriptstyle+};
(5.552361,1.438428) *[blue]{\scriptscriptstyle+};
(5.560575,1.441414) *[blue]{\scriptscriptstyle+};
(5.568789,1.441903) *[blue]{\scriptscriptstyle+};
(5.577002,1.446791) *[blue]{\scriptscriptstyle+};
(5.585216,1.448019) *[blue]{\scriptscriptstyle+};
(5.593429,1.471500) *[blue]{\scriptscriptstyle+};
(5.601643,1.472411) *[blue]{\scriptscriptstyle+};
(5.609856,1.475282) *[blue]{\scriptscriptstyle+};
(5.618070,1.477650) *[blue]{\scriptscriptstyle+};
(5.626283,1.478275) *[blue]{\scriptscriptstyle+};
(5.634497,1.480102) *[blue]{\scriptscriptstyle+};
(5.642710,1.498706) *[blue]{\scriptscriptstyle+};
(5.650924,1.499037) *[blue]{\scriptscriptstyle+};
(5.659138,1.501779) *[blue]{\scriptscriptstyle+};
(5.667351,1.366730) *[blue]{\scriptscriptstyle+};
(5.675565,1.394655) *[blue]{\scriptscriptstyle+};
(5.683778,1.394709) *[blue]{\scriptscriptstyle+};
(5.691992,1.400781) *[blue]{\scriptscriptstyle+};
(5.700205,1.401852) *[blue]{\scriptscriptstyle+};
(5.708419,1.402296) *[blue]{\scriptscriptstyle+};
(5.716632,1.404012) *[blue]{\scriptscriptstyle+};
(5.724846,1.421727) *[blue]{\scriptscriptstyle+};
(5.733060,1.426671) *[blue]{\scriptscriptstyle+};
(5.741273,1.432570) *[blue]{\scriptscriptstyle+};
(5.749487,1.432776) *[blue]{\scriptscriptstyle+};
(5.757700,1.436863) *[blue]{\scriptscriptstyle+};
(5.765914,1.438265) *[blue]{\scriptscriptstyle+};
(5.774127,1.439986) *[blue]{\scriptscriptstyle+};
(5.782341,1.476973) *[blue]{\scriptscriptstyle+};
(5.790554,1.430343) *[blue]{\scriptscriptstyle+};
(5.798768,1.430443) *[blue]{\scriptscriptstyle+};
(5.806982,1.431541) *[blue]{\scriptscriptstyle+};
(5.815195,1.431717) *[blue]{\scriptscriptstyle+};
(5.823409,1.439363) *[blue]{\scriptscriptstyle+};
(5.831622,1.441930) *[blue]{\scriptscriptstyle+};
(5.839836,1.442340) *[blue]{\scriptscriptstyle+};
(5.848049,1.442542) *[blue]{\scriptscriptstyle+};
(5.856263,1.445020) *[blue]{\scriptscriptstyle+};
(5.864476,1.448240) *[blue]{\scriptscriptstyle+};
(5.872690,1.458104) *[blue]{\scriptscriptstyle+};
(5.880903,1.463037) *[blue]{\scriptscriptstyle+};
(5.889117,1.471514) *[blue]{\scriptscriptstyle+};
(5.897331,1.493263) *[blue]{\scriptscriptstyle+};
(5.905544,1.496625) *[blue]{\scriptscriptstyle+};
(5.913758,1.441941) *[blue]{\scriptscriptstyle+};
(5.921971,1.441989) *[blue]{\scriptscriptstyle+};
(5.930185,1.442013) *[blue]{\scriptscriptstyle+};
(5.938398,1.442018) *[blue]{\scriptscriptstyle+};
(5.946612,1.445277) *[blue]{\scriptscriptstyle+};
(5.954825,1.448093) *[blue]{\scriptscriptstyle+};
(5.963039,1.472768) *[blue]{\scriptscriptstyle+};
(5.971253,1.472820) *[blue]{\scriptscriptstyle+};
(5.979466,1.474494) *[blue]{\scriptscriptstyle+};
(5.987680,1.475257) *[blue]{\scriptscriptstyle+};
(5.995893,1.478769) *[blue]{\scriptscriptstyle+};
(6.004107,1.495510) *[blue]{\scriptscriptstyle+};
(6.012320,1.502774) *[blue]{\scriptscriptstyle+};
(6.020534,1.504655) *[blue]{\scriptscriptstyle+};
(6.028747,1.507532) *[blue]{\scriptscriptstyle+};
(6.036961,1.425206) *[blue]{\scriptscriptstyle+};
(6.045175,1.431132) *[blue]{\scriptscriptstyle+};
(6.053388,1.456527) *[blue]{\scriptscriptstyle+};
(6.061602,1.457614) *[blue]{\scriptscriptstyle+};
(6.069815,1.457721) *[blue]{\scriptscriptstyle+};
(6.078029,1.459459) *[blue]{\scriptscriptstyle+};
(6.086242,1.462732) *[blue]{\scriptscriptstyle+};
(6.094456,1.471390) *[blue]{\scriptscriptstyle+};
(6.102669,1.486849) *[blue]{\scriptscriptstyle+};
(6.110883,1.487852) *[blue]{\scriptscriptstyle+};
(6.119097,1.492814) *[blue]{\scriptscriptstyle+};
(6.127310,1.497965) *[blue]{\scriptscriptstyle+};
(6.135524,1.498051) *[blue]{\scriptscriptstyle+};
(6.143737,1.505502) *[blue]{\scriptscriptstyle+};
(6.151951,1.538141) *[blue]{\scriptscriptstyle+};
(6.160164,1.426620) *[blue]{\scriptscriptstyle+};
(6.168378,1.448340) *[blue]{\scriptscriptstyle+};
(6.176591,1.449061) *[blue]{\scriptscriptstyle+};
(6.184805,1.449701) *[blue]{\scriptscriptstyle+};
(6.193018,1.450835) *[blue]{\scriptscriptstyle+};
(6.201232,1.454900) *[blue]{\scriptscriptstyle+};
(6.209446,1.476788) *[blue]{\scriptscriptstyle+};
(6.217659,1.478202) *[blue]{\scriptscriptstyle+};
(6.225873,1.479888) *[blue]{\scriptscriptstyle+};
(6.234086,1.483513) *[blue]{\scriptscriptstyle+};
(6.242300,1.485479) *[blue]{\scriptscriptstyle+};
(6.250513,1.487007) *[blue]{\scriptscriptstyle+};
(6.258727,1.493268) *[blue]{\scriptscriptstyle+};
(6.266940,1.518461) *[blue]{\scriptscriptstyle+};
(6.275154,1.526811) *[blue]{\scriptscriptstyle+};
(6.283368,1.456172) *[blue]{\scriptscriptstyle+};
(6.291581,1.483659) *[blue]{\scriptscriptstyle+};
(6.299795,1.486849) *[blue]{\scriptscriptstyle+};
(6.308008,1.489564) *[blue]{\scriptscriptstyle+};
(6.316222,1.492181) *[blue]{\scriptscriptstyle+};
(6.324435,1.499772) *[blue]{\scriptscriptstyle+};
(6.332649,1.513772) *[blue]{\scriptscriptstyle+};
(6.340862,1.513835) *[blue]{\scriptscriptstyle+};
(6.349076,1.518539) *[blue]{\scriptscriptstyle+};
(6.357290,1.522447) *[blue]{\scriptscriptstyle+};
(6.365503,1.522451) *[blue]{\scriptscriptstyle+};
(6.373717,1.527082) *[blue]{\scriptscriptstyle+};
(6.381930,1.543534) *[blue]{\scriptscriptstyle+};
(6.390144,1.554479) *[blue]{\scriptscriptstyle+};
(6.398357,1.569351) *[blue]{\scriptscriptstyle+};
(6.406571,1.459030) *[blue]{\scriptscriptstyle+};
(6.414784,1.461430) *[blue]{\scriptscriptstyle+};
(6.422998,1.489056) *[blue]{\scriptscriptstyle+};
(6.431211,1.494626) *[blue]{\scriptscriptstyle+};
(6.439425,1.499382) *[blue]{\scriptscriptstyle+};
(6.447639,1.500577) *[blue]{\scriptscriptstyle+};
(6.455852,1.500581) *[blue]{\scriptscriptstyle+};
(6.464066,1.518961) *[blue]{\scriptscriptstyle+};
(6.472279,1.524558) *[blue]{\scriptscriptstyle+};
(6.480493,1.524657) *[blue]{\scriptscriptstyle+};
(6.488706,1.526395) *[blue]{\scriptscriptstyle+};
(6.496920,1.546809) *[blue]{\scriptscriptstyle+};
(6.505133,1.551039) *[blue]{\scriptscriptstyle+};
(6.513347,1.556564) *[blue]{\scriptscriptstyle+};
(6.521561,1.590742) *[blue]{\scriptscriptstyle+};
(6.529774,1.448530) *[blue]{\scriptscriptstyle+};
(6.537988,1.450827) *[blue]{\scriptscriptstyle+};
(6.546201,1.462888) *[blue]{\scriptscriptstyle+};
(6.554415,1.475672) *[blue]{\scriptscriptstyle+};
(6.562628,1.481337) *[blue]{\scriptscriptstyle+};
(6.570842,1.488600) *[blue]{\scriptscriptstyle+};
(6.579055,1.504169) *[blue]{\scriptscriptstyle+};
(6.587269,1.507256) *[blue]{\scriptscriptstyle+};
(6.595483,1.512800) *[blue]{\scriptscriptstyle+};
(6.603696,1.513274) *[blue]{\scriptscriptstyle+};
(6.611910,1.518619) *[blue]{\scriptscriptstyle+};
(6.620123,1.527002) *[blue]{\scriptscriptstyle+};
(6.628337,1.577994) *[blue]{\scriptscriptstyle+};
(6.636550,1.587709) *[blue]{\scriptscriptstyle+};
(6.644764,1.646838) *[blue]{\scriptscriptstyle+};
(6.652977,1.428894) *[blue]{\scriptscriptstyle+};
(6.661191,1.457110) *[blue]{\scriptscriptstyle+};
(6.669405,1.457146) *[blue]{\scriptscriptstyle+};
(6.677618,1.457194) *[blue]{\scriptscriptstyle+};
(6.685832,1.457338) *[blue]{\scriptscriptstyle+};
(6.694045,1.460112) *[blue]{\scriptscriptstyle+};
(6.702259,1.463079) *[blue]{\scriptscriptstyle+};
(6.710472,1.490770) *[blue]{\scriptscriptstyle+};
(6.718686,1.502312) *[blue]{\scriptscriptstyle+};
(6.726899,1.504987) *[blue]{\scriptscriptstyle+};
(6.735113,1.517736) *[blue]{\scriptscriptstyle+};
(6.743326,1.523394) *[blue]{\scriptscriptstyle+};
(6.751540,1.537348) *[blue]{\scriptscriptstyle+};
(6.759754,1.558237) *[blue]{\scriptscriptstyle+};
(6.767967,1.608267) *[blue]{\scriptscriptstyle+};
(6.776181,1.485001) *[blue]{\scriptscriptstyle+};
(6.784394,1.485672) *[blue]{\scriptscriptstyle+};
(6.792608,1.497730) *[blue]{\scriptscriptstyle+};
(6.800821,1.514097) *[blue]{\scriptscriptstyle+};
(6.809035,1.517895) *[blue]{\scriptscriptstyle+};
(6.817248,1.519775) *[blue]{\scriptscriptstyle+};
(6.825462,1.519815) *[blue]{\scriptscriptstyle+};
(6.833676,1.522603) *[blue]{\scriptscriptstyle+};
(6.841889,1.522631) *[blue]{\scriptscriptstyle+};
(6.850103,1.539858) *[blue]{\scriptscriptstyle+};
(6.858316,1.546768) *[blue]{\scriptscriptstyle+};
(6.866530,1.550307) *[blue]{\scriptscriptstyle+};
(6.874743,1.554826) *[blue]{\scriptscriptstyle+};
(6.882957,1.557385) *[blue]{\scriptscriptstyle+};
(6.891170,1.576288) *[blue]{\scriptscriptstyle+};
(6.899384,1.496277) *[blue]{\scriptscriptstyle+};
(6.907598,1.496316) *[blue]{\scriptscriptstyle+};
(6.915811,1.496430) *[blue]{\scriptscriptstyle+};
(6.924025,1.498891) *[blue]{\scriptscriptstyle+};
(6.932238,1.499394) *[blue]{\scriptscriptstyle+};
(6.940452,1.503889) *[blue]{\scriptscriptstyle+};
(6.948665,1.526014) *[blue]{\scriptscriptstyle+};
(6.956879,1.526043) *[blue]{\scriptscriptstyle+};
(6.965092,1.532859) *[blue]{\scriptscriptstyle+};
(6.973306,1.533118) *[blue]{\scriptscriptstyle+};
(6.981520,1.534400) *[blue]{\scriptscriptstyle+};
(6.989733,1.535402) *[blue]{\scriptscriptstyle+};
(6.997947,1.539355) *[blue]{\scriptscriptstyle+};
(7.006160,1.541275) *[blue]{\scriptscriptstyle+};
(7.014374,1.557810) *[blue]{\scriptscriptstyle+};
(7.022587,1.458056) *[blue]{\scriptscriptstyle+};
(7.030801,1.466572) *[blue]{\scriptscriptstyle+};
(7.039014,1.467186) *[blue]{\scriptscriptstyle+};
(7.047228,1.487493) *[blue]{\scriptscriptstyle+};
(7.055441,1.488317) *[blue]{\scriptscriptstyle+};
(7.063655,1.488456) *[blue]{\scriptscriptstyle+};
(7.071869,1.490123) *[blue]{\scriptscriptstyle+};
(7.080082,1.497274) *[blue]{\scriptscriptstyle+};
(7.088296,1.521513) *[blue]{\scriptscriptstyle+};
(7.096509,1.527227) *[blue]{\scriptscriptstyle+};
(7.104723,1.535447) *[blue]{\scriptscriptstyle+};
(7.112936,1.556140) *[blue]{\scriptscriptstyle+};
(7.121150,1.560023) *[blue]{\scriptscriptstyle+};
(7.129363,1.607053) *[blue]{\scriptscriptstyle+};
(7.137577,1.667783) *[blue]{\scriptscriptstyle+};
(7.145791,1.502009) *[blue]{\scriptscriptstyle+};
(7.154004,1.509142) *[blue]{\scriptscriptstyle+};
(7.162218,1.519254) *[blue]{\scriptscriptstyle+};
(7.170431,1.533951) *[blue]{\scriptscriptstyle+};
(7.178645,1.534387) *[blue]{\scriptscriptstyle+};
(7.186858,1.534427) *[blue]{\scriptscriptstyle+};
(7.195072,1.538605) *[blue]{\scriptscriptstyle+};
(7.203285,1.545457) *[blue]{\scriptscriptstyle+};
(7.211499,1.563953) *[blue]{\scriptscriptstyle+};
(7.219713,1.571669) *[blue]{\scriptscriptstyle+};
(7.227926,1.591747) *[blue]{\scriptscriptstyle+};
(7.236140,1.603609) *[blue]{\scriptscriptstyle+};
(7.244353,1.606549) *[blue]{\scriptscriptstyle+};
(7.252567,1.640635) *[blue]{\scriptscriptstyle+};
(7.260780,1.667636) *[blue]{\scriptscriptstyle+};
(7.268994,1.476604) *[blue]{\scriptscriptstyle+};
(7.277207,1.481818) *[blue]{\scriptscriptstyle+};
(7.285421,1.505322) *[blue]{\scriptscriptstyle+};
(7.293634,1.508906) *[blue]{\scriptscriptstyle+};
(7.301848,1.509089) *[blue]{\scriptscriptstyle+};
(7.310062,1.516027) *[blue]{\scriptscriptstyle+};
(7.318275,1.529914) *[blue]{\scriptscriptstyle+};
(7.326489,1.535015) *[blue]{\scriptscriptstyle+};
(7.334702,1.535769) *[blue]{\scriptscriptstyle+};
(7.342916,1.537798) *[blue]{\scriptscriptstyle+};
(7.351129,1.540070) *[blue]{\scriptscriptstyle+};
(7.359343,1.559324) *[blue]{\scriptscriptstyle+};
(7.367556,1.569584) *[blue]{\scriptscriptstyle+};
(7.375770,1.575648) *[blue]{\scriptscriptstyle+};
(7.383984,1.632867) *[blue]{\scriptscriptstyle+};
(7.392197,1.467219) *[blue]{\scriptscriptstyle+};
(7.400411,1.483968) *[blue]{\scriptscriptstyle+};
(7.408624,1.486272) *[blue]{\scriptscriptstyle+};
(7.416838,1.486274) *[blue]{\scriptscriptstyle+};
(7.425051,1.486338) *[blue]{\scriptscriptstyle+};
(7.433265,1.489337) *[blue]{\scriptscriptstyle+};
(7.441478,1.492335) *[blue]{\scriptscriptstyle+};
(7.449692,1.506437) *[blue]{\scriptscriptstyle+};
(7.457906,1.510621) *[blue]{\scriptscriptstyle+};
(7.466119,1.511855) *[blue]{\scriptscriptstyle+};
(7.474333,1.521953) *[blue]{\scriptscriptstyle+};
(7.482546,1.544434) *[blue]{\scriptscriptstyle+};
(7.490760,1.548718) *[blue]{\scriptscriptstyle+};
(7.498973,1.569252) *[blue]{\scriptscriptstyle+};
(7.507187,1.571997) *[blue]{\scriptscriptstyle+};
(7.515400,1.493826) *[blue]{\scriptscriptstyle+};
(7.523614,1.496837) *[blue]{\scriptscriptstyle+};
(7.531828,1.499496) *[blue]{\scriptscriptstyle+};
(7.540041,1.508724) *[blue]{\scriptscriptstyle+};
(7.548255,1.526416) *[blue]{\scriptscriptstyle+};
(7.556468,1.527237) *[blue]{\scriptscriptstyle+};
(7.564682,1.530196) *[blue]{\scriptscriptstyle+};
(7.572895,1.532073) *[blue]{\scriptscriptstyle+};
(7.581109,1.532598) *[blue]{\scriptscriptstyle+};
(7.589322,1.540061) *[blue]{\scriptscriptstyle+};
(7.597536,1.553055) *[blue]{\scriptscriptstyle+};
(7.605749,1.553986) *[blue]{\scriptscriptstyle+};
(7.613963,1.561068) *[blue]{\scriptscriptstyle+};
(7.622177,1.591019) *[blue]{\scriptscriptstyle+};
(7.630390,1.595946) *[blue]{\scriptscriptstyle+};
(7.638604,1.487332) *[blue]{\scriptscriptstyle+};
(7.646817,1.491908) *[blue]{\scriptscriptstyle+};
(7.655031,1.491925) *[blue]{\scriptscriptstyle+};
(7.663244,1.493106) *[blue]{\scriptscriptstyle+};
(7.671458,1.521682) *[blue]{\scriptscriptstyle+};
(7.679671,1.522478) *[blue]{\scriptscriptstyle+};
(7.687885,1.525430) *[blue]{\scriptscriptstyle+};
(7.696099,1.526865) *[blue]{\scriptscriptstyle+};
(7.704312,1.548001) *[blue]{\scriptscriptstyle+};
(7.712526,1.548746) *[blue]{\scriptscriptstyle+};
(7.720739,1.550535) *[blue]{\scriptscriptstyle+};
(7.728953,1.575750) *[blue]{\scriptscriptstyle+};
(7.737166,1.605341) *[blue]{\scriptscriptstyle+};
(7.745380,1.607688) *[blue]{\scriptscriptstyle+};
(7.753593,1.619784) *[blue]{\scriptscriptstyle+};
(7.761807,1.560828) *[blue]{\scriptscriptstyle+};
(7.770021,1.560829) *[blue]{\scriptscriptstyle+};
(7.778234,1.560925) *[blue]{\scriptscriptstyle+};
(7.786448,1.563589) *[blue]{\scriptscriptstyle+};
(7.794661,1.563604) *[blue]{\scriptscriptstyle+};
(7.802875,1.589193) *[blue]{\scriptscriptstyle+};
(7.811088,1.591967) *[blue]{\scriptscriptstyle+};
(7.819302,1.596475) *[blue]{\scriptscriptstyle+};
(7.827515,1.597234) *[blue]{\scriptscriptstyle+};
(7.835729,1.597322) *[blue]{\scriptscriptstyle+};
(7.843943,1.598694) *[blue]{\scriptscriptstyle+};
(7.852156,1.601392) *[blue]{\scriptscriptstyle+};
(7.860370,1.617271) *[blue]{\scriptscriptstyle+};
(7.868583,1.618853) *[blue]{\scriptscriptstyle+};
(7.876797,1.624335) *[blue]{\scriptscriptstyle+};
(7.885010,1.580868) *[blue]{\scriptscriptstyle+};
(7.893224,1.581144) *[blue]{\scriptscriptstyle+};
(7.901437,1.586278) *[blue]{\scriptscriptstyle+};
(7.909651,1.596173) *[blue]{\scriptscriptstyle+};
(7.917864,1.600861) *[blue]{\scriptscriptstyle+};
(7.926078,1.609139) *[blue]{\scriptscriptstyle+};
(7.934292,1.610443) *[blue]{\scriptscriptstyle+};
(7.942505,1.611647) *[blue]{\scriptscriptstyle+};
(7.950719,1.613343) *[blue]{\scriptscriptstyle+};
(7.958932,1.614339) *[blue]{\scriptscriptstyle+};
(7.967146,1.616753) *[blue]{\scriptscriptstyle+};
(7.975359,1.624880) *[blue]{\scriptscriptstyle+};
(7.983573,1.630168) *[blue]{\scriptscriptstyle+};
(7.991786,1.636755) *[blue]{\scriptscriptstyle+};
(8.000000,1.639233) *[blue]{\scriptscriptstyle+};
(0.000000,0.950999) *[red]{\scriptscriptstyle\times};
(0.008214,0.953606) *[red]{\scriptscriptstyle\times};
(0.016427,0.957705) *[red]{\scriptscriptstyle\times};
(0.024641,0.957794) *[red]{\scriptscriptstyle\times};
(0.032854,0.957947) *[red]{\scriptscriptstyle\times};
(0.041068,0.963190) *[red]{\scriptscriptstyle\times};
(0.049281,0.964676) *[red]{\scriptscriptstyle\times};
(0.057495,0.970544) *[red]{\scriptscriptstyle\times};
(0.065708,0.992551) *[red]{\scriptscriptstyle\times};
(0.073922,0.992668) *[red]{\scriptscriptstyle\times};
(0.082136,0.997949) *[red]{\scriptscriptstyle\times};
(0.090349,0.999861) *[red]{\scriptscriptstyle\times};
(0.098563,1.002185) *[red]{\scriptscriptstyle\times};
(0.106776,1.010518) *[red]{\scriptscriptstyle\times};
(0.114990,1.096048) *[red]{\scriptscriptstyle\times};
(0.123203,0.978206) *[red]{\scriptscriptstyle\times};
(0.131417,0.979385) *[red]{\scriptscriptstyle\times};
(0.139630,0.982495) *[red]{\scriptscriptstyle\times};
(0.147844,1.024188) *[red]{\scriptscriptstyle\times};
(0.156057,1.027666) *[red]{\scriptscriptstyle\times};
(0.164271,1.032817) *[red]{\scriptscriptstyle\times};
(0.172485,1.040657) *[red]{\scriptscriptstyle\times};
(0.180698,1.045411) *[red]{\scriptscriptstyle\times};
(0.188912,1.058097) *[red]{\scriptscriptstyle\times};
(0.197125,1.066051) *[red]{\scriptscriptstyle\times};
(0.205339,1.069406) *[red]{\scriptscriptstyle\times};
(0.213552,1.074068) *[red]{\scriptscriptstyle\times};
(0.221766,1.080715) *[red]{\scriptscriptstyle\times};
(0.229979,1.112319) *[red]{\scriptscriptstyle\times};
(0.238193,1.177529) *[red]{\scriptscriptstyle\times};
(0.246407,1.014902) *[red]{\scriptscriptstyle\times};
(0.254620,1.016232) *[red]{\scriptscriptstyle\times};
(0.262834,1.016943) *[red]{\scriptscriptstyle\times};
(0.271047,1.020080) *[red]{\scriptscriptstyle\times};
(0.279261,1.020654) *[red]{\scriptscriptstyle\times};
(0.287474,1.024095) *[red]{\scriptscriptstyle\times};
(0.295688,1.024298) *[red]{\scriptscriptstyle\times};
(0.303901,1.028037) *[red]{\scriptscriptstyle\times};
(0.312115,1.055094) *[red]{\scriptscriptstyle\times};
(0.320329,1.057462) *[red]{\scriptscriptstyle\times};
(0.328542,1.058166) *[red]{\scriptscriptstyle\times};
(0.336756,1.061302) *[red]{\scriptscriptstyle\times};
(0.344969,1.073042) *[red]{\scriptscriptstyle\times};
(0.353183,1.076824) *[red]{\scriptscriptstyle\times};
(0.361396,1.093949) *[red]{\scriptscriptstyle\times};
(0.369610,1.022690) *[red]{\scriptscriptstyle\times};
(0.377823,1.024450) *[red]{\scriptscriptstyle\times};
(0.386037,1.025683) *[red]{\scriptscriptstyle\times};
(0.394251,1.063890) *[red]{\scriptscriptstyle\times};
(0.402464,1.076981) *[red]{\scriptscriptstyle\times};
(0.410678,1.098331) *[red]{\scriptscriptstyle\times};
(0.418891,1.105576) *[red]{\scriptscriptstyle\times};
(0.427105,1.108007) *[red]{\scriptscriptstyle\times};
(0.435318,1.110515) *[red]{\scriptscriptstyle\times};
(0.443532,1.123243) *[red]{\scriptscriptstyle\times};
(0.451745,1.124369) *[red]{\scriptscriptstyle\times};
(0.459959,1.130238) *[red]{\scriptscriptstyle\times};
(0.468172,1.154184) *[red]{\scriptscriptstyle\times};
(0.476386,1.157852) *[red]{\scriptscriptstyle\times};
(0.484600,1.231860) *[red]{\scriptscriptstyle\times};
(0.492813,1.039362) *[red]{\scriptscriptstyle\times};
(0.501027,1.042159) *[red]{\scriptscriptstyle\times};
(0.509240,1.043214) *[red]{\scriptscriptstyle\times};
(0.517454,1.078527) *[red]{\scriptscriptstyle\times};
(0.525667,1.080901) *[red]{\scriptscriptstyle\times};
(0.533881,1.089833) *[red]{\scriptscriptstyle\times};
(0.542094,1.089881) *[red]{\scriptscriptstyle\times};
(0.550308,1.091533) *[red]{\scriptscriptstyle\times};
(0.558522,1.095456) *[red]{\scriptscriptstyle\times};
(0.566735,1.095722) *[red]{\scriptscriptstyle\times};
(0.574949,1.115954) *[red]{\scriptscriptstyle\times};
(0.583162,1.117264) *[red]{\scriptscriptstyle\times};
(0.591376,1.126216) *[red]{\scriptscriptstyle\times};
(0.599589,1.126775) *[red]{\scriptscriptstyle\times};
(0.607803,1.172819) *[red]{\scriptscriptstyle\times};
(0.616016,1.024308) *[red]{\scriptscriptstyle\times};
(0.624230,1.028831) *[red]{\scriptscriptstyle\times};
(0.632444,1.035484) *[red]{\scriptscriptstyle\times};
(0.640657,1.045695) *[red]{\scriptscriptstyle\times};
(0.648871,1.065587) *[red]{\scriptscriptstyle\times};
(0.657084,1.073660) *[red]{\scriptscriptstyle\times};
(0.665298,1.074731) *[red]{\scriptscriptstyle\times};
(0.673511,1.081245) *[red]{\scriptscriptstyle\times};
(0.681725,1.111849) *[red]{\scriptscriptstyle\times};
(0.689938,1.119182) *[red]{\scriptscriptstyle\times};
(0.698152,1.119329) *[red]{\scriptscriptstyle\times};
(0.706366,1.129707) *[red]{\scriptscriptstyle\times};
(0.714579,1.144821) *[red]{\scriptscriptstyle\times};
(0.722793,1.173212) *[red]{\scriptscriptstyle\times};
(0.731006,1.314084) *[red]{\scriptscriptstyle\times};
(0.739220,1.052266) *[red]{\scriptscriptstyle\times};
(0.747433,1.052350) *[red]{\scriptscriptstyle\times};
(0.755647,1.058011) *[red]{\scriptscriptstyle\times};
(0.763860,1.092578) *[red]{\scriptscriptstyle\times};
(0.772074,1.095365) *[red]{\scriptscriptstyle\times};
(0.780287,1.096200) *[red]{\scriptscriptstyle\times};
(0.788501,1.104865) *[red]{\scriptscriptstyle\times};
(0.796715,1.132013) *[red]{\scriptscriptstyle\times};
(0.804928,1.133489) *[red]{\scriptscriptstyle\times};
(0.813142,1.135013) *[red]{\scriptscriptstyle\times};
(0.821355,1.139528) *[red]{\scriptscriptstyle\times};
(0.829569,1.173819) *[red]{\scriptscriptstyle\times};
(0.837782,1.176585) *[red]{\scriptscriptstyle\times};
(0.845996,1.178739) *[red]{\scriptscriptstyle\times};
(0.854209,1.180619) *[red]{\scriptscriptstyle\times};
(0.862423,1.039346) *[red]{\scriptscriptstyle\times};
(0.870637,1.039365) *[red]{\scriptscriptstyle\times};
(0.878850,1.047145) *[red]{\scriptscriptstyle\times};
(0.887064,1.083877) *[red]{\scriptscriptstyle\times};
(0.895277,1.084790) *[red]{\scriptscriptstyle\times};
(0.903491,1.087584) *[red]{\scriptscriptstyle\times};
(0.911704,1.102978) *[red]{\scriptscriptstyle\times};
(0.919918,1.104904) *[red]{\scriptscriptstyle\times};
(0.928131,1.119935) *[red]{\scriptscriptstyle\times};
(0.936345,1.123253) *[red]{\scriptscriptstyle\times};
(0.944559,1.125820) *[red]{\scriptscriptstyle\times};
(0.952772,1.127355) *[red]{\scriptscriptstyle\times};
(0.960986,1.134465) *[red]{\scriptscriptstyle\times};
(0.969199,1.136015) *[red]{\scriptscriptstyle\times};
(0.977413,1.171646) *[red]{\scriptscriptstyle\times};
(0.985626,1.107109) *[red]{\scriptscriptstyle\times};
(0.993840,1.107148) *[red]{\scriptscriptstyle\times};
(1.002053,1.108674) *[red]{\scriptscriptstyle\times};
(1.010267,1.108747) *[red]{\scriptscriptstyle\times};
(1.018480,1.112208) *[red]{\scriptscriptstyle\times};
(1.026694,1.130520) *[red]{\scriptscriptstyle\times};
(1.034908,1.154698) *[red]{\scriptscriptstyle\times};
(1.043121,1.156738) *[red]{\scriptscriptstyle\times};
(1.051335,1.158679) *[red]{\scriptscriptstyle\times};
(1.059548,1.163831) *[red]{\scriptscriptstyle\times};
(1.067762,1.173984) *[red]{\scriptscriptstyle\times};
(1.075975,1.185537) *[red]{\scriptscriptstyle\times};
(1.084189,1.187331) *[red]{\scriptscriptstyle\times};
(1.092402,1.189081) *[red]{\scriptscriptstyle\times};
(1.100616,1.196543) *[red]{\scriptscriptstyle\times};
(1.108830,1.107420) *[red]{\scriptscriptstyle\times};
(1.117043,1.111143) *[red]{\scriptscriptstyle\times};
(1.125257,1.121018) *[red]{\scriptscriptstyle\times};
(1.133470,1.123272) *[red]{\scriptscriptstyle\times};
(1.141684,1.123353) *[red]{\scriptscriptstyle\times};
(1.149897,1.143251) *[red]{\scriptscriptstyle\times};
(1.158111,1.146336) *[red]{\scriptscriptstyle\times};
(1.166324,1.146912) *[red]{\scriptscriptstyle\times};
(1.174538,1.154522) *[red]{\scriptscriptstyle\times};
(1.182752,1.157071) *[red]{\scriptscriptstyle\times};
(1.190965,1.161941) *[red]{\scriptscriptstyle\times};
(1.199179,1.188017) *[red]{\scriptscriptstyle\times};
(1.207392,1.188641) *[red]{\scriptscriptstyle\times};
(1.215606,1.217491) *[red]{\scriptscriptstyle\times};
(1.223819,1.268837) *[red]{\scriptscriptstyle\times};
(1.232033,1.127876) *[red]{\scriptscriptstyle\times};
(1.240246,1.128119) *[red]{\scriptscriptstyle\times};
(1.248460,1.131053) *[red]{\scriptscriptstyle\times};
(1.256674,1.142753) *[red]{\scriptscriptstyle\times};
(1.264887,1.166115) *[red]{\scriptscriptstyle\times};
(1.273101,1.166399) *[red]{\scriptscriptstyle\times};
(1.281314,1.171523) *[red]{\scriptscriptstyle\times};
(1.289528,1.173564) *[red]{\scriptscriptstyle\times};
(1.297741,1.179329) *[red]{\scriptscriptstyle\times};
(1.305955,1.180657) *[red]{\scriptscriptstyle\times};
(1.314168,1.198311) *[red]{\scriptscriptstyle\times};
(1.322382,1.203446) *[red]{\scriptscriptstyle\times};
(1.330595,1.210812) *[red]{\scriptscriptstyle\times};
(1.338809,1.211550) *[red]{\scriptscriptstyle\times};
(1.347023,1.220180) *[red]{\scriptscriptstyle\times};
(1.355236,1.101113) *[red]{\scriptscriptstyle\times};
(1.363450,1.116176) *[red]{\scriptscriptstyle\times};
(1.371663,1.136509) *[red]{\scriptscriptstyle\times};
(1.379877,1.137313) *[red]{\scriptscriptstyle\times};
(1.388090,1.138128) *[red]{\scriptscriptstyle\times};
(1.396304,1.138289) *[red]{\scriptscriptstyle\times};
(1.404517,1.147970) *[red]{\scriptscriptstyle\times};
(1.412731,1.178449) *[red]{\scriptscriptstyle\times};
(1.420945,1.183000) *[red]{\scriptscriptstyle\times};
(1.429158,1.205783) *[red]{\scriptscriptstyle\times};
(1.437372,1.215691) *[red]{\scriptscriptstyle\times};
(1.445585,1.235163) *[red]{\scriptscriptstyle\times};
(1.453799,1.268292) *[red]{\scriptscriptstyle\times};
(1.462012,1.301549) *[red]{\scriptscriptstyle\times};
(1.470226,1.365818) *[red]{\scriptscriptstyle\times};
(1.478439,1.117426) *[red]{\scriptscriptstyle\times};
(1.486653,1.117498) *[red]{\scriptscriptstyle\times};
(1.494867,1.121337) *[red]{\scriptscriptstyle\times};
(1.503080,1.121341) *[red]{\scriptscriptstyle\times};
(1.511294,1.128649) *[red]{\scriptscriptstyle\times};
(1.519507,1.136020) *[red]{\scriptscriptstyle\times};
(1.527721,1.156059) *[red]{\scriptscriptstyle\times};
(1.535934,1.157350) *[red]{\scriptscriptstyle\times};
(1.544148,1.159881) *[red]{\scriptscriptstyle\times};
(1.552361,1.159893) *[red]{\scriptscriptstyle\times};
(1.560575,1.169730) *[red]{\scriptscriptstyle\times};
(1.568789,1.170690) *[red]{\scriptscriptstyle\times};
(1.577002,1.177803) *[red]{\scriptscriptstyle\times};
(1.585216,1.229210) *[red]{\scriptscriptstyle\times};
(1.593429,1.254313) *[red]{\scriptscriptstyle\times};
(1.601643,1.154053) *[red]{\scriptscriptstyle\times};
(1.609856,1.156353) *[red]{\scriptscriptstyle\times};
(1.618070,1.157754) *[red]{\scriptscriptstyle\times};
(1.626283,1.157782) *[red]{\scriptscriptstyle\times};
(1.634497,1.157864) *[red]{\scriptscriptstyle\times};
(1.642710,1.159967) *[red]{\scriptscriptstyle\times};
(1.650924,1.163922) *[red]{\scriptscriptstyle\times};
(1.659138,1.176044) *[red]{\scriptscriptstyle\times};
(1.667351,1.188282) *[red]{\scriptscriptstyle\times};
(1.675565,1.188304) *[red]{\scriptscriptstyle\times};
(1.683778,1.195073) *[red]{\scriptscriptstyle\times};
(1.691992,1.195460) *[red]{\scriptscriptstyle\times};
(1.700205,1.206678) *[red]{\scriptscriptstyle\times};
(1.708419,1.207807) *[red]{\scriptscriptstyle\times};
(1.716632,1.232506) *[red]{\scriptscriptstyle\times};
(1.724846,1.129472) *[red]{\scriptscriptstyle\times};
(1.733060,1.131845) *[red]{\scriptscriptstyle\times};
(1.741273,1.133490) *[red]{\scriptscriptstyle\times};
(1.749487,1.164317) *[red]{\scriptscriptstyle\times};
(1.757700,1.167042) *[red]{\scriptscriptstyle\times};
(1.765914,1.171843) *[red]{\scriptscriptstyle\times};
(1.774127,1.178543) *[red]{\scriptscriptstyle\times};
(1.782341,1.201577) *[red]{\scriptscriptstyle\times};
(1.790554,1.205492) *[red]{\scriptscriptstyle\times};
(1.798768,1.206544) *[red]{\scriptscriptstyle\times};
(1.806982,1.207459) *[red]{\scriptscriptstyle\times};
(1.815195,1.217848) *[red]{\scriptscriptstyle\times};
(1.823409,1.241478) *[red]{\scriptscriptstyle\times};
(1.831622,1.249600) *[red]{\scriptscriptstyle\times};
(1.839836,1.284941) *[red]{\scriptscriptstyle\times};
(1.848049,1.141084) *[red]{\scriptscriptstyle\times};
(1.856263,1.142940) *[red]{\scriptscriptstyle\times};
(1.864476,1.144111) *[red]{\scriptscriptstyle\times};
(1.872690,1.153898) *[red]{\scriptscriptstyle\times};
(1.880903,1.159306) *[red]{\scriptscriptstyle\times};
(1.889117,1.179597) *[red]{\scriptscriptstyle\times};
(1.897331,1.180767) *[red]{\scriptscriptstyle\times};
(1.905544,1.182186) *[red]{\scriptscriptstyle\times};
(1.913758,1.186075) *[red]{\scriptscriptstyle\times};
(1.921971,1.189741) *[red]{\scriptscriptstyle\times};
(1.930185,1.191441) *[red]{\scriptscriptstyle\times};
(1.938398,1.219738) *[red]{\scriptscriptstyle\times};
(1.946612,1.223063) *[red]{\scriptscriptstyle\times};
(1.954825,1.258987) *[red]{\scriptscriptstyle\times};
(1.963039,1.262316) *[red]{\scriptscriptstyle\times};
(1.971253,1.171003) *[red]{\scriptscriptstyle\times};
(1.979466,1.201730) *[red]{\scriptscriptstyle\times};
(1.987680,1.203784) *[red]{\scriptscriptstyle\times};
(1.995893,1.205062) *[red]{\scriptscriptstyle\times};
(2.004107,1.205310) *[red]{\scriptscriptstyle\times};
(2.012320,1.208105) *[red]{\scriptscriptstyle\times};
(2.020534,1.212254) *[red]{\scriptscriptstyle\times};
(2.028747,1.240505) *[red]{\scriptscriptstyle\times};
(2.036961,1.244920) *[red]{\scriptscriptstyle\times};
(2.045175,1.251969) *[red]{\scriptscriptstyle\times};
(2.053388,1.253689) *[red]{\scriptscriptstyle\times};
(2.061602,1.258638) *[red]{\scriptscriptstyle\times};
(2.069815,1.265969) *[red]{\scriptscriptstyle\times};
(2.078029,1.266322) *[red]{\scriptscriptstyle\times};
(2.086242,1.278658) *[red]{\scriptscriptstyle\times};
(2.094456,1.155857) *[red]{\scriptscriptstyle\times};
(2.102669,1.182748) *[red]{\scriptscriptstyle\times};
(2.110883,1.182767) *[red]{\scriptscriptstyle\times};
(2.119097,1.182771) *[red]{\scriptscriptstyle\times};
(2.127310,1.182834) *[red]{\scriptscriptstyle\times};
(2.135524,1.185223) *[red]{\scriptscriptstyle\times};
(2.143737,1.186426) *[red]{\scriptscriptstyle\times};
(2.151951,1.187358) *[red]{\scriptscriptstyle\times};
(2.160164,1.219492) *[red]{\scriptscriptstyle\times};
(2.168378,1.224995) *[red]{\scriptscriptstyle\times};
(2.176591,1.226650) *[red]{\scriptscriptstyle\times};
(2.184805,1.229413) *[red]{\scriptscriptstyle\times};
(2.193018,1.256287) *[red]{\scriptscriptstyle\times};
(2.201232,1.278427) *[red]{\scriptscriptstyle\times};
(2.209446,1.300640) *[red]{\scriptscriptstyle\times};
(2.217659,1.181279) *[red]{\scriptscriptstyle\times};
(2.225873,1.181302) *[red]{\scriptscriptstyle\times};
(2.234086,1.182823) *[red]{\scriptscriptstyle\times};
(2.242300,1.184986) *[red]{\scriptscriptstyle\times};
(2.250513,1.185052) *[red]{\scriptscriptstyle\times};
(2.258727,1.185141) *[red]{\scriptscriptstyle\times};
(2.266940,1.187249) *[red]{\scriptscriptstyle\times};
(2.275154,1.188475) *[red]{\scriptscriptstyle\times};
(2.283368,1.202672) *[red]{\scriptscriptstyle\times};
(2.291581,1.215443) *[red]{\scriptscriptstyle\times};
(2.299795,1.218195) *[red]{\scriptscriptstyle\times};
(2.308008,1.219453) *[red]{\scriptscriptstyle\times};
(2.316222,1.224104) *[red]{\scriptscriptstyle\times};
(2.324435,1.251061) *[red]{\scriptscriptstyle\times};
(2.332649,1.254230) *[red]{\scriptscriptstyle\times};
(2.340862,1.151069) *[red]{\scriptscriptstyle\times};
(2.349076,1.151379) *[red]{\scriptscriptstyle\times};
(2.357290,1.157924) *[red]{\scriptscriptstyle\times};
(2.365503,1.172896) *[red]{\scriptscriptstyle\times};
(2.373717,1.188869) *[red]{\scriptscriptstyle\times};
(2.381930,1.193501) *[red]{\scriptscriptstyle\times};
(2.390144,1.195716) *[red]{\scriptscriptstyle\times};
(2.398357,1.207779) *[red]{\scriptscriptstyle\times};
(2.406571,1.225747) *[red]{\scriptscriptstyle\times};
(2.414784,1.227309) *[red]{\scriptscriptstyle\times};
(2.422998,1.230697) *[red]{\scriptscriptstyle\times};
(2.431211,1.235708) *[red]{\scriptscriptstyle\times};
(2.439425,1.235964) *[red]{\scriptscriptstyle\times};
(2.447639,1.242802) *[red]{\scriptscriptstyle\times};
(2.455852,1.282142) *[red]{\scriptscriptstyle\times};
(2.464066,1.165050) *[red]{\scriptscriptstyle\times};
(2.472279,1.166106) *[red]{\scriptscriptstyle\times};
(2.480493,1.200744) *[red]{\scriptscriptstyle\times};
(2.488706,1.204895) *[red]{\scriptscriptstyle\times};
(2.496920,1.205327) *[red]{\scriptscriptstyle\times};
(2.505133,1.209827) *[red]{\scriptscriptstyle\times};
(2.513347,1.211726) *[red]{\scriptscriptstyle\times};
(2.521561,1.212003) *[red]{\scriptscriptstyle\times};
(2.529774,1.236114) *[red]{\scriptscriptstyle\times};
(2.537988,1.237830) *[red]{\scriptscriptstyle\times};
(2.546201,1.238932) *[red]{\scriptscriptstyle\times};
(2.554415,1.247570) *[red]{\scriptscriptstyle\times};
(2.562628,1.272029) *[red]{\scriptscriptstyle\times};
(2.570842,1.277828) *[red]{\scriptscriptstyle\times};
(2.579055,1.279606) *[red]{\scriptscriptstyle\times};
(2.587269,1.166845) *[red]{\scriptscriptstyle\times};
(2.595483,1.169792) *[red]{\scriptscriptstyle\times};
(2.603696,1.173921) *[red]{\scriptscriptstyle\times};
(2.611910,1.206834) *[red]{\scriptscriptstyle\times};
(2.620123,1.207756) *[red]{\scriptscriptstyle\times};
(2.628337,1.211025) *[red]{\scriptscriptstyle\times};
(2.636550,1.212260) *[red]{\scriptscriptstyle\times};
(2.644764,1.212482) *[red]{\scriptscriptstyle\times};
(2.652977,1.217948) *[red]{\scriptscriptstyle\times};
(2.661191,1.222814) *[red]{\scriptscriptstyle\times};
(2.669405,1.223287) *[red]{\scriptscriptstyle\times};
(2.677618,1.244016) *[red]{\scriptscriptstyle\times};
(2.685832,1.246688) *[red]{\scriptscriptstyle\times};
(2.694045,1.248238) *[red]{\scriptscriptstyle\times};
(2.702259,1.278869) *[red]{\scriptscriptstyle\times};
(2.710472,1.148141) *[red]{\scriptscriptstyle\times};
(2.718686,1.151827) *[red]{\scriptscriptstyle\times};
(2.726899,1.170066) *[red]{\scriptscriptstyle\times};
(2.735113,1.194045) *[red]{\scriptscriptstyle\times};
(2.743326,1.202089) *[red]{\scriptscriptstyle\times};
(2.751540,1.214176) *[red]{\scriptscriptstyle\times};
(2.759754,1.221244) *[red]{\scriptscriptstyle\times};
(2.767967,1.228636) *[red]{\scriptscriptstyle\times};
(2.776181,1.229451) *[red]{\scriptscriptstyle\times};
(2.784394,1.233145) *[red]{\scriptscriptstyle\times};
(2.792608,1.239936) *[red]{\scriptscriptstyle\times};
(2.800821,1.255667) *[red]{\scriptscriptstyle\times};
(2.809035,1.276860) *[red]{\scriptscriptstyle\times};
(2.817248,1.310396) *[red]{\scriptscriptstyle\times};
(2.825462,1.364630) *[red]{\scriptscriptstyle\times};
(2.833676,1.168685) *[red]{\scriptscriptstyle\times};
(2.841889,1.168846) *[red]{\scriptscriptstyle\times};
(2.850103,1.195678) *[red]{\scriptscriptstyle\times};
(2.858316,1.195685) *[red]{\scriptscriptstyle\times};
(2.866530,1.197893) *[red]{\scriptscriptstyle\times};
(2.874743,1.198772) *[red]{\scriptscriptstyle\times};
(2.882957,1.201835) *[red]{\scriptscriptstyle\times};
(2.891170,1.202513) *[red]{\scriptscriptstyle\times};
(2.899384,1.202606) *[red]{\scriptscriptstyle\times};
(2.907598,1.208776) *[red]{\scriptscriptstyle\times};
(2.915811,1.220221) *[red]{\scriptscriptstyle\times};
(2.924025,1.239463) *[red]{\scriptscriptstyle\times};
(2.932238,1.242341) *[red]{\scriptscriptstyle\times};
(2.940452,1.245710) *[red]{\scriptscriptstyle\times};
(2.948665,1.296877) *[red]{\scriptscriptstyle\times};
(2.956879,1.193090) *[red]{\scriptscriptstyle\times};
(2.965092,1.217810) *[red]{\scriptscriptstyle\times};
(2.973306,1.219854) *[red]{\scriptscriptstyle\times};
(2.981520,1.226754) *[red]{\scriptscriptstyle\times};
(2.989733,1.227924) *[red]{\scriptscriptstyle\times};
(2.997947,1.227965) *[red]{\scriptscriptstyle\times};
(3.006160,1.229642) *[red]{\scriptscriptstyle\times};
(3.014374,1.234868) *[red]{\scriptscriptstyle\times};
(3.022587,1.235177) *[red]{\scriptscriptstyle\times};
(3.030801,1.238530) *[red]{\scriptscriptstyle\times};
(3.039014,1.255612) *[red]{\scriptscriptstyle\times};
(3.047228,1.255887) *[red]{\scriptscriptstyle\times};
(3.055441,1.265390) *[red]{\scriptscriptstyle\times};
(3.063655,1.291229) *[red]{\scriptscriptstyle\times};
(3.071869,1.307426) *[red]{\scriptscriptstyle\times};
(3.080082,1.199199) *[red]{\scriptscriptstyle\times};
(3.088296,1.199523) *[red]{\scriptscriptstyle\times};
(3.096509,1.199835) *[red]{\scriptscriptstyle\times};
(3.104723,1.202752) *[red]{\scriptscriptstyle\times};
(3.112936,1.221319) *[red]{\scriptscriptstyle\times};
(3.121150,1.223499) *[red]{\scriptscriptstyle\times};
(3.129363,1.232496) *[red]{\scriptscriptstyle\times};
(3.137577,1.234973) *[red]{\scriptscriptstyle\times};
(3.145791,1.236563) *[red]{\scriptscriptstyle\times};
(3.154004,1.238316) *[red]{\scriptscriptstyle\times};
(3.162218,1.239587) *[red]{\scriptscriptstyle\times};
(3.170431,1.244528) *[red]{\scriptscriptstyle\times};
(3.178645,1.275197) *[red]{\scriptscriptstyle\times};
(3.186858,1.285022) *[red]{\scriptscriptstyle\times};
(3.195072,1.286579) *[red]{\scriptscriptstyle\times};
(3.203285,1.173633) *[red]{\scriptscriptstyle\times};
(3.211499,1.210802) *[red]{\scriptscriptstyle\times};
(3.219713,1.216097) *[red]{\scriptscriptstyle\times};
(3.227926,1.217472) *[red]{\scriptscriptstyle\times};
(3.236140,1.221239) *[red]{\scriptscriptstyle\times};
(3.244353,1.221311) *[red]{\scriptscriptstyle\times};
(3.252567,1.224727) *[red]{\scriptscriptstyle\times};
(3.260780,1.226967) *[red]{\scriptscriptstyle\times};
(3.268994,1.228312) *[red]{\scriptscriptstyle\times};
(3.277207,1.246935) *[red]{\scriptscriptstyle\times};
(3.285421,1.247622) *[red]{\scriptscriptstyle\times};
(3.293634,1.250165) *[red]{\scriptscriptstyle\times};
(3.301848,1.266359) *[red]{\scriptscriptstyle\times};
(3.310062,1.285669) *[red]{\scriptscriptstyle\times};
(3.318275,1.318463) *[red]{\scriptscriptstyle\times};
(3.326489,1.175893) *[red]{\scriptscriptstyle\times};
(3.334702,1.209610) *[red]{\scriptscriptstyle\times};
(3.342916,1.213261) *[red]{\scriptscriptstyle\times};
(3.351129,1.215481) *[red]{\scriptscriptstyle\times};
(3.359343,1.216106) *[red]{\scriptscriptstyle\times};
(3.367556,1.220129) *[red]{\scriptscriptstyle\times};
(3.375770,1.223645) *[red]{\scriptscriptstyle\times};
(3.383984,1.239266) *[red]{\scriptscriptstyle\times};
(3.392197,1.247635) *[red]{\scriptscriptstyle\times};
(3.400411,1.249334) *[red]{\scriptscriptstyle\times};
(3.408624,1.252746) *[red]{\scriptscriptstyle\times};
(3.416838,1.266315) *[red]{\scriptscriptstyle\times};
(3.425051,1.281190) *[red]{\scriptscriptstyle\times};
(3.433265,1.304545) *[red]{\scriptscriptstyle\times};
(3.441478,1.317778) *[red]{\scriptscriptstyle\times};
(3.449692,1.193192) *[red]{\scriptscriptstyle\times};
(3.457906,1.196739) *[red]{\scriptscriptstyle\times};
(3.466119,1.211886) *[red]{\scriptscriptstyle\times};
(3.474333,1.232202) *[red]{\scriptscriptstyle\times};
(3.482546,1.235368) *[red]{\scriptscriptstyle\times};
(3.490760,1.237138) *[red]{\scriptscriptstyle\times};
(3.498973,1.238889) *[red]{\scriptscriptstyle\times};
(3.507187,1.239130) *[red]{\scriptscriptstyle\times};
(3.515400,1.267890) *[red]{\scriptscriptstyle\times};
(3.523614,1.268875) *[red]{\scriptscriptstyle\times};
(3.531828,1.275191) *[red]{\scriptscriptstyle\times};
(3.540041,1.275779) *[red]{\scriptscriptstyle\times};
(3.548255,1.279561) *[red]{\scriptscriptstyle\times};
(3.556468,1.307108) *[red]{\scriptscriptstyle\times};
(3.564682,1.314479) *[red]{\scriptscriptstyle\times};
(3.572895,1.201221) *[red]{\scriptscriptstyle\times};
(3.581109,1.206429) *[red]{\scriptscriptstyle\times};
(3.589322,1.211926) *[red]{\scriptscriptstyle\times};
(3.597536,1.237672) *[red]{\scriptscriptstyle\times};
(3.605749,1.237818) *[red]{\scriptscriptstyle\times};
(3.613963,1.241058) *[red]{\scriptscriptstyle\times};
(3.622177,1.241723) *[red]{\scriptscriptstyle\times};
(3.630390,1.250999) *[red]{\scriptscriptstyle\times};
(3.638604,1.260190) *[red]{\scriptscriptstyle\times};
(3.646817,1.273441) *[red]{\scriptscriptstyle\times};
(3.655031,1.278224) *[red]{\scriptscriptstyle\times};
(3.663244,1.283447) *[red]{\scriptscriptstyle\times};
(3.671458,1.286695) *[red]{\scriptscriptstyle\times};
(3.679671,1.290107) *[red]{\scriptscriptstyle\times};
(3.687885,1.327544) *[red]{\scriptscriptstyle\times};
(3.696099,1.184933) *[red]{\scriptscriptstyle\times};
(3.704312,1.187022) *[red]{\scriptscriptstyle\times};
(3.712526,1.219629) *[red]{\scriptscriptstyle\times};
(3.720739,1.219747) *[red]{\scriptscriptstyle\times};
(3.728953,1.223050) *[red]{\scriptscriptstyle\times};
(3.737166,1.226768) *[red]{\scriptscriptstyle\times};
(3.745380,1.233177) *[red]{\scriptscriptstyle\times};
(3.753593,1.261142) *[red]{\scriptscriptstyle\times};
(3.761807,1.267215) *[red]{\scriptscriptstyle\times};
(3.770021,1.269513) *[red]{\scriptscriptstyle\times};
(3.778234,1.272771) *[red]{\scriptscriptstyle\times};
(3.786448,1.290956) *[red]{\scriptscriptstyle\times};
(3.794661,1.293052) *[red]{\scriptscriptstyle\times};
(3.802875,1.339996) *[red]{\scriptscriptstyle\times};
(3.811088,1.413326) *[red]{\scriptscriptstyle\times};
(3.819302,1.191998) *[red]{\scriptscriptstyle\times};
(3.827515,1.197898) *[red]{\scriptscriptstyle\times};
(3.835729,1.199359) *[red]{\scriptscriptstyle\times};
(3.843943,1.227244) *[red]{\scriptscriptstyle\times};
(3.852156,1.228876) *[red]{\scriptscriptstyle\times};
(3.860370,1.231436) *[red]{\scriptscriptstyle\times};
(3.868583,1.234517) *[red]{\scriptscriptstyle\times};
(3.876797,1.266422) *[red]{\scriptscriptstyle\times};
(3.885010,1.267737) *[red]{\scriptscriptstyle\times};
(3.893224,1.269753) *[red]{\scriptscriptstyle\times};
(3.901437,1.278097) *[red]{\scriptscriptstyle\times};
(3.909651,1.283522) *[red]{\scriptscriptstyle\times};
(3.917864,1.312386) *[red]{\scriptscriptstyle\times};
(3.926078,1.339121) *[red]{\scriptscriptstyle\times};
(3.934292,1.412922) *[red]{\scriptscriptstyle\times};
(3.942505,1.200268) *[red]{\scriptscriptstyle\times};
(3.950719,1.239701) *[red]{\scriptscriptstyle\times};
(3.958932,1.239833) *[red]{\scriptscriptstyle\times};
(3.967146,1.240175) *[red]{\scriptscriptstyle\times};
(3.975359,1.246806) *[red]{\scriptscriptstyle\times};
(3.983573,1.265326) *[red]{\scriptscriptstyle\times};
(3.991786,1.272211) *[red]{\scriptscriptstyle\times};
(4.000000,1.272864) *[red]{\scriptscriptstyle\times};
(4.008214,1.280173) *[red]{\scriptscriptstyle\times};
(4.016427,1.282373) *[red]{\scriptscriptstyle\times};
(4.024641,1.286673) *[red]{\scriptscriptstyle\times};
(4.032854,1.310359) *[red]{\scriptscriptstyle\times};
(4.041068,1.314713) *[red]{\scriptscriptstyle\times};
(4.049281,1.328995) *[red]{\scriptscriptstyle\times};
(4.057495,1.351734) *[red]{\scriptscriptstyle\times};
(4.065708,1.203567) *[red]{\scriptscriptstyle\times};
(4.073922,1.203630) *[red]{\scriptscriptstyle\times};
(4.082136,1.203673) *[red]{\scriptscriptstyle\times};
(4.090349,1.206239) *[red]{\scriptscriptstyle\times};
(4.098563,1.209458) *[red]{\scriptscriptstyle\times};
(4.106776,1.209712) *[red]{\scriptscriptstyle\times};
(4.114990,1.212167) *[red]{\scriptscriptstyle\times};
(4.123203,1.214057) *[red]{\scriptscriptstyle\times};
(4.131417,1.239816) *[red]{\scriptscriptstyle\times};
(4.139630,1.245859) *[red]{\scriptscriptstyle\times};
(4.147844,1.255410) *[red]{\scriptscriptstyle\times};
(4.156057,1.295241) *[red]{\scriptscriptstyle\times};
(4.164271,1.298595) *[red]{\scriptscriptstyle\times};
(4.172485,1.328614) *[red]{\scriptscriptstyle\times};
(4.180698,1.347490) *[red]{\scriptscriptstyle\times};
(4.188912,1.197737) *[red]{\scriptscriptstyle\times};
(4.197125,1.199027) *[red]{\scriptscriptstyle\times};
(4.205339,1.199816) *[red]{\scriptscriptstyle\times};
(4.213552,1.234372) *[red]{\scriptscriptstyle\times};
(4.221766,1.237600) *[red]{\scriptscriptstyle\times};
(4.229979,1.241903) *[red]{\scriptscriptstyle\times};
(4.238193,1.242566) *[red]{\scriptscriptstyle\times};
(4.246407,1.244061) *[red]{\scriptscriptstyle\times};
(4.254620,1.259029) *[red]{\scriptscriptstyle\times};
(4.262834,1.271137) *[red]{\scriptscriptstyle\times};
(4.271047,1.283734) *[red]{\scriptscriptstyle\times};
(4.279261,1.289921) *[red]{\scriptscriptstyle\times};
(4.287474,1.303234) *[red]{\scriptscriptstyle\times};
(4.295688,1.306202) *[red]{\scriptscriptstyle\times};
(4.303901,1.355067) *[red]{\scriptscriptstyle\times};
(4.312115,1.244589) *[red]{\scriptscriptstyle\times};
(4.320329,1.244879) *[red]{\scriptscriptstyle\times};
(4.328542,1.248161) *[red]{\scriptscriptstyle\times};
(4.336756,1.248394) *[red]{\scriptscriptstyle\times};
(4.344969,1.250349) *[red]{\scriptscriptstyle\times};
(4.353183,1.251523) *[red]{\scriptscriptstyle\times};
(4.361396,1.256886) *[red]{\scriptscriptstyle\times};
(4.369610,1.267001) *[red]{\scriptscriptstyle\times};
(4.377823,1.283420) *[red]{\scriptscriptstyle\times};
(4.386037,1.289961) *[red]{\scriptscriptstyle\times};
(4.394251,1.290127) *[red]{\scriptscriptstyle\times};
(4.402464,1.302981) *[red]{\scriptscriptstyle\times};
(4.410678,1.317975) *[red]{\scriptscriptstyle\times};
(4.418891,1.320880) *[red]{\scriptscriptstyle\times};
(4.427105,1.324169) *[red]{\scriptscriptstyle\times};
(4.435318,1.233839) *[red]{\scriptscriptstyle\times};
(4.443532,1.257364) *[red]{\scriptscriptstyle\times};
(4.451745,1.260948) *[red]{\scriptscriptstyle\times};
(4.459959,1.262654) *[red]{\scriptscriptstyle\times};
(4.468172,1.265492) *[red]{\scriptscriptstyle\times};
(4.476386,1.270499) *[red]{\scriptscriptstyle\times};
(4.484600,1.274358) *[red]{\scriptscriptstyle\times};
(4.492813,1.292305) *[red]{\scriptscriptstyle\times};
(4.501027,1.292749) *[red]{\scriptscriptstyle\times};
(4.509240,1.295973) *[red]{\scriptscriptstyle\times};
(4.517454,1.299283) *[red]{\scriptscriptstyle\times};
(4.525667,1.321981) *[red]{\scriptscriptstyle\times};
(4.533881,1.329941) *[red]{\scriptscriptstyle\times};
(4.542094,1.355503) *[red]{\scriptscriptstyle\times};
(4.550308,1.414246) *[red]{\scriptscriptstyle\times};
(4.558522,1.259010) *[red]{\scriptscriptstyle\times};
(4.566735,1.259040) *[red]{\scriptscriptstyle\times};
(4.574949,1.261198) *[red]{\scriptscriptstyle\times};
(4.583162,1.262583) *[red]{\scriptscriptstyle\times};
(4.591376,1.262995) *[red]{\scriptscriptstyle\times};
(4.599589,1.266784) *[red]{\scriptscriptstyle\times};
(4.607803,1.294189) *[red]{\scriptscriptstyle\times};
(4.616016,1.297385) *[red]{\scriptscriptstyle\times};
(4.624230,1.298276) *[red]{\scriptscriptstyle\times};
(4.632444,1.300744) *[red]{\scriptscriptstyle\times};
(4.640657,1.301084) *[red]{\scriptscriptstyle\times};
(4.648871,1.305303) *[red]{\scriptscriptstyle\times};
(4.657084,1.306151) *[red]{\scriptscriptstyle\times};
(4.665298,1.310409) *[red]{\scriptscriptstyle\times};
(4.673511,1.335336) *[red]{\scriptscriptstyle\times};
(4.681725,1.201232) *[red]{\scriptscriptstyle\times};
(4.689938,1.204688) *[red]{\scriptscriptstyle\times};
(4.698152,1.210145) *[red]{\scriptscriptstyle\times};
(4.706366,1.211809) *[red]{\scriptscriptstyle\times};
(4.714579,1.239622) *[red]{\scriptscriptstyle\times};
(4.722793,1.244735) *[red]{\scriptscriptstyle\times};
(4.731006,1.248041) *[red]{\scriptscriptstyle\times};
(4.739220,1.249268) *[red]{\scriptscriptstyle\times};
(4.747433,1.251235) *[red]{\scriptscriptstyle\times};
(4.755647,1.285300) *[red]{\scriptscriptstyle\times};
(4.763860,1.298487) *[red]{\scriptscriptstyle\times};
(4.772074,1.350494) *[red]{\scriptscriptstyle\times};
(4.780287,1.368186) *[red]{\scriptscriptstyle\times};
(4.788501,1.386460) *[red]{\scriptscriptstyle\times};
(4.796715,1.393925) *[red]{\scriptscriptstyle\times};
(4.804928,1.237293) *[red]{\scriptscriptstyle\times};
(4.813142,1.237966) *[red]{\scriptscriptstyle\times};
(4.821355,1.241364) *[red]{\scriptscriptstyle\times};
(4.829569,1.270693) *[red]{\scriptscriptstyle\times};
(4.837782,1.270705) *[red]{\scriptscriptstyle\times};
(4.845996,1.270920) *[red]{\scriptscriptstyle\times};
(4.854209,1.273974) *[red]{\scriptscriptstyle\times};
(4.862423,1.277529) *[red]{\scriptscriptstyle\times};
(4.870637,1.296838) *[red]{\scriptscriptstyle\times};
(4.878850,1.305817) *[red]{\scriptscriptstyle\times};
(4.887064,1.314048) *[red]{\scriptscriptstyle\times};
(4.895277,1.318790) *[red]{\scriptscriptstyle\times};
(4.903491,1.319291) *[red]{\scriptscriptstyle\times};
(4.911704,1.392086) *[red]{\scriptscriptstyle\times};
(4.919918,1.417768) *[red]{\scriptscriptstyle\times};
(4.928131,1.250445) *[red]{\scriptscriptstyle\times};
(4.936345,1.253838) *[red]{\scriptscriptstyle\times};
(4.944559,1.260509) *[red]{\scriptscriptstyle\times};
(4.952772,1.260647) *[red]{\scriptscriptstyle\times};
(4.960986,1.266757) *[red]{\scriptscriptstyle\times};
(4.969199,1.282210) *[red]{\scriptscriptstyle\times};
(4.977413,1.289359) *[red]{\scriptscriptstyle\times};
(4.985626,1.292047) *[red]{\scriptscriptstyle\times};
(4.993840,1.293205) *[red]{\scriptscriptstyle\times};
(5.002053,1.295354) *[red]{\scriptscriptstyle\times};
(5.010267,1.295579) *[red]{\scriptscriptstyle\times};
(5.018480,1.321226) *[red]{\scriptscriptstyle\times};
(5.026694,1.330798) *[red]{\scriptscriptstyle\times};
(5.034908,1.332835) *[red]{\scriptscriptstyle\times};
(5.043121,1.359309) *[red]{\scriptscriptstyle\times};
(5.051335,1.255559) *[red]{\scriptscriptstyle\times};
(5.059548,1.258411) *[red]{\scriptscriptstyle\times};
(5.067762,1.258927) *[red]{\scriptscriptstyle\times};
(5.075975,1.288887) *[red]{\scriptscriptstyle\times};
(5.084189,1.288975) *[red]{\scriptscriptstyle\times};
(5.092402,1.292170) *[red]{\scriptscriptstyle\times};
(5.100616,1.293390) *[red]{\scriptscriptstyle\times};
(5.108830,1.297312) *[red]{\scriptscriptstyle\times};
(5.117043,1.298870) *[red]{\scriptscriptstyle\times};
(5.125257,1.300134) *[red]{\scriptscriptstyle\times};
(5.133470,1.300638) *[red]{\scriptscriptstyle\times};
(5.141684,1.306650) *[red]{\scriptscriptstyle\times};
(5.149897,1.326609) *[red]{\scriptscriptstyle\times};
(5.158111,1.329643) *[red]{\scriptscriptstyle\times};
(5.166324,1.330724) *[red]{\scriptscriptstyle\times};
(5.174538,1.233262) *[red]{\scriptscriptstyle\times};
(5.182752,1.236815) *[red]{\scriptscriptstyle\times};
(5.190965,1.240798) *[red]{\scriptscriptstyle\times};
(5.199179,1.247121) *[red]{\scriptscriptstyle\times};
(5.207392,1.254004) *[red]{\scriptscriptstyle\times};
(5.215606,1.258553) *[red]{\scriptscriptstyle\times};
(5.223819,1.272266) *[red]{\scriptscriptstyle\times};
(5.232033,1.274709) *[red]{\scriptscriptstyle\times};
(5.240246,1.283014) *[red]{\scriptscriptstyle\times};
(5.248460,1.290355) *[red]{\scriptscriptstyle\times};
(5.256674,1.313631) *[red]{\scriptscriptstyle\times};
(5.264887,1.314651) *[red]{\scriptscriptstyle\times};
(5.273101,1.318058) *[red]{\scriptscriptstyle\times};
(5.281314,1.341382) *[red]{\scriptscriptstyle\times};
(5.289528,1.415667) *[red]{\scriptscriptstyle\times};
(5.297741,1.244496) *[red]{\scriptscriptstyle\times};
(5.305955,1.245665) *[red]{\scriptscriptstyle\times};
(5.314168,1.245764) *[red]{\scriptscriptstyle\times};
(5.322382,1.246875) *[red]{\scriptscriptstyle\times};
(5.330595,1.255197) *[red]{\scriptscriptstyle\times};
(5.338809,1.257740) *[red]{\scriptscriptstyle\times};
(5.347023,1.268311) *[red]{\scriptscriptstyle\times};
(5.355236,1.280176) *[red]{\scriptscriptstyle\times};
(5.363450,1.286612) *[red]{\scriptscriptstyle\times};
(5.371663,1.286634) *[red]{\scriptscriptstyle\times};
(5.379877,1.287442) *[red]{\scriptscriptstyle\times};
(5.388090,1.324253) *[red]{\scriptscriptstyle\times};
(5.396304,1.328019) *[red]{\scriptscriptstyle\times};
(5.404517,1.330557) *[red]{\scriptscriptstyle\times};
(5.412731,1.355925) *[red]{\scriptscriptstyle\times};
(5.420945,1.269905) *[red]{\scriptscriptstyle\times};
(5.429158,1.270008) *[red]{\scriptscriptstyle\times};
(5.437372,1.271409) *[red]{\scriptscriptstyle\times};
(5.445585,1.273525) *[red]{\scriptscriptstyle\times};
(5.453799,1.276850) *[red]{\scriptscriptstyle\times};
(5.462012,1.286148) *[red]{\scriptscriptstyle\times};
(5.470226,1.304696) *[red]{\scriptscriptstyle\times};
(5.478439,1.311048) *[red]{\scriptscriptstyle\times};
(5.486653,1.311151) *[red]{\scriptscriptstyle\times};
(5.494867,1.321152) *[red]{\scriptscriptstyle\times};
(5.503080,1.323844) *[red]{\scriptscriptstyle\times};
(5.511294,1.338982) *[red]{\scriptscriptstyle\times};
(5.519507,1.344874) *[red]{\scriptscriptstyle\times};
(5.527721,1.345601) *[red]{\scriptscriptstyle\times};
(5.535934,1.347999) *[red]{\scriptscriptstyle\times};
(5.544148,1.281221) *[red]{\scriptscriptstyle\times};
(5.552361,1.294298) *[red]{\scriptscriptstyle\times};
(5.560575,1.316637) *[red]{\scriptscriptstyle\times};
(5.568789,1.316746) *[red]{\scriptscriptstyle\times};
(5.577002,1.316764) *[red]{\scriptscriptstyle\times};
(5.585216,1.316807) *[red]{\scriptscriptstyle\times};
(5.593429,1.318961) *[red]{\scriptscriptstyle\times};
(5.601643,1.319007) *[red]{\scriptscriptstyle\times};
(5.609856,1.325495) *[red]{\scriptscriptstyle\times};
(5.618070,1.327663) *[red]{\scriptscriptstyle\times};
(5.626283,1.334780) *[red]{\scriptscriptstyle\times};
(5.634497,1.347127) *[red]{\scriptscriptstyle\times};
(5.642710,1.358696) *[red]{\scriptscriptstyle\times};
(5.650924,1.378929) *[red]{\scriptscriptstyle\times};
(5.659138,1.379287) *[red]{\scriptscriptstyle\times};
(5.667351,1.237918) *[red]{\scriptscriptstyle\times};
(5.675565,1.262989) *[red]{\scriptscriptstyle\times};
(5.683778,1.263271) *[red]{\scriptscriptstyle\times};
(5.691992,1.267684) *[red]{\scriptscriptstyle\times};
(5.700205,1.271197) *[red]{\scriptscriptstyle\times};
(5.708419,1.279935) *[red]{\scriptscriptstyle\times};
(5.716632,1.287566) *[red]{\scriptscriptstyle\times};
(5.724846,1.298452) *[red]{\scriptscriptstyle\times};
(5.733060,1.304938) *[red]{\scriptscriptstyle\times};
(5.741273,1.319120) *[red]{\scriptscriptstyle\times};
(5.749487,1.330948) *[red]{\scriptscriptstyle\times};
(5.757700,1.344247) *[red]{\scriptscriptstyle\times};
(5.765914,1.348338) *[red]{\scriptscriptstyle\times};
(5.774127,1.421018) *[red]{\scriptscriptstyle\times};
(5.782341,1.509137) *[red]{\scriptscriptstyle\times};
(5.790554,1.268735) *[red]{\scriptscriptstyle\times};
(5.798768,1.303580) *[red]{\scriptscriptstyle\times};
(5.806982,1.303629) *[red]{\scriptscriptstyle\times};
(5.815195,1.305145) *[red]{\scriptscriptstyle\times};
(5.823409,1.305168) *[red]{\scriptscriptstyle\times};
(5.831622,1.306999) *[red]{\scriptscriptstyle\times};
(5.839836,1.308330) *[red]{\scriptscriptstyle\times};
(5.848049,1.308436) *[red]{\scriptscriptstyle\times};
(5.856263,1.310200) *[red]{\scriptscriptstyle\times};
(5.864476,1.326412) *[red]{\scriptscriptstyle\times};
(5.872690,1.340839) *[red]{\scriptscriptstyle\times};
(5.880903,1.341034) *[red]{\scriptscriptstyle\times};
(5.889117,1.341851) *[red]{\scriptscriptstyle\times};
(5.897331,1.353520) *[red]{\scriptscriptstyle\times};
(5.905544,1.383106) *[red]{\scriptscriptstyle\times};
(5.913758,1.306959) *[red]{\scriptscriptstyle\times};
(5.921971,1.310325) *[red]{\scriptscriptstyle\times};
(5.930185,1.320171) *[red]{\scriptscriptstyle\times};
(5.938398,1.338961) *[red]{\scriptscriptstyle\times};
(5.946612,1.340852) *[red]{\scriptscriptstyle\times};
(5.954825,1.343902) *[red]{\scriptscriptstyle\times};
(5.963039,1.347198) *[red]{\scriptscriptstyle\times};
(5.971253,1.351595) *[red]{\scriptscriptstyle\times};
(5.979466,1.353626) *[red]{\scriptscriptstyle\times};
(5.987680,1.356777) *[red]{\scriptscriptstyle\times};
(5.995893,1.358399) *[red]{\scriptscriptstyle\times};
(6.004107,1.361156) *[red]{\scriptscriptstyle\times};
(6.012320,1.379442) *[red]{\scriptscriptstyle\times};
(6.020534,1.381409) *[red]{\scriptscriptstyle\times};
(6.028747,1.381916) *[red]{\scriptscriptstyle\times};
(6.036961,1.281767) *[red]{\scriptscriptstyle\times};
(6.045175,1.286623) *[red]{\scriptscriptstyle\times};
(6.053388,1.312955) *[red]{\scriptscriptstyle\times};
(6.061602,1.313141) *[red]{\scriptscriptstyle\times};
(6.069815,1.316244) *[red]{\scriptscriptstyle\times};
(6.078029,1.316265) *[red]{\scriptscriptstyle\times};
(6.086242,1.317748) *[red]{\scriptscriptstyle\times};
(6.094456,1.317868) *[red]{\scriptscriptstyle\times};
(6.102669,1.327836) *[red]{\scriptscriptstyle\times};
(6.110883,1.346911) *[red]{\scriptscriptstyle\times};
(6.119097,1.350181) *[red]{\scriptscriptstyle\times};
(6.127310,1.370795) *[red]{\scriptscriptstyle\times};
(6.135524,1.383007) *[red]{\scriptscriptstyle\times};
(6.143737,1.423153) *[red]{\scriptscriptstyle\times};
(6.151951,1.537499) *[red]{\scriptscriptstyle\times};
(6.160164,1.290744) *[red]{\scriptscriptstyle\times};
(6.168378,1.325026) *[red]{\scriptscriptstyle\times};
(6.176591,1.328205) *[red]{\scriptscriptstyle\times};
(6.184805,1.328858) *[red]{\scriptscriptstyle\times};
(6.193018,1.334114) *[red]{\scriptscriptstyle\times};
(6.201232,1.341066) *[red]{\scriptscriptstyle\times};
(6.209446,1.347480) *[red]{\scriptscriptstyle\times};
(6.217659,1.359355) *[red]{\scriptscriptstyle\times};
(6.225873,1.371139) *[red]{\scriptscriptstyle\times};
(6.234086,1.371141) *[red]{\scriptscriptstyle\times};
(6.242300,1.374354) *[red]{\scriptscriptstyle\times};
(6.250513,1.385867) *[red]{\scriptscriptstyle\times};
(6.258727,1.400767) *[red]{\scriptscriptstyle\times};
(6.266940,1.415987) *[red]{\scriptscriptstyle\times};
(6.275154,1.494093) *[red]{\scriptscriptstyle\times};
(6.283368,1.319705) *[red]{\scriptscriptstyle\times};
(6.291581,1.321927) *[red]{\scriptscriptstyle\times};
(6.299795,1.322980) *[red]{\scriptscriptstyle\times};
(6.308008,1.331069) *[red]{\scriptscriptstyle\times};
(6.316222,1.353510) *[red]{\scriptscriptstyle\times};
(6.324435,1.356518) *[red]{\scriptscriptstyle\times};
(6.332649,1.356765) *[red]{\scriptscriptstyle\times};
(6.340862,1.358180) *[red]{\scriptscriptstyle\times};
(6.349076,1.369161) *[red]{\scriptscriptstyle\times};
(6.357290,1.369386) *[red]{\scriptscriptstyle\times};
(6.365503,1.389256) *[red]{\scriptscriptstyle\times};
(6.373717,1.392555) *[red]{\scriptscriptstyle\times};
(6.381930,1.395837) *[red]{\scriptscriptstyle\times};
(6.390144,1.421660) *[red]{\scriptscriptstyle\times};
(6.398357,1.426293) *[red]{\scriptscriptstyle\times};
(6.406571,1.325167) *[red]{\scriptscriptstyle\times};
(6.414784,1.337966) *[red]{\scriptscriptstyle\times};
(6.422998,1.355853) *[red]{\scriptscriptstyle\times};
(6.431211,1.358499) *[red]{\scriptscriptstyle\times};
(6.439425,1.358523) *[red]{\scriptscriptstyle\times};
(6.447639,1.359600) *[red]{\scriptscriptstyle\times};
(6.455852,1.366162) *[red]{\scriptscriptstyle\times};
(6.464066,1.370935) *[red]{\scriptscriptstyle\times};
(6.472279,1.371002) *[red]{\scriptscriptstyle\times};
(6.480493,1.374212) *[red]{\scriptscriptstyle\times};
(6.488706,1.388278) *[red]{\scriptscriptstyle\times};
(6.496920,1.391249) *[red]{\scriptscriptstyle\times};
(6.505133,1.391250) *[red]{\scriptscriptstyle\times};
(6.513347,1.426427) *[red]{\scriptscriptstyle\times};
(6.521561,1.427524) *[red]{\scriptscriptstyle\times};
(6.529774,1.317475) *[red]{\scriptscriptstyle\times};
(6.537988,1.323267) *[red]{\scriptscriptstyle\times};
(6.546201,1.325840) *[red]{\scriptscriptstyle\times};
(6.554415,1.325874) *[red]{\scriptscriptstyle\times};
(6.562628,1.351018) *[red]{\scriptscriptstyle\times};
(6.570842,1.351298) *[red]{\scriptscriptstyle\times};
(6.579055,1.352991) *[red]{\scriptscriptstyle\times};
(6.587269,1.356587) *[red]{\scriptscriptstyle\times};
(6.595483,1.359027) *[red]{\scriptscriptstyle\times};
(6.603696,1.367466) *[red]{\scriptscriptstyle\times};
(6.611910,1.392419) *[red]{\scriptscriptstyle\times};
(6.620123,1.396793) *[red]{\scriptscriptstyle\times};
(6.628337,1.400667) *[red]{\scriptscriptstyle\times};
(6.636550,1.421095) *[red]{\scriptscriptstyle\times};
(6.644764,1.478670) *[red]{\scriptscriptstyle\times};
(6.652977,1.298250) *[red]{\scriptscriptstyle\times};
(6.661191,1.298269) *[red]{\scriptscriptstyle\times};
(6.669405,1.304939) *[red]{\scriptscriptstyle\times};
(6.677618,1.332228) *[red]{\scriptscriptstyle\times};
(6.685832,1.335541) *[red]{\scriptscriptstyle\times};
(6.694045,1.335634) *[red]{\scriptscriptstyle\times};
(6.702259,1.338889) *[red]{\scriptscriptstyle\times};
(6.710472,1.365592) *[red]{\scriptscriptstyle\times};
(6.718686,1.368997) *[red]{\scriptscriptstyle\times};
(6.726899,1.377077) *[red]{\scriptscriptstyle\times};
(6.735113,1.380716) *[red]{\scriptscriptstyle\times};
(6.743326,1.422788) *[red]{\scriptscriptstyle\times};
(6.751540,1.439413) *[red]{\scriptscriptstyle\times};
(6.759754,1.470277) *[red]{\scriptscriptstyle\times};
(6.767967,1.476411) *[red]{\scriptscriptstyle\times};
(6.776181,1.355425) *[red]{\scriptscriptstyle\times};
(6.784394,1.358475) *[red]{\scriptscriptstyle\times};
(6.792608,1.359453) *[red]{\scriptscriptstyle\times};
(6.800821,1.365054) *[red]{\scriptscriptstyle\times};
(6.809035,1.392242) *[red]{\scriptscriptstyle\times};
(6.817248,1.397448) *[red]{\scriptscriptstyle\times};
(6.825462,1.397499) *[red]{\scriptscriptstyle\times};
(6.833676,1.398336) *[red]{\scriptscriptstyle\times};
(6.841889,1.398393) *[red]{\scriptscriptstyle\times};
(6.850103,1.401722) *[red]{\scriptscriptstyle\times};
(6.858316,1.403665) *[red]{\scriptscriptstyle\times};
(6.866530,1.404626) *[red]{\scriptscriptstyle\times};
(6.874743,1.420786) *[red]{\scriptscriptstyle\times};
(6.882957,1.426943) *[red]{\scriptscriptstyle\times};
(6.891170,1.429509) *[red]{\scriptscriptstyle\times};
(6.899384,1.365256) *[red]{\scriptscriptstyle\times};
(6.907598,1.368660) *[red]{\scriptscriptstyle\times};
(6.915811,1.371621) *[red]{\scriptscriptstyle\times};
(6.924025,1.374752) *[red]{\scriptscriptstyle\times};
(6.932238,1.383997) *[red]{\scriptscriptstyle\times};
(6.940452,1.388331) *[red]{\scriptscriptstyle\times};
(6.948665,1.397721) *[red]{\scriptscriptstyle\times};
(6.956879,1.397831) *[red]{\scriptscriptstyle\times};
(6.965092,1.399030) *[red]{\scriptscriptstyle\times};
(6.973306,1.401882) *[red]{\scriptscriptstyle\times};
(6.981520,1.406043) *[red]{\scriptscriptstyle\times};
(6.989733,1.419801) *[red]{\scriptscriptstyle\times};
(6.997947,1.429525) *[red]{\scriptscriptstyle\times};
(7.006160,1.431076) *[red]{\scriptscriptstyle\times};
(7.014374,1.434188) *[red]{\scriptscriptstyle\times};
(7.022587,1.329804) *[red]{\scriptscriptstyle\times};
(7.030801,1.345603) *[red]{\scriptscriptstyle\times};
(7.039014,1.363156) *[red]{\scriptscriptstyle\times};
(7.047228,1.372615) *[red]{\scriptscriptstyle\times};
(7.055441,1.378966) *[red]{\scriptscriptstyle\times};
(7.063655,1.399819) *[red]{\scriptscriptstyle\times};
(7.071869,1.402026) *[red]{\scriptscriptstyle\times};
(7.080082,1.417532) *[red]{\scriptscriptstyle\times};
(7.088296,1.437433) *[red]{\scriptscriptstyle\times};
(7.096509,1.442662) *[red]{\scriptscriptstyle\times};
(7.104723,1.443068) *[red]{\scriptscriptstyle\times};
(7.112936,1.444176) *[red]{\scriptscriptstyle\times};
(7.121150,1.461537) *[red]{\scriptscriptstyle\times};
(7.129363,1.476477) *[red]{\scriptscriptstyle\times};
(7.137577,1.482762) *[red]{\scriptscriptstyle\times};
(7.145791,1.357981) *[red]{\scriptscriptstyle\times};
(7.154004,1.368467) *[red]{\scriptscriptstyle\times};
(7.162218,1.393973) *[red]{\scriptscriptstyle\times};
(7.170431,1.400096) *[red]{\scriptscriptstyle\times};
(7.178645,1.404292) *[red]{\scriptscriptstyle\times};
(7.186858,1.406277) *[red]{\scriptscriptstyle\times};
(7.195072,1.425960) *[red]{\scriptscriptstyle\times};
(7.203285,1.428299) *[red]{\scriptscriptstyle\times};
(7.211499,1.428841) *[red]{\scriptscriptstyle\times};
(7.219713,1.428875) *[red]{\scriptscriptstyle\times};
(7.227926,1.428929) *[red]{\scriptscriptstyle\times};
(7.236140,1.431682) *[red]{\scriptscriptstyle\times};
(7.244353,1.431731) *[red]{\scriptscriptstyle\times};
(7.252567,1.431816) *[red]{\scriptscriptstyle\times};
(7.260780,1.433274) *[red]{\scriptscriptstyle\times};
(7.268994,1.332769) *[red]{\scriptscriptstyle\times};
(7.277207,1.338255) *[red]{\scriptscriptstyle\times};
(7.285421,1.339063) *[red]{\scriptscriptstyle\times};
(7.293634,1.342288) *[red]{\scriptscriptstyle\times};
(7.301848,1.342366) *[red]{\scriptscriptstyle\times};
(7.310062,1.372538) *[red]{\scriptscriptstyle\times};
(7.318275,1.401498) *[red]{\scriptscriptstyle\times};
(7.326489,1.401736) *[red]{\scriptscriptstyle\times};
(7.334702,1.407982) *[red]{\scriptscriptstyle\times};
(7.342916,1.418053) *[red]{\scriptscriptstyle\times};
(7.351129,1.423247) *[red]{\scriptscriptstyle\times};
(7.359343,1.430620) *[red]{\scriptscriptstyle\times};
(7.367556,1.451355) *[red]{\scriptscriptstyle\times};
(7.375770,1.454266) *[red]{\scriptscriptstyle\times};
(7.383984,1.467705) *[red]{\scriptscriptstyle\times};
(7.392197,1.307420) *[red]{\scriptscriptstyle\times};
(7.400411,1.314768) *[red]{\scriptscriptstyle\times};
(7.408624,1.338220) *[red]{\scriptscriptstyle\times};
(7.416838,1.341104) *[red]{\scriptscriptstyle\times};
(7.425051,1.341354) *[red]{\scriptscriptstyle\times};
(7.433265,1.341730) *[red]{\scriptscriptstyle\times};
(7.441478,1.354057) *[red]{\scriptscriptstyle\times};
(7.449692,1.360697) *[red]{\scriptscriptstyle\times};
(7.457906,1.374293) *[red]{\scriptscriptstyle\times};
(7.466119,1.377919) *[red]{\scriptscriptstyle\times};
(7.474333,1.409848) *[red]{\scriptscriptstyle\times};
(7.482546,1.412935) *[red]{\scriptscriptstyle\times};
(7.490760,1.413106) *[red]{\scriptscriptstyle\times};
(7.498973,1.438510) *[red]{\scriptscriptstyle\times};
(7.507187,1.477815) *[red]{\scriptscriptstyle\times};
(7.515400,1.385485) *[red]{\scriptscriptstyle\times};
(7.523614,1.385591) *[red]{\scriptscriptstyle\times};
(7.531828,1.388768) *[red]{\scriptscriptstyle\times};
(7.540041,1.389484) *[red]{\scriptscriptstyle\times};
(7.548255,1.394896) *[red]{\scriptscriptstyle\times};
(7.556468,1.396005) *[red]{\scriptscriptstyle\times};
(7.564682,1.397666) *[red]{\scriptscriptstyle\times};
(7.572895,1.398536) *[red]{\scriptscriptstyle\times};
(7.581109,1.400528) *[red]{\scriptscriptstyle\times};
(7.589322,1.409347) *[red]{\scriptscriptstyle\times};
(7.597536,1.417582) *[red]{\scriptscriptstyle\times};
(7.605749,1.423849) *[red]{\scriptscriptstyle\times};
(7.613963,1.430055) *[red]{\scriptscriptstyle\times};
(7.622177,1.453385) *[red]{\scriptscriptstyle\times};
(7.630390,1.458358) *[red]{\scriptscriptstyle\times};
(7.638604,1.349741) *[red]{\scriptscriptstyle\times};
(7.646817,1.354456) *[red]{\scriptscriptstyle\times};
(7.655031,1.381230) *[red]{\scriptscriptstyle\times};
(7.663244,1.382651) *[red]{\scriptscriptstyle\times};
(7.671458,1.384477) *[red]{\scriptscriptstyle\times};
(7.679671,1.385520) *[red]{\scriptscriptstyle\times};
(7.687885,1.388928) *[red]{\scriptscriptstyle\times};
(7.696099,1.394595) *[red]{\scriptscriptstyle\times};
(7.704312,1.397840) *[red]{\scriptscriptstyle\times};
(7.712526,1.416504) *[red]{\scriptscriptstyle\times};
(7.720739,1.419333) *[red]{\scriptscriptstyle\times};
(7.728953,1.423204) *[red]{\scriptscriptstyle\times};
(7.737166,1.459544) *[red]{\scriptscriptstyle\times};
(7.745380,1.477427) *[red]{\scriptscriptstyle\times};
(7.753593,1.488042) *[red]{\scriptscriptstyle\times};
(7.761807,1.415031) *[red]{\scriptscriptstyle\times};
(7.770021,1.419741) *[red]{\scriptscriptstyle\times};
(7.778234,1.420564) *[red]{\scriptscriptstyle\times};
(7.786448,1.427595) *[red]{\scriptscriptstyle\times};
(7.794661,1.446378) *[red]{\scriptscriptstyle\times};
(7.802875,1.446413) *[red]{\scriptscriptstyle\times};
(7.811088,1.446528) *[red]{\scriptscriptstyle\times};
(7.819302,1.450724) *[red]{\scriptscriptstyle\times};
(7.827515,1.451880) *[red]{\scriptscriptstyle\times};
(7.835729,1.452411) *[red]{\scriptscriptstyle\times};
(7.843943,1.454457) *[red]{\scriptscriptstyle\times};
(7.852156,1.455365) *[red]{\scriptscriptstyle\times};
(7.860370,1.463159) *[red]{\scriptscriptstyle\times};
(7.868583,1.463874) *[red]{\scriptscriptstyle\times};
(7.876797,1.476842) *[red]{\scriptscriptstyle\times};
(7.885010,1.461346) *[red]{\scriptscriptstyle\times};
(7.893224,1.461844) *[red]{\scriptscriptstyle\times};
(7.901437,1.482014) *[red]{\scriptscriptstyle\times};
(7.909651,1.483481) *[red]{\scriptscriptstyle\times};
(7.917864,1.483674) *[red]{\scriptscriptstyle\times};
(7.926078,1.484717) *[red]{\scriptscriptstyle\times};
(7.934292,1.486002) *[red]{\scriptscriptstyle\times};
(7.942505,1.488764) *[red]{\scriptscriptstyle\times};
(7.950719,1.492407) *[red]{\scriptscriptstyle\times};
(7.958932,1.515794) *[red]{\scriptscriptstyle\times};
(7.967146,1.516381) *[red]{\scriptscriptstyle\times};
(7.975359,1.517797) *[red]{\scriptscriptstyle\times};
(7.983573,1.524831) *[red]{\scriptscriptstyle\times};
(7.991786,1.528270) *[red]{\scriptscriptstyle\times};
(8.000000,1.549829) *[red]{\scriptscriptstyle\times};
\endxy
}
\caption{Skylake cycles
  for the CSIDH-1024 action
  using {\tt velusqrt-asm}.
}
\label{action-cycles1024}
\end{figure}

\begin{figure}[t]
\centerline{
\xy <1.1cm,0cm>:<0cm,4cm>::
(0,0.554923); (8,0.554923) **[blue]@{-};
(8.1,0.554923) *[blue]{\rlap{440727}};
(0,0.620536); (8,0.620536) **[blue]@{-};
(8.1,0.620536) *[blue]{\rlap{461234}};
(0,0.689878); (8,0.689878) **[blue]@{-};
(8.1,0.689878) *[blue]{\rlap{483944}};
(0,0.446422); (8,0.446422) **[red]@{-};
(-0.1,0.446422) *[red]{\llap{408797}};
(0,0.510479); (8,0.510479) **[red]@{-};
(-0.1,0.510479) *[red]{\llap{427357}};
(0,0.583128); (8,0.583128) **[red]@{-};
(-0.1,0.583128) *[red]{\llap{449428}};
(-0.008811,0.200772); (-0.008811,0.942323) **[lightgray]@{-};
(0.114537,0.200772); (0.114537,0.942323) **[lightgray]@{-};
(0.237885,0.200772); (0.237885,0.942323) **[lightgray]@{-};
(0.361233,0.200772); (0.361233,0.942323) **[lightgray]@{-};
(0.484581,0.200772); (0.484581,0.942323) **[lightgray]@{-};
(0.607930,0.200772); (0.607930,0.942323) **[lightgray]@{-};
(0.731278,0.200772); (0.731278,0.942323) **[lightgray]@{-};
(0.854626,0.200772); (0.854626,0.942323) **[lightgray]@{-};
(0.977974,0.200772); (0.977974,0.942323) **[lightgray]@{-};
(1.101322,0.200772); (1.101322,0.942323) **[lightgray]@{-};
(1.224670,0.200772); (1.224670,0.942323) **[lightgray]@{-};
(1.348018,0.200772); (1.348018,0.942323) **[lightgray]@{-};
(1.471366,0.200772); (1.471366,0.942323) **[lightgray]@{-};
(1.594714,0.200772); (1.594714,0.942323) **[lightgray]@{-};
(1.718062,0.200772); (1.718062,0.942323) **[lightgray]@{-};
(1.841410,0.200772); (1.841410,0.942323) **[lightgray]@{-};
(1.964758,0.200772); (1.964758,0.942323) **[lightgray]@{-};
(2.088106,0.200772); (2.088106,0.942323) **[lightgray]@{-};
(2.211454,0.200772); (2.211454,0.942323) **[lightgray]@{-};
(2.334802,0.200772); (2.334802,0.942323) **[lightgray]@{-};
(2.458150,0.200772); (2.458150,0.942323) **[lightgray]@{-};
(2.581498,0.200772); (2.581498,0.942323) **[lightgray]@{-};
(2.704846,0.200772); (2.704846,0.942323) **[lightgray]@{-};
(2.828194,0.200772); (2.828194,0.942323) **[lightgray]@{-};
(2.951542,0.200772); (2.951542,0.942323) **[lightgray]@{-};
(3.074890,0.200772); (3.074890,0.942323) **[lightgray]@{-};
(3.198238,0.200772); (3.198238,0.942323) **[lightgray]@{-};
(3.321586,0.200772); (3.321586,0.942323) **[lightgray]@{-};
(3.444934,0.200772); (3.444934,0.942323) **[lightgray]@{-};
(3.568282,0.200772); (3.568282,0.942323) **[lightgray]@{-};
(3.691630,0.200772); (3.691630,0.942323) **[lightgray]@{-};
(3.814978,0.200772); (3.814978,0.942323) **[lightgray]@{-};
(3.938326,0.200772); (3.938326,0.942323) **[lightgray]@{-};
(4.061674,0.200772); (4.061674,0.942323) **[lightgray]@{-};
(4.185022,0.200772); (4.185022,0.942323) **[lightgray]@{-};
(4.308370,0.200772); (4.308370,0.942323) **[lightgray]@{-};
(4.431718,0.200772); (4.431718,0.942323) **[lightgray]@{-};
(4.555066,0.200772); (4.555066,0.942323) **[lightgray]@{-};
(4.678414,0.200772); (4.678414,0.942323) **[lightgray]@{-};
(4.801762,0.200772); (4.801762,0.942323) **[lightgray]@{-};
(4.925110,0.200772); (4.925110,0.942323) **[lightgray]@{-};
(5.048458,0.200772); (5.048458,0.942323) **[lightgray]@{-};
(5.171806,0.200772); (5.171806,0.942323) **[lightgray]@{-};
(5.295154,0.200772); (5.295154,0.942323) **[lightgray]@{-};
(5.418502,0.200772); (5.418502,0.942323) **[lightgray]@{-};
(5.541850,0.200772); (5.541850,0.942323) **[lightgray]@{-};
(5.665198,0.200772); (5.665198,0.942323) **[lightgray]@{-};
(5.788546,0.200772); (5.788546,0.942323) **[lightgray]@{-};
(5.911894,0.200772); (5.911894,0.942323) **[lightgray]@{-};
(6.035242,0.200772); (6.035242,0.942323) **[lightgray]@{-};
(6.158590,0.200772); (6.158590,0.942323) **[lightgray]@{-};
(6.281938,0.200772); (6.281938,0.942323) **[lightgray]@{-};
(6.405286,0.200772); (6.405286,0.942323) **[lightgray]@{-};
(6.528634,0.200772); (6.528634,0.942323) **[lightgray]@{-};
(6.651982,0.200772); (6.651982,0.942323) **[lightgray]@{-};
(6.775330,0.200772); (6.775330,0.942323) **[lightgray]@{-};
(6.898678,0.200772); (6.898678,0.942323) **[lightgray]@{-};
(7.022026,0.200772); (7.022026,0.942323) **[lightgray]@{-};
(7.145374,0.200772); (7.145374,0.942323) **[lightgray]@{-};
(7.268722,0.200772); (7.268722,0.942323) **[lightgray]@{-};
(7.392070,0.200772); (7.392070,0.942323) **[lightgray]@{-};
(7.515419,0.200772); (7.515419,0.942323) **[lightgray]@{-};
(7.638767,0.200772); (7.638767,0.942323) **[lightgray]@{-};
(7.762115,0.200772); (7.762115,0.942323) **[lightgray]@{-};
(7.885463,0.200772); (7.885463,0.942323) **[lightgray]@{-};
(8.008811,0.200772); (8.008811,0.942323) **[lightgray]@{-};
(-0.008811,0.200772); (-0.008811,0.942323) **[lightgray]@{-};
(0.114537,0.200772); (0.114537,0.942323) **[lightgray]@{-};
(0.237885,0.200772); (0.237885,0.942323) **[lightgray]@{-};
(0.361233,0.200772); (0.361233,0.942323) **[lightgray]@{-};
(0.484581,0.200772); (0.484581,0.942323) **[lightgray]@{-};
(0.607930,0.200772); (0.607930,0.942323) **[lightgray]@{-};
(0.731278,0.200772); (0.731278,0.942323) **[lightgray]@{-};
(0.854626,0.200772); (0.854626,0.942323) **[lightgray]@{-};
(0.977974,0.200772); (0.977974,0.942323) **[lightgray]@{-};
(1.101322,0.200772); (1.101322,0.942323) **[lightgray]@{-};
(1.224670,0.200772); (1.224670,0.942323) **[lightgray]@{-};
(1.348018,0.200772); (1.348018,0.942323) **[lightgray]@{-};
(1.471366,0.200772); (1.471366,0.942323) **[lightgray]@{-};
(1.594714,0.200772); (1.594714,0.942323) **[lightgray]@{-};
(1.718062,0.200772); (1.718062,0.942323) **[lightgray]@{-};
(1.841410,0.200772); (1.841410,0.942323) **[lightgray]@{-};
(1.964758,0.200772); (1.964758,0.942323) **[lightgray]@{-};
(2.088106,0.200772); (2.088106,0.942323) **[lightgray]@{-};
(2.211454,0.200772); (2.211454,0.942323) **[lightgray]@{-};
(2.334802,0.200772); (2.334802,0.942323) **[lightgray]@{-};
(2.458150,0.200772); (2.458150,0.942323) **[lightgray]@{-};
(2.581498,0.200772); (2.581498,0.942323) **[lightgray]@{-};
(2.704846,0.200772); (2.704846,0.942323) **[lightgray]@{-};
(2.828194,0.200772); (2.828194,0.942323) **[lightgray]@{-};
(2.951542,0.200772); (2.951542,0.942323) **[lightgray]@{-};
(3.074890,0.200772); (3.074890,0.942323) **[lightgray]@{-};
(3.198238,0.200772); (3.198238,0.942323) **[lightgray]@{-};
(3.321586,0.200772); (3.321586,0.942323) **[lightgray]@{-};
(3.444934,0.200772); (3.444934,0.942323) **[lightgray]@{-};
(3.568282,0.200772); (3.568282,0.942323) **[lightgray]@{-};
(3.691630,0.200772); (3.691630,0.942323) **[lightgray]@{-};
(3.814978,0.200772); (3.814978,0.942323) **[lightgray]@{-};
(3.938326,0.200772); (3.938326,0.942323) **[lightgray]@{-};
(4.061674,0.200772); (4.061674,0.942323) **[lightgray]@{-};
(4.185022,0.200772); (4.185022,0.942323) **[lightgray]@{-};
(4.308370,0.200772); (4.308370,0.942323) **[lightgray]@{-};
(4.431718,0.200772); (4.431718,0.942323) **[lightgray]@{-};
(4.555066,0.200772); (4.555066,0.942323) **[lightgray]@{-};
(4.678414,0.200772); (4.678414,0.942323) **[lightgray]@{-};
(4.801762,0.200772); (4.801762,0.942323) **[lightgray]@{-};
(4.925110,0.200772); (4.925110,0.942323) **[lightgray]@{-};
(5.048458,0.200772); (5.048458,0.942323) **[lightgray]@{-};
(5.171806,0.200772); (5.171806,0.942323) **[lightgray]@{-};
(5.295154,0.200772); (5.295154,0.942323) **[lightgray]@{-};
(5.418502,0.200772); (5.418502,0.942323) **[lightgray]@{-};
(5.541850,0.200772); (5.541850,0.942323) **[lightgray]@{-};
(5.665198,0.200772); (5.665198,0.942323) **[lightgray]@{-};
(5.788546,0.200772); (5.788546,0.942323) **[lightgray]@{-};
(5.911894,0.200772); (5.911894,0.942323) **[lightgray]@{-};
(6.035242,0.200772); (6.035242,0.942323) **[lightgray]@{-};
(6.158590,0.200772); (6.158590,0.942323) **[lightgray]@{-};
(6.281938,0.200772); (6.281938,0.942323) **[lightgray]@{-};
(6.405286,0.200772); (6.405286,0.942323) **[lightgray]@{-};
(6.528634,0.200772); (6.528634,0.942323) **[lightgray]@{-};
(6.651982,0.200772); (6.651982,0.942323) **[lightgray]@{-};
(6.775330,0.200772); (6.775330,0.942323) **[lightgray]@{-};
(6.898678,0.200772); (6.898678,0.942323) **[lightgray]@{-};
(7.022026,0.200772); (7.022026,0.942323) **[lightgray]@{-};
(7.145374,0.200772); (7.145374,0.942323) **[lightgray]@{-};
(7.268722,0.200772); (7.268722,0.942323) **[lightgray]@{-};
(7.392070,0.200772); (7.392070,0.942323) **[lightgray]@{-};
(7.515419,0.200772); (7.515419,0.942323) **[lightgray]@{-};
(7.638767,0.200772); (7.638767,0.942323) **[lightgray]@{-};
(7.762115,0.200772); (7.762115,0.942323) **[lightgray]@{-};
(7.885463,0.200772); (7.885463,0.942323) **[lightgray]@{-};
(8.008811,0.200772); (8.008811,0.942323) **[lightgray]@{-};
(0.000000,0.336839) *[blue]{\scriptscriptstyle+};
(0.017621,0.352664) *[blue]{\scriptscriptstyle+};
(0.035242,0.367170) *[blue]{\scriptscriptstyle+};
(0.052863,0.372284) *[blue]{\scriptscriptstyle+};
(0.070485,0.420247) *[blue]{\scriptscriptstyle+};
(0.088106,0.452490) *[blue]{\scriptscriptstyle+};
(0.105727,0.455993) *[blue]{\scriptscriptstyle+};
(0.123348,0.321259) *[blue]{\scriptscriptstyle+};
(0.140969,0.360593) *[blue]{\scriptscriptstyle+};
(0.158590,0.389706) *[blue]{\scriptscriptstyle+};
(0.176211,0.396934) *[blue]{\scriptscriptstyle+};
(0.193833,0.401048) *[blue]{\scriptscriptstyle+};
(0.211454,0.410591) *[blue]{\scriptscriptstyle+};
(0.229075,0.417553) *[blue]{\scriptscriptstyle+};
(0.246696,0.351474) *[blue]{\scriptscriptstyle+};
(0.264317,0.374440) *[blue]{\scriptscriptstyle+};
(0.281938,0.374841) *[blue]{\scriptscriptstyle+};
(0.299559,0.427306) *[blue]{\scriptscriptstyle+};
(0.317181,0.427470) *[blue]{\scriptscriptstyle+};
(0.334802,0.438414) *[blue]{\scriptscriptstyle+};
(0.352423,0.451112) *[blue]{\scriptscriptstyle+};
(0.370044,0.387947) *[blue]{\scriptscriptstyle+};
(0.387665,0.389776) *[blue]{\scriptscriptstyle+};
(0.405286,0.393218) *[blue]{\scriptscriptstyle+};
(0.422907,0.418809) *[blue]{\scriptscriptstyle+};
(0.440529,0.421106) *[blue]{\scriptscriptstyle+};
(0.458150,0.425184) *[blue]{\scriptscriptstyle+};
(0.475771,0.433487) *[blue]{\scriptscriptstyle+};
(0.493392,0.427181) *[blue]{\scriptscriptstyle+};
(0.511013,0.444490) *[blue]{\scriptscriptstyle+};
(0.528634,0.460907) *[blue]{\scriptscriptstyle+};
(0.546256,0.469049) *[blue]{\scriptscriptstyle+};
(0.563877,0.472973) *[blue]{\scriptscriptstyle+};
(0.581498,0.492479) *[blue]{\scriptscriptstyle+};
(0.599119,0.502914) *[blue]{\scriptscriptstyle+};
(0.616740,0.400815) *[blue]{\scriptscriptstyle+};
(0.634361,0.425935) *[blue]{\scriptscriptstyle+};
(0.651982,0.440630) *[blue]{\scriptscriptstyle+};
(0.669604,0.441246) *[blue]{\scriptscriptstyle+};
(0.687225,0.457174) *[blue]{\scriptscriptstyle+};
(0.704846,0.479220) *[blue]{\scriptscriptstyle+};
(0.722467,0.479386) *[blue]{\scriptscriptstyle+};
(0.740088,0.473659) *[blue]{\scriptscriptstyle+};
(0.757709,0.477104) *[blue]{\scriptscriptstyle+};
(0.775330,0.481585) *[blue]{\scriptscriptstyle+};
(0.792952,0.485083) *[blue]{\scriptscriptstyle+};
(0.810573,0.488028) *[blue]{\scriptscriptstyle+};
(0.828194,0.526019) *[blue]{\scriptscriptstyle+};
(0.845815,0.574185) *[blue]{\scriptscriptstyle+};
(0.863436,0.475106) *[blue]{\scriptscriptstyle+};
(0.881057,0.484788) *[blue]{\scriptscriptstyle+};
(0.898678,0.497171) *[blue]{\scriptscriptstyle+};
(0.916300,0.502548) *[blue]{\scriptscriptstyle+};
(0.933921,0.508006) *[blue]{\scriptscriptstyle+};
(0.951542,0.522300) *[blue]{\scriptscriptstyle+};
(0.969163,0.577170) *[blue]{\scriptscriptstyle+};
(0.986784,0.425620) *[blue]{\scriptscriptstyle+};
(1.004405,0.428550) *[blue]{\scriptscriptstyle+};
(1.022026,0.483794) *[blue]{\scriptscriptstyle+};
(1.039648,0.485856) *[blue]{\scriptscriptstyle+};
(1.057269,0.505352) *[blue]{\scriptscriptstyle+};
(1.074890,0.561924) *[blue]{\scriptscriptstyle+};
(1.092511,0.588869) *[blue]{\scriptscriptstyle+};
(1.110132,0.477878) *[blue]{\scriptscriptstyle+};
(1.127753,0.491436) *[blue]{\scriptscriptstyle+};
(1.145374,0.505396) *[blue]{\scriptscriptstyle+};
(1.162996,0.525675) *[blue]{\scriptscriptstyle+};
(1.180617,0.540067) *[blue]{\scriptscriptstyle+};
(1.198238,0.588888) *[blue]{\scriptscriptstyle+};
(1.215859,0.656591) *[blue]{\scriptscriptstyle+};
(1.233480,0.510341) *[blue]{\scriptscriptstyle+};
(1.251101,0.511960) *[blue]{\scriptscriptstyle+};
(1.268722,0.514174) *[blue]{\scriptscriptstyle+};
(1.286344,0.546502) *[blue]{\scriptscriptstyle+};
(1.303965,0.591456) *[blue]{\scriptscriptstyle+};
(1.321586,0.603394) *[blue]{\scriptscriptstyle+};
(1.339207,0.619019) *[blue]{\scriptscriptstyle+};
(1.356828,0.519482) *[blue]{\scriptscriptstyle+};
(1.374449,0.526990) *[blue]{\scriptscriptstyle+};
(1.392070,0.536680) *[blue]{\scriptscriptstyle+};
(1.409692,0.546703) *[blue]{\scriptscriptstyle+};
(1.427313,0.550579) *[blue]{\scriptscriptstyle+};
(1.444934,0.557565) *[blue]{\scriptscriptstyle+};
(1.462555,0.558261) *[blue]{\scriptscriptstyle+};
(1.480176,0.488138) *[blue]{\scriptscriptstyle+};
(1.497797,0.521238) *[blue]{\scriptscriptstyle+};
(1.515419,0.537743) *[blue]{\scriptscriptstyle+};
(1.533040,0.542506) *[blue]{\scriptscriptstyle+};
(1.550661,0.563454) *[blue]{\scriptscriptstyle+};
(1.568282,0.568094) *[blue]{\scriptscriptstyle+};
(1.585903,0.655776) *[blue]{\scriptscriptstyle+};
(1.603524,0.482384) *[blue]{\scriptscriptstyle+};
(1.621145,0.535990) *[blue]{\scriptscriptstyle+};
(1.638767,0.540748) *[blue]{\scriptscriptstyle+};
(1.656388,0.541016) *[blue]{\scriptscriptstyle+};
(1.674009,0.619438) *[blue]{\scriptscriptstyle+};
(1.691630,0.635819) *[blue]{\scriptscriptstyle+};
(1.709251,0.684313) *[blue]{\scriptscriptstyle+};
(1.726872,0.525942) *[blue]{\scriptscriptstyle+};
(1.744493,0.529714) *[blue]{\scriptscriptstyle+};
(1.762115,0.530104) *[blue]{\scriptscriptstyle+};
(1.779736,0.536205) *[blue]{\scriptscriptstyle+};
(1.797357,0.539147) *[blue]{\scriptscriptstyle+};
(1.814978,0.543855) *[blue]{\scriptscriptstyle+};
(1.832599,0.551426) *[blue]{\scriptscriptstyle+};
(1.850220,0.492352) *[blue]{\scriptscriptstyle+};
(1.867841,0.526119) *[blue]{\scriptscriptstyle+};
(1.885463,0.536633) *[blue]{\scriptscriptstyle+};
(1.903084,0.571305) *[blue]{\scriptscriptstyle+};
(1.920705,0.588360) *[blue]{\scriptscriptstyle+};
(1.938326,0.599889) *[blue]{\scriptscriptstyle+};
(1.955947,0.600460) *[blue]{\scriptscriptstyle+};
(1.973568,0.553891) *[blue]{\scriptscriptstyle+};
(1.991189,0.560138) *[blue]{\scriptscriptstyle+};
(2.008811,0.570677) *[blue]{\scriptscriptstyle+};
(2.026432,0.575809) *[blue]{\scriptscriptstyle+};
(2.044053,0.585793) *[blue]{\scriptscriptstyle+};
(2.061674,0.599188) *[blue]{\scriptscriptstyle+};
(2.079295,0.601199) *[blue]{\scriptscriptstyle+};
(2.096916,0.540728) *[blue]{\scriptscriptstyle+};
(2.114537,0.540867) *[blue]{\scriptscriptstyle+};
(2.132159,0.545352) *[blue]{\scriptscriptstyle+};
(2.149780,0.547861) *[blue]{\scriptscriptstyle+};
(2.167401,0.559929) *[blue]{\scriptscriptstyle+};
(2.185022,0.563360) *[blue]{\scriptscriptstyle+};
(2.202643,0.574741) *[blue]{\scriptscriptstyle+};
(2.220264,0.513113) *[blue]{\scriptscriptstyle+};
(2.237885,0.524793) *[blue]{\scriptscriptstyle+};
(2.255507,0.529281) *[blue]{\scriptscriptstyle+};
(2.273128,0.538568) *[blue]{\scriptscriptstyle+};
(2.290749,0.539541) *[blue]{\scriptscriptstyle+};
(2.308370,0.540216) *[blue]{\scriptscriptstyle+};
(2.325991,0.559280) *[blue]{\scriptscriptstyle+};
(2.343612,0.546173) *[blue]{\scriptscriptstyle+};
(2.361233,0.553259) *[blue]{\scriptscriptstyle+};
(2.378855,0.574789) *[blue]{\scriptscriptstyle+};
(2.396476,0.580570) *[blue]{\scriptscriptstyle+};
(2.414097,0.580695) *[blue]{\scriptscriptstyle+};
(2.431718,0.587989) *[blue]{\scriptscriptstyle+};
(2.449339,0.604150) *[blue]{\scriptscriptstyle+};
(2.466960,0.565249) *[blue]{\scriptscriptstyle+};
(2.484581,0.568259) *[blue]{\scriptscriptstyle+};
(2.502203,0.572146) *[blue]{\scriptscriptstyle+};
(2.519824,0.573097) *[blue]{\scriptscriptstyle+};
(2.537445,0.580409) *[blue]{\scriptscriptstyle+};
(2.555066,0.588773) *[blue]{\scriptscriptstyle+};
(2.572687,0.613673) *[blue]{\scriptscriptstyle+};
(2.590308,0.554923) *[blue]{\scriptscriptstyle+};
(2.607930,0.557621) *[blue]{\scriptscriptstyle+};
(2.625551,0.577270) *[blue]{\scriptscriptstyle+};
(2.643172,0.580332) *[blue]{\scriptscriptstyle+};
(2.660793,0.586456) *[blue]{\scriptscriptstyle+};
(2.678414,0.590476) *[blue]{\scriptscriptstyle+};
(2.696035,0.610508) *[blue]{\scriptscriptstyle+};
(2.713656,0.567039) *[blue]{\scriptscriptstyle+};
(2.731278,0.568441) *[blue]{\scriptscriptstyle+};
(2.748899,0.575971) *[blue]{\scriptscriptstyle+};
(2.766520,0.590620) *[blue]{\scriptscriptstyle+};
(2.784141,0.593491) *[blue]{\scriptscriptstyle+};
(2.801762,0.599388) *[blue]{\scriptscriptstyle+};
(2.819383,0.617217) *[blue]{\scriptscriptstyle+};
(2.837004,0.535174) *[blue]{\scriptscriptstyle+};
(2.854626,0.558503) *[blue]{\scriptscriptstyle+};
(2.872247,0.585860) *[blue]{\scriptscriptstyle+};
(2.889868,0.586533) *[blue]{\scriptscriptstyle+};
(2.907489,0.611009) *[blue]{\scriptscriptstyle+};
(2.925110,0.648742) *[blue]{\scriptscriptstyle+};
(2.942731,0.653898) *[blue]{\scriptscriptstyle+};
(2.960352,0.543153) *[blue]{\scriptscriptstyle+};
(2.977974,0.545319) *[blue]{\scriptscriptstyle+};
(2.995595,0.569886) *[blue]{\scriptscriptstyle+};
(3.013216,0.574234) *[blue]{\scriptscriptstyle+};
(3.030837,0.589802) *[blue]{\scriptscriptstyle+};
(3.048458,0.614496) *[blue]{\scriptscriptstyle+};
(3.066079,0.622221) *[blue]{\scriptscriptstyle+};
(3.083700,0.518156) *[blue]{\scriptscriptstyle+};
(3.101322,0.547575) *[blue]{\scriptscriptstyle+};
(3.118943,0.549242) *[blue]{\scriptscriptstyle+};
(3.136564,0.561439) *[blue]{\scriptscriptstyle+};
(3.154185,0.565977) *[blue]{\scriptscriptstyle+};
(3.171806,0.614339) *[blue]{\scriptscriptstyle+};
(3.189427,0.622356) *[blue]{\scriptscriptstyle+};
(3.207048,0.562409) *[blue]{\scriptscriptstyle+};
(3.224670,0.562921) *[blue]{\scriptscriptstyle+};
(3.242291,0.593182) *[blue]{\scriptscriptstyle+};
(3.259912,0.598813) *[blue]{\scriptscriptstyle+};
(3.277533,0.617715) *[blue]{\scriptscriptstyle+};
(3.295154,0.620777) *[blue]{\scriptscriptstyle+};
(3.312775,0.623776) *[blue]{\scriptscriptstyle+};
(3.330396,0.532294) *[blue]{\scriptscriptstyle+};
(3.348018,0.547736) *[blue]{\scriptscriptstyle+};
(3.365639,0.576825) *[blue]{\scriptscriptstyle+};
(3.383260,0.580216) *[blue]{\scriptscriptstyle+};
(3.400881,0.599296) *[blue]{\scriptscriptstyle+};
(3.418502,0.628522) *[blue]{\scriptscriptstyle+};
(3.436123,0.669329) *[blue]{\scriptscriptstyle+};
(3.453744,0.569984) *[blue]{\scriptscriptstyle+};
(3.471366,0.578385) *[blue]{\scriptscriptstyle+};
(3.488987,0.581676) *[blue]{\scriptscriptstyle+};
(3.506608,0.582495) *[blue]{\scriptscriptstyle+};
(3.524229,0.589866) *[blue]{\scriptscriptstyle+};
(3.541850,0.603768) *[blue]{\scriptscriptstyle+};
(3.559471,0.654359) *[blue]{\scriptscriptstyle+};
(3.577093,0.596783) *[blue]{\scriptscriptstyle+};
(3.594714,0.621743) *[blue]{\scriptscriptstyle+};
(3.612335,0.634057) *[blue]{\scriptscriptstyle+};
(3.629956,0.643317) *[blue]{\scriptscriptstyle+};
(3.647577,0.654167) *[blue]{\scriptscriptstyle+};
(3.665198,0.660488) *[blue]{\scriptscriptstyle+};
(3.682819,0.662588) *[blue]{\scriptscriptstyle+};
(3.700441,0.565103) *[blue]{\scriptscriptstyle+};
(3.718062,0.566896) *[blue]{\scriptscriptstyle+};
(3.735683,0.572864) *[blue]{\scriptscriptstyle+};
(3.753304,0.573988) *[blue]{\scriptscriptstyle+};
(3.770925,0.579280) *[blue]{\scriptscriptstyle+};
(3.788546,0.590297) *[blue]{\scriptscriptstyle+};
(3.806167,0.603698) *[blue]{\scriptscriptstyle+};
(3.823789,0.595784) *[blue]{\scriptscriptstyle+};
(3.841410,0.613327) *[blue]{\scriptscriptstyle+};
(3.859031,0.623527) *[blue]{\scriptscriptstyle+};
(3.876652,0.658545) *[blue]{\scriptscriptstyle+};
(3.894273,0.660107) *[blue]{\scriptscriptstyle+};
(3.911894,0.673912) *[blue]{\scriptscriptstyle+};
(3.929515,0.678958) *[blue]{\scriptscriptstyle+};
(3.947137,0.575216) *[blue]{\scriptscriptstyle+};
(3.964758,0.576087) *[blue]{\scriptscriptstyle+};
(3.982379,0.582662) *[blue]{\scriptscriptstyle+};
(4.000000,0.585347) *[blue]{\scriptscriptstyle+};
(4.017621,0.599083) *[blue]{\scriptscriptstyle+};
(4.035242,0.624004) *[blue]{\scriptscriptstyle+};
(4.052863,0.664173) *[blue]{\scriptscriptstyle+};
(4.070485,0.595373) *[blue]{\scriptscriptstyle+};
(4.088106,0.623414) *[blue]{\scriptscriptstyle+};
(4.105727,0.624687) *[blue]{\scriptscriptstyle+};
(4.123348,0.626679) *[blue]{\scriptscriptstyle+};
(4.140969,0.633868) *[blue]{\scriptscriptstyle+};
(4.158590,0.654820) *[blue]{\scriptscriptstyle+};
(4.176211,0.667287) *[blue]{\scriptscriptstyle+};
(4.193833,0.607071) *[blue]{\scriptscriptstyle+};
(4.211454,0.616204) *[blue]{\scriptscriptstyle+};
(4.229075,0.617512) *[blue]{\scriptscriptstyle+};
(4.246696,0.620536) *[blue]{\scriptscriptstyle+};
(4.264317,0.634147) *[blue]{\scriptscriptstyle+};
(4.281938,0.665669) *[blue]{\scriptscriptstyle+};
(4.299559,0.700537) *[blue]{\scriptscriptstyle+};
(4.317181,0.549219) *[blue]{\scriptscriptstyle+};
(4.334802,0.576664) *[blue]{\scriptscriptstyle+};
(4.352423,0.592825) *[blue]{\scriptscriptstyle+};
(4.370044,0.602077) *[blue]{\scriptscriptstyle+};
(4.387665,0.654717) *[blue]{\scriptscriptstyle+};
(4.405286,0.667980) *[blue]{\scriptscriptstyle+};
(4.422907,0.679775) *[blue]{\scriptscriptstyle+};
(4.440529,0.602622) *[blue]{\scriptscriptstyle+};
(4.458150,0.608383) *[blue]{\scriptscriptstyle+};
(4.475771,0.628491) *[blue]{\scriptscriptstyle+};
(4.493392,0.632699) *[blue]{\scriptscriptstyle+};
(4.511013,0.633642) *[blue]{\scriptscriptstyle+};
(4.528634,0.643530) *[blue]{\scriptscriptstyle+};
(4.546256,0.662133) *[blue]{\scriptscriptstyle+};
(4.563877,0.615912) *[blue]{\scriptscriptstyle+};
(4.581498,0.617831) *[blue]{\scriptscriptstyle+};
(4.599119,0.624659) *[blue]{\scriptscriptstyle+};
(4.616740,0.635364) *[blue]{\scriptscriptstyle+};
(4.634361,0.638741) *[blue]{\scriptscriptstyle+};
(4.651982,0.698055) *[blue]{\scriptscriptstyle+};
(4.669604,0.706782) *[blue]{\scriptscriptstyle+};
(4.687225,0.596455) *[blue]{\scriptscriptstyle+};
(4.704846,0.597027) *[blue]{\scriptscriptstyle+};
(4.722467,0.601779) *[blue]{\scriptscriptstyle+};
(4.740088,0.613038) *[blue]{\scriptscriptstyle+};
(4.757709,0.683720) *[blue]{\scriptscriptstyle+};
(4.775330,0.709079) *[blue]{\scriptscriptstyle+};
(4.792952,0.746138) *[blue]{\scriptscriptstyle+};
(4.810573,0.636710) *[blue]{\scriptscriptstyle+};
(4.828194,0.654246) *[blue]{\scriptscriptstyle+};
(4.845815,0.655721) *[blue]{\scriptscriptstyle+};
(4.863436,0.662573) *[blue]{\scriptscriptstyle+};
(4.881057,0.665350) *[blue]{\scriptscriptstyle+};
(4.898678,0.668522) *[blue]{\scriptscriptstyle+};
(4.916300,0.688339) *[blue]{\scriptscriptstyle+};
(4.933921,0.597809) *[blue]{\scriptscriptstyle+};
(4.951542,0.608942) *[blue]{\scriptscriptstyle+};
(4.969163,0.618784) *[blue]{\scriptscriptstyle+};
(4.986784,0.640402) *[blue]{\scriptscriptstyle+};
(5.004405,0.646673) *[blue]{\scriptscriptstyle+};
(5.022026,0.648505) *[blue]{\scriptscriptstyle+};
(5.039648,0.651255) *[blue]{\scriptscriptstyle+};
(5.057269,0.621774) *[blue]{\scriptscriptstyle+};
(5.074890,0.627305) *[blue]{\scriptscriptstyle+};
(5.092511,0.646384) *[blue]{\scriptscriptstyle+};
(5.110132,0.653088) *[blue]{\scriptscriptstyle+};
(5.127753,0.664003) *[blue]{\scriptscriptstyle+};
(5.145374,0.664377) *[blue]{\scriptscriptstyle+};
(5.162996,0.669991) *[blue]{\scriptscriptstyle+};
(5.180617,0.607323) *[blue]{\scriptscriptstyle+};
(5.198238,0.608317) *[blue]{\scriptscriptstyle+};
(5.215859,0.623221) *[blue]{\scriptscriptstyle+};
(5.233480,0.630371) *[blue]{\scriptscriptstyle+};
(5.251101,0.672087) *[blue]{\scriptscriptstyle+};
(5.268722,0.688252) *[blue]{\scriptscriptstyle+};
(5.286344,0.700345) *[blue]{\scriptscriptstyle+};
(5.303965,0.643853) *[blue]{\scriptscriptstyle+};
(5.321586,0.646058) *[blue]{\scriptscriptstyle+};
(5.339207,0.649152) *[blue]{\scriptscriptstyle+};
(5.356828,0.649897) *[blue]{\scriptscriptstyle+};
(5.374449,0.664037) *[blue]{\scriptscriptstyle+};
(5.392070,0.677116) *[blue]{\scriptscriptstyle+};
(5.409692,0.680738) *[blue]{\scriptscriptstyle+};
(5.427313,0.643462) *[blue]{\scriptscriptstyle+};
(5.444934,0.655898) *[blue]{\scriptscriptstyle+};
(5.462555,0.664744) *[blue]{\scriptscriptstyle+};
(5.480176,0.665526) *[blue]{\scriptscriptstyle+};
(5.497797,0.679313) *[blue]{\scriptscriptstyle+};
(5.515419,0.680099) *[blue]{\scriptscriptstyle+};
(5.533040,0.756827) *[blue]{\scriptscriptstyle+};
(5.550661,0.626205) *[blue]{\scriptscriptstyle+};
(5.568282,0.626738) *[blue]{\scriptscriptstyle+};
(5.585903,0.632585) *[blue]{\scriptscriptstyle+};
(5.603524,0.649039) *[blue]{\scriptscriptstyle+};
(5.621145,0.650216) *[blue]{\scriptscriptstyle+};
(5.638767,0.669904) *[blue]{\scriptscriptstyle+};
(5.656388,0.720891) *[blue]{\scriptscriptstyle+};
(5.674009,0.659928) *[blue]{\scriptscriptstyle+};
(5.691630,0.660174) *[blue]{\scriptscriptstyle+};
(5.709251,0.661351) *[blue]{\scriptscriptstyle+};
(5.726872,0.663536) *[blue]{\scriptscriptstyle+};
(5.744493,0.663621) *[blue]{\scriptscriptstyle+};
(5.762115,0.665687) *[blue]{\scriptscriptstyle+};
(5.779736,0.709114) *[blue]{\scriptscriptstyle+};
(5.797357,0.625766) *[blue]{\scriptscriptstyle+};
(5.814978,0.653311) *[blue]{\scriptscriptstyle+};
(5.832599,0.653329) *[blue]{\scriptscriptstyle+};
(5.850220,0.662649) *[blue]{\scriptscriptstyle+};
(5.867841,0.666217) *[blue]{\scriptscriptstyle+};
(5.885463,0.670221) *[blue]{\scriptscriptstyle+};
(5.903084,0.683939) *[blue]{\scriptscriptstyle+};
(5.920705,0.616223) *[blue]{\scriptscriptstyle+};
(5.938326,0.648870) *[blue]{\scriptscriptstyle+};
(5.955947,0.671188) *[blue]{\scriptscriptstyle+};
(5.973568,0.672655) *[blue]{\scriptscriptstyle+};
(5.991189,0.684014) *[blue]{\scriptscriptstyle+};
(6.008811,0.686099) *[blue]{\scriptscriptstyle+};
(6.026432,0.729139) *[blue]{\scriptscriptstyle+};
(6.044053,0.673330) *[blue]{\scriptscriptstyle+};
(6.061674,0.673330) *[blue]{\scriptscriptstyle+};
(6.079295,0.674153) *[blue]{\scriptscriptstyle+};
(6.096916,0.696130) *[blue]{\scriptscriptstyle+};
(6.114537,0.697551) *[blue]{\scriptscriptstyle+};
(6.132159,0.699155) *[blue]{\scriptscriptstyle+};
(6.149780,0.744867) *[blue]{\scriptscriptstyle+};
(6.167401,0.671604) *[blue]{\scriptscriptstyle+};
(6.185022,0.683792) *[blue]{\scriptscriptstyle+};
(6.202643,0.689896) *[blue]{\scriptscriptstyle+};
(6.220264,0.698681) *[blue]{\scriptscriptstyle+};
(6.237885,0.706111) *[blue]{\scriptscriptstyle+};
(6.255507,0.727528) *[blue]{\scriptscriptstyle+};
(6.273128,0.731447) *[blue]{\scriptscriptstyle+};
(6.290749,0.686959) *[blue]{\scriptscriptstyle+};
(6.308370,0.689878) *[blue]{\scriptscriptstyle+};
(6.325991,0.692737) *[blue]{\scriptscriptstyle+};
(6.343612,0.710913) *[blue]{\scriptscriptstyle+};
(6.361233,0.711178) *[blue]{\scriptscriptstyle+};
(6.378855,0.733253) *[blue]{\scriptscriptstyle+};
(6.396476,0.764692) *[blue]{\scriptscriptstyle+};
(6.414097,0.691897) *[blue]{\scriptscriptstyle+};
(6.431718,0.700960) *[blue]{\scriptscriptstyle+};
(6.449339,0.743320) *[blue]{\scriptscriptstyle+};
(6.466960,0.760508) *[blue]{\scriptscriptstyle+};
(6.484581,0.766923) *[blue]{\scriptscriptstyle+};
(6.502203,0.806797) *[blue]{\scriptscriptstyle+};
(6.519824,0.819344) *[blue]{\scriptscriptstyle+};
(6.537445,0.679262) *[blue]{\scriptscriptstyle+};
(6.555066,0.679667) *[blue]{\scriptscriptstyle+};
(6.572687,0.705194) *[blue]{\scriptscriptstyle+};
(6.590308,0.712660) *[blue]{\scriptscriptstyle+};
(6.607930,0.723302) *[blue]{\scriptscriptstyle+};
(6.625551,0.756016) *[blue]{\scriptscriptstyle+};
(6.643172,0.775888) *[blue]{\scriptscriptstyle+};
(6.660793,0.691635) *[blue]{\scriptscriptstyle+};
(6.678414,0.697617) *[blue]{\scriptscriptstyle+};
(6.696035,0.709279) *[blue]{\scriptscriptstyle+};
(6.713656,0.712672) *[blue]{\scriptscriptstyle+};
(6.731278,0.713834) *[blue]{\scriptscriptstyle+};
(6.748899,0.716303) *[blue]{\scriptscriptstyle+};
(6.766520,0.717330) *[blue]{\scriptscriptstyle+};
(6.784141,0.720141) *[blue]{\scriptscriptstyle+};
(6.801762,0.729830) *[blue]{\scriptscriptstyle+};
(6.819383,0.731456) *[blue]{\scriptscriptstyle+};
(6.837004,0.737638) *[blue]{\scriptscriptstyle+};
(6.854626,0.752968) *[blue]{\scriptscriptstyle+};
(6.872247,0.755483) *[blue]{\scriptscriptstyle+};
(6.889868,0.766705) *[blue]{\scriptscriptstyle+};
(6.907489,0.724490) *[blue]{\scriptscriptstyle+};
(6.925110,0.743041) *[blue]{\scriptscriptstyle+};
(6.942731,0.743885) *[blue]{\scriptscriptstyle+};
(6.960352,0.755793) *[blue]{\scriptscriptstyle+};
(6.977974,0.761705) *[blue]{\scriptscriptstyle+};
(6.995595,0.767776) *[blue]{\scriptscriptstyle+};
(7.013216,0.771162) *[blue]{\scriptscriptstyle+};
(7.030837,0.706391) *[blue]{\scriptscriptstyle+};
(7.048458,0.710893) *[blue]{\scriptscriptstyle+};
(7.066079,0.714493) *[blue]{\scriptscriptstyle+};
(7.083700,0.728957) *[blue]{\scriptscriptstyle+};
(7.101322,0.729070) *[blue]{\scriptscriptstyle+};
(7.118943,0.738448) *[blue]{\scriptscriptstyle+};
(7.136564,0.763117) *[blue]{\scriptscriptstyle+};
(7.154185,0.716613) *[blue]{\scriptscriptstyle+};
(7.171806,0.762703) *[blue]{\scriptscriptstyle+};
(7.189427,0.767237) *[blue]{\scriptscriptstyle+};
(7.207048,0.791972) *[blue]{\scriptscriptstyle+};
(7.224670,0.808404) *[blue]{\scriptscriptstyle+};
(7.242291,0.866642) *[blue]{\scriptscriptstyle+};
(7.259912,0.872529) *[blue]{\scriptscriptstyle+};
(7.277533,0.732347) *[blue]{\scriptscriptstyle+};
(7.295154,0.753513) *[blue]{\scriptscriptstyle+};
(7.312775,0.765775) *[blue]{\scriptscriptstyle+};
(7.330396,0.773816) *[blue]{\scriptscriptstyle+};
(7.348018,0.784074) *[blue]{\scriptscriptstyle+};
(7.365639,0.793197) *[blue]{\scriptscriptstyle+};
(7.383260,0.809892) *[blue]{\scriptscriptstyle+};
(7.400881,0.755292) *[blue]{\scriptscriptstyle+};
(7.418502,0.765792) *[blue]{\scriptscriptstyle+};
(7.436123,0.799295) *[blue]{\scriptscriptstyle+};
(7.453744,0.815103) *[blue]{\scriptscriptstyle+};
(7.471366,0.831012) *[blue]{\scriptscriptstyle+};
(7.488987,0.838019) *[blue]{\scriptscriptstyle+};
(7.506608,0.883154) *[blue]{\scriptscriptstyle+};
(7.524229,0.767333) *[blue]{\scriptscriptstyle+};
(7.541850,0.796983) *[blue]{\scriptscriptstyle+};
(7.559471,0.809681) *[blue]{\scriptscriptstyle+};
(7.577093,0.815187) *[blue]{\scriptscriptstyle+};
(7.594714,0.815450) *[blue]{\scriptscriptstyle+};
(7.612335,0.825441) *[blue]{\scriptscriptstyle+};
(7.629956,0.888388) *[blue]{\scriptscriptstyle+};
(7.647577,0.796711) *[blue]{\scriptscriptstyle+};
(7.665198,0.798098) *[blue]{\scriptscriptstyle+};
(7.682819,0.837017) *[blue]{\scriptscriptstyle+};
(7.700441,0.840857) *[blue]{\scriptscriptstyle+};
(7.718062,0.862667) *[blue]{\scriptscriptstyle+};
(7.735683,0.865805) *[blue]{\scriptscriptstyle+};
(7.753304,0.871399) *[blue]{\scriptscriptstyle+};
(7.770925,0.823940) *[blue]{\scriptscriptstyle+};
(7.788546,0.855979) *[blue]{\scriptscriptstyle+};
(7.806167,0.865420) *[blue]{\scriptscriptstyle+};
(7.823789,0.867470) *[blue]{\scriptscriptstyle+};
(7.841410,0.899707) *[blue]{\scriptscriptstyle+};
(7.859031,0.935875) *[blue]{\scriptscriptstyle+};
(7.876652,0.942323) *[blue]{\scriptscriptstyle+};
(7.894273,0.852963) *[blue]{\scriptscriptstyle+};
(7.911894,0.854185) *[blue]{\scriptscriptstyle+};
(7.929515,0.875646) *[blue]{\scriptscriptstyle+};
(7.947137,0.886113) *[blue]{\scriptscriptstyle+};
(7.964758,0.902632) *[blue]{\scriptscriptstyle+};
(7.982379,0.919833) *[blue]{\scriptscriptstyle+};
(8.000000,0.920771) *[blue]{\scriptscriptstyle+};
(0.000000,0.255489) *[red]{\scriptscriptstyle\times};
(0.017621,0.265549) *[red]{\scriptscriptstyle\times};
(0.035242,0.267720) *[red]{\scriptscriptstyle\times};
(0.052863,0.294508) *[red]{\scriptscriptstyle\times};
(0.070485,0.301326) *[red]{\scriptscriptstyle\times};
(0.088106,0.307063) *[red]{\scriptscriptstyle\times};
(0.105727,0.313950) *[red]{\scriptscriptstyle\times};
(0.123348,0.227560) *[red]{\scriptscriptstyle\times};
(0.140969,0.227881) *[red]{\scriptscriptstyle\times};
(0.158590,0.255182) *[red]{\scriptscriptstyle\times};
(0.176211,0.309227) *[red]{\scriptscriptstyle\times};
(0.193833,0.310914) *[red]{\scriptscriptstyle\times};
(0.211454,0.313563) *[red]{\scriptscriptstyle\times};
(0.229075,0.376246) *[red]{\scriptscriptstyle\times};
(0.246696,0.200772) *[red]{\scriptscriptstyle\times};
(0.264317,0.208559) *[red]{\scriptscriptstyle\times};
(0.281938,0.212187) *[red]{\scriptscriptstyle\times};
(0.299559,0.240347) *[red]{\scriptscriptstyle\times};
(0.317181,0.296690) *[red]{\scriptscriptstyle\times};
(0.334802,0.304912) *[red]{\scriptscriptstyle\times};
(0.352423,0.383722) *[red]{\scriptscriptstyle\times};
(0.370044,0.279491) *[red]{\scriptscriptstyle\times};
(0.387665,0.280549) *[red]{\scriptscriptstyle\times};
(0.405286,0.280822) *[red]{\scriptscriptstyle\times};
(0.422907,0.286995) *[red]{\scriptscriptstyle\times};
(0.440529,0.297047) *[red]{\scriptscriptstyle\times};
(0.458150,0.314836) *[red]{\scriptscriptstyle\times};
(0.475771,0.348406) *[red]{\scriptscriptstyle\times};
(0.493392,0.316706) *[red]{\scriptscriptstyle\times};
(0.511013,0.325966) *[red]{\scriptscriptstyle\times};
(0.528634,0.334366) *[red]{\scriptscriptstyle\times};
(0.546256,0.339780) *[red]{\scriptscriptstyle\times};
(0.563877,0.345881) *[red]{\scriptscriptstyle\times};
(0.581498,0.354956) *[red]{\scriptscriptstyle\times};
(0.599119,0.360934) *[red]{\scriptscriptstyle\times};
(0.616740,0.285955) *[red]{\scriptscriptstyle\times};
(0.634361,0.288181) *[red]{\scriptscriptstyle\times};
(0.651982,0.352958) *[red]{\scriptscriptstyle\times};
(0.669604,0.366551) *[red]{\scriptscriptstyle\times};
(0.687225,0.375386) *[red]{\scriptscriptstyle\times};
(0.704846,0.386403) *[red]{\scriptscriptstyle\times};
(0.722467,0.555355) *[red]{\scriptscriptstyle\times};
(0.740088,0.379070) *[red]{\scriptscriptstyle\times};
(0.757709,0.384695) *[red]{\scriptscriptstyle\times};
(0.775330,0.386660) *[red]{\scriptscriptstyle\times};
(0.792952,0.390862) *[red]{\scriptscriptstyle\times};
(0.810573,0.393815) *[red]{\scriptscriptstyle\times};
(0.828194,0.439783) *[red]{\scriptscriptstyle\times};
(0.845815,0.450056) *[red]{\scriptscriptstyle\times};
(0.863436,0.364841) *[red]{\scriptscriptstyle\times};
(0.881057,0.372562) *[red]{\scriptscriptstyle\times};
(0.898678,0.374978) *[red]{\scriptscriptstyle\times};
(0.916300,0.382213) *[red]{\scriptscriptstyle\times};
(0.933921,0.421616) *[red]{\scriptscriptstyle\times};
(0.951542,0.428407) *[red]{\scriptscriptstyle\times};
(0.969163,0.452265) *[red]{\scriptscriptstyle\times};
(0.986784,0.299803) *[red]{\scriptscriptstyle\times};
(1.004405,0.342547) *[red]{\scriptscriptstyle\times};
(1.022026,0.351779) *[red]{\scriptscriptstyle\times};
(1.039648,0.368485) *[red]{\scriptscriptstyle\times};
(1.057269,0.451650) *[red]{\scriptscriptstyle\times};
(1.074890,0.503980) *[red]{\scriptscriptstyle\times};
(1.092511,0.506876) *[red]{\scriptscriptstyle\times};
(1.110132,0.357954) *[red]{\scriptscriptstyle\times};
(1.127753,0.371414) *[red]{\scriptscriptstyle\times};
(1.145374,0.377257) *[red]{\scriptscriptstyle\times};
(1.162996,0.381593) *[red]{\scriptscriptstyle\times};
(1.180617,0.399962) *[red]{\scriptscriptstyle\times};
(1.198238,0.459180) *[red]{\scriptscriptstyle\times};
(1.215859,0.459715) *[red]{\scriptscriptstyle\times};
(1.233480,0.409816) *[red]{\scriptscriptstyle\times};
(1.251101,0.433547) *[red]{\scriptscriptstyle\times};
(1.268722,0.442230) *[red]{\scriptscriptstyle\times};
(1.286344,0.444512) *[red]{\scriptscriptstyle\times};
(1.303965,0.449697) *[red]{\scriptscriptstyle\times};
(1.321586,0.470289) *[red]{\scriptscriptstyle\times};
(1.339207,0.476088) *[red]{\scriptscriptstyle\times};
(1.356828,0.416534) *[red]{\scriptscriptstyle\times};
(1.374449,0.424424) *[red]{\scriptscriptstyle\times};
(1.392070,0.426744) *[red]{\scriptscriptstyle\times};
(1.409692,0.437782) *[red]{\scriptscriptstyle\times};
(1.427313,0.446563) *[red]{\scriptscriptstyle\times};
(1.444934,0.469525) *[red]{\scriptscriptstyle\times};
(1.462555,0.470264) *[red]{\scriptscriptstyle\times};
(1.480176,0.411672) *[red]{\scriptscriptstyle\times};
(1.497797,0.419765) *[red]{\scriptscriptstyle\times};
(1.515419,0.433558) *[red]{\scriptscriptstyle\times};
(1.533040,0.454358) *[red]{\scriptscriptstyle\times};
(1.550661,0.466615) *[red]{\scriptscriptstyle\times};
(1.568282,0.476834) *[red]{\scriptscriptstyle\times};
(1.585903,0.515568) *[red]{\scriptscriptstyle\times};
(1.603524,0.371485) *[red]{\scriptscriptstyle\times};
(1.621145,0.382140) *[red]{\scriptscriptstyle\times};
(1.638767,0.394159) *[red]{\scriptscriptstyle\times};
(1.656388,0.427424) *[red]{\scriptscriptstyle\times};
(1.674009,0.444254) *[red]{\scriptscriptstyle\times};
(1.691630,0.454737) *[red]{\scriptscriptstyle\times};
(1.709251,0.526099) *[red]{\scriptscriptstyle\times};
(1.726872,0.427059) *[red]{\scriptscriptstyle\times};
(1.744493,0.429021) *[red]{\scriptscriptstyle\times};
(1.762115,0.432649) *[red]{\scriptscriptstyle\times};
(1.779736,0.432799) *[red]{\scriptscriptstyle\times};
(1.797357,0.439676) *[red]{\scriptscriptstyle\times};
(1.814978,0.441409) *[red]{\scriptscriptstyle\times};
(1.832599,0.485585) *[red]{\scriptscriptstyle\times};
(1.850220,0.424367) *[red]{\scriptscriptstyle\times};
(1.867841,0.445797) *[red]{\scriptscriptstyle\times};
(1.885463,0.455499) *[red]{\scriptscriptstyle\times};
(1.903084,0.464978) *[red]{\scriptscriptstyle\times};
(1.920705,0.473195) *[red]{\scriptscriptstyle\times};
(1.938326,0.493641) *[red]{\scriptscriptstyle\times};
(1.955947,0.598397) *[red]{\scriptscriptstyle\times};
(1.973568,0.420010) *[red]{\scriptscriptstyle\times};
(1.991189,0.426444) *[red]{\scriptscriptstyle\times};
(2.008811,0.442535) *[red]{\scriptscriptstyle\times};
(2.026432,0.452880) *[red]{\scriptscriptstyle\times};
(2.044053,0.454934) *[red]{\scriptscriptstyle\times};
(2.061674,0.455253) *[red]{\scriptscriptstyle\times};
(2.079295,0.466939) *[red]{\scriptscriptstyle\times};
(2.096916,0.424575) *[red]{\scriptscriptstyle\times};
(2.114537,0.440332) *[red]{\scriptscriptstyle\times};
(2.132159,0.445246) *[red]{\scriptscriptstyle\times};
(2.149780,0.454386) *[red]{\scriptscriptstyle\times};
(2.167401,0.456249) *[red]{\scriptscriptstyle\times};
(2.185022,0.466695) *[red]{\scriptscriptstyle\times};
(2.202643,0.485626) *[red]{\scriptscriptstyle\times};
(2.220264,0.409780) *[red]{\scriptscriptstyle\times};
(2.237885,0.415762) *[red]{\scriptscriptstyle\times};
(2.255507,0.440800) *[red]{\scriptscriptstyle\times};
(2.273128,0.441044) *[red]{\scriptscriptstyle\times};
(2.290749,0.451671) *[red]{\scriptscriptstyle\times};
(2.308370,0.478848) *[red]{\scriptscriptstyle\times};
(2.325991,0.516919) *[red]{\scriptscriptstyle\times};
(2.343612,0.439495) *[red]{\scriptscriptstyle\times};
(2.361233,0.449028) *[red]{\scriptscriptstyle\times};
(2.378855,0.461553) *[red]{\scriptscriptstyle\times};
(2.396476,0.463061) *[red]{\scriptscriptstyle\times};
(2.414097,0.471108) *[red]{\scriptscriptstyle\times};
(2.431718,0.489128) *[red]{\scriptscriptstyle\times};
(2.449339,0.494689) *[red]{\scriptscriptstyle\times};
(2.466960,0.435572) *[red]{\scriptscriptstyle\times};
(2.484581,0.442064) *[red]{\scriptscriptstyle\times};
(2.502203,0.448179) *[red]{\scriptscriptstyle\times};
(2.519824,0.466841) *[red]{\scriptscriptstyle\times};
(2.537445,0.468166) *[red]{\scriptscriptstyle\times};
(2.555066,0.469844) *[red]{\scriptscriptstyle\times};
(2.572687,0.478516) *[red]{\scriptscriptstyle\times};
(2.590308,0.441395) *[red]{\scriptscriptstyle\times};
(2.607930,0.446422) *[red]{\scriptscriptstyle\times};
(2.625551,0.458924) *[red]{\scriptscriptstyle\times};
(2.643172,0.477335) *[red]{\scriptscriptstyle\times};
(2.660793,0.480296) *[red]{\scriptscriptstyle\times};
(2.678414,0.484478) *[red]{\scriptscriptstyle\times};
(2.696035,0.492181) *[red]{\scriptscriptstyle\times};
(2.713656,0.436560) *[red]{\scriptscriptstyle\times};
(2.731278,0.451006) *[red]{\scriptscriptstyle\times};
(2.748899,0.451882) *[red]{\scriptscriptstyle\times};
(2.766520,0.467046) *[red]{\scriptscriptstyle\times};
(2.784141,0.469049) *[red]{\scriptscriptstyle\times};
(2.801762,0.473409) *[red]{\scriptscriptstyle\times};
(2.819383,0.480338) *[red]{\scriptscriptstyle\times};
(2.837004,0.447911) *[red]{\scriptscriptstyle\times};
(2.854626,0.452335) *[red]{\scriptscriptstyle\times};
(2.872247,0.455443) *[red]{\scriptscriptstyle\times};
(2.889868,0.457409) *[red]{\scriptscriptstyle\times};
(2.907489,0.474459) *[red]{\scriptscriptstyle\times};
(2.925110,0.478616) *[red]{\scriptscriptstyle\times};
(2.942731,0.525882) *[red]{\scriptscriptstyle\times};
(2.960352,0.446239) *[red]{\scriptscriptstyle\times};
(2.977974,0.450510) *[red]{\scriptscriptstyle\times};
(2.995595,0.466065) *[red]{\scriptscriptstyle\times};
(3.013216,0.471808) *[red]{\scriptscriptstyle\times};
(3.030837,0.476824) *[red]{\scriptscriptstyle\times};
(3.048458,0.489406) *[red]{\scriptscriptstyle\times};
(3.066079,0.520300) *[red]{\scriptscriptstyle\times};
(3.083700,0.414897) *[red]{\scriptscriptstyle\times};
(3.101322,0.432828) *[red]{\scriptscriptstyle\times};
(3.118943,0.444038) *[red]{\scriptscriptstyle\times};
(3.136564,0.445140) *[red]{\scriptscriptstyle\times};
(3.154185,0.460397) *[red]{\scriptscriptstyle\times};
(3.171806,0.472301) *[red]{\scriptscriptstyle\times};
(3.189427,0.550152) *[red]{\scriptscriptstyle\times};
(3.207048,0.423664) *[red]{\scriptscriptstyle\times};
(3.224670,0.453537) *[red]{\scriptscriptstyle\times};
(3.242291,0.455632) *[red]{\scriptscriptstyle\times};
(3.259912,0.460148) *[red]{\scriptscriptstyle\times};
(3.277533,0.474663) *[red]{\scriptscriptstyle\times};
(3.295154,0.490823) *[red]{\scriptscriptstyle\times};
(3.312775,0.505982) *[red]{\scriptscriptstyle\times};
(3.330396,0.401783) *[red]{\scriptscriptstyle\times};
(3.348018,0.408618) *[red]{\scriptscriptstyle\times};
(3.365639,0.420427) *[red]{\scriptscriptstyle\times};
(3.383260,0.440258) *[red]{\scriptscriptstyle\times};
(3.400881,0.445723) *[red]{\scriptscriptstyle\times};
(3.418502,0.515070) *[red]{\scriptscriptstyle\times};
(3.436123,0.613830) *[red]{\scriptscriptstyle\times};
(3.453744,0.461731) *[red]{\scriptscriptstyle\times};
(3.471366,0.462736) *[red]{\scriptscriptstyle\times};
(3.488987,0.465100) *[red]{\scriptscriptstyle\times};
(3.506608,0.482029) *[red]{\scriptscriptstyle\times};
(3.524229,0.514541) *[red]{\scriptscriptstyle\times};
(3.541850,0.522598) *[red]{\scriptscriptstyle\times};
(3.559471,0.549968) *[red]{\scriptscriptstyle\times};
(3.577093,0.486920) *[red]{\scriptscriptstyle\times};
(3.594714,0.496033) *[red]{\scriptscriptstyle\times};
(3.612335,0.499160) *[red]{\scriptscriptstyle\times};
(3.629956,0.512048) *[red]{\scriptscriptstyle\times};
(3.647577,0.515060) *[red]{\scriptscriptstyle\times};
(3.665198,0.520310) *[red]{\scriptscriptstyle\times};
(3.682819,0.528775) *[red]{\scriptscriptstyle\times};
(3.700441,0.455814) *[red]{\scriptscriptstyle\times};
(3.718062,0.477626) *[red]{\scriptscriptstyle\times};
(3.735683,0.481733) *[red]{\scriptscriptstyle\times};
(3.753304,0.519243) *[red]{\scriptscriptstyle\times};
(3.770925,0.520682) *[red]{\scriptscriptstyle\times};
(3.788546,0.554762) *[red]{\scriptscriptstyle\times};
(3.806167,0.564794) *[red]{\scriptscriptstyle\times};
(3.823789,0.494590) *[red]{\scriptscriptstyle\times};
(3.841410,0.495654) *[red]{\scriptscriptstyle\times};
(3.859031,0.497679) *[red]{\scriptscriptstyle\times};
(3.876652,0.521044) *[red]{\scriptscriptstyle\times};
(3.894273,0.523355) *[red]{\scriptscriptstyle\times};
(3.911894,0.540639) *[red]{\scriptscriptstyle\times};
(3.929515,0.544670) *[red]{\scriptscriptstyle\times};
(3.947137,0.470552) *[red]{\scriptscriptstyle\times};
(3.964758,0.472373) *[red]{\scriptscriptstyle\times};
(3.982379,0.475189) *[red]{\scriptscriptstyle\times};
(4.000000,0.483172) *[red]{\scriptscriptstyle\times};
(4.017621,0.483646) *[red]{\scriptscriptstyle\times};
(4.035242,0.491025) *[red]{\scriptscriptstyle\times};
(4.052863,0.566809) *[red]{\scriptscriptstyle\times};
(4.070485,0.493985) *[red]{\scriptscriptstyle\times};
(4.088106,0.513373) *[red]{\scriptscriptstyle\times};
(4.105727,0.523445) *[red]{\scriptscriptstyle\times};
(4.123348,0.533480) *[red]{\scriptscriptstyle\times};
(4.140969,0.539951) *[red]{\scriptscriptstyle\times};
(4.158590,0.543215) *[red]{\scriptscriptstyle\times};
(4.176211,0.549479) *[red]{\scriptscriptstyle\times};
(4.193833,0.481678) *[red]{\scriptscriptstyle\times};
(4.211454,0.482886) *[red]{\scriptscriptstyle\times};
(4.229075,0.491880) *[red]{\scriptscriptstyle\times};
(4.246696,0.502093) *[red]{\scriptscriptstyle\times};
(4.264317,0.514046) *[red]{\scriptscriptstyle\times};
(4.281938,0.537564) *[red]{\scriptscriptstyle\times};
(4.299559,0.559916) *[red]{\scriptscriptstyle\times};
(4.317181,0.460522) *[red]{\scriptscriptstyle\times};
(4.334802,0.467394) *[red]{\scriptscriptstyle\times};
(4.352423,0.485564) *[red]{\scriptscriptstyle\times};
(4.370044,0.493811) *[red]{\scriptscriptstyle\times};
(4.387665,0.494214) *[red]{\scriptscriptstyle\times};
(4.405286,0.590811) *[red]{\scriptscriptstyle\times};
(4.422907,0.622880) *[red]{\scriptscriptstyle\times};
(4.440529,0.488261) *[red]{\scriptscriptstyle\times};
(4.458150,0.488889) *[red]{\scriptscriptstyle\times};
(4.475771,0.510479) *[red]{\scriptscriptstyle\times};
(4.493392,0.523960) *[red]{\scriptscriptstyle\times};
(4.511013,0.539551) *[red]{\scriptscriptstyle\times};
(4.528634,0.547677) *[red]{\scriptscriptstyle\times};
(4.546256,0.565513) *[red]{\scriptscriptstyle\times};
(4.563877,0.482687) *[red]{\scriptscriptstyle\times};
(4.581498,0.510229) *[red]{\scriptscriptstyle\times};
(4.599119,0.523539) *[red]{\scriptscriptstyle\times};
(4.616740,0.529468) *[red]{\scriptscriptstyle\times};
(4.634361,0.537654) *[red]{\scriptscriptstyle\times};
(4.651982,0.557876) *[red]{\scriptscriptstyle\times};
(4.669604,0.569394) *[red]{\scriptscriptstyle\times};
(4.687225,0.426801) *[red]{\scriptscriptstyle\times};
(4.704846,0.486721) *[red]{\scriptscriptstyle\times};
(4.722467,0.513753) *[red]{\scriptscriptstyle\times};
(4.740088,0.535499) *[red]{\scriptscriptstyle\times};
(4.757709,0.577041) *[red]{\scriptscriptstyle\times};
(4.775330,0.577199) *[red]{\scriptscriptstyle\times};
(4.792952,0.594214) *[red]{\scriptscriptstyle\times};
(4.810573,0.535383) *[red]{\scriptscriptstyle\times};
(4.828194,0.537730) *[red]{\scriptscriptstyle\times};
(4.845815,0.550763) *[red]{\scriptscriptstyle\times};
(4.863436,0.550911) *[red]{\scriptscriptstyle\times};
(4.881057,0.553744) *[red]{\scriptscriptstyle\times};
(4.898678,0.554589) *[red]{\scriptscriptstyle\times};
(4.916300,0.564944) *[red]{\scriptscriptstyle\times};
(4.933921,0.483578) *[red]{\scriptscriptstyle\times};
(4.951542,0.503237) *[red]{\scriptscriptstyle\times};
(4.969163,0.505698) *[red]{\scriptscriptstyle\times};
(4.986784,0.509939) *[red]{\scriptscriptstyle\times};
(5.004405,0.512824) *[red]{\scriptscriptstyle\times};
(5.022026,0.564677) *[red]{\scriptscriptstyle\times};
(5.039648,0.619097) *[red]{\scriptscriptstyle\times};
(5.057269,0.518589) *[red]{\scriptscriptstyle\times};
(5.074890,0.536368) *[red]{\scriptscriptstyle\times};
(5.092511,0.541442) *[red]{\scriptscriptstyle\times};
(5.110132,0.557356) *[red]{\scriptscriptstyle\times};
(5.127753,0.570861) *[red]{\scriptscriptstyle\times};
(5.145374,0.577992) *[red]{\scriptscriptstyle\times};
(5.162996,0.586600) *[red]{\scriptscriptstyle\times};
(5.180617,0.498949) *[red]{\scriptscriptstyle\times};
(5.198238,0.527778) *[red]{\scriptscriptstyle\times};
(5.215859,0.530297) *[red]{\scriptscriptstyle\times};
(5.233480,0.569705) *[red]{\scriptscriptstyle\times};
(5.251101,0.574538) *[red]{\scriptscriptstyle\times};
(5.268722,0.583128) *[red]{\scriptscriptstyle\times};
(5.286344,0.612548) *[red]{\scriptscriptstyle\times};
(5.303965,0.542951) *[red]{\scriptscriptstyle\times};
(5.321586,0.546532) *[red]{\scriptscriptstyle\times};
(5.339207,0.547015) *[red]{\scriptscriptstyle\times};
(5.356828,0.547262) *[red]{\scriptscriptstyle\times};
(5.374449,0.548012) *[red]{\scriptscriptstyle\times};
(5.392070,0.570291) *[red]{\scriptscriptstyle\times};
(5.409692,0.571777) *[red]{\scriptscriptstyle\times};
(5.427313,0.539190) *[red]{\scriptscriptstyle\times};
(5.444934,0.541026) *[red]{\scriptscriptstyle\times};
(5.462555,0.562220) *[red]{\scriptscriptstyle\times};
(5.480176,0.564437) *[red]{\scriptscriptstyle\times};
(5.497797,0.567244) *[red]{\scriptscriptstyle\times};
(5.515419,0.576699) *[red]{\scriptscriptstyle\times};
(5.533040,0.603252) *[red]{\scriptscriptstyle\times};
(5.550661,0.518867) *[red]{\scriptscriptstyle\times};
(5.568282,0.524355) *[red]{\scriptscriptstyle\times};
(5.585903,0.544380) *[red]{\scriptscriptstyle\times};
(5.603524,0.563223) *[red]{\scriptscriptstyle\times};
(5.621145,0.578095) *[red]{\scriptscriptstyle\times};
(5.638767,0.581955) *[red]{\scriptscriptstyle\times};
(5.656388,0.631350) *[red]{\scriptscriptstyle\times};
(5.674009,0.535575) *[red]{\scriptscriptstyle\times};
(5.691630,0.539690) *[red]{\scriptscriptstyle\times};
(5.709251,0.549383) *[red]{\scriptscriptstyle\times};
(5.726872,0.564655) *[red]{\scriptscriptstyle\times};
(5.744493,0.569057) *[red]{\scriptscriptstyle\times};
(5.762115,0.579309) *[red]{\scriptscriptstyle\times};
(5.779736,0.594220) *[red]{\scriptscriptstyle\times};
(5.797357,0.479651) *[red]{\scriptscriptstyle\times};
(5.814978,0.516045) *[red]{\scriptscriptstyle\times};
(5.832599,0.535628) *[red]{\scriptscriptstyle\times};
(5.850220,0.538621) *[red]{\scriptscriptstyle\times};
(5.867841,0.553849) *[red]{\scriptscriptstyle\times};
(5.885463,0.593038) *[red]{\scriptscriptstyle\times};
(5.903084,0.633676) *[red]{\scriptscriptstyle\times};
(5.920705,0.514195) *[red]{\scriptscriptstyle\times};
(5.938326,0.533693) *[red]{\scriptscriptstyle\times};
(5.955947,0.535914) *[red]{\scriptscriptstyle\times};
(5.973568,0.555898) *[red]{\scriptscriptstyle\times};
(5.991189,0.561217) *[red]{\scriptscriptstyle\times};
(6.008811,0.565191) *[red]{\scriptscriptstyle\times};
(6.026432,0.699667) *[red]{\scriptscriptstyle\times};
(6.044053,0.564261) *[red]{\scriptscriptstyle\times};
(6.061674,0.565197) *[red]{\scriptscriptstyle\times};
(6.079295,0.572327) *[red]{\scriptscriptstyle\times};
(6.096916,0.576158) *[red]{\scriptscriptstyle\times};
(6.114537,0.579290) *[red]{\scriptscriptstyle\times};
(6.132159,0.597139) *[red]{\scriptscriptstyle\times};
(6.149780,0.605757) *[red]{\scriptscriptstyle\times};
(6.167401,0.551340) *[red]{\scriptscriptstyle\times};
(6.185022,0.554595) *[red]{\scriptscriptstyle\times};
(6.202643,0.613261) *[red]{\scriptscriptstyle\times};
(6.220264,0.621765) *[red]{\scriptscriptstyle\times};
(6.237885,0.623190) *[red]{\scriptscriptstyle\times};
(6.255507,0.635194) *[red]{\scriptscriptstyle\times};
(6.273128,0.636259) *[red]{\scriptscriptstyle\times};
(6.290749,0.556516) *[red]{\scriptscriptstyle\times};
(6.308370,0.578800) *[red]{\scriptscriptstyle\times};
(6.325991,0.581309) *[red]{\scriptscriptstyle\times};
(6.343612,0.599661) *[red]{\scriptscriptstyle\times};
(6.361233,0.603274) *[red]{\scriptscriptstyle\times};
(6.378855,0.620643) *[red]{\scriptscriptstyle\times};
(6.396476,0.630837) *[red]{\scriptscriptstyle\times};
(6.414097,0.601582) *[red]{\scriptscriptstyle\times};
(6.431718,0.603720) *[red]{\scriptscriptstyle\times};
(6.449339,0.623898) *[red]{\scriptscriptstyle\times};
(6.466960,0.646749) *[red]{\scriptscriptstyle\times};
(6.484581,0.647404) *[red]{\scriptscriptstyle\times};
(6.502203,0.649428) *[red]{\scriptscriptstyle\times};
(6.519824,0.652314) *[red]{\scriptscriptstyle\times};
(6.537445,0.562286) *[red]{\scriptscriptstyle\times};
(6.555066,0.569011) *[red]{\scriptscriptstyle\times};
(6.572687,0.609831) *[red]{\scriptscriptstyle\times};
(6.590308,0.651028) *[red]{\scriptscriptstyle\times};
(6.607930,0.657481) *[red]{\scriptscriptstyle\times};
(6.625551,0.670124) *[red]{\scriptscriptstyle\times};
(6.643172,0.687619) *[red]{\scriptscriptstyle\times};
(6.660793,0.569076) *[red]{\scriptscriptstyle\times};
(6.678414,0.587503) *[red]{\scriptscriptstyle\times};
(6.696035,0.588299) *[red]{\scriptscriptstyle\times};
(6.713656,0.603565) *[red]{\scriptscriptstyle\times};
(6.731278,0.623352) *[red]{\scriptscriptstyle\times};
(6.748899,0.626486) *[red]{\scriptscriptstyle\times};
(6.766520,0.676478) *[red]{\scriptscriptstyle\times};
(6.784141,0.624828) *[red]{\scriptscriptstyle\times};
(6.801762,0.638469) *[red]{\scriptscriptstyle\times};
(6.819383,0.642905) *[red]{\scriptscriptstyle\times};
(6.837004,0.645305) *[red]{\scriptscriptstyle\times};
(6.854626,0.668013) *[red]{\scriptscriptstyle\times};
(6.872247,0.671568) *[red]{\scriptscriptstyle\times};
(6.889868,0.678099) *[red]{\scriptscriptstyle\times};
(6.907489,0.619050) *[red]{\scriptscriptstyle\times};
(6.925110,0.628646) *[red]{\scriptscriptstyle\times};
(6.942731,0.635640) *[red]{\scriptscriptstyle\times};
(6.960352,0.635751) *[red]{\scriptscriptstyle\times};
(6.977974,0.646408) *[red]{\scriptscriptstyle\times};
(6.995595,0.665147) *[red]{\scriptscriptstyle\times};
(7.013216,0.673918) *[red]{\scriptscriptstyle\times};
(7.030837,0.599289) *[red]{\scriptscriptstyle\times};
(7.048458,0.601097) *[red]{\scriptscriptstyle\times};
(7.066079,0.617129) *[red]{\scriptscriptstyle\times};
(7.083700,0.628301) *[red]{\scriptscriptstyle\times};
(7.101322,0.629868) *[red]{\scriptscriptstyle\times};
(7.118943,0.633081) *[red]{\scriptscriptstyle\times};
(7.136564,0.696875) *[red]{\scriptscriptstyle\times};
(7.154185,0.645256) *[red]{\scriptscriptstyle\times};
(7.171806,0.671429) *[red]{\scriptscriptstyle\times};
(7.189427,0.672051) *[red]{\scriptscriptstyle\times};
(7.207048,0.689946) *[red]{\scriptscriptstyle\times};
(7.224670,0.713130) *[red]{\scriptscriptstyle\times};
(7.242291,0.715214) *[red]{\scriptscriptstyle\times};
(7.259912,0.717857) *[red]{\scriptscriptstyle\times};
(7.277533,0.646473) *[red]{\scriptscriptstyle\times};
(7.295154,0.666557) *[red]{\scriptscriptstyle\times};
(7.312775,0.673520) *[red]{\scriptscriptstyle\times};
(7.330396,0.673843) *[red]{\scriptscriptstyle\times};
(7.348018,0.733745) *[red]{\scriptscriptstyle\times};
(7.365639,0.737116) *[red]{\scriptscriptstyle\times};
(7.383260,0.759210) *[red]{\scriptscriptstyle\times};
(7.400881,0.621631) *[red]{\scriptscriptstyle\times};
(7.418502,0.631536) *[red]{\scriptscriptstyle\times};
(7.436123,0.650844) *[red]{\scriptscriptstyle\times};
(7.453744,0.673593) *[red]{\scriptscriptstyle\times};
(7.471366,0.674723) *[red]{\scriptscriptstyle\times};
(7.488987,0.699350) *[red]{\scriptscriptstyle\times};
(7.506608,0.869711) *[red]{\scriptscriptstyle\times};
(7.524229,0.663223) *[red]{\scriptscriptstyle\times};
(7.541850,0.663888) *[red]{\scriptscriptstyle\times};
(7.559471,0.679664) *[red]{\scriptscriptstyle\times};
(7.577093,0.688816) *[red]{\scriptscriptstyle\times};
(7.594714,0.713752) *[red]{\scriptscriptstyle\times};
(7.612335,0.724487) *[red]{\scriptscriptstyle\times};
(7.629956,0.757168) *[red]{\scriptscriptstyle\times};
(7.647577,0.690724) *[red]{\scriptscriptstyle\times};
(7.665198,0.696540) *[red]{\scriptscriptstyle\times};
(7.682819,0.721510) *[red]{\scriptscriptstyle\times};
(7.700441,0.722113) *[red]{\scriptscriptstyle\times};
(7.718062,0.723398) *[red]{\scriptscriptstyle\times};
(7.735683,0.723608) *[red]{\scriptscriptstyle\times};
(7.753304,0.746915) *[red]{\scriptscriptstyle\times};
(7.770925,0.728071) *[red]{\scriptscriptstyle\times};
(7.788546,0.734268) *[red]{\scriptscriptstyle\times};
(7.806167,0.755107) *[red]{\scriptscriptstyle\times};
(7.823789,0.758386) *[red]{\scriptscriptstyle\times};
(7.841410,0.804775) *[red]{\scriptscriptstyle\times};
(7.859031,0.845540) *[red]{\scriptscriptstyle\times};
(7.876652,0.847545) *[red]{\scriptscriptstyle\times};
(7.894273,0.717658) *[red]{\scriptscriptstyle\times};
(7.911894,0.717807) *[red]{\scriptscriptstyle\times};
(7.929515,0.758036) *[red]{\scriptscriptstyle\times};
(7.947137,0.758198) *[red]{\scriptscriptstyle\times};
(7.964758,0.759613) *[red]{\scriptscriptstyle\times};
(7.982379,0.808289) *[red]{\scriptscriptstyle\times};
(8.000000,0.848734) *[red]{\scriptscriptstyle\times};
\endxy
}
\caption{Multiplication counts
  for the CSIDH-512 action
  using {\tt velusqrt-asm}.
}
\label{action-mults512}
\end{figure}

\begin{figure}[t]
\centerline{
\xy <1.1cm,0cm>:<0cm,4cm>::
(0,0.694323); (8,0.694323) **[blue]@{-};
(8.1,0.694323) *[blue]{\rlap{647250}};
(0,0.759855); (8,0.759855) **[blue]@{-};
(8.1,0.759855) *[blue]{\rlap{677328}};
(0,0.814809); (8,0.814809) **[blue]@{-};
(8.1,0.814809) *[blue]{\rlap{703626}};
(0,0.485254); (8,0.485254) **[red]@{-};
(-0.1,0.485254) *[red]{\llap{559933}};
(0,0.545165); (8,0.545165) **[red]@{-};
(-0.1,0.545165) *[red]{\llap{583675}};
(0,0.601739); (8,0.601739) **[red]@{-};
(-0.1,0.601739) *[red]{\llap{607018}};
(-0.008811,0.264203); (-0.008811,1.001536) **[lightgray]@{-};
(0.114537,0.264203); (0.114537,1.001536) **[lightgray]@{-};
(0.237885,0.264203); (0.237885,1.001536) **[lightgray]@{-};
(0.361233,0.264203); (0.361233,1.001536) **[lightgray]@{-};
(0.484581,0.264203); (0.484581,1.001536) **[lightgray]@{-};
(0.607930,0.264203); (0.607930,1.001536) **[lightgray]@{-};
(0.731278,0.264203); (0.731278,1.001536) **[lightgray]@{-};
(0.854626,0.264203); (0.854626,1.001536) **[lightgray]@{-};
(0.977974,0.264203); (0.977974,1.001536) **[lightgray]@{-};
(1.101322,0.264203); (1.101322,1.001536) **[lightgray]@{-};
(1.224670,0.264203); (1.224670,1.001536) **[lightgray]@{-};
(1.348018,0.264203); (1.348018,1.001536) **[lightgray]@{-};
(1.471366,0.264203); (1.471366,1.001536) **[lightgray]@{-};
(1.594714,0.264203); (1.594714,1.001536) **[lightgray]@{-};
(1.718062,0.264203); (1.718062,1.001536) **[lightgray]@{-};
(1.841410,0.264203); (1.841410,1.001536) **[lightgray]@{-};
(1.964758,0.264203); (1.964758,1.001536) **[lightgray]@{-};
(2.088106,0.264203); (2.088106,1.001536) **[lightgray]@{-};
(2.211454,0.264203); (2.211454,1.001536) **[lightgray]@{-};
(2.334802,0.264203); (2.334802,1.001536) **[lightgray]@{-};
(2.458150,0.264203); (2.458150,1.001536) **[lightgray]@{-};
(2.581498,0.264203); (2.581498,1.001536) **[lightgray]@{-};
(2.704846,0.264203); (2.704846,1.001536) **[lightgray]@{-};
(2.828194,0.264203); (2.828194,1.001536) **[lightgray]@{-};
(2.951542,0.264203); (2.951542,1.001536) **[lightgray]@{-};
(3.074890,0.264203); (3.074890,1.001536) **[lightgray]@{-};
(3.198238,0.264203); (3.198238,1.001536) **[lightgray]@{-};
(3.321586,0.264203); (3.321586,1.001536) **[lightgray]@{-};
(3.444934,0.264203); (3.444934,1.001536) **[lightgray]@{-};
(3.568282,0.264203); (3.568282,1.001536) **[lightgray]@{-};
(3.691630,0.264203); (3.691630,1.001536) **[lightgray]@{-};
(3.814978,0.264203); (3.814978,1.001536) **[lightgray]@{-};
(3.938326,0.264203); (3.938326,1.001536) **[lightgray]@{-};
(4.061674,0.264203); (4.061674,1.001536) **[lightgray]@{-};
(4.185022,0.264203); (4.185022,1.001536) **[lightgray]@{-};
(4.308370,0.264203); (4.308370,1.001536) **[lightgray]@{-};
(4.431718,0.264203); (4.431718,1.001536) **[lightgray]@{-};
(4.555066,0.264203); (4.555066,1.001536) **[lightgray]@{-};
(4.678414,0.264203); (4.678414,1.001536) **[lightgray]@{-};
(4.801762,0.264203); (4.801762,1.001536) **[lightgray]@{-};
(4.925110,0.264203); (4.925110,1.001536) **[lightgray]@{-};
(5.048458,0.264203); (5.048458,1.001536) **[lightgray]@{-};
(5.171806,0.264203); (5.171806,1.001536) **[lightgray]@{-};
(5.295154,0.264203); (5.295154,1.001536) **[lightgray]@{-};
(5.418502,0.264203); (5.418502,1.001536) **[lightgray]@{-};
(5.541850,0.264203); (5.541850,1.001536) **[lightgray]@{-};
(5.665198,0.264203); (5.665198,1.001536) **[lightgray]@{-};
(5.788546,0.264203); (5.788546,1.001536) **[lightgray]@{-};
(5.911894,0.264203); (5.911894,1.001536) **[lightgray]@{-};
(6.035242,0.264203); (6.035242,1.001536) **[lightgray]@{-};
(6.158590,0.264203); (6.158590,1.001536) **[lightgray]@{-};
(6.281938,0.264203); (6.281938,1.001536) **[lightgray]@{-};
(6.405286,0.264203); (6.405286,1.001536) **[lightgray]@{-};
(6.528634,0.264203); (6.528634,1.001536) **[lightgray]@{-};
(6.651982,0.264203); (6.651982,1.001536) **[lightgray]@{-};
(6.775330,0.264203); (6.775330,1.001536) **[lightgray]@{-};
(6.898678,0.264203); (6.898678,1.001536) **[lightgray]@{-};
(7.022026,0.264203); (7.022026,1.001536) **[lightgray]@{-};
(7.145374,0.264203); (7.145374,1.001536) **[lightgray]@{-};
(7.268722,0.264203); (7.268722,1.001536) **[lightgray]@{-};
(7.392070,0.264203); (7.392070,1.001536) **[lightgray]@{-};
(7.515419,0.264203); (7.515419,1.001536) **[lightgray]@{-};
(7.638767,0.264203); (7.638767,1.001536) **[lightgray]@{-};
(7.762115,0.264203); (7.762115,1.001536) **[lightgray]@{-};
(7.885463,0.264203); (7.885463,1.001536) **[lightgray]@{-};
(8.008811,0.264203); (8.008811,1.001536) **[lightgray]@{-};
(-0.008811,0.264203); (-0.008811,1.001536) **[lightgray]@{-};
(0.114537,0.264203); (0.114537,1.001536) **[lightgray]@{-};
(0.237885,0.264203); (0.237885,1.001536) **[lightgray]@{-};
(0.361233,0.264203); (0.361233,1.001536) **[lightgray]@{-};
(0.484581,0.264203); (0.484581,1.001536) **[lightgray]@{-};
(0.607930,0.264203); (0.607930,1.001536) **[lightgray]@{-};
(0.731278,0.264203); (0.731278,1.001536) **[lightgray]@{-};
(0.854626,0.264203); (0.854626,1.001536) **[lightgray]@{-};
(0.977974,0.264203); (0.977974,1.001536) **[lightgray]@{-};
(1.101322,0.264203); (1.101322,1.001536) **[lightgray]@{-};
(1.224670,0.264203); (1.224670,1.001536) **[lightgray]@{-};
(1.348018,0.264203); (1.348018,1.001536) **[lightgray]@{-};
(1.471366,0.264203); (1.471366,1.001536) **[lightgray]@{-};
(1.594714,0.264203); (1.594714,1.001536) **[lightgray]@{-};
(1.718062,0.264203); (1.718062,1.001536) **[lightgray]@{-};
(1.841410,0.264203); (1.841410,1.001536) **[lightgray]@{-};
(1.964758,0.264203); (1.964758,1.001536) **[lightgray]@{-};
(2.088106,0.264203); (2.088106,1.001536) **[lightgray]@{-};
(2.211454,0.264203); (2.211454,1.001536) **[lightgray]@{-};
(2.334802,0.264203); (2.334802,1.001536) **[lightgray]@{-};
(2.458150,0.264203); (2.458150,1.001536) **[lightgray]@{-};
(2.581498,0.264203); (2.581498,1.001536) **[lightgray]@{-};
(2.704846,0.264203); (2.704846,1.001536) **[lightgray]@{-};
(2.828194,0.264203); (2.828194,1.001536) **[lightgray]@{-};
(2.951542,0.264203); (2.951542,1.001536) **[lightgray]@{-};
(3.074890,0.264203); (3.074890,1.001536) **[lightgray]@{-};
(3.198238,0.264203); (3.198238,1.001536) **[lightgray]@{-};
(3.321586,0.264203); (3.321586,1.001536) **[lightgray]@{-};
(3.444934,0.264203); (3.444934,1.001536) **[lightgray]@{-};
(3.568282,0.264203); (3.568282,1.001536) **[lightgray]@{-};
(3.691630,0.264203); (3.691630,1.001536) **[lightgray]@{-};
(3.814978,0.264203); (3.814978,1.001536) **[lightgray]@{-};
(3.938326,0.264203); (3.938326,1.001536) **[lightgray]@{-};
(4.061674,0.264203); (4.061674,1.001536) **[lightgray]@{-};
(4.185022,0.264203); (4.185022,1.001536) **[lightgray]@{-};
(4.308370,0.264203); (4.308370,1.001536) **[lightgray]@{-};
(4.431718,0.264203); (4.431718,1.001536) **[lightgray]@{-};
(4.555066,0.264203); (4.555066,1.001536) **[lightgray]@{-};
(4.678414,0.264203); (4.678414,1.001536) **[lightgray]@{-};
(4.801762,0.264203); (4.801762,1.001536) **[lightgray]@{-};
(4.925110,0.264203); (4.925110,1.001536) **[lightgray]@{-};
(5.048458,0.264203); (5.048458,1.001536) **[lightgray]@{-};
(5.171806,0.264203); (5.171806,1.001536) **[lightgray]@{-};
(5.295154,0.264203); (5.295154,1.001536) **[lightgray]@{-};
(5.418502,0.264203); (5.418502,1.001536) **[lightgray]@{-};
(5.541850,0.264203); (5.541850,1.001536) **[lightgray]@{-};
(5.665198,0.264203); (5.665198,1.001536) **[lightgray]@{-};
(5.788546,0.264203); (5.788546,1.001536) **[lightgray]@{-};
(5.911894,0.264203); (5.911894,1.001536) **[lightgray]@{-};
(6.035242,0.264203); (6.035242,1.001536) **[lightgray]@{-};
(6.158590,0.264203); (6.158590,1.001536) **[lightgray]@{-};
(6.281938,0.264203); (6.281938,1.001536) **[lightgray]@{-};
(6.405286,0.264203); (6.405286,1.001536) **[lightgray]@{-};
(6.528634,0.264203); (6.528634,1.001536) **[lightgray]@{-};
(6.651982,0.264203); (6.651982,1.001536) **[lightgray]@{-};
(6.775330,0.264203); (6.775330,1.001536) **[lightgray]@{-};
(6.898678,0.264203); (6.898678,1.001536) **[lightgray]@{-};
(7.022026,0.264203); (7.022026,1.001536) **[lightgray]@{-};
(7.145374,0.264203); (7.145374,1.001536) **[lightgray]@{-};
(7.268722,0.264203); (7.268722,1.001536) **[lightgray]@{-};
(7.392070,0.264203); (7.392070,1.001536) **[lightgray]@{-};
(7.515419,0.264203); (7.515419,1.001536) **[lightgray]@{-};
(7.638767,0.264203); (7.638767,1.001536) **[lightgray]@{-};
(7.762115,0.264203); (7.762115,1.001536) **[lightgray]@{-};
(7.885463,0.264203); (7.885463,1.001536) **[lightgray]@{-};
(8.008811,0.264203); (8.008811,1.001536) **[lightgray]@{-};
(0.000000,0.548622) *[blue]{\scriptscriptstyle+};
(0.017621,0.548622) *[blue]{\scriptscriptstyle+};
(0.035242,0.548622) *[blue]{\scriptscriptstyle+};
(0.052863,0.554471) *[blue]{\scriptscriptstyle+};
(0.070485,0.556616) *[blue]{\scriptscriptstyle+};
(0.088106,0.615324) *[blue]{\scriptscriptstyle+};
(0.105727,0.617893) *[blue]{\scriptscriptstyle+};
(0.123348,0.496838) *[blue]{\scriptscriptstyle+};
(0.140969,0.499746) *[blue]{\scriptscriptstyle+};
(0.158590,0.503626) *[blue]{\scriptscriptstyle+};
(0.176211,0.507495) *[blue]{\scriptscriptstyle+};
(0.193833,0.530420) *[blue]{\scriptscriptstyle+};
(0.211454,0.540746) *[blue]{\scriptscriptstyle+};
(0.229075,0.546502) *[blue]{\scriptscriptstyle+};
(0.246696,0.526277) *[blue]{\scriptscriptstyle+};
(0.264317,0.531496) *[blue]{\scriptscriptstyle+};
(0.281938,0.535742) *[blue]{\scriptscriptstyle+};
(0.299559,0.567037) *[blue]{\scriptscriptstyle+};
(0.317181,0.576290) *[blue]{\scriptscriptstyle+};
(0.334802,0.576844) *[blue]{\scriptscriptstyle+};
(0.352423,0.605366) *[blue]{\scriptscriptstyle+};
(0.370044,0.523005) *[blue]{\scriptscriptstyle+};
(0.387665,0.524808) *[blue]{\scriptscriptstyle+};
(0.405286,0.558694) *[blue]{\scriptscriptstyle+};
(0.422907,0.564351) *[blue]{\scriptscriptstyle+};
(0.440529,0.564412) *[blue]{\scriptscriptstyle+};
(0.458150,0.566162) *[blue]{\scriptscriptstyle+};
(0.475771,0.603008) *[blue]{\scriptscriptstyle+};
(0.493392,0.555492) *[blue]{\scriptscriptstyle+};
(0.511013,0.558285) *[blue]{\scriptscriptstyle+};
(0.528634,0.560108) *[blue]{\scriptscriptstyle+};
(0.546256,0.562919) *[blue]{\scriptscriptstyle+};
(0.563877,0.595215) *[blue]{\scriptscriptstyle+};
(0.581498,0.605461) *[blue]{\scriptscriptstyle+};
(0.599119,0.672665) *[blue]{\scriptscriptstyle+};
(0.616740,0.544856) *[blue]{\scriptscriptstyle+};
(0.634361,0.579407) *[blue]{\scriptscriptstyle+};
(0.651982,0.581144) *[blue]{\scriptscriptstyle+};
(0.669604,0.601126) *[blue]{\scriptscriptstyle+};
(0.687225,0.622987) *[blue]{\scriptscriptstyle+};
(0.704846,0.625654) *[blue]{\scriptscriptstyle+};
(0.722467,0.627588) *[blue]{\scriptscriptstyle+};
(0.740088,0.561060) *[blue]{\scriptscriptstyle+};
(0.757709,0.561676) *[blue]{\scriptscriptstyle+};
(0.775330,0.566735) *[blue]{\scriptscriptstyle+};
(0.792952,0.592450) *[blue]{\scriptscriptstyle+};
(0.810573,0.595890) *[blue]{\scriptscriptstyle+};
(0.828194,0.596060) *[blue]{\scriptscriptstyle+};
(0.845815,0.607205) *[blue]{\scriptscriptstyle+};
(0.863436,0.622497) *[blue]{\scriptscriptstyle+};
(0.881057,0.622553) *[blue]{\scriptscriptstyle+};
(0.898678,0.629615) *[blue]{\scriptscriptstyle+};
(0.916300,0.631013) *[blue]{\scriptscriptstyle+};
(0.933921,0.637202) *[blue]{\scriptscriptstyle+};
(0.951542,0.664731) *[blue]{\scriptscriptstyle+};
(0.969163,0.722510) *[blue]{\scriptscriptstyle+};
(0.986784,0.613888) *[blue]{\scriptscriptstyle+};
(1.004405,0.625740) *[blue]{\scriptscriptstyle+};
(1.022026,0.645960) *[blue]{\scriptscriptstyle+};
(1.039648,0.649466) *[blue]{\scriptscriptstyle+};
(1.057269,0.661011) *[blue]{\scriptscriptstyle+};
(1.074890,0.693610) *[blue]{\scriptscriptstyle+};
(1.092511,0.696837) *[blue]{\scriptscriptstyle+};
(1.110132,0.623431) *[blue]{\scriptscriptstyle+};
(1.127753,0.679205) *[blue]{\scriptscriptstyle+};
(1.145374,0.689688) *[blue]{\scriptscriptstyle+};
(1.162996,0.689742) *[blue]{\scriptscriptstyle+};
(1.180617,0.697663) *[blue]{\scriptscriptstyle+};
(1.198238,0.699922) *[blue]{\scriptscriptstyle+};
(1.215859,0.733543) *[blue]{\scriptscriptstyle+};
(1.233480,0.635518) *[blue]{\scriptscriptstyle+};
(1.251101,0.671604) *[blue]{\scriptscriptstyle+};
(1.268722,0.689952) *[blue]{\scriptscriptstyle+};
(1.286344,0.702165) *[blue]{\scriptscriptstyle+};
(1.303965,0.717296) *[blue]{\scriptscriptstyle+};
(1.321586,0.720134) *[blue]{\scriptscriptstyle+};
(1.339207,0.738874) *[blue]{\scriptscriptstyle+};
(1.356828,0.581199) *[blue]{\scriptscriptstyle+};
(1.374449,0.613970) *[blue]{\scriptscriptstyle+};
(1.392070,0.621102) *[blue]{\scriptscriptstyle+};
(1.409692,0.624613) *[blue]{\scriptscriptstyle+};
(1.427313,0.628256) *[blue]{\scriptscriptstyle+};
(1.444934,0.691813) *[blue]{\scriptscriptstyle+};
(1.462555,0.692150) *[blue]{\scriptscriptstyle+};
(1.480176,0.660792) *[blue]{\scriptscriptstyle+};
(1.497797,0.664262) *[blue]{\scriptscriptstyle+};
(1.515419,0.664262) *[blue]{\scriptscriptstyle+};
(1.533040,0.667724) *[blue]{\scriptscriptstyle+};
(1.550661,0.667724) *[blue]{\scriptscriptstyle+};
(1.568282,0.682652) *[blue]{\scriptscriptstyle+};
(1.585903,0.695112) *[blue]{\scriptscriptstyle+};
(1.603524,0.652470) *[blue]{\scriptscriptstyle+};
(1.621145,0.655961) *[blue]{\scriptscriptstyle+};
(1.638767,0.660924) *[blue]{\scriptscriptstyle+};
(1.656388,0.662187) *[blue]{\scriptscriptstyle+};
(1.674009,0.687419) *[blue]{\scriptscriptstyle+};
(1.691630,0.689348) *[blue]{\scriptscriptstyle+};
(1.709251,0.690665) *[blue]{\scriptscriptstyle+};
(1.726872,0.667311) *[blue]{\scriptscriptstyle+};
(1.744493,0.675732) *[blue]{\scriptscriptstyle+};
(1.762115,0.676082) *[blue]{\scriptscriptstyle+};
(1.779736,0.677650) *[blue]{\scriptscriptstyle+};
(1.797357,0.709308) *[blue]{\scriptscriptstyle+};
(1.814978,0.711629) *[blue]{\scriptscriptstyle+};
(1.832599,0.717877) *[blue]{\scriptscriptstyle+};
(1.850220,0.694323) *[blue]{\scriptscriptstyle+};
(1.867841,0.696619) *[blue]{\scriptscriptstyle+};
(1.885463,0.696619) *[blue]{\scriptscriptstyle+};
(1.903084,0.722070) *[blue]{\scriptscriptstyle+};
(1.920705,0.726818) *[blue]{\scriptscriptstyle+};
(1.938326,0.737754) *[blue]{\scriptscriptstyle+};
(1.955947,0.737780) *[blue]{\scriptscriptstyle+};
(1.973568,0.689098) *[blue]{\scriptscriptstyle+};
(1.991189,0.689098) *[blue]{\scriptscriptstyle+};
(2.008811,0.689098) *[blue]{\scriptscriptstyle+};
(2.026432,0.694615) *[blue]{\scriptscriptstyle+};
(2.044053,0.702662) *[blue]{\scriptscriptstyle+};
(2.061674,0.726110) *[blue]{\scriptscriptstyle+};
(2.079295,0.732762) *[blue]{\scriptscriptstyle+};
(2.096916,0.686487) *[blue]{\scriptscriptstyle+};
(2.114537,0.686487) *[blue]{\scriptscriptstyle+};
(2.132159,0.687670) *[blue]{\scriptscriptstyle+};
(2.149780,0.693297) *[blue]{\scriptscriptstyle+};
(2.167401,0.695466) *[blue]{\scriptscriptstyle+};
(2.185022,0.702048) *[blue]{\scriptscriptstyle+};
(2.202643,0.766112) *[blue]{\scriptscriptstyle+};
(2.220264,0.665095) *[blue]{\scriptscriptstyle+};
(2.237885,0.696410) *[blue]{\scriptscriptstyle+};
(2.255507,0.700619) *[blue]{\scriptscriptstyle+};
(2.273128,0.700779) *[blue]{\scriptscriptstyle+};
(2.290749,0.726271) *[blue]{\scriptscriptstyle+};
(2.308370,0.726376) *[blue]{\scriptscriptstyle+};
(2.325991,0.729692) *[blue]{\scriptscriptstyle+};
(2.343612,0.690732) *[blue]{\scriptscriptstyle+};
(2.361233,0.693465) *[blue]{\scriptscriptstyle+};
(2.378855,0.725405) *[blue]{\scriptscriptstyle+};
(2.396476,0.728123) *[blue]{\scriptscriptstyle+};
(2.414097,0.735624) *[blue]{\scriptscriptstyle+};
(2.431718,0.754755) *[blue]{\scriptscriptstyle+};
(2.449339,0.761404) *[blue]{\scriptscriptstyle+};
(2.466960,0.681096) *[blue]{\scriptscriptstyle+};
(2.484581,0.681903) *[blue]{\scriptscriptstyle+};
(2.502203,0.682041) *[blue]{\scriptscriptstyle+};
(2.519824,0.707229) *[blue]{\scriptscriptstyle+};
(2.537445,0.709079) *[blue]{\scriptscriptstyle+};
(2.555066,0.711750) *[blue]{\scriptscriptstyle+};
(2.572687,0.757126) *[blue]{\scriptscriptstyle+};
(2.590308,0.666554) *[blue]{\scriptscriptstyle+};
(2.607930,0.692278) *[blue]{\scriptscriptstyle+};
(2.625551,0.697329) *[blue]{\scriptscriptstyle+};
(2.643172,0.700821) *[blue]{\scriptscriptstyle+};
(2.660793,0.702578) *[blue]{\scriptscriptstyle+};
(2.678414,0.726101) *[blue]{\scriptscriptstyle+};
(2.696035,0.767127) *[blue]{\scriptscriptstyle+};
(2.713656,0.658207) *[blue]{\scriptscriptstyle+};
(2.731278,0.667279) *[blue]{\scriptscriptstyle+};
(2.748899,0.690337) *[blue]{\scriptscriptstyle+};
(2.766520,0.723812) *[blue]{\scriptscriptstyle+};
(2.784141,0.723995) *[blue]{\scriptscriptstyle+};
(2.801762,0.727186) *[blue]{\scriptscriptstyle+};
(2.819383,0.937576) *[blue]{\scriptscriptstyle+};
(2.837004,0.719319) *[blue]{\scriptscriptstyle+};
(2.854626,0.719714) *[blue]{\scriptscriptstyle+};
(2.872247,0.719897) *[blue]{\scriptscriptstyle+};
(2.889868,0.721158) *[blue]{\scriptscriptstyle+};
(2.907489,0.755011) *[blue]{\scriptscriptstyle+};
(2.925110,0.778259) *[blue]{\scriptscriptstyle+};
(2.942731,0.785984) *[blue]{\scriptscriptstyle+};
(2.960352,0.706553) *[blue]{\scriptscriptstyle+};
(2.977974,0.707324) *[blue]{\scriptscriptstyle+};
(2.995595,0.720957) *[blue]{\scriptscriptstyle+};
(3.013216,0.770706) *[blue]{\scriptscriptstyle+};
(3.030837,0.801268) *[blue]{\scriptscriptstyle+};
(3.048458,0.829277) *[blue]{\scriptscriptstyle+};
(3.066079,0.835423) *[blue]{\scriptscriptstyle+};
(3.083700,0.736782) *[blue]{\scriptscriptstyle+};
(3.101322,0.739585) *[blue]{\scriptscriptstyle+};
(3.118943,0.742744) *[blue]{\scriptscriptstyle+};
(3.136564,0.746792) *[blue]{\scriptscriptstyle+};
(3.154185,0.756579) *[blue]{\scriptscriptstyle+};
(3.171806,0.782471) *[blue]{\scriptscriptstyle+};
(3.189427,0.783930) *[blue]{\scriptscriptstyle+};
(3.207048,0.705859) *[blue]{\scriptscriptstyle+};
(3.224670,0.739239) *[blue]{\scriptscriptstyle+};
(3.242291,0.747173) *[blue]{\scriptscriptstyle+};
(3.259912,0.750441) *[blue]{\scriptscriptstyle+};
(3.277533,0.771738) *[blue]{\scriptscriptstyle+};
(3.295154,0.781484) *[blue]{\scriptscriptstyle+};
(3.312775,0.843998) *[blue]{\scriptscriptstyle+};
(3.330396,0.734224) *[blue]{\scriptscriptstyle+};
(3.348018,0.746609) *[blue]{\scriptscriptstyle+};
(3.365639,0.751297) *[blue]{\scriptscriptstyle+};
(3.383260,0.752194) *[blue]{\scriptscriptstyle+};
(3.400881,0.754554) *[blue]{\scriptscriptstyle+};
(3.418502,0.778919) *[blue]{\scriptscriptstyle+};
(3.436123,0.799419) *[blue]{\scriptscriptstyle+};
(3.453744,0.675719) *[blue]{\scriptscriptstyle+};
(3.471366,0.682715) *[blue]{\scriptscriptstyle+};
(3.488987,0.691527) *[blue]{\scriptscriptstyle+};
(3.506608,0.706539) *[blue]{\scriptscriptstyle+};
(3.524229,0.709875) *[blue]{\scriptscriptstyle+};
(3.541850,0.745283) *[blue]{\scriptscriptstyle+};
(3.559471,0.748300) *[blue]{\scriptscriptstyle+};
(3.577093,0.735462) *[blue]{\scriptscriptstyle+};
(3.594714,0.737797) *[blue]{\scriptscriptstyle+};
(3.612335,0.769947) *[blue]{\scriptscriptstyle+};
(3.629956,0.770049) *[blue]{\scriptscriptstyle+};
(3.647577,0.771826) *[blue]{\scriptscriptstyle+};
(3.665198,0.775895) *[blue]{\scriptscriptstyle+};
(3.682819,0.843982) *[blue]{\scriptscriptstyle+};
(3.700441,0.749090) *[blue]{\scriptscriptstyle+};
(3.718062,0.752436) *[blue]{\scriptscriptstyle+};
(3.735683,0.757740) *[blue]{\scriptscriptstyle+};
(3.753304,0.778272) *[blue]{\scriptscriptstyle+};
(3.770925,0.782427) *[blue]{\scriptscriptstyle+};
(3.788546,0.832532) *[blue]{\scriptscriptstyle+};
(3.806167,0.841937) *[blue]{\scriptscriptstyle+};
(3.823789,0.716750) *[blue]{\scriptscriptstyle+};
(3.841410,0.743360) *[blue]{\scriptscriptstyle+};
(3.859031,0.746586) *[blue]{\scriptscriptstyle+};
(3.876652,0.746637) *[blue]{\scriptscriptstyle+};
(3.894273,0.775868) *[blue]{\scriptscriptstyle+};
(3.911894,0.789122) *[blue]{\scriptscriptstyle+};
(3.929515,0.807759) *[blue]{\scriptscriptstyle+};
(3.947137,0.725608) *[blue]{\scriptscriptstyle+};
(3.964758,0.758438) *[blue]{\scriptscriptstyle+};
(3.982379,0.763492) *[blue]{\scriptscriptstyle+};
(4.000000,0.768146) *[blue]{\scriptscriptstyle+};
(4.017621,0.790536) *[blue]{\scriptscriptstyle+};
(4.035242,0.790737) *[blue]{\scriptscriptstyle+};
(4.052863,0.796899) *[blue]{\scriptscriptstyle+};
(4.070485,0.720051) *[blue]{\scriptscriptstyle+};
(4.088106,0.726419) *[blue]{\scriptscriptstyle+};
(4.105727,0.755390) *[blue]{\scriptscriptstyle+};
(4.123348,0.788310) *[blue]{\scriptscriptstyle+};
(4.140969,0.788483) *[blue]{\scriptscriptstyle+};
(4.158590,0.794056) *[blue]{\scriptscriptstyle+};
(4.176211,0.843313) *[blue]{\scriptscriptstyle+};
(4.193833,0.741961) *[blue]{\scriptscriptstyle+};
(4.211454,0.751781) *[blue]{\scriptscriptstyle+};
(4.229075,0.771235) *[blue]{\scriptscriptstyle+};
(4.246696,0.774199) *[blue]{\scriptscriptstyle+};
(4.264317,0.784983) *[blue]{\scriptscriptstyle+};
(4.281938,0.809067) *[blue]{\scriptscriptstyle+};
(4.299559,0.903665) *[blue]{\scriptscriptstyle+};
(4.317181,0.720432) *[blue]{\scriptscriptstyle+};
(4.334802,0.723449) *[blue]{\scriptscriptstyle+};
(4.352423,0.723762) *[blue]{\scriptscriptstyle+};
(4.370044,0.753480) *[blue]{\scriptscriptstyle+};
(4.387665,0.759499) *[blue]{\scriptscriptstyle+};
(4.405286,0.759855) *[blue]{\scriptscriptstyle+};
(4.422907,0.763197) *[blue]{\scriptscriptstyle+};
(4.440529,0.718872) *[blue]{\scriptscriptstyle+};
(4.458150,0.733858) *[blue]{\scriptscriptstyle+};
(4.475771,0.751879) *[blue]{\scriptscriptstyle+};
(4.493392,0.751879) *[blue]{\scriptscriptstyle+};
(4.511013,0.772485) *[blue]{\scriptscriptstyle+};
(4.528634,0.777693) *[blue]{\scriptscriptstyle+};
(4.546256,0.786734) *[blue]{\scriptscriptstyle+};
(4.563877,0.740789) *[blue]{\scriptscriptstyle+};
(4.581498,0.740789) *[blue]{\scriptscriptstyle+};
(4.599119,0.740789) *[blue]{\scriptscriptstyle+};
(4.616740,0.776616) *[blue]{\scriptscriptstyle+};
(4.634361,0.810490) *[blue]{\scriptscriptstyle+};
(4.651982,0.902849) *[blue]{\scriptscriptstyle+};
(4.669604,0.914484) *[blue]{\scriptscriptstyle+};
(4.687225,0.767276) *[blue]{\scriptscriptstyle+};
(4.704846,0.773642) *[blue]{\scriptscriptstyle+};
(4.722467,0.795971) *[blue]{\scriptscriptstyle+};
(4.740088,0.795995) *[blue]{\scriptscriptstyle+};
(4.757709,0.797366) *[blue]{\scriptscriptstyle+};
(4.775330,0.806817) *[blue]{\scriptscriptstyle+};
(4.792952,0.854494) *[blue]{\scriptscriptstyle+};
(4.810573,0.766739) *[blue]{\scriptscriptstyle+};
(4.828194,0.766739) *[blue]{\scriptscriptstyle+};
(4.845815,0.766739) *[blue]{\scriptscriptstyle+};
(4.863436,0.769964) *[blue]{\scriptscriptstyle+};
(4.881057,0.773994) *[blue]{\scriptscriptstyle+};
(4.898678,0.782467) *[blue]{\scriptscriptstyle+};
(4.916300,0.804958) *[blue]{\scriptscriptstyle+};
(4.933921,0.759484) *[blue]{\scriptscriptstyle+};
(4.951542,0.760176) *[blue]{\scriptscriptstyle+};
(4.969163,0.765959) *[blue]{\scriptscriptstyle+};
(4.986784,0.785325) *[blue]{\scriptscriptstyle+};
(5.004405,0.789587) *[blue]{\scriptscriptstyle+};
(5.022026,0.790561) *[blue]{\scriptscriptstyle+};
(5.039648,0.790687) *[blue]{\scriptscriptstyle+};
(5.057269,0.754947) *[blue]{\scriptscriptstyle+};
(5.074890,0.756944) *[blue]{\scriptscriptstyle+};
(5.092511,0.756944) *[blue]{\scriptscriptstyle+};
(5.110132,0.758710) *[blue]{\scriptscriptstyle+};
(5.127753,0.785843) *[blue]{\scriptscriptstyle+};
(5.145374,0.795322) *[blue]{\scriptscriptstyle+};
(5.162996,0.864658) *[blue]{\scriptscriptstyle+};
(5.180617,0.760034) *[blue]{\scriptscriptstyle+};
(5.198238,0.763273) *[blue]{\scriptscriptstyle+};
(5.215859,0.787144) *[blue]{\scriptscriptstyle+};
(5.233480,0.790499) *[blue]{\scriptscriptstyle+};
(5.251101,0.821753) *[blue]{\scriptscriptstyle+};
(5.268722,0.821875) *[blue]{\scriptscriptstyle+};
(5.286344,0.864963) *[blue]{\scriptscriptstyle+};
(5.303965,0.741051) *[blue]{\scriptscriptstyle+};
(5.321586,0.789652) *[blue]{\scriptscriptstyle+};
(5.339207,0.814788) *[blue]{\scriptscriptstyle+};
(5.356828,0.819701) *[blue]{\scriptscriptstyle+};
(5.374449,0.824126) *[blue]{\scriptscriptstyle+};
(5.392070,0.833574) *[blue]{\scriptscriptstyle+};
(5.409692,0.875988) *[blue]{\scriptscriptstyle+};
(5.427313,0.723419) *[blue]{\scriptscriptstyle+};
(5.444934,0.756272) *[blue]{\scriptscriptstyle+};
(5.462555,0.759648) *[blue]{\scriptscriptstyle+};
(5.480176,0.759648) *[blue]{\scriptscriptstyle+};
(5.497797,0.762610) *[blue]{\scriptscriptstyle+};
(5.515419,0.769429) *[blue]{\scriptscriptstyle+};
(5.533040,0.881171) *[blue]{\scriptscriptstyle+};
(5.550661,0.763161) *[blue]{\scriptscriptstyle+};
(5.568282,0.776049) *[blue]{\scriptscriptstyle+};
(5.585903,0.788824) *[blue]{\scriptscriptstyle+};
(5.603524,0.798243) *[blue]{\scriptscriptstyle+};
(5.621145,0.798343) *[blue]{\scriptscriptstyle+};
(5.638767,0.806011) *[blue]{\scriptscriptstyle+};
(5.656388,0.844900) *[blue]{\scriptscriptstyle+};
(5.674009,0.791939) *[blue]{\scriptscriptstyle+};
(5.691630,0.820224) *[blue]{\scriptscriptstyle+};
(5.709251,0.820469) *[blue]{\scriptscriptstyle+};
(5.726872,0.821988) *[blue]{\scriptscriptstyle+};
(5.744493,0.828455) *[blue]{\scriptscriptstyle+};
(5.762115,0.884541) *[blue]{\scriptscriptstyle+};
(5.779736,0.885246) *[blue]{\scriptscriptstyle+};
(5.797357,0.771438) *[blue]{\scriptscriptstyle+};
(5.814978,0.773262) *[blue]{\scriptscriptstyle+};
(5.832599,0.773414) *[blue]{\scriptscriptstyle+};
(5.850220,0.783000) *[blue]{\scriptscriptstyle+};
(5.867841,0.809460) *[blue]{\scriptscriptstyle+};
(5.885463,0.843664) *[blue]{\scriptscriptstyle+};
(5.903084,0.872856) *[blue]{\scriptscriptstyle+};
(5.920705,0.748577) *[blue]{\scriptscriptstyle+};
(5.938326,0.755076) *[blue]{\scriptscriptstyle+};
(5.955947,0.778373) *[blue]{\scriptscriptstyle+};
(5.973568,0.780969) *[blue]{\scriptscriptstyle+};
(5.991189,0.787000) *[blue]{\scriptscriptstyle+};
(6.008811,0.791051) *[blue]{\scriptscriptstyle+};
(6.026432,0.809962) *[blue]{\scriptscriptstyle+};
(6.044053,0.781515) *[blue]{\scriptscriptstyle+};
(6.061674,0.783827) *[blue]{\scriptscriptstyle+};
(6.079295,0.802783) *[blue]{\scriptscriptstyle+};
(6.096916,0.806794) *[blue]{\scriptscriptstyle+};
(6.114537,0.813092) *[blue]{\scriptscriptstyle+};
(6.132159,0.814809) *[blue]{\scriptscriptstyle+};
(6.149780,0.865493) *[blue]{\scriptscriptstyle+};
(6.167401,0.773842) *[blue]{\scriptscriptstyle+};
(6.185022,0.797984) *[blue]{\scriptscriptstyle+};
(6.202643,0.798009) *[blue]{\scriptscriptstyle+};
(6.220264,0.810517) *[blue]{\scriptscriptstyle+};
(6.237885,0.816866) *[blue]{\scriptscriptstyle+};
(6.255507,0.829224) *[blue]{\scriptscriptstyle+};
(6.273128,0.890937) *[blue]{\scriptscriptstyle+};
(6.290749,0.736778) *[blue]{\scriptscriptstyle+};
(6.308370,0.769002) *[blue]{\scriptscriptstyle+};
(6.325991,0.780755) *[blue]{\scriptscriptstyle+};
(6.343612,0.800695) *[blue]{\scriptscriptstyle+};
(6.361233,0.802284) *[blue]{\scriptscriptstyle+};
(6.378855,0.802803) *[blue]{\scriptscriptstyle+};
(6.396476,0.835178) *[blue]{\scriptscriptstyle+};
(6.414097,0.780428) *[blue]{\scriptscriptstyle+};
(6.431718,0.813869) *[blue]{\scriptscriptstyle+};
(6.449339,0.815196) *[blue]{\scriptscriptstyle+};
(6.466960,0.820473) *[blue]{\scriptscriptstyle+};
(6.484581,0.824556) *[blue]{\scriptscriptstyle+};
(6.502203,0.832888) *[blue]{\scriptscriptstyle+};
(6.519824,0.853946) *[blue]{\scriptscriptstyle+};
(6.537445,0.800885) *[blue]{\scriptscriptstyle+};
(6.555066,0.808465) *[blue]{\scriptscriptstyle+};
(6.572687,0.826831) *[blue]{\scriptscriptstyle+};
(6.590308,0.836457) *[blue]{\scriptscriptstyle+};
(6.607930,0.841339) *[blue]{\scriptscriptstyle+};
(6.625551,0.886679) *[blue]{\scriptscriptstyle+};
(6.643172,0.926650) *[blue]{\scriptscriptstyle+};
(6.660793,0.793671) *[blue]{\scriptscriptstyle+};
(6.678414,0.800370) *[blue]{\scriptscriptstyle+};
(6.696035,0.828175) *[blue]{\scriptscriptstyle+};
(6.713656,0.832289) *[blue]{\scriptscriptstyle+};
(6.731278,0.835492) *[blue]{\scriptscriptstyle+};
(6.748899,0.855910) *[blue]{\scriptscriptstyle+};
(6.766520,0.870864) *[blue]{\scriptscriptstyle+};
(6.784141,0.804576) *[blue]{\scriptscriptstyle+};
(6.801762,0.807225) *[blue]{\scriptscriptstyle+};
(6.819383,0.808165) *[blue]{\scriptscriptstyle+};
(6.837004,0.817101) *[blue]{\scriptscriptstyle+};
(6.854626,0.872181) *[blue]{\scriptscriptstyle+};
(6.872247,0.872299) *[blue]{\scriptscriptstyle+};
(6.889868,0.930828) *[blue]{\scriptscriptstyle+};
(6.907489,0.777249) *[blue]{\scriptscriptstyle+};
(6.925110,0.802578) *[blue]{\scriptscriptstyle+};
(6.942731,0.805798) *[blue]{\scriptscriptstyle+};
(6.960352,0.808962) *[blue]{\scriptscriptstyle+};
(6.977974,0.850851) *[blue]{\scriptscriptstyle+};
(6.995595,0.858478) *[blue]{\scriptscriptstyle+};
(7.013216,0.969451) *[blue]{\scriptscriptstyle+};
(7.030837,0.867875) *[blue]{\scriptscriptstyle+};
(7.048458,0.900575) *[blue]{\scriptscriptstyle+};
(7.066079,0.904789) *[blue]{\scriptscriptstyle+};
(7.083700,0.923845) *[blue]{\scriptscriptstyle+};
(7.101322,0.926965) *[blue]{\scriptscriptstyle+};
(7.118943,0.932235) *[blue]{\scriptscriptstyle+};
(7.136564,0.972634) *[blue]{\scriptscriptstyle+};
(7.154185,0.839494) *[blue]{\scriptscriptstyle+};
(7.171806,0.842392) *[blue]{\scriptscriptstyle+};
(7.189427,0.868722) *[blue]{\scriptscriptstyle+};
(7.207048,0.872884) *[blue]{\scriptscriptstyle+};
(7.224670,0.874608) *[blue]{\scriptscriptstyle+};
(7.242291,0.878776) *[blue]{\scriptscriptstyle+};
(7.259912,0.953379) *[blue]{\scriptscriptstyle+};
(7.277533,0.825159) *[blue]{\scriptscriptstyle+};
(7.295154,0.829086) *[blue]{\scriptscriptstyle+};
(7.312775,0.830789) *[blue]{\scriptscriptstyle+};
(7.330396,0.831395) *[blue]{\scriptscriptstyle+};
(7.348018,0.833706) *[blue]{\scriptscriptstyle+};
(7.365639,0.859310) *[blue]{\scriptscriptstyle+};
(7.383260,0.876374) *[blue]{\scriptscriptstyle+};
(7.400881,0.859035) *[blue]{\scriptscriptstyle+};
(7.418502,0.862061) *[blue]{\scriptscriptstyle+};
(7.436123,0.862061) *[blue]{\scriptscriptstyle+};
(7.453744,0.901834) *[blue]{\scriptscriptstyle+};
(7.471366,0.906683) *[blue]{\scriptscriptstyle+};
(7.488987,0.930336) *[blue]{\scriptscriptstyle+};
(7.506608,0.935886) *[blue]{\scriptscriptstyle+};
(7.524229,0.849361) *[blue]{\scriptscriptstyle+};
(7.541850,0.871812) *[blue]{\scriptscriptstyle+};
(7.559471,0.876484) *[blue]{\scriptscriptstyle+};
(7.577093,0.876625) *[blue]{\scriptscriptstyle+};
(7.594714,0.876625) *[blue]{\scriptscriptstyle+};
(7.612335,0.877477) *[blue]{\scriptscriptstyle+};
(7.629956,0.915037) *[blue]{\scriptscriptstyle+};
(7.647577,0.819345) *[blue]{\scriptscriptstyle+};
(7.665198,0.855954) *[blue]{\scriptscriptstyle+};
(7.682819,0.859220) *[blue]{\scriptscriptstyle+};
(7.700441,0.883208) *[blue]{\scriptscriptstyle+};
(7.718062,0.886207) *[blue]{\scriptscriptstyle+};
(7.735683,0.921061) *[blue]{\scriptscriptstyle+};
(7.753304,0.945829) *[blue]{\scriptscriptstyle+};
(7.770925,0.817572) *[blue]{\scriptscriptstyle+};
(7.788546,0.825725) *[blue]{\scriptscriptstyle+};
(7.806167,0.843461) *[blue]{\scriptscriptstyle+};
(7.823789,0.847388) *[blue]{\scriptscriptstyle+};
(7.841410,0.874805) *[blue]{\scriptscriptstyle+};
(7.859031,0.935575) *[blue]{\scriptscriptstyle+};
(7.876652,1.001536) *[blue]{\scriptscriptstyle+};
(7.894273,0.865814) *[blue]{\scriptscriptstyle+};
(7.911894,0.868896) *[blue]{\scriptscriptstyle+};
(7.929515,0.868900) *[blue]{\scriptscriptstyle+};
(7.947137,0.873463) *[blue]{\scriptscriptstyle+};
(7.964758,0.896936) *[blue]{\scriptscriptstyle+};
(7.982379,0.922046) *[blue]{\scriptscriptstyle+};
(8.000000,0.950849) *[blue]{\scriptscriptstyle+};
(0.000000,0.316386) *[red]{\scriptscriptstyle\times};
(0.017621,0.316386) *[red]{\scriptscriptstyle\times};
(0.035242,0.316386) *[red]{\scriptscriptstyle\times};
(0.052863,0.326805) *[red]{\scriptscriptstyle\times};
(0.070485,0.331972) *[red]{\scriptscriptstyle\times};
(0.088106,0.367150) *[red]{\scriptscriptstyle\times};
(0.105727,0.386008) *[red]{\scriptscriptstyle\times};
(0.123348,0.264203) *[red]{\scriptscriptstyle\times};
(0.140969,0.266688) *[red]{\scriptscriptstyle\times};
(0.158590,0.282351) *[red]{\scriptscriptstyle\times};
(0.176211,0.310262) *[red]{\scriptscriptstyle\times};
(0.193833,0.313539) *[red]{\scriptscriptstyle\times};
(0.211454,0.358408) *[red]{\scriptscriptstyle\times};
(0.229075,0.444500) *[red]{\scriptscriptstyle\times};
(0.246696,0.271793) *[red]{\scriptscriptstyle\times};
(0.264317,0.282039) *[red]{\scriptscriptstyle\times};
(0.281938,0.310762) *[red]{\scriptscriptstyle\times};
(0.299559,0.321114) *[red]{\scriptscriptstyle\times};
(0.317181,0.354737) *[red]{\scriptscriptstyle\times};
(0.334802,0.355952) *[red]{\scriptscriptstyle\times};
(0.352423,0.420992) *[red]{\scriptscriptstyle\times};
(0.370044,0.298347) *[red]{\scriptscriptstyle\times};
(0.387665,0.309557) *[red]{\scriptscriptstyle\times};
(0.405286,0.311061) *[red]{\scriptscriptstyle\times};
(0.422907,0.350593) *[red]{\scriptscriptstyle\times};
(0.440529,0.358911) *[red]{\scriptscriptstyle\times};
(0.458150,0.391702) *[red]{\scriptscriptstyle\times};
(0.475771,0.424766) *[red]{\scriptscriptstyle\times};
(0.493392,0.307411) *[red]{\scriptscriptstyle\times};
(0.511013,0.308600) *[red]{\scriptscriptstyle\times};
(0.528634,0.311047) *[red]{\scriptscriptstyle\times};
(0.546256,0.359735) *[red]{\scriptscriptstyle\times};
(0.563877,0.390764) *[red]{\scriptscriptstyle\times};
(0.581498,0.398531) *[red]{\scriptscriptstyle\times};
(0.599119,0.411537) *[red]{\scriptscriptstyle\times};
(0.616740,0.306732) *[red]{\scriptscriptstyle\times};
(0.634361,0.355233) *[red]{\scriptscriptstyle\times};
(0.651982,0.373211) *[red]{\scriptscriptstyle\times};
(0.669604,0.397663) *[red]{\scriptscriptstyle\times};
(0.687225,0.400221) *[red]{\scriptscriptstyle\times};
(0.704846,0.401333) *[red]{\scriptscriptstyle\times};
(0.722467,0.438463) *[red]{\scriptscriptstyle\times};
(0.740088,0.338787) *[red]{\scriptscriptstyle\times};
(0.757709,0.355887) *[red]{\scriptscriptstyle\times};
(0.775330,0.368640) *[red]{\scriptscriptstyle\times};
(0.792952,0.390137) *[red]{\scriptscriptstyle\times};
(0.810573,0.400267) *[red]{\scriptscriptstyle\times};
(0.828194,0.401090) *[red]{\scriptscriptstyle\times};
(0.845815,0.423269) *[red]{\scriptscriptstyle\times};
(0.863436,0.368157) *[red]{\scriptscriptstyle\times};
(0.881057,0.371969) *[red]{\scriptscriptstyle\times};
(0.898678,0.405614) *[red]{\scriptscriptstyle\times};
(0.916300,0.434353) *[red]{\scriptscriptstyle\times};
(0.933921,0.446576) *[red]{\scriptscriptstyle\times};
(0.951542,0.450826) *[red]{\scriptscriptstyle\times};
(0.969163,0.478405) *[red]{\scriptscriptstyle\times};
(0.986784,0.414206) *[red]{\scriptscriptstyle\times};
(1.004405,0.422427) *[red]{\scriptscriptstyle\times};
(1.022026,0.438777) *[red]{\scriptscriptstyle\times};
(1.039648,0.441393) *[red]{\scriptscriptstyle\times};
(1.057269,0.445592) *[red]{\scriptscriptstyle\times};
(1.074890,0.483101) *[red]{\scriptscriptstyle\times};
(1.092511,0.490660) *[red]{\scriptscriptstyle\times};
(1.110132,0.444365) *[red]{\scriptscriptstyle\times};
(1.127753,0.448301) *[red]{\scriptscriptstyle\times};
(1.145374,0.468291) *[red]{\scriptscriptstyle\times};
(1.162996,0.484350) *[red]{\scriptscriptstyle\times};
(1.180617,0.491107) *[red]{\scriptscriptstyle\times};
(1.198238,0.492151) *[red]{\scriptscriptstyle\times};
(1.215859,0.492307) *[red]{\scriptscriptstyle\times};
(1.233480,0.408777) *[red]{\scriptscriptstyle\times};
(1.251101,0.410048) *[red]{\scriptscriptstyle\times};
(1.268722,0.410048) *[red]{\scriptscriptstyle\times};
(1.286344,0.415056) *[red]{\scriptscriptstyle\times};
(1.303965,0.419658) *[red]{\scriptscriptstyle\times};
(1.321586,0.461542) *[red]{\scriptscriptstyle\times};
(1.339207,0.502252) *[red]{\scriptscriptstyle\times};
(1.356828,0.408144) *[red]{\scriptscriptstyle\times};
(1.374449,0.416594) *[red]{\scriptscriptstyle\times};
(1.392070,0.425481) *[red]{\scriptscriptstyle\times};
(1.409692,0.451570) *[red]{\scriptscriptstyle\times};
(1.427313,0.465261) *[red]{\scriptscriptstyle\times};
(1.444934,0.496335) *[red]{\scriptscriptstyle\times};
(1.462555,0.576370) *[red]{\scriptscriptstyle\times};
(1.480176,0.429742) *[red]{\scriptscriptstyle\times};
(1.497797,0.435607) *[red]{\scriptscriptstyle\times};
(1.515419,0.442591) *[red]{\scriptscriptstyle\times};
(1.533040,0.474459) *[red]{\scriptscriptstyle\times};
(1.550661,0.474522) *[red]{\scriptscriptstyle\times};
(1.568282,0.485254) *[red]{\scriptscriptstyle\times};
(1.585903,0.505944) *[red]{\scriptscriptstyle\times};
(1.603524,0.432342) *[red]{\scriptscriptstyle\times};
(1.621145,0.439376) *[red]{\scriptscriptstyle\times};
(1.638767,0.470719) *[red]{\scriptscriptstyle\times};
(1.656388,0.470813) *[red]{\scriptscriptstyle\times};
(1.674009,0.473434) *[red]{\scriptscriptstyle\times};
(1.691630,0.486457) *[red]{\scriptscriptstyle\times};
(1.709251,0.516809) *[red]{\scriptscriptstyle\times};
(1.726872,0.438626) *[red]{\scriptscriptstyle\times};
(1.744493,0.443726) *[red]{\scriptscriptstyle\times};
(1.762115,0.456924) *[red]{\scriptscriptstyle\times};
(1.779736,0.478519) *[red]{\scriptscriptstyle\times};
(1.797357,0.496843) *[red]{\scriptscriptstyle\times};
(1.814978,0.500853) *[red]{\scriptscriptstyle\times};
(1.832599,0.508048) *[red]{\scriptscriptstyle\times};
(1.850220,0.475676) *[red]{\scriptscriptstyle\times};
(1.867841,0.481113) *[red]{\scriptscriptstyle\times};
(1.885463,0.514211) *[red]{\scriptscriptstyle\times};
(1.903084,0.515844) *[red]{\scriptscriptstyle\times};
(1.920705,0.515874) *[red]{\scriptscriptstyle\times};
(1.938326,0.522302) *[red]{\scriptscriptstyle\times};
(1.955947,0.526091) *[red]{\scriptscriptstyle\times};
(1.973568,0.473252) *[red]{\scriptscriptstyle\times};
(1.991189,0.475116) *[red]{\scriptscriptstyle\times};
(2.008811,0.482998) *[red]{\scriptscriptstyle\times};
(2.026432,0.510296) *[red]{\scriptscriptstyle\times};
(2.044053,0.512664) *[red]{\scriptscriptstyle\times};
(2.061674,0.525550) *[red]{\scriptscriptstyle\times};
(2.079295,0.555014) *[red]{\scriptscriptstyle\times};
(2.096916,0.480048) *[red]{\scriptscriptstyle\times};
(2.114537,0.488212) *[red]{\scriptscriptstyle\times};
(2.132159,0.499912) *[red]{\scriptscriptstyle\times};
(2.149780,0.526565) *[red]{\scriptscriptstyle\times};
(2.167401,0.536677) *[red]{\scriptscriptstyle\times};
(2.185022,0.559235) *[red]{\scriptscriptstyle\times};
(2.202643,0.559235) *[red]{\scriptscriptstyle\times};
(2.220264,0.445189) *[red]{\scriptscriptstyle\times};
(2.237885,0.473080) *[red]{\scriptscriptstyle\times};
(2.255507,0.484997) *[red]{\scriptscriptstyle\times};
(2.273128,0.512470) *[red]{\scriptscriptstyle\times};
(2.290749,0.525335) *[red]{\scriptscriptstyle\times};
(2.308370,0.566815) *[red]{\scriptscriptstyle\times};
(2.325991,0.587557) *[red]{\scriptscriptstyle\times};
(2.343612,0.479590) *[red]{\scriptscriptstyle\times};
(2.361233,0.518442) *[red]{\scriptscriptstyle\times};
(2.378855,0.524653) *[red]{\scriptscriptstyle\times};
(2.396476,0.529903) *[red]{\scriptscriptstyle\times};
(2.414097,0.542070) *[red]{\scriptscriptstyle\times};
(2.431718,0.563785) *[red]{\scriptscriptstyle\times};
(2.449339,0.581264) *[red]{\scriptscriptstyle\times};
(2.466960,0.501724) *[red]{\scriptscriptstyle\times};
(2.484581,0.501910) *[red]{\scriptscriptstyle\times};
(2.502203,0.505388) *[red]{\scriptscriptstyle\times};
(2.519824,0.513377) *[red]{\scriptscriptstyle\times};
(2.537445,0.518371) *[red]{\scriptscriptstyle\times};
(2.555066,0.547677) *[red]{\scriptscriptstyle\times};
(2.572687,0.594021) *[red]{\scriptscriptstyle\times};
(2.590308,0.502975) *[red]{\scriptscriptstyle\times};
(2.607930,0.526212) *[red]{\scriptscriptstyle\times};
(2.625551,0.528041) *[red]{\scriptscriptstyle\times};
(2.643172,0.533224) *[red]{\scriptscriptstyle\times};
(2.660793,0.539335) *[red]{\scriptscriptstyle\times};
(2.678414,0.598920) *[red]{\scriptscriptstyle\times};
(2.696035,0.612933) *[red]{\scriptscriptstyle\times};
(2.713656,0.389746) *[red]{\scriptscriptstyle\times};
(2.731278,0.428716) *[red]{\scriptscriptstyle\times};
(2.748899,0.435159) *[red]{\scriptscriptstyle\times};
(2.766520,0.506863) *[red]{\scriptscriptstyle\times};
(2.784141,0.517356) *[red]{\scriptscriptstyle\times};
(2.801762,0.518140) *[red]{\scriptscriptstyle\times};
(2.819383,0.625869) *[red]{\scriptscriptstyle\times};
(2.837004,0.510408) *[red]{\scriptscriptstyle\times};
(2.854626,0.513807) *[red]{\scriptscriptstyle\times};
(2.872247,0.528544) *[red]{\scriptscriptstyle\times};
(2.889868,0.548360) *[red]{\scriptscriptstyle\times};
(2.907489,0.548419) *[red]{\scriptscriptstyle\times};
(2.925110,0.559614) *[red]{\scriptscriptstyle\times};
(2.942731,0.563983) *[red]{\scriptscriptstyle\times};
(2.960352,0.513630) *[red]{\scriptscriptstyle\times};
(2.977974,0.513721) *[red]{\scriptscriptstyle\times};
(2.995595,0.528939) *[red]{\scriptscriptstyle\times};
(3.013216,0.529089) *[red]{\scriptscriptstyle\times};
(3.030837,0.540241) *[red]{\scriptscriptstyle\times};
(3.048458,0.567095) *[red]{\scriptscriptstyle\times};
(3.066079,0.581421) *[red]{\scriptscriptstyle\times};
(3.083700,0.479034) *[red]{\scriptscriptstyle\times};
(3.101322,0.480834) *[red]{\scriptscriptstyle\times};
(3.118943,0.519773) *[red]{\scriptscriptstyle\times};
(3.136564,0.556042) *[red]{\scriptscriptstyle\times};
(3.154185,0.557601) *[red]{\scriptscriptstyle\times};
(3.171806,0.561035) *[red]{\scriptscriptstyle\times};
(3.189427,0.563495) *[red]{\scriptscriptstyle\times};
(3.207048,0.482683) *[red]{\scriptscriptstyle\times};
(3.224670,0.521393) *[red]{\scriptscriptstyle\times};
(3.242291,0.526777) *[red]{\scriptscriptstyle\times};
(3.259912,0.529116) *[red]{\scriptscriptstyle\times};
(3.277533,0.563165) *[red]{\scriptscriptstyle\times};
(3.295154,0.577040) *[red]{\scriptscriptstyle\times};
(3.312775,0.599887) *[red]{\scriptscriptstyle\times};
(3.330396,0.447672) *[red]{\scriptscriptstyle\times};
(3.348018,0.450591) *[red]{\scriptscriptstyle\times};
(3.365639,0.487319) *[red]{\scriptscriptstyle\times};
(3.383260,0.487381) *[red]{\scriptscriptstyle\times};
(3.400881,0.495258) *[red]{\scriptscriptstyle\times};
(3.418502,0.515733) *[red]{\scriptscriptstyle\times};
(3.436123,0.568808) *[red]{\scriptscriptstyle\times};
(3.453744,0.452993) *[red]{\scriptscriptstyle\times};
(3.471366,0.457158) *[red]{\scriptscriptstyle\times};
(3.488987,0.460965) *[red]{\scriptscriptstyle\times};
(3.506608,0.496473) *[red]{\scriptscriptstyle\times};
(3.524229,0.505053) *[red]{\scriptscriptstyle\times};
(3.541850,0.620152) *[red]{\scriptscriptstyle\times};
(3.559471,0.634261) *[red]{\scriptscriptstyle\times};
(3.577093,0.526131) *[red]{\scriptscriptstyle\times};
(3.594714,0.533740) *[red]{\scriptscriptstyle\times};
(3.612335,0.537530) *[red]{\scriptscriptstyle\times};
(3.629956,0.565551) *[red]{\scriptscriptstyle\times};
(3.647577,0.571238) *[red]{\scriptscriptstyle\times};
(3.665198,0.575946) *[red]{\scriptscriptstyle\times};
(3.682819,0.579805) *[red]{\scriptscriptstyle\times};
(3.700441,0.519786) *[red]{\scriptscriptstyle\times};
(3.718062,0.522468) *[red]{\scriptscriptstyle\times};
(3.735683,0.529289) *[red]{\scriptscriptstyle\times};
(3.753304,0.541113) *[red]{\scriptscriptstyle\times};
(3.770925,0.556466) *[red]{\scriptscriptstyle\times};
(3.788546,0.565088) *[red]{\scriptscriptstyle\times};
(3.806167,0.565673) *[red]{\scriptscriptstyle\times};
(3.823789,0.484759) *[red]{\scriptscriptstyle\times};
(3.841410,0.523625) *[red]{\scriptscriptstyle\times};
(3.859031,0.529061) *[red]{\scriptscriptstyle\times};
(3.876652,0.531037) *[red]{\scriptscriptstyle\times};
(3.894273,0.563024) *[red]{\scriptscriptstyle\times};
(3.911894,0.571839) *[red]{\scriptscriptstyle\times};
(3.929515,0.603454) *[red]{\scriptscriptstyle\times};
(3.947137,0.495956) *[red]{\scriptscriptstyle\times};
(3.964758,0.503725) *[red]{\scriptscriptstyle\times};
(3.982379,0.503725) *[red]{\scriptscriptstyle\times};
(4.000000,0.556368) *[red]{\scriptscriptstyle\times};
(4.017621,0.579170) *[red]{\scriptscriptstyle\times};
(4.035242,0.579402) *[red]{\scriptscriptstyle\times};
(4.052863,0.599725) *[red]{\scriptscriptstyle\times};
(4.070485,0.505944) *[red]{\scriptscriptstyle\times};
(4.088106,0.514721) *[red]{\scriptscriptstyle\times};
(4.105727,0.544161) *[red]{\scriptscriptstyle\times};
(4.123348,0.551705) *[red]{\scriptscriptstyle\times};
(4.140969,0.556746) *[red]{\scriptscriptstyle\times};
(4.158590,0.584830) *[red]{\scriptscriptstyle\times};
(4.176211,0.622052) *[red]{\scriptscriptstyle\times};
(4.193833,0.530193) *[red]{\scriptscriptstyle\times};
(4.211454,0.534385) *[red]{\scriptscriptstyle\times};
(4.229075,0.564100) *[red]{\scriptscriptstyle\times};
(4.246696,0.564129) *[red]{\scriptscriptstyle\times};
(4.264317,0.574915) *[red]{\scriptscriptstyle\times};
(4.281938,0.600912) *[red]{\scriptscriptstyle\times};
(4.299559,0.601941) *[red]{\scriptscriptstyle\times};
(4.317181,0.502972) *[red]{\scriptscriptstyle\times};
(4.334802,0.537368) *[red]{\scriptscriptstyle\times};
(4.352423,0.537547) *[red]{\scriptscriptstyle\times};
(4.370044,0.541118) *[red]{\scriptscriptstyle\times};
(4.387665,0.549920) *[red]{\scriptscriptstyle\times};
(4.405286,0.590600) *[red]{\scriptscriptstyle\times};
(4.422907,0.629273) *[red]{\scriptscriptstyle\times};
(4.440529,0.528551) *[red]{\scriptscriptstyle\times};
(4.458150,0.532324) *[red]{\scriptscriptstyle\times};
(4.475771,0.536297) *[red]{\scriptscriptstyle\times};
(4.493392,0.539901) *[red]{\scriptscriptstyle\times};
(4.511013,0.561091) *[red]{\scriptscriptstyle\times};
(4.528634,0.565544) *[red]{\scriptscriptstyle\times};
(4.546256,0.631060) *[red]{\scriptscriptstyle\times};
(4.563877,0.517445) *[red]{\scriptscriptstyle\times};
(4.581498,0.555451) *[red]{\scriptscriptstyle\times};
(4.599119,0.559007) *[red]{\scriptscriptstyle\times};
(4.616740,0.559095) *[red]{\scriptscriptstyle\times};
(4.634361,0.568577) *[red]{\scriptscriptstyle\times};
(4.651982,0.606783) *[red]{\scriptscriptstyle\times};
(4.669604,0.610355) *[red]{\scriptscriptstyle\times};
(4.687225,0.532998) *[red]{\scriptscriptstyle\times};
(4.704846,0.534492) *[red]{\scriptscriptstyle\times};
(4.722467,0.540632) *[red]{\scriptscriptstyle\times};
(4.740088,0.542057) *[red]{\scriptscriptstyle\times};
(4.757709,0.574211) *[red]{\scriptscriptstyle\times};
(4.775330,0.577835) *[red]{\scriptscriptstyle\times};
(4.792952,0.586275) *[red]{\scriptscriptstyle\times};
(4.810573,0.531396) *[red]{\scriptscriptstyle\times};
(4.828194,0.538978) *[red]{\scriptscriptstyle\times};
(4.845815,0.550275) *[red]{\scriptscriptstyle\times};
(4.863436,0.572536) *[red]{\scriptscriptstyle\times};
(4.881057,0.574661) *[red]{\scriptscriptstyle\times};
(4.898678,0.583719) *[red]{\scriptscriptstyle\times};
(4.916300,0.588735) *[red]{\scriptscriptstyle\times};
(4.933921,0.505108) *[red]{\scriptscriptstyle\times};
(4.951542,0.528386) *[red]{\scriptscriptstyle\times};
(4.969163,0.565668) *[red]{\scriptscriptstyle\times};
(4.986784,0.576919) *[red]{\scriptscriptstyle\times};
(5.004405,0.578456) *[red]{\scriptscriptstyle\times};
(5.022026,0.584121) *[red]{\scriptscriptstyle\times};
(5.039648,0.602383) *[red]{\scriptscriptstyle\times};
(5.057269,0.532661) *[red]{\scriptscriptstyle\times};
(5.074890,0.537291) *[red]{\scriptscriptstyle\times};
(5.092511,0.549812) *[red]{\scriptscriptstyle\times};
(5.110132,0.570128) *[red]{\scriptscriptstyle\times};
(5.127753,0.573794) *[red]{\scriptscriptstyle\times};
(5.145374,0.575046) *[red]{\scriptscriptstyle\times};
(5.162996,0.610195) *[red]{\scriptscriptstyle\times};
(5.180617,0.524781) *[red]{\scriptscriptstyle\times};
(5.198238,0.534044) *[red]{\scriptscriptstyle\times};
(5.215859,0.564732) *[red]{\scriptscriptstyle\times};
(5.233480,0.569523) *[red]{\scriptscriptstyle\times};
(5.251101,0.597789) *[red]{\scriptscriptstyle\times};
(5.268722,0.608608) *[red]{\scriptscriptstyle\times};
(5.286344,0.612178) *[red]{\scriptscriptstyle\times};
(5.303965,0.505749) *[red]{\scriptscriptstyle\times};
(5.321586,0.543852) *[red]{\scriptscriptstyle\times};
(5.339207,0.546373) *[red]{\scriptscriptstyle\times};
(5.356828,0.558934) *[red]{\scriptscriptstyle\times};
(5.374449,0.581236) *[red]{\scriptscriptstyle\times};
(5.392070,0.628319) *[red]{\scriptscriptstyle\times};
(5.409692,0.634716) *[red]{\scriptscriptstyle\times};
(5.427313,0.541235) *[red]{\scriptscriptstyle\times};
(5.444934,0.571451) *[red]{\scriptscriptstyle\times};
(5.462555,0.576626) *[red]{\scriptscriptstyle\times};
(5.480176,0.593292) *[red]{\scriptscriptstyle\times};
(5.497797,0.610575) *[red]{\scriptscriptstyle\times};
(5.515419,0.621949) *[red]{\scriptscriptstyle\times};
(5.533040,0.657903) *[red]{\scriptscriptstyle\times};
(5.550661,0.544987) *[red]{\scriptscriptstyle\times};
(5.568282,0.545165) *[red]{\scriptscriptstyle\times};
(5.585903,0.548481) *[red]{\scriptscriptstyle\times};
(5.603524,0.593246) *[red]{\scriptscriptstyle\times};
(5.621145,0.622042) *[red]{\scriptscriptstyle\times};
(5.638767,0.622099) *[red]{\scriptscriptstyle\times};
(5.656388,0.629079) *[red]{\scriptscriptstyle\times};
(5.674009,0.527033) *[red]{\scriptscriptstyle\times};
(5.691630,0.532280) *[red]{\scriptscriptstyle\times};
(5.709251,0.568412) *[red]{\scriptscriptstyle\times};
(5.726872,0.572198) *[red]{\scriptscriptstyle\times};
(5.744493,0.602925) *[red]{\scriptscriptstyle\times};
(5.762115,0.602953) *[red]{\scriptscriptstyle\times};
(5.779736,0.612263) *[red]{\scriptscriptstyle\times};
(5.797357,0.531434) *[red]{\scriptscriptstyle\times};
(5.814978,0.533648) *[red]{\scriptscriptstyle\times};
(5.832599,0.536008) *[red]{\scriptscriptstyle\times};
(5.850220,0.572776) *[red]{\scriptscriptstyle\times};
(5.867841,0.585833) *[red]{\scriptscriptstyle\times};
(5.885463,0.612909) *[red]{\scriptscriptstyle\times};
(5.903084,0.644975) *[red]{\scriptscriptstyle\times};
(5.920705,0.525545) *[red]{\scriptscriptstyle\times};
(5.938326,0.529029) *[red]{\scriptscriptstyle\times};
(5.955947,0.529356) *[red]{\scriptscriptstyle\times};
(5.973568,0.563312) *[red]{\scriptscriptstyle\times};
(5.991189,0.570118) *[red]{\scriptscriptstyle\times};
(6.008811,0.617700) *[red]{\scriptscriptstyle\times};
(6.026432,0.671019) *[red]{\scriptscriptstyle\times};
(6.044053,0.549775) *[red]{\scriptscriptstyle\times};
(6.061674,0.555733) *[red]{\scriptscriptstyle\times};
(6.079295,0.556295) *[red]{\scriptscriptstyle\times};
(6.096916,0.557261) *[red]{\scriptscriptstyle\times};
(6.114537,0.563190) *[red]{\scriptscriptstyle\times};
(6.132159,0.595048) *[red]{\scriptscriptstyle\times};
(6.149780,0.635803) *[red]{\scriptscriptstyle\times};
(6.167401,0.553459) *[red]{\scriptscriptstyle\times};
(6.185022,0.578064) *[red]{\scriptscriptstyle\times};
(6.202643,0.598620) *[red]{\scriptscriptstyle\times};
(6.220264,0.601300) *[red]{\scriptscriptstyle\times};
(6.237885,0.628736) *[red]{\scriptscriptstyle\times};
(6.255507,0.632287) *[red]{\scriptscriptstyle\times};
(6.273128,0.645257) *[red]{\scriptscriptstyle\times};
(6.290749,0.544453) *[red]{\scriptscriptstyle\times};
(6.308370,0.548185) *[red]{\scriptscriptstyle\times};
(6.325991,0.553297) *[red]{\scriptscriptstyle\times};
(6.343612,0.555004) *[red]{\scriptscriptstyle\times};
(6.361233,0.591646) *[red]{\scriptscriptstyle\times};
(6.378855,0.596945) *[red]{\scriptscriptstyle\times};
(6.396476,0.690942) *[red]{\scriptscriptstyle\times};
(6.414097,0.568419) *[red]{\scriptscriptstyle\times};
(6.431718,0.568915) *[red]{\scriptscriptstyle\times};
(6.449339,0.574617) *[red]{\scriptscriptstyle\times};
(6.466960,0.602416) *[red]{\scriptscriptstyle\times};
(6.484581,0.607136) *[red]{\scriptscriptstyle\times};
(6.502203,0.613230) *[red]{\scriptscriptstyle\times};
(6.519824,0.644791) *[red]{\scriptscriptstyle\times};
(6.537445,0.570740) *[red]{\scriptscriptstyle\times};
(6.555066,0.608407) *[red]{\scriptscriptstyle\times};
(6.572687,0.609854) *[red]{\scriptscriptstyle\times};
(6.590308,0.614437) *[red]{\scriptscriptstyle\times};
(6.607930,0.622443) *[red]{\scriptscriptstyle\times};
(6.625551,0.652124) *[red]{\scriptscriptstyle\times};
(6.643172,0.653065) *[red]{\scriptscriptstyle\times};
(6.660793,0.574511) *[red]{\scriptscriptstyle\times};
(6.678414,0.579178) *[red]{\scriptscriptstyle\times};
(6.696035,0.643969) *[red]{\scriptscriptstyle\times};
(6.713656,0.643997) *[red]{\scriptscriptstyle\times};
(6.731278,0.647480) *[red]{\scriptscriptstyle\times};
(6.748899,0.686182) *[red]{\scriptscriptstyle\times};
(6.766520,0.697827) *[red]{\scriptscriptstyle\times};
(6.784141,0.570900) *[red]{\scriptscriptstyle\times};
(6.801762,0.607484) *[red]{\scriptscriptstyle\times};
(6.819383,0.621693) *[red]{\scriptscriptstyle\times};
(6.837004,0.624480) *[red]{\scriptscriptstyle\times};
(6.854626,0.677765) *[red]{\scriptscriptstyle\times};
(6.872247,0.687847) *[red]{\scriptscriptstyle\times};
(6.889868,0.691597) *[red]{\scriptscriptstyle\times};
(6.907489,0.595809) *[red]{\scriptscriptstyle\times};
(6.925110,0.595952) *[red]{\scriptscriptstyle\times};
(6.942731,0.599496) *[red]{\scriptscriptstyle\times};
(6.960352,0.608831) *[red]{\scriptscriptstyle\times};
(6.977974,0.609093) *[red]{\scriptscriptstyle\times};
(6.995595,0.635511) *[red]{\scriptscriptstyle\times};
(7.013216,0.734664) *[red]{\scriptscriptstyle\times};
(7.030837,0.637476) *[red]{\scriptscriptstyle\times};
(7.048458,0.639170) *[red]{\scriptscriptstyle\times};
(7.066079,0.649712) *[red]{\scriptscriptstyle\times};
(7.083700,0.670500) *[red]{\scriptscriptstyle\times};
(7.101322,0.670582) *[red]{\scriptscriptstyle\times};
(7.118943,0.673543) *[red]{\scriptscriptstyle\times};
(7.136564,0.686662) *[red]{\scriptscriptstyle\times};
(7.154185,0.601445) *[red]{\scriptscriptstyle\times};
(7.171806,0.610823) *[red]{\scriptscriptstyle\times};
(7.189427,0.637216) *[red]{\scriptscriptstyle\times};
(7.207048,0.640637) *[red]{\scriptscriptstyle\times};
(7.224670,0.672339) *[red]{\scriptscriptstyle\times};
(7.242291,0.675403) *[red]{\scriptscriptstyle\times};
(7.259912,0.734428) *[red]{\scriptscriptstyle\times};
(7.277533,0.593801) *[red]{\scriptscriptstyle\times};
(7.295154,0.612772) *[red]{\scriptscriptstyle\times};
(7.312775,0.648157) *[red]{\scriptscriptstyle\times};
(7.330396,0.655343) *[red]{\scriptscriptstyle\times};
(7.348018,0.658716) *[red]{\scriptscriptstyle\times};
(7.365639,0.736297) *[red]{\scriptscriptstyle\times};
(7.383260,0.750002) *[red]{\scriptscriptstyle\times};
(7.400881,0.650493) *[red]{\scriptscriptstyle\times};
(7.418502,0.651979) *[red]{\scriptscriptstyle\times};
(7.436123,0.678153) *[red]{\scriptscriptstyle\times};
(7.453744,0.687477) *[red]{\scriptscriptstyle\times};
(7.471366,0.689970) *[red]{\scriptscriptstyle\times};
(7.488987,0.703446) *[red]{\scriptscriptstyle\times};
(7.506608,0.717550) *[red]{\scriptscriptstyle\times};
(7.524229,0.647490) *[red]{\scriptscriptstyle\times};
(7.541850,0.660906) *[red]{\scriptscriptstyle\times};
(7.559471,0.671830) *[red]{\scriptscriptstyle\times};
(7.577093,0.692438) *[red]{\scriptscriptstyle\times};
(7.594714,0.704088) *[red]{\scriptscriptstyle\times};
(7.612335,0.727646) *[red]{\scriptscriptstyle\times};
(7.629956,0.730801) *[red]{\scriptscriptstyle\times};
(7.647577,0.625020) *[red]{\scriptscriptstyle\times};
(7.665198,0.648125) *[red]{\scriptscriptstyle\times};
(7.682819,0.649172) *[red]{\scriptscriptstyle\times};
(7.700441,0.654371) *[red]{\scriptscriptstyle\times};
(7.718062,0.694980) *[red]{\scriptscriptstyle\times};
(7.735683,0.783463) *[red]{\scriptscriptstyle\times};
(7.753304,0.786197) *[red]{\scriptscriptstyle\times};
(7.770925,0.624258) *[red]{\scriptscriptstyle\times};
(7.788546,0.625352) *[red]{\scriptscriptstyle\times};
(7.806167,0.626472) *[red]{\scriptscriptstyle\times};
(7.823789,0.631284) *[red]{\scriptscriptstyle\times};
(7.841410,0.662921) *[red]{\scriptscriptstyle\times};
(7.859031,0.668782) *[red]{\scriptscriptstyle\times};
(7.876652,0.834938) *[red]{\scriptscriptstyle\times};
(7.894273,0.601739) *[red]{\scriptscriptstyle\times};
(7.911894,0.635669) *[red]{\scriptscriptstyle\times};
(7.929515,0.637559) *[red]{\scriptscriptstyle\times};
(7.947137,0.668979) *[red]{\scriptscriptstyle\times};
(7.964758,0.711612) *[red]{\scriptscriptstyle\times};
(7.982379,0.772456) *[red]{\scriptscriptstyle\times};
(8.000000,0.785540) *[red]{\scriptscriptstyle\times};
\endxy
}
\caption{Multiplication counts
  for the CSIDH-1024 action
  using {\tt velusqrt-asm}.
}
\label{action-mults1024}
\end{figure}

The specific graphs are as follows:
\begin{itemize}
\item
CSIDH-512 and CSURF-512 cycles
using {\tt velusqrt-magma}:
\cref{fig: action-magma}
and \cref{fig: csurf-magma} respectively,
with $E=7$ and $K=63$.
\item
CSIDH-512 and CSURF-512 cycles
using {\tt velusqrt-flint}:
\cref{fig:csidh-c}
and
\cref{fig:csurf-c} respectively,
with $E=15$ and $K=65$.
The new $\ell$-isogeny algorithm
speeds up CSIDH-512 by approximately 5\%,
and CSURF-512 by approximately 3\%.
\item
CSIDH-512 and CSIDH-1024 cycles
using {\tt velusqrt-asm}:
\cref{action-cycles512}
and
\cref{action-cycles1024} respectively,
with $E=15$ and $K=65$.
The new $\ell$-isogeny algorithm
speeds up CSIDH-512 by approximately 1\%,
and CSIDH-1024 by approximately 8\%
(on top of the assembly-language speedup
mentioned above).
\item
CSIDH-512 and CSIDH-1024 multiplication counts
using {\tt velusqrt-asm}:
\cref{action-mults512}
and
\cref{action-mults1024} respectively,
with $E=7$ and $K=65$.
The new $\ell$-isogeny algorithm
saves approximately 8\% and 16\% respectively.
\end{itemize}

Finally,
we evaluated the performance
of the {\tt velusqrt-julia}
implementation of B-SIDH using the
prime above.
We only measured the time needed for computing the
isogeny defined by a random point of order $p+1$ or $p-1$, and to
evaluate it at three random points.
This simulates the workload of the first stage of B-SIDH;
  the second stage does not need to evaluate the isogeny at any point,
  and is thus considerably cheaper.
We estimate that the other costs occurring in B-SIDH, such as the
double-point Montgomery ladder, are negligible compared to these.

Using our algorithm, an isogeny of degree $p-1$ is evaluated in about
0.56 seconds, whereas it takes approximately 2 seconds to evaluate it
using the conventional algorithms. %
More remarkably, we can evaluate an isogeny of degree $p+1$ in
approximately 10 seconds,
whereas the naive approach
(in one experiment)
takes 6.5 minutes.

\subsection{Techniques to save time
  inside the $\ell$-isogeny algorithm}
Here is a brief survey,
based on a more detailed analysis
of the results above,
of ways to reduce
the cost of evaluating an $\ell$-isogeny
and computing the new curve coefficient:
\begin{itemize}
\item
  Instead of computing
  the coefficients of the quadratic polynomial
  $Q(\alpha)
  = F_0(\alpha,x([j]P))Z^2
  + F_1(\alpha,x([j]P))Z
  + F_2(\alpha,x([j]P))
  $
  separately for each $\alpha$,
  merge the four computations of
  $Q(\alpha),Q(1/\alpha),Q(1),Q(-1)$
  across the four computations
  of $h_S(\alpha),h_S(1/\alpha),h_S(1),h_S(-1)$.
\item
  Observe that $Q(1)$ and $Q(-1)$
  are self-reciprocal quadratics.
  Speed up multiplication of
  self-reciprocal polynomials,
  exploiting the similarity of this problem
  to the problem of
  multiplying half-size polynomials.
\item
  Use scaled remainder trees
  (see~\cite{2003/bostan},
  \cite{2004/bostan},
  and~\cite{2004/bernstein-scaledmod})
  whenever those are faster than
  traditional unscaled remainder trees.
  Precompute the product trees
  used inside these remainder trees.
\item
  Speed up polynomial divisions,
  especially by precomputing reciprocals
  of tree nodes.
  See~\cite{2011/harvey}
  for various techniques to save time in reciprocals;
  most of these techniques
  are not incorporated into our current software.
\item
  Speed up polynomial multiplications,
  including polynomial multiplications
  that produce only a stretch of output coefficients.
  Divisions use ``low'' and ``high'' products,
  and scaled remainder trees
  use ``middle'' products.
\item
  Merge reductions across field multiplications:
  e.g., compute $ab+cd$
  by first adding the unreduced products
  $ab$ and $cd$ and then reducing the sum.
  This needs a more complicated field API;
  our current software does not do this.
\end{itemize}

It is well known that
multiplying two $n$-coefficient polynomials
costs just $2n-1$ field multiplications
(in large characteristic),
since one can interpolate
the product from its values at $0,1,\dots,2n-2$.
Divisions by various positive integers
inside this interpolation
can be replaced by multiplications
by various positive integers,
and thus by additions,
since isogeny outputs are represented projectively.
However,
more work is required
to optimize polynomial multiplication
in metrics that go beyond multiplications.
There is an extensive literature on this topic,
including many techniques
not used in our current software.

At a lower level,
reducing the cycles for field operations is helpful
in any cycle-counting metric.
At a higher level,
using one or more field divisions could be helpful
if divisions are fast enough
compared to the size of $\ell$.
Furthermore,
$\#I$ and $\#J$
should be chosen in light of the costs
of all of these operations.
Presumably the optimal $\#I/\#J$
converges to a constant as $\ell\to\infty$,
but it is not at all obvious what this constant is
for any particular metric.

\end{document}